\documentclass[11pt]{article}

\pdfoutput=1

\usepackage{graphicx}   % --> LaTex =>PDF con epstodf
\usepackage{epstopdf}           % --> LaTex =>PDF con epstodf

\usepackage{graphicx,epsf, epsfig, amssymb}% Include figure files
\usepackage{bm}
\usepackage{longtable}
\usepackage{color}
\usepackage{enumitem}

\usepackage{float}
\usepackage{epsfig, amssymb,graphicx} %include figure files
\usepackage{epstopdf}
\usepackage{amsmath, amsfonts}

\usepackage{amsfonts,amsmath,amssymb,mathrsfs}
\usepackage{soul}
\usepackage{graphicx}
\usepackage{amssymb}
\usepackage{amsmath}
\usepackage{enumerate}
\usepackage{amsfonts}
\usepackage[section]{placeins}
\usepackage[colorlinks=true,linktocpage=true,linkcolor=blue,citecolor=blue]{hyperref}
\usepackage{tensor}
\usepackage{multirow}

\newcommand{\be}{\begin{equation}}
\newcommand{\ee}{\end{equation}}
\newcommand{\bea}{\begin{eqnarray}}
\newcommand{\eea}{\end{eqnarray}}

\newcommand{\dmoff}[1]{}

\newcommand{\pa}{\partial}

\def\beq{\begin{eqnarray}}
\def\eeq{\end{eqnarray}}
\newcommand{\nn}{\nonumber}

\def\th{\vartheta}
\def\lp{{l+1}}
\def\lm{{l-1}}
\def\lpp{{l+2}}
\def\lmm{{l-2}}
\newcommand{\beqa}{\begin{eqnarray}}
\newcommand{\eeqa}{\end{eqnarray}}

\newcommand{\reoff}[1]{}

\newcommand{\Rmnum}[1]{\expandafter\@slowromancap\romannumeral #1@}

\numberwithin{equation}{section} %For numbering equations according to section

\begin{document}

\pagestyle{plain}
\setcounter{page}{1}

\begin{titlepage}

\begin{center}
\vspace*{-1cm} \today 
% {}~{} \hfill  MAD-TH-11-03 \\

\vskip 2.0cm

{\huge {\bf Superradiance}}

%{\huge {\bf Black-hole superradiance: \\ {\large from fundamental physics to astrophysics}}}

\vskip 14mm

{\large  {\bf Richard Brito,$^{1}$ Vitor Cardoso,$^{2}$ Paolo Pani,$^{1}$}}

\vspace{0.5 cm}

 ${}^1$ {\it Dipartimento di Fisica, ``Sapienza'' Universit\`a di Roma \& Sezione INFN Roma1, P.A. Moro 5, 00185, Roma, 
Italy}
%  \medskip

${}^2$ {\it CENTRA, Departamento de F\'{\i}sica, Instituto Superior
   T\'ecnico, Universidade de Lisboa, Av.~Rovisco Pais 1, 1049
   Lisboa, Portugal.}

{\tt  richard.brito@roma1.infn.it,\, vitor.cardoso@tecnico.ulisboa.pt, \, paolo.pani@uniroma1.it}

\vspace{5mm}

{\bf Abstract}
\end{center}
 \noindent
Superradiance is a radiation enhancement process that involves dissipative systems. With a 60 year-old history, superradiance has played a prominent role in optics, quantum mechanics and especially in relativity and astrophysics. 
In General Relativity, black-hole superradiance is permitted by the ergoregion, that allows for energy, 
charge and angular momentum extraction from the vacuum, even at the classical level. Stability of the spacetime is 
enforced by the event horizon, where negative energy-states are dumped. Black-hole superradiance is intimately connected 
to the black-hole area theorem, Penrose process, tidal forces, and even Hawking radiation, which can be interpreted as a 
quantum version of black-hole superradiance.
Various mechanisms (as diverse as massive fields, magnetic fields, anti-de Sitter boundaries, nonlinear interactions, etc...) can confine the amplified radiation and give rise to strong instabilities. These ``black-hole bombs'' have applications in searches of dark matter and of physics beyond the Standard Model, are associated to the threshold of formation of new black hole solutions that evade the no-hair theorems, can be studied in the laboratory by devising analog models of gravity, and might even provide a holographic description of spontaneous symmetry breaking and superfluidity through the gauge-gravity duality.
This work is meant to provide a unified picture of this multifaceted subject. We focus on the recent developments in the field, and work out a number of novel examples and applications, ranging from fundamental physics to astrophysics.

\noindent

\vskip 0.2cm
\noindent
%{\small PACS numbers: 
% 11.25.Tq, %Gauge/string duality
% 04.70.-s, %Physics of black holes
% 11.25.-w, %Strings and branes
% 41.60.-m, %Radiation by moving charges
% 04.25.Nx %Post-Newtonian approximation; perturbation theory; related approximations 
%} 

\end{titlepage}

\tableofcontents

%%%%%%%%%%%%%%%%%%%%%%%%%%%%%%%%%%%%%%%%%%%%%%%%%%%%%%%%%%%%%%%%%%%%%%%%%%%%%%%%%%%%%
\clearpage
\newpage
\section*{Notation and conventions}
%%%%%%%%%%%%%%%%%%%%%%%%%%%%%%%%%%%%%%%%%%%%%%%%%%%%%%%%%%%%%%%%%%%%%%%%%%%%%%%%%%%%%

Unless otherwise and explicitly stated, we use geometrized units where
$G=c=1$, so that energy and time have units of length. We also adopt the
$(-+++\dots)$ convention for the metric.  For reference, the following is a list of
symbols that are used often throughout the text.

\begin{table}[ht]
\begin{tabular}{ll}
  $\varphi$             & Azimuthal coordinate.\\
  $\vartheta$           & Angular coordinate.\\
  $m$                   & Azimuthal number with respect to the axis of rotation, $|m|\leq l$.\\ 
  $l$                   & Integer angular number, related to the eigenvalue $A_{lm}=l(l+1)$ in four spacetime dimensions.\\ 
  $s$                   & Spin of the field.\\
  $\omega$              & Fourier transform variable. The time dependence of any field is $\sim e^{-i\omega t}$.  \\
                        & For stable spacetimes, ${\rm Im}(\omega)<0$. \\
  $\omega_R,\,\omega_I$ & Real and imaginary part of the quasinormal mode frequencies.\\
  ${\cal R}, {\cal I}$  & Amplitude of reflected and incident waves, which characterize a wavefunction $\Phi$. \\
  $Z_{slm}$             & Amplification factor of fluxes for a wave with spin $s$ and harmonic indices $(l,m)$. For scalar fields,\\
                        & $Z_{0lm}=|{\cal R}|^2/|{\cal I}|^2-1$ with the asymptotic expansion at spatial infinity, $\Phi\sim {\cal R}e^{i\omega t}+{\cal I}e^{-i\omega t}$.\\
                        & Occasionally, when clear from the context, we will omit the indices $s$ and $l$ and simply write $Z_{m}$.\\
  $n$                   & Overtone numbers of the eigenfrequencies.\\
                        & We conventionally start counting from a ``fundamental mode'' with $n=0$.\\
  $D$                   & Total number of spacetime dimensions (we always consider one timelike\\
                        & and $D-1$ spacelike dimensions).\\
  $L$                   & Curvature radius of (A)dS spacetime, related to the negative \\
                        & cosmological constant $\Lambda$ in the Einstein equations ($G_{\mu\nu}+\Lambda g_{\mu\nu}=0$).\\
  $L^2$                 &$=\mp(D-2)(D-1)/(2\Lambda)$ is the curvature radius of anti-de Sitter (- sign) or de Sitter.\\
  $M$                   & Mass of the black-hole spacetime. \\
  $r_+$                 & Radius of the black-hole event horizon in the chosen coordinates.\\
  $\Omega_{\rm H}$      & Angular velocity of a zero-angular momentum observer at the black hole horizon,\\
			& as measured by a static observer at infinity.\\
 $\mu_{S,\,V,\,T}$      & Mass parameter of the (scalar, vector or tensor) field. \\
			& In geometric units, the field mass is $\mu_{S,V,T}\hbar$, respectively.\\
  $a$                   & Kerr rotation parameter: $a = J/M \in [0,M]$.\\
  $g_{\alpha \beta}$    & Spacetime metric; Greek indices run from 0 to $D-1$. \\
  $Y_{lm}$		& Spherical harmonics, orthonormal with respect to the integral over the 2-sphere.\\
  $S_{slm}$		& Spin-weighted spheroidal harmonics. 

%%%%%%%%%%%%%%%%%%%%%%%%%%%%%%%%%%%%%
\end{tabular}
\end{table}
%

%%%%%%%%%%%%%%%%%%%%%%%%%%%%%%%%%%%%%%%%%%%%%%%%%%%%%%%%%%%%%%%%%%%%%%%%%%%%%%%%%%%%%
\clearpage
\newpage
\section*{Acronyms}
%%%%%%%%%%%%%%%%%%%%%%%%%%%%%%%%%%%%%%%%%%%%%%%%%%%%%%%%%%%%%%%%%%%%%%%%%%%%%%%%%%%%%
% \begin{table}[ht]
\begin{tabular}{ll}
ADM    & Arnowitt-Deser-Misner                                \\
AGN    & Active Galactic Nuclei				\\
AdS    & Anti-de Sitter                                       \\
BH     & Black hole                                           \\
CFT    & Conformal field theory                                 \\
% EoS    & Equation of state                                    \\
% ER     & Ergoregion                                           \\
EM     & Electromagnetic                                      \\
GR     & General Relativity                                   \\ 
GW     & Gravitational Wave                                   \\
% HEP    & High Energy Physics                                  \\
% LHC    & Large Hadron Collider                               \\
LIGO   & Laser Interferometric Gravitational Wave Observatory \\
LISA   & Laser Interferometer Space Antenna   \\
ODE    & Ordinary differential equation                       \\
% NR     & Numerical Relativity                                 \\
NS     & Neutron star                                        \\
PDE    & Partial differential equation                        \\
PSD    & Power spectral density  \\ 
QCD    & Quantum Chromodynamics                               \\
% PN     & Post-Newtonian                                       \\
QNM    & Quasinormal mode                                     \\
RN     & Reissner-Nordstr\"om				      \\
% RHIC   & Relativistic Heavy Ion Collider                      \\
ZAMO   & Zero Angular Momentum Observer
\end{tabular}
% \end{table}
%
% \bigskip

\clearpage
\newpage

%%%%%%%%%%%%%%%%%%%%%%%%%%%%%%%%%%%%%%%%%%%%%%%%%%%%%%%%%%%%%
\clearpage
\newpage
\begin{quote}
\raggedleft
 \textit{Macroscopic objects, as we see them all around us, are governed by a variety of forces, derived from a variety of approximations to a variety of physical theories. In contrast, the only elements in the construction of black holes are our basic concepts of space and time. They are, thus, almost by definition, the most perfect macroscopic objects there are in the universe.} \\ -- Subrahmanyan Chandrasekhar \\
\end{quote}
\section{Prologue}
\label{sec:prologue}
%%%%%%%%%%%%%%%%%%%%%%%%%%%%%%%%%%%%%%%%%%%%%%%%%%%%%%%%%%%%%

Radiation-enhancement processes have a long history that can be traced back to the dawn of quantum mechanics, when Klein showed that the Dirac equation allows for electrons to be transmitted even in classically forbidden regions~\cite{Kleinparadox}.
In 1954, Dicke introduced the concept of {\it superradiance}, standing for a collective phenomena
whereby radiation is amplified by coherence of emitters~\cite{Dicke:1954zz}. In 1971 Zel'dovich showed that scattering of radiation off rotating absorbing surfaces
results, under certain conditions, in waves with a larger amplitude~\cite{zeldovich1,zeldovich2}. This phenomenon
is now widely known also as (rotational) superradiance and requires that
the incident radiation, assumed monochromatic of frequency $\omega$, satisfies
\be
\omega <m\Omega\,,\label{eq:superradiance_condition}
\ee
with $m$ the azimuthal number with respect to the rotation axis and $\Omega$ the angular velocity of the body.
Rotational superradiance belongs to a wider class of classical problems
displaying stimulated or spontaneous energy emission, such as the Vavilov-Cherenkov effect, the anomalous 
Doppler effect, and other examples of ``superluminal motion''. When quantum effects were incorporated,
it was argued that rotational superradiance would become a spontaneous process and that rotating bodies -- including black holes (BHs)--
would slow down by spontaneous emission of photons satisfying~\eqref{eq:superradiance_condition}.
In parallel, similar conclusions were reached when analyzing BH superradiance from a thermodynamic viewpoint~\cite{Bekenstein:1973mi,Bekenstein:1998nt}.
From a historic perspective, the first studies of BH superradiance
played a decisive role in the discovery of BH evaporation~\cite{Hawking:1974sw,Hawking:book}.

Interest in BH superradiance was recently revived in different areas, including astrophysics and high-energy physics (via the gauge/gravity duality), and fundamental issues in General Relativity (GR). 
Superradiant instabilities can be used to constrain the mass of ultralight degrees of freedom~\cite{Arvanitaki:2009fg,Arvanitaki:2010sy,Pani:2012vp,Brito:2013wya}, with important applications to dark-matter searches and to physics beyond the Standard Model. BH superradiance is also associated to the existence of new asymptotically flat, hairy BH solutions~\cite{Herdeiro:2014goa} and to phase transitions between spinning or charged black objects in asymptotically anti-de Sitter (AdS) spacetime~\cite{Cardoso:2004hs,Dias:2011tj,Dias:2011at} or in higher dimensions~\cite{Shibata:2010wz}. Finally, superradiance is fundamental in deciding the stability of BHs and the fate of the gravitational collapse in confining geometries.
In fact, the strong connection between some recent applications and the original phenomenon of superradiance has not always been fully recognized. This is the case, for instance, of holographic models of superfluids~\cite{Hartnoll:2008vx}, which hinge on a spontaneous symmetry breaking of charged BHs in AdS spacetime~\cite{Gubser:2008px}. 
In global AdS, the associated phase transition can be interpreted in terms of \emph{superradiant instability} of a 
Reissner-Nordstrom AdS BH triggered by a charged scalar field~\cite{Hartnoll:2011fn,Dias:2011tj}. 
 
With the exception of the outstanding -- but focused-- work by Bekenstein and Schiffer~\cite{Bekenstein:1998nt}, a 
proper overview on superradiance, including various aspects of wave propagation in BH spacetimes, does not exist. We 
hope to fill this gap with the current work. 
In view of the multifaceted nature of this subject, we wish to present a unified treatment where various aspects of superradiance in flat spacetime are connected to their counterparts in curved spacetime, with particular emphasis on the superradiant amplification by BHs. 
In addition, we wish to review various applications of BH superradiance which have been developed in the last decade. 
These developments embrace different communities (e.g., gravity, particle physics, string 
theorists, experimentalists), and our scope is to present a concise treatment that can be fruitful for the reader who 
is not familiar with the specific area. As will become clear throughout this work, some of these topics are far from 
being fully explored. We hope this study will serve as a guide for the exciting developments lying ahead.

\subsection*{Preface to second edition}

As the first version of this work was being published, the field experienced a phase transition.
To name but a few, gravitational-wave (GW) observatories made the first-ever direct detections of BH binaries; the 
numerical evolution of massive fields around spinning BHs was reported; long-baseline interferometry produced the 
first-ever images of supermassive BHs, and detected motion close to their horizon. These novel observations made it 
possible to search for \emph{direct} signatures of superradiant instabilities around BHs. Furthermore, superradiance 
was {\it measured} in the laboratory in BH-analog systems. 
The topic is more timely than ever and urged us to write a second edition, where a number of typos and some wrong statements were corrected. We hope that this updated 
revision reflects all the main developments and the excitement of the last years.

%%%%%%%%%%%%%%%%%%%%%%%%%%%%%%%%%%%%%%%%%%%%%%%%%%%%%%%%%%%%%%%%%%%%%%%%%%%%%%%
\clearpage
\newpage
\section{Milestones}
%%%%%%%%%%%%%%%%%%%%%%%%%%%%%%%%%%%%%%%%%%%%%%%%%%%%%%%%%%%%%%%%%%%%%%%%%%%%%%%
The term superradiance was coined by Dicke in 1954~\cite{Dicke:1954zz}, but studies on related phenomena date back to at least 1947 with the pioneering work of Ginzburg and Frank~\cite{Ginzburg:1947} on the ``anomalous'' Doppler effect. It is impossible to summarize all the important work
in the field in this work, but we think it is both useful and pedagogical to have a chronogram
of some of the most relevant milestones. We will keep this list -- necessarily incomplete and necessarily biased --
confined mostly to the realm of General Relativity (GR), although we can't help 
making a reference to some of the breakthrough work in other areas.
A more complete set of references can be found in the rest of this work.
\begin{itemize}[noitemsep]

\item[1899] In his book ``Electromagnetic Theory'', Oliver Heaviside discusses the motion of a charged body moving faster than light in a medium. Remarkably (since the electron had not been discovered yet), this was a precursor of the Vavilov-Cherenkov effect.

\item[1915] Einstein develops GR~\cite{Einstein:1916vd}.
 
\item[1916] Shortly afterwards, Schwarzschild~\cite{Schwarzschild:1916uq} and Droste~\cite{Droste:2017} derive the first solution of
Einstein's equations, describing the gravitational field generated by a point mass. Most of the subtleties and implications of this solution will only be understood many years later.

\item[1920s] In order to unify electromagnetism with GR, Kaluza
and Klein propose a model in which the spacetime has five dimensions, one of
which is compactified on a circle~\cite{Kaluza:1921tu,Klein:1926tv}.

\item[1929] In his studies of the Dirac equation, Klein finds that electrons can ``cross'' a potential barrier without the exponential damping expected from nonrelativistic quantum tunneling~\cite{Kleinparadox}. This process was soon dubbed \emph{Klein paradox} by Sauter. The expression was later used to describe an incorrectly obtained phenomenon of fermion superradiance (Klein's original work correctly shows that no superradiance occurs for fermions). An interesting historical account of these events is given by Manogue~\cite{Manogue1988}.
 
\item[1931] Chandrasekhar derives an upper limit for white dwarf masses,
above which electron degeneracy pressure cannot sustain the star~\cite{Chandrasekhar:1931ih}. 
The Chandrasekhar limit was subsequently extended to neutron stars~(NS) by Oppenheimer and 
Volkoff~\cite{Oppenheimer:1939ne}.

\item[1934] Vavilov and Cherenkov discover spontaneous emission from a charge moving uniformly and superluminally in a dielectric. The effect was interpreted theoretically by Tamm and Frank in 1937~\cite{TammFrank}. In 1958, Tamm, Frank and Cherenkov receive the Nobel prize in physics for these studies.
 
\item[1937] Kapitska discovers superfluidity in liquid helium.
 
\item[1945] Ginzburg and Frank discuss transition radiation~\cite{Ginzburg:1945zz}.
 
\item[1947] Ginzburg and Frank discover an ``anomalous Doppler effect''~\cite{Ginzburg:1947}: the emission of radiation 
by a system moving faster than the phase velocity of EM waves in a medium and followed by the \emph{excitation} (rather 
than by the standard de-excitation) to a higher energy level.

\item[1947] Pierce describes a ``traveling wave tube amplifier'', where an electron beam {\it extracts} energy
from an EM wave traveling at a speed less than its vacuum value. 
The EM wave is forced to slow down using an helix electrode, a spiral of wire around the electron 
beam~\cite{PierceBook,webpagepierce}.

\item[1953] Smith and Purcell experimentally show that motion near finite-size objects induces radiation emission, or ``diffraction radiation''~\cite{Smith:1953sq}.

\item[1954] Dicke coins the term ``superradiance'' in the context of coherent emission in quantum optics~\cite{Dicke:1954zz}. The first high-resolution measurement of superradiance using coherent synchrotron radiation was recently achieved~\cite{Billinghurst:2013kna}. 
 
\item[1957] Regge and Wheeler~\cite{Regge:1957rw} analyze a special class of gravitational perturbations of the Schwarzschild
geometry. This effectively marks the birth of BH perturbation theory.

\item[1958] Finkelstein understands that the $r=2M$ surface of
the Schwarzschild geometry is not a singularity but a horizon~\cite{Finkelstein:1958zz}. The so-called ``golden age of GR'' begins: in a few
years there would be an enormous progress in the understanding of GR and of its solutions.

\item[1962] Newman and Penrose~\cite{Newman:1961qr} develop a
formalism to study gravitational radiation using spin coefficients. 

\item[1963] Kerr~\cite{Kerr:1963ud} discovers the mathematical
solution of Einstein's field equations describing rotating BHs. In the same
year, Schmidt identifies the first quasar~\cite{Schmidt:1963}. Quasars are now believed to
be supermassive BHs, described by the Kerr solution.

\item[1964] The UHURU orbiting X-ray observatory makes the
first surveys of the X-ray sky discovering over 300 X-ray ``stars''.
One of these X-ray sources, Cygnus X-1, is soon accepted as the first plausible
stellar-mass BH candidate (see e.g. Ref.~\cite{Bolton:1972}).

\item[1967] Wheeler~\cite{Ruffini:1971,wheeler} coins the term
\textit{black hole} (see the April 2009 issue of {\em Physics Today},
Ref.~\cite{Wheeler:1998vs} for a fascinating, first-person historical
account, and a recent overview with new and interesting details~\cite{Herdeiro:2018ldf}).

\item[1969] Hawking's singularity theorems imply that collapse
of ordinary matter leads, under generic conditions, to spacetime singularities.
In the same year Penrose conjectures that these singularities, where quantum gravitational effects
become important, are generically contained within BHs, the so-called \textit{Cosmic Censorship Conjecture} 
\cite{Penrose:1969,Wald:1997wa}.

 \item[1969] Penrose shows that the existence of an ergoregion allows to extract energy and angular momentum from a Kerr BH and to amplify energy in particle collisions~\cite{Penrose:1969}.
 
 \item[1970] Zerilli~\cite{Zerilli:1970se,Zerilli:1971wd}
extends the Regge-Wheeler analysis to general perturbations of a Schwarzschild
BH, reducing the problem to the study of a pair of
Schr\"odinger-like equations, and computing the
gravitational radiation emitted by infalling test particles.

 \item[1970] Vishveshwara~\cite{vish} studies numerically the
scattering of GWs by BHs: at late times the
waveform consists of damped sinusoids, now called ringdown waves. The latter coincide with the BH quasinormal modes (QNMs)~\cite{Nollert:1999ji,Kokkotas:1999bd,Ferrari:2007dd,Berti:2009kk,Konoplya:2011qq}.

 \item[1971] Zeldovich shows that dissipative rotating bodies amplify
incident waves~\cite{zeldovich1,zeldovich2}. In the same study, quantum spontaneous
pair creation by rotating bodies is also predicted, which effectively is a precursor to
Hawking's result on BH evaporation. Misner explored some of the physics associated with energy extraction~\cite{Misner:1972kx}. Aspects of the quantization procedure of test fields in the Kerr geometry
were further independently elaborated by Starobinski~\cite{Starobinski:1973,Starobinski2:1973} and Deruelle and collaborators~\cite{Deruelle:1974zy,Deruelle:1975yz}.

 \item[1972-1974] Teukolsky~\cite{Teukolsky:1972my} decouples and
separates the equations for perturbations in the Kerr geometry using the
Newman-Penrose formalism~\cite{Newman:1961qr}. In the same year, Teukolsky and Press discuss quantitatively the superradiant scattering from a spinning BH~\cite{Teukolsky:1974yv}. They predict that, if confined, superradiance can give rise to \emph{BH bombs} and \emph{floating orbits} around spinning BHs~\cite{Press:1972zz}. This work introduces the term ``superradiance'' for the first time, in connection to Zel'dovich classical process of energy amplification.

\item[1973] Unruh independently separates the massless spin-1/2 equations on a Kerr background and proves the absence of superradiance for massless fermions~\cite{Unruh:1973,Toth:2015cda}. The result was later generalized to massive fermions by Chandrasekhar and by Iyer and Kumar~\cite{Chandra:1976,Iyer:1978,Dolan:2015eua}. 

\item[1975] Using quantum field theory in curved space and building on Zeldovich's 1971 result, Hawking
finds that BHs have a thermal emission~\cite{Hawking:1974sw}. This result is one
of the most important links between general relativity, quantum mechanics and thermodynamics.

\item[1977] Blandford and Znajek propose a mechanism to extract energy from rotating BHs immersed in force-free magnetic fields~\cite{Blandford:1977ds}.
This is thought to be one of the main mechanisms behind jet formation.

\item[1976-1980] Damour, Deruelle and Ruffini discover that superradiance triggers an instability of the Kerr BH solution against massive scalar fields~\cite{Damour:1976kh}. The study is then formalized by Detweiler~\cite{Detweiler:1980uk} and by Zouros and Eardley~\cite{Zouros:1979iw}. 
 
\item[1978] Friedman~\cite{1978CMaPh..63..243F} shows that horizonless spacetimes with ergoregions
are unstable. Similar result is shown simultaneously for quantum fields~\cite{Sato:1978ue}.

\item[1983] Chandrasekhar's monograph~\cite{Chandra} summarizes the
state of the art in BH perturbation theory, elucidating connections between
different formalisms.

\item[1985] Leaver~\cite{Leaver:1985ax,leJMP,Leaver:1986gd}
provides the most accurate method to date to compute BH QNMs
using continued fraction representations of the relevant wavefunctions.
Recently, accurate spectral methods have been developed to handle partial differential equations (PDEs)~\cite{Cardoso:2013pza}.

\item[1986] McClintock and Remillard~\cite{McClintock:1986}
show that the X-ray nova A0620-00 contains a compact object of mass almost
certainly larger than $3M_\odot$, paving the way for the identification of
many more stellar-mass BH candidates. 

\item[1986] Myers and Perry construct higher-dimensional
rotating, topologically spherical, BH solutions~\cite{Myers:1986un}.

 \item[1992] In a series of papers, Kojima develops the theory of linear perturbations of a slowly-rotating, relativistic star~\cite{Kojima:1992ie,1993PThPh..90..977K,1993ApJ...414..247K}. 

 \item[1998] Maldacena formulates the
AdS/CFT duality conjecture~\cite{Maldacena:1997re}. Shortly afterward, the papers by Gubser, Klebanov, Polyakov~\cite{Gubser:1998bc} and
Witten~\cite{Witten:1998qj} establish a concrete quantitative recipe for the
duality. The AdS/CFT era begins. In the same year, the correspondence is generalized to
nonconformal theories in a variety of approaches. The terms ``gauge/string duality'', ``gauge/gravity duality'' and ``holography'' appear, referring to these generalized settings (we refer to Ref.~\cite{Aharony:1999ti} for a review).

\item[1999] Banks and Fischler~\cite{Banks:1999gd} show that in
braneworld scenarios BHs could be
produced in particle accelerators. Shortly after, it is proposed to look
for BH production in the LHC and in ultra high-energy cosmic
rays~\cite{Dimopoulos:2001hw,Giddings:2001bu}.

\item[2001] Emparan and Reall provide the first example of a stationary asymptotically flat vacuum solution 
with an event horizon of nonspherical topology: the ``black ring''~\cite{Emparan:2001wn}.

\item[2003] In a series of papers
\cite{Kodama:2003jz,Ishibashi:2003ap,Kodama:2003kk}, Kodama and Ishibashi
extend the Regge-Wheeler-Zerilli formalism to higher dimensions.

\item[2004] Small, rapidly spinning Kerr-AdS BHs are found to be linearly unstable because of the AdS boundary providing a natural confinement mechanism for superradiant radiation~\cite{Cardoso:2004hs}. Rigorous growth-rate estimates for generic initial data are provided in Ref.~\cite{Shlapentokh-Rothman:2013ysa}.

\item [2005-2009] The D1-D5 system is used as a toy model to understand the microscopic origin of superradiant scattering~\cite{Cardoso:2005gj,Dias:2007nj}. For horizonless geometries, ergoregion instabilities lead precisely to Hawking radiation~\cite{Chowdhury:2007jx,Chowdhury:2008bd}.

\item[2008] Gubser proposes a spontaneous symmetry breaking mechanism, giving an effective mass to charged scalars in AdS~\cite{Gubser:2008px}.
Shortly afterwards, Hartnoll, Herzog and Horowitz provide a nonlinear realization of the mechanism, building the holographic analog of a superfluid~\cite{Hartnoll:2008vx}. Depending on the magnitude of the induced mass, tachyonic or superradiant instabilities may be triggered in BH spacetimes~\cite{Hartnoll:2011fn,Dias:2010ma,Dias:2011tj,Basu:2010uz}.
 
\item[2009] The string-axiverse scenario is proposed, where a number of ultralight degrees of freedom -- prone to superradiant instabilities around spinning BHs -- are conjectured to exist ~\cite{Arvanitaki:2009fg}. Precision measurements of mass and spin of BHs, together with GW observations, may be used to explore some of the consequences of this scenario~\cite{Arvanitaki:2009fg,Arvanitaki:2010sy}. Such searches were later shown to provide constraints on the QCD axion~\cite{Arvanitaki:2014wva}.

\item[2011] Superradiant instabilities are shown to drive AdS BHs to hairy configurations~\cite{Dias:2011tj,Dias:2011at}.

\item[2011] Floating orbits around Kerr BHs are predicted in scalar-tensor theories as a generic outcome of superradiant amplification of scalar waves~\cite{Cardoso:2011xi}.

\item[2012] Rotating Kerr BHs are shown to be linearly unstable against massive vector field perturbations~\cite{Pani:2012vp,Pani:2012bp,Witek:2012tr} and massive tensor field perturbations~\cite{Brito:2013wya}. Competitive bounds on the photon and graviton mass are derived from the observations of spinning BHs~\cite{PDG}.

\item[2013] Superradiance is shown to occur at full nonlinear level~\cite{East:2013mfa}.

\item[2014] Asymptotically flat, hairy BH solutions are constructed analytically~\cite{Hod:2012px} and numerically~\cite{Herdeiro:2014goa,Herdeiro:2015waa,Herdeiro:2016tmi}. These are thought to be one possible end-state of superradiant instabilities for complex scalar or vector fields. The superradiance threshold of the standard Kerr solution marks the onset of a phase transition towards a hairy BH. 

\item[2014] Reissner-Nordstrom de Sitter BHs are found to be unstable against charged scalar perturbations~\cite{Zhu:2014sya}. The unstable modes satisfy the superradiant condition~\cite{Konoplya:2014lha}. 

\item[2014] The adiabatic evolution of the superradiant instability of the Kerr spacetime is simulated, in the presence of an accretion disk and of GW emission~\cite{Brito:2014wla}. Growth of a scalar cloud and subsequent depletion through GW emission is reported. 

\item[2016] The LIGO-Virgo Collaboration announces the first direct detection of GWs, generated by a pair of merging 
BHs~\cite{Abbott:2016blz}. This historical event was followed by the detection of a binary NS merger, both in the GW 
and EM window~\cite{TheLIGOScientific:2017qsa}. An overview of the field and prospects for the future can be found in Ref.~\cite{Barack:2018yly}.

\item[2016] Post-merger continuous GW signals from merging BHs are proposed as a smoking gun of ultralight fields~\cite{Arvanitaki:2016qwi}. 

\item[2016] Superradiance from BHs is derived using effective field theory techniques~\cite{Endlich:2016jgc}.

\item[2016] The first observation of rotational superradiance in the laboratory is reported~\cite{Torres:2016iee}.

\item[2016] Superradiance from binaries or triple systems is shown to yield observational signatures in GW 
signals~\cite{Rosa:2016bli}.

\item[2017] The nonlinear evolution of the superradiant instability of a massive vector is reported. Mass extraction from the BH and growth of a vector ``cloud'' are observed~\cite{East:2017ovw}. 

\item[2017] Superradiance is reported for EM waves impinging on a rotating conducting star~\cite{Cardoso:2017kgn}. The 
mechanism is used to constrain invisible, massive vector fields.

\item[2017] Superradiance around rotating BHs is computed for massive vector fields using a large Compton-wavelength approximation~\cite{Baryakhtar:2017ngi}. Results are subsequently computed numerically, at linearized level~\cite{Cardoso:2018tly}.

\item[2017] The nonlinear growth, saturation of, and gravitational radiation from the instability of a massive vector 
field around a spinning BH is reported~\cite{East:2018glu}.

\item[2018] First constraints on light fields from the absence of a GW stochastic background~\cite{Brito:2017wnc,Brito:2017zvb,Tsukada:2018mbp,LIGOScientific:2019vic}, and null results for continuous searches~\cite{Brito:2017wnc,Brito:2017zvb,Isi:2018pzk,DAntonio:2018sff,Palomba:2019vxe,Dergachev:2019wqa,Sun:2019mqb,Zhu:2020tht}.

\item[2018] The nonlinear evolution of the superradiant instability of spinning BHs in four-dimensional AdS space is reported~\cite{Chesler:2018txn}.

\item[2018] The equations of motion for a massive vector on a Kerr background are shown to be separable~\cite{Frolov:2018ezx}.

\item[2018] The GRAVITY instrument reports detection of an orbiting hotspot close to the innermost stable orbit
of the BH at the center of our galaxy~\cite{2018A&A...618L..10G}.

\item[2018] The influence of a companion on superradiant clouds is studied. Level mixing and resonances are reported analytically~\cite{Baumann:2018vus,Zhang:2018kib,Baumann:2019ztm}, and verified numerically, including cloud disruption for large tides~\cite{Cardoso:2020hca}. Cloud is shown to mediate angular momentum exchange, which can lead to floating or sinking orbits~\cite{Zhang:2018kib,Baumann:2019ztm}.

\item[2019] The Event Horizon Telescope produces the first direct imaging of a supermassive BH 
in M87 galaxy~\cite{Akiyama:2019cqa}.

\item[2019] Energy extraction from binaries~\cite{Bernard:2019nkv,Wong:2019kru} and isolated, moving~\cite{Cardoso:2019dte} BHs is 
reported.

\end{itemize}

\newpage

%%%%%%%%%%%%%%%%%%%%%%%%%%%%%%%%%%%%%%%%%%%%%%%%%%%%%%%%%%%%%%%%%%%%%%%%%%%%%%%%%%%%%%%%%%%%%%%%%%%%%%%%%%%%%%%%%%%%%%%%%%%%%%%%%%

%%%%%%%%%%%%%%%%%%%%%%%%%%%%%%%%%%%%%%%%%%%%%%%%%%%%%%%%%%%%%%%%%%%%%%
\clearpage
\newpage
\section{Superradiance in flat spacetime}\label{sec:faces}
%%%%%%%%%%%%%%%%%%%%%%%%%%%%%%%%%%%%%%%%%%%%%%%%%%%%%%%%%%%%%%%%%%%%%%
%%%%%%%%%%%%%%%%%%%%%%%%%%%%%%%%%%%%%%%%%%%%%%%%%%%%%%%%%%%%%%%%%%%%%%%%%%%%%%%%%%%%%%%%
\subsection{Klein paradox: the first example of superradiance} \label{sec:KleinParadox}
%%%%%%%%%%%%%%%%%%%%%%%%%%%%%%%%%%%%%%%%%%%%%%%%%%%%%%%%%%%%%%%%%%%%%%%%%%%%%%%%%%%%%%%%
The first treatment of what came to be known as the Klein paradox can be traced back to the original paper by Klein~\cite{Kleinparadox}, who pioneered studies of Dirac's equation in the presence of a step potential. He showed that an electron beam propagating in a region with a large enough potential barrier $V$ can emerge without the exponential damping expected from nonrelativistic quantum tunneling processes. When trying to understand if such a result was an artifact of the step-potential used by Klein, Sauter found that the essentials of the process were independent on the details of the potential barrier, although the probability of transmission decreases with decreasing slope~\cite{Sauter:1931zz}. 
{\it This} phenomenon was originally dubbed ``Klein paradox'' by Sauter\footnote{The Klein paradox as we understand it today has an interesting history. Few years after Klein's original study (written in German), the expression {\it Klein paradox} appeared in some British literature in relation with \emph{fermionic superradiance}: due to some confusion (and probably because Klein's paper didn't have an English translation), some authors wrongly interpreted Klein's results as if the fermionic current reflected by the potential barrier could be greater than the incident current. This result was due to an incorrect evaluation of the reflected and transmitted wave's group velocities, although Klein --~following suggestions by Niels Bohr~-- had the correct calculation in the original work~\cite{Manogue1988}. Although not explicitly mentioned by Klein, this phenomenon can actually happen for bosonic fields~\cite{Winter:1959AJP} and it goes under the name of \emph{superradiant} scattering.} in 1931~\cite{Sauter:1931zz}.

Further studies by Hund in 1941~\cite{Hund:1941zz}, now dealing with a charged scalar field and the Klein-Gordon equation, showed that the step potential could give rise to the production of pairs of charged particles when the potential is sufficiently strong. Hund tried -- but failed -- to derive the same result for fermions. 
It is quite interesting to note that this result can be seen as a precursor of the modern quantum field theory results 
of Schwinger~\cite{Schwinger:1951nm} and Hawking~\cite{Hawking:1974sw} who showed that spontaneous pair production is 
possible in the presence of strong EM and gravitational fields for both bosons and fermions. In fact we know today that 
the resolution of the ``old'' Klein paradox is due to the creation of particle--antiparticle pairs at the barrier, which 
explains the undamped transmitted part.

In the remaining of this section we present a simple treatment of bosonic and fermionic scattering, to illustrate these phenomena. 

%%%%%%%%%%%%%%%%%%%%%%%%%%%%%%%%%%%%%%%%%%%%%%%%%%%%%%%%%%%%
\subsubsection{Bosonic scattering}
%%%%%%%%%%%%%%%%%%%%%%%%%%%%%%%%%%%%%%%%%%%%%%%%%%%%%%%%%%%%
Consider a massless scalar field $\Phi$ minimally coupled to an EM potential $A_{\mu}$ in $(1+1)$--dimensions, governed 
by the Klein-Gordon equation
\be\label{paradox_scalar}
\Phi_{;\mu}^{\phantom{;\mu};\mu}=0\,,
\ee
where we defined $\Phi_{;\mu}\equiv (\partial_{\mu}-i e A_{\mu})\Phi$ and $e$ is the charge of the scalar field. For simplicity we consider an external potential $A^{\mu}=\left\{A_0(x),0\right\}$, with the asymptotic behavior
\bea~
A_0\to \left\{\begin{array}{l}
        0 \quad \text{ as}\,\, x\to-\infty \\
        V \quad \text{as}\,\, x\to+\infty
       \end{array}\right. \,
\,.\label{potential_Klein}
\eea

With the ansatz $\Phi=e^{-i\omega t}f(x)$, Eq.~\eqref{paradox_scalar} can be separated yielding the ordinary differential equation (ODE)
\be\label{ODE_boson_paradox}
\frac{d^2f}{dx^2}+\left(\omega-e A_0\right)^2f=0\,.
\ee

Consider a beam of particles coming from $-\infty$ and scattering off the potential with reflection and transmission amplitudes $\mathcal{R}$ and $\mathcal{T}$ respectively. With these boundary conditions, the solution to Eq.~\eqref{paradox_scalar} behaves asymptotically as
\bea
\label{sol_boson1}
f_{\rm in}(x)&=&\mathcal{I} e^{i\omega x}+\mathcal{R} e^{-i\omega x}\,, \quad \text{as}\,\, x\to-\infty\,,\nn\\
f_{\rm in}(x)&=&\mathcal{T} e^{ikx}\,, \quad \text{as}\,\, x\to+\infty\,,
\eea
where
\be\label{disp_relation}
k=\pm(\omega-e V)\,.
\ee
To define the sign of $\omega$ and $k$ we must look at the wave's group velocity. We require the incoming and the transmitted part of the waves to have positive group velocity, $\partial\omega/\partial k>0$, so that they travel from the left to the right in the $x$--direction. Hence, we take $\omega>0$ and the plus sign in~\eqref{disp_relation}.

The reflection and transmission coefficients depend on the specific shape of the potential $A_0$. However one can easily show that the Wronskian
\be\label{wronskian}
W=\tilde{f}_1 \frac{d\tilde{f}_2}{dx}-\tilde{f}_2\frac{d\tilde{f}_1}{dx}\,,
\ee
between two independent solutions, $\tilde{f}_1$ and $\tilde{f}_2$, of~\eqref{ODE_boson_paradox} is conserved.
From the equation~\eqref{ODE_boson_paradox} on the other hand, if $f$ is a solution then its complex conjugate $f^*$ is another linearly independent solution. Evaluating the Wronskian~\eqref{wronskian}, or equivalently, the particle current density, for the solution~\eqref{sol_boson1} and its complex conjugate we find
\bea\label{reflection_boson}
\left|\mathcal{R}\right|^2=|\mathcal{I}|^2-\frac{\omega-eV}{\omega}\left|\mathcal{T}\right|^2\,.
\eea
Thus, for
\be\label{super_klein} 0<\omega<e V\,,
\ee
it is possible to have superradiant amplification of the reflected current, i.e, $\left|\mathcal{R}\right|>|\mathcal{I}|$. 
There are other exactly solvable potentials which also display superradiance explicitly, but we will not discuss them here~\cite{Puente:2019qin}.
%%%%%%%%%%%%%%%%%%%%%%%%%%%%%%%%%%%%%%%%%%
\subsubsection{Fermionic scattering}
%%%%%%%%%%%%%%%%%%%%%%%%%%%%%%%%%%%%%%%%%%
Now let us consider the Dirac equation for a spin-$\frac{1}{2}$ massless fermion $\Psi$, minimally coupled to the same 
EM potential $A_{\mu}$ as in Eq.~\eqref{potential_Klein},
\be\label{paradox_Dirac}
\gamma^{\mu}\Psi_{;\mu}=0\,,
\ee
where $\gamma^{\mu}$ are the four Dirac matrices satisfying the anticommutation relation $\{\gamma^{\mu},\gamma^{\nu}\}=2g^{\mu\nu}$. 
The solution to~\eqref{paradox_Dirac} takes the form $\Psi=e^{-i\omega t}\chi(x)$, where $\chi$ is a two-spinor given by
\be
\chi=\begin{pmatrix}f_1(x)\\f_2(x)\end{pmatrix}\,.
\ee
Using the representation
\be\label{rep}
\gamma^0=\begin{pmatrix}i&0\\0&-i\end{pmatrix}\,,\,\,\gamma^1=\begin{pmatrix}0&i\\-i&0\end{pmatrix}\,,
\ee
the functions $f_1$ and $f_2$ satisfy the system of equations:
\bea
df_1/dx-i(\omega-eA_0)f_2&=&0\,,\nn\\
df_2/dx-i(\omega-eA_0)f_1&=&0\,.
\eea
One set of solutions can be once more formed by the `in' modes, representing a flux of particles coming from $x\to-\infty$ being partially reflected (with reflection amplitude $|\mathcal{R}|^2$) and partially transmitted at the barrier
\bea\label{sol_fermion1}
\left(f_1^{\rm in},f_2^{\rm in}\right)=\left\{\begin{array}{l}
                                               \left(\mathcal{I}e^{i\omega x}-\mathcal{R} e^{-i\omega x},\mathcal{I} e^{i\omega x}+\mathcal{R} e^{-i\omega x}\right) \quad \text{as} \,\, x\to-\infty\\
                                               \left(\mathcal{T}e^{ikx},\mathcal{T}e^{ikx}\right) \quad \hspace{2.9cm} \text{as}\,\, x\to+\infty\
                                              \end{array}\right.\,.
\eea

On the other hand, the conserved current associated with the Dirac equation~\eqref{paradox_Dirac} is given by
$j^{\mu}=-e\Psi^{\dagger}\gamma^{0}\gamma^{\mu}\Psi$ and, by equating the latter at $x\to-\infty$ and $x\to+\infty$, we find some general relations between the reflection and the transmission coefficients, in particular, 
\bea\label{reflection_fermion}
\left|\mathcal{R}\right|^2&=&|\mathcal{I}|^2-\left|\mathcal{T}\right|^2\,.
\eea
Therefore, $\left|\mathcal{R}\right|^2\leq |\mathcal{I}|^2$ for any frequency, showing that there is no superradiance for fermions. The same kind of relation can be found for massive fields~\cite{Manogue1988}. 

The difference between fermions and bosons comes from the intrinsic properties of these two kinds of particles. Fermions 
have positive definite current densities and bounded transmission amplitudes $0\leq \left|\mathcal{T}\right|^2\leq 
|\mathcal{I}|^2$, while for bosons the current density can change its sign as it is partially transmitted and the 
transmission amplitude can be negative, $-\infty < \frac{\omega-eV}{\omega}\left|\mathcal{T}\right|^2\leq 
|\mathcal{I}|^2$. From the quantum field theory point of view one can understand this process as a spontaneous pair 
production phenomenon due to the presence of a strong EM field~(see e.g.~\cite{Manogue1988}). The number of fermionic 
pairs produced spontaneously in a given state is limited by the Pauli exclusion principle, while such limitation does 
not exist for bosons.  
%%%%%%%%%%%%%%%%%%%%%%%%%%%%%%%%%%%%%%%%%%%%%%%
\subsection{Superradiance and pair creation}
%%%%%%%%%%%%%%%%%%%%%%%%%%%%%%%%%%%%%%%%%%%%%%%
To understand how pair creation is related to superradiance consider the potential used in the Klein paradox. Take a superradiant mode obeying Eq.~\eqref{super_klein} and $\mathcal{P}\leq 1$ to be the probability for spontaneous production of a single particle-antiparticle pair. The average number $\bar{n}$ of bosonic and fermionic pairs in a given state follows the Bose-Einstein and the Fermi-Dirac distributions, respectively,~\cite{Hansen:1980nc}
\be
\bar{n}_{{\rm B},{\rm F}}=\frac{1}{{1}/{\mathcal{P}}\mp 1}\,, \label{BE-FD}
\ee
where the minus sign refers to bosons, whereas the plus sign in the equation above is dictated by the Pauli exclusion principle, which allows only one fermionic pair to be produced in the same state.

Now, by using a second quantization procedure, the number of pairs produced in a given state for bosons and fermions in the superradiant region~\eqref{super_klein} is~\cite{Manogue1988}
\be
\bar{n}_{\rm B}=\left|\frac{\omega-eV}{\omega}\right|\left|\mathcal{T}\right|^2\,,\qquad 
\bar{n}_{\rm F}=\left|\mathcal{T}\right|^2\,. \label{eq:nb}
\ee
From Eq.~\eqref{BE-FD} we see that $0\leq\bar{n}_{\rm F}\leq 1$ while $\bar{n}_{\rm B}\to \infty$ when $\mathcal{P}\to 1$ and $\bar{n}_{\rm B}\to 0$ when $\mathcal{P}\to 0$. Equations~\eqref{BE-FD},~\eqref{eq:nb} and~\eqref{reflection_boson} show that $|{\cal R}|^2\to |\mathcal{I}|^2$ as ${\cal P}\to0$, so that superradiance is possible only when $\mathcal{P}\neq 0$, i.e. superradiance occurs due to spontaneous pair creation. On the other hand, we also see that the bounded value for the amplification factor in fermions is due to the Pauli exclusion principle. 

Although superradiance and spontaneous pair production in strong fields are related phenomena, they are nevertheless 
distinct. Indeed, pair production can occur without superradiance and it can occur whenever is kinematically allowed. On 
the other hand, superradiance is enough to ensure that bosonic spontaneous pair emission will occur. This is a well 
known result in BH physics. For example, in Sec.~\ref{sec:BHsuperradiance} we shall see that even nonrotating BHs do not 
allow for superradiance, but nonetheless emit Hawking radiation~\cite{Hawking:1974sw}, the latter can be considered as 
the gravitational analogue of pair production in strong EM fields.

To examine the question of energy conservation in this process, let us follow the following thought experiment~\cite{Winter:1959AJP}. Consider a battery connected to two boxes, such that a potential $V$ increase occurs between an outer grounded box and an inner box. An absorber is placed at the end of the inner box, which absorbs all particles incident on it. Let us consider an incident superradiant massless bosonic wave with charge $e$ and energy $\omega<eV$. From~\eqref{reflection_boson} we see that
\be
\left|\mathcal{R}\right|^2-\frac{eV-\omega}{\omega}\left|\mathcal{T}\right|^2=|{\cal I}|^2\,,
\ee
The minus sign in front of $\frac{eV-\omega}{\omega}\left|\mathcal{T}\right|^2$ is a consequence of the fact that the current for bosons is not positive definite, and ``negatively'' charged waves have a negative current density. Since more particles are reflected than incident we can also picture the process in the following way: all particles incident on the potential barrier are reflected, however the incident beam stimulates pair creation at the barrier, which emits particles and antiparticles. Particles join the reflected beam, while the negative transmitted current can be interpreted as a flow of antiparticles with charge $-e$.
All the particles incident with energy $\omega$ are reflected back with energy $\omega$ and in addition, because of pair creation, more particles with charge $e$ and energy $\omega$ join the beam. For each additional particle another one with charge $-e$ is transmitted to the box and transmits its energy to the absorber, delivering a kinetic energy $e V-\omega$. To keep the potential of the inner box at $V$, the battery loses an amount of stored energy equal to $eV$. The total change of the system, battery plus boxes, is therefore $E_{\rm diss}=-\omega$, for each particle with energy $\omega$ that is created to join the beam. 

Now, imagine exactly the same experiment but $\omega>eV$, when superradiance does not occur, and $\left|\mathcal{R}\right|^2\leq |{\cal I}|^2$. In this case the kinetic energy delivered to the absorber is $\omega-eV$. An amount of energy $eV$ is given to the battery and the system battery plus boxes gains a total energy $\omega$. By energy conservation the reflected beam must have energy $-\omega$, which we can interpret as being due to the fact that the reflected beam is composed by antiparticles and the transmitted beam by particles.

Although the result might seem evident from the energetic point of view, we see that superradiance is connected to dissipation within the system. As we will see in the rest, this fact is a very generic feature of superradiance.

If we repeat the same experiment for fermions we see from~\eqref{reflection_fermion} that $\left|\mathcal{R}\right|^2+\left|\mathcal{T}\right|^2=|{\cal I}|^2$.
Since the current density for fermions is positive definite the flux across the potential barrier must be positive and, therefore, the flux in the reflected wave must be less than the incident wave. Since fewer particles are reflected than transmitted, then by energy conservation the total energy given to the battery-boxes system must be positive and given by $\omega$. Thus the reflected beam has a negative energy $-\omega$, which can be interpreted as being due the production of antiparticles. In this case the kinetic energy delivered to the absorber will always be $|\omega-eV|$.  
%%%%%%%%%%%%%%%%%%%%%%%%%%%%%%%%%%%%%%%%%%%%%%%%%%%%%%%%%%%%%%%%%%%%%%%%%%%%%%
\subsection{Superradiance and spontaneous emission by a moving object~\label{subsec:spontaneous}}
%%%%%%%%%%%%%%%%%%%%%%%%%%%%%%%%%%%%%%%%%%%%%%%%%%%%%%%%%%%%%%%%%%%%%%%%%%%%%%
As counterintuitive as it can appear at first sight, in fact superradiance can be understood purely kinematically in terms of Lorentz transformations. Consider an object moving with velocity $\mathbf{v}_i$ (with respect to the laboratory frame) and emitting/absorbing a photon. Let the initial $4$-momentum of the object be $p^\mu_i=(E_i,\mathbf{p}_i)$ and the final one be $p^\mu_f=(E_f,\mathbf{p}_f)$ with $E_f=E_i\mp\hbar \omega$ and $\mathbf{p}_f=\mathbf{p}_i\mp\hbar \mathbf{k}$, where $(\hbar\omega,\hbar\mathbf{k})$ is the $4$-momentum of the emitted/absorbed photon, respectively. The object's rest mass can be computed by using Lorentz transformations to go to the comoving frame,
%%%
\begin{equation}
 {\cal E}_i=\gamma_i(E_i-\mathbf{v}_i\cdot\mathbf{p}_i)\,,
\end{equation}
and similarly for ${\cal E}_f$, where $\gamma_i=1/\sqrt{1-\mathbf{v}^2_i}$. Assuming $\mathbf{v}_f=\mathbf{v}_i+\delta\mathbf{v}$, to zeroth order in the recoil term $\delta\mathbf{v}$ the increase of the rest mass reads
%%%
\begin{equation}
 \Delta {\cal E}\equiv {\cal E}_f-{\cal E}_i=\mp\gamma_i\hbar(\omega-\mathbf{v}_i\cdot\mathbf{k})+{\cal O}(\delta\mathbf{v})\,, \label{deltaE}
\end{equation}
%%%
where the minus and plus signs refer to emission and absorption of the photon, respectively. Therefore, if the object is in its fundamental state (${\cal E}_i<{\cal E}_f$), the emission of a photon can only occur when the Ginzburg-Frank condition is satisfied, namely~\cite{Ginzburg:1945zz,Ginzburg:1947}
%%%
\begin{equation}
 \omega(k)-\mathbf{v}_i\cdot\mathbf{k}<0\,, \label{SR_spontaneous}
\end{equation}
%%%
where $k=|\mathbf{k}|$ and $\omega(k)$ is given by the photon's dispersion relation. In vacuum, $\omega(k)=k$ so that the equation above can never be fulfilled. This reflects the obvious fact that Lorentz invariance forbids a particle in its ground state to emit a photon in vacuum. However, spontaneous emission can occur any time the dispersion relation allows for $\omega<k$. For example, suppose that the particle emits a massive wave whose dispersion relation is $\omega=\sqrt{\mu^2+k^2}$, where $\mu$ is the mass of the emitted radiation. For modes with $\mu\ll k$, Eq.~\eqref{SR_spontaneous} reads
%%%
\begin{equation}
 \frac{\mu^2}{2k^2}<v_i\cos\vartheta-1\leq0\,,
\end{equation}
%%%
where $\mathbf{v}_i\cdot \mathbf{k}=v_i k \cos\vartheta$. Hence, only unphysical radiation with $\mu^2<0$ can be spontaneously radiated, this fact being related to the so-called \emph{tachyonic instability} and it is relevant for those theories that predict radiation with an effective mass $\mu$ through nonminimal couplings (e.g. this happens in scalar-tensor theories of gravity~\cite{Will:2005va} and it is associated to so-called spontaneous scalarization~\cite{Damour:1993hw}).

Another relevant example occurs when the object is travelling through an isotropic dielectric that is transparent to radiation. In this case $\omega=k/n(\omega)$ where $n(\omega)=1/v_{\rm ph}$ is the medium's refractive index and $v_{\rm ph}$ is the phase velocity of radiation in the medium. In this case Eq.~\eqref{SR_spontaneous} reads
%%%
\begin{equation}
 \cos\vartheta>\frac{v_{\rm ph}}{v_i}\,.
\end{equation}
%%%
Therefore, if the object's speed is smaller than the phase velocity of radiation, no spontaneous emission can occur, whereas in the opposite case spontaneous superradiance occurs when $\vartheta<\vartheta_c=\cos^{-1}(v_{\rm ph}/v_i)$. This phenomenon was dubbed anomalous Doppler effect~\cite{Ginzburg:1945zz,Ginzburg:1947}. The angle $\vartheta_c$ defines the angle of coherent scattering, i.e. a photon incident with an angle $\vartheta_c$ can be absorbed and re-emitted along the same direction without changing the object motion, even when the latter is structureless, i.e. when ${\cal E}_i={\cal E}_f$.

As discussed in Ref.~\cite{Bekenstein:1998nt}, spontaneous superradiance is not only a simple consequence of Lorentz invariance, but it also follows from thermodynamical arguments. Indeed, for a finite body that absorbs nearly monochromatic radiation, the second law of thermodynamics implies
%%%
\begin{equation}
 (\omega-\mathbf{v}_i\cdot\mathbf{k})a(\omega)>0\,,
\end{equation}
%%%
where $a(\omega)$ is the characteristic absorptivity of the body. Hence, the superradiance condition is associated with a negative absorptivity, that is, superradiance is intimately connected to \emph{dissipation} within the system.
%%%%%%%%%%%%%%%%%%%%%%%%%%%%%%%%%%%%%%%%%%%%%%%%%%%%%%%%%%%%%%%%%%%%%%%%%
\subsubsection{Cherenkov emission and superradiance\label{sec:cherenkov}}
%%%%%%%%%%%%%%%%%%%%%%%%%%%%%%%%%%%%%%%%%%%%%%%%%%%%%%%%%%%%%%%%%%%%%%%%%
The emission of radiation by a charge moving superluminally relative to the phase velocity of radiation in a dielectric --~also known as the Vavilov-Cherenkov effect~-- has a simple interpretation in terms of spontaneous superradiance~\cite{Ginzburg:2005}. A point charge has no internal structure, so $\Delta {\cal E}=0$ in Eq.~\eqref{deltaE}. Such condition can only be fulfilled when the charge moves faster than the phase velocity of radiation in the dielectric and it occurs when photons are emitted with an angle
%%%
\begin{equation}
 \vartheta_c=\cos^{-1}(v_{\rm ph}/v_i)\,.
\end{equation}
%%%
In general, $v_{\rm ph}=v_{\rm ph}(\omega)$ and radiation at different frequencies will be emitted in different 
directions. In case of a dielectric with zero dispersivity, the refraction index is independent from $\omega$ and the 
front of the photons emitted during the charge's motion forms a cone with opening angle $\pi-2\vartheta_c$. Such cone is 
the EM counterpart of the Mach cone that characterizes a shock wave produced by supersonic motion as will be discussed 
in Sec.~\ref{sec:shock_waves}. 
%%%

%%%%%%%%%%%%%%%%%%%%%%%%%%%%%%%%%%%%%%%%%%%%%%%%%%%%%%%%%%%%%%%%%%%%%%%%%%%%%%%%%%%%%%%%%%%%%%%%%%%%%%%%%%%%%%%%%%%%%%%%%%%%%%%
\subsubsection{Cherenkov radiation by neutral particles}
%%%%%%%%%%%%%%%%%%%%%%%%%%%%%%%%%%%%%%%%%%%%%%%%%%%%%%%%%%%%%%%%%%%%%%%%%%%%%%%%%%%%%%%%%%%%%%%%%%%%%%%%%%%%%%%%%%%%%%%%%%%%%%%
In their seminal work, Ginzburg and Frank also studied the anomalous Doppler effect occurring when a charge moves through a pipe drilled into a dielectric~\cite{Ginzburg:1945zz,Ginzburg:1947}. More recently, Bekenstein and Schiffer have generalized this effect to the case of a \emph{neutral} object which sources a large gravitational potential (e.g. a neutral BH) moving through a dielectric~\cite{Bekenstein:1998nt}. As we now briefly discuss, this effect is similar to Cherenkov emission, although it occurs even in presence of neutral particles.

Consider first a neutral massive object with mass $M$ surrounded by a ionized, two-component plasma of electrons and positively-charged nuclei\footnote{
Because we want to use thermodynamic equilibrium at the same temperature $T$, it is physically more transparent to work with a plasma than with a dielectric, as done instead in Ref.~\cite{Bekenstein:1998nt}.}. It was realized by Milne and Eddington that
in hydrostatic and thermodynamic equilibrium, an electric field necessarily develops to keep protons and electrons from separating completely~\cite{Milne,Eddington,1979ApJ...229..694M}. In equilibrium, the partial pressure $P_{\rm e, N}$ of electrons and nuclei is, respectively 
\be
\frac{\partial \log P_{\rm e, N}}{\partial r}=-\frac{m_{\rm e, N} g}{kT}-\frac{eE}{kT}\,,
\ee
where $m_{\rm e, N}$ is the mass of an electron and of the nucleon, $k$ is the Boltzmann constant, $T$ the temperature of the plasma
and $g$ the local gravitational acceleration.
Equality of the pressure gradient -- achieved when electrons and protons are separated -- happens for an electric field
\be
eE=\frac{(m_{\rm N}-m_{\rm e})g}{2}\sim \frac{m_{\rm N}g}{2}\,.
\ee

Consider now the same neutral massive object traveling through the ionized plasma. As we saw, the gravitational pull of the object will polarize the plasma because the positively charged nuclei are attracted more than the electrons. The polarization cloud is associated with an electric dipole field $\mathbf{E}$ that balances the gravitational force $\mathbf{g}$ and that acts as source of superradiant photons. This follows by thermodynamical arguments, even neglecting the entropy increase due to possible accretion~\cite{Bekenstein:1998nt}. The superradiant energy in this case comes from the massive object kinetic energy. Thus, the effect predicts that the object slows down because of superradiant emission of photons in the dielectric. 

In fact, the effect can be mapped into a Cherenkov process by noting that, in order to balance the gravitational pull, $e\mathbf{E}\sim-m_N \mathbf{g}$. Poisson equation then implies~\cite{Bekenstein:1998nt}
\be
\nabla\cdot\mathbf{E}=4\pi G\frac{M m_N}{e}\delta(r-r_0)\,,
\ee
where $r_0$ is the massive object position and for clarity we have restored the factor $G$. This equation is equivalent to that of an electric field sourced by a pointlike charge 
\begin{equation}
 Q=\frac{G m_N M}{e}\sim 5\times 10^4 A \left(\frac{M}{10^{17}{\rm g}}\right)e\,. \label{QcherenkovBH}
\end{equation}
where $A$ is the mass number of the atoms. Assuming that the plasma relaxation time is short enough, such effective charge will emit Cherenkov radiation whenever the Ginzburg-Frank condition~\eqref{SR_spontaneous} is met. 
Note that, modulo accretion issues which are not relevant to us here, the above derivation is equally valid for BHs. As already noted in Ref.~\cite{Bekenstein:1998nt} a primordial BH with $M\sim 10^{17}{\rm g}$ moving fast through a dielectric would Cherenkov radiate just like an elementary particle with charge $Q\sim 5\times 10^4e$. In particular, the Frank-Tamm formula for the energy $dE$ emitted per unit length $dx$ and per unit of frequency $d\omega$ reads
\be
dE=\frac{Q^2}{4\pi} \omega \mu(\omega)\left(1-\frac{1}{\beta^2 n^2(\omega)}\right)d\omega dx \,.
\ee
where $\mu$ and $n$ are the permeability and the refraction index of the medium, respectively, and $\beta=v/c$. Therefore, the total power reads
%%%
\begin{equation}
 \dot{E}_{\rm rad}=\frac{c Q^2}{4\pi}\int d\omega \mu(\omega)\omega\left(1-\frac{1}{\beta^2 n^2(\omega)}\right) \lesssim \frac{Q^2}{8\pi \epsilon_0 c}\omega_c^2
\end{equation}
%%%
where the integral is taken over the Cherenkov regime. In the last step we assumed $\mu(\omega)\approx \mu_0=1/(\epsilon_0 c^2)$ and $\beta\sim1$. The upper limit is expressed in terms of a cutoff frequency which depends solely on the plasma's properties $\omega_c \lesssim 2\pi c/a_0$, where $a_0$ is Bohr's radius. As a result of this energy emission, the BH slows down on a time scale
%%%
\begin{equation}
 \tau_{\rm C}\sim \frac{M}{\dot{E}_{\rm rad}}\sim \frac{2\epsilon_0}{\pi}\frac{M^2}{Q^2}\frac{a_0^2}{Mc}\sim 10^{12}\left(\frac{10^{17}{\rm g}}{M}\right) {\rm yr}
\end{equation}
where we have used Eq.~\eqref{QcherenkovBH}. Therefore, the effect is negligible for primordial BHs~\cite{Carr:2009jm} which were originally considered in Ref.~\cite{Bekenstein:1998nt}, but it might be relevant for more massive BHs travelling at relativistic velocities in a plasma with short relaxation time.
%%%

%%%%%%%%%%%%%%%%%%%%%%%%%%%%%%%%%%%%%%%%%%%%%%%%%%%%%%%%%%%%%%%%%%%%%%%%%%%%%%%%%%%%%%%
\subsubsection{Superradiance in superfluids and superconductors\label{sec:superfluid}}
%%%%%%%%%%%%%%%%%%%%%%%%%%%%%%%%%%%%%%%%%%%%%%%%%%%%%%%%%%%%%%%%%%%%%%%%%%%%%%%%%%%%%%%
Another example of linear superradiance in flat spacetime is related to superfluids\footnote{In the context of the gauge-gravity duality, the holographic dual of a superfluid is also a superradiant state, cf. Sec.~\ref{sec:breaking}.}~\cite{Bekenstein:1998nt}. Superfluids can flow through pipes with no friction when their speed is below a critical value known as Landau critical speed~\cite{Landau:Statistical}. If the fluid moves faster than the Landau critical speed, quasiparticle production in the fluid becomes energetically convenient at expenses of the fluid kinetic energy.

This process can be understood in terms of linear superradiance similarly to the Cherenkov effect previously discussed. In the fluid rest frame, consider a quasiparticle (e.g. a phonon) with frequency $\omega(\mathbf{k})$ and wavenumber $\mathbf{k}$. In this frame, the walls of the channel move with velocity $\mathbf{v}$ relative to the fluid. Therefore, the quantity
$\omega-\mathbf{v}\cdot\mathbf{k}$ is the analog of the Ginzburg-Frank condition~\eqref{SR_spontaneous} and becomes negative when 
%%%
\begin{equation}\label{landau_vel}
 v>v_c\equiv {\rm min}\frac{\omega(\mathbf{k})}{|\mathbf{k}|}\,,
\end{equation}
%%%
where $\omega(\mathbf{k})$ gives the dispersion relation of the quasiparticle. As discussed above, in this configuration it is energetically favorable to create a quasiparticle mode. This quasiparticles formation contributes a component which is not superfluid because its energy can be dissipated in various channels.

The same kind of reasoning can be used to predict the critical current flowing through a superconductor above which superconductivity is disrupted. Supercurrents are carried by Cooper pairs that move through a solid lattice with no resistance. However, whenever the kinetic energy of the current carriers exceeds the binding energy of a Cooper pair, it is energetically more favorable for the electrons in a pair to separate, with these broken pairs behaving as quasiparticles. Consider a superconductor, taken to be at zero temperature for simplicity, with supercurrent density $J=n q v_d$, where $n$ is the current carrier density, $q$ is the carrier charge and  $v_d$ is the drift velocity of the carriers measured in the frame of the solid lattice. In the rest frame of the superconductor ``fluid'', a quasiparticle created due to the scattering of a current carrier with the solid lattice has minimum momentum given by $2\hbar k_F$, where $k_F$ is the Fermi momentum of the electrons in the pair, and an energy $2\Delta_0$ which is the minimum energy needed to broke a Cooper pair at zero temperature. Landau arguments then predicts that to break a Cooper pair, i.e., to spontaneously emit a quasiparticle, the drift velocity must be given by
%%%
\begin{equation}\label{landau_vel_conductor}
 v_d>{\rm min}\frac{\omega(\mathbf{k})}{|\mathbf{k}|}\equiv\frac{\Delta_0}{\hbar k_F}\,.
\end{equation}
%%%
This in turn can be used to estimate the critical magnetic field above which superconductivity is broken. Take, for example, a circular superconductor with radius $R$, carrying a current density $J$. The magnetic field at the surface of the superconductor is then given by $H=JR/2$. The critical current density $J_c=n q \Delta_0/\hbar k_F$, then predicts that the critical magnetic field strength is given by $H_c=J_cR/2$ (see e.g. Ref.~\cite{HughesSuper}).

%%%%%%%%%%%%%%%%%%%%%%%%%%%%%%%%%%%%%%%%%%%%%%%%%%%%%%%%%%%%%%%%%%%%%%
\subsection{Sound amplification by shock waves}\label{sec:shock_waves}
%%%%%%%%%%%%%%%%%%%%%%%%%%%%%%%%%%%%%%%%%%%%%%%%%%%%%%%%%%%%%%%%%%%%%%

%%%%%%%%%%%%%%%%%%%%%%%%%%%%%%%%%%%%%%%%%%%%%%%%%%%%%%%%%%%%%%%%%%%%%%
\subsubsection{Sonic ``booms''}
%%%%%%%%%%%%%%%%%%%%%%%%%%%%%%%%%%%%%%%%%%%%%%%%%%%%%%%%%%%%%%%%%%%%%%
Curiously, very familiar phenomena can be understood from the point of view of superradiance. One of the most striking examples is the ``sonic boom'' originating from the supersonic motion of objects in a fluid. 

Imagine a structureless solid object traveling through a quiescent fluid with speed $v_i>c_s$ where $c_s$ is the speed of sound in the medium. Since the object is structureless then $\Delta {\cal E}=0$ in Eq.~\eqref{deltaE}, and in analogy with the Vavilov-Cherenkov effect we see that the object will emit \emph{phonons} with dispersion relation $\omega=c_s k$, when their angle with respect to the object's velocity satisfy
%%%
\begin{equation}
 \vartheta_M=\cos^{-1}(c_s/v_i)\,.
\end{equation}
%%%
Due to the supersonic motion of the object the emitted phonons will form a shock wave in the form of a cone, known as the Mach cone, with an opening angle $\pi-2\vartheta_M$~\cite{Landau:book_fluids}.

If there is any sound wave present in the fluid which satisfy the Ginzburg-Frank condition~\eqref{SR_spontaneous}, it will be superradiantly amplified as the object overtakes them. In the fluid's rest frame the wave fronts will propagate with an angle
%%%
\begin{equation}
 \cos\vartheta>c_s/v_i\,,
\end{equation}
%%%
which means that they are emitted inside the Mach cone and the cone surface marks the transition between the superradiant and non-superradiant regions. Thus the ``sonic boom'' associated with the supersonic motion in a fluid can be understood as a superradiant amplification of sound waves.

Although very different in spirit, the effects we discussed can be all explained in terms of spontaneous superradiance, and they just follow from energy and momentum conservation and by considering the emission in the comoving frame. As we shall discuss in the Sec.~\ref{sec:rotSR}, this guiding principle turns out to be extremely useful also in the case of rotational superradiance.

%%%%%%%%%%%%%%%%%%%%%%%%%%%%%%%%%%%%%%%%%%%%%%%%%%%%%%%%%%%%%%%%%%%%%%%%%%%%%%%%%%%%%%%%%
\subsubsection{Superradiant amplification at discontinuities~\label{super_discontinuity}}
%%%%%%%%%%%%%%%%%%%%%%%%%%%%%%%%%%%%%%%%%%%%%%%%%%%%%%%%%%%%%%%%%%%%%%%%%%%%%%%%%%%%%%%%%
A second instructive example concerning superradiant amplification by shock waves refers to sound waves at a discontinuity.
Consider an ideal fluid, locally irrotational (vorticity free), barotropic and inviscid. Focus now on small propagating disturbances --~i.e., sound waves~-- such that $\vec{v} = \vec{v_0}+\delta \vec{v}$, where $\vec{v}$ is the velocity of the perturbed fluid. Then, by linearizing the Navier-Stokes equations around the background flow,
it can be shown that small irrotational perturbations $\delta \vec{v} = -\nabla \Phi$ are described by the Klein-Gordon equation~\cite{Unruh:1980cg,Visser:1997ux}\footnote{this formal equivalence will prove useful later on when discussing analogue BHs.},
\begin{equation}
\Box \Phi = 0\,,
\label{KG}
\end{equation}
where the box operator is defined in the effective spacetime 
\begin{eqnarray}
g^{\mu \nu}\equiv \frac{1}{\rho c_s} \left[
\begin{array}{ccc}
-1 &\vdots&-v_0^{j}\\
\hdots \hdots &.&\hdots \hdots \hdots\\
-v_0^{i}&\vdots& (c_s^2 \delta _{ij}-v_0^iv_0^j)
\end{array}
\right]\,. \label{metricvisserinv}
\end{eqnarray}
and where $\rho(r)$ and $c_s(r)$ are the density of the fluid and the local speed of sound, respectively.
The effective geometry on which sound waves propagate is dictated solely by the background velocity $v_0$
and local speed of sound $c$. The (perturbed) fluid velocity and pressure can be expressed in terms of the master field $\Phi$ as
\beq
\delta\vec{v}&=&-\vec{\nabla} \Phi\,,\\
\delta P&=&\rho_0\left(\frac{\partial \Phi}{\partial t}+\vec{v_0}\cdot\vec{\nabla} \Phi\right)\,.\label{inversion_sound}
\eeq

We consider now a very simple example worked out by in Ref.~\cite{Ribner:1957} (and reproduced also in Landau and Lifshitz monograph~\cite{Landau:book_fluids}), where the normal to the discontinuity lies on the $z=0$ plane. Suppose that the surface of discontinuity separates a medium ``2'' at rest ($z<0$) from a medium ``1'' moving with velocity $\vec{v}_0=v_x\equiv v$ along the $x-$axis.
The scattering of a sound wave in medium 2 gives rise in medium 1 to a transmitted wave with the form\footnote{The slightly unorthodox normalization of the transmitted wave was chosen so that the final result for the amplification factor exactly matches Landau and Lifshitz's result, in their formalism.}
\be
\Phi_1=\frac{\omega}{\omega-k_x v_0}{\cal T}e^{ik_x\,x+ik_y\,y+ik\,z-i\omega\,t}\,.
\ee
The equation of motion~\eqref{KG} forces the dispersion relation
\be
(\omega-v_0 k_x)^2=c_s^2(k_x^2+k_y^2+k^2)\,.
\ee
In medium 2, the incident wave gets reflected, and has the general form
\be
\Phi_2={\cal I} e^{ik_x\,x+ik_y\,y+ik_z\,z-i\omega\,t}+{\cal R}e^{ik_x\,x+ik_y\,y-ik_z\,z-i\omega\,t}\,.
\ee

There are two boundary conditions relevant for this problem. The pressure must be continuous at the interface, yielding the condition
\be
{\cal R}+{\cal I}={\cal T}\,.
\ee
Finally, the vertical displacement $\zeta(x,t)$ of the fluid particles at the interface must also be continuous.
The derivative $\partial \zeta/\partial t$ is the rate of change of the surface coordinate $\zeta$ for a given $x$.
Since the fluid velocity component normal to the surface of discontinuity is equal to the rate of displacement of the surface itself, we have
\be
\partial \zeta/\partial t=\delta v_z-v_0\partial \zeta/\partial x\,.
\ee
Assuming for the displacement $\zeta$ the same harmonic dependence as we took for $\Phi$, we then have the second condition
\be
\frac{k}{\left(\omega-v_0k_x\right)^2}{\cal T}=\frac{k_z}{\omega^2}\left({\cal I}-{\cal R}\right)\,.
\ee
The sign of $k$ is as yet undetermined, and it is fixed by the requirement that the velocity of the refracted wave is away from the discontinuity, i.e., $\partial\omega/\partial k=c_s^2k/(\omega-v_0k_x)>0$. It can be verified that for $v_0>2c_s$ superradiant amplification of the reflected waves (${\cal R}>{\cal I}$) is possible, provided that $k<0$ and consequently that $\omega-v_0k_x<0$ ~\cite{Ribner:1957,Landau:book_fluids}. The energy carried away is supposedly being drawn from the whole of the medium ``1'' in motion, although a verification of this would require nonlinearities to be taken into account. Such nonlinear results have not been presented in the original work~\cite{Ribner:1957,Landau:book_fluids}; in the context of BH superradiance, we show in Section~\ref{backreaction_charged} that superradiance does result in mass (and charge) loss 
from the (BH) medium, at nonlinear order in the fluctuation.

This example considers compressible fluids and sound waves, but it can be shown that similar energy extraction mechanisms
are at play for waves in {\it incompressible} stratified fluids with shear flow~\cite{Booker:1967,Jones:1968,McKenzie:1972}.
An intuitive explanation in terms of negative-energy states is given in Ref.~\cite{McKenzie:1972}.
%%%%%%%%%%%%%%%%%%%%%%%%%%%%%%%%%%%%%%%%%%%%%%%%%%%%%%%%%%%%%%
\subsection{Rotational superradiance\label{sec:rotSR}}
%%%%%%%%%%%%%%%%%%%%%%%%%%%%%%%%%%%%%%%%%%%%%%%%%%%%%%%%%%%%%%

%%%%%%%%%%%%%%%%%%%%%%%%%%%%%%%%%%%%%%%%%%%%%%%%%%%%%%%%%%%%%%%%%%%%%%%%%%%%%%%%%%%%%%%%%%%%%%%%%%%%%%%%%
\subsubsection{Thermodynamics and dissipation: Zel'dovich and Bekenstein's argument\label{sec:Zeldovich}}
%%%%%%%%%%%%%%%%%%%%%%%%%%%%%%%%%%%%%%%%%%%%%%%%%%%%%%%%%%%%%%%%%%%%%%%%%%%%%%%%%%%%%%%%%%%%%%%%%%%%%%%%%
One important aspect of the previous examples is that the linear velocity 
of the medium from which the energy is drawn exceeds the phase velocity of the corresponding
oscillations~\cite{zeldovich2}. It is clearly impossible to extend such process to waves in vacuum and in plane geometry, because it would require
superluminal velocities, as already pointed out. However, in a cylindrical or spherical geometry
the angular phase velocity of an $m-$pole wave ($m$ is an azimuthal number, specified in more detail below),
is $\omega/m$. If the body is assumed to rotate with angular velocity $\Omega$, then amplification is in principle possible for waves satisfying condition \eqref{eq:superradiance_condition}, $\omega<m\Omega$, if the previous example is faithful.

It should be also clear from all the previous examples that rotating bodies with internal degrees of freedom
(where energy can be dumped into) display superradiance.
Two different arguments can be made in order to show this rigorously~\cite{zeldovich2,Bekenstein:1998nt}.

The first is of a thermodynamic origin. Consider an axi-symmetric macroscopic body
rotating rigidly with constant angular velocity about its symmetry axis.
Assume also the body has reached equilibrium, with well defined entropy $S$,
rest mass $M$ and temperature $T$.
Suppose now that a wavepacket with frequency $(\omega,\omega+d\omega)$ and azimuthal number $m$
is incident upon this body, with a power $P_m(\omega)d\omega$. Radiation with a specific
frequency and azimuthal number carries angular momentum
at a rate $(m/\omega)P_m(\omega) d\omega$ (c.f. Appendix \ref{appendix_energyangularmomentum}). Neglecting any spontaneous emission by the body
(of thermal or any other origin), the latter will absorb a fraction $Z_{m}$ of the incident energy and angular momentum,
\be
\frac{dE}{dt}=Z_{m} P_md\omega\,,\quad \frac{dJ}{dt}=Z_{m} \frac{m}{\omega} P_m d\omega\,.
\ee
Notice that the assumption of axi-symmetry is crucial. No precession occurs during the interaction,
and no Doppler shifts are involved. This implies that both the frequency and multipolarity of the incident
and scattered wave are the same, as assumed in the equations above.
Now, in the frame co-rotating with the body, the change in energy is simply~\cite{Landau:Statistical}
\be
dE_0=dE-\Omega dJ=dE\left(1-\frac{m\Omega}{\omega}\right)\,,\label{energy_transform}
\ee
and thus the absorption process is followed by an increase in entropy, $dS=dE_0/T$, of
\be
\frac{dS}{dt}=\frac{\omega-m\Omega}{\omega\,T}Z_{m}\,P_m(\omega)d\omega\,.
\ee
Finally, the second law of thermodynamics demands that 
\be
(\omega-m\Omega)Z_{m}>0\,,
\ee
and superradiance ($Z_{m}<0$) follows in the superradiant regime $\omega-m\Omega<0$. A similar argument within a quantum mechanics context also leads to the same conclusion~\cite{Alicki:2017bgh}.

Next, consider Zel'dovich's original ``dynamical'' argument, and take for definiteness a scalar field $\Phi$,
governed in vacuum by the Lorentz-invariant Klein-Gordon equation, $\Box \Phi=0$.
An absorbing medium breaks Lorentz invariance. Assume that, in a coordinate system in which
the medium is at rest, the absorption is characterized by a parameter $\alpha$ as
\be
\Box \Phi+\alpha\frac{\partial \Phi}{\partial t}=0\,. \label{KGdiss}
\ee
The $\Box$ term is Lorentz-invariant, but if the frequency in the accelerated frame is $\omega$ and the field behaves as $e^{-i\omega t+im\varphi}$ in the inertial frame the azimuthal coordinate is $\varphi=\varphi'-\Omega t$, and hence the frequency is $\omega'=\omega-m\Omega$. In other words, the effective damping parameter $\alpha \omega'$ becomes negative in the superradiant regime and the medium amplifies -- rather than absorbing-- radiation.

A very appealing classical example of rotational (EM) superradiance is worked out in some detail for the 
original model by Zel'dovich~\cite{zeldovich2,Bekenstein:1998nt}.

%%%%%%%%%%%%%%%%%%%%%%%%%%%%%%%%%%%%%%%%%%%%%%%
\subsubsection{EFT approach}
%%%%%%%%%%%%%%%%%%%%%%%%%%%%%%%%%%%%%%%%%%%%%%%
The thermodynamic and dynamical ``Lorentz-violating'' construction of Sec.~\ref{sec:Zeldovich} can be re-stated in a 
language more familiar to quantum mechanics~\cite{Endlich:2016jgc,Porto:2016pyg}. In this framework, one considers a spinning object 
(along the z-axis, say) interacting with a particle (for definiteness we consider a spin-0 particle) of energy $\omega$ 
and with $m$ as eigenvalue of the angular momentum along the $z$ direction. For a spinning object in an initial state 
$X_i$, interaction with the particle will cause it to transition with some probability to state $X_f$. The total probability $P$ for absorption of this particle is a sum over all final states
\be
P_{\rm abs}=\sum_{X_f}\frac{|\left<X_f;0|S|X_i;\omega,l,m\right>}{\left<\omega,l,m|\omega,l,m\right>}\,.
\ee
Here, $l,m$ define the angular states of the particle, the states of the object and of the vacuum are normalized, and $S$ encodes the interaction physics
\be
S=Te^{-i\int dt H_{\rm int}(t)}\,,
\ee
where $T$ is the time-ordering operator. 

The interaction Hamiltonian $H_{\rm int}(t)$ contains all possible composite operators ${\cal O}$, which encode all of the microscopic degrees of freedom of the spinning object.
As we noted previously, in the thermodynamic approach and in others, it is these dissipative degrees of freedom which 
are ultimately responsible for superradiance. The simplest possible composite operator ${\cal O}_I$ results in an 
Hamiltonian which involves rotation through the rotation matrix $R$
\be
H_{\rm int}=\partial^I \phi R_I^J{\cal O}_J\,,
\ee
and gives rise to an absorption cross section which becomes negative in the superradiant regime~\cite{Endlich:2016jgc,Porto:2016pyg}. These general results should be thought of as complementing the approach in Section~\ref{sec:Zeldovich}, and a similar discussion of superradiance
arising when a quantum field interacts with a rotating heat bath can be found in Ref.~\cite{Alicki:2017bgh}.

Below, we present three examples, one of which can also potentially be implemented in the laboratory.
We end with a well-known Newtonian system where energy transfer akin to superradiance is active.
%%%%%%%%%%%%%%%%%%%%%%%%%%%%%%%%%%%%%%%%%%%%%%%%%%%%%%%%%%%%%%
\subsubsection{Example 1. Scalar waves}
%%%%%%%%%%%%%%%%%%%%%%%%%%%%%%%%%%%%%%%%%%%%%%%%%%%%%%%%%%%%%%
%
\begin{figure}
%\begin{center}
\centerline{\includegraphics[height=7 cm]{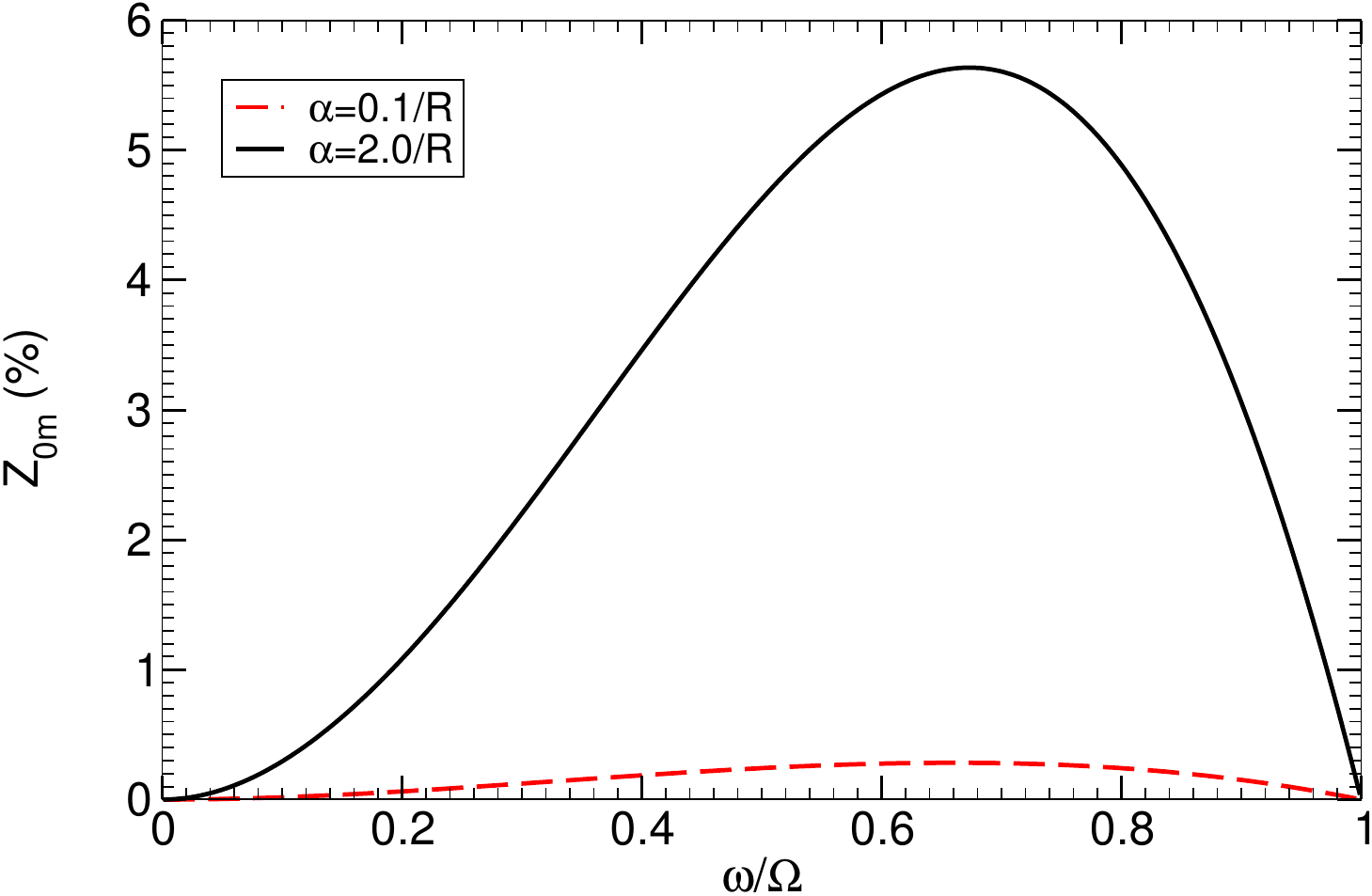}}
\caption{Amplification values $Z_{0m}=|{\cal R}|^2/|{\cal I}|^2-1$ of the scalar toy model for $m=1$, $\Omega R=0.5$ and $\alpha R=0.1,2$. } \label{Fig:scalar_toy}
%\end{center}
\end{figure}
Let us work out explicitly the case of a rotating cylinder in $(r,z,\varphi)$ spatial coordinates with a dissipative surface at $r=R$. For simplicity
the scalar is assumed to be independent of $z$, $\Phi=\phi(r)e^{-i\omega t+im\varphi }$. From what we said, by using Eq.~\eqref{KGdiss}, the problem can be modelled by
\be
\frac{1}{r}\left(r\phi'\right)'+\left(\omega^2-i\alpha (\omega-m\Omega) \delta(r-R)-\frac{m^2}{r^2}\right)\phi=0\,,
\ee
which can be solved analytically in terms of Bessel functions,
\be
\phi=\left\{\begin{array}{l}
             C_0J_m(\omega r)  \qquad \hspace{2cm} r<R \\
             C_1J_m(\omega r)+C_2Y_m(\omega r) \qquad r>R
            \end{array}\right.
\,.
\ee
The constants $C_1,\,C_2$ can be determined by continuity at $r=R$ along with the jump implied by the delta function. At infinity the solution is a superposition of ingoing and outgoing waves, $\phi\to {\cal I} e^{-i\omega r}/(r\omega)^{3/2}+{\cal R} e^{i\omega r}/(r\omega)^{3/2}$, where the constants ${\cal I}$ and ${\cal R}$ can be expressed in terms of $C_1$ and $C_2$. 
Figure~\ref{Fig:scalar_toy} shows a typical amplification factor $Z_{0m}\equiv |{\cal R}|^2/|{\cal I}|^2-1$ (in percentage) for $m=1$, $\Omega R=0.5$ and $\alpha R=0.1,2$.

%%%%%%%%%%%%%%%%%%%%%%%%%%%%%%%%%%%%%%%%%%%%%%%%%%%%%%%%%%%%%%%%%%%%%%%%%%%%%%%%%%%%%%%%%%%%%%%%%%%%%%%%%%%%%%%%%%%%%%%%
\subsubsection{Example 2. Amplification of sound and surface waves at the surface of a spinning cylinder\label{sec:lab}}
%%%%%%%%%%%%%%%%%%%%%%%%%%%%%%%%%%%%%%%%%%%%%%%%%%%%%%%%%%%%%%%%%%%%%%%%%%%%%%%%%%%%%%%%%%%%%%%%%%%%%%%%%%%%%%%%%%%%%%%%
A second example concerns amplification of {\it sound waves} at the surface of a rotating cylinder of radius $R$, 
but can also be directly used with {\it surface gravity waves}~\cite{Cardoso:2016zvz}. 
As we discussed in Section~\ref{super_discontinuity}, sound waves propagate in moving fluids as a massless scalar field in curved spacetime,
with an effective geometry dictated by the background fluid flow~\eqref{metricvisserinv}.

We focus here on fluids at rest, so that the effective metric is Minkowskian,
$ds^2=\frac{\rho}{c_s}\left(-c_s^2dt^2+dr^2+r^2d\vartheta^2+dz^2\right)$ in cylindrical coordinates.
Coincidentally, exactly the same equation of motion governs small gravity waves in a shallow basin~\cite{Schutzhold:2002rf},
thus the results below apply equally well to sound and gravity waves\footnote{Notice that Ref.~\cite{Schutzhold:2002rf} always implicitly
assumes a nontrivial background flow and the presence of a horizon in the effective geometry. In contrast, in our setup this is not required. 
All that it needs is a rotating boundary.}.

Solutions to Eq.~\eqref{KG} are better studied using the cylindrical symmetry of the effective background metric. 
In particular, we may decompose the field $\Phi$ in terms of azimuthal modes,
\beq
\Phi(t,r,\vartheta, z) = \frac{\phi(r)}{\sqrt{r\rho(r)}}  e^{-i\omega t+i m \vartheta+i k z}\,, \label{modesum}
\eeq
and we get
\begin{eqnarray}
\frac{d^2\phi}{dr^2}+\left[\frac{\omega^2}{c_s^2}-k^2-\frac{m^2-1/4}{r^2}-\frac{\rho '}{2 r \rho }+\frac{\rho'^2}{4 \rho^2}-\frac{\rho ''}{2 \rho}\right]\phi(r)=0\,.
\label{radialsound}
\end{eqnarray} 
For simplicity, let us focus on $k=0$ modes and assume that the density and the speed of sound are constant, so that the last three terms in the potential above vanishes and the background metric can be cast in Minkowski form. In this case, Eq.~\eqref{radialsound} admits the general solution $\phi=C_1\sqrt{r}J_m(\omega r/c_s)+C_2\sqrt{r}Y_m(\omega r/c_s)$. The constants $C_1$ and $C_2$ are related to the amplitude of the ingoing and outgoing wave at infinity, i.e., asymptotically one has
\be
\phi \sim{\cal I}e^{-i\omega r}+{\cal R}e^{i\omega r}=\sqrt{\frac{c_s}{2\pi\omega}}\left((C_1-iC_2)e^{i(\omega r/c_s-m\pi/2-\pi/4)}+(C_1+iC_2)e^{-i(\omega r/c_s-m\pi/2-\pi/4)}\right)\,.
\ee
The ratio ${\cal R}/\cal I$ can be computed by imposing appropriate boundary conditions. For nonrotating cylinders the latter read~\cite{Lax:1948}
\begin{equation}
\left(\frac{r\Phi'}{\Phi}\right)_{r=R}=-\frac{i\rho \omega R}{\Upsilon}\,, \label{BCsound}
\end{equation}
in terms of the original perturbation function, where $\Upsilon$ is the impedance of the cylinder material. As explained before, when the cylinder rotates uniformly with angular velocity $\Omega$, it is sufficient to transform to a new angular coordinate $\vartheta'=\vartheta+\Omega t$ which effectively amounts to the replacement of $\omega$ with $\omega-m\Omega$ in the boundary condition~\eqref{BCsound}. An empirical impedance model for fibrous porous materials was developed in Ref.~\cite{Delany:1970}, yielding a universal function of the flow resistance $\sigma$ and frequency of the waves,
\be
\Upsilon=\rho c_s\left[1+0.0511\left(2\pi\sigma/\omega\,\, {\rm kg}^{-1}{\rm m}^3\right)^{0.75}-i0.0768\left(2\pi\sigma/\omega\,\, {\rm kg}^{-1}{\rm m}^3\right)^{0.73}\right]\,.
\ee
Typical values at frequencies $\omega \sim 1000 {\rm s}^{-1}$ are $\Upsilon\sim \rho c_s(1-0.2i)$~\cite{Delany:1970}.

We will define the amplification factor $Z_{m}$ to be
\be
Z_{m}=|{\cal R}|^2/|{\cal I}|^2-1\,.
\ee
Notice that, from \eqref{inversion_sound}, the amplification factor measures the gain in pressure.
Using Eq.~\eqref{BCsound} and the exact solution of Eq.~\eqref{radialsound}, the final result for the amplification factor reads
%%%
\begin{equation}
Z_{m}=\left|\frac{i \tilde{\omega} \tilde{\Upsilon} J_{m-1}-2 (\tilde{\omega}-1) J_m-i \tilde{\omega} \tilde{\Upsilon} J_{m+1}+\tilde{\omega} \tilde{\Upsilon} Y_{m-1}+2 i(\tilde\omega-1) Y_m-\tilde{\omega} \tilde{\Upsilon} Y_{m+1}}{\tilde{\omega} \tilde{\Upsilon} J_{m-1}+2 i (\tilde{\omega}-1) J_{m}-\tilde{\omega} \tilde{\Upsilon} J_{m+1}+i \tilde{\omega} \tilde{\Upsilon} Y_{m-1}-2(\tilde\omega-1) Y_{m}-i \tilde{\omega} \tilde{\Upsilon} Y_{m+1}}\right|^2-1\,,\label{sigmasound}
\end{equation}
%%%
where we have defined the dimensionless quantities $\tilde{\Upsilon}=\Upsilon/(\rho c_s)$,  $\tilde{\omega}=\omega/(m\Omega)$ and we indicate $J_i=J_i(\omega R/c_s)$ and $Y_i=Y_i(\omega R/c_s)$ for short. Note that the argument of the Bessel functions reads $m\tilde{\omega}v/c_s$, where $v$ is the linear velocity at the cylinder's surface. Therefore, the amplification factor does not depend on the fluid density and it only depends on the dimensionless quantities $v/c_s$ and $\tilde{\omega}$.  Although not evident from Eq.~\eqref{sigmasound}, $Z_{m}=0$ when $\tilde\omega=1$ and it is positive (i.e. there is superradiant amplification) for $\tilde\omega<1$, for any $v/c_s$.

\begin{figure}[hbt]
\begin{center}
\begin{tabular}{cc}
\epsfig{file=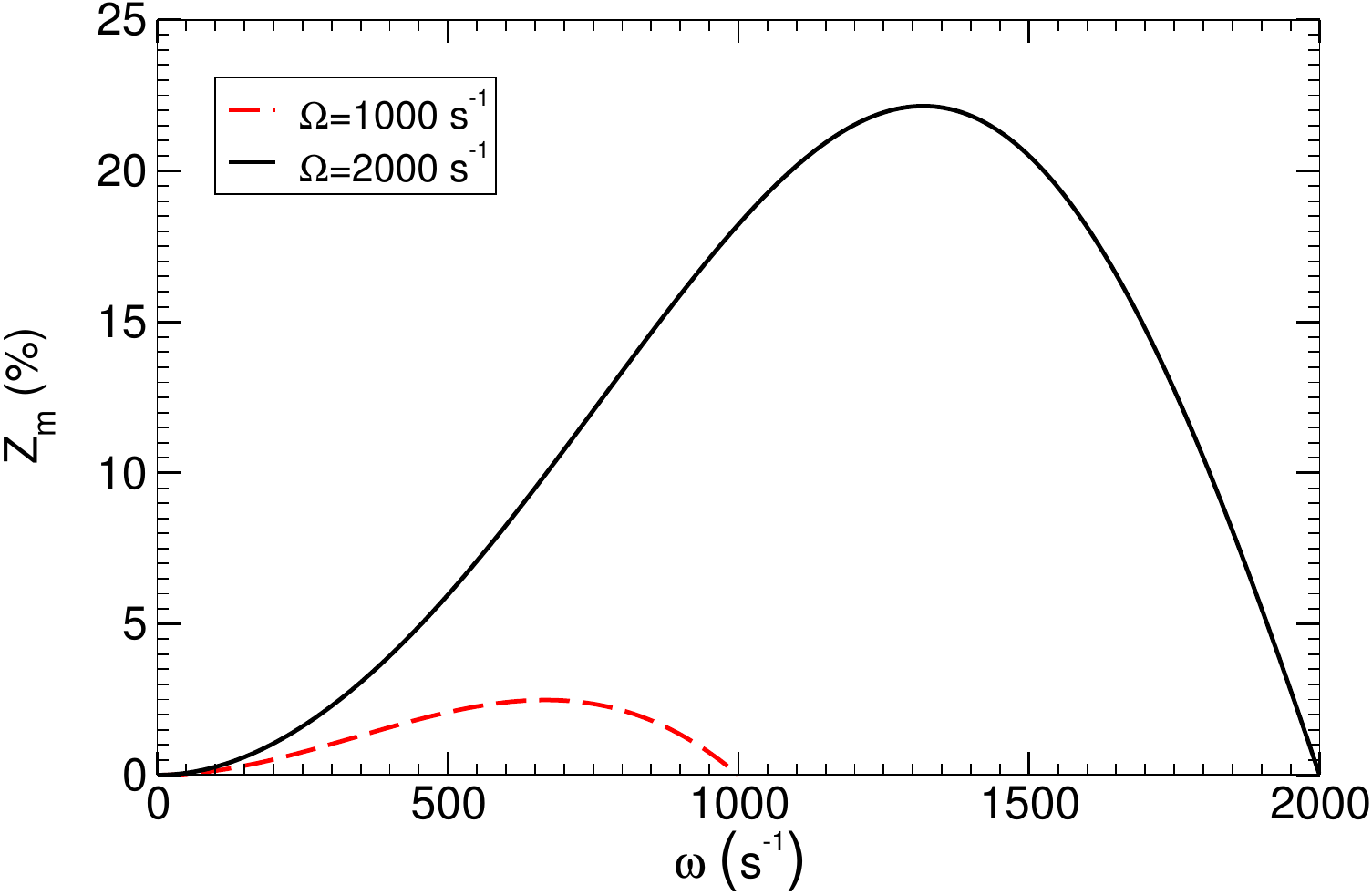,width=7.5cm,angle=0,clip=true}&
\epsfig{file=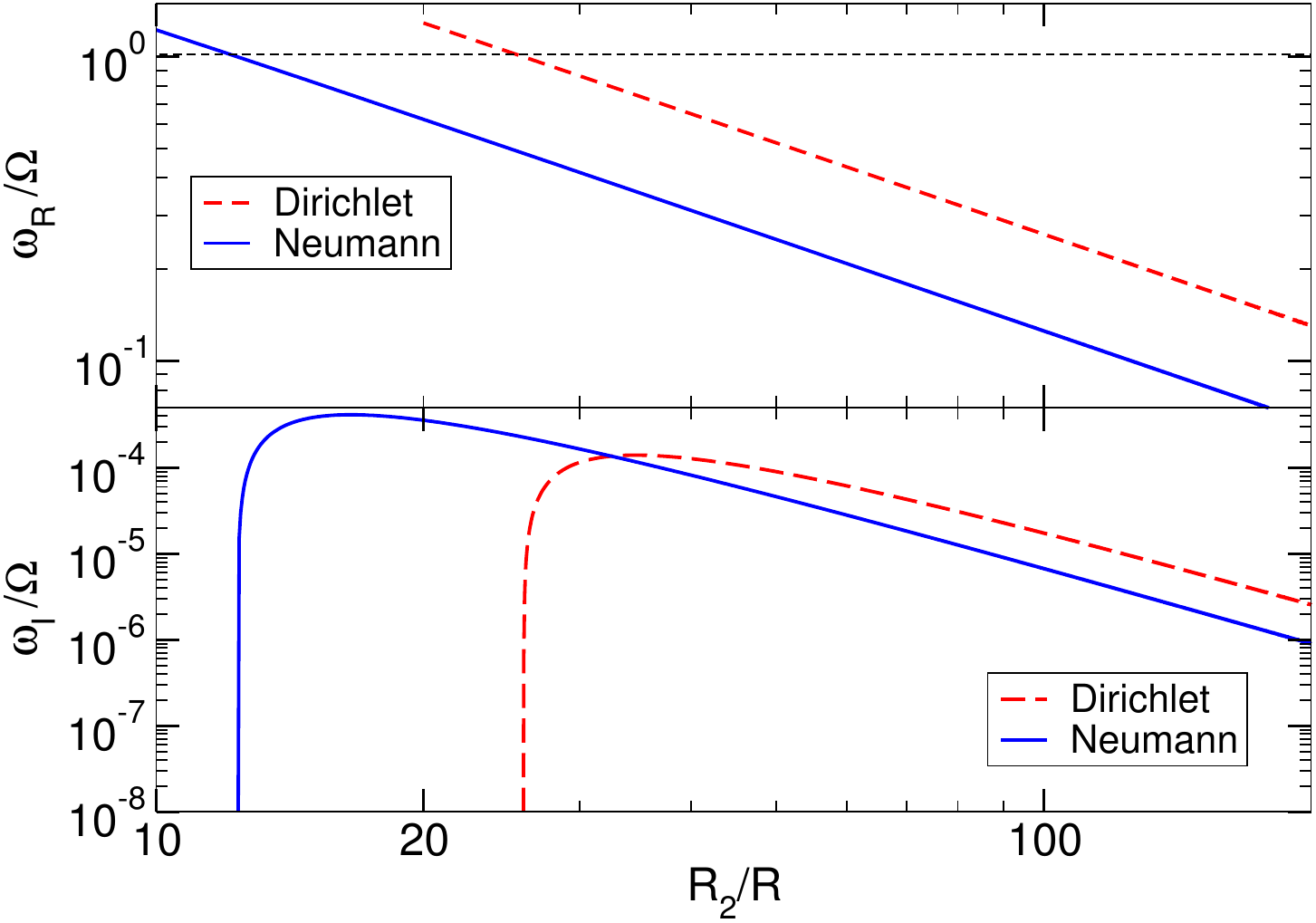,width=7.0cm,angle=0,clip=true}
\end{tabular}
\caption{Left panel: Amplification values $Z_{m}$ of acoustic waves for $m=1,\,R=10\,{\rm cm}$ and $\Omega=1000,\,2000\,{\rm s}^{-1}$. Right panel: fundamental unstable mode for the ``acoustic bomb'', a rotating cylinder with radius $R$ enclosed in a cylindrical cavity at distance $R_2$. In this example we set $m=1$ and $v/c_s\approx 0.147$. Note that the mode becomes unstable $(\omega_I>0)$ precisely when the superradiance condition $\omega_R<\Omega$ is fulfilled.
\label{Fig:acoustic}}
\end{center}
\end{figure}

As a point of principle, let us use a typical value for the impedance, $\tilde{\Upsilon}\approx (1-0.2i)$, to compute the amplification of sound waves in {\it air} within this setup.
We take $\Omega=1000,\,2000\,{\rm s}^{-1}$ and a cylinder with radius $R=10\,{\rm cm}$,
corresponding to linear velocities at the cylinder surface of the order of $v=100,\,200\,{\rm m}\,{\rm s}^{-1}$, but below the sound speed. The (percentage) results are shown in Fig.~\ref{Fig:acoustic},
and can be close to 100$\%$ amplification for large enough cylinder angular velocity. Note the result only depends on the combination $\Omega R/c_s$, which can be tweaked to obtain the optimal experimental configuration.

Another interesting application is to build an ``acoustic bomb'', similar in spirit with the ``BH bombs'' that we discuss in Sec.~\ref{sec:bombs}. In other words, by confining the superradiant modes near the rotating cylinder we can amplify the superradiant extraction of energy and trigger an instability. In this simple setup, confinement can be achieved by placing a cylindrical reflecting surface at some distance $R_2$ (note that this configuration is akin to the ``perfect mirror'' used by Press and Teukolsky to create what they called a BH bomb~\cite{Press:1972zz}). The details of the instability depend quantitatively on the outer boundary, specifically on its acoustic impedance. We will not perform a thorough parameter search, but focus on two extreme cases: Dirichlet and Neumann conditions.
Imposing the boundary conditions at $r=R_2$, we obtain the equation that defines the (complex) eigenfrequencies of the problem analytically,
%%%
\begin{eqnarray}
&&\tilde{\omega} \tilde{\Upsilon}\left[\hat{J}_m (Y_{m-1}-Y_{m+1})+ \hat{Y}_m (J_{m+1}- J_{m-1})\right]+2 i (\tilde{\omega}-1) \left[\hat{J}_m Y_m- J_m \hat{Y}_m \right]=0\,,\\
&&\tilde{\omega} \tilde{\Upsilon} \left[\left(J_{m-1}-J_{m+1}\right) \left(\hat{Y}_{m+1}-\hat{Y}_{m-1}\right)+\hat{J}_{m-1} \left(Y_{m-1}-Y_{m+1}\right)+\hat{J}_{m+1} \left(Y_{m+1}-Y_{m-1}\right)\right]\nn\\
&&+2i (\tilde{\omega}-1) \left[J_m \left(\hat{Y}_{m+1}-\hat{Y}_{m-1}\right)+\hat{J}_{m-1} Y_m-\hat{J}_{m+1} Y_m\right]=0\,,
\end{eqnarray}
%%%
for Dirichlet ($\Phi(r=R_2)=0$) and Neumann ($\Phi'(r=R_2)=0$) conditions, respectively. In the equations above, we have further defined $\hat{J}_i=J_i(\omega R_2/c_s)$ and $\hat{Y}_i=Y_i(\omega R_2/c_s)$ for short. In both cases the eigenmode equation only depends on the ratio $R_2/R$, $\tilde{\omega}$ and $v/c_s$. 
Neumann conditions, $\Phi'(r=R_2)=0$, mimic rigid outer boundaries. The fundamental eigenfrequencies $\omega=\omega_R+i\omega_I$ for these two cases are shown in the right panel of Fig.~\ref{Fig:acoustic} as functions of the mirror position $R_2/R$. Within our conventions, the modes are unstable when the imaginary part is positive (because of the time dependence $e^{-i\omega t}$). As expected, the modes become unstable when $\omega_R<m\Omega$, i.e. when the superradiance condition is satisfied. In the example shown in Fig.~\ref{Fig:acoustic}, the maximum instability occurs for $R_2\sim 30 R$ and corresponds to a very short instability time scale, 
%%%
\begin{equation}
 \tau\equiv\frac{1}{\omega_I}\sim 10\left(\frac{1000\,{\rm Hz}}{\Omega}\right)\,{\rm s}\,.
\end{equation}
Although our model is extremely simple, these results suggest the interesting prospect of detecting sound-wave superradiance amplification and ``acoustic bomb'' instabilities in the laboratory.
Complementary results can be found in Ref.~\cite{Cardoso:2016zvz}.

Finally, note that an alternative to make the system unstable is to have the fluid confined within a {\it single, rotating} absorbing cylinder. We find however, that in this particular setup
the instability only sets in for supersonic cylinder surface velocities, presumably harder to achieve experimentally.

%%%%%%%%%%%%%%%%%%%%%%%%%%%%%%%%%%%%%%%%%%%%%%%%%%%%%%%%%%%%
\subsubsection{Example 3. Tidal heating and acceleration\label{sec:tides}}
%%%%%%%%%%%%%%%%%%%%%%%%%%%%%%%%%%%%%%%%%%%%%%%%%%%%%%%%%%%%
%
\begin{figure}
\begin{center}
\begin{tabular}{cc}
\epsfig{file=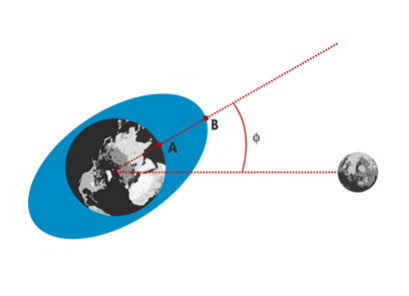,width=0.7\textwidth,angle=0,clip=true}
\end{tabular}
\caption{Tides on the Earth caused by our moon (as seen by a frame anchored on the Moon). The tidal forces create a bulge on Earth's ocean surface, which leads Moon's orbital position by a constant angle $\phi$. Earth rotates faster than the Moon in its orbit, thus a point $A$ on the surface of the Earth will differentially rotate with respect to the oceans, causing dissipation of energy and decrease of Earth's rotation period.
\label{fig:tides}}
\end{center}
\end{figure}
Although the processes we have discussed so far all involve radiation, it is possible to extract 
energy away from rotating bodies even in the absence of waves\footnote{This statement can be disputed however, since the phenomenon we discuss in the following does involve time retardation effects and is therefore intimately associated with wave phenomena. For further arguments that tides are in fact GWs, see Ref.~\cite{Goswami:2019fyk}.}.
A prime example concerns \emph{``tidal heating''} and consequent tidal acceleration, which is most commonly known to occur in the Earth-Moon system.

As explained by George Darwin back in 1880~\cite{Darwin:1880} (see also Refs.~\cite{Hut:1981,Verbunt} which are excellent overviews of the topic), tides are caused by differential forces on the oceans, which raise tidal bulges on them, as depicted in Figure~\ref{fig:tides}.
Because Earth rotates with angular velocity $\Omega_{\rm Earth}$, these bulges are not exactly aligned with the Earth-Moon direction. In fact, because Earth rotates faster than the Moon's orbital motion ($\Omega_{\rm Earth}>\Omega$), the bulges {\it lead} the Earth-Moon direction by a constant angle. This angle would be zero if friction were absent, and the magnitude of the angle depends on the amount of friction. Friction between the ocean and the Earth's crust slows down Earth's rotation by roughly $\dot{\Omega}_{\rm Earth}\sim -5.6\times 10^{-22}/s^2$, about $0.002s$ per century. Conservation of angular momentum of the entire system lifts the Moon into a higher orbit with a longer period and larger semi-major axis. Lunar ranging experiments have measured the magnitude of this tidal acceleration to be about $\dot{a}=3.82 {\rm cm}/{\rm yr}$~\cite{Dickey:1994zz}.

%%%%%%%%%%%%%%%%%%%%%%%%%%%%%%%%%%%%%%%%%%%%%%%%%%%%%%%%%%%%%%%%%%%%%%%%%%%%%%%%%%%%%%%%
\paragraph{Tidal acceleration and superradiance in the ``Newtonian'' approximation}
%%%%%%%%%%%%%%%%%%%%%%%%%%%%%%%%%%%%%%%%%%%%%%%%%%%%%%%%%%%%%%%%%%%%%%%%%%%%%%%%%%%%%%%%
Let us consider a generic power-law interaction between a central body of gravitational mass $M$ and radius $R$ and its moon with mass $m_p$ at a distance $r_0$. The magnitude is (in this section we re-insert factors of $G$ and $c$ for clarity)
\be
F=\frac{GMm_p}{r_0^{n}}\,, \label{NewtonGeneral}
\ee
and Newton's law is recovered for $n=2$. The tidal acceleration in $M$ is given by
\be
a_{\rm tidal}=\frac{nGm_p}{R^n}\left(\frac{R}{r_0}\right)^{n+1}=n g_M\left(\frac{R}{r_0}\right)^{n+1}\frac{m_p}{M}\,,
\ee
where $g_M$ is the surface gravity on $M$. This acceleration causes tidal bulges of height $h$ and mass $\mu$ to be raised on $M$. These can be estimated by equating the specific energy of the tidal field, $E_{\rm tidal}\sim a_{\rm tidal} R$, with the specific gravitational energy, $E_G\sim g_M h$, needed to lift a unit mass from the surface of $M$ to a distance $h$. We get
\be
\frac{h}{R}=n\left(\frac{R}{r_0}\right)^{n+1}\frac{m_p}{M}\,,
\ee
which corresponds to a bulge mass of approximately $\mu=\frac{\kappa}{4}n m_p \left({R}/{r_0}\right)^{n+1}$,
where $\kappa$ is a constant of order 1, which encodes the details of Earth's internal structure.
Without dissipation, the position angle $\phi$ in Figure~\ref{fig:tides} is $\phi=0$, while the tidal bulge is aligned with moon's motion. Dissipation contributes a constant, small, time lag $\tau$ such that the lag angle is $\phi=(\Omega_{\rm Earth}-\Omega)\tau$.

With these preliminaries, a trivial extension of the results of Ref.~\cite{Hut:1981} yields a tangential tidal force on $M$, assuming a circular orbit for the moon, 
\be
F_{\vartheta}\sim \frac{n(n+1)G\kappa}{2}m_p^2\frac{R^{n+3}}{r_0^{2n+3}}(\Omega_{\rm Earth}-\Omega)\tau\,. \label{torquetides}
\ee
%
% This perturbation exerts a torque $N=r_0F_{\vartheta}$. 
The change in orbital energy over one orbit is related to the torque $r_0F_{\vartheta}$ and reads
$\int_0^{2\pi}r_0F_{\vartheta}\Omega/2\pi d\vartheta=\Omega r_0 F_{\vartheta}$. Thus, we get
\be
\dot{E}_{\rm orbital}=\frac{n(n+1)G\kappa m_p^2}{2}\frac{R^{n+3}}{r_0^{2n+2}}\Omega(\Omega_{\rm Earth}-\Omega)\tau\,, \label{tidesGeneral}
\ee
and, for gravitational forces obeying Gauss's law ($n=2$), the latter reduces to
\be
\dot{E}_{\rm orbital}=3G\kappa m_p^2\frac{R^{5}}{r_0^{6}}\Omega(\Omega_{\rm Earth}-\Omega)\tau\,.\label{tidesNewt}
\ee

Summarizing, tidal heating extracts energy and angular momentum from the Earth. Conservation of both these quantities then requires the moon
to slowly spiral outwards. It can be shown that tidal acceleration works in any number of spacetime dimensions and with 
other fields (scalar or EM)~\cite{Cardoso:2012zn,Brito:2012gw}.

This and the previous examples make it clear that any rotating object should be prone to energy extraction and 
superradiance, provided that some negative-energy states are available, which usually mean that some dissipation 
mechanism of any sort is at work when the system is non-rotating.
When the tidally distorted object is a BH, negative energies are naturally provided by the ergoregion. The event horizon --~as we discuss in the next section~-- behaves in many respects as a viscous one-way membrane~\cite{MembraneParadigm}, providing the dissipation in the non-rotating limit, and stabilizing the rotating system~\cite{Vicente:2018mxl}. Interestingly, by substituting $\Omega_{\rm Earth}\to\Omega_{\rm H}$ in Eq.~\eqref{tidesNewt}, setting $\kappa\sim1/3\approx{\cal O}(1)$, and with the simple argument that the only relevant dissipation time scale in the BH case is the light-crossing time $\tau\sim M$, Eq.~\eqref{tidesNewt} was found to agree~\cite{Cardoso:2012zn} with the exact result for BH tidal heating obtained through BH perturbation theory~\cite{Hartle:1973zz,Hartle:1974gy,Poisson:2009di,Binnington:2009bb,Glampedakis:2013jya}.

%%%%%%%%%%%%%%%%%%%%%%%%%%%%%%%%%%%%%%%%%%%%%%%%%%%%%%%%%%%%%%%%%%%%%%%%%
\clearpage
\newpage
\section{Superradiance in black hole physics}\label{sec:BHsuperradiance}
%%%%%%%%%%%%%%%%%%%%%%%%%%%%%%%%%%%%%%%%%%%%%%%%%%%%%%%%%%%%%%%%%%%%%%%%%
As discussed in the previous section, superradiance requires \emph{dissipation}. The latter can emerge in various forms, e.g. viscosity, friction, turbulence, radiative cooling, etc. All these forms of dissipation are associated with some medium or some matter field that provides the \emph{arena} for superradiance. It is thus truly remarkable that --~when spacetime is \emph{curved}~-- superradiance can also occur in vacuum, even at the classical level. In this section we discuss in detail \emph{BH superradiance}, which is the main topic of this work. 

BHs are classical vacuum solutions of essentially any relativistic (metric) theory of gravity, including Einstein General Theory of Relativity.
Despite their simplicity, BHs are probably the most fascinating predictions of GR and enjoy some extremely nontrivial 
properties. The most important property (which also defines the very concept of BH) is the existence of an event 
horizon, a boundary in spacetime which causally disconnects the interior from the exterior. Among 
the various properties of BH event horizons, the one that is most relevant for the present discussion is that BHs behave 
in many respects as a viscous one-way membrane in flat spacetime. This is the so-called BH membrane 
paradigm~\cite{MembraneParadigm}. Thus, the existence of an event horizon provides vacuum with an intrinsic dissipative 
mechanism. Perhaps the second most relevant property of BHs is the existence of {\it ergoregions}, regions close to the 
horizon where timelike particles can have negative energies. Further details are given below in 
Section~\ref{subsub:ergoregion}. As we shall see, the very existence of ergoregions allows to extract energy from the 
vacuum, basically in any relativistic theory of gravity, while the horizon works to stabilize the system.

While most of our discussion is largely model- and theory-independent, for calculation purposes we will be dealing with the Kerr-Newman family of BHs~\cite{Newman:1965my},
which describes the most general stationary electrovacuum solution of the Einstein-Maxwell theory~\cite{uniqueness}. We will be specially interested in two different spacetimes
which display superradiance of different nature, the uncharged Kerr and the nonrotating charged BH geometry.

%%%%%%%%%%%%%%%%%%%%%%%%%%%%%%%%%%%%%%%%%%%%%%%%%%%%%%%%%%%%%%%%%%%%%%%%%%%%%%%%%%%%%%%%%%%%%%%%%%
\subsection{Action, equations of motion and black hole spacetimes}\label{sec:kerr}
%%%%%%%%%%%%%%%%%%%%%%%%%%%%%%%%%%%%%%%%%%%%%%%%%%%%%%%%%%%%%%%%%%%%%%%%%%%%%%%%%%%%%%%%%%%%%%%%%%
We consider a generic action involving one complex, charged massive scalar $\Psi$ and a massive vector
field $A_{\mu}$ with mass $m_S = \mu_S \hbar$ and $m_V=\mu_V\hbar$,
respectively,
\beq
S&=&\int d^4x \sqrt{-g} \left( \frac{R-2\Lambda}{\kappa} - \frac{1}{4}F^{\mu\nu}F_{\mu\nu}- \frac{\mu_V^2}{2}A_{\nu}A^{\nu}
-\frac{1}{2}g^{\mu\nu}\Psi^{\ast}_{,\mu}\Psi^{}_{,\nu} - \frac{\mu_S^2}{2}\Psi^{\ast}\Psi %-\frac{\mu_S^2}{2}\Phi^2 
\right)\nonumber\\
&+&\int d^4x \sqrt{-g}\left(i\frac{q}{2}A_{\mu}\left(\Psi\nabla^{\mu}\Psi^{\ast}-\Psi^{\ast}\nabla^{\mu}\Psi\right)-\frac{q^2}{2}A_{\mu}A^{\mu}\Psi\Psi^{\ast}
\right)+S_M       
\,.\label{eq:MFaction}
\eeq
where $\kappa=16\pi$, $\Lambda$ is the cosmological constant, $F_{\mu\nu} \equiv
\nabla_{\mu}A_{\nu} - \nabla_{\nu} A_{\mu}$ is the Maxwell tensor, and $S_M$ is the standard matter action that we neglect henceforth.
More generic actions could include a coupling between the scalar and vector sector, 
and also higher-order self-interaction terms. However, most of the work on BH superradiance is framed
in the above theory and we therefore restrict our discussion to this scenario.
The resulting equations of motion are
\begin{subequations}
\label{eq:MFEoMgen}
\begin{eqnarray}
  \label{eq:MFEoMScalar}
  &&\left(\nabla_{\mu}-iqA_{\mu}\right)\left(\nabla^{\mu}-iqA^{\mu}\right)\Psi =\mu_S^2\Psi
			\,,\\
  \label{eq:MFEoMVector}
  &&\nabla_{\mu} F^{\mu\nu} =
      \mu_V^2A^\nu+q^2\Psi\Psi^{\ast}A^{\nu}-i\frac{q}{2}\left(\Psi\nabla^{\nu}\Psi^{\ast}-\Psi^{\ast}\nabla^{\nu}\Psi\right) \,,\\
  \label{eq:MFEoMTensor}
  &&\frac{1}{\kappa} \left(R^{\mu \nu} - \frac{1}{2}g^{\mu\nu}R+\Lambda g^{\mu\nu}\right) =
      - \frac{1}{8}F^{\alpha\beta}F_{\alpha\beta}g^{\mu\nu}+\frac{1}{2}F^{\mu}_{\,\,\alpha}F^{\nu\alpha}
      - \frac{1}{4}\mu_V^2A_{\alpha}A^{\alpha}g^{\mu\nu}+\frac{\mu_V^2}{2}A^{\mu}A^{\nu}
      \nonumber\\
   && -\frac{1}{4}g^{\mu\nu}\left( \Psi^{\ast}_{,\alpha}\Psi^{,\alpha}+{\mu_S^2}\Psi^{\ast}\Psi\right)
      +\frac{1}{4}\left( \Psi^{\ast,\mu}\Psi^{,\nu}+ \Psi^{,\mu}\Psi^{\ast,\nu} \right)-i\frac{q}{2}A^{\mu}\left(\Psi\nabla^{\nu}\Psi^{\ast}-\Psi^{\ast}\nabla^{\nu}\Psi\right) \nonumber\\
   &&-\frac{q^2}{4}g^{\mu\nu}\Psi\Psi^{\ast}A_{\alpha}A^{\alpha}+\frac{q^2}{2} \Psi\Psi^{\ast}A^{\mu}A^{\nu}+i\frac{q}{4}g^{\mu\nu}A_{\alpha}\left(\Psi\nabla^{\alpha}\Psi^{\ast}-\Psi^{\ast}\nabla^{\alpha}\Psi\right)\,.
\end{eqnarray}
\end{subequations}

These equations describe the fully nonlinear evolution of
the system. For the most part of our work, we will specialize to perturbation theory, i.e. we consider $A^\mu$ and $\Psi$ to be small --~say of order ${\cal O}(\epsilon)$~-- and include their backreaction on the metric only perturbatively. Because the stress-energy tensor is quadratic in the fields, to order ${\cal O}(\epsilon)$ the gravitational sector is described by the standard Einstein equations in vacuum, $R_{\mu\nu}=0$, so that the scalar and Maxwell field propagate on a Kerr-Newman geometry. Backreaction on the metric appears at order ${\cal O}(\epsilon^2)$ in the fields.
We consider two particular cases and focus on the following background geometries:
%%%%%%%%%%%%%%%%%%%%%%%%%%%%%%%%%%%%%%%%%%%%%%%%%%%%%%%%%%%%%%%%%%%%%%%%%%%%%%%%%%%%%%%%%%%%%%%%%%
\subsubsection{Static, charged backgrounds}
%%%%%%%%%%%%%%%%%%%%%%%%%%%%%%%%%%%%%%%%%%%%%%%%%%%%%%%%%%%%%%%%%%%%%%%%%%%%%%%%%%%%%%%%%%%%%%%%%%
For static backgrounds, the uniqueness theorem~\cite{uniqueness} guarantees that the only regular, asymptotically flat solution necessarily has $\psi=0$ and belongs to the Reissner-Nordstr\"om (RN) family of charged BHs. In the presence of a cosmological constant, $\Lambda\neq0$, other solutions exist, some of them are in fact allowed by superradiant mechanisms, as we shall discuss. For definiteness, we focus for the most part of our work on the fundamental family of RN-(A)dS solution, described by the metric 
\be
ds^2=-fdt^2+fdr^2+r^2d\vartheta^2+r^2\sin^2\vartheta d\varphi^2\,, 
\ee
where
\be
f(r)=1-\frac{2M}{r}+\frac{Q^2}{r^2}-\frac{\Lambda }{3}r^2\,, \label{RNLambda}
\ee
and the background vector potential $A_{\mu}=(Q/r,0,0,0)$, where $M$ and $Q$ are the mass and the charge of the BH, respectively.
When $\Lambda=0$ the spacetime is asymptotically flat and the roots of $f(r)$ determine the event horizon, located at $r_+=M+\sqrt{M^2-Q^2}$, and a Cauchy horizon at $r_-=M-\sqrt{M^2-Q^2}$. 
In this case the electrostatic potential at the horizon is $\Phi_H=Q/r_+$. When $\Lambda>0$, the spacetime is asymptotically de Sitter (dS) and the function $f(r)$ has a further positive root which defines the cosmological horizon $r_c$, whereas when $\Lambda<0$ the spacetime is asymptotically anti-de Sitter (AdS) and the function $f(r)$ has only two positive roots.

Fluctuations of order ${\cal O}(\epsilon)$ in the scalar field in this background induce changes in the spacetime geometry and in the vector potential which are of order ${\cal O}(\epsilon^2)$, and therefore to leading order can be studied on a fixed RN-(A)dS geometry. 
This is done in Section~\ref{sec:superradiance_charged} below.

%%%%%%%%%%%%%%%%%%%%%%%%%%%%%%%%%%%%%%%%%%%%%%%%%%%%%%%%%%%%%%%%%%%%%%%%%%%%%%%%%%%%%%%%%%%%%%%%%%
\subsubsection{Spinning, neutral backgrounds}
%%%%%%%%%%%%%%%%%%%%%%%%%%%%%%%%%%%%%%%%%%%%%%%%%%%%%%%%%%%%%%%%%%%%%%%%%%%%%%%%%%%%%%%%%%%%%%%%%%
For neutral backgrounds $A_{\mu}=0$ to zeroth order, and the uniqueness theorems guarantee that the scalar field is trivial and the only regular,
asymptotically flat solution to the background equations is given by the Kerr family of spinning BHs. Because we also wish to consider the effect of a cosmological constant, we will enlarge it to the Kerr-(A)dS family of spinning BHs,
which in standard Boyer-Lindquist coordinates reads (for details on the Kerr spacetime, we refer the reader to the monograph~\cite{Wiltshire:2009zza})

\begin{eqnarray} 
ds^2=-\frac{\Delta_r}{\rho^2}\left ( dt-\frac{a}{\Sigma}\sin^2\vartheta \,d\varphi\right )^2
+\frac{\rho^2}{\Delta_r}\,dr^2
+\frac{\rho^2}{\Delta_{\vartheta}}\,d\vartheta^2
+\frac{\Delta_\vartheta}{\rho^2} \sin^2\vartheta\left (
a\,dt-\frac{r^2+a^2}{\Sigma} \,d\varphi\right )^2 \,, \label{metricKerrLambda}
\end{eqnarray}
with
\begin{eqnarray}
& & \Delta_r=\left (r^2+a^2\right )\left (1-\frac{\Lambda}{3}{r^2}
\right )-2Mr\,, \qquad  \Sigma=1+\frac{\Lambda}{3}{a^2}\,,\nonumber \\
& &\Delta_{\vartheta}= 1+\frac{\Lambda}{3}{a^2}\cos^2\vartheta\,, \qquad
\rho^2=r^2+a^2 \cos^2\vartheta  \,.  \label{metric parameters}
\end{eqnarray}
This metric describes the gravitational field of a spinning BH with mass $M/\Sigma^2$ and angular
momentum $J=a M/\Sigma^2$. When $\Lambda=0$, the roots of $\Delta$ determine the event horizon, located at $r_+=M+\sqrt{M^2-a^2}$, and a Cauchy horizon at $r_-=M-\sqrt{M^2-a^2}$. The static surface $g_{tt}=0$ defines the ergosphere given by $r_{\rm ergo}=M+\sqrt{M^2-a^2\cos^2\vartheta}$.
As in the static case, when $\Lambda>0$ the spacetime possesses also a cosmological horizon.

A fundamental parameter of a spinning BH is the angular velocity of its event horizon, which for the Kerr-(A)dS solution is given by
\begin{equation}
\Omega_{\rm H}=\frac{a}{r_+^2+a^2}\left ( 1+\frac{\Lambda}{3}{a^2}\right ) \,.  \label{Omega}
\end{equation}
The area and the temperature of the BH event horizon respectively read
\begin{eqnarray}
A_H=\frac{4\pi  (r_+^2+a^2)}{\Sigma} \,,\qquad  T_H=\frac{r_+\left(1-\frac{\Lambda}{3} a^2-{\Lambda}r_+^2 -a^2 r_+^{-2}\right)}{4\pi (r_+^2+a^2)}\,.  \label{area_temperature}
\end{eqnarray}
%%%

%%%%%%%%%%%%%%%%%%%%%%%%%%%%%%%%%%%%%%%%%%%%%%%%%%%%%%%%%%%%%%%%%%%%%%%%%%%%%%%%%%%%%%%%%%%%%%%%%%%%%%%%%%%%%%%
\subsubsection{Geodesics and frame dragging in the Kerr geometry}\label{sec:ZAMO}
%%%%%%%%%%%%%%%%%%%%%%%%%%%%%%%%%%%%%%%%%%%%%%%%%%%%%%%%%%%%%%%%%%%%%%%%%%%%%%%%%%%%%%%%%%%%%%%%%%%%%%%%%%%%%%%
The motion of free pointlike particles in the equatorial plane of this geometry is described by the following geodesic equations~\cite{Bardeen:1972fi,Chandra},
\beq
\dot{t}&=&\frac{1}{\Delta}\left[\left(r^2+a^2+\frac{2a^2M}{r}\right)E-\frac{2aM}{r}\,L\right]\,,\label{geodesics:kerrt}\\
\dot{\varphi}&=&\frac{1}{\Delta}\left[\frac{2aM}{r}\,E+\left(1-\frac{2M}{r}\right)L\right]\,,\label{geodesics:kerrphi}\\
r^2\dot{r}^2&=&r^2E^2+\frac{2M}{r}(aE-L)^2+(a^2E^2-L^2)-\delta_1\Delta\,,\label{geodesics:kerr}
\eeq
where $\delta_1=1,0$ for timelike and null geodesics, respectively, and the dot denotes differentiation with respect to the geodesic's affine parameter. The first two equations follow from the symmetry of the Kerr background under time translations and rotations, while the last equation is simply the defining relation for timelike and null geodesics. A more thorough analysis
of the geodesics of the Kerr geometry can be found in the classic work by Bardeen {\it et al}~\cite{Bardeen:1972fi} or in Chandrasekhar's book~\cite{Chandra}. The conserved quantities $E,\,L$ are, respectively, the energy and angular momentum per unit rest mass of the object undergoing geodesic motion (or the energy and angular momentum for massless particles).

%%%%%%%%%%%%%%%%%%%%%%%%%%%%%%%%%%%%%%%%%%%%%%%%%%%%%%%%%%%%%%%%%%%%%%%%%%%%%%%%%%%%%%%%%%%%%%%%%%%%%%%%%%%%%%%
%\subsubsection{Frame dragging}
%%%%%%%%%%%%%%%%%%%%%%%%%%%%%%%%%%%%%%%%%%%%%%%%%%%%%%%%%%%%%%%%%%%%%%%%%%%%%%%%%%%%%%%%%%%%%%%%%%%%%%%%%%%%%%%
%
\begin{figure}
\centerline{\includegraphics[height=6 cm] {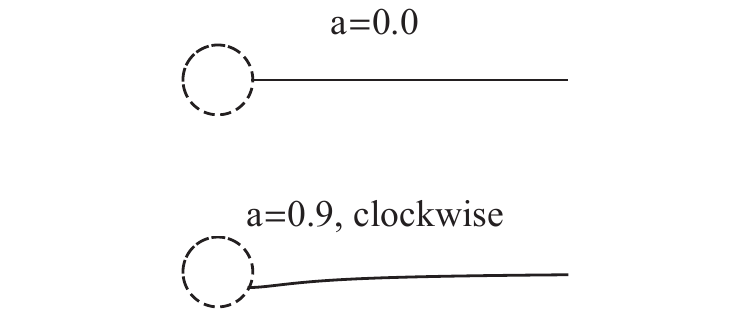}}
\caption{Frame dragging effects: sketch of the trajectory of a zero-angular-momentum observer as it falls into a BH.
The BH is either static (upper panel) or rotating clockwise (lower panel).
The infall into a rotating BH is drag along the BH's sense of rotation.} \label{fig:framedragging}
\end{figure}
Consider an observer with timelike four-velocity which falls into the BH with zero angular momentum. This observer is known as the ZAMO (Zero Angular Momentum Observer). From Eqs.~\eqref{geodesics:kerrt} and \eqref{geodesics:kerrphi} with $L=0$, we get the following angular velocity, as measured at infinity,
\be
\Omega \equiv \frac{\dot{\varphi}}{\dot{t}}=-\frac{g_{t\varphi}}{g_{\varphi\varphi}}=\frac{2Mar}{r^4+r^2a^2+2a^2Mr}\,.
\ee
At infinity $\Omega=0$ consistent with the fact that these are zero angular momentum observers. However, $\Omega\neq0$ at any finite distance and at the horizon one finds 
\be
\Omega_{\rm H}^{\rm ZAMO}=\frac{a}{2Mr_+}\,.
\ee
Thus, observers are frame-dragged and forced to co-rotate with the geometry. This phenomenon is depicted in Fig.\ref{fig:framedragging}, where we sketch the trajectory of a ZAMO in a nonrotating and rotating BH background. 
%%%%%%%%%%%%%%%%%%%%%%%%%%%%%%%%%%%%%%%%%%%%%%%%%%%%%%%%%%%%%%%%%%%%%%%%%%%%%%%%%%%%%%%%%%%%%%%%%%%%%%%%%%%%%%%
\subsubsection{The ergoregion\label{subsub:ergoregion}}
\begin{figure}
\begin{center}
\begin{tabular}{cc}
\epsfig{file=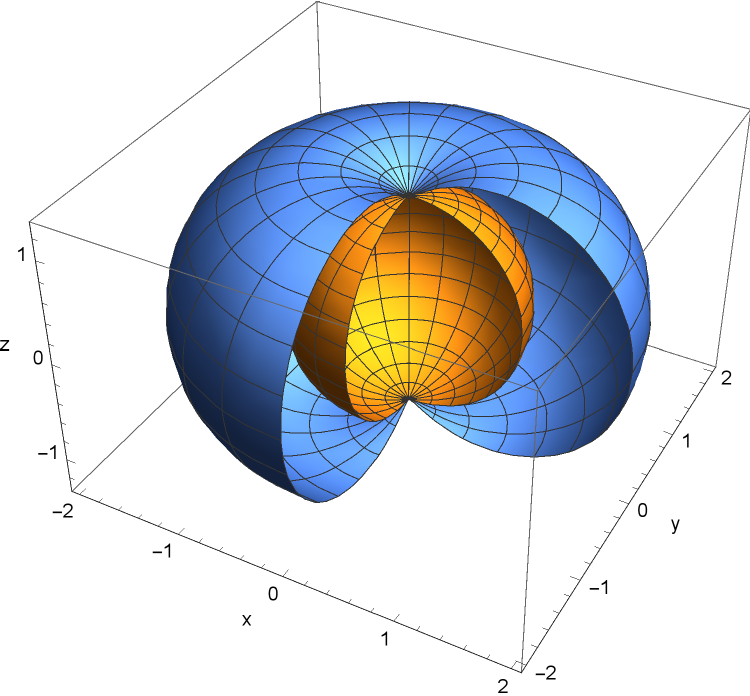,width=0.7\textwidth,angle=0,clip=true}
\end{tabular}
\caption{The ergosphere of a Kerr BH is shown together with the horizon for a nearly-extremal BH with $a\sim 0.999M$. The coordinates $(x,y,z)$ are similar to standard Cartesian-coordinate but obtained from the Boyer-Lindquist coordinates.
\label{fig:ERKerr}}
\end{center}
\end{figure}

%%%%%%%%%%%%%%%%%%%%%%%%%%%%%%%%%%%%%%%%%%%%%%%%%%%%%%%%%%%%%%%%%%%%%%%%%%%%%%%%%%%%%%%%%%%%%%%%%%%%%%%%%%%%%%%
The Kerr geometry is also endowed with an infinite-redshift surface outside the horizon. These points
define the ergosurface and are the roots of $g_{tt}=0$. The ergosurface exterior to the event horizon is located at
\be
r_{\rm ergo}=M+\sqrt{M^2-a^2\cos^2\vartheta}\,.\label{ergoKerr}
\ee
In particular, it is defined by $r=2M$ at the equator and $r=r_+$ at the poles. The region between the event horizon and the ergosurface is the ergoregion.
The ergosurface is an infinite-redshift surface, in the sense that any light ray emitted from the ergosurface will be infinitely redshifted when observed at infinity. The ergosphere of a Kerr BH is shown in Fig.~\ref{fig:ERKerr}.

The ergosurface is also the static limit, as no static observer is allowed inside the ergoregion.
Indeed, the Killing vector $\xi^{\mu}=(1,0,0,0)$ becomes spacelike in the ergoregion $\xi^{\mu}\xi^{\mu}g_{\mu\nu}=g_{tt}>0$.
We define a static observer as an observer (i.e., a timelike curve) with tangent vector proportional to
$\xi^{\mu}$. The coordinates $(r,\vartheta,\phi)$ are constant along this wordline.
Such an observer cannot exist inside the ergoregion, because $\xi^{\mu}$ is spacelike there. In other words, an observer
cannot stay still, but is forced to rotate with the BH.

Let's consider this in more detail, taking a stationary observer at constant $(r,\vartheta)$, with four-velocity
\be
v^{\mu}=(\dot{t},0,0,\dot{\varphi})=\dot{t}(1,0,0,\Omega)\,,
\ee
This observer can exist provided its orbit is time-like, which implies $v^2<0$. This translates in a necessary condition for an existence of a stationary observer, which reads
\be
g_{tt}+2\Omega g_{t\varphi}+\Omega^2g_{\varphi\varphi}<0\,.
\ee
Let's consider the zeroes of the above. We have
\be
\Omega_{\pm}=\frac{-g_{t\varphi}\pm\sqrt{g_{t\varphi}^2-g_{tt}g_{\varphi\varphi}}}{g_{\varphi\varphi}}=\frac{-g_{t\varphi}\pm\sqrt{\Delta}\sin\vartheta}{g_{\varphi\varphi}}\,.
\ee
Thus, a stationary observer cannot exist when $r_-<r<r_+$. In general, the allowed range of $\Omega$ is $\Omega_-\leq\Omega\leq\Omega_+$.
On the outer horizon, we have $\Omega_-=\Omega_+$ and the only possible stationary observer on the horizon has 
\be
\Omega=-\frac{g_{t\varphi}}{g_{\varphi\varphi}}=\Omega_{\rm H}\,,
\ee
which coincides with the angular velocity of a ZAMO at the event horizon. Note also that a static observer is a stationary observer with $\Omega=0$. Indeed, it is easy to check that $\Omega_-$ changes sign at the static limit, i.e. $\Omega=0$ is not allowed within the ergoregion.

%%%%%%%%%%%%%%%%%%%%%%%%%%%%%%%%%%%%%%%%%%%%%%%%%%%%%%%%%%%%%%%%%%%%%%%%%%%%%%%%%%%%%%%%%%%%%%%%%%%%%%%%
\subsubsection{Intermezzo: stationary and axisymmetric black holes have an ergoregion}\label{sec:ERhor}
%%%%%%%%%%%%%%%%%%%%%%%%%%%%%%%%%%%%%%%%%%%%%%%%%%%%%%%%%%%%%%%%%%%%%%%%%%%%%%%%%%%%%%%%%%%%%%%%%%%%%%%%
At this point it is instructive to take one step back and try to understand what are the minimal ingredients for the existence of an ergoregion in a BH spacetime. Indeed, in many applications it would be useful to disentangle the role of the ergoregion from that of the horizon. Unfortunately, this cannot be done because, as we now prove, the existence of an event horizon in a stationary and axisymmetric spacetime automatically implies the existence of an ergoregion~\cite{Cardoso:2012zn}.

Let us consider the most general stationary and axisymmetric metric\footnote{We also require the spacetime to be invariant under the ``circularity condition'', $t\to-t$ and $\varphi\to-\varphi$, which implies $g_{t\vartheta}=g_{t\varphi}=g_{r\vartheta}=g_{r\varphi}=0$~\cite{Chandra}. While the circularity condition follows from Einstein and Maxwell equations in electrovacuum, it might not hold true in modified gravities or for exotic matter fields.}:
\begin{equation}
ds^2=g_{tt} dt^2+g_{rr}dr^2+2g_{t\varphi}dtd\varphi+g_{\varphi\varphi} d\varphi^2+g_{\vartheta\vartheta}d\vartheta^2\,, 
\end{equation}
%%%
where $g_{ij}$ are functions of $r$ and $\vartheta$ only. The event horizon is the locus $r_+=r_+(\vartheta)$ defined as the largest root of the lapse function:
\begin{equation}
N_{r=r_+}\equiv\left( g_{t\varphi}^2-g_{\varphi\varphi}g _{tt}\right)_{r=r_+}=0\,. \label{defhor}
\end{equation}
%%%
In a region outside the horizon $N>0$, whereas $N<0$ inside the horizon. As we discussed, the boundary of the ergoregion, $r_{\rm ergo}=r_{\rm ergo}(\vartheta)$, is defined by $\left. g _{tt}\right|_{r=r_{\rm ergo}}=0$, and $g_{tt}<0$ in a region outside the ergoregion, whereas $g_{tt}>0$ inside the ergoregion.
From Eq.~(\ref{defhor}) we get, at the horizon, 
\begin{equation}
 \left.g_{tt}\right|_{r=r_+}= \left.\frac{g_{t\varphi}^2}{g_{\varphi\varphi}}\right|_{r=r_+}\geq0\,,
\end{equation}
%%%%%
where, in the last inequality, we assumed no closed timelike curves outside the horizon, i.e. $g_{\varphi\varphi}>0$. The inequality is saturated only when the gyromagnetic term vanishes, $\left.g_{t\varphi}\right|_{r=r_+}=0$.
On the other hand, at asymptotic infinity $g_{tt}\to-1$. Therefore, by continuity, there must exist a region $r_{\rm ergo}(\vartheta)$ such that $r_+\leq r_{\rm ergo}<\infty$ and where the function $g_{tt}$ vanishes.
This proves that an ergoregion necessarily exists in the spacetime of a stationary and axisymmetric BH. As a by-product, we showed that the boundaries of the ergoregion (i.e. the ergosphere) must lay outside the horizon or coincide with it, $r_{\rm ergo}\geq r_+$. In the case of a static and spherically symmetric spacetime, $g_{t\varphi}\equiv0$ and the ergosphere coincides with the horizon.

%%%%%%%%%%%%%%%%%%%%%%%%%%%%%%%%%%%%%%%%%%%%%%%%%%%%%%%%%%%%%%%%%%%%%%%
\subsection{Area theorem implies superradiance}\label{areaimpliesSR}
%%%%%%%%%%%%%%%%%%%%%%%%%%%%%%%%%%%%%%%%%%%%%%%%%%%%%%%%%%%%%%%%%%%%%%%
It was realized by Bekenstein that BH superradiance can be naturally understood using the classical laws of BH mechanics~\cite{Bekenstein:1973mi}.
In fact, given these laws, the argument in Section~\ref{sec:rotSR} can be applied {\it ipsis verbis}.
The first law relates the changes in mass $M$, angular momentum $J$, horizon area $A_H$ and charge $Q$, of a stationary BH when it is perturbed. To first order, the variations of these quantities in the vacuum case satisfy
\be
\delta M = \frac{k}{8\pi}\delta A_H + \Omega_{\rm H} \delta J+\Phi_H \delta Q\,,\label{first_law_BH}
\ee
with $k\equiv 2\pi T_H$ the BH surface gravity, $\Omega_{\rm H}$ the angular velocity of the horizon \eqref{Omega} and $\Phi_H$ is the electrostatic potential at the horizon~\cite{Wald:1999vt}.
The first law can be shown to be quite generic, holding for a class of field equations derived from a diffeomorphism covariant Lagrangian with the form ${\cal L}(g_{ab};\,R_{abcd};\,\nabla_{a}R_{bcde},...;\,\psi,\,\nabla_{a}\psi,.......)$.
The second law of BH mechanics states that, {\it if matter obeys the weak energy condition}~\cite{Bekenstein:1973mi,Unruh:1973,Chandra} (see also the discussion in Sec.~\ref{DiracKerr} for a counterexample with fermions), then $\delta A_H\geq0$.
Whether or not the second law can be generalized to
arbitrary theories is an open question, but it seems to hinge on energy conditions~\cite{Wald:1993nt,Iyer:1994ys}.

For the sake of the argument, let us consider a neutral BH, $\Phi=0$.
The ratio of angular momentum flux $L$ to energy $E$ of a wave with frequency $\omega$ and azimuthal number $m$ is $L/E=m/\omega$ (see Appendix~\ref{appendix_energyangularmomentum}). Thus, interaction with the BH causes it to change its angular momentum as
\be\label{delta_J}
\delta J/\delta M=m/\omega\,.
\ee
Substitution in the first law of BH mechanics \eqref{first_law_BH} yields
\be\label{delta_M}
\delta M=\frac{\omega k}{8\pi}\frac{\delta A_H}{\omega-m\Omega_{\rm H}}\,.
\ee
Finally, the second law of BH thermodynamics, $\delta A_H\geq0$, implies that waves with $\omega<m \Omega_{\rm H}$ extract energy from the horizon, $\delta M<0$.

Likewise, the interaction between a static charged BH and a wave with charge $q$ causes a change in the BH charge as
\be\label{delta_Q}
\delta Q/\delta M=q/\omega\,,
\ee
and therefore in this case Eq.~\eqref{delta_M} reads
\be\label{delta_M2}
\delta M=\frac{\omega k}{8\pi}\frac{\delta A_H}{\omega-q\Phi_H}\,.
\ee

This argument holds in GR in various circumstances, but note that it assumes that the wave is initially ingoing at infinity
and that the matter fields obey the weak energy condition.
The latter condition is violated for fermions in asymptotically flat spacetimes (cf. Sec.~\ref{DiracKerr} below), while the former
needs to be carefully analyzed in asymptotically de Sitter spacetimes where a subtlety arises at the cosmological horizon~\cite{Tachizawa:1992ue}. 

These results can be generalized to any test field, possibly charged, propagating on a Kerr-Newman or 
Kerr-Newman-AdS spacetime, with a stress-energy tensor satisfying the null energy condition at the event horizon and 
appropriate boundary conditions at infinity~\cite{Natario:2016bay}. Under those assumptions it was shown that BH 
thermodynamics does not allow to overspin/overcharge an extremal Kerr-Newman BH, nor to violate the weak cosmic 
censorship~\cite{Natario:2016bay}.

%%%%%%%%%%%%%%%%%%%%%%%%%%%%%%%%%%%%%%%%%%%%%%%%%%%%%%%%%%%%%%%%%%%%%%%%%%%%%%%%%%%%%%%%%%%%%%
\subsection{Energy extraction {\it from} black holes: the Penrose process}\label{sec:energyextraction}
%%%%%%%%%%%%%%%%%%%%%%%%%%%%%%%%%%%%%%%%%%%%%%%%%%%%%%%%%%%%%%%%%%%%%%%%%%%%%%%%%%%%%%%%%%%%%%
Despite being classically perfect absorbers, BHs can be used as a ``catalyst'' to extract 
the rest energy of a particle or even as an energy reservoir themselves, if they are spinning or charged.

\begin{figure}
\centerline{\includegraphics[height=0.8\textwidth]{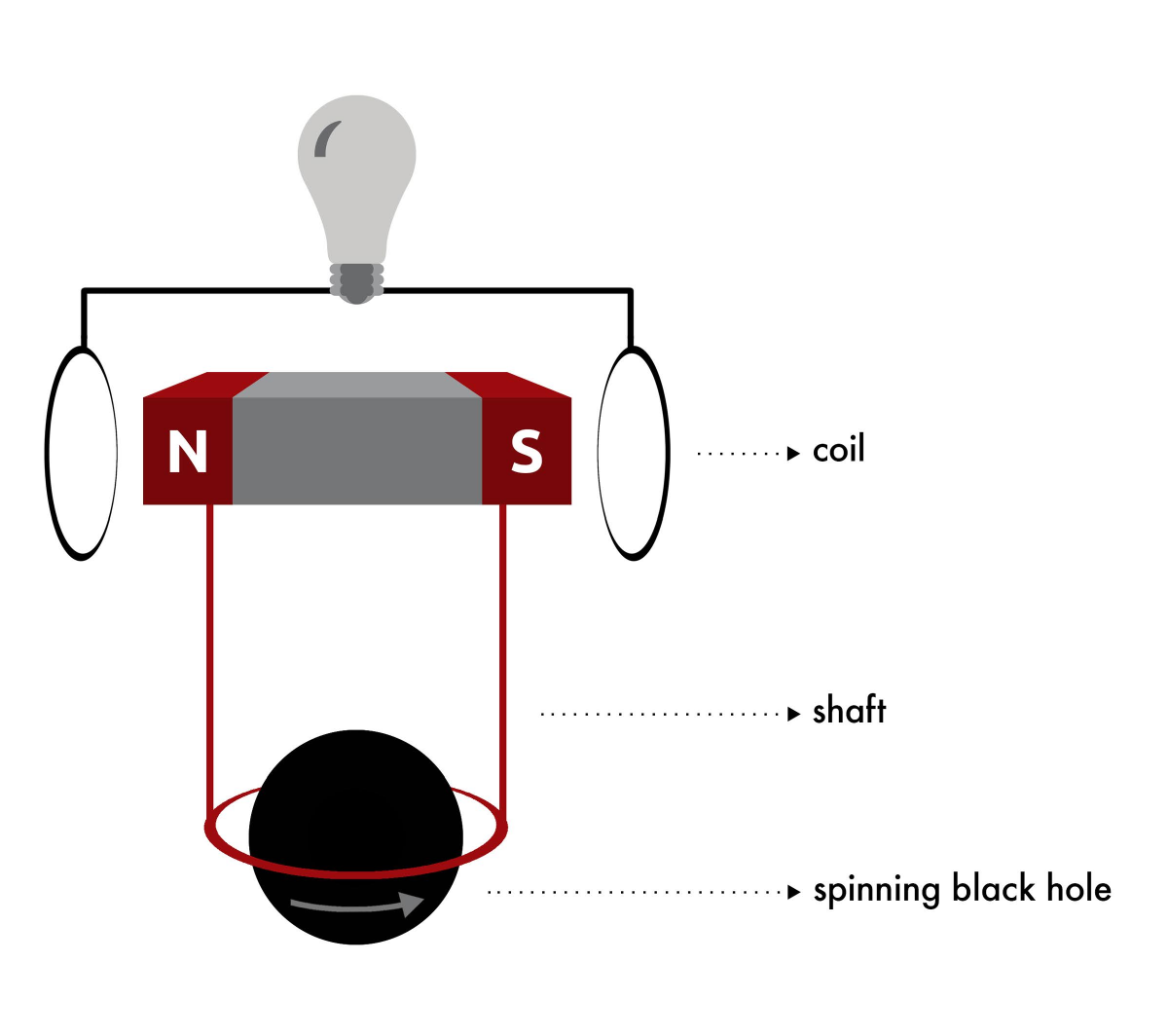}}
\caption{Cartoon of a BH-powered circuit. Two shafts are rigidly attached to a ring, which is inside the ergoregion. The ring and therefore the shafts, are forced to rotate with the BH, turning the magnet at the other end of the shafts end over end, thereby producing a current in any closed circuit. Adapted from a diagram by Dan Watson~\cite{DanWatson}.} \label{fig:generator}
\end{figure}
Classical energy extraction {\it with} BHs works in exactly the same way as in Newtonian mechanics,
by converting into useful work the binding energy of an object orbiting around another. Let's take for simplicity a point particle
of mass $\mu$ around a much more massive body of mass $M$. In Newtonian mechanics, the maximum energy that can be converted in this way
is given by the potential difference between infinity and the surface of the planet, ${\rm Work}/(\mu c^2)=GM/(c^2R)$, where $R$ is the planet's radius. A similar result holds true when the planet is replaced by a BH; for a nonrotating BH, all the object's mass energy can be extracted as useful work as the particle is lowered towards the BH, as the Newtonian calculation suggests!
Notice that in the previous example, what one accomplished was to trade binding energy with useful work, no energy was extracted from the BH itself.

Ways to extract energy from BHs make use of the existence of the ergoregion whose boundary is also a static limit:
all observers are dragged along with the spacetime and cannot remain at rest with respect to distant observers.
The ergoregion is the chief responsible for allowing energy extraction from vacuum, spinning BHs. Just like in 
the spinning cylinder example of Section~\ref{sec:rotSR}, it provides a ``contact surface'' with the impinging object.
A cartoonish application of this property to extract energy is depicted in Fig.~\ref{fig:generator}. 
Quantitative estimates of energy extraction from BHs were first made in a simpler context, which we now discuss.

%%%%%%%%%%%%%%%%%%%%%%%%%%%%%%%%%%%%%%%%%%%%%%%%%%%%%%%%%%%%%%%%%%%%%%%
\subsubsection{The original Penrose process\label{ec:penrose_original}}
%%%%%%%%%%%%%%%%%%%%%%%%%%%%%%%%%%%%%%%%%%%%%%%%%%%%%%%%%%%%%%%%%%%%%%%

%
\begin{figure}
\begin{center}
\begin{tabular}{cc}
\epsfig{file=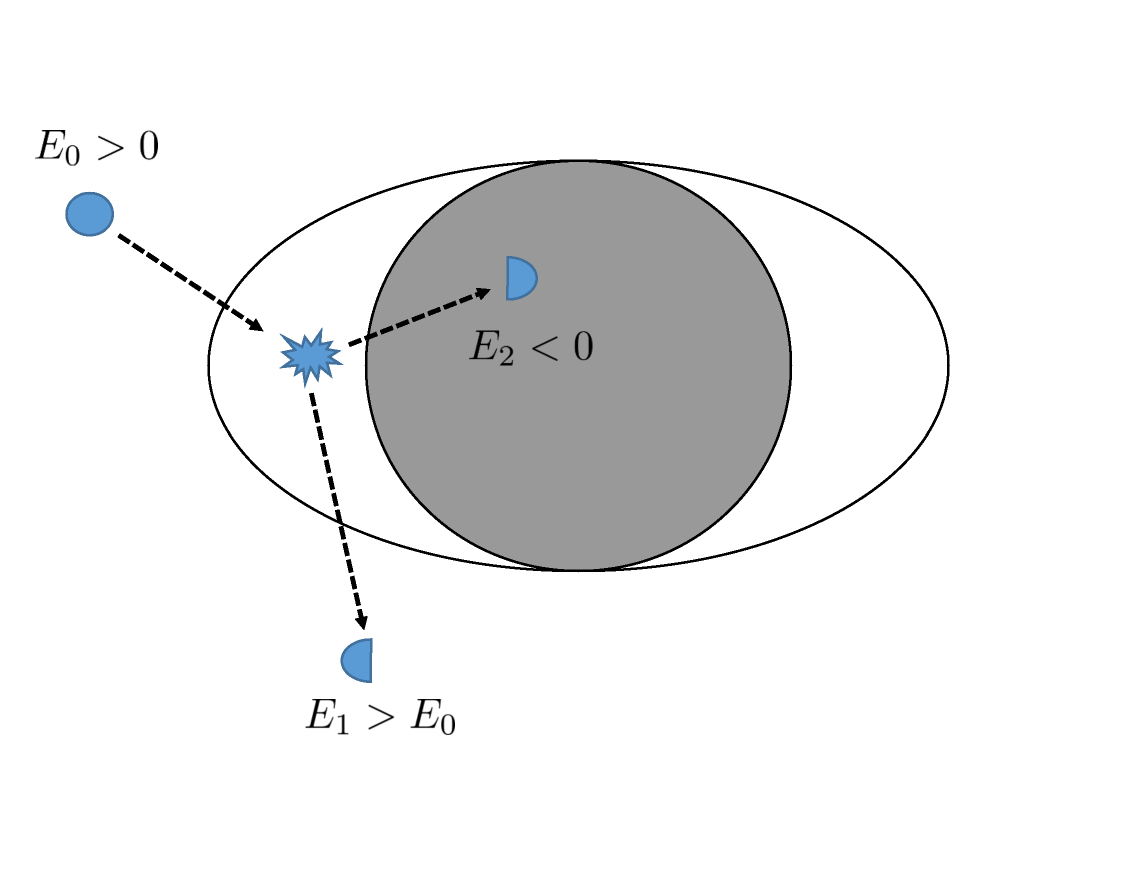,width=0.7\textwidth,angle=0,clip=true}
\end{tabular}
\caption{Pictorial view of the original Penrose processes. A particle with energy $E_0$ decays inside the ergosphere into two particles, one with negative energy $E_2<0$ which falls into the BH, while the second particle escapes to infinity with an energy higher than the original particle, $E_1>E_0$.
\label{penrose_fig1}}
\end{center}
\end{figure}

The possibility to extract energy \emph{from} a spinning BH was first quantified by Roger Penrose~\cite{Penrose:1969} some years before the discovery of BH superradiance, and it is related to the fact that the energy of a particle within the ergoregion, as perceived by an observer at infinity, can be negative. Penrose conceived the following \emph{gedankenexperiment}.
Consider the Kerr geometry~\eqref{metricKerrLambda} with vanishing cosmological constant.
Penrose's thought experiment consists on a particle of rest mass $\mu_{\rm 0}$, at rest at infinity, decaying into two identical particles
each with rest mass $\mu_{\rm fin}$ (Penrose considered these two to be photons, we will keep it generic) at a turning point in its (geodesic) trajectory. Because the particle is initially at rest, the conserved dimensionless energy parameter is $E^{(0)}={\cal E}^{(0)}/\mu_{\rm 0}=1$, and we denote the conserved energy and angular momentum parameters of the two decay-products by $(E^{(1)},L^{(1)})=({\cal E}^{(1)}/\mu_{\rm fin},{\cal L}^{(1)}/\mu_{\rm fin})$ and $(E^{(2)},L^{(2)})=({\cal E}^{(2)}/\mu_{\rm fin},{\cal L}^{(2)}/\mu_{\rm fin})$. Here ${\cal E},\,{\cal L}$ are the physical dimensionful energy and angular momentum of the particles.
From \eqref{geodesics:kerr}, the turning point condition, $\dot{r}(r=r_0)=0$, immediately gives
\beq
L^{(0)}&=&\frac{1}{r_0-2M}\left(-2aM+\sqrt{2Mr_0\Delta}\right)\,,\\
L^{(1), \,(2)}&=&\frac{- 2aM\, E^{(1),\,(2)}\pm\sqrt{r_0\Delta\left(2M+r(\left(E^{(1),\,(2)}\right)^2-1)\right)}}{r_0-2M}\,.
\eeq
Imposing conservation of energy and angular momentum,
\be
{\cal E}^{(1)}+{\cal E}^{(2)}={\cal E}^{(0)}=\mu_{0}\,,\qquad {\cal L}^{(1)}+{\cal L}^{(2)}={\cal L}^{(0)}\,,
\ee
one gets finally,
\be
{\cal E}^{(1)}=\frac{\mu_0}{2}\left(1\pm\sqrt{\frac{2M(1-4\mu_{\rm fin}^2/\mu_{0}^2)}{r_0}}\right)\,,\qquad {\cal E}^{(2)}=\frac{\mu_0}{2}\left(1\mp\sqrt{\frac{2M(1-4\mu_{\rm fin}^2/\mu_{0}^2)}{r_0}}\right)\,.
\ee
It is thus clear that one of the decay products will have an energy larger than the incoming particle. This is schematically shown in Fig.~\ref{penrose_fig1}.
How much larger, depends on the details of the break-up process and is encoded in the quantity $0<1-4\mu_{\rm fin}^2/\mu_{0}^2<1$.
That is, there will be a gain in energy at infinity provided that the turning point satisfies $r_0<2M(1-4\mu_{\rm fin}^2/\mu_{0}^2)<2M$ or, in other words, provided that the decay takes place between the ergosurface and the event horizon.

The maximum gain of energy is obtained when the decay takes place at the horizon and reads
\be
\eta_{\rm max}=\frac{{\cal E}^{(1)}}{{\cal E}^{(0)}}=\frac{1}{2}\left(\sqrt{\frac{2M(1-4\mu_{\rm fin}^2/\mu_{0}^2)}{r_+}}+1\right)\,.
\ee
As we noted, the efficiency depends on the details of the process. The maximum efficiency occurs for conversion into photons, such that $\mu_{\rm fin}^2/\mu_{0}^2=0$, and for which we recover Penrose's result $2{\cal E}^{(1)}/{\cal E}^{(0)}=\left(\sqrt{2M/r_+}+1\right)$.

In this latter case, it is possible to show that the negative-energy photon is doomed to fall into the horizon~\cite{Contopoulos1984}, decreasing the BH mass and angular momentum by $\delta E$ and $\delta L$ but in such a way that the irreducible mass, $M_{\rm irr}=\sqrt{Mr_+/2}$, actually increases~\cite{Chandra}. Furthermore, a generic condition on the energy and angular momentum of the infalling particle can be computed as follows. In the ZAMO frame the energy flux across the horizon is given by 
\be\label{energy_hole_zamo}
\delta E_{H}= -\int_{r_+} d\Sigma_\mu T^{\mu}_{\nu}n^{\nu} \propto \delta E-\Omega_{\rm H} \delta L\,,
\ee
where $T_{\mu\nu}$ is a generic stress-energy tensor of the matter/radiation crossing the horizon, $n^{\mu}=\xi^{\mu}_{(t)}+\Omega_{\rm H}\xi^{\mu}_{(\varphi)}$, $\xi^{\mu}_{(t)}\equiv \partial^{\mu}t$ is the time Killing vector, $\xi^{\mu}_{(\varphi)}\equiv \partial^{\mu}\varphi$ is the axial Killing vector (see Appendix~\ref{appendix_energyangularmomentum}),  while $E$ and $L$ are the (conserved) specific energy and angular momentum of the particle crossing the horizon. Since the locally measured energy must be positive,  assuming $\delta E$ and $\delta L$ are small, it follows that  
\be\label{penrose_condition}
E_{H}\propto E-\Omega_{\rm H} L>0\, \implies \, \Omega_{\rm H} L<E\,.
\ee 
The result above applies to any form of energy and angular momentum crossing the horizon and is related to the area theorem. In addition, if the infalling particle has a negative energy, the bound above implies that $J<0$, i.e. the negative-energy particle must be counter-rotating. The original process conceived by Penrose makes use of pointlike particles, but energy extraction is also possible with extended objects, namely rigidly rotating string around a Kerr BH~\cite{Kinoshita:2016lqd}.
%%%%%%%%%%%%%%%%%%%%%%%%%%%%%%%%%%%%%%%%%%%%%%%
\subsubsection{The Newtonian carousel analogy}
%%%%%%%%%%%%%%%%%%%%%%%%%%%%%%%%%%%%%%%%%%%%%%%
%
\begin{figure}
\begin{center}
\begin{tabular}{cc}
\epsfig{file=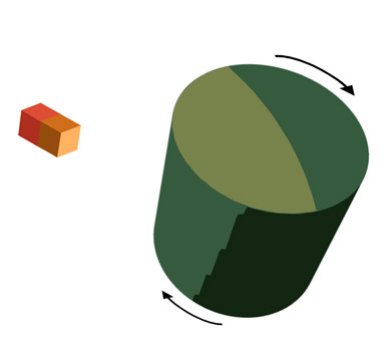,width=0.48\textwidth,angle=0,clip=true}&
\epsfig{file=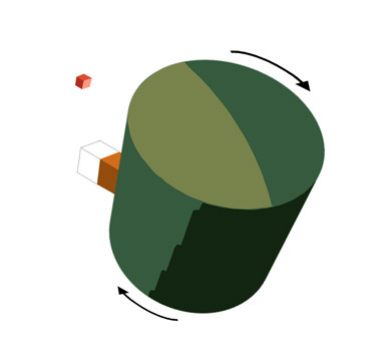,width=0.48\textwidth,angle=0,clip=true}
\end{tabular}
\caption{The carousel analogy of the Penrose process. A body falls nearly from rest into a rotating cylinder, whose surface is sprayed with glue. At the surface the body is forced to co-rotate with the cylinder (analog therefore of the BH ergosphere, the surface beyond which no observer can remain stationary with respect to infinity). The negative energy states of the ergoregion are played by the potential energy associated with the sticky surface. If now half the object (in reddish) is detached from the first half (yellowish), it will reach infinity with more (kinetic) energy than it had initially, extracting rotational energy out of the system. 
\label{fig:carrousel}}
\end{center}
\end{figure}
A simple Newtonian, non-relativistic analog of the Penrose process is the ``carousel process'' depicted in Fig.~\ref{fig:carrousel}. The process consists of two steps. In the first step a point particle collides with a rotating thin cylinder with a ``sticky'' surface and angular velocity $\Omega_i$. The calculations will be done in the inertial frame centered at the cylinder's original axis. For simplicity, we assume the collision to be completely inelastic and we take the particle's mass $m$ to be much smaller than the mass $M$ of the cylinder so that, to first approximation, after the collision the particle is absorbed by the cylinder without changing its shape. Furthermore, consider the particle to have a velocity $v_{\rm in}$ perpendicular to the axis of rotation of the cylinder and with a zero impact parameter. Because of the sticky surface, after the collision the particle is forced to co-rotate with the cylinder. In the second step a fraction $\epsilon$ of the initial mass is ejected from the surface of the cylinder. We want to understand under which conditions the ejected particle has an energy larger than the initial one.

Conservation of angular and linear momenta implies that, after the collision with the cylinder, the linear and angular velocities of the cylinder respectively are
%%%
\begin{equation}
 v_f=\frac{m}{M+m}v_{\rm in}\,,\qquad \Omega_f=\frac{M}{M+m}\Omega_i\,.
\end{equation}
%%%
Because in this example the impact parameter vanishes, the particle has zero angular momentum and the angular velocity of the cylinder decreases. After the collision, the particle is stuck to the surface. Let a fraction $\epsilon$ of the initial mass be ejected at the radial direction forming an angle $\vartheta$ with the initial direction of the particle (and in the same direction of the angular velocity of the disk). Then, the components of the particle's velocity in the collision plane, $v_{\rm out}=(v_x,v_y)$, read
%%%
\begin{equation}
 v_x=-\Omega_f R \cos\vartheta\,,\qquad v_y=v_f+\Omega_f R\sin\vartheta\,,
\end{equation}
%%%
where $R$ is the radius of the cylinder. Finally, we can compare the final energy of the ejected particle, $K_{\rm out}=\epsilon m v_{\rm out}^2/2$, with the initial energy $K_{\rm in}=m v_{\rm in}^2/2$. To first order in the mass ratio $m/M$, the efficiency reads
%%%
\begin{equation}
 \eta\equiv\frac{K_{\rm out}}{K_{\rm in}}=1+\left(\epsilon\frac{R^2\Omega_i^2}{v_{\rm in}^2}-1\right)+2\epsilon \frac{ R \Omega_i}{v_{\rm in}}\left(\sin\vartheta-\frac{R\Omega_i}{v_{\rm in}}\right)\frac{m}{M}+{\cal O}\left[\left(\frac{m}{M}\right)^2\right]\,.
\end{equation}
%%%
Interestingly, the efficiency does not depend on the angle $\vartheta$ to lowest order in the mass ratio. When $m\ll M$ the energy of the ejected particle is larger than the initial kinetic energy provided
\begin{equation}
 \Omega_i>\frac{v_{\rm in}}{\sqrt{\epsilon} R}\,.
\end{equation}

Thus, the rotating ``sticky'' surface plays the same role as the BH ergosphere. The perfectly inelastic collision is the analog of the frame-dragging effect according to which no observer within the ergoregion can remain stationary with respect to infinity. The negative energy states of the ergoregion are played by the potential energy associated with the sticky surface.

%%%%%%%%%%%%%%%%%%%%%%%%%%%%%%%%%%%%%%%%%%%%%%%%
\subsubsection{Penrose's process: energy limits}
%%%%%%%%%%%%%%%%%%%%%%%%%%%%%%%%%%%%%%%%%%%%%%%%

We have seen already that the energy gain provided by the Penrose mechanism is modest, at least for equal-rest-mass fragments.
Still open however, is the possibility that the efficiency, or that the Lorentz factor of one of the fragments, is large
for some situations\footnote{This possibility was at some stage considered of potential interest for the physics of jets emitted by quasars.}.
Strong limits on the energy that can be extracted from the Penrose process can be obtained~\cite{Bardeen:1972fi,Wald:1974kya}:
consider a particle with four-velocity $U^{\mu}$ and conserved energy parameter $E$ that breaks up and emits a fragment with energy $E'$ and four-velocity $u^{\mu}$. We want to impose limits on $E'$, given the three-velocity of the fragment $\vec{v}$ as measured in the rest frame of the incident body.
Suppose that the breakup occurs in a spacetime with a Killing vector $\xi^{\mu}=\partial^{\mu} t$ which is timelike at infinity. In the laboratory frame we define an orthonormal tetrad, $e_{(\alpha)}^{\mu}$, where $e_{(0)}^{\mu}=U^{\mu}$. The four-velocity of the fragment in the locally flat space is given by
\be
u^{(\alpha)}=\frac{dx^{(\alpha)}}{d\tau}=\gamma \frac{dx^{(\alpha)}}{dx^{(0)}}\,,
\ee
where $\gamma=dx^{(0)}/d\tau=\left(1-v^2\right)^{-1/2}$ and $v^2=v^{(i)}v_{(i)}$. In the frame defined by $e_{(\alpha)}^{\mu}$ we can write $u^{\mu}=e_{(\alpha)}^{\mu}u^{(\alpha)}=\gamma(U^{\mu}+v^{(i)} e_{(i)}^{\mu})$ and $\xi^{\mu}=\xi^{(0)}U^{\mu}+\xi^{(i)}e_{(i)}^{\mu}$ (with $i=1,2,3$).
We then have
\be\label{energy_wald}
E=-\xi_{\mu}U^{\mu}=-\xi_{(0)}=-\xi^{\mu}U_{\mu}=-\xi^{(0)}\,, \quad g_{tt}=\xi^{\mu}\xi_{\mu}=-E^2+\xi^2\,,
\ee
where $\xi^2=\xi^{(i)}\xi_{(i)}$. The energy of the ejected particle reads
\be
E'=-\xi_{\mu}u^{\mu}=\gamma\left(E+v^{(i)} \xi_{(i)}\right)=\gamma\left(E+v \xi\cos\vartheta\right)\,,
\ee
where $\vartheta$ is the angle between the fragment velocity $v^{(i)}$ and $\xi_{(i)}$.
Using~\eqref{energy_wald} we can write
\be
E'=\gamma E+\gamma v\left(E^2+g_{tt}\right)^{1/2}\cos\vartheta,
\ee
which implies the inequality
\be
\gamma E-\gamma v\left(E^2+g_{tt}\right)^{1/2} \leq E' \leq \gamma E+\gamma v\left(E^2+g_{tt}\right)^{1/2}\,. \label{Waldineq}
\ee
In the Kerr metric~\eqref{metricKerrLambda}, $g_{tt}$ is always less than 1 outside the horizon;
furthermore, realistic configurations of matter outside BHs are likely to be well approximated with circular geodesics, for which the maximum possible energy is $E=1/\sqrt{3}$~\cite{Bardeen:1972fi}. Thus, for $E'$ to be negative, or equivalently, for the Penrose process to be {\it possible}, it is necessary that
\be
v>\frac{E}{\sqrt{E^2+1}}=\frac{1}{2}\,.
\ee
This means that the disintegration process must convert most of the rest mass energy of the initial body into kinetic energy for any extraction of energy to become possible. In other words, the breakup process itself is relativistic. Such conclusion might be avoided if one is willing to accept the existence of naked singularities or wormholes, where $g_{tt}$ can in principle become very large.

It is interesting to note that the inequality~\eqref{Waldineq} applies also in flat space, where $g_{tt}=-1$. In this case the bound reads
\be
\gamma E-\gamma v\left(E^2-1\right)^{1/2} \leq E' \leq \gamma E+\gamma v\left(E^2-1\right)^{1/2}\,.
\ee
We conclude that (i) there is no great gain compared to what could be achieved from a breakup process in flat space and (ii)
the left-hand-side can never become negative, as expected. 
 
Bardeen \emph{et al.} also showed that similar limits can be derived by following two particles which collide at some point inside the ergoregion~\cite{Bardeen:1972fi}. Following similar steps to the ones we discussed above, they computed a lower bound on the magnitude of the relative three-velocity $w$ between them, obtaining $w\geq 1/2$, in agreement with Ref.~\cite{Wald:1974kya}. This leads to the conclusion that for the Penrose process to be possible, the particles must first acquire relativistic energies through some other mechanism.

In its simplest incarnation, energy extraction from spinning BHs in vacuum is not efficient enough to explain highly-energetic phenomena such as the emission of relativistic jets from quasars. However, in the presence of magnetic fields the limits discussed above can be lowered significantly for charged particles~\cite{1985JApA....6...85B,Wagh:1986tsa,1986ApJ...301.1018W}, or as we discuss in Section~\ref{super_penrose}, the situation can change completely by considering a variant of the Penrose process known as the {\em collisional Penrose process}.

%%%%%%%%%%%%%%%%%%%%%%%%%%%%%%%%%%%%%%%%%%%%%%%%%%%%%%%%%%%%%%%%%%%%%%%%%%%%%%%%%%%
\subsubsection{The Penrose process in generic spacetimes\label{sec:penrosegeneral}}
%%%%%%%%%%%%%%%%%%%%%%%%%%%%%%%%%%%%%%%%%%%%%%%%%%%%%%%%%%%%%%%%%%%%%%%%%%%%%%%%%%%

The overall picture discussed above for the Penrose's extraction of energy from a Kerr BH can be actually generalized to any stationary and axisymmetric spacetime with an ergoregion. Consider a massive particle with specific energy $E^{(0)}$ at infinity, falling along the equatorial plane and finally decaying into two photons within the ergoregion. In such circumstances, one photon can have negative energy, $E^{(1)}<0$, so that by energy conservation the second photon must have $E^{(2)}>E^{(0)}$. In the case of a Kerr BH the negative-energy photon is forced to fall into the horizon~\cite{Contopoulos1984}, whereas the other can escape to infinity with an energy excess compensated by the BH angular momentum. In fact, as shown by Chandrasekhar~\cite{Chandra}, the process can be also understood in terms of the BH area theorem, i.e. energy extraction is related to the property that the surface area of a BH never decreases in a continuous process.

Let us start by repeating the essentials of the Penrose process in a generic stationary, axisymmetric spacetime. Focusing on equatorial motion, the line element~\eqref{genericspinningmetric} can be simplified as 
\begin{equation}
 ds^2=g_{tt}(r)dt^2+g_{rr}(r)dr^2+g_{\varphi\varphi}(r) d\varphi^2+2g_{t\varphi}(r)dtd\varphi\,,
\end{equation}
where all metric coefficients are evaluated at $\vartheta=\pi/2$. Generalizing the geodesics analysis presented in Sec.~\ref{ec:penrose_original}, it is easy to show that a massive particle in this spacetime has a negative energy if and only if it is counter-rotating (i.e. its angular momentum along the rotation axis is negative, $L<0$) and 
\begin{equation}
g_{tt}\left(1+\frac{g_{\varphi\varphi}}{L^2}\right)<\frac{g_{t\varphi}^2}{L^2}\,. \label{ergocondition}
\end{equation}
Because the right-hand side of the equation above is positive and regularity of the spacetime requires $g_{\varphi\varphi}>0$, the condition above implies $g_{tt}>0$, i.e. that the negative-energy particle is confined within the ergoregion. Likewise, for a particle with specific energy $E^{(0)}=1$ decaying into two particles with specific energies $E^{(1)}$ and $E^{(2)}$ at its turning point and rest masses $\mu_{\rm fin}$ each, we obtain
\be
{\cal E}^{(1)}=\frac{\mu_0}{2}\left(1\pm\sqrt{(1+g_{tt})(1-4\mu_{\rm fin}^2/\mu_{0}^2)}\right)\,,\qquad {\cal E}^{(2)}=\frac{\mu_0}{2}\left(1\mp\sqrt{(1+g_{tt})(1-4\mu_{\rm fin}^2/\mu_{0}^2)}\right)\,.
\ee
The efficiency reads
\begin{equation}
\eta=\frac{{\cal E}^{(1)}}{{\cal E}^{(0)}}=\frac{1}{2}\left[\sqrt{(1+g_{tt})(1-4\mu_{\rm fin}^2/\mu_{0}^2)}+1\right]\,, 
\label{efficiencyPenrose}
\end{equation}
and is limited by the maximum value of $|g_{tt}|$. The latter must be finite to ensure regularity of the geometry\footnote{Interestingly, in the case of a naked singularity large-curvature regions become accessible to outside observers and $g_{tt}$ can be arbitrarily large. This suggests that the Penrose effects around spinning naked singularities can be very efficient. It is also possible that rotating wormholes are prone to
efficient Penrose-like processes, although to the best of our knowledge a detailed investigation has not been performed.} and this limits the efficiency of Penrose's process, in addition to the bounds discussed above for the case of a Kerr BH. Crucially, this derivation does \emph{not} assume the existence of an event horizon and is valid for any stationary and axisymmetric spacetime. This is analogous to what happens with superradiance~\cite{Vicente:2018mxl}.

While in the case of a Kerr BH the negative-energy particle is doomed to fall into the BH~\cite{Contopoulos1984}, if the spacetime does not possess an event horizon Eq.~\eqref{ergocondition} requires that the negative-energy particle be confined within the ergoregion. In this case there are two possibilities: (i) the particle does not interact with the rotating object and it remains in orbital motion in the region $g_{tt}>0$, or (ii) the particle is absorbed by the object and transfers its negative energy and angular momentum through other (nongravitational) mechanisms. 
As we will see in Sec.~\ref{sec:ergoregioninstability} the former possibility is related to the so-called ergoregion instability.

We showed that the Penrose mechanism extends trivially to generic axi-symmetric
stationary spacetimes. Specifically, it has been studied for
rotating wormholes~\cite{Teo:1998dp}, BHs in other theories of gravity such as the ``Horava-Lifshitz'' gravity BH~\cite{Abdujabbarov:2011af}, Kerr-NUT BHs~\cite{Abdujabbarov:2011uv}, BHs with a global monopole~\cite{Chen:2013vja}, charged rotating BHs in Einstein-Maxwell axion-dilaton coupled gravity~\cite{Ganguly:2014pwa}, and to arbitrarily ``deformed'' Kerr BHs~\cite{Liu:2012qe}, where it was shown that the maximum energy gain can be several times larger than for a Kerr BH.

The efficiency of the Penrose mechanism was also studied in the context of higher-dimensional physics,
for higher dimensional BHs and black rings~\cite{Nozawa:2005eu}, to the five-dimensional supergravity rotating BH~\cite{Prabhu:2009ju}, 
and even to arbitrarily deformed BHs~\cite{Ghosh:2013ona}. Finally, the astrophysically more relevant Penrose process for a Kerr BH immersed in a magnetic field, was studied in Refs.~\cite{1986ApJ...307...38P,Wagh:1986tsa,1986ApJ...301.1018W,Wagh:1989zqa} where it was shown that the maximum efficiency could be up to ten times larger than in a vacuum Kerr BH.

%%%%%%%%%%%%%%%%%%%%%%%%%%%%%%%%%%%%%%%%%%%%%%%%%%%%%%%%%%%%%%%%%%%%%%%%%%%%%%%%%%%%%%%%%%%%%%%
\subsubsection{The collisional Penrose process: ultra-high-energy debris}\label{super_penrose}
%%%%%%%%%%%%%%%%%%%%%%%%%%%%%%%%%%%%%%%%%%%%%%%%%%%%%%%%%%%%%%%%%%%%%%%%%%%%%%%%%%%%%%%%%%%%%%%

%
\begin{figure*}[hbt]
\begin{center}
\begin{tabular}{cc}
\epsfig{file=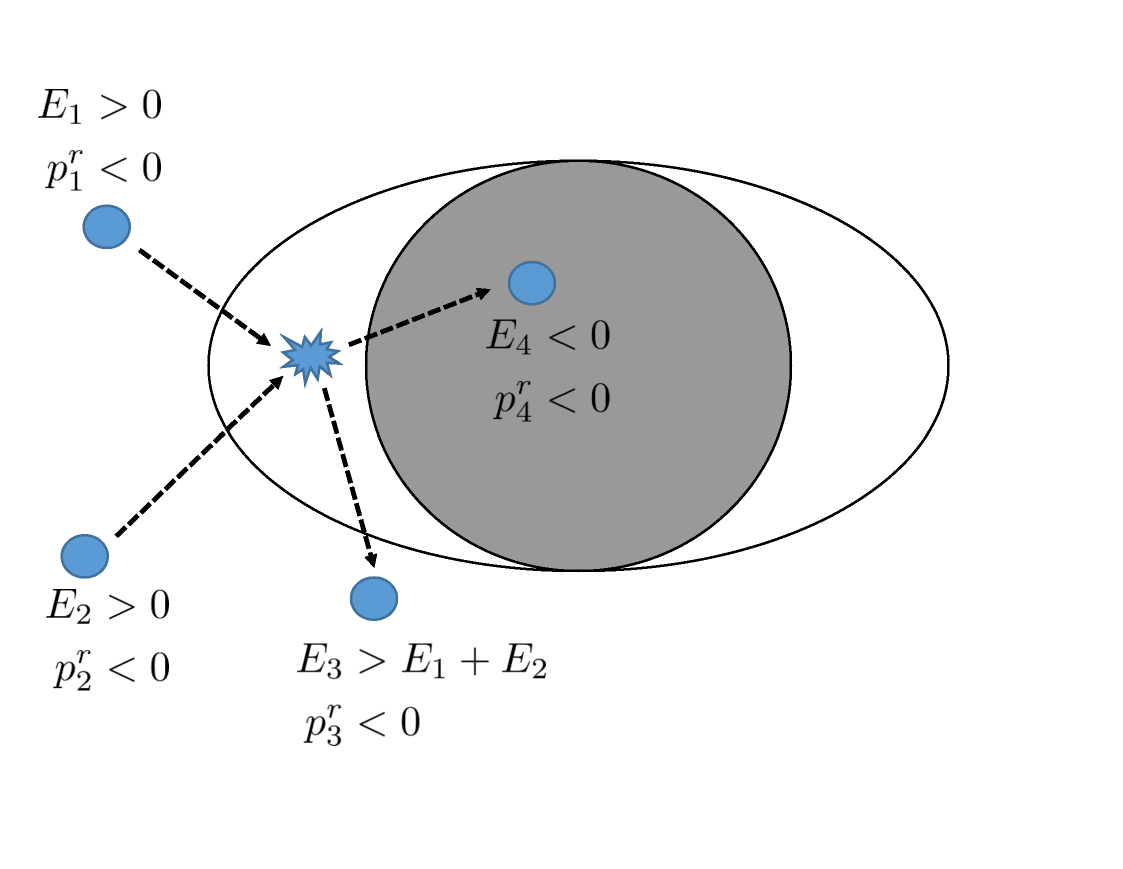,width=0.48\textwidth,angle=0,clip=true}&
\epsfig{file=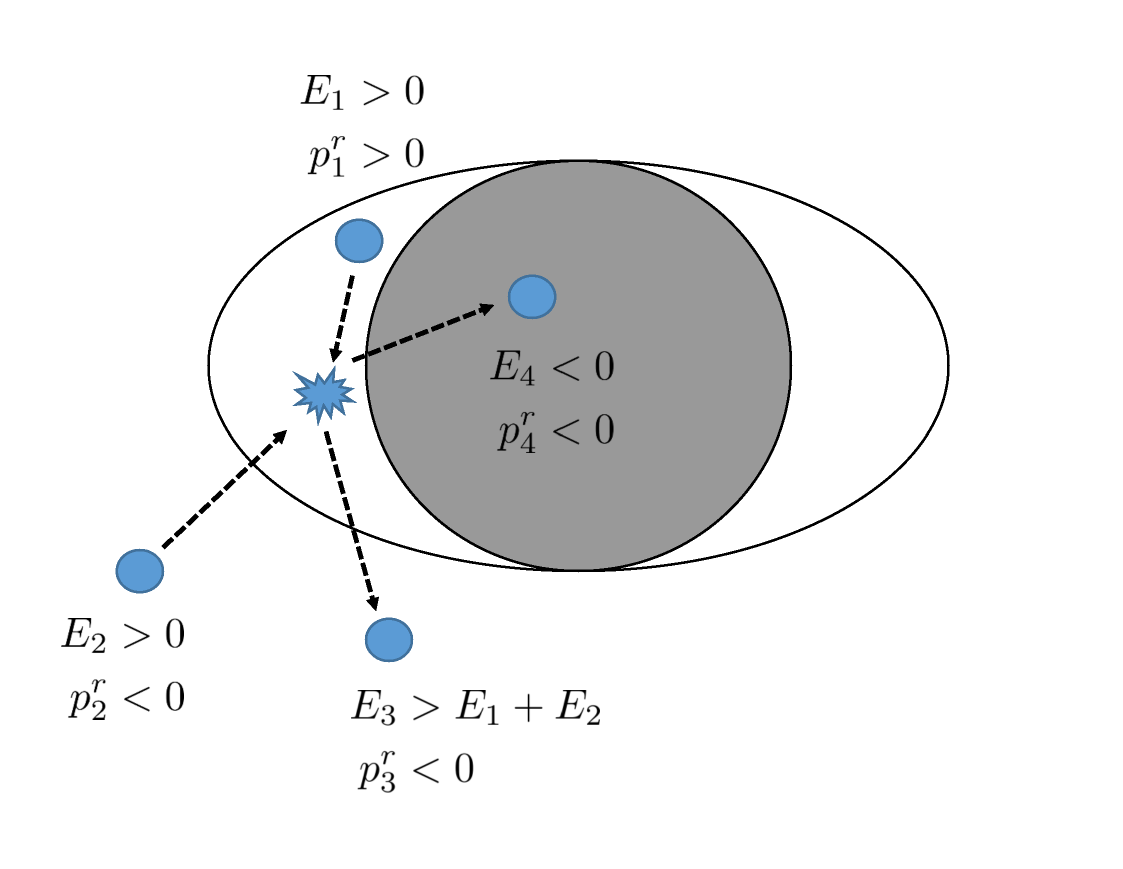,width=0.48\textwidth,angle=0,clip=true}
\end{tabular}
\caption{Pictorial view of the different collisional Penrose processes. Left: initial particles with ingoing radial momentum ($p_1^r<0$ and $p_2^r<0$). Particle 3 has initial ingoing radial momentum, but eventually finds a turning point and escapes to infinity. The maximum efficiency for this was shown to be quite modest $\eta\sim 1.5$~\cite{Piran:1977dm,Harada:2012ap,Bejger:2012yb,Zaslavskii:2012yp}. Right: initial particles with $p_1^r>0$ and $p_2^r<0$. In this case particle 1 must have $p_1^r>0$ inside the ergosphere. For this process the efficiency can be unbound for extremal BHs~\cite{Schnittman:2014zsa,Berti:2014lva}.\label{penrose_fig2}}
\end{center}
\end{figure*}

\begin{figure*}[hbt]
\begin{center}
\begin{tabular}{cc}
\epsfig{file=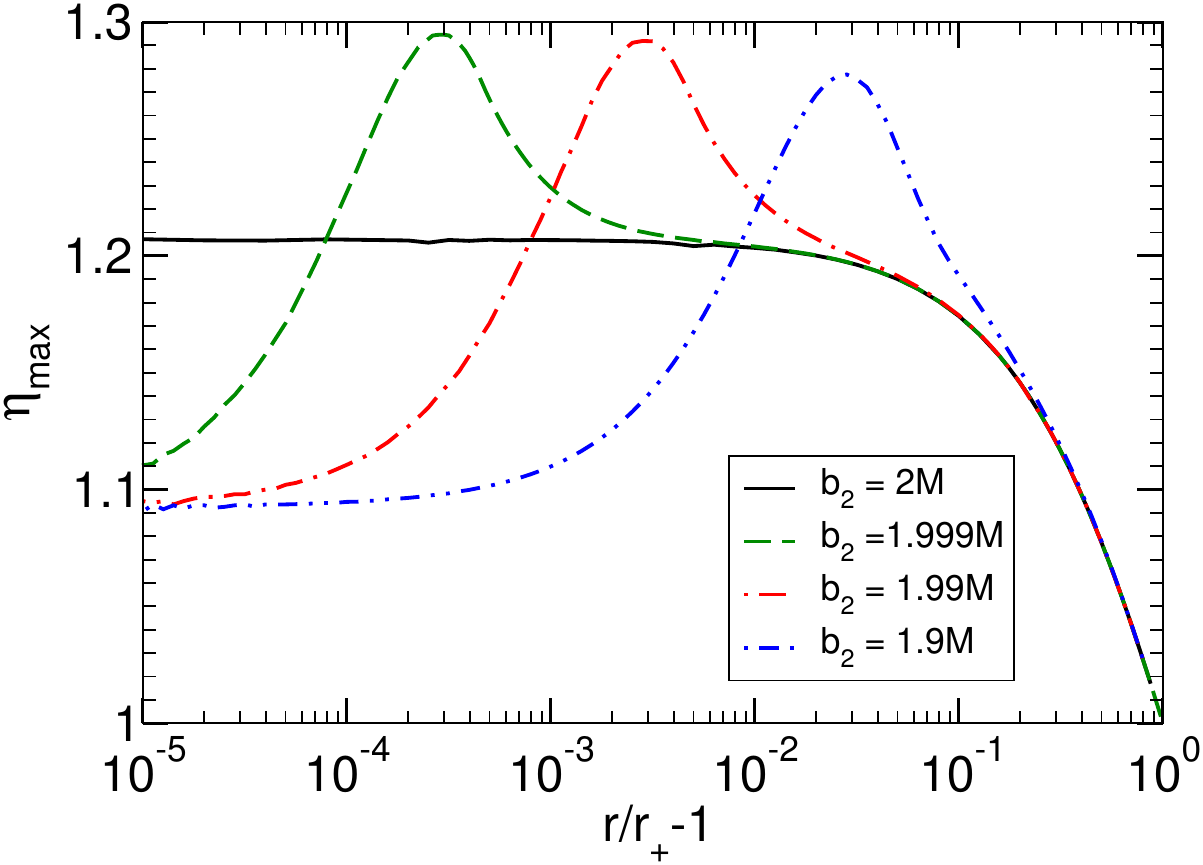,width=0.48\textwidth,angle=0,clip=true}&
\epsfig{file=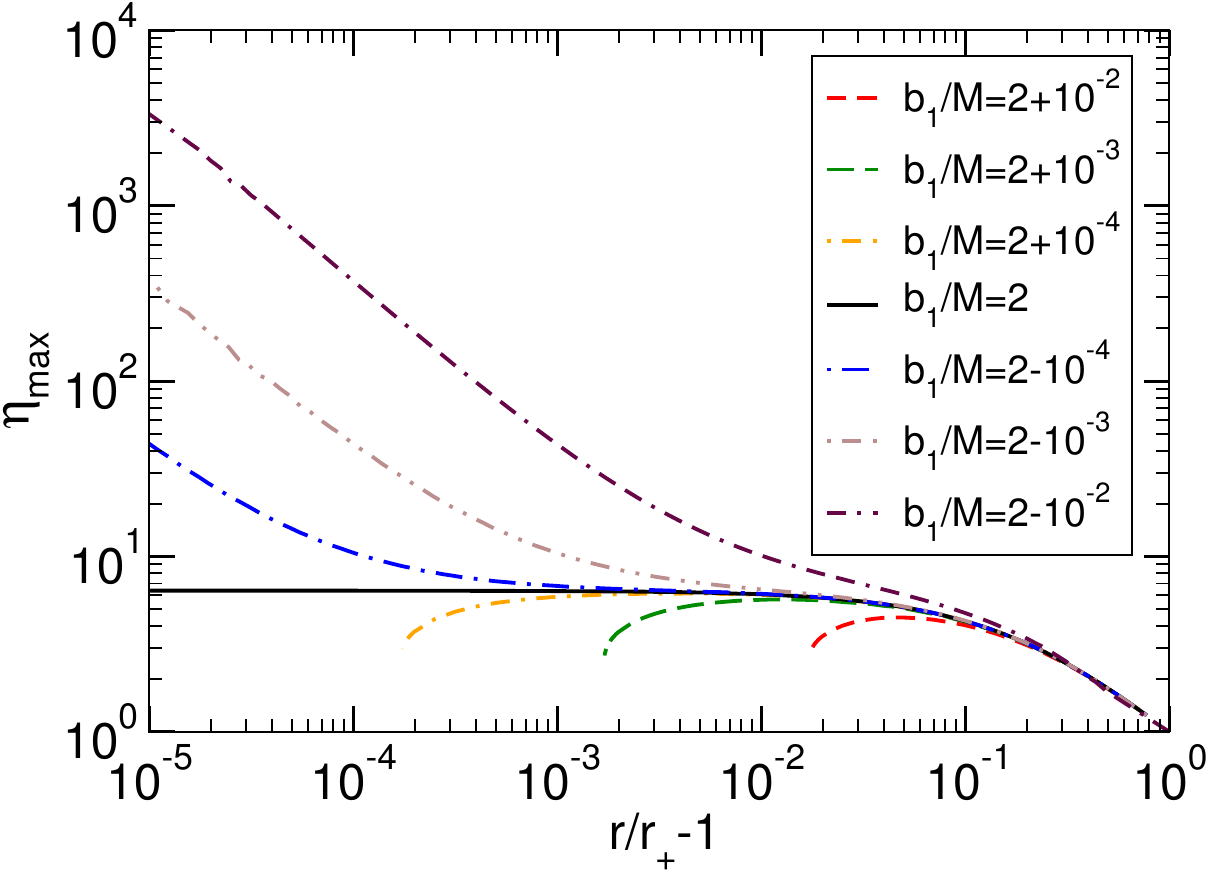,width=0.48\textwidth,angle=0,clip=true}
\end{tabular}
\caption{\label{fig:penrose_schnittman} Left: Maximum efficiency $\eta_{\rm
    max}$ for the collision of equal-energy particles as a
  function of the radius at which the reaction occurs, for
  $p_1^r<0$, $p_2^r<0$, $L_1/E_1\equiv b_1=2M$, and an extremal BH ($a=M$). The case $b_1=b_2=2M$ corresponds to the decay of a single particle into two photons discussed in Sec.~\ref{ec:penrose_original}. The maximum efficiency for this case is $\eta_{\rm max}\sim 1.3$, as shown in Ref.~\cite{Bejger:2012yb}.
	Right: Same, but for $p_1^r>0$, $p_2^r<0$ and $b_2=-2(1+\sqrt{2})M$. The curves for $b_1>2M$ terminate at the turning point
  of particle 1. The process considered in Ref.~\cite{Schnittman:2014zsa} corresponds to the case $b_1\geq 2M$, while Ref.~\cite{Berti:2014lva} extended these results to the case $b_1<2M$. From~\cite{Berti:2014lva}.}
\end{center}
\end{figure*}

A variant of the Penrose process which might be astrophysically more promising is the {\em collisional Penrose process}, first proposed in 1975~\cite{1975ApJ...196L.107P} and studied in detail in Ref.~\cite{Piran:1977dm}. The process consists of two particles 1 and 2 colliding with four-momenta $p^{\mu}_1$ and $p^{\mu}_2$ at some Boyer-Lindquist coordinate position $r$, and resulting in the emission of two bodies 3 and 4 with four-momenta $p^{\mu}_3$ and $p^{\mu}_4$. This process was mostly studied in the equatorial plane where the geodesic equations are given by Eqs.~\eqref{geodesics:kerrt}--\eqref{geodesics:kerr}. In the local `lab' reference frame, the four-momentum is $p^{\mu}=\dot{x}^{\mu}$ for massless particles, while for massive particles we can choose the geodesic's affine parameter to be $\tau/\mu$ ($\tau$ being the proper time and $\mu$ the particle rest mass), so that $p_{\mu}p^{\mu}=-\mu^2$. Using \eqref{geodesics:kerrt}--\eqref{geodesics:kerr} and imposing the local conservation of four-momentum
\be
p^{\mu}_1+p^{\mu}_2=p^{\mu}_3+p^{\mu}_4\,,
\ee
it is possible to numerically compute the ratio $\eta$ 
between the energy of the post-collision escaping particle 3 and the energy of the colliding particles, $\eta\equiv {\cal E}_3/({\cal E}_1+{\cal E}_2)$. Imposing that the initial particles have {\it ingoing} radial momentum ($p_1^r<0$ and $p_2^r<0$) and that particle 3 can escape and reach an observer at infinity, it was shown that the process would result in modest maximum efficiencies ($\eta\lesssim 1.5$) for the escaping particle, where the precise upper bound depends on the nature of the colliding particles~\cite{Piran:1977dm,Harada:2012ap,Bejger:2012yb,Zaslavskii:2012yp}. However, recently, Schnittman~\cite{Schnittman:2014zsa} found the surprising result that one could achieve much higher energy gains ($\eta \lesssim 15$) by allowing one of the colliding particles (say, particle 1) to rebound at a turning point, so it has {\it outgoing} radial momentum ($p_1^r>0$) when it collides with the incoming particle 2. This outgoing momentum favors ejection of a high-energy particle after the collision. A schematic view of the two processes is shown in Fig.~\ref{penrose_fig2}. This was further extended in Ref.~\cite{Berti:2014lva}, with the striking conclusion that particle collisions in the vicinity of rapidly rotating BHs could, in principle, reach arbitrarily high efficiencies. They allowed for one of the particles to have outgoing radial momentum but with angular momentum $L_1<2E_1 M$, such that this particle cannot come from infinity but is still kinematically allowed to be created inside the ergosphere by previous scattering events (however see Ref.~\cite{Leiderschneider:2015ika} for a particular case where there is no energy amplification taking into account multiple scattering). These results are summarized in Fig.~\ref{fig:penrose_schnittman}.

In principle multiple scattering events can also be used to increase the efficiency of any possible collisional Penrose process. The energy of particles that cannot escape to infinity may be substantially larger than the energy of those that can, and even if these particles are unable to escape themselves, they may collide with other particles and give rise to high-energy collision products that may escape and be detected at infinity. This may lead to very large efficiencies, even away from $a=M$~\cite{Schnittman:2014zsa,Berti:2014lva}, which can even increase further when the particles involved in the collision are spinning~\cite{Maeda:2018hfi}. However, whether these processes play a role in the production of observable gamma rays or ultra-high-energy cosmic rays is still an open problem.

%

%%%%%%%%%%%%%%%%%%%%%%%%%%%%%%%%%%%%%%%%%%%%%%%%%%%%%%%%%%%%%%%
\subsection{The ABC of black hole superradiance}\label{sec:ABC}
%%%%%%%%%%%%%%%%%%%%%%%%%%%%%%%%%%%%%%%%%%%%%%%%%%%%%%%%%%%%%%%
In this section we introduce the theory of superradiant scattering of test fields on a BH background. 
Fluctuations of order ${\cal O}(\epsilon)$ in the scalar or vector field in a given background induce changes in the spacetime geometry of order ${\cal O}(\epsilon^2)$, and therefore to leading order can be studied on a fixed BH geometry. 
Before entering in the details of the problem, it is instructive to consider a model that captures the basic ingredients of superradiant scattering in curved spacetime. For simplicity, we assume asymptotic flatness.

Let us assume that the spacetime is stationary and axisymmetric~\footnote{A special feature of vacuum stationary GR solutions is their \emph{axisymmetry}~\cite{Hawking:1971vc}. This simplifies considerably the treatment of superradiant instabilities in GR, as it excludes mixing between modes with different azimuthal number $m$.}.
As we shall see, in this case various types of 
perturbations propagating on fixed BH metrics can be expressed in terms of a single master variable $\psi$, after a 
Fourier-decomposition and harmonic expansion of the time-domain fields. For example,
the Klein-Gordon equation for a minimally coupled charged scalar field in this curved spacetime was given in Eq.~\eqref{eq:MFEoMScalar}. Using the ansatz 
\be
\Psi(t,r,\vartheta,\varphi)=\int d\omega \sum_{lm} e^{-i\omega t}Y_{lm}(\vartheta,\varphi) \frac{\psi(r)}{r}\,,\label{eq:Fourier}
\ee
the Klein-Gordon equation above can be written in the Schroedinger-like form
\begin{equation}
\frac{d^2 \psi}{dr_*^2}+V_{\rm eff} \psi=0\,,\label{wave} 
\end{equation}
where the potential $V_{\rm eff}(r)$ is model dependent and encodes the curvature of the background and the properties of the test fields. The coordinate $r_*$ maps the region $r\in[r_+,\infty[$ to the entire real axis. Given the symmetries of the background, we consider a scattering experiment of a monochromatic wave with frequency $\omega$ and azimuthal and time dependence $e^{-i\omega t+im\varphi}$. Assuming $V_{\rm eff}$ is constant at the boundaries, Eq.\eqref{wave} has the following asymptotic behavior 
\begin{equation}
 \psi \sim\left\{
\begin{array}{ll}
{\cal T}e^{-i k_H r_*}+{\cal O}e^{i k_H r_*} & {\rm as}\ r\rightarrow r_+ \,, \\
{\cal R}e^{i k_\infty r_*}+ {\cal I}e^{-i k_\infty r_*}& {\rm as}\ r\rightarrow \infty\,.
\end{array}
\right. \label{bound2}
\end{equation}
where $r_+$ is the horizon radius in some chosen coordinates, $k_H^2=V_{\rm eff}(r\to r_+)$ and $k_\infty^2=V_{\rm eff}(r\to \infty)$.
These boundary conditions correspond to an incident wave of amplitude ${\cal I}$ from spatial infinity giving rise to a reflected wave of amplitude ${\cal R}$ and a transmitted wave of amplitude ${\cal T}$ at the horizon. 
The ${\cal O}$ term describes a putative outgoing flux across the surface at $r=r_+$. Although the presence of a horizon and a well-posed Cauchy problem would imply ${\cal O}\equiv0$, here we shall generically keep this term, in order to allow for a nonvanishing outgoing flux in absence of an event horizon.

Let us assume that the potential is real\footnote{As we shall discuss, this condition holds for scalar 
perturbations of spinning and charged BHs, but it does not hold in other cases of interest, for example for 
EM and gravitational perturbations satisfying the Teukolsky equation for a Kerr BH. In the latter cases, 
it is convenient to make a change of variables by introducing the Detweiler's 
function, which can be chosen such that the effective potential is real~\cite{1977RSPSA.352..381D,Maggio:2018ivz}.}. 
Then, since the background is stationary, the field equations are invariant under the transformations $t\to -t$ and 
$\omega\to-\omega$. Thus, there exists another solution $\bar\psi$ to Eq.~(\ref{wave}) which satisfies the complex 
conjugate boundary conditions. 
The solutions $\psi$ and $\bar\psi$ are linearly independent and standard theory of ODEs tells us that their Wronskian is independent of $r_*$. Thus, the Wronskian evaluated near the horizon, $W= -2i k_H\left(|{\cal T}|^2-|{\cal O}|^2\right)$, must equal the one evaluated at infinity, $W=2i k_\infty(|{\cal R}|^2-|{\cal I}|^2)$, so that 
%%%%%
\begin{equation}
 |{\cal R}|^2=|{\cal I}|^2-\frac{k_H}{k_\infty}\left(|{\cal T}|^2-|{\cal O}|^2\right)\,,\label{reflectivity}
\end{equation}
%%%
independently from the details of the potential in the wave equation.

In the case of a one-way membrane boundary conditions at the horizon, i.e. ${\cal O}=0$, one gets $|{\cal R}|^2<|{\cal I}|^2$ when $k_H/k_\infty>0$, as is to be expected for scattering off perfect absorbers. However, for $k_H/k_\infty<0$, the wave is superradiantly amplified, $|{\cal R}|^2>|{\cal I}|^2$~\cite{Teukolsky:1974yv}.

The analysis above is classical, in the sense that the fields are not quantized. Superradiance from BHs generalizes to the quantum world.
For example, the quantum analogue of the classical process of superradiance for a massless charged scalar field on a static charged BH space-time was studied in Ref.~\cite{Balakumar:2020gli}.

%%%%%%%%%%%%%%%%%%%%%%%%%%%%%%%%%%%%%%%%%%%%%%%%%%%%%%%%%%%%%%%%%%%%%%%
\subsection{Horizons are unrelated to superradiance}\label{sec:nohorSR}
%%%%%%%%%%%%%%%%%%%%%%%%%%%%%%%%%%%%%%%%%%%%%%%%%%%%%%%%%%%%%%%%%%%%%%%
Boundary conditions are important for any problem. One might be tempted to conclude, as has been done in the past, that 
without ingoing boundary conditions at the horizon, no superradiant scattering can 
occur~\cite{zeldovich1,zeldovich2,Bekenstein:1973mi,Richartz:2009mi,Cardoso:2012zn}. In fact, in absence of a horizon 
(for example in the case of rotating perfect-fluid stars), regularity boundary conditions must be imposed at the center 
of the object. By applying the same argument as above, the Wronskian at the center vanishes, which implies $|{\cal 
R}|^2=|{\cal I}|^2$. One could conclude (erroneously) that superradiance is absent.  It is thus generally believed 
(erroneously) that replacing an horizon by a hard surface would lead to no superradiance in BH geometries.

 Such conclusions are {\it wrong} and could be anticipated with pure causality arguments: since waves  take  an
infinite  (coordinate) time to reach the horizon, it has no causal contact with the region where ``dynamics  is  
happening.'' Thus, boundary conditions should  be  irrelevant for the occurrence of 
superradiance~\cite{Vicente:2018mxl}. Pulses of waves can be amplified with or without horizons, as can be easily shown 
with time-domain evolutions~\cite{Vicente:2018mxl}; these results are supported also by a formal study of the Cauchy problem with highly oscillatory initial data, with some initial energy~\cite{Eskin:2015ssa}. It can be shown that the time evolution splits the solution into two parts: one with the negative energy traveling inwards (towards the event horizon if there is one), and the second part, with positive energy escaping, under some conditions, to infinity. In summary, there is strong evidence that horizons are unrelated to the phenomenon of superradiance.

The two results are compatible by noting that the Fourier-decomposition assumes a stationary profile for the incoming and scattered waves, and requires integrable functions. To arrive at expressions like \eqref{bound2}-
\eqref{reflectivity}, the Fourier decomposition \eqref{eq:Fourier} was used, but it describes {\it only stable 
solutions} $\Psi(t,r,\vartheta,\varphi)$. In fact, as we discuss in Section~\ref{sec:ERpartial} below, when 
the boundary conditions are those of regularity at the center, but ergoregions exist, then an instability is triggered. 
The initial development is precisely that of superradiant amplification, but the  field grows exponentially with time, 
and is not captured by a Fourier analysis. 

To summarize, superradiance requires friction, which for BHs is provided by the ergoregion. When ergoregions are present without horizons, superradiance exists, but comes hand in hand with an instability.
%%%%%%%%%%%%%%%%%%%%%%%%%%%%%%%%%%%%%%%%%%%%%%%%%%%%%%%%%%%%%%%%%%%%%%%%%%%%%%%%%%%%%%%%%%
\subsection{Superradiance from charged static black holes}\label{sec:superradiance_charged}
%%%%%%%%%%%%%%%%%%%%%%%%%%%%%%%%%%%%%%%%%%%%%%%%%%%%%%%%%%%%%%%%%%%%%%%%%%%%%%%%%%%%%%%%%%%
From the discussion of the previous section, it is clear that BH superradiance also occurs for electrically charged waves scattered by a static, charged BHs whenever (cf. Eq.~\eqref{delta_M2})
\begin{equation}
\omega-q\Phi_H<0\,. \label{SRRNthermo}
\end{equation}
Because the background is spherically symmetric, this type of superradiance is simpler to treat and in this section we start our analysis with this simpler case.
%
%%%%%%%%%%%%%%%%%%%%%%%%%%%%%%%%%%%%%%%%%%%%%%%%%%%%%%%%%%%
\subsubsection{Linearized analysis: amplification factors}
%%%%%%%%%%%%%%%%%%%%%%%%%%%%%%%%%%%%%%%%%%%%%%%%%%%%%%%%%%%
The problem can be investigated at linearized level by considering a charged scalar field $\Psi$ propagating on a RN 
background, which is defined by Eq.~\eqref{RNLambda} with $\Lambda=0$. Using the harmonic expansion and 
Fourier-decomposition of Eq.~\eqref{eq:Fourier}, the Klein-Gordon equation for a minimally coupled scalar can be written 
in the Schroedinger-like form~\eqref{wave} with the potential
\begin{equation}
V_{\rm eff}(r)=\omega^2-f\left(\frac{l(l+1)}{r^2}+\frac{f'(r)}{r}+\mu_S^2\right)-\frac{2qQ\omega}{r}+\frac{q^2Q^2}{r^2}\,,\label{VRN}
\end{equation}
where $r$ is defined in terms of $r_*$ through $dr/dr_*=f=(r-r_+)(r-r_-)/r^2$.
We can compute the reflectivity of a scattering experiment as done at the beginning of Sec.~\ref{sec:ABC}. In this specific case $k_H=\omega-q\Phi_H=\omega-qQ/r_+$ and $k_\infty=\sqrt{\omega^2-\mu_S^2}$. Equation~\eqref{reflectivity} then reduces to
%%%
\begin{equation}
 |{\cal R}|^2=|{\cal I}|^2-\frac{\omega-qQ/r_+}{\sqrt{\omega^2-\mu_S^2}}|{\cal T}|^2\,.\label{reflectivityRN}
\end{equation}
%%%
This equation shows that only waves with $\omega>\mu_S$ propagate to infinity and that superradiant scattering occurs, $|{\cal R}|^2>|{\cal I}|^2$, whenever $\omega<qQ/r_+$, which coincides with the condition~\eqref{SRRNthermo} derived from thermodynamical arguments.

The amplification factor for each frequency can be computed by integrating numerically the wave equation (cf. available {\scshape Mathematica}\textsuperscript{\textregistered} notebook in Appendix~\ref{app:codes}). Figure~\ref{fig:SR_charged} shows the amplification factor as a function of the frequency for monopole, $l=0$, waves
and different BH $Q$ and field $q$ charge parameters.
\begin{figure*}[hbt]
\begin{center}
\begin{tabular}{cc}
\epsfig{file=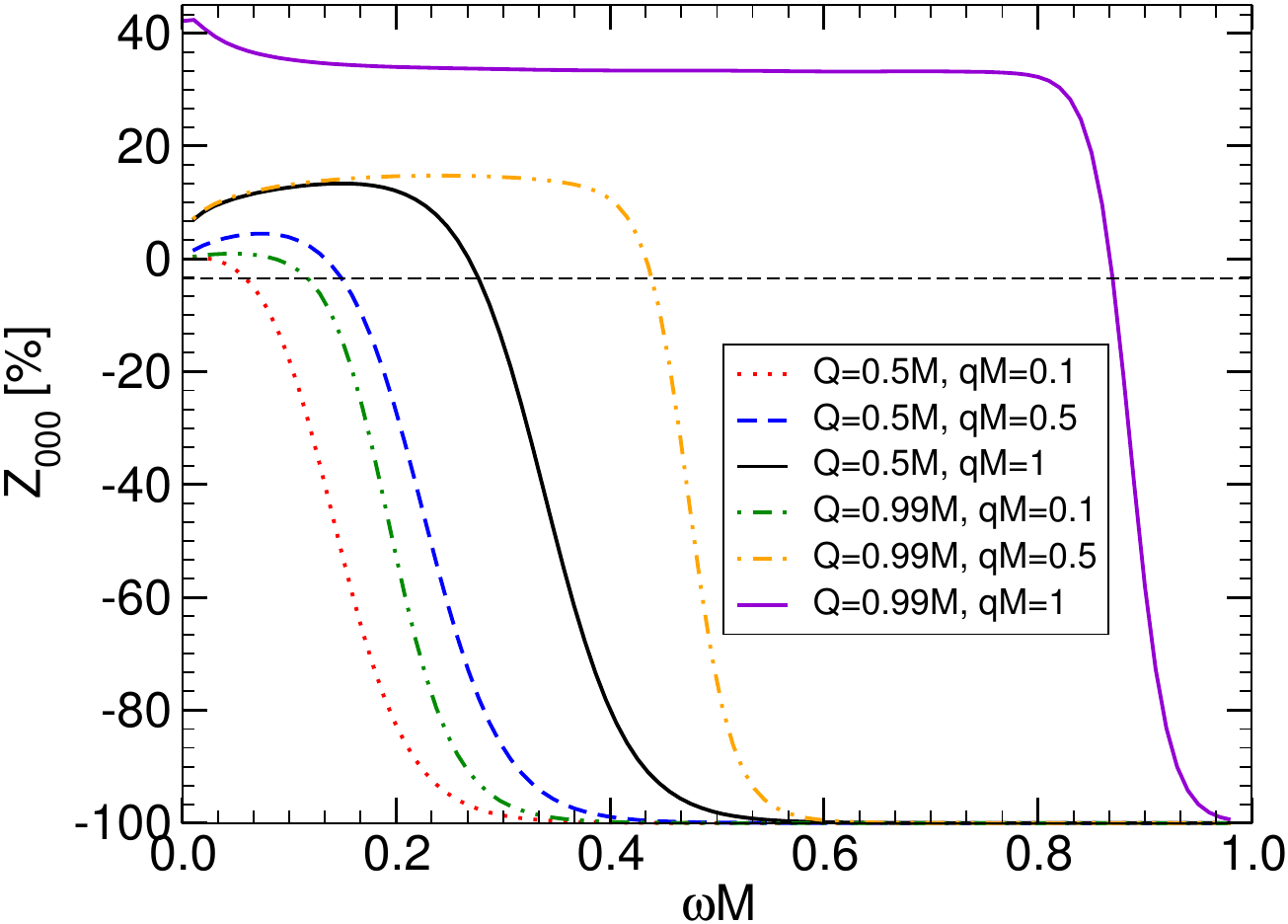,width=0.7\textwidth,angle=0,clip=true}
\end{tabular}
\caption{Amplification factor $Z_{000}=|{\cal R}|^2/|{\cal I}|^2-1$ as a function of the frequency for a  massless bosonic wave with $l=0$ and charge $q$ scattered off a RN BH with charge $Q$ and mass $M$. The threshold of superradiance, $Z_{000}>0$, occurs when $\omega=qQ/r_+$.
\label{fig:SR_charged}}
\end{center}
\end{figure*}
The amplification factor can be as high as $40\%$ for nearly extreme BHs, substantially larger than the amplification factors
of scalar fields in Kerr backgrounds, as we will see. Note also that the critical threshold for superradiance to occur, $Z_{000}>0$, is to numerical accuracy described by condition \eqref{SRRNthermo}. The amplification factor is proportional to $Qq$ at intermediate values, but tends to $100\%$ at large values of $q$. We find that at large $qM$, the amplification factor satisfies
\be
Z_{000}\sim 100-\frac{80}{Qq} \quad(\%)\,.\label{maxcharge}
\ee

A detailed analysis in the time-domain has also recently been performed in Ref.~\cite{DiMenza:2014vpa}. Their results agree with the frequency-domain computation here presented and show indications that the maximum energy gain is always finite, independently of the initial conditions, in accord with the linear stability of the (sub-extremal) RN geometry. These results, in particular \eqref{maxcharge}, are fully consistent with an analytical, small-frequency expansion for the amplification factors~\cite{Richartz:2011vf}. 
An interesting result concerning charge relates to the possibility of attaining a negative total absorption cross section, which implies in particular planar scalar waves can be superradiantly amplified by BHs~\cite{Benone:2015bst,Benone:2019all}.

As we shall see in the next sections, the existence of superradiance for static charged BHs is a crucial ingredient for interesting applications in the context of the gauge/gravity duality. For example the spontaneous symmetry breaking mechanism near a RN-AdS BH~\cite{Gubser:2008px}, and applications therein related to holographic models of superconductors~\cite{Hartnoll:2008vx}, all hinge on this superradiant phenomenon.
%%%%%%%%%%%%%%%%%%%%%%%%%%%%%%%%%%%%%%%%%%%%%%%%%%%%%%%%%%%%%%%%%%%%%%%%%%%%%%%%%%%%%%%%%%%%%%%
\subsubsection{Backreaction on the geometry: mass and charge loss\label{backreaction_charged}}
%%%%%%%%%%%%%%%%%%%%%%%%%%%%%%%%%%%%%%%%%%%%%%%%%%%%%%%%%%%%%%%%%%%%%%%%%%%%%%%%%%%%%%%%%%%%%%%
Superradiant scattering seems to imply that energy is being extracted from the background which -- at linearized
order where superradiance is observed -- is kept fixed. This is not a particularity of superradiant scattering from BHs,
but rather a very generic property. We will now show
that when backreaction effects are included, both the mass and charge of the BH decrease. 

Take a spherically symmetric, linearized charged scalar field
\be
\Psi=\epsilon \frac{\psi(t,r)}{r}\,, 
\ee
where now we explicitly introduced a bookkeeping parameter $\epsilon$ to help keep track of the expansion order.
When allowed to propagate in a RN background, such field  introduces backreactions in both the geometry and vector potential
which are both of order $\epsilon^2$,
\be
A_{\mu}=\left(\frac{Q}{r}+\epsilon^2\frac{Q_t(t,r)}{r},\epsilon^2\frac{\int{dt Q_r(t,r)}}{r^2},0,0\right)\,,
\ee
where the form of the perturbation quantities $Q_t(t,r),\,Q_r(t,r)$ at order ${\cal O}(\epsilon^2)$ was chosen so that the radial electric field ${\cal E}_r$ at large distances is 
\be
r^2{\cal E}_r=Q+\epsilon^2\left(Q_r(t,r)-rQ_t'(t,r)+Q_t(t,r)\right)\,,
\ee
and therefore the charge flux can be obtained via Gauss's law to be
\be
\dot{Q}_{\rm tot}=\epsilon^2\left(\dot{Q}_r-r\dot{Q}_t'+\dot{Q}_t\right)\,.\label{Qtot_gauss}
\ee

Likewise, the metric gets ${\cal O}(\epsilon^2)$ corrections of the form
\be
ds^2=-\left(f-\epsilon^2\frac{2\mu(t,r)}{r}\right)dt^2+\left(f-\epsilon^2\frac{2\mu(t,r)}{r}-\epsilon^2\frac{X(t,r)}{r}\right)^{-1}dr^2+r^2d\Omega^2\,,
\ee
with $\mu$ the mass loss (or gain) induced by the scalar field.
At large distances, we know from the previous analysis of the scalar field equation at order ${\cal O}(\epsilon)$
that the solutions are oscillatory. Let the solutions at large distances be 
\be
\psi \sim f(t-r)+g(t+r)\,,
\ee
where the first term represents an outgoing wave and the second an ingoing wave. The field equations yield a vanishing $\dot X(t,r)$ at large distances, whereas the $(t,r)$ component of Einstein's equations yields
\be
2\frac{r}{f}\dot{\mu}=r\left[\left(\psi^*\right)'\dot{\psi}+\psi'\dot{\psi}^*\right]-\psi\dot{\psi}^*-\psi^*\dot{\psi}-iqQ\left[\psi\left(\psi^*\right)'-\psi^*\psi'\right]\,,
\ee
where the first term on the r.h.s dominates at large distance. Because $\left(\psi^*\right)'\dot{\psi}+\psi'\dot{\psi}^*=2g'(g^*)'-2f'(f^*)'$, we obtain
\be
\dot{\mu}\sim g'(g^*)'-f'(f^*)'\,,\label{super_back_ener}
\ee
where now primes stand for derivative with respect to the argument ($(t-r)$ and $(t+r)$ for $f$ and $g$ respectively).
In other words, for $f'>g'$ -- which can be seen to be the condition for superradiance at order ${\cal O}(\epsilon)$ --
the mass of the BH does decrease at order ${\cal O}(\epsilon^2)$. From the scalar field stress-tensor, which can be read off from \eqref{eq:MFEoMTensor}, the energy flux at infinity can be computed using only the linearized result and reads
\be
\dot{E}_{\infty}=-\left(g'(g^*)'-f'(f^*)'\right)\,.
\ee
In other words, equation \eqref{super_back_ener} tells us that the BH looses or gains mass at a rate which matches
{\it exactly} the energy dissipated or ingoing at infinity, respectively and which is evaluated using only the linearized quantities.
This is an important consistency result and shows that the energy for superradiant amplification does come -- at the nonlinear level -- from the medium, in this case the BH. For monochromatic scalar waves, $\psi\sim {\cal I}e^{-i\omega (t+r)}+{\cal R}e^{-i\omega (t-r)}$ at large distances, one gets
\be
\dot{\mu}=-\omega^2\left(|{\cal R}|^2-|{\cal I}|^2\right)\,,
\ee
indicating that superradiance extracts mass.

Finally, the $r$ component of Maxwell's equations~\eqref{eq:MFEoMVector} yields
\be
qf\left[\psi^*\psi'-\psi(\psi^*)'\right]+2i\left[\dot{Q}_r-r\dot{Q}'_t+\dot{Q}_t\right]=0\,.
\ee
From \eqref{Qtot_gauss} this can be re-written as
\be
2\dot{Q}_{\rm tot}=iqf\left[\psi^*\psi'-\psi(\psi^*)'\right]\,,
\ee
which leads to loss of charge at order ${\cal O}(\epsilon^2)$ whenever the superradiance condition for the scalar field is satisfied
at order ${\cal O}(\epsilon^2)$. For monochromatic scalar waves, $\psi\sim {\cal I} e^{-i\omega (t+r)}+{\cal R}e^{-i\omega (t-r)}$ at large distances, one finds
\be
\dot{Q}_{\rm tot}=-\omega q\left(|{\cal R}|^2-|{\cal I}|^2\right)\,.
\ee

One can now use the first law of BH mechanics \eqref{first_law_BH} to find
\be
\dot{A}_H=\frac{8\pi}{k} \left(\dot{M}-\Phi_H\dot{Q} \right)=-\frac{8\pi}{k} \omega \left(\omega-q\Phi_H \right)\left(|{\cal R}|^2-|{\cal I}|^2\right)\,.
\ee
In the superradiant regime, $|{\cal R}|^2-|{\cal I}|^2>0$ but a necessary condition is that $\omega-q\Phi_H<0$ thus yielding a positive area increase.
Outside the superradiant regime $\omega-q\Phi_H>0$ but there is no amplification and $|{\cal R}|^2-|{\cal I}|^2<0$. In conclusion, the area always increases in agreement with the second law of BH mechanics.

\begin{figure*}[hbt]
\begin{center}
\begin{tabular}{cc}
\epsfig{file=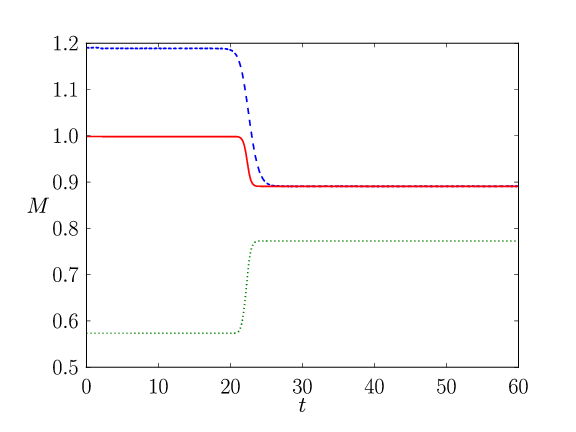,width=0.5\textwidth,angle=0,clip=true}
\epsfig{file=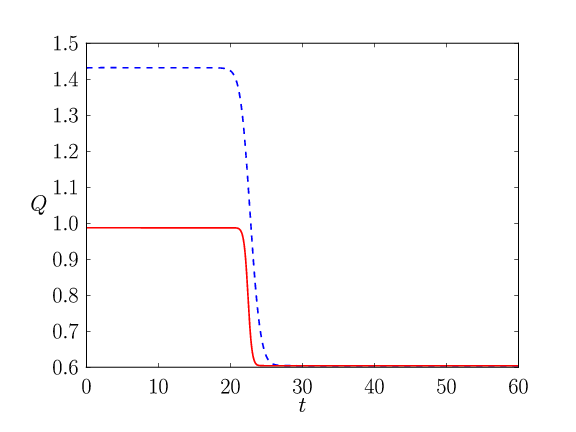,width=0.5\textwidth,angle=0,clip=true}
\end{tabular}
\caption{Time evolution of the BH mass and charge for the superradiant scattering of a massless scalar field around a BH with initial charge $Q = 0.987M$. The charge of the scalar field and the initial data for the simulation are chosen in such a way that one expects superradiant scattering to occur. Left: Evolution of the BH mass (solid red), Bondi mass (dashed blue) and irreducible mass (dotted green). Right: Evolution of the BH charge at the horizon (solid red) and total charge measured at future null infinity (dashed blue). From~\cite{Baake:2016oku}.
\label{fig:SR_charged2}}
\end{center}
\end{figure*}
These conclusions agree with fully nonlinear studies of the superradiant scattering of charged scalar fields around 
charged BHs. This was first studied in Ref.~\cite{Baake:2016oku} where numerical simulations of massless and massive 
charged scalar fields scattering off a charged BH were performed. The authors considered initial data representing an 
ingoing spherically symmetric scalar wave packet scattering off a charged BH and found that superradiance indeed occurs 
at the full nonlinear level leading to a decrease of mass and charge of the BH. Their simulation also confirms that the 
BH area always increases during the process. An example of such superradiant mass and charge extraction is shown in 
Fig.~\ref{fig:SR_charged2}, for an initially nearly-extremal charged BH. In those simulations more than $10\%$ of the BH 
mass is lost in the process whereas the horizon charge decreases by more than $40\%$. Thus, charged superradiance occurs 
at the full nonlinear level.

%%%%%%%%%%%%%%%%%%%%%%%%%%%%%%%%%%%%%%%%%%%%%%%%%%%%%
\subsection{Superradiance from rotating black holes}
%%%%%%%%%%%%%%%%%%%%%%%%%%%%%%%%%%%%%%%%%%%%%%%%%%%%%

Here we introduce the superradiant scattering of rotating BHs. We focus on the asymptotically-flat case and consider the geometry~\eqref{metricKerrLambda} with $\Lambda=0$.

%%%%%%%%%%%%%%%%%%%%%%%%%%%%%%%%%%%%%%%%%%%%%%%%%%%%%%%%%%%%%%%%%%%%%%%%%%%%%%%%%%%%%%
\subsubsection{Bosonic and fermionic fields in the Kerr geometry}\label{sec:Teukolsky}
%%%%%%%%%%%%%%%%%%%%%%%%%%%%%%%%%%%%%%%%%%%%%%%%%%%%%%%%%%%%%%%%%%%%%%%%%%%%%%%%%%%%%%
%
\begin{table}[htb]
% \scriptsize
\centering \caption{Wavefunction $\psi$ for each value of the spin weight-$s$. The spin coefficient is given by 
$\rho\equiv-1/(r-ia\cos\vartheta)$. The quantities $\phi_0$, $\phi_2$, $\Psi_0$ and $\Psi_4$ are Newman-Penrose 
scalars~\cite{Newman:1961qr} describing EM and gravitational perturbations, respectively. The quantities $\chi_0$ and 
$\chi_1$ denote components of the Dirac spinor along dyad legs.} 
\vskip 12pt
\begin{tabular}{|c|| c c c c |}
\hline%
$s$      & 0  & ($1/2$, $-1/2$) & ($1$, $-1$) & ($2$, $-2$)\\
$\psi$  & $\Phi$  & ($\chi_0$,$\rho^{-1}\chi_1$)	&($\phi_0$,$\rho^{-2}\phi_2$) &($\Psi_0$,$\rho^{-4}\Psi_4$)\\
\hline
\end{tabular}
\label{tab:NP_scalars}
\end{table}
The wave equation for linearized fluctuations around the Kerr geometry was studied by Teukolsky, Press and collaborators in great detail~\cite{Teukolsky:1972my,Teukolsky:1973ha,Teukolsky:1974yv,Press:1973zz}. 
Following Carter's unexpected result on the separability of the Hamilton-Jacobi equation for the geodesics in a Kerr 
geometry~\cite{Carter:1968rr}, he also noted that the analogue scalar field equation was separable~\cite{Carter:1968ks}, 
as was explicitly shown in Ref.~\cite{Brill:1972xj}. In a breakthrough work (see Ref.~\cite{Teukolsky:2014vca} for a 
first-person historical account), it was shown that linearized perturbations of the Kerr geometry could be described 
with a single master equation, describing ``probe'' scalar ($s=0$), massless Dirac ($s=\pm 1/2$), EM ($s=\pm 1$) and 
gravitational ($s=\pm 2$) fields in a Kerr background~\cite{Teukolsky:1972my}. The master equation reads
\bea
&&\left[\frac{\left(r^2+a^2\right)^2}{\Delta}-a^2\sin^2\vartheta\right]\frac{\partial\psi^2}{\partial t^2}+\frac{4Mar}{\Delta}\frac{\partial\psi^2}{\partial t\partial\varphi}+
\left[\frac{a^2}{\Delta}-\frac{1}{\sin^2\vartheta}\right]\frac{\partial\psi^2}{\partial\varphi^2}\nn\\
&&-\Delta^{-s}\frac{\partial}{\partial r}\left(\Delta^{s+1} \frac{\partial \psi}{\partial r}\right)-\frac{1}{\sin\vartheta}\frac{\partial}{\partial \vartheta}\left(\sin\vartheta\frac{\partial\psi}{\partial \vartheta}\right)-2s\left[\frac{a(r-M)}{\Delta}+\frac{i\cos\vartheta}{\sin^2\vartheta}\right]\frac{\partial\psi}{\partial\varphi}\nn\\
&&-2s\left[\frac{M(r^2-a^2)}{\Delta}-r-ia\cos\vartheta\right]\frac{\partial\psi}{\partial t}+\left(s^2\cot^2\vartheta-s\right)\psi=0\,,
\eea
where $s$ is the field's spin weight, and the field quantity $\psi$ is directly related to Newman-Penrose quantities as shown in Table~\ref{tab:NP_scalars}.
By Fourier transforming $\psi(t,r,\vartheta,\varphi)$ and using the ansatz
\be\label{teu_eigen}
\psi=\frac{1}{2\pi}\int d\omega e^{-i\omega t}e^{im\varphi}S(\vartheta)R(r)\,,
\ee
Teukolsky found separated ODEs for the radial and angular part, which read, respectively
\be\label{teu_radial}
\Delta^{-s}\frac{d}{dr}\left(\Delta^{s+1}\frac{dR}{dr}\right)+\left(\frac{K^2-2is(r-M)K}{\Delta}+4is\omega r-\lambda\right)R=0\,,
\ee
and
{\small
\be
\frac{1}{\sin\vartheta}\frac{d}{d\vartheta}\left(\sin\vartheta\frac{dS}{d\vartheta}\right)+
\left(a^2\omega^2\cos^2\vartheta-\frac{m^2}{\sin^2\vartheta}-2a\omega s\cos\vartheta
-\frac{2m s\cos\vartheta}{\sin^2\vartheta}-s^2\cot^2\vartheta+s+A_{slm}\right)S=0\,, \label{spheroidal}
\ee}
where $K\equiv (r^2+a^2)\omega-am$ and $\lambda\equiv A_{slm}+a^2\omega^2-2am\omega$. Together with the orthonormality condition
\be\label{sphe_norm}
\int_0^{2\pi}\int_0^{\pi}|S|^2\sin\vartheta d\vartheta d\varphi=1\,,
\ee
the solutions to the angular equation~\eqref{spheroidal} are known as spin-weighted spheroidal harmonics $e^{im\varphi}S\equiv S_{slm}(a\omega,\vartheta,\varphi)$. When $a\omega=0$ they reduce to the spin-weighted spherical harmonics $Y_{slm}(\vartheta,\varphi)$~\cite{Goldberg1967}.
For small $a\omega$ the angular eigenvalues are (cf. Ref.~\cite{Berti:2005gp} for higher-order terms)
\be
A_{slm}=l(l+1)-s(s+1)+\mathcal{O}(a^2\omega^2)\,.
\ee
The computation of the eigenvalues for generic spin can only be done numerically~\cite{Berti:2005gp}. 

Besides these equations, to have complete information about the gravitational and EM fluctuations, we need to find the 
relative normalization between $\phi_0$ and $\phi_2$ for EM fields and between $\Psi_0$ and $\Psi_4$ for 
gravitational perturbations. This was done in Refs.~\cite{Teukolsky:1974yv,1973ZhETF..65....3S,Staro2} assuming the 
normalization condition~\eqref{sphe_norm} and using what is now known as the Teukolsky-Starobinsky identities (see 
also~\cite{Chandra} for details).

Defining the tortoise coordinate $r_*$ as $dr/dr_*=\Delta/(r^2+a^2)$, Eq.~\eqref{teu_radial} has the following asymptotic solutions
\be\label{bc_kerr}
R_{slm}\sim \mathcal{T}\Delta^{-s} e^{-ik_H r_*}+\mathcal{O} e^{ik_H r_*}\,, {\rm as}\quad r \to r_+\,,\quad 
R_{slm}\sim \mathcal{I}\frac{e^{-i\omega r}}{r}+\mathcal{R}\frac{e^{i\omega r}}{r^{2s+1}}\,, {\rm as} \quad r \to \infty\,,
\ee
where $k_H=\omega-m\Omega_{\rm H}$ and $\Omega_{\rm H}=a/(2M r_+)$ is the angular velocity of the BH horizon. Regularity at the horizon requires purely ingoing boundary conditions, i.e., $\mathcal{O}=0$ (see Section 3 in Ref.~\cite{Berti:2009kk} for a careful discussion of boundary conditions).

%%%%%%%%%%%%%%%%%%%%%%%%%%%%%%%%%%%%%%%%%%%%%%%%%%%%%%%%%%%%%%%%%%%%%%%%%%%%%%
\subsubsection{Energy fluxes of bosonic fields at infinity and on the horizon} 
%%%%%%%%%%%%%%%%%%%%%%%%%%%%%%%%%%%%%%%%%%%%%%%%%%%%%%%%%%%%%%%%%%%%%%%%%%%%%%

The perturbation equations~\eqref{teu_radial} and~\eqref{spheroidal} and their asymptotic behavior~\eqref{bc_kerr} can be used to define the energy fluxes that the fields carry through the horizon and to infinity. The expressions for the energy fluxes were computed in Ref.~\cite{Teukolsky:1974yv}, to which we refer the reader for further details.  
The total energy fluxes at infinity per unit solid angle for scalar $s=0$ and EM $s=\pm 1$ are given by (see 
Appendix~\ref{appendix_energyangularmomentum}):
\be
\frac{d^2E}{dtd\Omega}=\lim_{r\to +\infty} r^2 T^r_{\phantom{r}t}\,,
\ee
where $T_{\mu\nu}$ is the stress-energy tensor of the test field.
For the scalar case $s=0$, one has
\be
\frac{dE_{\rm out}}{dt}=\frac{\omega^2}{2}\left|\mathcal{R}\right|^2\,,
\quad \frac{dE_{\rm in}}{dt}=\frac{\omega^2}{2}\left|\mathcal{I}\right|^2\,,\label{flux_scalar}
\ee
whereas, for the EM case $s\pm 1$,
\be
\frac{d^2E_{\rm out}}{dtd\Omega}=\lim_{r\to +\infty} \frac{r^2}{2\pi}\left|\phi_2\right|^2\,,\quad
\frac{d^2E_{\rm in}}{dtd\Omega}=\lim_{r\to +\infty} \frac{r^2}{8\pi}\left|\phi_0\right|^2\,.%\label{flux_elec}
\ee
From these definitions it can be shown that the fluxes, valid for $s=1$, are given by
\be
\frac{dE_{\rm out}}{dt}=\frac{4\omega^4}{B^2}\left|\mathcal{R}\right|^2\,,\quad
\frac{dE_{\rm in}}{dt}=\frac{1}{4}\left|\mathcal{I}\right|^2\,,\label{flux_elec}
\ee
where $B^2=Q^2+4ma\omega-4a^2\omega^2$ and $Q=\lambda+s(s+1)$. The corresponding fluxes for $s=-1$ can be found using the Teukolsky-Starobinsky identities and can be obtained from the above relations by doing the transformation: $\mathcal{I}\to -(8\omega^2/B) \mathcal{I}$ and $\mathcal{R}\to -B/(2\omega^2)\mathcal{R}$.
Finally, for gravitational perturbations $s\pm 2$ the fluxes can be computed using the effective stress-energy tensor 
for linearized GWs~\cite{Misner:1974qy}. In terms of the Weyl scalars they are given by 
\be
\frac{d^2E_{\rm out}}{dtd\Omega}=\lim_{r\to +\infty} \frac{r^2}{4\pi\omega^2}\left|\Psi_4\right|^2\,,\quad
\frac{d^2E_{\rm in}}{dtd\Omega}=\lim_{r\to +\infty} \frac{r^2}{64\pi\omega^2}\left|\Psi_0\right|^2\,,
\ee
which can be shown to give for $s=2$,
\be
\frac{dE_{\rm out}}{dt}
=\frac{8\omega^6}{|C|^2}\left|\mathcal{R}\right|^2\,,\quad
\frac{dE_{\rm in}}{dt}=\frac{1}{32\omega^2}\left|\mathcal{I}\right|^2\,,\label{flux_grav}
\ee
where $\left|C\right|^2=B^2\left[(Q-2)^2+36a\omega m-36a^2\omega^2)\right]+(2Q-1)(96a^2\omega^2-48a\omega m)+144\omega^2(M^2-a^2)$. For $s=-2$ the fluxes can be found once again using the Teukolsky-Starobinsky identities and can be obtained from the above relations by doing the transformation: $\mathcal{I}\to (64\omega^4/C) \mathcal{I}$ and $\mathcal{R}\to C^*/(4\omega^4)\mathcal{R}$.

The flux at the horizon for $s=0, \pm 1$ can be computed evaluating the change in energy of the hole. As showed in Appendix~\ref{appendix_energyangularmomentum} it is given by
\be
\frac{d^2E_{\rm hole}}{dtd\Omega}=\frac{\omega}{k_H}2Mr_+ T^{\mu\nu}n_{\mu}n_{\nu}\,,
\ee
where $n_{\mu}$ is an inward unit vector, normal to the horizon surface.

Using~\eqref{bc_kerr}, one finds for the scalar case
\be\label{fluxhor_sca}
\frac{d^2E_{\rm hole}}{dtd\Omega}=M r_+ \omega k_H\frac{S_{0lm}^2(\vartheta)}{2\pi}\left|\mathcal{T}\right|^2\,,
\ee
whereas, the EM case for $s=1$ gives 
\be\label{fluxhor_elec}
\frac{d^2E_{\rm hole}}{dtd\Omega}=\frac{\omega}{8M r_+ k_H}\frac{S_{1lm}^2(\vartheta)}{2\pi}\left|\mathcal{T}\right|^2\,.
\ee
The case $s=-1$ can be obtained doing the transformation $B\mathcal{T}\to -32ik_H M^2r_+^2(-ik_H+2\epsilon)\, \mathcal{T}$, where $\epsilon=\sqrt{M^2-a^2}/(4Mr_+)$.

For gravitational perturbations one can use the first law of BH mechanics~\eqref{first_law_BH} to find the flux at the horizon~\cite{Hawking:1972hy}. The rate of change of the area can be found from Eq.~\eqref{delta_M}. Since $\delta M=\delta E_{\rm hole}$ we find
\be\label{area_rate}
\frac{d^2A}{dtd\Omega}=\frac{16\pi r_+k_H}{(M^2-a^2)^{1/2}\omega}\frac{d^2E_{\rm hole}}{dtd\Omega}\,.
\ee
We can also show that~\cite{Hawking:1972hy}
\be\label{area_rate_HH}
\frac{d^2A}{dtd\Omega}=\frac{2Mr_+\Delta^4}{16(r^2+a^2)^4\epsilon(k_H^2+4\epsilon^2)}\left|\Psi_0\right|^2\,,
\ee
Equating~\eqref{area_rate} with~\eqref{area_rate_HH} at the horizon, we find for $s=2$ 
\be
\frac{d^2E_{\rm hole}}{dtd\Omega}=\frac{S_{2lm}^2(\vartheta)}{2\pi}\frac{\omega}{32k_H(k_H^2+4\epsilon^2)(2Mr_+)^3}\left|\mathcal{T}\right|^2\,.\label{fluxhor_grav}
\ee
whereas the corresponding for $s=-2$ can be found doing the transformation $C\mathcal{T}\to 64(2Mr_+)^4 ik_H (k_H^2+4\epsilon^2)(-ik_H+4\epsilon)\,\mathcal{T}$. 

From Eqs.~\eqref{fluxhor_sca},~\eqref{fluxhor_elec} and~\eqref{fluxhor_grav} one can see that if the superradiance condition is met, $k_H<0$, the energy flux at the horizon is negative, i.e. energy (and angular momentum) are extracted from the BH.
%
%%%
\subsubsection{Amplification factors}\label{sec:amplcoeff}
%%%
For any scattering process experiment, energy conservation implies that
\be
\frac{dE_{\rm in}}{dt}-\frac{dE_{\rm out}}{dt}=\frac{dE_{\rm hole}}{dt}\,.
\ee
This equation relates the asymptotic coefficients $\mathcal{R}$, $\mathcal{I}$ and $\mathcal{T}$, which can be used to check the consistency of numerical computations {\it a posteriori}. Using Eqs.~\eqref{fluxhor_sca},~\eqref{fluxhor_elec} and~\eqref{fluxhor_grav}, it is also clear that when energy is extracted from the BH, $k_H<0 \implies \frac{dE_{\rm hole}}{dt}<0$, there is superradiance,$\frac{dE_{\rm in}}{dt}<\frac{dE_{\rm out}}{dt}$, as it should by energy conservation.
Finally, from the energy fluxes at infinity one can define the quantity
\be
Z_{slm}=\frac{dE_{\rm out}}{dE_{\rm in}}-1\,,
\ee
which, depending on whether the superradiance condition is met or not, provides the amplification or the absorption factor for a bosonic wave of generic spin $s$ and quantum numbers $(l,m)$ scattered off a Kerr BH. Using Eqs.~\eqref{flux_scalar},~\eqref{flux_elec} and~\eqref{flux_grav} we find
\be\label{def:sigma}
Z_{slm}=
 \left\{ 
 \begin{array}{l l}
\frac{|{\cal R}|^2}{|{\cal I}|^2}-1\,, &\quad \text{if  } s=0\,,\\
\frac{|{\cal R}|^2}{|{\cal I}|^2}\left(\frac{16\omega^4}{B^2}\right)^{\pm 1}-1\,, &\quad \text{if  } s=\pm 1\,,\\ 
\frac{|{\cal R}|^2}{|{\cal I}|^2}\left(\frac{256\omega^8}{\left|C\right|^2}\right)^{\pm 1}-1\,, &\quad \text{if  } s=\pm 2\,.
\end{array} \right.
\ee
From the symmetries of the differential equations~\eqref{teu_radial} and~\eqref{spheroidal}, one can prove the following relation
%%%
\begin{equation}
 Z_{slm}(\omega)=Z_{sl-m}(-\omega) \,.\label{symmetry}
\end{equation}
%%%
This symmetry relation can be used to fix the sign of $\omega$. In other words, if the full dependence on $m$ is known for a given $(s,l)$ and $\omega>0$, the corresponding amplification factor for $-\omega$ follows immediately from Eq.~\eqref{symmetry}. Thus, the amplification factor $Z_{slm}$ in the entire real $\omega$-axis can be obtained by only looking at $\omega>0$. In the following we will exploit these symmetries when computing superradiant amplification factors numerically.

%%%%%%%%%%%%%%%%%%%%%%%%%%%%%%%%%%%%%%%%%%%%%%%%%%%%%%%%%%%%%%%%%%%
\subsubsection{Dirac fields on the Kerr geometry} \label{DiracKerr}
%%%%%%%%%%%%%%%%%%%%%%%%%%%%%%%%%%%%%%%%%%%%%%%%%%%%%%%%%%%%%%%%%%%
The absence of superradiance for massless Dirac fields was proved in 1973, through the separation of the massless spin-1/2 equations on a Kerr background~\cite{Unruh:1973}. In 1976, the separation of variables was extended to massive Dirac particles~\cite{Chandra:1976}, a result soon generalized to the Kerr-Newman geometry~\cite{Page:1976,Lee:1977}. In 1978, these results were used to show that
generic massive Dirac fields do not exhibit superradiant scattering in the Kerr BH background geometry~\cite{Iyer:1978} (thereby correcting a previous analysis~\cite{Martellini:1977qf}).
The Dirac equation in curved spacetime is
\begin{eqnarray}
 \gamma^\mu\nabla_\mu\psi+i\mu_e \psi&=&0\,, \label{Dirac}
\end{eqnarray}
where $\left[\gamma^\mu,\gamma^\nu\right]=2 g^{\mu\nu}$, $\nabla_\mu\psi=\partial_\mu\psi-\Gamma_\mu\psi$, $\nabla_\mu\bar\psi=\partial_\mu\bar\psi+\bar\psi\Gamma_\mu$, $\bar\psi=\psi^\dagger\gamma^0$ is the Dirac adjoint, $\Gamma_\mu$ is the spinor affine connection~\cite{Chandra} and $\mu_e$ is the fermion mass. The Dirac equation can be separated on a Kerr background using the ansatz
%%%%
\begin{equation}
 \psi=\left(\frac{R_{-}S_{-}}{\sqrt{2}\rho^*},\frac{R_{+}S_{+}}{\sqrt{\Delta}},-\frac{R_{+}S_{-}}{\sqrt{\Delta}},-\frac{R_{-}S_{+}}{\sqrt{2}\rho}\right)^T e^{-i\omega t}e^{im\varphi}\,, \label{reprDirac}
\end{equation}
%%%%
where $\rho=r+ia\cos\vartheta$. The functions $R_{\pm}(r)$ and $S_{\pm}(\vartheta)$ satisfy a system of first-order differential equations, which can be reduced to the following second-order form~\cite{Page:1976}
%%%%
\begin{eqnarray}
 &&\sqrt{\Delta}\frac{d}{dr}\left(\sqrt{\Delta}\frac{dR_-}{dr}\right)-\frac{i \mu_e \Delta}{\sqrt{\lambda}+i\mu_e r}\frac{dR_-}{dr}\nn\\
 &&+\left[\frac{K^2+i(r-M)K}{\Delta}-2i\omega r-\frac{\mu_e K}{\sqrt{\lambda}+i\mu_e r}-\mu_e^2 r^2-\lambda\right]R_-=0\,,  \label{radDirac}\\
 %%%
 &&\frac{1}{\sin\vartheta}\frac{d}{d\vartheta}\left(\sin\vartheta\frac{dS_-}{d\vartheta}\right)+\frac{a\mu_e \sin\vartheta}{\sqrt{\lambda}+a\mu_e\cos\vartheta}\frac{dS_-}{d\vartheta}+\nn\\
 &&+\left[a^2\omega^2\cos^2\vartheta-\frac{m^2}{\sin^2\vartheta}+a\omega\cos\vartheta+
\frac{m\cos\vartheta}{\sin^2\vartheta}-\frac{\cot^2\vartheta}{4}-\frac{1}{2}+\lambda-2am\omega-a^2\omega^2\right.\nn\\
&&\left.+\frac{a\mu_e(1/2\cos\vartheta+a\omega\sin^2\vartheta-m)}{\sqrt{\lambda}+a\mu_e\cos\vartheta}-a^2\mu_e^2\cos^2\vartheta\right]S_-=0\,,
\end{eqnarray}
%%%%
and $R_{+}$ and $S_{+}$ can be obtained once $R_-$ and $S_-$ are known~\cite{Chandra:1976}. The equations above were extended by Page to the case of Kerr-Newman metric and they reduce to Teukolsky's equations~\eqref{teu_radial} and \eqref{spheroidal} when $\mu_e=0$ and setting $s=-1/2$. 
Near the horizon, the radial functions behave as
%%%%
\begin{equation}
 R_{\pm}(r)\to A_{\pm}\Delta^{\frac{1\mp1}{4}}e^{-i k_H r_*}\,,
\end{equation}
%%%
so that $R_{-}$ is vanishing at the horizon. Although the asymptotic solution exhibits the usual $k_H$ term that appears due to the BH rotation relative to the reference frame (cf. Eq.~\eqref{bc_kerr}), in this case superradiance is forbidden to occur, as we now discuss.

Absence of superradiance is a direct consequence of the properties of the stress-energy tensor for fermions. Dirac's equation~\eqref{Dirac} is associated with a conserved current
\begin{equation}
 J^\mu=\bar\psi\gamma^\mu\psi\,,
\end{equation}
whose conservation, $\nabla_\mu J^\mu=0$, implies that the net number current flowing down the horizon is always positive
\begin{equation}
\frac{dN}{dt}=-\int d\vartheta d\varphi\sqrt{-g}J^r=\pi\sum_{lm}|A_{+}|^2\int d\vartheta \sin\vartheta(|S_{+}|^2+|S_{-}|^2)\,,
\end{equation}
where the last step follows from the representation~\eqref{reprDirac} and the orthonormality of the eigenfunctions, Eq.~\eqref{sphe_norm}~\cite{Iyer:1978}. From the equation above, it is clear that $dN/dt>0$, i.e. there is no net flux coming {\it from} the horizon, for any frequency. Indeed, using the stress-energy tensor for a Dirac field, it is easy to show that the net energy flow across the horizon per unit time and solid angle is $\sim\omega dN/dt$, signaling the absence of energy and angular momentum extraction for fermions. 

The same conclusion can be obtained by studying the reflection and transmission coefficients in the scattering of a fermionic wave off a Kerr BH. Chandrasekhar showed that Eq.~\eqref{radDirac} can be written as a Schroedinger-like equation in modified tortoise coordinates~\cite{Chandra}. Using the homogeneity of the Wronskian, the same analysis performed at the beginning of Sec.~\ref{sec:ABC}, allows to relate the reflection coefficient ${\cal R}$ and the transmission coefficient ${\cal T}$ as 
\begin{equation}
 |{\cal R}|^2=|{\cal I}|^2-\frac{\omega}{\sqrt{\omega^2-\mu_e^2/2}}|{\cal T}|^2\,.
\end{equation}
The reflection coefficient is always less than unity, showing that superradiance cannot occur.

As discussed in Sec.~\ref{areaimpliesSR}, at the classical level superradiant amplification is a consequence of Hawking's area theorem~\cite{Hawking:1971tu,Bekenstein:1973mi}. It might appear that the absence of superradiance for fermions is at odds with this fact. However, as already pointed out in the original analysis~\cite{Unruh:1973} and later generalized to generic charged massive fermions interacting with Kerr-Newman BHs~\cite{Wagh:1986cz}, the stress-energy tensor for fermions does not satisfy the weak energy condition, $T_{\mu\nu}t^\mu t^\nu>0$ for any timelike vector $t^\mu$, which is one of the assumptions behind Hawking's theorem.

%%%%%%%%%%%%%%%%%%%%%%%%%%%%%%%%%%%%%%%%%%%%%%%%%%%%%%%%%%%%%%%%%%%%%%%%%%%%%%%%%%%
\subsubsection{Linearized analysis: analytic vs numerics}\label{sec:super_anavsnum}
%%%%%%%%%%%%%%%%%%%%%%%%%%%%%%%%%%%%%%%%%%%%%%%%%%%%%%%%%%%%%%%%%%%%%%%%%%%%%%%%%%%
The amplification factors $Z_{slm}$ for a bosonic wave of generic spin $s$ and quantum numbers $(l,m)$ (cf. Eq.~\eqref{def:sigma}) scattered off a Kerr BH can be computed by integrating numerically the Teukolsky equations presented in Sec.~\ref{sec:Teukolsky}. When the superradiance condition is not fulfilled, the same computation provides the absorption cross section of a spinning BH.
Remarkably, the problem was also solved analytically in the low-frequency regime~\cite{Starobinski:1973,Starobinski2:1973}. Using matching-asymptotic techniques (see Appendix~\ref{appendix_super_ana}), the authors showed that in the low-frequency regime 
%%%
\begin{eqnarray}
 Z_{slm} &=& Z_{0lm}\left[\frac{(l-s)!(l+s)!}{(l!)^2}\right]^2 \,, \label{sigma}\\
 %%%
 Z_{0lm} &=&-8M r_+(\omega-m\Omega_{\rm H})\omega^{2l+1}\left(r_+-r_-\right)^{2l}\left[\frac{(l!)^2}{(2l)!(2l+1)!!}\right]^2\prod_{k=1}^l\left[1+\frac{M^2}{k^2}\left(\frac{\omega-m\Omega_{\rm H}}{\pi r_+ T_H}\right)^2\right]\,, \nn\\ \label{sigma0}
\end{eqnarray}
%%%
where $T_H=(r_+-r_-)/(4\pi r_+^2)$ is the BH temperature and $Z_{0lm}$ is the amplification factor for scalar waves. The 
formulas above are valid for any spin $a\leq M$ provided $\omega M\ll1$. The superradiant condition is independent of 
the spin of the field and $Z_{slm}>0$ whenever $\omega<m\Omega_{\rm H}$ for any $l$ and $s$. In addition, 
Eq.~\eqref{sigma} shows that: (i) the amplification factor is independent of the spin of the field when $l\gg 2s^2$, and 
(ii) in the low-frequency limit the amplification of EM waves is only a factor $4$ larger than that of 
scalar waves (this maximum is obtained when $l=m=1$), whereas the amplification of GWs is a factor $36$ larger than 
that of scalar waves for $l=m=2$.

Defining $\alpha=1-\omega/(m\Omega_{\rm H})$, the equations above predict $Z_{slm}\propto \alpha$ when $|\alpha|\ll(r_+-r_-)/(am)$, and the exact coefficient can be extracted from Eqs.~\eqref{sigma} and \eqref{sigma0}. Thus, in this regime $Z_{slm}$ is linear and continuous in $\omega-m\Omega_{\rm H}$ near the threshold. 
Furthermore, the amplification is largest at $\omega_{\rm max}\sim (2l+1)/(2l+2)m\Omega_{\rm H}$, independently of $s$.

\begin{figure*}[hbt]
\begin{center}
\begin{tabular}{cc}
\epsfig{file=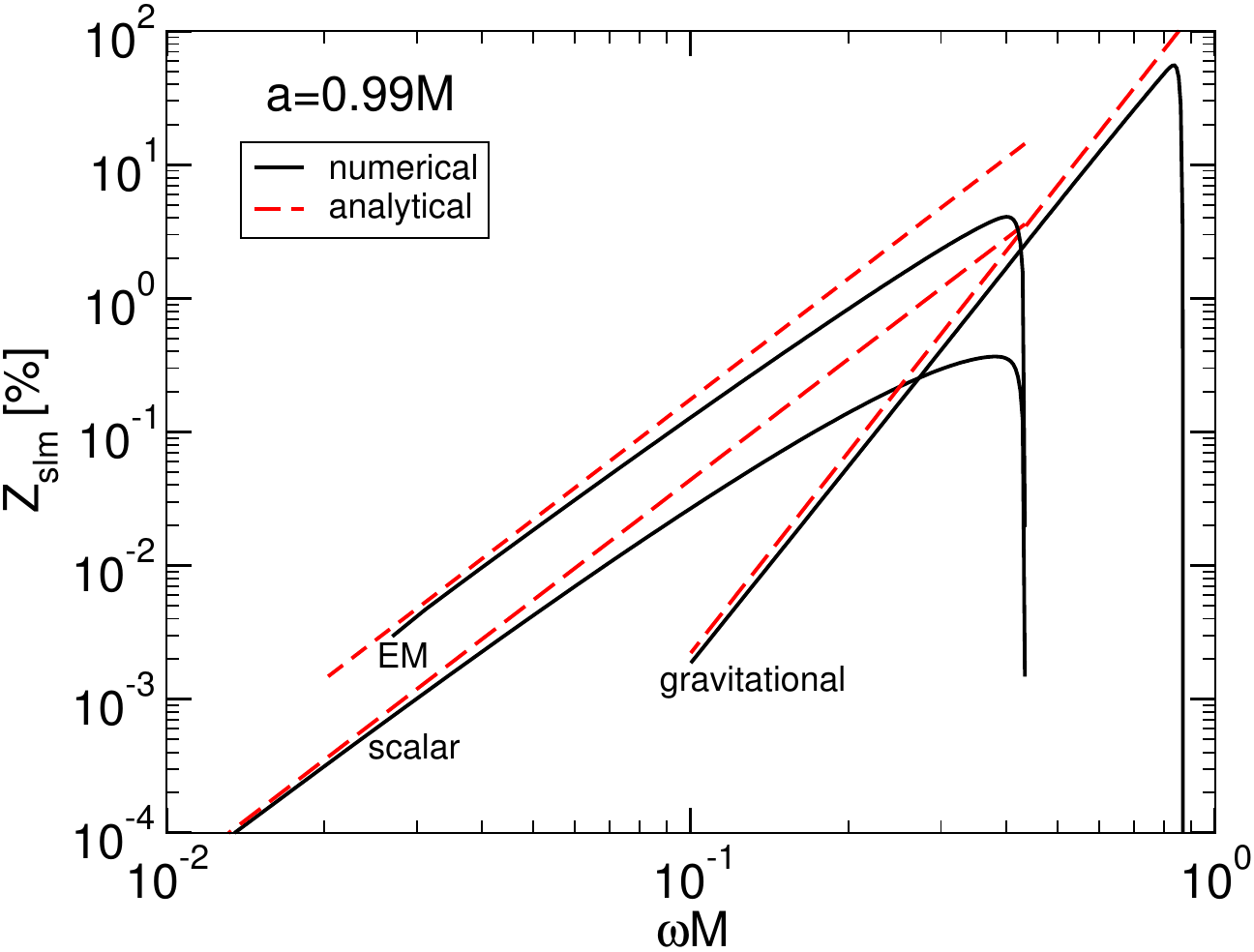,width=0.48\textwidth,angle=0,clip=true}&
\epsfig{file=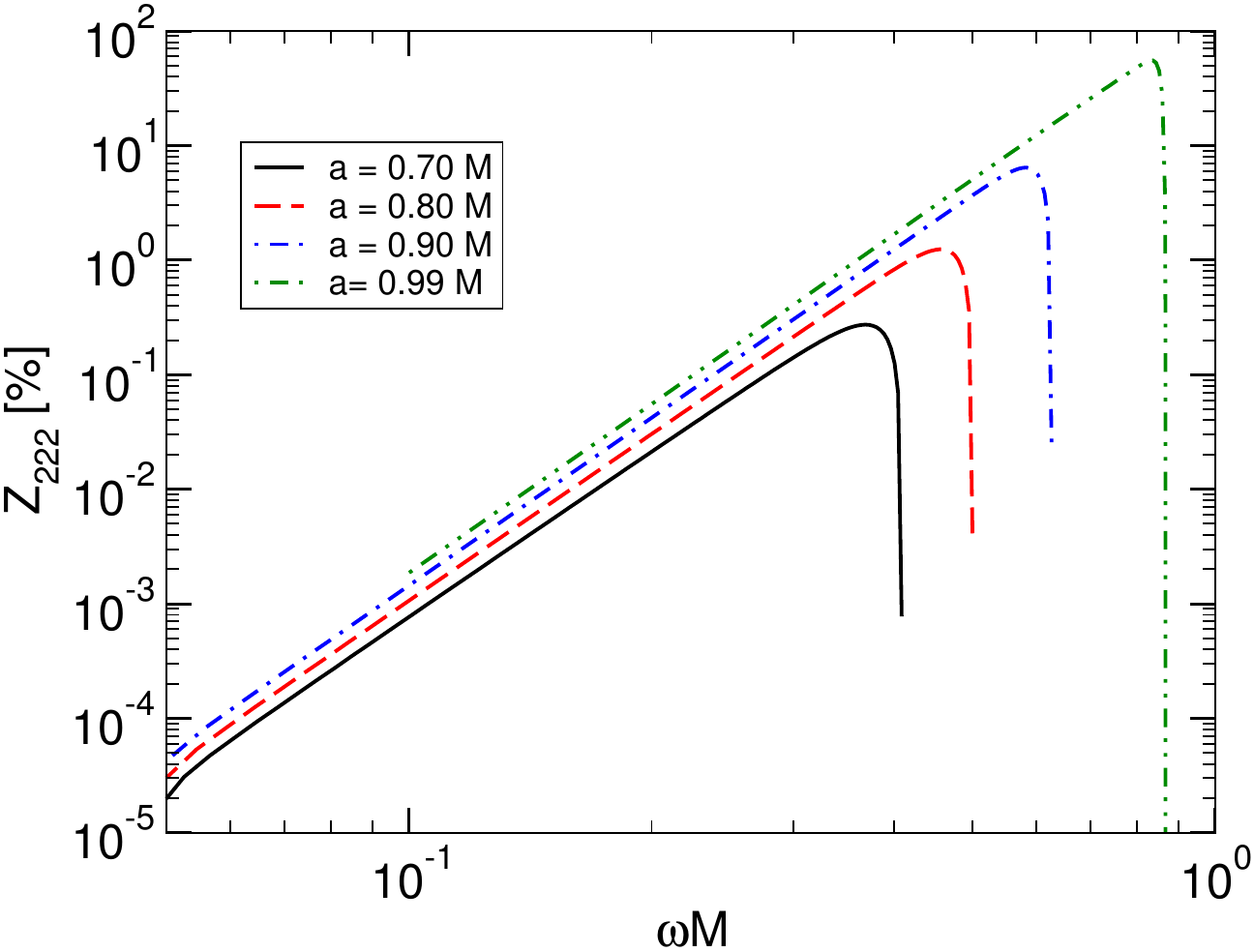,width=0.48\textwidth,angle=0,clip=true}
\end{tabular}
\caption{Left: Amplification factor $Z_{slm}$ as a function of the frequency $\omega$ of a wave scattered off a Kerr BH 
with spin parameter $a=0.99M$ obtained by solving numerically the Teukolsky equations and compared to the analytical 
result in the low-frequency limit. We consider scalar and EM waves with $l=m=1$ and GWs with $l=m=2$. 
Superradiance, $Z_{slm}>0$, occurs when $0<\omega<m\Omega_{\rm H}$ in all cases.
Right: The amplification factor for GWs and for different values of the BH spin.
\label{fig:Acomparison}}
\end{center}
\end{figure*}

With the further assumption $\omega\ll m \Omega_{\rm H}$, Eq.~\eqref{sigma} reduces to
\begin{eqnarray}
 Z_{slm}&=&8 r_+^2 T_H \omega ^{2 l+1} (r_+-r_-)^{2 l} \left[\frac{\Gamma (1+l-s) \Gamma (1+l+s)}{(2 l+1)!!\Gamma (l+1) \Gamma (2 l+1)}\right]^2 \times\nn\\
 &&\sinh \left(\frac{m \Omega_{\rm H}}{r_+ T_H}\right) \Gamma \left(l-\frac{i m \Omega_{\rm H}}{\pi  r_+ T_H}+1\right) \Gamma \left(l+\frac{i m \Omega_{\rm H}}{\pi  r_+ T_H}+1\right)\,.\label{sigma2}
\end{eqnarray}
which, although not reproducing the threshold behavior $Z_{slm}\to0$ as $\omega\to m\Omega_{\rm H}$, reproduces well the exact numerical results even at moderately large frequencies, whereas the full equation~\eqref{sigma} breaks down before.
A comparison between the low-frequency analytical result~\eqref{sigma2} and the exact result obtained by solving the 
Teukolsky equation numerically (the {\scshape Mathematica}\textsuperscript{\textregistered} notebook to compute this 
factor and data tables are publicly available at~\cite{webpage}, cf. Appendix~\ref{app:codes}) is presented in the left 
panel of Fig.~\ref{fig:Acomparison} for scalar, EM and GWs scattered off a nearly-extremal BH with 
$a=0.99M$. 
In this figure we only focus on the superradiant regime, $0<\omega<m\Omega_{\rm H}$. Data files of the amplification factors in the entire parameter spaces are provided in a supplementary file (cf. Appendix~\ref{app:codes}).

Equations \eqref{sigma} and \eqref{sigma2} break down when $\omega M\sim1$, a condition which is generically fulfilled near the superradiant threshold $\omega\sim m\Omega_{\rm H}$ (equivalently $\alpha\sim0$) and in the quasi-extremal limit, even at low $m$. In fact, it is clear from Fig.~\ref{fig:Acomparison} that the low-frequency limit~\eqref{sigma2} generically overestimates the amplification factors. 
The behavior near the threshold has been also studied analytically using a different matching asymptotic technique~\cite{Starobinski:1973}. 
In the extremal case, $a=M$, defining $\delta^2=2m^2-A_{slm}-(s+1/2)^2$ [we recall that $A_{slm}$ are the eigenvalues of the spin-$s$ spheroidal harmonics which satisfy Eq.~\eqref{spheroidal}], when $\delta^2<0$ one finds~\cite{Starobinski2:1973}
\begin{equation}
Z_{slm}=4\,{S_\alpha} |\delta|^2 \left(2m^2|\alpha|\right)^{2|\delta|}\frac{|\Gamma(1/2+s+|\delta|+im)|^2|\Gamma(1/2-s+|\delta|+im)|^2}{\Gamma(1+2|\delta|)^4} e^{\pi n[1-{S_\alpha}]}\,, \label{extreme_negative}
\end{equation}
where $S_\alpha=\text{sgn}(\alpha)$, and
\begin{eqnarray}
 Z_{slm}^{-1}&=& \frac{{S_\alpha}e^{\pi m[{S_\alpha}-1]}}{\sinh(2\pi\delta)^2}\left\{\cosh[\pi(m-\delta)]^2 e^{\pi\delta[{S_\alpha}-1]}+
 \cosh[\pi(m+\delta)]^2 e^{-\pi\delta[{S_\alpha}-1]}\right.\nn\\
 &&\left.-2\cosh[\pi(m-\delta)]\cosh[\pi(m+\delta)] \cos[\gamma_0-2\delta\log(2m^2|\alpha|)]\right\}   \,, \label{extreme_positive}
\end{eqnarray}
for $\delta^2>0$ and $|\alpha|\ll m^{-4}\max(1,|\alpha|^2)$. In the equation above
%%%%
\begin{equation}
 \gamma_0=4\arg[\Gamma(1+2i\delta)]+2\arg[\Gamma(1/2+s+im-i\delta)]+2\arg[\Gamma(1/2+s-im-i\delta)]\,.
\end{equation}
%%%%

Note that the condition $\delta^2>0$ is satisfied by almost all modes~\cite{Cardoso:2004hh}, for example it is satisfied for $s=1$ for any $l=m\geq1$ and for $s=2$ for any $l=m\geq2$, i.e. for the cases that correspond to the largest amplification.
The behaviors described by Eq.~\eqref{extreme_negative} and Eq.~\eqref{extreme_positive} are quite different. When $\delta^2<0$, $Z_{slm}$ is continuous and monotonic near $\alpha\sim 0$, whereas when $\delta^2>0$ it displays an infinite number of oscillations as $\alpha\to 0$ in the region $|\alpha|\ll1/m^2$ (provided $\delta\ll1$). Remarkably, as understood already in Ref.~\cite{Starobinski:1973}, these oscillations are related to the existence of quasi-stationary bound states near the event horizon of a nearly-extremal Kerr BH. These quasi-bound states have been computed in Refs.~\cite{Detweiler:1980gk,Andersson:1999wj,Yang:2012pj}. 

When $\delta^2>0$, the oscillations have a small amplitude and --~except for the exceptional case $m=1$ and $\pi\delta\lesssim1$~-- can be ignored. In such case, for $\alpha>0$ one finds
\begin{equation}
 Z_{slm}\sim e^{2\pi(\delta-m)}\,,
\end{equation}
and the amplification factor is discontinuous near the superradiant threshold. Finally, when $\alpha<0$ we have $\min(Z_{slm})=-1$, i.e. there are regions of the parameter space in which the reflectivity is zero and the BH is totally transparent~\cite{Starobinski:1973,Starobinski2:1973}.

Equations~\eqref{extreme_negative} and Eq.~\eqref{extreme_positive} are also valid in the quasi-extremal limit, $a\sim M$, provided $m<\sqrt{M/(M-a)}$ and $(r_+-r_-)/(am)\ll|\alpha|\ll1/m^2$. Since when $|\alpha|\ll (r_+-r_-)/(am)$ the amplification factor is described by Eq.~\eqref{sigma}, near the threshold $Z_{slm}\propto\alpha$ and it is continuous for any $a<M$.
Note however that there exists a regime which is not captured by the formulas above, namely when $a\sim M$ and $\omega\sim m\Omega_{\rm H}$ such that $\alpha\ll (r_+-r_-)/(am)$. Describing this regime analytically requires more sophisticated matching techniques. Various analytical treatments of the Teukolsky's equation can be found in Refs.~\cite{Hod:2013zza,Hod:2012px,Hod:2012bw,Hod:2012zzb,Hod:2014dda} and they are in agreement with the exact results. A representative example of the dependence of $Z_{slm}$ with the BH spin is presented in the right panel of Fig.~\ref{fig:Acomparison}.

The maximum amplification factors are about $0.4\%$, $4.4\%$ and $138\%$ for scattering of massless scalar, 
EM and GWs, respectively, and for the minimum value of $l=m$ allowed (namely $l=m=1$ for scalar and 
EM waves and $l=m=2$ for GWs). As evident from Fig.~\ref{fig:Acomparison}, the maximum 
amplification occurs for BHs with $a\approx M$ and very close to the superradiant threshold, $\omega\sim m\Omega_{\rm 
H}$. Indeed, near the threshold the curve becomes very steep (with a steepness that increases with the BH spin) and it 
attains a maximum right before reaching $\omega=m\Omega_{\rm H}$ where superradiance stops. Detailed tables of the 
amplification factors for scalar, EM and GWs for various parameters are provided in accompanying data 
files (cf. Appendix~\ref{app:codes}).

\begin{figure*}[hbt]
\begin{center}
\begin{tabular}{cc}
\epsfig{file=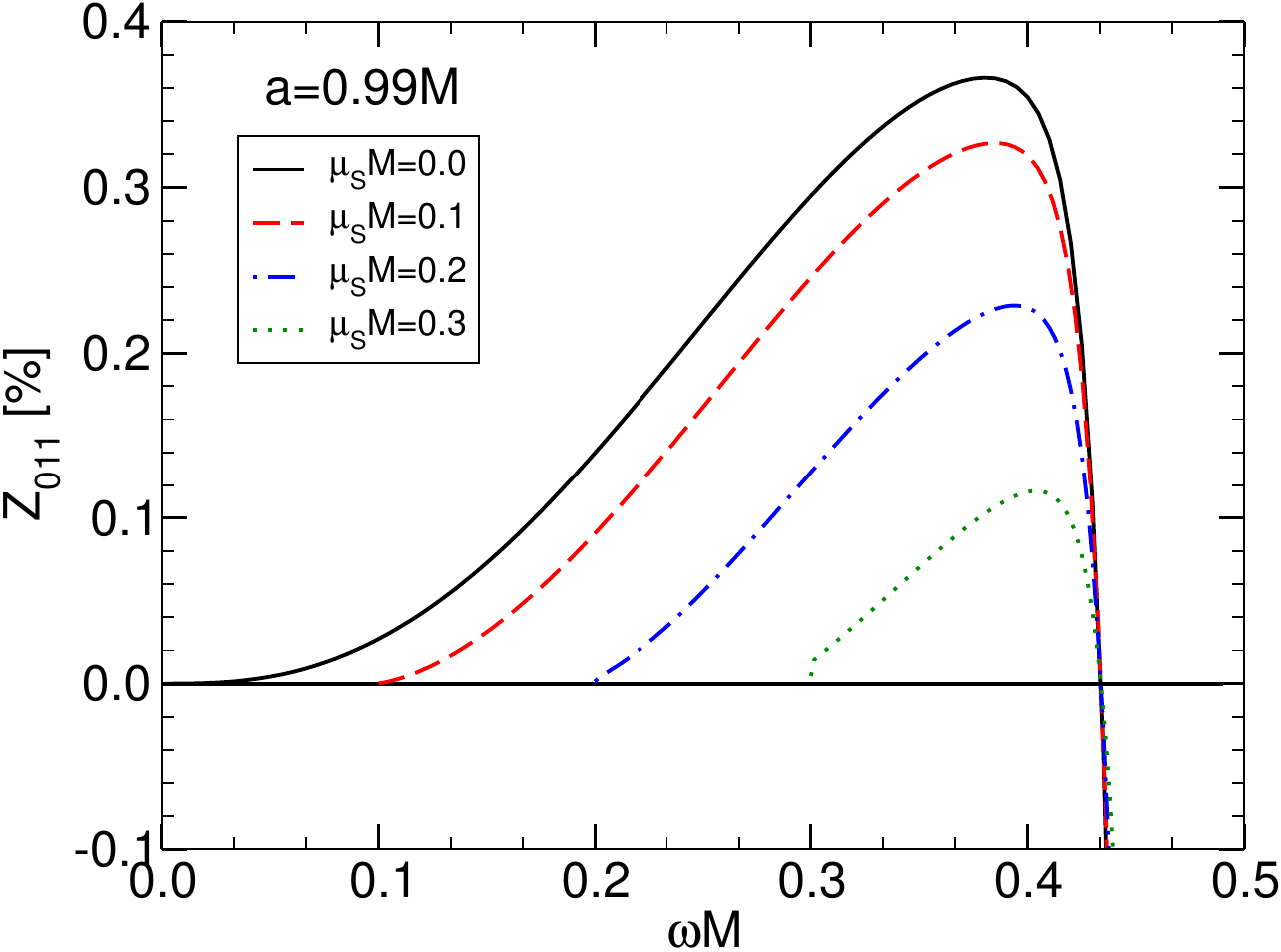,width=0.7\textwidth,angle=0,clip=true}
\end{tabular}
\caption{Amplification factor $Z_{0lm}=|{\cal R}|^2/|{\cal I}|^2-1$ as a function of the frequency for a  massive scalar field with $l=m=1$ and mass $\mu_S$ scattered off a Kerr BH with angular momentum parameter $a=0.99M$. Superradiance, $Z_{0lm}>0$, occurs when $\mu_S<\omega<m\Omega_{\rm H}$.
\label{fig:SR_massive}}
\end{center}
\end{figure*}
The previous analysis concerns massless fields, but the extension to massive fields is, in principle, straightforward. As an example, we show in Fig.~\ref{fig:SR_massive} the amplification factors
of a massive scalar field -- with mass $\mu_S \bar{h}$ --in the background of a Kerr BH. It is clear from Eq.~\eqref{eq:MFEoMScalar} that no propagation is possible for energies $\omega<\mu_S$.
Thus, superradiance can also occur for massive waves as long as the condition $\mu_S<\omega<m\Omega_{\rm H}$ is satisfied. Waves with $\omega<\mu_S$ are trapped near the horizon and are exponentially suppressed at infinity.
Figure~\ref{fig:SR_massive} shows that superradiance is {\it less} pronounced for massive fields; the larger the field 
mass $\mu_S$, the smaller the amplification factors are.

%%%%%%%%%%%%%%%%%%%%%%%%%%%%%%%%%%%%%%%%%%%%%%%%%%%%%%%%%%%%%%%%%%%%%%%%%%%%%%%%
\subsubsection{Scattering of plane waves}\label{sec:planewaves}
%%%%%%%%%%%%%%%%%%%%%%%%%%%%%%%%%%%%%%%%%%%%%%%%%%%%%%%%%%%%%%%%%%%%%%%%%%%%%%%%
%
\begin{figure*}[hbt]
\begin{center}
\begin{tabular}{cc}
\epsfig{file=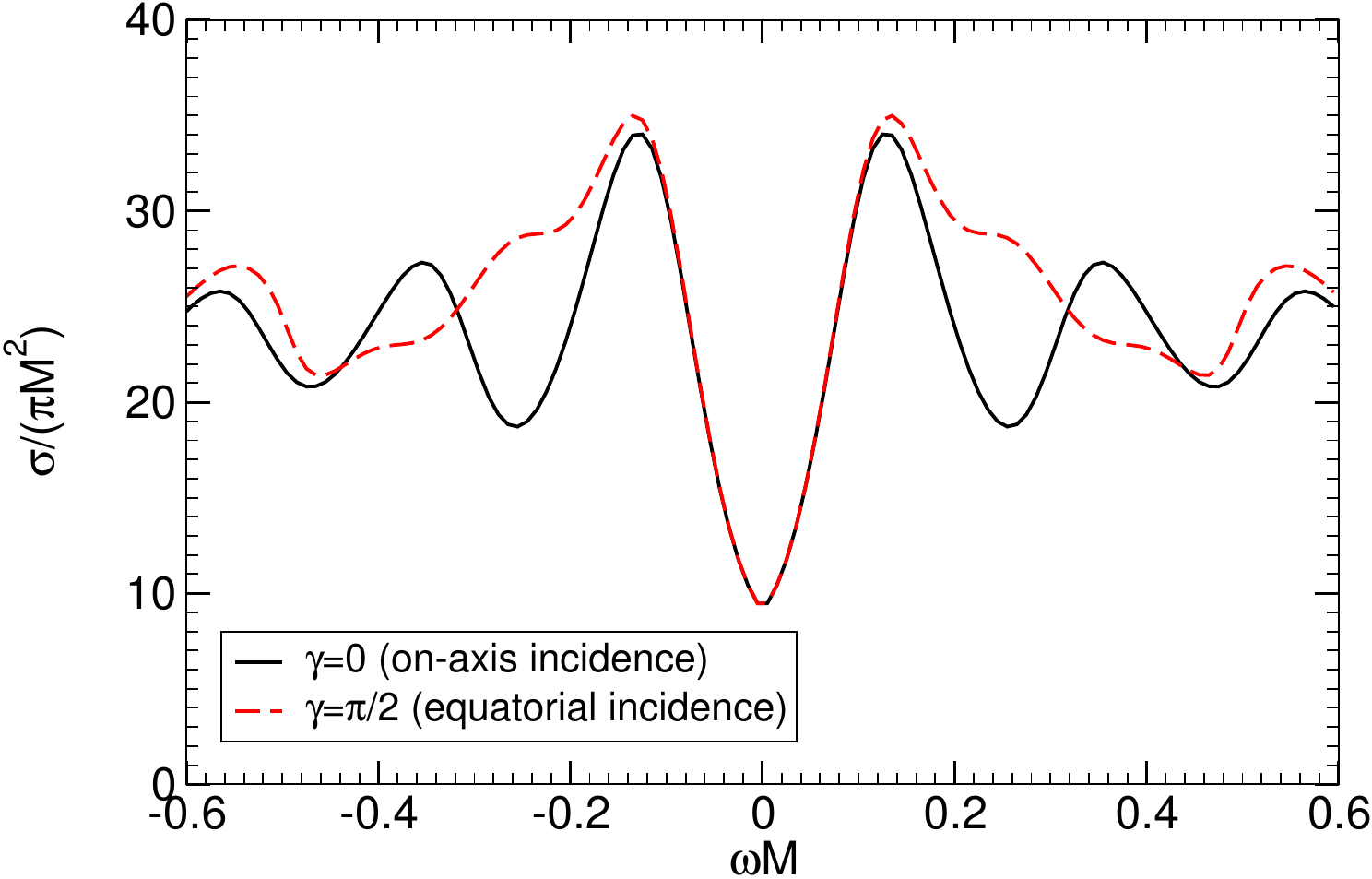,width=0.49\textwidth,angle=0,clip=true}&
\epsfig{file=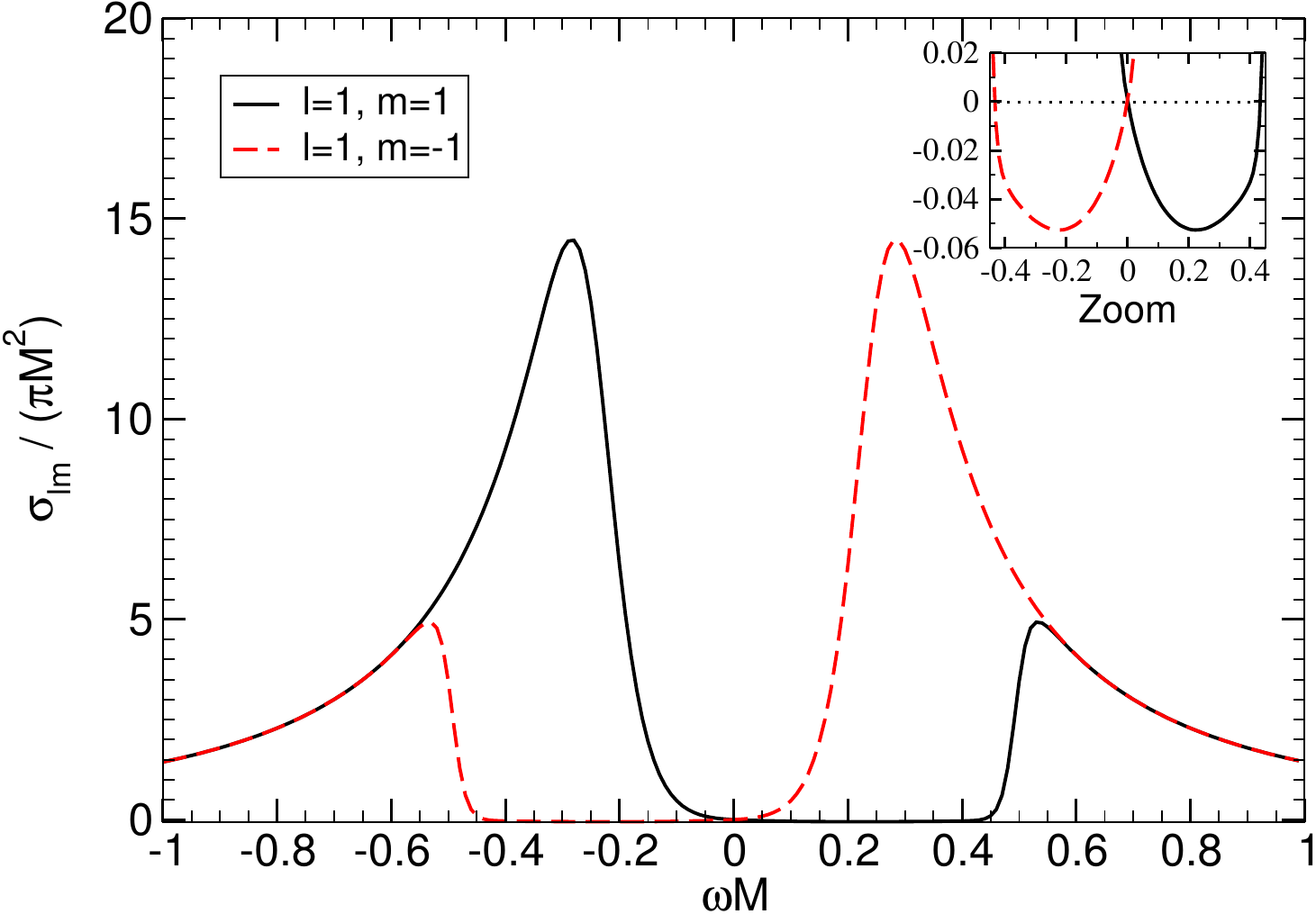,width=0.45\textwidth,angle=0,clip=true}
\end{tabular}
\caption{Absorption cross-section of a scalar plane wave incident on a rotating Kerr BH ($a/M=0.99$)
along the axis and equator. The left panel shows that the absorption cross-section is always positive, i.e., plane waves
are never superradiantly amplified. However, as expected some partial waves are indeed subjected to superradiance, as the right panel 
shows. 
\label{fig:cross_scalar}}
\end{center}
\end{figure*}
Generically, the field scattering off a BH is a superposition of multipoles. Of particular interest for a variety of applications is a field which is a plane wave at infinity. The multipolar expansion of a plane wave is straightforward to perform~\cite{Unruh:1976fm}.

%%%%%%%%%%%%%%%%%%%%%%%%
\paragraph{Scalar waves}
%%%%%%%%%%%%%%%%%%%%%%%%
Let us focus on a massless scalar field, and assume without loss of generality that there is an incoming monochromatic plane wave propagating along the $(\sin\gamma,0,\cos\gamma)$-direction. The absorption cross section $\sigma$ of a spinning BH can then be computed as~\cite{Unruh:1976fm,Macedo:2013afa}
\be
\sigma=\frac{4\pi^2}{\omega^2}\sum_{lm}\sigma_{lm}\equiv\sum_{lm}\left|S_{0lm}(\gamma)\right|^2\left(1-\frac{|{\cal R}|^2}{|{\cal I}|^2}\right)=-\frac{4\pi^2}{\omega^2}\sum_{lm}\left|S_{0lm}(\gamma)\right|^2\,Z_{0lm}\,,\label{cross_scalar}
\ee
where we used the asymptotic behavior as defined in \eqref{bc_kerr}. In other words, once the amplification factors have been computed for any $l$ and $m$, the cross-section is trivial to obtain. 

The results for two extreme cases --~incidence along the equatorial ($\gamma=\pi/2$) and axial ($\gamma=0$) directions~-- are summarized in Fig.~\ref{fig:cross_scalar}
for a rapidly spinning BH with $a/M=0.99$. Because $S_{0lm}(0)=0$ unless $m=0$, the cross-section for waves incident along the axial direction simplifies as
%%%
\begin{equation}
 \sigma(\gamma=0)=-\frac{4\pi^2}{\omega^2}\sum_{l=0}^\infty\left|S_{0l0}(0)\right|^2\,Z_{0l0}\,.\label{sigma_scalar_gamma0}
\end{equation}
%%%
For generic incidence angles, the total cross-section is symmetric along the $\omega=0$ axis, as could be anticipated 
from the general symmetry properties of the wave equation, cf. Eq.~\eqref{symmetry}. The first important conclusion is 
that plane scalar waves are never superradiantly amplified, or in other words, the absorption cross-section is positive 
for all values of frequency $\omega$. As might be expected from the general equation \eqref{cross_scalar}, because the 
amplification factor can become positive, some of the partial cross-sections $\sigma_{lm}$ {\it can} become negative, as 
shown in the right panel of Fig.~\ref{fig:cross_scalar} for the $l=|m|=1$ modes~\cite{Macedo:2013afa}. This result is true for neutral fields in a rotating BH background;
it can be shown, however, that  the scattering of a charged scalar by a charged BH can give rise to {\it negative} total absorption cross section, and that therefore planar scalar waves can be superradiantly amplified by BH~\cite{Benone:2015bst,Benone:2019all}.

%%%%%%%%%%%%%%%%%%%%%%%%%%%%%%%%
\paragraph{Electromagnetic waves}
%%%%%%%%%%%%%%%%%%%%%%%%%%%%%%%%
Scattering of monochromatic EM plane waves off a Kerr BH was studied in detail in 
Ref.~\cite{Leite:2018mon}. The numerical results agree with previous analytical approximations in the low- and 
high-frequency regimes.
For incidence along the axis of symmetry of a Kerr 
BH~\cite{Leite:2018mon},
\be
\sigma(\gamma=0)=\frac{4\pi^2}{\omega^2}\sum_{l=1}^{\infty}\left|S_{1l1}(0)\right|^2\,Z_{1l1}\,,\label{
sigma_EM_gamma0}
\ee
where the $Z_{1l1}$ are the amplification factors studied previously (see also Eq.~\eqref{sigma_grav_gamma0} below for 
the gravitational case.)

At variance with the scalar case, there exists a narrow parameter window (at low frequencies and moderate incidence 
angles) in which superradiant emission in the $l=1$ mode can exceed absorption in the non-superradiant modes. In other 
words, a planar EM wave can be superradiantly amplified by a spinning BH.
This effect might have observational consequences, for example in binary pulsar 
systems~\cite{Rosa:2015hoa,Rosa:2016bli}.

%%%%%%%%%%%%%%%%%%%%%%%%%%%%%%%%
\paragraph{Gravitational waves}
%%%%%%%%%%%%%%%%%%%%%%%%%%%%%%%%

%
\begin{figure*}[hbt]
\begin{center}
\begin{tabular}{c}
\epsfig{file=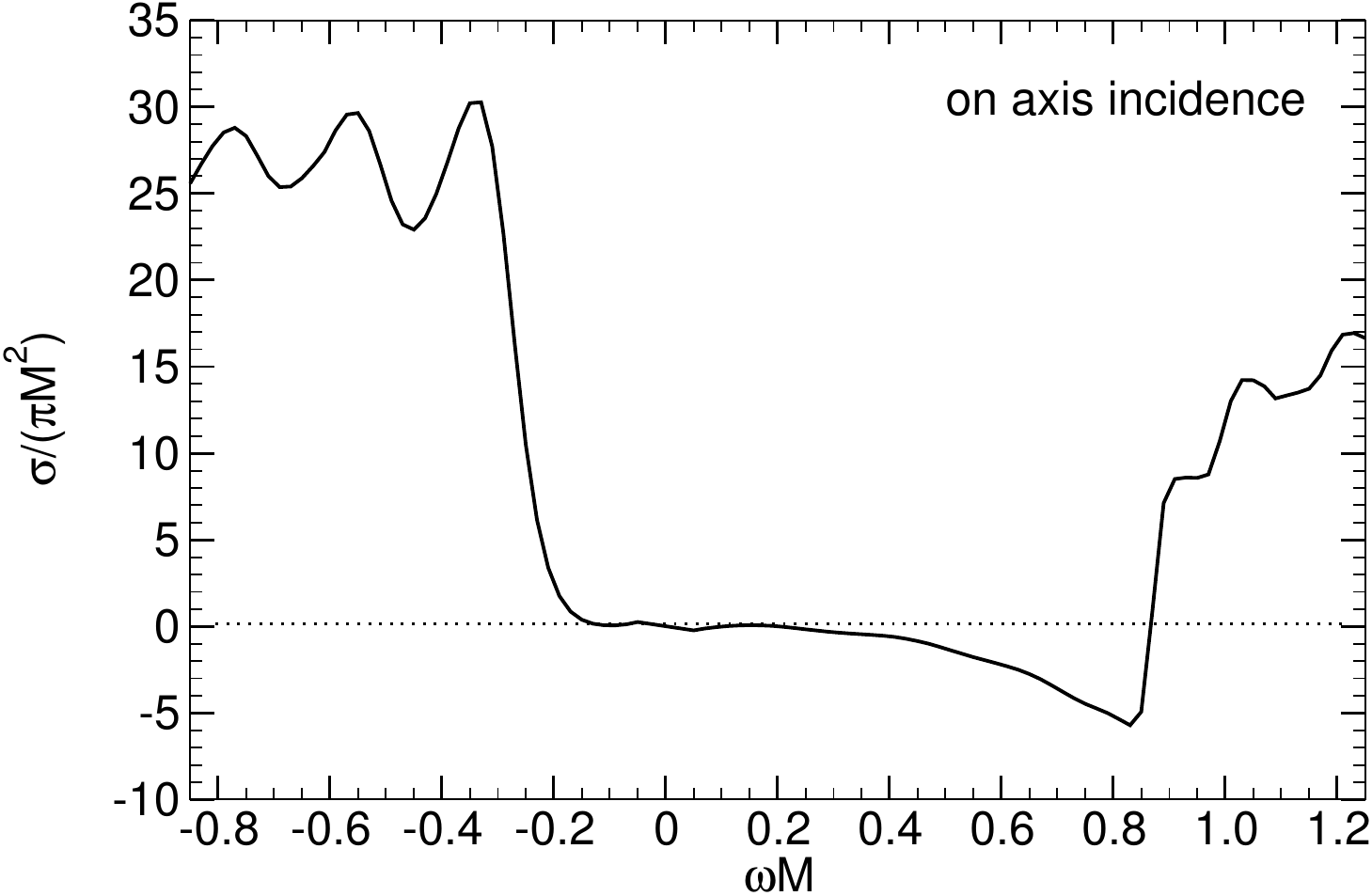,width=0.7\textwidth,angle=0,clip=true}
\end{tabular}
\caption{Absorption cross-section of a gravitational plane wave incident on a rotating Kerr BH ($a/M=0.99$)
along the rotation axis. The figure shows that counter-rotating ($\omega<0$) plane waves are more absorbed than co-rotating waves ($\omega>0$),
and that in the superradiant regime plane waves are amplified.
\label{fig:cross_grav}}
\end{center}
\end{figure*}
The scattering of plane GWs off rotating BHs is an important, decades-old problem~\cite{Chrzanowski:1976jb,Matzner:1977dn,1978ApJS...36..451M,1988sfbh.book.....F,Dolan:2008kf}. One of the important differences with respect to scalar waves, is that the symmetry along the $\omega=0$ axis is lost. In fact, for scattering along the symmetry axis of a Kerr BH, the low-frequency differential scattering cross reads
\be
M^{-2}\frac{d\sigma}{d\Omega}\approx \frac{\cos^8(\vartheta/2)}{\sin^4(\vartheta/2)}\left(1-4a\omega\sin^2(\vartheta/2)\right)+\frac{\sin^8(\vartheta/2)}{\sin^4(\vartheta/2)}\left(1+4a\omega\sin^2(\vartheta/2)\right)\,.
\ee
Thus, waves of different sign of $\omega$ are scattered differently from a rotating BH, generically inducing nontrivial polarization
on the scattered field.

The absorption cross-section of GWs off rotating BHs can be obtained in a similar fashion to those of scalar waves.
One finds, for incidence along the axis of symmetry of a Kerr BH~\cite{Chrzanowski:1976jb,Matzner:1977dn,1978ApJS...36..451M,1988sfbh.book.....F,Dolan:2008kf},
\be
\sigma(\gamma=0)=\frac{4\pi^2}{\omega^2}\sum_{l=2}^{\infty}\left|S_{2l2}(0)\right|^2\,Z_{2l2}\,,
\label{sigma_grav_gamma0}
\ee
where again $Z_{2l2}$ are the amplification factors studied previously\footnote{Note that a planar tensor wave along $\gamma=0$ in Cartesian coordinates will have a $\sin2\varphi$ modulation when transformed to spherical coordinates, in which the multipolar decomposition is performed. This explains why Eq.~\eqref{sigma_grav_gamma0} depends only on $|m|=2$ and on a sum over all multipolar indices $l\leq2$. Likewise, an EM wave along $\gamma=0$ would be modulated by $\sin\varphi$ and its cross-section would only depend on $|m|=1$, whereas the cross-section~\eqref{sigma_scalar_gamma0} for a scalar wave along $\gamma=0$ only depends on $m=0$.} (see Fig.~\ref{fig:Acomparison}). 
Because the amplification of GWs can be two orders of magnitude larger than that of scalars,
the cross-section for scattering of plane waves can now become {\it negative}. Thus, similarly to the EM 
case, plane GWs {\it can} be superradiantly
amplified. This is shown in Fig.~\ref{fig:cross_grav}, from which two features stand out: negative-frequency waves -- or waves counter-rotating with respect to the BH -- are always absorbed. On the other hand, positive-frequency waves (which co-rotate with the BH) are amplified in the superradiant regime.

Generically, a plane wave is a superposition of positive and negative-frequencies. For linearly polarized waves, for example,
one can easily show that the net effect always results in absorption~\cite{Dolan:2008kf}.

Recently, the scattering of plane waves off a Kerr BH has been analyzed in the context of superradiant amplification of 
the radiation from a BH-pulsar system~\cite{Rosa:2015hoa}. In this case, the pulsar's GW and EM luminosities show a 
characteristic modulation, which is due to superradiant scattering and depends on the pulsar position relative to the 
BH.

%%%%%%%%%%%%%%%%%%%%%%%%%%%%%%%%
\paragraph{Acoustic geometries}
%%%%%%%%%%%%%%%%%%%%%%%%%%%%%%%%
The scattering of sounds waves off acoustic BH geometries, in particular the one discussed in Section~\ref{sec:SR_analog}
was studied recently~\cite{Oliveira:2010zzb,Hod:2017lho}. Clear hints of superradiance were found, manifested as negative partial absorption ``lengths''
(as this is a $(2+1)$-dimensional geometry) for co-rotating modes at low frequencies.

%%%%%%%%%%%%%%%%%%%%%%%%%%%%%%%%%%%%%%%%%%%%%%%%%%%%%%%%%%%%%%%%%%%%%%%%%%%%%%%%%%%%%%%%%%%%%%%%%%%%%%%
\subsubsection{Nonlinear superradiant scattering from spinning black holes}\label{sec:super_nonlinear}
%%%%%%%%%%%%%%%%%%%%%%%%%%%%%%%%%%%%%%%%%%%%%%%%%%%%%%%%%%%%%%%%%%%%%%%%%%%%%%%%%%%%%%%%%%%%%%%%%%%%%%%
%
\begin{figure}[htbp]
\centerline{\includegraphics[width=0.7\textwidth]{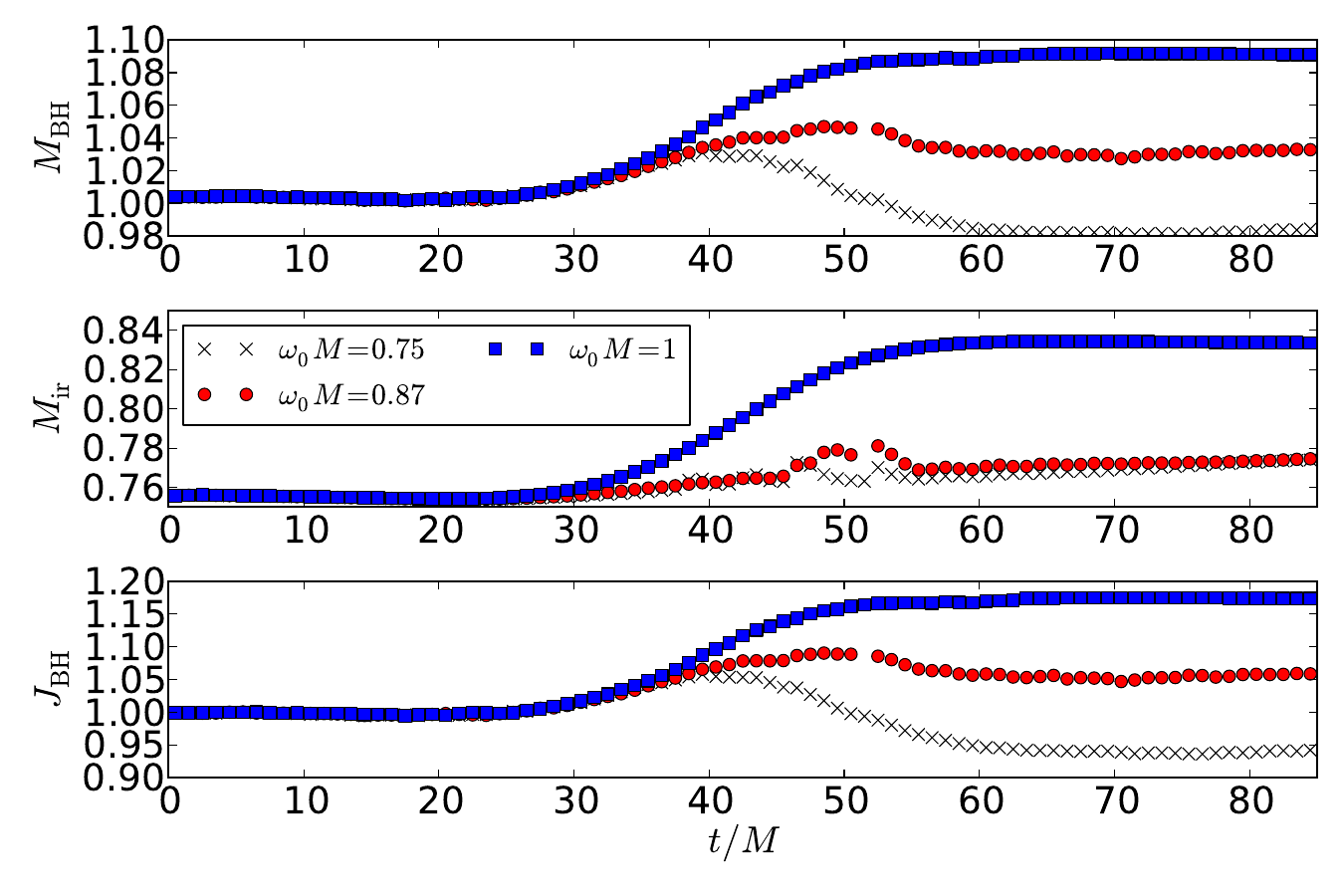}}
\caption{Evolution of a highly spinning BH ($a/M=0.99$) during interaction with
different frequency GW packets, each with initial mass $\approx 0.1M$.  Shown
(in units where $M=1$) are the mass, irreducible mass, and angular momentum of
the BH as inferred from AH properties. From Ref.~\cite{East:2013mfa}.}
\label{fig:ah_fig}
\end{figure}
In Section~\ref{backreaction_charged} we showed that when backreaction effects are taken into account then superradiance of charged fields does indeed extract mass and charge away from the BH. Fully nonlinear studies of superradiance are extremely scarse, with the notable exceptions of Ref.~\cite{Baake:2016oku} for charged BHs and Ref.~\cite{East:2013mfa} for spinning BHs.
The authors of Ref.~\cite{East:2013mfa} performed nonlinear scattering experiments, constructing initial data
representing a BH with dimensionless spin $a/M=0.99$, and an incoming
quadrupolar GW packet. Their results are summarized in Fig.~\ref{fig:ah_fig}, for three different wavepacket frequencies,
$M\omega=0.75,\,0.87,\,1$ (note that only the first is in the superradiant regime~\eqref{eq:superradiance_condition}). 
The wavepackets carry roughly $10\%$ of the spacetime's total mass. These results confirm that low frequency radiation does
extract mass and spin from the BH (both the mass $M_{\rm BH}$ and spin $J_{\rm BH}$ of the BH decrease for the superradiant wavepacket with $M\omega=0.75$), and that nonlinear results agree quantitatively with linear predictions for small wavepacket amplitudes~\cite{Teukolsky:1974yv}. To summarize, although further studies would certainly be interesting, superradiance is confirmed at full nonlinear level for rotating BHs.

%%%%%%%%%%%%%%%%%%%%%%%%%%%%%%%%%%%%%%%%%%%%%%%%%%%%%%%%
\subsection{Superradiance from stars} \label{sec:SRNS}
%%%%%%%%%%%%%%%%%%%%%%%%%%%%%%%%%%%%%%%%%%%%%%%%%%%%%%%%
As is clear from the entire discussion and from the classical examples of Section~\ref{sec:faces}, rotation and a 
dissipation channel are enough to trigger superradiance. As such, ordinary stars are also prone to superradiant 
amplification. A formal proof of this was recently produced for stars in GR~\cite{Richartz:2013unq}.

Explicit calculations require a modeling of dissipation. Ref.~\cite{Cardoso:2015zqa} developed a toy model similar to 
that adopted by Zeldovich in his original study (see also Ref.~\cite{Glampedakis:2013jya} who studied 
the correspondence between superradiance and tidal friction on viscous Newtonian anisotropic stars).
The toy model assumes the modified Klein-Gordon equation~\eqref{KGdiss} {\it inside} the star and in a co-rotating 
frame~\cite{zeldo1}. The term proportional to $\alpha$ in Eq.~\eqref{KGdiss} is added to break Lorentz invariance, and 
describes absorption on a time scale $\tau \sim 1/\alpha$. 
Following Zeldovich, if the frequency in the accelerated frame is $\omega$ and the field behaves as $e^{-i\omega 
t+im\varphi}$, then in the inertial frame the azimuthal coordinate is $\varphi=\varphi'-\Omega t$, and hence the 
frequency is $\omega'=\omega-m\Omega$ (see also Sec.~\ref{sec:rotSR}). In other words, the effective damping parameter 
$\alpha \omega'$ becomes negative in the superradiant regime and the medium amplifies -- rather than absorbing-- 
radiation~\cite{zeldovich1,zeldovich2}.

The constant $\alpha$ appearing in Eq.~\eqref{KGdiss} can be related to more physical 
parameters describing the microscopic details of the absorption process~\cite{Cardoso:2015zqa}. 
%%%
Indeed, this toy model was recently used to model the dissipation of weakly-interacting scalar fields within a pulsar, 
where the role of the dissipation is played by the interaction between these fields and neutrons~\cite{Kaplan:2019ako}.

Going beyond the above toy model, Ref.~\cite{Cardoso:2017kgn} studied superradiant scattering from conducting spinning 
compact stars and the superradiant instability in a model of massive vector fields with kinetic mixing terms with 
ordinary photons. They considered the theory
%
% \begin{widetext}
\be
S =\int d^4x \sqrt{-g} 
 \left( \frac{R}{16\pi} - \frac{1}{4}F^{\mu\nu}F_{\mu\nu} - \frac{\mu_V^2}{2}A_{\nu}A^{\nu}+4\pi  j^\mu 
A_\mu\right)+S_{\rm matter}\,,\label{action}
\ee
% \end{widetext}
%
which is a particular case of Eq.~\eqref{eq:MFaction}, with an extra term accounting for the standard coupling between 
the vector field $A_\mu$ and the current $j^\mu$, and with $S_{\rm matter}$ describing the action for the conducting 
fluid of the star.

The field equations were solved in the linearized regime, assuming a small vector field. The background Einstein's 
equations describe a spinning uncharged, rotating star made of material with conductivity $\sigma$ and proper charge 
density $\rho_{\rm EM}$. The star is characterized by a mass $M$, radius $R$, and angular velocity 
$\Omega\ll \Omega_K$, where $\Omega_K=\sqrt{M/R^3}$ is the Keplerian frequency.
To linear order in the spin, an ansatz for the metric is
\begin{equation}
ds^2=-F(r) dt^2+\frac{dr^2}{B(r)}-2r^2 \zeta(r) \sin^2\vartheta dt d\varphi 
+r^2d\Omega^2\,,
\end{equation}
where the radial coefficients $B$, $F$, and $\zeta$ are obtained by solving the background 
Einstein's equations.

The coupling between the vector and the material is given by the constitutive Ohm's law, which 
in covariant form reads~\cite{Bekenstein:1998nt},
\be
j^{\alpha}=\sigma F^{\alpha\beta} u_{\beta} +\rho_{\rm EM} u^\alpha\,,
\label{eq:conductivitycurrent}
\ee
where all quantities are computed in the frame of the material whose 4-velocity is $u^{\alpha}$. 

The linearized Maxwell/Proca equations are solved to linear order in the star's spin, i.e. to ${\cal 
O}(\Omega/\Omega_K)$. The result for the axial sector of the perturbations is particularly simple and can be written 
as a single, second-order, differential equation
\beq
&&\frac{d^2 a}{dr_*^2}+\left(\omega^2-2m\omega\zeta(r)-V\right)a=0\,, \label{axial_final}\\
&&V=F\left(\frac{l(l+1)}{r^2}+\mu_V^2-\frac{4i\pi\sigma (\omega-m\Omega)}{\sqrt{F}}\right)\,.
\eeq
where $dr/dr_*=\sqrt{BF}$. Note the combination $\omega-m\Omega$ in the last term of the 
above effective potential, which arises 
naturally from the field equations. 

\begin{figure}[th]
  \includegraphics[width=0.48\textwidth]{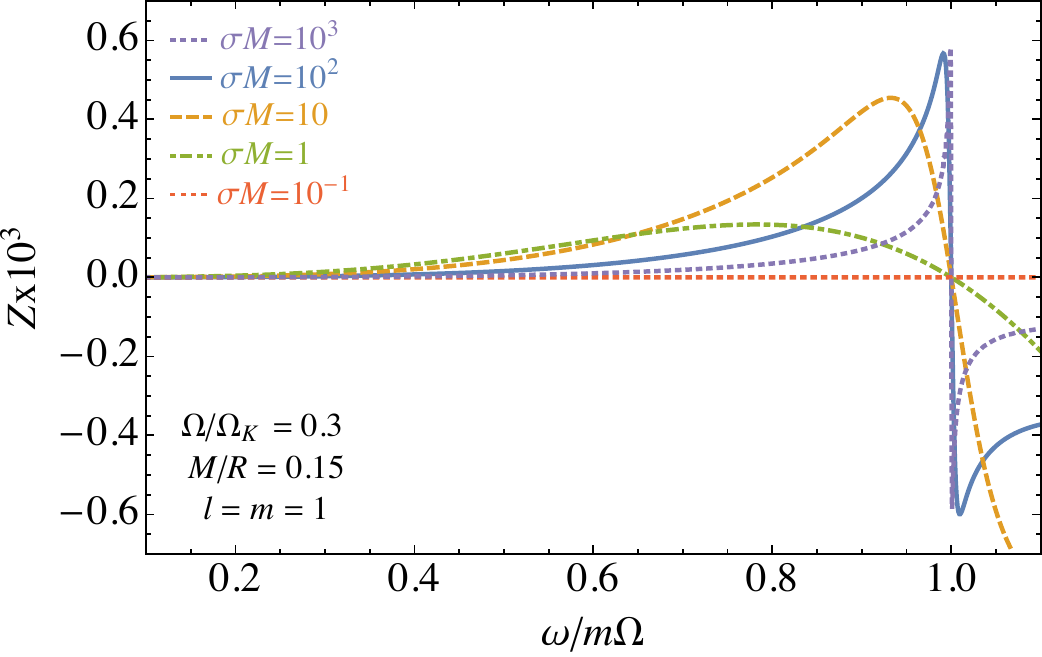}
  \includegraphics[width=0.48\textwidth]{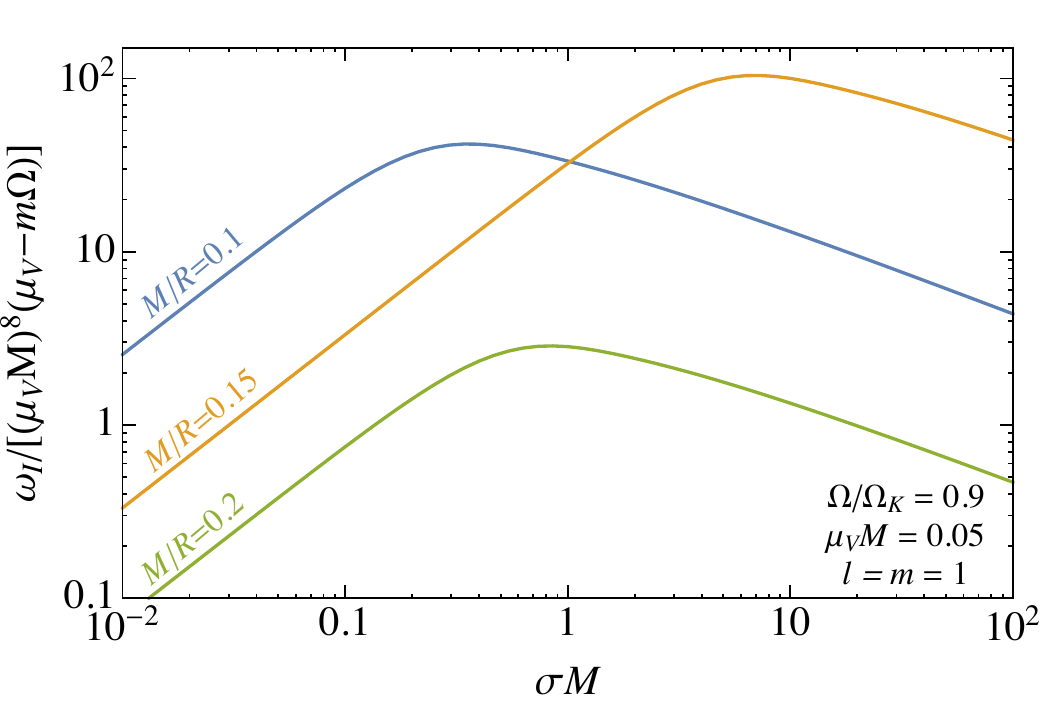}
\caption{Left: Amplification factor for axial vector $l=m=1$ modes as a function of the frequency, for 
a moderately-spinning star ($\Omega=0.3\,\Omega_K$) with compactness $M/R=0.15$ and for different values of the 
conductivity $\sigma$.
Right: The prefactor in square brackets of Eq.~\eqref{wIaxial} as a function of $\sigma M$ and for different values of 
the compactness at fixed $\Omega/\Omega_K=0.9$. A fit of the numerical data is consistent with Eq.~\eqref{wIaxial} with 
$(\alpha_1,\alpha_2)\sim(39,0.13)$, $(429,11)$ and $(4.2,0.48)$ for $M/R=0.1,0.15,0.2$, respectively.
From Ref.~\cite{Cardoso:2017kgn}.
} \label{fig:SRstar}
\end{figure}

The superradiant scattering factor computed in Ref.~\cite{Cardoso:2017kgn} is shown in the left 
panel of Fig.~\ref{fig:SRstar}. As expected, $Z>0$ when the superradiant condition is satisfied, $\omega<m\Omega$. The 
amplification factor grows with $\sigma$, until it saturates in the large-$\sigma$ limit displaying a sharp maximum at 
$\omega\lesssim m\Omega$. The amplification grows with the compactness and with the spin of the object.

An analytical approximation of the amplification factor in the Newtonian limit and for small 
conductivity reads~\cite{Cardoso:2017kgn}
\be
Z\equiv -\frac{2^{1-2l}\pi^2}{\Gamma[l+3/2]\Gamma[l+5/2]}\sigma R^2\left(\omega -m\Omega\right)(\omega 
R)^{2l+1}\,.
\ee
The above expression agrees remarkably well with the exact numerical result up to $M/R\sim 0.2$ and for $\sigma M\ll1$.
Furthermore, it agrees {\it exactly}~\cite{Cardoso:2017kgn} with the results that can be obtained in flat space by 
invoking the membrane paradigm~\cite{MembraneParadigm}, where horizons are endowed with a surface conductivity of 
$1/4\pi$.

For large conductivities, in the Newtonian limit the numerical results are well approximated by
\be
Z=k_l\frac{\left(\omega R\right)^{2l+1}}{\sqrt{\sigma R^2\left(m\Omega-\omega\right)+c_l}}\left[1+\frac{1}{2\sigma 
R^2\left(m\Omega -\omega\right)}\right]^{-1}\,,\label{large_sigma_Z}
\ee
in the superradiant regime, with $k_1\sim 0.78,\,k_2\sim0.09$ and $c_1\sim2, c_2\sim 25$. 
The amplification factor is peaked at $\omega-m\Omega \sim 1/(\sigma R)$, and bounded. The analytical expression above 
is not very accurate close to the peak of the amplification factor. For $\sigma M\gg1$, 
the $l=m=1$ peak is well described by
\be
Z_{\rm max} \sim (0.48-0.78 M/R) (\Omega R)^3\,,
\ee
where, interestingly, the prefactor decreases at large compactness.

It can be shown that a finite bulk conductivity results in an effective resistivity in the stellar magnetosphere of neutron stars. In the context of axionic fields, the axion can mix with photon modes which superradiantly scatter off the magnetosphere, extracting rotational energy~\cite{Day:2019bbh}. Thus, superradiance in stars also arises naturally in other standard model extensions.

%%%%%%%%%%%%%%%%%%%%%%%%%%%%%%%%%%%%%%%%%%%%%%%%%%%%%%%%%%%%%%%%%%%%%%%%%%%%%%%%%%
\subsection{Superradiance in analogue black hole geometries} \label{sec:SR_analog}
%%%%%%%%%%%%%%%%%%%%%%%%%%%%%%%%%%%%%%%%%%%%%%%%%%%%%%%%%%%%%%%%%%%%%%%%%%%%%%%%%%
%
\begin{figure*}[hbt]
\begin{center}
\begin{tabular}{cc}
\epsfig{file=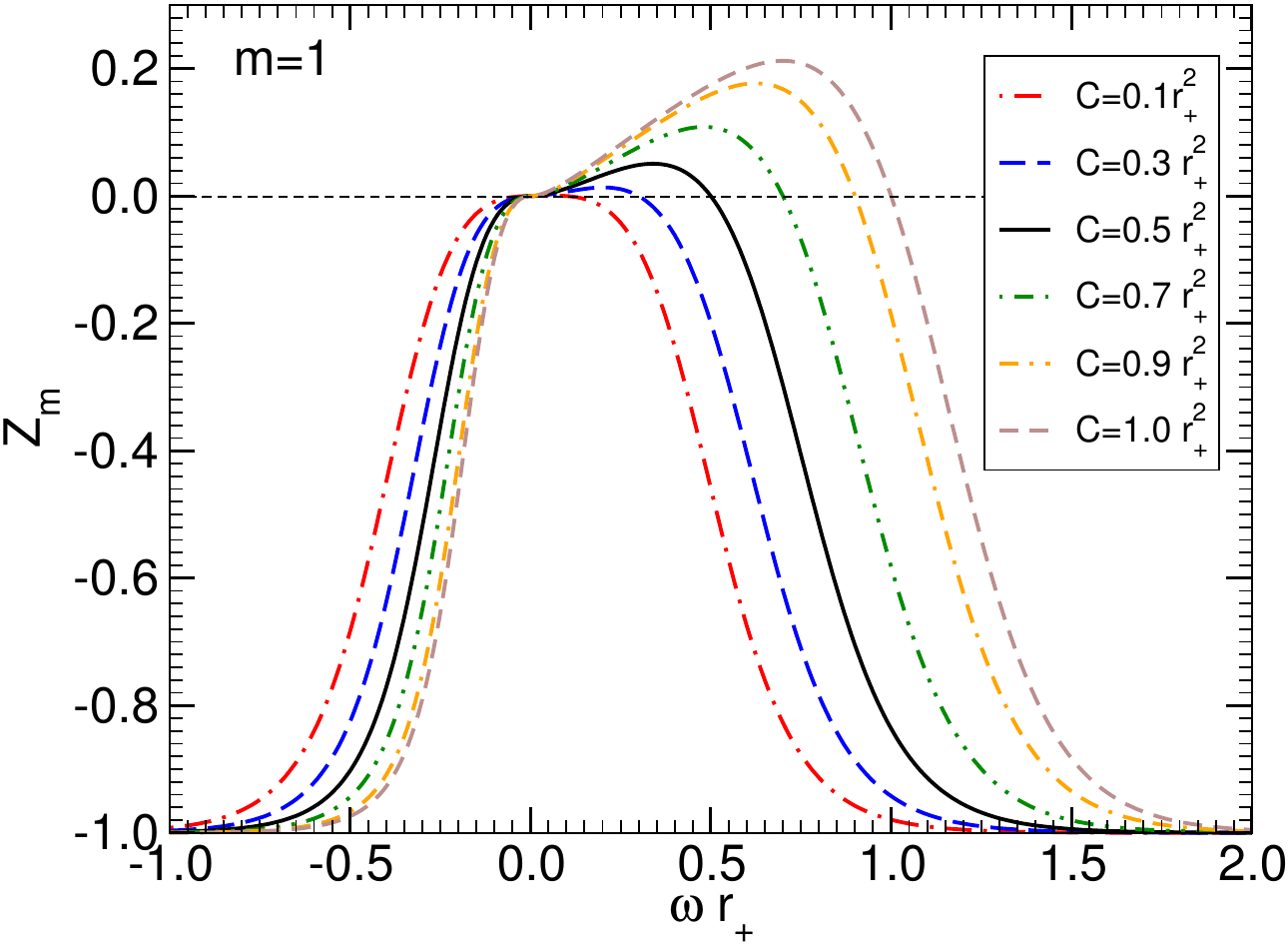,width=0.48\textwidth,angle=0,clip=true}&
\epsfig{file=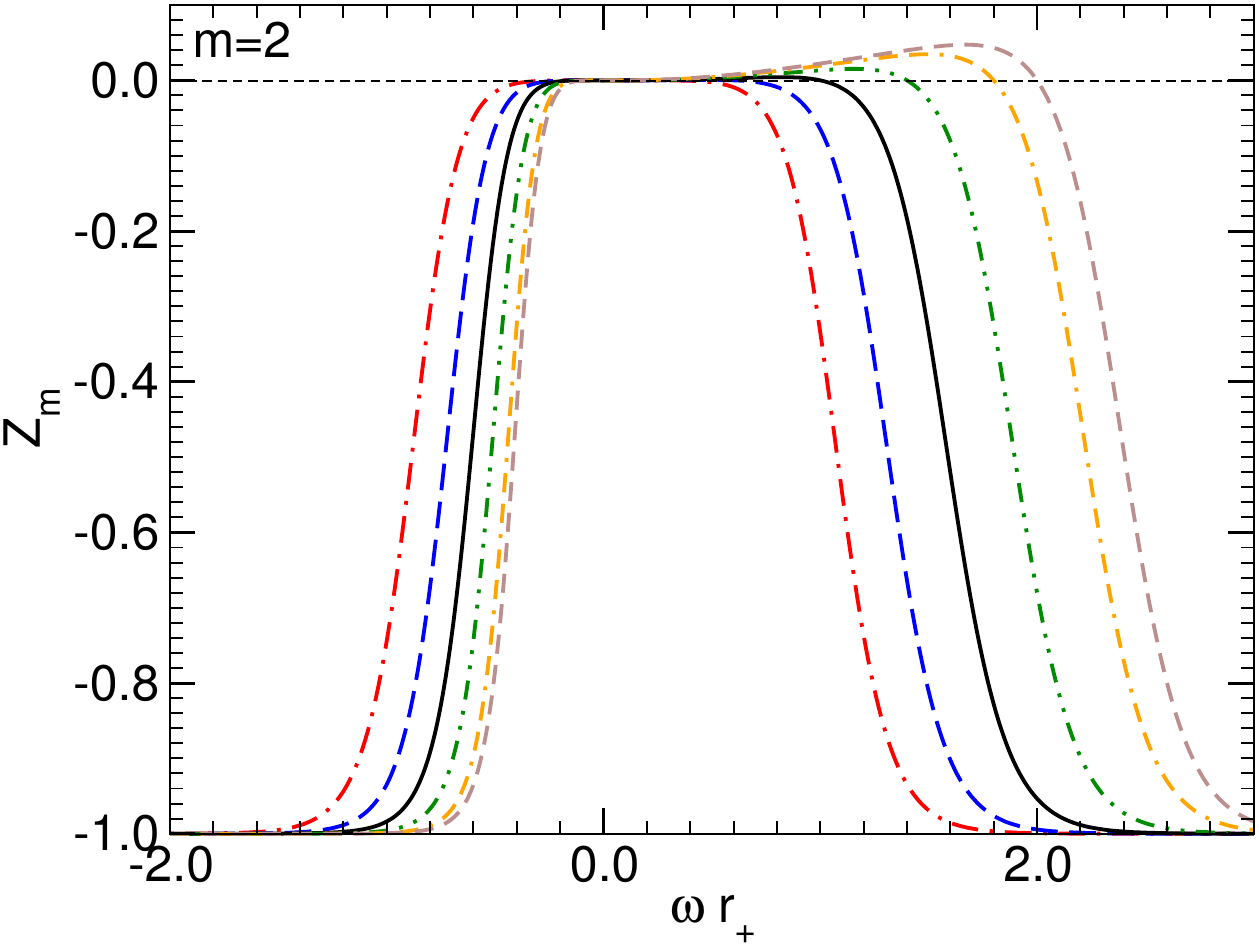,width=0.48\textwidth,angle=0,clip=true}
\end{tabular}
\caption{Amplification factor $Z_{m}$ for the draining vortex as a
function of $\omega$ for $m=1$ (left panel) and $m=2$ (right
panel), for a unit-amplitude incident wave. Results are normalized by the effective horizon $r_+$.
Each curve corresponds to a different value of rotation $C$, as
indicated. Adapted from Ref.~\cite{Berti:2004ju}.\label{fig:superradiantfactor}}
\end{center}
\end{figure*}
The construction outlined in Sec.~\ref{sec:shock_waves} established a formal equivalence between the propagation of sound waves
and the Klein Gordon equation in an effective, curved spacetime. Under certain conditions, a horizon in the effective metric
is present, when the local fluid velocity surpasses the local sound speed. This object is usually called an acoustic BH or ``dumb hole'' (cf. Ref.~\cite{Barcelo:2005fc} for a review).
Superradiance in acoustic BH geometries was studied in some detail for the two-dimensional draining geometry (``draining bathtub vortex''), described by a two dimensional fluid flow
\be
{\vec v}=\frac{-A {\vec r}+C {\vec\phi}}{r}\,,
\ee
in polar coordinates, where ${\vec r}$ and ${\vec\phi}$ are orthogonal unit basis vectors.
The flow above is that of an ideal fluid, which is locally irrotational (vorticity free), barotropic and inviscid.
The quantity $A$ thus measures the flow radial speed and the circulation $C$ measures its angular speed.
In these setups the notion of horizon and ergospheres is very intuitive: the effective spacetime has an acoustic horizon at the point where the radial speed is equal to the local sound speed, $r_+=Ac^{-1}$ and an ergosurface at
the location where the total speed equal the speed of sound, $r_{\rm ergo}^2=c^{-2}(A^2+C^2)$.

With the following coordinate transformation~\cite{Berti:2004ju},
\begin{eqnarray}
dt&\rightarrow& d\tilde{t}=dt-\frac{Ar}{r^2c^2-A^2}dr\\
d\phi&\rightarrow& d\tilde{\phi}=d\phi-\frac{CA}{r(r^2c^2-A^2)}dr\,,
\label{coordtransf}
\end{eqnarray}
the effective metric \eqref{metricvisserinv} takes the form
\be
ds^2=-\left(1-\frac{A^2+C^2}{c^2r^2} \right)c^2 d\tilde{t}^2 + \left(1-\frac{A^2}{c^2r^2} \right )^{-1}dr^2-2C d\tilde{\phi}d\tilde{t}+r^2d\tilde{\phi}^2\,.\label{metric2}
\ee

Superradiance was studied in this effective acoustic spacetime in the frequency domain, by studying the amplification factors~\cite{Berti:2004ju}. 
For an incident wave of amplitude ${\cal I}$, the reflection coefficients are shown in Fig.~\ref{fig:superradiantfactor}, the amplification factors for fluxes are $Z_{m}=\left|{\cal R}\right|^2-|{\cal I}|^2$;
the reflection coefficient depends only on the dimensionless parameter $C/A$~\cite{Berti:2004ju} and therefore without loss of generality one can set $A=c=1$.
the amplification factor grows with rotation parameter $C$, albeit slowly (the numerics indicate a logarithmic growth at large $C$). At a moderately large value of $C=1$,
the peak amplification factor for $m=1$ modes is 21.2 \%. Amplification factors higher that 100\% are extremely hard to achieve, which might be connected to entropy bounds, see Section
\ref{open4} for a further discussions on this.
Superradiant wave scattering for the same geometry was analyzed in the time domain in Ref.~\cite{Cherubini:2005dha}. These studies were complemented by a low-frequency analysis~\cite{Lepe:2004kv} and by an energy flux analysis~\cite{Choy:2005qx}.

Recently, Ref.~\cite{Richartz:2014lda} considered a similar, but slightly more realistic draining geometry taking into account the varying depth of water.
Superradiance in this analog system depends now on two parameters, and can be as large as 60\% or higher.

Analogue geometries can be realized outside acoustic setups, and include Bose-Einstein condensates for instance~\cite{Barcelo:2005fc}.
Superradiant scattering of sound wave fluctuations from vortex excitations of Bose-Einstein condensates
was considered in Refs.~\cite{Federici:2005ty,Ghazanfari:2014fta}. Bose-Einstein condensates are also interesting models for 
dark-matter halos and boson stars self-gravitating scalar fields~\cite{Liebling:2012fv}; in this context, a gravitational analogue description also displays superradiant scattering~\cite{Kuhnel:2014bga}.

It has also been realized that effective photon-photon interactions of a monochromatic laser beam propagating in a nonlinear medium lead to a collective behaviour of the many photon system, leading to a superfluid behaviour~\cite{PhysRevLett.69.1644,Vocke:15}. These fluids are naturally prone to mimic curved spacetimes as we described.
Kerr-like geometries and superradiance in such setups has been reported recently~\cite{PhysRevA.80.065802,Prain:2019jqk}.

%%%%%%%%%%%%%%%%%%%%%%%%%%%%%%%%%%%%%%%%%%%%%%%%%%%%%%%%%%%%%%%%%%%%%%%%%%%%%%%%%%%%%%%%%%%%%%%%%%%%%%%%%%%%%%%%%
\subsection{The experimental observation of superradiance\label{sec:lab2}}
%%%%%%%%%%%%%%%%%%%%%%%%%%%%%%%%%%%%%%%%%%%%%%%%%%%%%%%%%%%%%%%%%%%%%%%%%%%%%%%%%%%%%%%%%%%%%%%%%%%%%%%%%%%%%%%%%
%
\begin{figure}
\begin{center}
\begin{tabular}{cc}
\epsfig{file=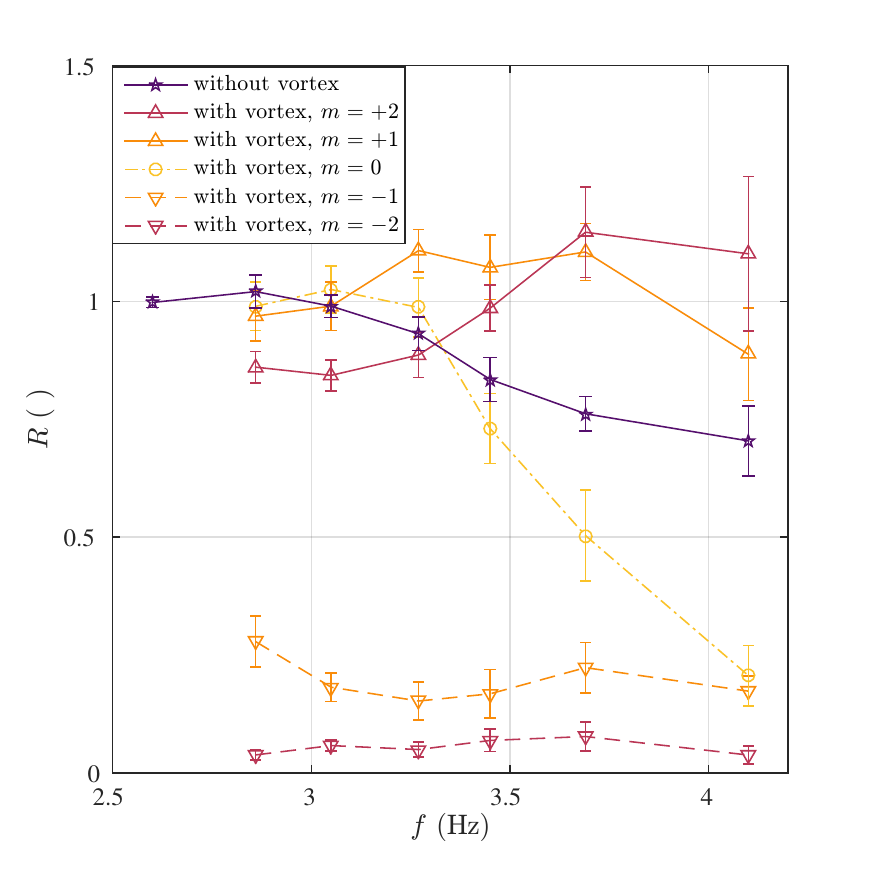,width=0.5\textwidth,angle=0,clip=true}
\end{tabular}
\caption{Reflection coefficients of surface waves in a tank where water was made to rotate in a ``vortex''. The results are shown for various frequencies and various azimuthal number $m$.  The purple line (star points) shows the reflection coefficients of a plane wave in standing water of the same height. There is a significant damping for frequencies above $3\,{\rm Hz}$. There is statistical evidence for amplification of positive $m$ modes at frequencies $3.5-4\,{\rm Hz}$. Taken from Ref.~\cite{Torres:2016iee}.
\label{fig:lab_super}}
\end{center}
\end{figure}
The example above concerned sound or surface waves on a background fluid at rest. The energy was being transferred 
from a spinning cylinder to the waves through local interactions between these two media. One can do away with the 
material cylinder if the fluid is made to rotate at a speed faster than the local sound speed. This will force the sound 
to co-rotate with the fluid, providing an interaction mechanism. Such situation is equivalent to the appearance of 
ergoregions in curved spacetime, which are the essential ingredient for superradiance.
If an inner core is made where the fluid is dumped into, then such apparatus can be made to mimic a rotating BH 
spacetime. Such a simple experimental setup was realized recently~\cite{Torres:2016iee}.

Surface waves were setup on a water tank, wherein water was pumped continuously at the corners, and made to
drain through a hole. This established a stationary, rotating draining flow that satisfied the conditions above.
The free-surface waves were measured with a high speed sensor. These waves were subsequently decomposed in the basis $e^{im\phi}$ (as in the previous section), with $m$ an integer number. The results are summarized in Fig.~\ref{fig:lab_super} and clearly indicate superradiance of co-rotating (positive $m$) modes in a certain frequency range. This is the first experimental support for the existence of rotational superradiance.

%%%%%%%%%%%%%%%%%%%%%%%%%%%%%%%%%%%%%%%%%%%%%%%%%%%%%%%%%%%%%%%%%%%%%%%%%%%
\subsection{Superradiance in higher dimensional spacetimes}
%%%%%%%%%%%%%%%%%%%%%%%%%%%%%%%%%%%%%%%%%%%%%%%%%%%%%%%%%%%%%%%%%%%%%%%%%%%
With the exception of the boosted black string just discussed, we have so far only considered BH superradiance in 4-dimensional spacetimes. Generalization to higher dimensions can be done along the same lines\footnote{There are no gravitational degrees of freedom in less than 4 dimensions, and a BH solution only exists for a negative cosmological constant, the so-called BTZ solution~\cite{Banados:1992wn}. This solution has some similarities with the Kerr-AdS metric and, as we shall discuss in Section~\ref{sec:super_nonAF}, superradiance does not occur when reflective boundary conditions at infinity are imposed~\cite{Ortiz:2011wd}.}. The multitude of black objects in higher dimensions makes this an interesting and relatively unexplored subject (for a review on BHs in higher dimensions see Ref.~\cite{Emparan:2008eg}).

% %%%%%%%%%%%%%%%%%%%%%%%%%%%%%%%%%%%%%%%%%%%%%%%%%%%%%%%%%%%%%
% \subsubsection{Rotational superradiance in higher dimensions}
% %%%%%%%%%%%%%%%%%%%%%%%%%%%%%%%%%%%%%%%%%%%%%%%%%%%%%%%%%%%%%
From the rigidity theorem, a stationary $D$--dimensional BH must be axisymmetric~\cite{Hawking:1971vc,Hollands:2006rj}, meaning that it must have $D-3$ rotational Killing vectors in addition to the time translation Killing vector. Thus, to study superradiance in higher-dimensions, one must take into account that there exist at most $D-3$ rotation axis. The condition for superradiance in the background of a five-dimensional, topologically spherical BH was computed in Ref.~\cite{Frolov:2002xf}; using the area theorem this condition was generalized to arbitrary dimensions for Myers-Perry BHs with a single angular momentum parameter~\cite{Jung:2005cn} and finally with multiple angular momentum parameters in Ref.~\cite{Jung:2005nf}. More recently the condition was computed for asymptotically flat rotating BHs with generic spacetime dimension and horizon topology using a Wronskian approach~\cite{Kodama:2007sf}. The generalized superradiance condition~\eqref{eq:superradiance_condition} is given by
\be~\label{eq:superradiance_D_dimensions}
\omega<\sum_{i=1}^{i\leq D-3} m_i \Omega_{\rm H}^i\,, 
\ee
where $m_i$ is a set of integers, corresponding to the azimuthal numbers with respect to the different rotation axis, and $\Omega_{\rm H}^i$ represents the multi-component angular velocity of the horizon.

Amplification factors for a scalar field where computed for Myers-Perry BH with a single angular momentum parameter in Refs.~\cite{Brito:2012gj,Creek:2007pw,Casals:2008pq}. They showed that the superradiant amplification is less efficient in higher dimensions and the maximum amplification factor decreases with the dimension of the spacetime; for a doubly spinning Myer-Perry BH in 5D, the amplification factors were computed in Ref.~\cite{Jung:2005pk}. 
Motivated by extra dimensional models which predicted the possibility of creating micro BHs in particle accelerators such as the LHC, amplification factors for a singly-spinning higher-dimensional Myers-Perry BH induced on an asymptotically flat 4--dimensional brane were also computed. This was done for spin-0 particles~\cite{Harris:2005jx,Ida:2005ax,Creek:2007sy} and spin-1 fields~\cite{Casals:2005sa}. Superradiant amplification on the brane was shown to be much larger than in the $D$--dimensional bulk and to be greatly enhanced compared to the 4--dimensional Kerr BH case. 

Interesting tidal effects related to the superradiant energy extraction in higher-dimensions were shown to occur in Refs.~\cite{Cardoso:2012zn,Brito:2012gw}. As first suggested in Ref.~\cite{Cardoso:2012zn} and later confirmed~\cite{Brito:2012gw}, the energy extracted by superradiant scalar waves generated by the circular motion of a point particle around a singly-spinning Myers-Perry BH could be higher than the energy lost to infinity through the emission of scalar waves, in contrast to the 4--dimensional case, where the BH energy absorption (or extraction) is negligible compared to the energy emitted to infinity~\cite{Poisson:1994yf}.

%%%%%%%%%%%%%%%%%%%%%%%%%%%%%%%%%%%%%%%%%%%%%%%%%%%%%%%%%%%%%%%%%%%%%%%%%%%%%%%%%%%%%%
\subsection{Superradiance in nonasymptotically flat spacetimes}\label{sec:super_nonAF}
%%%%%%%%%%%%%%%%%%%%%%%%%%%%%%%%%%%%%%%%%%%%%%%%%%%%%%%%%%%%%%%%%%%%%%%%%%%%%%%%%%%%%%
The literature on superradiance amplification from BHs in Einstein's theory with a cosmological constant is limited. The dS and the AdS cases behave in a completely different way: when the cosmological constant $\Lambda>0$ new effects related to the presence of a dS cosmological horizon can occur, whereas when $\Lambda<0$ the AdS boundary can effectively confine superradiantly-amplified waves thus providing the arena for BH bomb instabilities. The latter effect is discussed at length in Sec.~\ref{sec:bombs} so in this section we focus only on the superradiant amplification, neglecting possible instabilities that it might trigger.

%%%%%%%%%%%%%%%%%%%%%%%%%%%%%%%%%%%%%%%%%%%%%%%%%%%%%%%%
\paragraph{Extracting energy from dS BHs}
%%%%%%%%%%%%%%%%%%%%%%%%%%%%%%%%%%%%%%%%%%%%%%%%%%%%%%%%
Superradiance of Kerr-dS BHs has also been studied~\cite{Tachizawa:1992ue}. Extending the analysis of Sec.~\ref{sec:ABC}, the radial Teukolsky equation can be solved in the asymptotic regions and the solution reads as in Eq.~\eqref{bound2} with
\begin{equation}
 k_H=\omega-m\Omega_{\rm H}\,,\qquad k_\infty =\omega-m\Omega_c\,,
\end{equation}
where $\Omega_c$ is the angular velocity of the cosmological horizon at $r=r_c$. Imposing ${\cal O}=0$ at the event horizon, Eq.~\eqref{reflectivity} takes the form
%%%
\begin{equation}
 |{\cal R}|^2=|{\cal I}|^2-\frac{\omega-m\Omega_{\rm H}}{\omega-m\Omega_c} |{\cal T}|^2 \,,\label{reflectivity_dS}
\end{equation}
%%%
and therefore superradiance occurs only when
%%%
\begin{equation}
 m\Omega_c<\omega<m \Omega_{\rm H}\,. \label{superradiance_dS}
\end{equation}
%%%
Although the range of superradiant frequencies is smaller than in the asymptotically flat case, the maximum superradiance amplification is slightly larger for positive values of $\Lambda$~\cite{Tachizawa:1992ue}.

On a more formal account, Ref.~\cite{Georgescu:2014yea} has proved asymptotic completeness for a class of Klein-Gordon equations which allow for superradiance, including the scalar equation on a Kerr-dS BH (see also references in Ref.~\cite{Georgescu:2014yea} for recent formal development on the local energy for the wave equation on the Kerr-dS metric).

Finally, an interesting effect related to dS superradiance was recently discovered in Ref.~\cite{Zhu:2014sya}. There, it was shown that RN-dS BHs are linearly unstable to spherical, charged scalar perturbations. The unstable modes were subsequently found to satisfy a superradiance condition analog to Eq.~\eqref{superradiance_dS} for static charged dS BHs~\cite{Konoplya:2014lha}. 

%%%%%%%%%%%%%%%%%%%%%%%%%%%%%%%%%%%%%%%%%%%%%%%%%%%%%%%%%%%%%%%%%%%%%%%%%%%
\paragraph{Extracting energy from black holes in AdS backgrounds}
%%%%%%%%%%%%%%%%%%%%%%%%%%%%%%%%%%%%%%%%%%%%%%%%%%%%%%%%%%%%%%%%%%%%%%%%%%%
AdS spacetime is not globally hyperbolic, so fields which satisfy a hyperbolic wave equation on AdS might not have a 
well-defined dynamics. Nonetheless scalar, vector and GWs propagating on AdS can be shown to possess some conserved 
energy, and their dynamics correspond to that defined by choosing some positive, self-adjoint wave 
operator~\cite{Ishibashi:2004wx}. Such formal analysis also determines all possible boundary conditions that can be 
imposed at AdS infinity.

These boundary conditions are indeed crucial for superradiance. It was shown that, using ``reflective'' boundary conditions (i.e. either Dirichlet or Neumann) at timelike infinity, all modes of a scalar field on a Kerr-Newman-AdS BH are \emph{not} superradiant whereas, for ``transparent'' boundary conditions, the presence of superradiance depends on the definition of positive frequencies, which is subtle in AdS~\cite{Winstanley:2001nx}.
For those BHs having a globally timelike Killing vector, a natural definition of positive frequency implies absence of superradiance. This is to be contrasted with the situation  in asymptotically
flat space previously discussed, where superradiance occurs regardless of the definition of positive frequency. This result has important implications for constructing a quantum field theory on a BH background in AdS.

Nonetheless, even at the classical level, the issue of boundary conditions in rotating AdS spacetimes is subtle. Imposing that the perturbations conserve the symmetries of asymptotically global AdS, a set of Robin boundary conditions for the Teukolsky equation of a Kerr-AdS BH was found~\cite{Dias:2013sdc} (cf. also Ref.~\cite{Cardoso:2013pza} for some applications). Furthermore, in a scattering experiment the boundary conditions at infinity should allow for a nonvanishing flux, thus corresponding to the ``transparent'' case discussed above. A thorough analysis of this problem was recently performed in~\cite{Jorge:2014kra}, where it was shown that superradiance occurs for AdS BHs in any spacetime dimension whenever transmittive boundary conditions are allowed at the AdS boundaries.

%%%%%%%%%%%%%%%%%%%%%%%%%%%%%%%%%%%%%%%%%%%%%%%%%%%%%%%%
\subsection{Superradiance beyond General Relativity}
%%%%%%%%%%%%%%%%%%%%%%%%%%%%%%%%%%%%%%%%%%%%%%%%%%%%%%%%
From our previous discussion, it is clear that superradiance is not a prerogative of BHs in GR but it would occur in any gravitational theory that admits BH solutions. Indeed, the analysis of Sec.~\ref{sec:ABC} only requires the presence of an event horizon and an asymptotically-flat spacetime. Clearly, the details of the superradiance amplification would depend on the specific BH geometry and on the wave dynamics in the modified theory and an interesting problem is to understand whether superradiance can be stronger in modified theories of gravity.

Extended theories of gravity usually predict novel BH solutions which reduce to the Kerr metric in the GR limit (see 
e.g. Refs.~\cite{Yunes:2013dva,Barausse:2014tra,Berti:2015itd} for reviews). On the other hand, constructing rotating 
metrics in closed form is usually very challenging and most solutions are known analytically only in the slow-rotation 
limit~\cite{Pani:2011gy} or fully numerically~\cite{Kleihaus:2011tg}. To the best of our knowledge, no studies of 
superradiance amplification in these spacetimes is available to date. However, at least for the slowly-rotating BH 
solutions predicted in quadratic gravity~\cite{Pani:2011gy}, the deformations from the Kerr geometry tend to 
\emph{decrease} the proper volume of the ergoregion. This suggests that at least the background geometry would 
contribute to decrease the amplification factor. A simpler analysis is to focus on theories which admit the same BH 
solutions as GR~\cite{Psaltis:2007cw,Berti:2015itd} but for which wave propagation is different. In some of these 
theories the superradiance amplification has been shown to lead to ``BH-bomb 
instabilities''~\cite{Myung:2011we,Myung:2013oca} analog to those discussed in Sec.~\ref{sec:bombs} below. Theories 
where the background BH spacetime is identical to GR will generically lead to superradiance;
such is the case for a neutral scalar in ``MOG'' theory~\cite{Wondrak:2018fza}. As a toy model for non-local 
gravitational theories, the superradiant scattering of a non-local scalar field interacting with a rotating cylinder, in 
analogy with Zel'dovich's original thought experiment~\cite{zeldovich1,zeldovich2}, was studied 
in Ref.~\cite{Frolov:2018bak}.

Another strategy consists in considering phenomenological nonKerr geometries which are not necessarily solutions of any 
specific theory~\cite{Johannsen:2011dh,Cardoso:2014rha}. However, the lack of an underlying theory prevents to study the dynamics of GWs and only test fields propagating in a fixed background can be analyzed. Even in this case, the 
separability properties of the Kerr metric are generically lost and even the Klein-Gordon equation might not be 
separable. Probably because of these technicalities, superradiance in such geometries has not been studied to date. On 
the other hand, the Penrose process in a restricted class of such metrics was studied in Ref.~\cite{Liu:2012qe}, showing 
that the maximum energy gain can be several times larger than for a Kerr BH.

Finally, superradiance amplification of test fields propagating on some exact solutions of Einstein's equations which represent spinning geometries other than Kerr were analyzed in~\cite{Bini:2003sy,Bini:2014kga,Khodadi:2020cht}. Although strictly speaking these geometries are GR solutions, they possess peculiar matter fields and they might be considered as modified BH solutions. Exceptions concern rotating dilatonic BHs~\cite{Koga:1994np}. These are truly new geometries arising from the compactification of higher dimensions, and lead to strong superradiant effects.

%%%%%%%%%%%%%%%%%%%%%%%%%%%%%%%%%%%%%%%%%%%%%%%%%%%%%%%%%%%%%%%%%%%%%%%%%%%%%%%%%%%%%%%%%%%%%%%%%%%%%%%
\paragraph{Superradiance of black holes surrounded by matter in scalar-tensor theories\label{sec:super_ST}}
%%%%%%%%%%%%%%%%%%%%%%%%%%%%%%%%%%%%%%%%%%%%%%%%%%%%%%%%%%%%%%%%%%%%%%%%%%%%%%%%%%%%%%%%%%%%%%%%%%%%%%%
In the context of scalar-tensor theories, superradiance amplification from spinning BHs has been investigated in Refs.~\cite{Cardoso:2013fwa,Cardoso:2013opa,Dima:2020rzg}, which showed that the presence of matter may drastically affect the amplification of scalar waves. In these theories the Klein-Gordon equation on a Kerr BH surrounded by matter takes the form $[\square-\mu_{\rm eff}^2]\Psi=0$, where the effective mass term $\mu_{\rm eff}$ depends on the specific scalar-tensor theory and it is proportional to the trace of the stress-energy tensor. 

\begin{center}
\begin{figure}[ht]
\begin{center}
\begin{tabular}{cc}
\epsfig{file=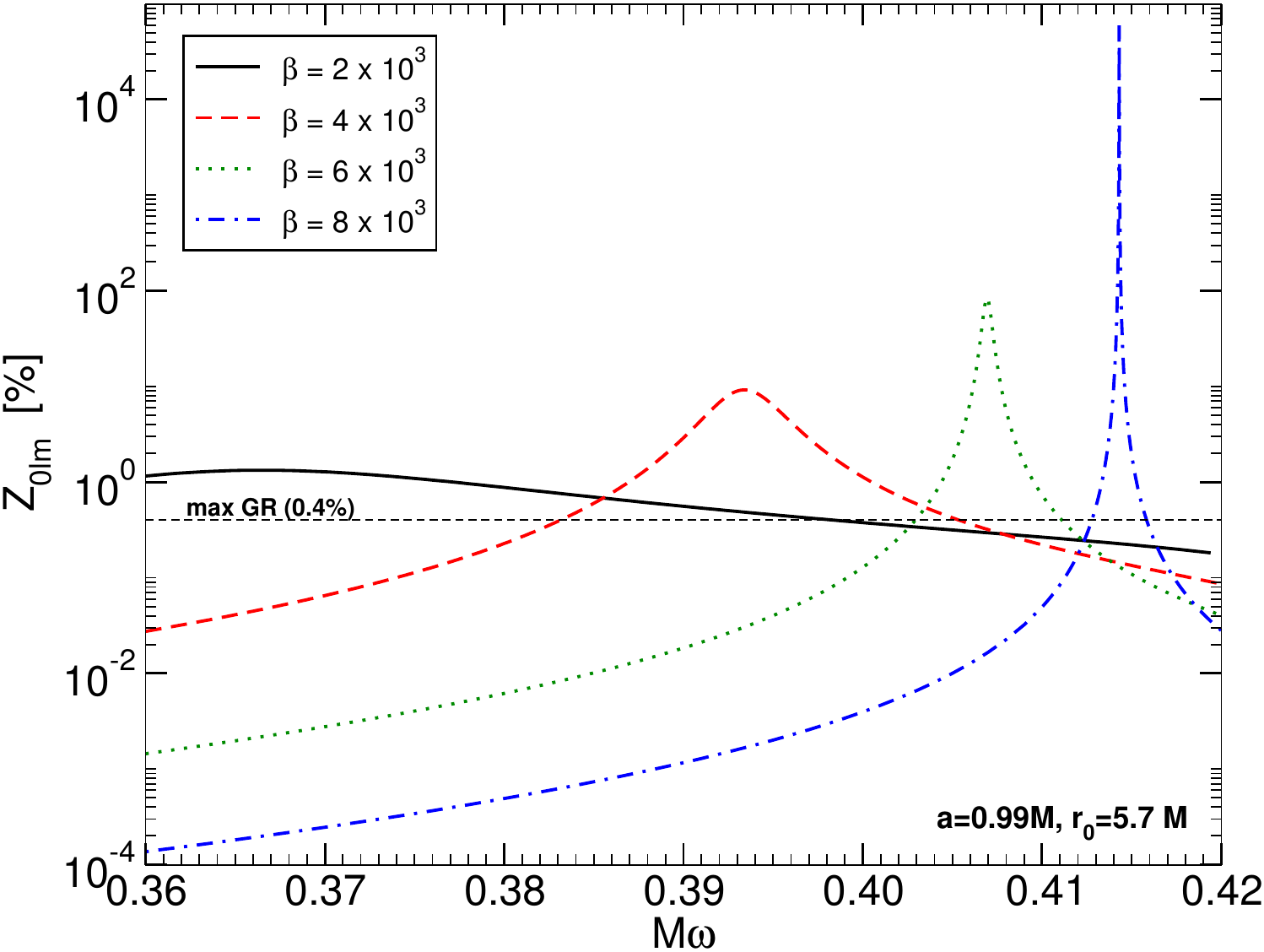,width=0.7\textwidth,angle=0,clip=true}
\end{tabular}
\end{center}
\caption{\label{fig:STamplification}
Percentage superradiant amplification factor, $Z_{0lm}=10^2\left(|{\cal R}|^2-|{\cal I}|^2\right)$, for a scalar wave scattered off a Kerr BH with $a=0.99M$ in a matter profile~\eqref{separable} with ${\cal G}=\beta\Theta[r - r_0](r-r_0)/r^3$ as a function of the wave frequency $\omega$ in a generic scalar tensor theory (where $\beta$ parametrizes the scalar-tensor coupling~\cite{Cardoso:2013opa}). 
As a reference, the horizontal line corresponds to the maximum amplification for scalar waves in vacuum, $Z_{0lm}^{\rm max}\approx 0.4\%$.
This example refer to $r_0=5.7 M$, but similar results hold for other choices of $r_0$ and for different matter profiles. The resonances correspond to the excitation of a new class of \emph{stable} QNMs.
}
\end{figure}
\end{center}

Figure~\ref{fig:STamplification} shows a representative example of superradiance amplification for a specific matter profile, namely
\begin{equation}
\mu_{\rm eff}^2(r,\vartheta)=\frac{2{\cal G}(r)}{a^2+2r^2+a^2\cos2\vartheta}\,.\label{separable}
\end{equation}
This choice simplifies the treatment significantly because the corresponding Teukolsky equation is separable and the problem can be solved with standard methods. More general mass distributions can be handled with spectral methods~\cite{Dima:2020rzg}.
For small coupling, the standard GR results are recovered, with a maximum amplification of about$~0.4\%$. On the other hand, as the scalar-tensor coupling to matter increases, the amplification factor can exceed the standard value by orders of magnitude. This is due to the appearance of resonances at specific frequencies $\omega=\omega_{\rm res}$ that depend on the parameters of the model. In some cases, the amplification factor can increase by six orders of magnitude or more, even in regions of the parameter space which are phenomenologically allowed~\cite{Cardoso:2013opa}. Understanding the astrophysical implications of such huge amplification may be used to constrain the parameter space of scalar-tensor theories. 

The presence of Breit-Wigner resonances~\cite{ChandraFerrari91} in the amplification factor has been interpreted in terms of very long-lived QNMs with $\omega_R\sim \omega_{\rm res}$ and $\omega_I\ll\omega_R$~\cite{Cardoso:2013fwa,Cardoso:2013opa}. Such long-lived modes  are associated to trapping by potential barriers (see also Sec.~\ref{sec:ERlonglived} for a related problem) and they also exist in the case of \emph{massive} scalar perturbations of Kerr BHs~\cite{Detweiler:1980uk}, but in that case the potential well extends to infinity, so that waves whose frequency is larger than the mass are exponentially suppressed and no superradiant amplification can occur, as previously discussed. Nonminimal scalar-matter interactions 
in scalar-tensor gravity produce an \emph{effective} scalar mass which is localized near the BH and vanishes at large distances. This effective mass can trap long-lived modes and, nonetheless, allows for propagation of scalar waves to infinity. This allows for a new class of long-lived QNMs of Kerr BHs surrounded by matter. Correspondingly to the excitation of these modes, the superradiant gain factor is resonantly amplified\footnote{It would be interesting to understand the large amplification of the superradiance energy in terms of violation of some energy condition due to the effective coupling that appears in scalar-tensor theories.}.

%%%%%%%%%%%%%%%%%%%%%%%%%%%%%%%%%%%%%%%%%%%%%%%%%%%%%%%%%%%%%%%%%%%%%%%%%%%%%%%%%%%%%%%%%%%%%%%%%%%%
\subsection{Microscopic description of superradiance and the Kerr/CFT duality} \label{sec:KerrCFT}
%%%%%%%%%%%%%%%%%%%%%%%%%%%%%%%%%%%%%%%%%%%%%%%%%%%%%%%%%%%%%%%%%%%%%%%%%%%%%%%%%%%%%%%%%%%%%%%%%%%%

It was shown by Hawking that when quantum effects are taken into account BHs emit thermal radiation with the expected number of emitted particles given by~\cite{Hawking:1974sw}
\be\label{Hawking_rate}
\left\langle N\right\rangle=-\frac{Z_{slm}}{e^{(\omega-m\Omega_{\rm H})/T_{\rm H}}\pm 1}\,,
\ee
where the minus sign is for bosons and the plus sign for fermions and $Z_{slm}$ is the absorption/amplification factor given by Eq.~\eqref{sigma}, whereas the same factor for fermions can be found in Ref.~\cite{Page:1976df} (in this case, as discussed in Sec.~\ref{DiracKerr}, we always have $Z_{slm}<0$). In the extremal limit $T_{\rm H}\to 0$ there is only emission in the superradiant regime $\omega<m\Omega_{\rm H}$ with a rate $\mp Z_{slm}$, where here the minus and the plus signs are for fermions and bosons, respectively. This clearly shows that when the BH temperature is different from zero, thermal Hawking radiation and spontaneous superradiant emission are strongly mixed. 
In fact, as discussed in detail in Ref.~\cite{Dai:2010xp}, the power spectrum of Hawking radiation comprises two terms: a black body term and a greybody term. The former is associated to the probability that a certain particle is thermally produced near the horizon, whereas the latter term term modifies
the thermal radiation due to the existence of the potential barrier created by the BH. 
While the probability of Hawking emission (for both bosons and fermions) can be greatly amplified by spin-spin interactions~\cite{Dai:2010xp}, one can show that superradiance affects only the greybody term. Therefore, Hawking fermion emission can also be amplified
by the BH rotation even if fermions do not experience superradiance amplification.

In the extremal limit, the only modes that are spontaneously emitted are superradiant. This was used in Refs.~\cite{Dias:2007nj,Bredberg:2009pv} to investigate the microscopic description of superradiance within a string theory and gauge/gravity duality context. These studies --~which are closely related to the program aiming to account for the microscopic degrees of freedom of BHs~-- have been met with a moderate degree of success.

In Ref.~\cite{Dias:2007nj} the authors were able to account for superradiant effects in a certain extremal BH background (more specifically the D1-D5-P BH solutions of type IIB supergravity), where the ${\rm AdS}_3/{\rm CFT}_2$ duality applies. In their picture the superradiant spontaneous emission was modeled as being due to the weak interaction between left and right-moving modes in the CFT. From this picture they argued that the superradiant bound~\eqref{eq:superradiance_condition} follows directly from the Fermi-Dirac statistics of the spin-carrying degrees of freedom in the dual CFT. More importantly, they showed that the superradiant emission rates agree in both sides of the duality. 
In the future it would be interesting to extend this study to other systems, and recover completely the superradiant amplification factors from the microscopic description.

Another important step was done within the so-called Kerr/CFT duality~\cite{Guica:2008mu} (see also Ref.~\cite{Compere:2012jk} for a recent review). The Kerr/CFT duality conjectures that the near-horizon extremal Kerr BH is holographically dual to a chiral left-moving (half of a) two-dimensional CFT with central charge $c=12J/\hbar$, where $J$ is the angular momentum of the extremal Kerr BH. In this picture the asymptotically flat region is removed from the spacetime and the CFT lives in the timelike boundary of the resulting spacetime\footnote{The geometry used in the original Kerr/CFT duality is the so-called near-horizon extreme Kerr ``NHEK'' geometry found by Bardeen and Horowitz~\cite{Bardeen:1999px} which is not asymptotically flat but resembles ${\rm AdS}_3$. That this geometry could have a dual CFT description was first pointed out in Ref.~\cite{Bardeen:1999px}.}. In Ref.~\cite{Bredberg:2009pv} the authors attempted to reproduce the superradiant scattering of a scalar field in a near-extremal Kerr geometry in terms of a dual two-dimensional CFT in which the BH corresponds to a thermal state while the scalar field is dual to a specific operator. They successfully showed that the amplification factor~\eqref{sigma} could be reproduced by the two point function of the operator dual to the scalar field. 
The analysis and results should however be taken with caution, as the boundary conditions --fundamental for the analysis -- were shown to be inconsistent with the field equations~\cite{Amsel:2009et,Dias:2009ex}.

%%%%%%%%%%%%%%%%%%%%%%%%%%%%%%%%%%%%%%%%%%%%%%%%%%%%%%%%%%%%%%%%%%%%%%%
\subsection{Boosted black holes: energy extraction without ergoregions}
%%%%%%%%%%%%%%%%%%%%%%%%%%%%%%%%%%%%%%%%%%%%%%%%%%%%%%%%%%%%%%%%%%%%%%%

We have so far been focusing exclusively on spinning BHs, and studying energy extraction from vacuum
by using the effective coupling to the ergoregion. It is easy to see that energy can also be extracted from moving BHs,
even if non-spinning~\cite{Bernard:2019nkv,Cardoso:2019dte}.
Consider a BH of mass $M$ and a high-frequency photon, described by null geodesics in the BH spacetime, with a large impact parameter $b\gg M$ and moving in the $-z$ direction.
The photon's incoming energy is $E_i$ in the frame where the BH is moving in the $+z$ direction with velocity $v$.
A boost in the $+z$ direction brings us to the BH frame, and blueshifts the wave to $E_1=\sqrt{(1+v/c)/(1-v/c)}E_i$.
In this frame, the photon is deflected by the Einstein angle $\alpha=4GM/(bc^2)$. Now boost back to the $-z$ direction, where due to relativistic aberration the angle with the $z-$axis is $\alpha'\sim \alpha\sqrt{(1+v/c)/(1-v/c)}$, and the frequency is now
$E_f=E_1/(\gamma(1+v\cos\alpha'/c))$. One finds the weak field energy amplification for such photons
\be
E_f^{\rm weak}=\left(1+\frac{8M^2v}{b^2c^4(c-v)}\right)E_i\,.\label{deflection_weak}
\ee

If a plane wave is passing through, one can see that the $1/r$ nature of the gravitational potential causes the total extracted energy to diverge; this phenomenon is akin to the divergence of the scattering cross section of the Coulomb potential~\cite{Merzbacher}. 
For a body of size $R_{\rm min}$ moving in a plane wave of density $\rho$ and extent $R_{\rm max}$, we find the total energy loss per second
\be
d E/dt=-\frac{16\pi M^2v^2\rho}{c^2(c-v)}\log{\left(R_{\rm max}/R_{\rm min}\right)}\,.
\ee
This is a generic result: moving objects will slow-down. In strong-field regions, photons can even turnaround.
In this case, a trivial change of frames and consequent blueshift yields
\be
E_f^{\rm peak}=\frac{1+v/c}{1-v/c}\,E_i\,,\label{max_amp}
\ee
for the energy gained by the photon during the process. This is also the blueshift by photons reflecting off a mirror moving with velocity $v$.

\begin{figure}
\centering
\begin{tabular}{c}
\includegraphics[width=10cm,height=10cm,keepaspectratio]{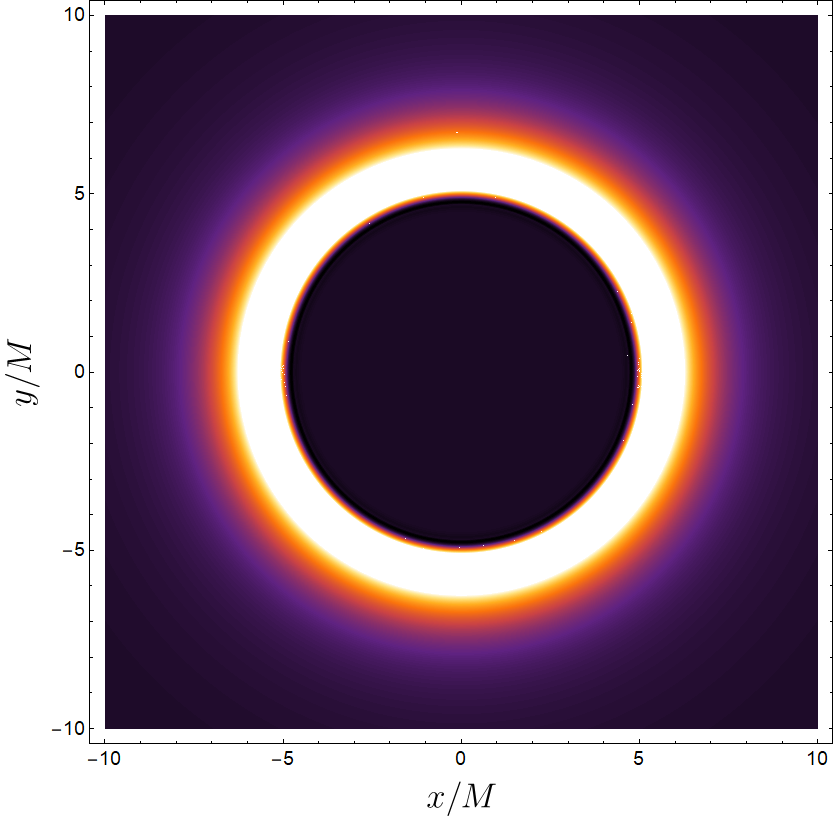} 
\end{tabular}
\caption{Appearance of a BH moving in a bath of cold (and counter-moving) radiation. The BH is moving along the $z$-axis towards us at a speed $v=0.9$. The colors denote energy flux intensity on a screen placed a short distance away from the BH. The peak energy flux is ten times larger than that of the environment. The bright ring has width $\sim M$ for all boost velocities $v$. For very large $v$ even a randomly-moving gas of photons will leave a similar observational imprint, since counter-moving photons will be red-shifted away. From Ref.~\cite{Cardoso:2019dte}.}
\label{fig:image}
\end{figure}
The large amplification for strongly-deflected photons implies that a rapidly moving BH looks peculiar. Downstream photons are deflected and blueshifted upstream. Thus a rapidly moving BH in a cold gas of radiation will be surrounded by a bright ring of thickness $\sim M$. A possible image of a moving BH is shown in Fig.~\ref{fig:image}. For a stellar-mass BH moving at velocities $v\sim 0.9996$ through the universe, the ambient microwave cosmic background will produce a kilometer-sized ring (locally $\sim 5000$ times hotter and brighter than the CMB) in the visible spectrum.

The overall result of energy transfer to external radiation echoes that of the inverse Compton scattering for fast-moving electrons in a radiation field~\cite{Beckmann,Cardoso:2019dte}. In this process, a nearly-isotropic radiation field is seen as extremely anisotropic to the individual ultrarelativistic electrons. Relativistic aberration causes the ambient photons to approach nearly head-on; Thomson scattering of this highly anisotropic radiation reduces the electron's kinetic energy and converts it into inverse-Compton radiation by upscattering radio photons into optical or $X$-ray photons. BHs are natural absorbers, but the universal -- and strong, close to the horizon -- pull of gravity can turn them also into overall amplifiers.

It is a question of nomenclature, or semantics, whether or not this process is superradiant.
We note that neither horizons nor ergoregions play a role. It is not a constant-frequency process either, and it does not 
entail mass-loss from the intervening objects, merely kinetic energy. 

Finally, we note that energy extraction from moving BHs may also be achieved using particles in non-geodesic motion or external magnetic fields~\cite{Penna:2015qta}.

%%%%%%%%%%%%%%%%%%%%%%%%%%%%%%%%%%%%%%%%%%%%%%%%%%%%%%%%%%%%%%%%%%%%%%
\subsection{Boosted black strings: ergoregions without superradiance}
%%%%%%%%%%%%%%%%%%%%%%%%%%%%%%%%%%%%%%%%%%%%%%%%%%%%%%%%%%%%%%%%%%%%%%
In the previous sections, we saw that superradiance is generically caused
by a medium moving faster than the speed of the interaction in the medium (for example the Cherenkov effect of Section~\ref{sec:cherenkov}
or sound amplification at discontinuities explained in Section~\ref{super_discontinuity}), or when the ``angular velocity of the medium''
is larger than the angular phase velocity of the interaction (an example was discussed in Section~\ref{sec:rotSR}, another is provided by the topic of this work, rotating BHs). These considerations seem to forbid gravitational superradiance for linear motion.
However, there are simple gravitational systems with ergoregions whose only motion is linear: consider a black string in five-dimensional spacetime\footnote{This example was suggested to us by Luis Lehner
and Frans Pretorius.},
\be
ds^2=-f(r)dt^2+\frac{dr^2}{f(r)}+r^2d\Omega_2^2+dz^2\,,
\ee
where $f(r)=1-2M/r$. Now boost the spacetime along the $z-$direction with boost $v=\tanh\beta$ and get 
\cite{Hovdebo:2006jy}
\be\label{boostedBH}
ds^2=-dt^2+\frac{dr^2}{f(r)}+r^2d\Omega_2^2+dz^2+(1-f(r))\cosh^2\beta\left(dt+\tanh\beta dz\right)^2\,,
\ee
It is easy to check that this solution has an event horizon at $r=2M$ and a ``momentum'' ergosurface at $r=2M\cosh^2\beta$\footnote{We follow the terminology of Dias, Emparan and Maccarrone who, in a completely different context, arrived at conclusions very similar to ours, see Section 2.4 in Ref.~\cite{Dias:2007nj}.}. Since this solution is just a non-boosted black string as seen by a boosted observer, it is clear that no superradiant amplification nor Penrose processes are possible. Let us show how this comes about.

\paragraph{Superradiance:} 
Consider perturbations of the metric~\eqref{boostedBH} due to a scalar field $\Psi$. Using the ansatz
\be
\Psi=\frac{\psi(r)}{r}Y(\vartheta,\varphi)e^{-i(\omega t+k z)}\,,
\ee
where $Y(\vartheta,\varphi)$ are the spherical harmonics, the radial function $\psi$ follows a Schroedinger-type equation of the form~\eqref{wave}. The solution $\psi$ has the asymptotic behavior given by~\eqref{bound2} with $k_H=\omega \cosh\beta-k\sinh\beta$ and $k_\infty=\sqrt{\omega^2-k^2}$, from which condition~\eqref{reflectivity} follows. Now, at first sight one could be led to think that superradiance is possible whenever the following condition is met:
\be\label{boosted_cond}
k_H<0\implies \omega<k v\,.
\ee
However the boundary condition at infinity also implies $|\omega|>|k|$. Since $-1<v<1$ we can see that the condition~\eqref{boosted_cond} is never met and, as expected, superradiance does not occur in this geometry. In other words, the potentially dangerous modes are redshifted away.

\paragraph{Penrose process:}

To understand why the Penrose process is not possible consider the negative energy particle that falls into the BH with energy and total linear momentum given by $\mathcal{E}<0$ and $p$, respectively. Denoting the particle's linear momentum along the $z$-direction by $p_z$, from arguments similar to those leading to Eq.~\eqref{penrose_condition} it follows that (note that $v_{\rm H}=-v$ is the velocity of a zero linear momentum observer at the horizon)
\be
\mathcal{E}+p_z v\geq 0 \implies |\mathcal{E}|\leq |p_z v|\,. 
\ee
The first condition also implies that, for negative energy particles and $0<v<1$, $p_z>0$. Moreover, since $0<v<1$, we have
\be\label{boosted_condition}
|\mathcal{E}|<p_z\,.
\ee
On the other hand, any particle must satisfy the relation
\be
\mathcal{E}^2=p^2+m^2\geq p_z^2 \implies |\mathcal{E}|\geq |p_z|\,.
\ee
Therefore, energy extraction is impossible because the inequality~\eqref{boosted_condition} is never satisfied for a negative energy particle\footnote{This simple proof was suggested to us by Roberto Emparan.}.

The absence of the Penrose process can also be understood through an analysis of geodesic motion. Let us focus on zero angular momentum trajectories for simplicity.
Geodesics in the spacetime~\eqref{boostedBH} are then described by the equations of motion,
\beq
&\dot{t}-(1-f(r))\cosh^2\beta(\dot{t}+\tanh\beta \dot{z})=E\,,\label{geodesics:time_boost}\\
&\dot{z}+(1-f(r))\cosh^2\beta(\dot{t}+\tanh\beta \dot{z})\tanh\beta=P\,,\\
&\dot{r}^2=(E^2-P^2)\left(1-\frac{M}{r}\right)-f(r)\delta_1+\frac{M\left[\left(E^2+P^2\right)\cosh 2\beta+2 E P\sinh 2\beta\right]}{r}\,,\label{geodesics:boost}
\eeq
where $E,P$ are the (conserved) energy and linear momentum per unit rest mass.
In the Penrose process, the breakup occurs at a turning point inside the ergoregion and with negative energy, $E<0$. From \eqref{geodesics:boost}, the turning point condition, $\dot{r}(r=r_0)=0$, gives
\beq
E&=&\frac{-P M \sinh 2\beta+\sqrt{f(r_0)r_0\left[\delta_1(r_0+2M(\cosh^2\beta-1))+P^2r_0\right]}}{r_0+2M(\cosh^2\beta-1)}\,,\label{energy_boost}\\
P&=&\frac{E M \sinh 2\beta\pm\sqrt{f(r_0)r_0\left[\delta_1(2M\cosh^2\beta-r_0)+E^2 r_0\right]}}{r_0-2M\cosh^2\beta}\,,\label{momentum_boost}
\eeq
where $E$ has been chosen such that when $r_0\to \infty$ we have $E>0$. It is clear from \eqref{energy_boost} that for $E<0$ we need $P\tanh\beta>0$ and
\beq
P M \sinh 2\beta&>&\sqrt{f(r_0)r_0\left[\delta_1(r_0+2M(\cosh^2\beta-1))+P^2 r_0\right]} \implies\nn\\
P^2 M^2 \sinh^2 2\beta&>&f(r_0)r_0\left[\delta_1(r_0+2M(\cosh^2\beta-1))+P^2 r_0\right] \implies\nn\\
P^2(r-2M\cosh^2\beta)&<&(2M-r)\delta_1<0 \,\implies \, r-2M\cosh^2\beta<0\,.
\eeq
Thus the particle needs to be inside the ergosphere to have a negative energy.	
For the Penrose process to occur we also need the positive energy fragment to be able to travel back to infinity. When $r\to\infty$ we have
\be
\dot{r}^2=E^2-P^2-\delta_1\,,\quad r\to\infty\,.
\ee
This means that only when $E^2-P^2-\delta_1>0$ is motion from $r_0$ to infinity allowed. Eq.~\eqref{geodesics:boost} however, says that there is only one turning point satisfying $\dot{r}(r=r_0)=0$, given by
\be
r_0=\frac{2M\left[(P\cosh\beta+E\sinh\beta)^2+\delta_1\right]}{P^2+\delta_1-E^2}\,.
\ee
The condition that $r_0>0$ implies that the particles are not allowed to escape to infinity since $E^2-P^2-\delta_1<0$. In fact since there is only one turning point and at the horizon we have
\be
\dot{r}^2=(E\cosh\beta+P\sinh\beta)^2 \,,\quad r\to 2M\,,
\ee
which is always positive, both particles are forced to fall into the horizon and there is no extraction of energy from the BH.

%%%%%%%%%%%%%%%%%%%%%%%%%%%%%%%%%%%%%%%%%%%%%%%%%%%%%%%%%%%%%%%%%%%%%%%%%%%%%%%%%%%%%
\subsection{Open issues\label{open4}}
%%%%%%%%%%%%%%%%%%%%%%%%%%%%%%%%%%%%%%%%%%%%%%%%%%%%%%%%%%%%%%%%%%%%%%%%%%%%%%%%%%%%%

The following is an incomplete list of the issues that are, in our opinion, not completely understood and which would merit further study.

\begin{itemize}

\item Superradiance in a BH-pulsar system has been recently discussed in Refs.~\cite{Rosa:2015hoa,Rosa:2016bli}, 
showing that superradiance of GWs from the pulsar can produce a peculiar modulation of the pulsar's GW luminosity at 
the percent level. Whether or not such effect is observationally important clearly deserves further study.

\item Is there a fundamental bound on superradiant amplification?
All the examples we have dealt with so far share a common denominator: the amplification factors $Z_{slm}\lesssim 100\%$~
\footnote{The only exception to this rule concerns BHs surrounded by matter coupled to scalar fields, where the amplification factors
can become unbounded (see Section~\ref{sec:super_ST}). Because the laws of BH mechanics will be different, these fall outside
the scope of this discussion.}. There are in fact suggestions that such bound also holds in some acoustic BH geometries~\cite{Choy:2005qx}.
Such relatively small amplification factors may be a consequence of a more fundamental principle at play.
Hints of such principle can be found with the following reasoning.
Recall that the area law for rotating BHs can be written as \eqref{first_law_BH} or, in terms of Bekenstein-Hawking entropy $S_H$, as
$\delta M=\frac{\omega k}{2\pi}\frac{\delta S_H}{\omega-m\Omega_{\rm H}}$. We can write this explicitly in terms of the amplification factor,
by considering that a wavepacket of energy $\delta E$ was thrown into the BH,
\be
Z_{m}\equiv \frac{-\delta M}{\delta E}=\frac{\omega k}{2\pi}\frac{\delta S_H/\delta E}{m\Omega_{\rm H}-\omega}\,.
\ee
It is clear that the BH mass decreases in the superradiant regime simply because the BH entropy must increase. 
This version of the first law doesn't immediately impose upper limits on the amplification factors, but that one should exist
follows from Bekenstein's entropy bound for any infalling matter~\cite{Bekenstein:1980jp},
\be
\delta S\leq 2\pi R\,\delta\,E\,,\label{Bek_bound}
\ee
where $R$ is the size of the object and $E$ its energy. This implies that
\begin{equation}
Z_{m}\leq k r_+ \frac{\omega}{m\Omega_{\rm H}-\omega}\,,\label{aiai}
\end{equation}
leading to a competitive bound on the amplification factor for small frequencies. Such bound becomes weaker close to the superradiant threshold. It is possible that a more refined argument can strengthen the bound in this regime as well.

This analysis is over-simplified\footnote{We thank Shahar Hod for drawing our attention to this point.}. In particular, the Bekenstein entropy bound \eqref{Bek_bound} is valid only 
for systems with a fixed radius; in general, there will be charge-dependent corrections. These may be important, as the bound \eqref{aiai} predicts that $Z_m \to 0$ 
when $\omega\to0$, in conflict with Fig.~\ref{fig:SR_charged} (see, in particular, the purple curve).

\item The scattering of massive Dirac waves off a Kerr BH has some connection with the original Klein paradox. Indeed, Chandrasekhar suggested that the effective potential for the Schroedinger-like problem can display some singularities outside the horizon in a certain region of the parameter space and in the case of rotation~\cite{Chandra}. If the potential is discontinuous, the transmission coefficient would be prone to the Klein paradox, as discussed in Sec.~\ref{sec:KleinParadox}. To the best of our knowledge a quantitative analysis of this phenomenon has not been performed yet. 

\item An outstanding open issue is the systematic calculation of the absorption cross-section of rotating BHs
for generic angles of incidence. In particular, a generalization of the low-frequency formulas available for 
GWs~\cite{Dolan:2008kf} to lower spin-fields would certainly be of interest, as well as thorough numerical studies. 
A generic computation for monochromatic EM plane waves scattered off a Kerr BH was recently done in 
Ref.~\cite{Leite:2018mon}.

\item Nonlinear effects and induced superradiance. The effect of nonlinear couplings have practically been ignored in all existing literature
on BH superradiance. Interesting effects could include induced superradiant-like effects in fermions when coupled to bosonic fields,
or mass-like effects when higher-order self-interaction terms are taken into account for boson fields. The backreaction of superradiant waves on the metric has been investigated only very recently, see Sec.~\ref{sec:super_nonlinear}.

\item Sound waves in matter outside {\it gravitational BHs} can feel an effective geometry with sonic horizons and ergoregions (differing from the true gravitational event horizon or ergoregions, which may be generically absent) and be subjected to superradiant effects. Although this is one more example of superradiance in analogue models, it is one with potentially important applications in astrophysical environments and may even lead to superradiant instabilities, c.f. Sec.~\ref{sec:bombs} and Refs.~\cite{Das:2004wf,Das:2006an}) (see also Ref.~\cite{Chaverra:2015aya} for a recent related realization in the case of nonspinning BHs).

\item Superradiance from BHs in modified theories of gravity has been studied only in a few 
cases~\cite{Wondrak:2018fza,Frolov:2018bak,Khodadi:2020cht}. At linearized level this requires having a stationary, axisymmetric BH 
solution and solving the modified wave dynamics in this background. Catalogs of interesting gravity theories and 
corresponding BH solutions can be found in the review~\cite{Berti:2015itd}.

\item Superradiance and non-axisymmetric spacetimes. All vacuum stationary solutions of Einstein's equations are also 
axisymmetric. This simplifies the treatment of superradiance considerably, because mode-mixing between different 
azimuthal numbers are avoided. However, this property can be broken in other theories, in non-stationary 
configurations, or by the presence of matter. Whether or not mode mixing would quench or favor superradiance is 
an interesting open problem. A specially relevant system is a binary of compact objects: does it superradiate or amplify incoming radiation? Preliminary results suggest a positive answer~\cite{Bernard:2019nkv,Wong:2019kru,Wong:2020qom}, but further work is necessary to clarify this issue.
\end{itemize}

%%%%%%%%%%%%%%%%%%%%%%%%%%%%%%%%%%%%%%%%%%%%%%%%%%%%%%%%%%%%%%%%%%%%%%
\clearpage
\newpage
\section{Black holes \& superradiant instabilities}\label{sec:bombs}
%%%%%%%%%%%%%%%%%%%%%%%%%%%%%%%%%%%%%%%%%%%%%%%%%%%%%%%%%%%%%%%%%%%%%%
Superradiant amplification lends itself to extraction of energy from BHs, but can also
be looked at as the chief cause of a number of important instabilities in BH spacetimes.
Some of these instabilities lead to hairy BH solutions, whereas others extract rotational
energy from the BH, spinning it down.
%%%%%%%%%%%%%%%%%%%%%%%%%%%%%%%%%%%%%%%%%%%%%%%%%%%%%%%%%%%%%%%%%%%%
\subsection{No black hole fission processes}
%%%%%%%%%%%%%%%%%%%%%%%%%%%%%%%%%%%%%%%%%%%%%%%%%%%%%%%%%%%%%%%%%%%%
One intriguing way of de-stabilizing a BH cluster using superradiance
is akin to more familiar fission processes. These however can be shown -- as we now do -- not to occur
for BH clusters.
\begin{figure}[ht]
\begin{center}
\epsfig{file=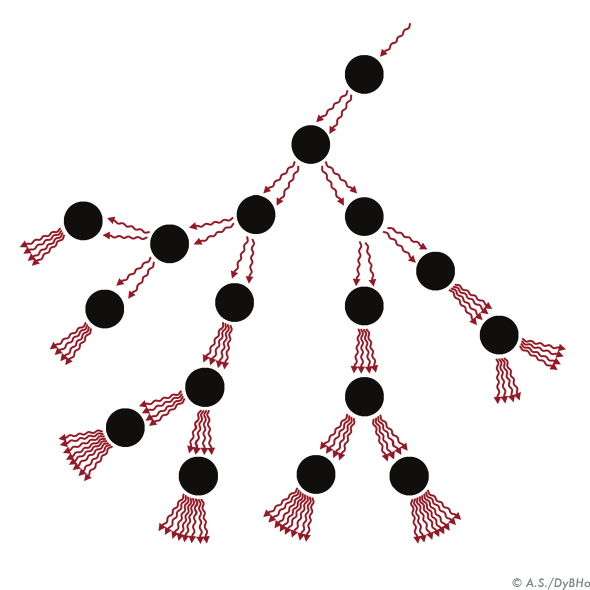,width=0.7\textwidth,angle=0}
\caption{Scheme of the hypothetical chain reaction in a cluster of rotating BHs. The incident
arrow denotes an incident wave on the rotating BH, which is then amplified and exits with larger amplitude, before interacting with other BHs.
The superradiantly scattered wave interacts with other BHs, in an exponential cascade.
\label{fig:chainreaction}}
\end{center}
\end{figure}
Take a cluster of rotating BHs, as in Fig.~\ref{fig:chainreaction}, and send in a low-frequency photon. If the cluster is appropriately built, it would seem {\it possible in principle} that the photon is successively amplified as it scatters off, leading to an exponential cascade. This kind of process is identical to the way fission bombs work, where neutrons play the role of our wave.

It was pointed out by Press and Teukolsky~\cite{Teukolsky:1974yv} that such a process could not occur for Kerr BHs, as the entire cluster would have to be contained in its own Schwarzschild radius. Let us see how this works in a generic $D$-dimensional setting. We take a cluster of $N$ rotating BHs of size $L$, and total mass $NM_{BH}$, where $M_{BH}$ is the mass of each individual BH. Assuming all the conditions are ideal, the process can only work if the mean free path $l$ of a photon (or any other boson field) is smaller than the size of the cluster,
\be
l<L\,.
\ee
Now, the mean free path is $l=\frac{1}{n\sigma}$, where $n$ is the BH number density in the cluster and $\sigma$ is an absorption cross section. The absorption cross-section could be negative if a plane wave is amplified upon incidence on a rotating BH (this happens for certain polarizations and angles of incidence only, see Section~\ref{sec:planewaves}).
Even in such case, it is at most of order the BH area. These two properties are very important. That the cross-section scales with the area can be seen on purely dimensional arguments and it holds true for all BH spacetimes we know of. A negative total cross-section is necessary to guarantee that whatever way the boson is scattered it will {\it on the average} be superradiantly amplified. In other words, we require that a plane wave is subjected to superradiance\footnote{Note that say, an $l=m=1$ mode is a sum of modes with respect to some other coordinate frame, where the following BH scatterer is sitting.}.
To summarize,
\be
\sigma \sim V_{D-2}r_+^{D-2}\,,
\ee
where $V_{D-2}=\pi^{D/2-1}/\Gamma[D/2]$ is the volume of a unit $(D-3)$ sphere. Thus, up to factors of order unity, the condition for fission would amount to $L^{D-2}/(N r_+^{D-2})<1$ or equivalently
\be
\frac{NM_{BH}}{L^{D-3}}>\frac{L}{r_+}\,.
\ee
This last condition is stating that the cluster lies within its own Schwarzschild radius, making the fission process impossible even in the most idealized scenario.

%%%%%%%%%%%%%%%%%%%%%%%%%%%%%%%%%%%%%%%%%%%%%%%%%%%%%%%%%%%%%%%%%%%%%%%%%%%%%%%%%%%%%%%%%%%%%%%%%%%%%%%%
\subsection{Spinning black holes in confining geometries are unstable}\label{sec:model}
%%%%%%%%%%%%%%%%%%%%%%%%%%%%%%%%%%%%%%%%%%%%%%%%%%%%%%%%%%%%%%%%%%%%%%%%%%%%%%%%%%%%%%%%%%%%%%%%%%%%%%%%
%
\begin{figure}[ht]
\begin{center}
\epsfig{file=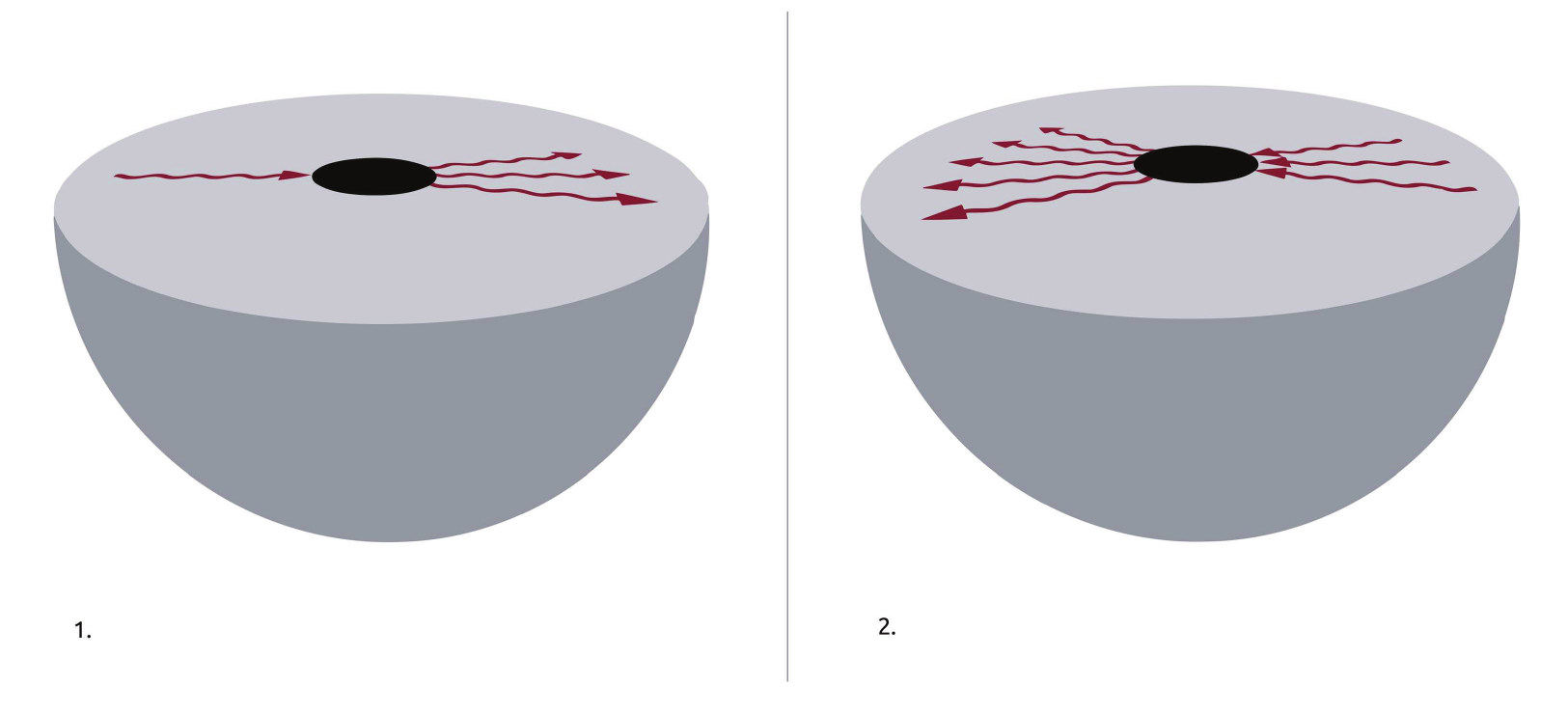,width=0.9\textwidth,angle=0}
\caption{Scheme of a confined rotating BH, and how an initially small fluctuation-- the single red arrow -- grows by successive reflections at the confining wall and amplifications by superradiance in the ergoregion.
\label{fig:bhbomb}}
\end{center}
\end{figure}
Fission-like processes don't work, but it was recognized early on that confinement will generically turn
superradiant amplification into an instability mechanism. The idea is very simple and is depicted in Fig.~\ref{fig:bhbomb}: superradiance amplifies
any incoming pulse, and the amplification process occurs near the ergoregion. If the pulse is now confined
(say, by a perfectly reflecting mirror at some distance) it is ``forced'' to interact -- and be amplified --
numerous times, giving rise to an exponentially increasing amplitude, i.e. to an instability. 

The details of the confinement are irrelevant and a simple picture in terms of a small perfect absorber immersed in a confining box can predict a number of features. A confining box supports stationary, \emph{normal} modes. Once a small BH is placed inside, one expects that the normal modes will become quasinormal and acquire a small imaginary part, describing absorption -- or amplification -- at the horizon of the small BH. Thus, it seems that one can separate the two scales -- BH and box size -- and describe quantitatively the system in this way~\cite{Brito:2014nja}.

Normal modes supported by a box have a wavelength comparable to the box size, in other words a frequency $\omega_R\sim1/r_0$. For small BHs, $M/r_0\ll 1$, we then have $M\omega\ll 1$, i.e., we are in the 
low-frequency limit. In this limit, the equation for wave propagation can be solved via matched asymptotics~\cite{Cardoso:2005mh}, similar to what is discussed in Appendix~\ref{appendix_super_ana}. 
Let $\mathcal{A}$ denote the absorption probability at the horizon of a rotating BH (which can be computed analytically in the small frequency regime~\cite{Cardoso:2005mh,1973ZhETF..64...48S,Staro1,1973ZhETF..65....3S,Staro2}).
By definition, a wave with initial amplitude $A_0$ is scattered with amplitude $A=A_0\left(1-|\mathcal{A}|^2\right)$ after one interaction with the BH.
In the superradiant regime $|\mathcal{A}|^2<0$.
Consider now a wave trapped inside the box and undergoing a large number of reflections. After a time $t$ the wave interacted $N=t/r_0$ times with the BH, and its amplitude changed to
$A=A_0\left(1-|\mathcal{A}|^2\right)^N\sim A_0\left(1-N|\mathcal{A}|^2\right)$. We then get
\be
A(t)=A_0\left(1-t|\mathcal{A}|^2/r_0\right)\,. \label{ampl0}
\ee
The net effect of this small absorption at the event horizon is to add a small imaginary part to the frequency, $\omega=\omega_R+i\omega_I$ (with $|\omega_I|\ll\omega_R$). 
In this limit, $A(t)\sim A_0 e^{-|\omega_I| t}\sim A_0(1-|\omega_I| t)$. Thus we immediately get that
\be
\omega_I=|\mathcal{A}|^2/r_0\,. \label{ampl}
\ee

For example, for a non-rotating BH~\cite{Cardoso:2005mh}
\beq
\left|\mathcal{A}\right|^2&=& 4\pi \left(\frac{M\omega_R}{2}\right)^{2+2l} \frac{\Gamma^2[1+l+s]\Gamma^2[1+l-s]}{\Gamma^2[1+2l]\Gamma^2[l+3/2]}\label{crosssection}\\
&\sim&\left({M}/{r_0}\right)^{2l+2}\ll1\,
\eeq
where $s=0,2$ for scalar and gravitational fields. Comparing with Eq.~\eqref{ampl}, we obtain
\be\label{absor_wIm}
M\omega_I\sim-(M/r_0)^{2l+3}\,. 
\ee
% % 

When the BH is rotating, the arguments leading to Eq.~\eqref{energy_transform} indicate that rotation can be taken into account by multiplying the previous result
by the superradiant factor $1-m\Omega/\omega$. In fact, low-frequency waves co-rotating with the BH are amplified by superradiance. Starobinsky has shown that, at least for moderate spin, the result in Eq.~\eqref{crosssection} still holds with the substitution~\cite{1973ZhETF..64...48S,Staro1,1973ZhETF..65....3S,Staro2}
\be
\omega^{2l+2}\to (\omega-m\Omega_{\rm H})\omega^{2l+1}\,,\label{absor_wIm_rot}
\ee
where we recall that $\Omega_{\rm H}$ is the horizon angular velocity. 

In other words, this intuitive picture immediately predicts that confined rotating BHs are {\it generically}
unstable and estimates the growth rate. The dependence of the growth rate on the confining radius $r_0$ is estimated to be independent on the spin of the field, and this behavior is observed in a variety of systems. The details need, of course, a careful consideration of the corresponding perturbation equations; nevertheless such conclusions hold for several different 
scenarios~\cite{Cardoso:2004nk,Cardoso:2004hs,Cardoso:2013pza,Cardoso:2005vk,Brito:2014nja}, as we discuss in more detail in the next sections.
%%%%%%%%%%%%%%%%%%%%%%%%%%%%%%%%%%%%%%%%%%%%%%%%%%%%%%%%%%%%%%%%%%%%%%%%%%%%%%%%%%%%%%%%%%%
\subsection{Superradiant instabilities: time-domain evolutions versus an eigenvalue search}
%%%%%%%%%%%%%%%%%%%%%%%%%%%%%%%%%%%%%%%%%%%%%%%%%%%%%%%%%%%%%%%%%%%%%%%%%%%%%%%%%%%%%%%%%%%
At linearized level BH superradiant instabilities are associated with perturbations of a fixed BH background which grow exponentially in time. Because the background is typically stationary, a Fourier-domain analysis proves to be very convenient. In a stationary and axisymmetric background, a given perturbation $\Psi(t,r,\vartheta,\varphi)$ can be Fourier transformed as 
\begin{equation}
 \Psi(t,r,\vartheta,\varphi)=\frac{1}{2\pi}\sum_m\int d\omega \tilde\Psi_m(\omega,r,\vartheta) e^{-i\omega t} e^{im\varphi}\,, \label{Fourier}
\end{equation}
%%%
and the perturbation function $\tilde\Psi_m$ will satisfy a set of PDEs in the variables $r$ and $\vartheta$. For the special case of a Kerr BH and for most types of fields, such PDEs can be miraculously separated using spheroidal harmonics (cf. Sec.~\ref{sec:Teukolsky} and Ref.~\cite{Kalnins-separability} for a proof of separability using Killing-Yano tensors), whereas in more generic settings other methods have to be used~\cite{Pani:2013pma}.

In any case, the system of equations for $\tilde\Psi_m$ together with suitable boundary conditions at the BH horizon (discussed already in Section~\ref{sec:Teukolsky}, and more thoroughly in Section 3 in Ref.~\cite{Berti:2009kk}) and at spatial infinity define an eigenvalue problem for the frequency $\omega$. Due to the boundary conditions at the BH horizon and at spatial infinity,
the eigenfrequencies (or quasinormal modes) are generically complex, $\omega=\omega_R+i\omega_I$~\cite{Berti:2009kk}.

In the rest of this section we discuss various superradiant instabilities obtained by solving the corresponding perturbation equations in the frequency domain and finding the complex eigenspectrum. Through Eq.~\eqref{Fourier}, an instability corresponds to an eigenfrequency with $\omega_I>0$ and the instability time scale is $\tau\equiv1/\omega_I$. In the case of superradiant modes this always occurs when the real part of the frequency satisfies the superradiant condition, e.g. $\omega_R<m\Omega_{\rm H}$ for a spinning BH. Although QNMs do not form a complete basis, they correspond to poles of the corresponding Green's function, and play 
an important part in the time-domain problem~\cite{Berti:2009kk}, as they arise in the contour-integration of \eqref{Fourier}.
A complementary approach consists in solving the perturbation equations directly in the time domain, by evolving an initially-small field and monitoring its energy-density as a function of time. As we will discuss, this approach has been used recently to study BH superradiance and its development. When both time-domain and frequency-domain computations are available, they yield consistent results~\cite{Dolan:2012yt,Witek:2012tr}.

%%%%%%%%%%%%%%%%%%%%%%%%%%%%%%%%%%%%%%%%%%%%%%%%%%%%%%%%%%%%%%%%%%%%%%%%%%%%%%
\subsection{Black holes enclosed in a mirror}\label{sec:mirror}
%%%%%%%%%%%%%%%%%%%%%%%%%%%%%%%%%%%%%%%%%%%%%%%%%%%%%%%%%%%%%%%%%%%%%%%%%%%%%%
%%%%%%%%%%%%%%%%%%%%%%%%%%%%%%%%%%%%%%%%%%%%%%%%%%%%%%%%%%%%%%%%%%%%%%%%%%%%%%
\subsubsection{Rotating black-hole bombs}
%%%%%%%%%%%%%%%%%%%%%%%%%%%%%%%%%%%%%%%%%%%%%%%%%%%%%%%%%%%%%%%%%%%%%%%%%%%%%%
%%%%%%%%%%%%%%%%%%%%%%%%%%%%%%%%%%%
\paragraph{Closed mirrors}
%%%%%%%%%%%%%%%%%%%%%%%%%%%%%%%%%%%
%
\begin{figure*}[hbt]
\begin{center}
\begin{tabular}{cc}
\epsfig{file=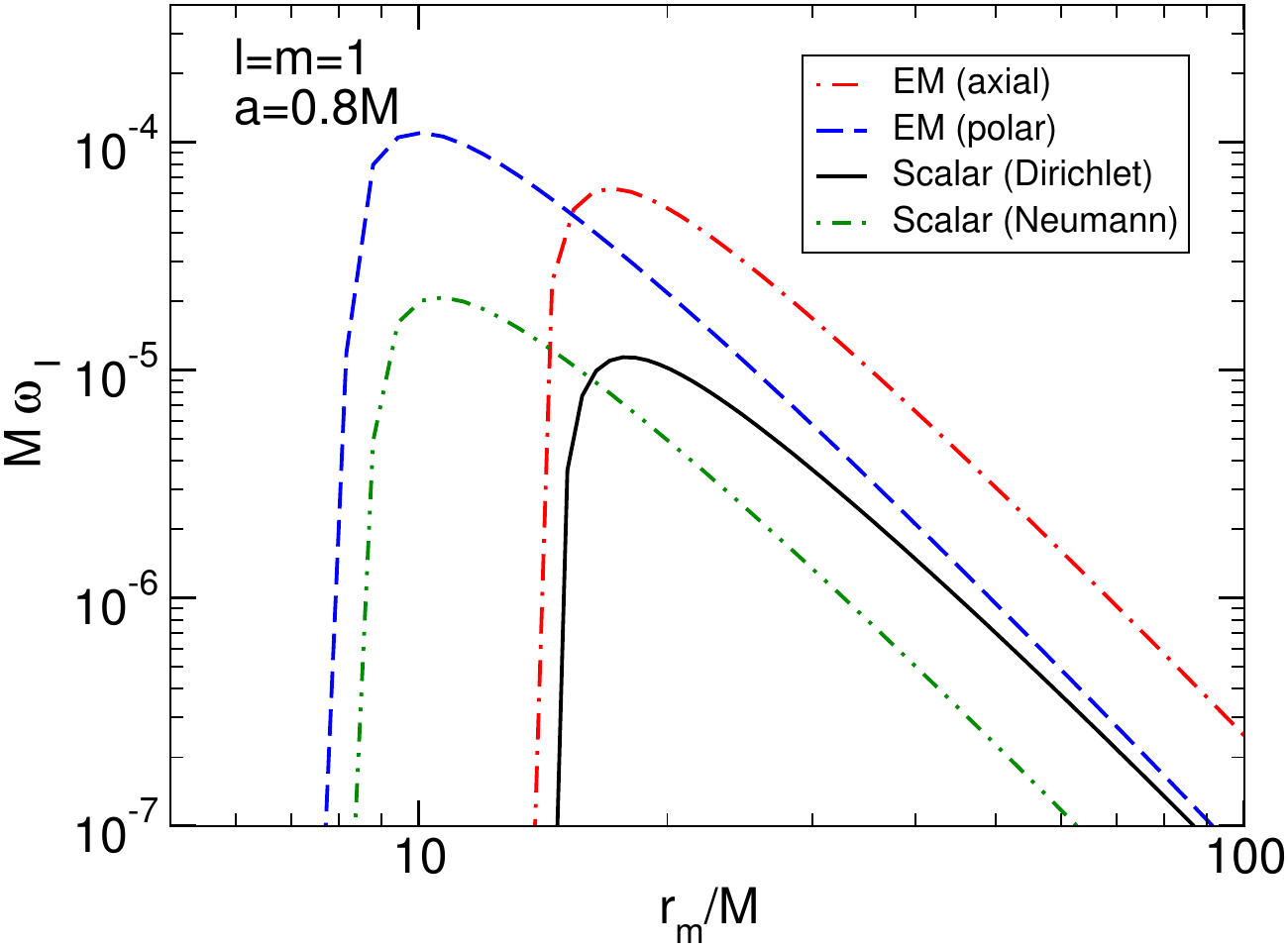,width=0.48\textwidth,angle=0,clip=true}&
\epsfig{file=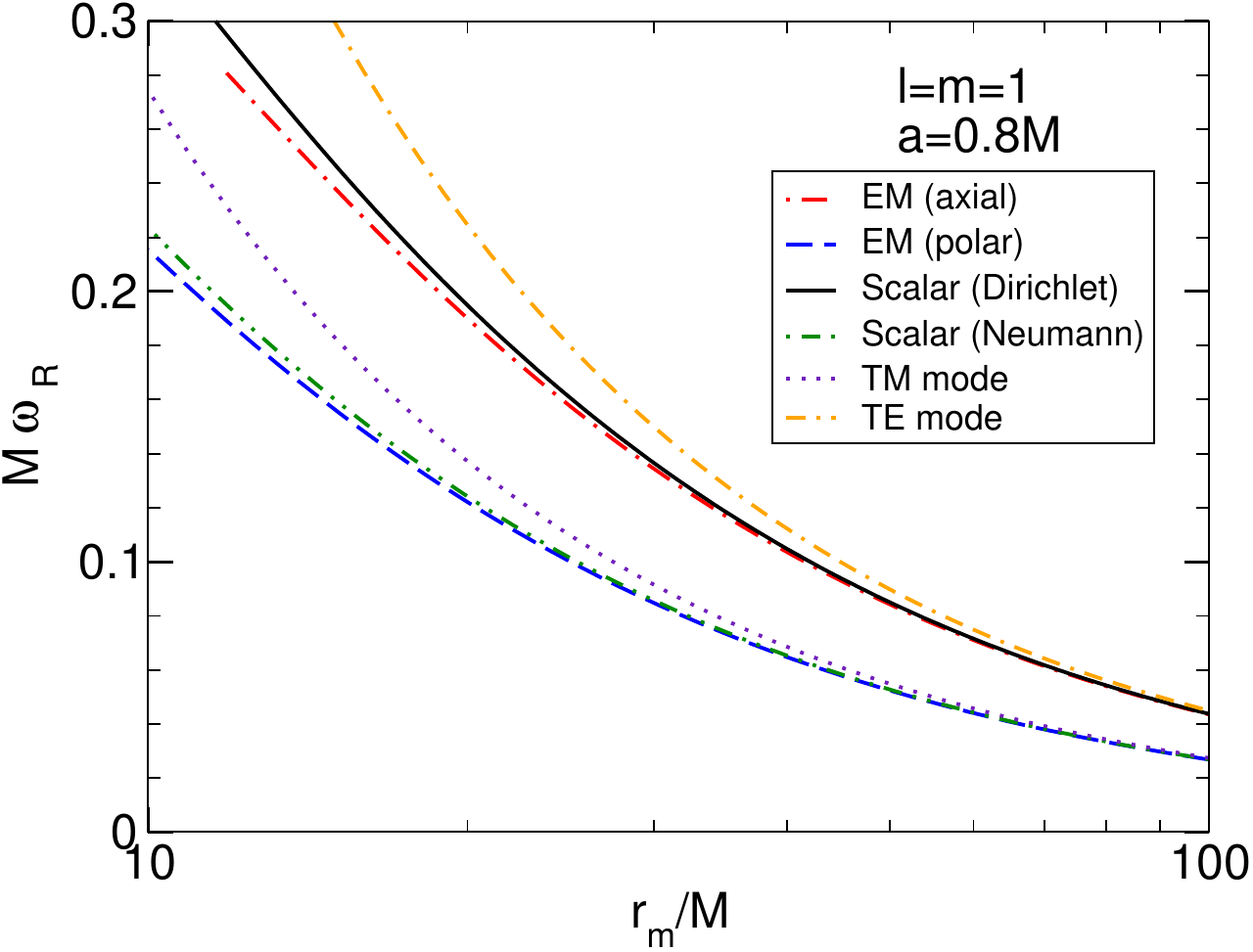,width=0.48\textwidth,angle=0,clip=true}
\end{tabular}
\caption{Fundamental ($n=0$) QNM frequency for scalar and EM perturbations of a confined Kerr BH as a function of the 
mirror's location $r_m$, for $l = m = 1$ and $a=0.8M$. For $r_m$ larger than a critical value the modes are unstable. We 
show the two different polarizations for the EM BH bomb compared to the modes of a scalar field for 
Dirichlet and Neumann boundary conditions at the boundary. For comparison we also show the flat space transverse 
electric (TE) and transverse magnetic (TM) modes inside a resonant cavity, as computed in Eqs.~\eqref{TE_modes} 
and~\eqref{TM_modes} of Appendix~\ref{app:EM_BCs}.\label{Fig:BQNM}}
\end{center}
\end{figure*}
One of the first conceptual experiments related to BH superradiance concerns a spinning BH surrounded by a perfectly reflecting mirror~\cite{zeldovich2,Press:1972zz,Cardoso:2004nk}.
As discussed in the previous section, confinement turns this system unstable against superradiant modes\footnote{
Any initial fluctuation grows exponentially, as we argued previously, leading to an ever increasing field density and pressure inside the mirror. The exponentially increasing pressure eventually disrupts the confining mirror,
leading to an ``explosion,'' and to this system being termed a \emph{black-hole bomb}~\cite{Press:1972zz}.}. A perfectly reflecting wall is an artificial way of confining fluctuations,
but is a useful guide to other more realistic and complex systems.

For scalars, the relevant equation \eqref{teu_radial} can be solved imposing suitable in-going or regularity boundary conditions at the horizon (discussed in Section \ref{sec:Teukolsky})
and a no-flux condition at the mirror boundary $r=r_m$ in Boyer-Lindquist coordinates. The latter can be realized in two different ways: either with Dirichlet $R(r_m)=0$ (see Ref.~\cite{Cardoso:2004nk} for a full analysis of this case) or Neumann $R'(r_m)=0$ conditions for the corresponding Teukolsky master wavefunction. Generic Robin boundary conditions can also be considered~\cite{Ferreira:2017tnc}.

The more realistic situation of EM waves trapped by a conducting spherical surface is also slightly more involved and is 
explained in Appendix~\ref{app:EM_BCs}.
The appropriate boundary conditions are that the electric field is tangent to the conductor and that the magnetic field is orthogonal to it in the mirror's frame~\cite{Jackson,King:1977}.
We find that the relevant boundary conditions at $r=r_m$ are
\beq
&&\partial_r R_{-1}=\frac{-i\Delta\left[\pm B+A_{-1lm}+\omega  \left(a^2 \omega -2 a m+2 i
   r\right)\right]}{2 \Delta\left(a^2 \omega -a m+r^2 \omega \right)}R_{-1}\nonumber\\
&&		+\frac{\left(a^2 \omega -a m+r^2 \omega \right) \left(2 i a^2 \omega -2 i a m+2 M+2 i r^2 \omega+ \partial_r\Delta-2  r\right)}{2 \Delta\left(a^2 \omega -a m+r^2 \omega \right)}R_{-1}\,,\label{cond1_a}
\eeq
where we have defined $B=\sqrt{(A_{-1lm}+a^2\omega^2-2am\omega)^2+4ma\omega-4a^2\omega^2}$ and $R_{-1}$ is a radial Teukolsky function defined in Appendix~\ref{app:EM_BCs}.
The perturbations can be written in terms of two Newman-Penrose scalars, $\phi_2$ and $\phi_0$, which are two linearly 
dependent complex functions. This explains the existence of two different boundary conditions, as would have been 
expected given the two degrees of freedom of EM fields. For $a=0$ we recover the condition~\eqref{cond1} when using the 
minus sign, while for the plus sign we recover the condition~\eqref{cond2}; accordingly, we label these modes as axial 
and polar modes, respectively.

\begin{figure}
\epsfig{file=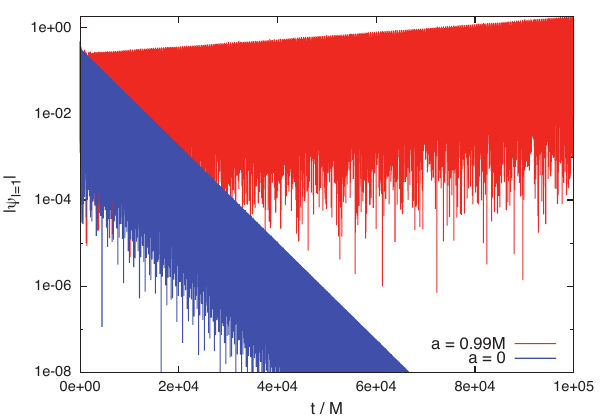,width=0.48\textwidth,angle=0,clip=true}
\epsfig{file=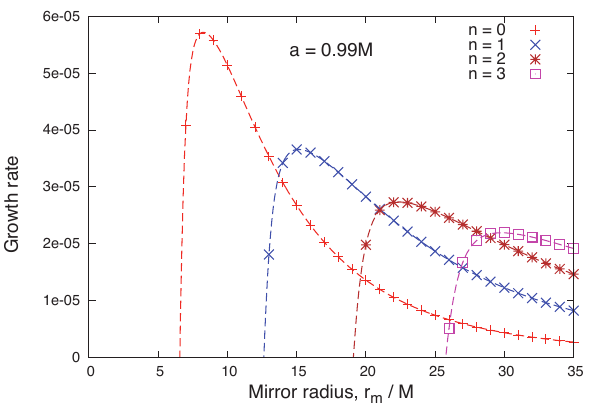,width=0.48\textwidth,angle=0,clip=true}
 \caption{Left: time evolution of a scalar field $\Psi$ obeying Dirichlet conditions at some boundary, on a logarithmic scale, up to $t=10^5M$. The envelope of the Kerr (Schwarzschild) field grows (decays) exponentially. Right: the growth rate $M\omega_I$ of the first few exponentially-growing modes, as a function of mirror radius $r_m$. The points show the growth rates calculated from time-domain data, using runs up to $t=10^5M$. The lines show the growth rates found in the frequency-domain analysis~\cite{Cardoso:2004nk}. From Ref.~\cite{Dolan:2012yt}.}
\label{fig:timeseries2}
\end{figure}
The boundary conditions described above are only satisfied for a discrete number of QNM eigenfrequencies $\omega$. Our results for the characteristic frequencies are shown in
Fig.~\ref{Fig:BQNM} for $l=m=1$ and $a=0.8M$. As the generic argument of the previous Section~\ref{sec:model} anticipated, confined BHs develop an instability, i.e.
some of the characteristic frequencies satisfy $\omega_I>0$\footnote{We recall that the time-dependence of the field is $\sim e^{-i\omega t}$, and a positive imaginary component
of the frequency signals an instability.}. Figure~\ref{Fig:BQNM} (left panel) shows that the time scale dependence on 
$r_m$ is the same for EM and scalar fluctuations,
as predicted in Section~\ref{sec:model}. Note that the EM growth rates $1/\omega_I$ are about one order of magnitude 
smaller than those of scalar fields. This is consistent with the fact that the maximum superradiant amplification factor 
for vector fields is approximately one order of magnitude larger than those of scalars, as shown in 
Fig.~\ref{fig:Acomparison}.

As also anticipated with the heuristic argument in the previous section, the instability time scale grows with $r_m^{2l+2}$
and the oscillation frequency $\omega_R$ is inversely proportional to the mirror position and reduces to the flat space result when $r_m\gg M$. Thus, for very small $r_m$ the superradiant condition $\omega<m\Omega_{\rm H}$ is violated and the superradiant instability is quenched. An analytic understanding of the onset of the instability is provided in Ref.~\cite{Hod:2014uoa}; generic analytic studies can be found in
Refs.~\cite{Cardoso:2004nk,Hod:2009cp,Rosa:2009ei,Hod:2009cw}. 
In the limit of very large cavity radius $r_m/M$ our results reduce to the TE and TM modes of a spherical cavity in flat space~\cite{Jackson} (see also Appendix~\ref{app:EM_BCs}).

These findings are fully corroborated by a time-domain analysis, summarized in Fig.~\ref{fig:timeseries2} for the case of a scalar field with
Dirichlet conditions at $r=r_m$~\cite{Dolan:2012yt}. The exponential growth of the scalar field for rapidly spinning BHs is apparent.
A full nonlinear evolution of the Einstein-Klein-Gordon system in a confining mirror was recently performed~\cite{Witek:2010qc}; the results were promising but numerically unstable on time scales too short to observe superradiant-induced growth of the scalar, or to probe the end-state of the instability. This remains an open issue to date.

The original investigation on BH bombs suggested that the setup could be used by advanced civilizations as an energy source~\cite{Press:1972zz}.
The idea is simple, and amounts to cutting small holes in the mirror (or making it semipermeable in some other way) such that one can extract (say) electromagnetic energy in a nearly stationary regime.
More recently, the functional dependence of such flux on the mirror radius and BH spin was observed to parallel that of the Blandford-Znajek mechanism (see Section~\ref{sec:BF} below) for extraction of energy via magnetic fields~\cite{Wilson-Gerow:2015esa}.

Rotating BHs surrounded by artificial mirrors were studied also in the context of higher dimensional BHs, with similar conclusions~\cite{Lee:2011ez,Aliev:2014aba}. Finally, as discussed in Sections~\ref{super_discontinuity} and~\ref{sec:lab}, realizations of effective BH metrics in the laboratory are possible through the use of acoustic setups. In this context, BH bombs were shown to be unstable on short, and possibly experimentally accessible, time scales~\cite{Berti:2004ju}. A possible end-state of the instability are ``distorted geometries'', which were recently discussed at the linearized level~\cite{Benone:2014nla}. Note however, that the boundary conditions used in both these references are very special and correspond to a highly absorbing boundary.
A discussion of general boundary conditions for acoustic geometries can be found elsewhere~\cite{Oliveira:2014oja}.

%%%%%%%%%%%%%%%%%%%%%%%%%%%%%%%%%%%%%%%%%
\paragraph{Accretion disks: open mirrors}
%%%%%%%%%%%%%%%%%%%%%%%%%%%%%%%%%%%%%%%%%

The BH bomb scenario discussed previously can serve as a model to describe astrophysical BHs surrounded by plasmas or 
accretion disks. Ionized matter is a good low-frequency EM waves reflector (see Section~\ref{plasma-triggered} below) 
and can thus play the role of the mirror (this was first realized by Teukolsky~\cite{teukolskythesis} and it is 
discussed in more detail in Secs.~\ref{plasma-triggered} and \ref{sec:astro_plasma}.). A very important question which 
still needs clarification concerns the effectiveness of the instability in these realistic situations. The matter 
surrounding the BH comes under the form of thin or thick accretion disks and not as spherically shaped mirrors. 
Confining the field along some angular direction means forbidding low angular eigenvalue modes, implying that only 
higher-angular eigenvalue modes (with longer time scales, cf. Eq.~\eqref{absor_wIm_rot}) are 
unstable~\cite{Putten,Aguirre:1999zn}.

Although the geometrical constraint imposed by accretion disks does not completely quench the instability, it can be 
argued that absorption effects at the mirror could~\cite{Aguirre:1999zn}. Consider an optimistic setup for which the EM 
wave is amplified by $\sim 1\%$ each time that it interacts with the BH~\cite{Teukolsky:1974yv}. A positive net gain 
only ensues if the wall has a reflection coefficient of $99\%$ or higher. 
On the other hand, this argument assumes that the mirror itself does not amplify the waves. But if it is rotating, it may too contribute to further amplification (an interesting example of amplification induced by a rotating cylinder is discussed in Ref.~\cite{Bekenstein:1998nt}).
Clearly, further and more realistic studies need to be made before any conclusion is reached about the effectiveness of ``BH bomb'' mechanisms in astrophysical settings.

%%%%%%%%%%%%%%%%%%%%%%%%%%%%%%%%%%%%%%%%%%%%%%%%%%%%%%%%%%%%%%%%%
\subsubsection{Charged black-hole bombs}\label{sec:chargedbombs}
%%%%%%%%%%%%%%%%%%%%%%%%%%%%%%%%%%%%%%%%%%%%%%%%%%%%%%%%%%%%%%%%%
 
As shown in Section.~\ref{sec:superradiance_charged}, charged fields can also be amplified through superradiance in a charged BH background. In complete analogy with the rotating BH bomb, one may also consider building a charged BH bomb. Numerical studies of charged BH bombs, at linear level in the field amplitude, were done in Refs.~\cite{Herdeiro:2013pia,Degollado:2013bha,Dias:2018zjg}, both in the frequency and time-domain whereas analytic studies were done in Refs.~\cite{Hod:2013fvl,Li:2014gfg}. It was shown that in the limit $q Q\to\infty$ and for a mirror in the near-horizon region, the characteristic frequency follows a linear scaling $\omega_I\propto q Q/r_+$, implying that the instability growth time scale $\tau_{\rm ins}\equiv 1/\omega_I$ can be made arbitrarily small by increasing $q$. In Refs.~\cite{Li:2014fna,Li:2015mqa} these results were extended to a charged massless scalar field in the background of a charged 
stringy BH with mirror-like boundary conditions.

Although astrophysical BHs are thought to be neutral due to quantum effects and plasma neutralization, this system is 
interesting from a conceptual point of view: the very short instability time scale (as compared to the very long time 
scales involved in the rotating case) makes it a very promising testbed for fully nonlinear studies following the 
development of the instability of BHs in a cavity. Such nonlinear evolutions were recently accomplished~\cite{Sanchis-Gual:2015lje,Sanchis-Gual:2016tcm}. The setup mimics the original BH bomb, and consists on an asymptotically flat, spherically symmetric spacetime with a surface at some (fixed coordinate) distance where the charged scalar obeys 
Dirichlet conditions~\cite{Sanchis-Gual:2015lje,Sanchis-Gual:2016tcm}. The framework allows the linearized instability to be followed through the nonlinear stage, and there is strong evidence 
in favour of a final state where a BH is surrounded by a charged condensate. These are in fact static, nonlinear solutions of the field equations which are entropically preferred~\cite{Dolan:2015dha}.

%%%%%%%%%%%%%%%%%%%%%%%%%%%%%%%%%%%%%%%%%%%%%%%%%%%%%%%%%%%
\subsection{Black holes in AdS backgrounds}\label{sec:AdS}
%%%%%%%%%%%%%%%%%%%%%%%%%%%%%%%%%%%%%%%%%%%%%%%%%%%%%%%%%%%
Black holes in anti-de Sitter backgrounds behave as BHs in a box, as the AdS boundary is
timelike and is may confine fluctuations. One way to see this is through the analysis of timelike geodesics:
no timelike particle is able to reach spatial infinity, and therefore AdS backgrounds can be looked at as 
really a confining system. Another intuition into these spacetimes is brought about by
following radial null geodesics in the geometry~\eqref{metricKerrLambda}
with a negative cosmological constant and zero mass. According to \eqref{geodesics:kerrt}-\eqref{geodesics:kerr}, these are governed by
\be
dr/dt=(r^2/L^2+1)\,,
\ee
where $L$ is the curvature radius of AdS spacetime, related to the negative
cosmological constant in the Einstein equations through 
\be
L^2=-3/\Lambda\,.
\ee
In other words, an observer at the origin measures a finite time
$t=\pi L/2$ for a light ray to travel from the origin to the AdS boundary at $r=\infty$. 
This short result teaches us that boundary conditions at spatial infinity are crucial to determine the evolution of the system.

In view of the above, rotating or charged BHs in anti-de Sitter are expected to behave as the ``BH bombs'' previously described:
for small BH size -- relative to the AdS curvature radius -- one expects superradiant instabilities, whereas for ``large'' BHs
the resonant frequencies are too large and outside the superradiant threshold. Alternatively, slowly rotating BHs with
$\Omega_{\rm H} L<1$ are expected to be stable whereas rapidly spinning BHs are expected to be 
unstable~\cite{Hawking:1999dp}. 
%%%%%%%%%%%%%%%%%%%%%%%%%%%%%%%%%%%%%%%%%%%%%%%%%%%%%%%%%%%%%%%%%%%%%%%%%%%%%%%%%%%%%%%%%%%%%%%%%%%%%%%%%
\subsubsection{Instability of small Kerr-AdS black holes and new black-hole solutions}\label{sec:KerrAdS}
%%%%%%%%%%%%%%%%%%%%%%%%%%%%%%%%%%%%%%%%%%%%%%%%%%%%%%%%%%%%%%%%%%%%%%%%%%%%%%%%%%%%%%%%%%%%%%%%%%%%%%%%%

%
\begin{figure}[ht]
\begin{center}
 \epsfig{file=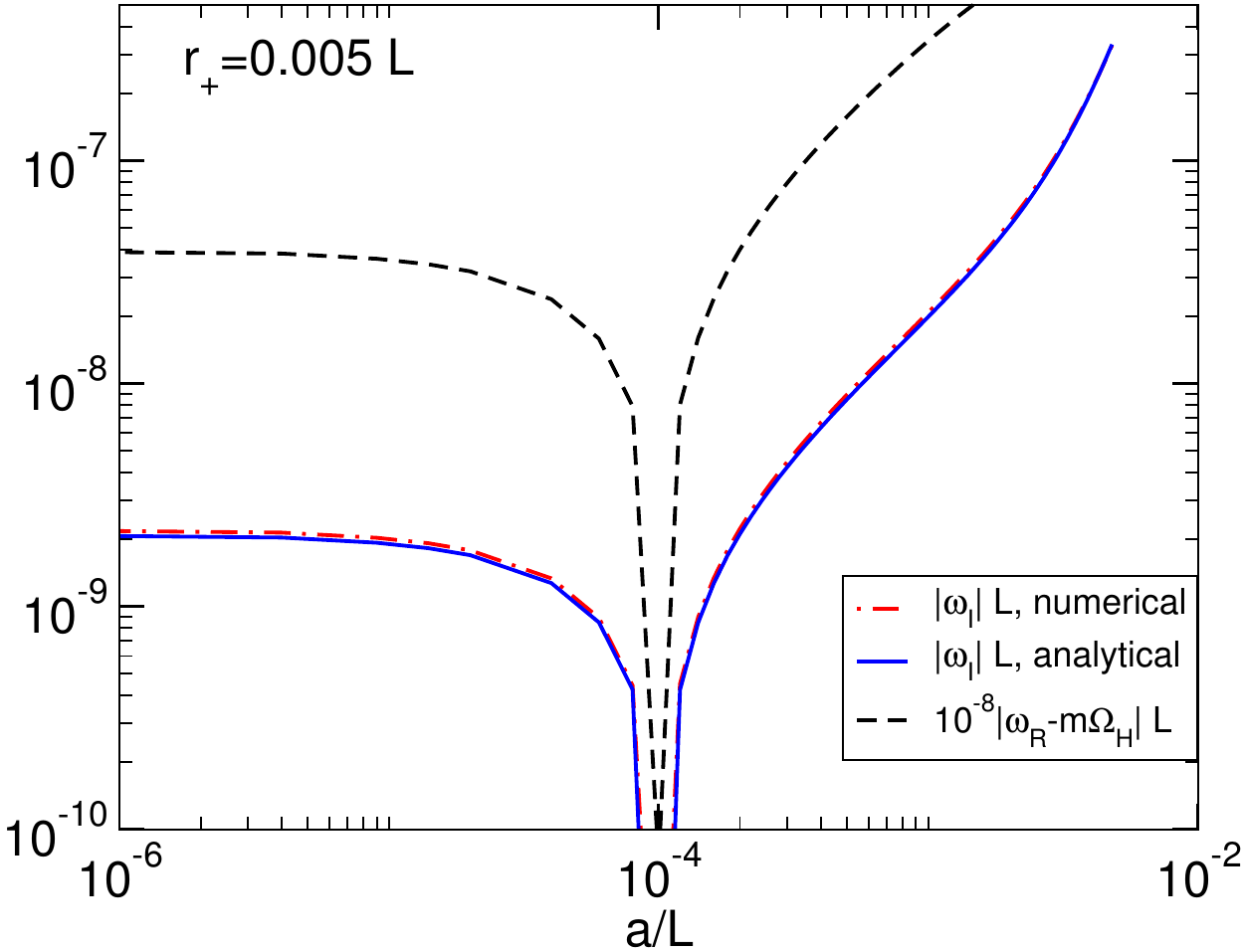,width=0.7\textwidth,angle=0,clip=true}
\end{center}
\caption{Details of the superradiant instability against dipole ($l=m=1$) scalar fields. We consider a spinning Kerr-AdS BH with $r_+/L=0.005$ (in Boyer-Lindquist coordinates).
The red dashed-dotted line represent numerical data points for the growth-rate $\omega_I$, and the solid blue curve is the analytical prediction~\eqref{adS_ana_super}.
Note that for rotation rates smaller than $a/L\sim 10^{-4}$ the perturbations become {\it stable}; as the dashed black line shows, this is also the critical point for superradiance, at which $\omega_R=m\Omega_{\rm H}$.}
\label{fig:KerrAdS_scalar}
\end{figure}
The previous arguments were shown to be correct in a series of works, starting with the proof that ``large''
Kerr-AdS BHs are stable~\cite{Hawking:1999dp}. Small Kerr-AdS BHs were subsequently shown to be mode-unstable against
scalar-field fluctuations~\cite{Cardoso:2004hs,Cardoso:2006wa,Uchikata:2009zz}. For small BHs, i.e. for $r_+/L\ll 1$,
the characteristic frequencies will be a deformation of the pure-AdS spectrum $L\omega=l+3+2n$~\cite{Burgess:1984ti}.
A matched asymptotic expansion method yields the eigenfrequencies~\cite{Cardoso:2004hs,Uchikata:2009zz},
\be
L\omega=l+3+2n-i\sigma\left(\frac{l+3+2n}{L}-m\Omega_{\rm H}\right)\frac{(r_+^2+a^2)(r_+-r_-)^{2l}}{\pi L^{2(l+1)}}\,,\label{adS_ana_super}
\ee
where
\be
\sigma=\frac{(l!)^2(l+2+n)!}{(2l+1)!(2l)!n!}\frac{2^{l+3}(2l+1+2n)!!}{(2l-1)!!(2l+1)!!(2n+3)!!}\prod_{k=1}^{l}(k^2+4\varpi^2)\,,
\ee
and
\be
\varpi=\left(\frac{l+3+2n}{L}-m\Omega_{\rm H}\right)\frac{r_+^2+a^2}{r_+-r_-}\,.
\ee
Here, $r_-$ is the smallest root of $\Delta$ in \eqref{metric parameters} and $\Omega_{\rm H}$ was defined in \eqref{Omega}.
The numerical solution of the eigenvalue problem was first considered in Ref.~\cite{Uchikata:2009zz} and agrees remarkably well with 
the analytical result \eqref{adS_ana_super}. As an example we used a direct integration, shooting method to determine numerically the eigenvalues
for $r_+/L=0.005$, the results are summarized in Fig.~\ref{fig:KerrAdS_scalar}, where we also show the analytical prediction.
At low rotations the imaginary component of the fundamental eigenfrequency is negative, $\omega_I<0$, signalling a stable spacetime.
As soon as the superradiance condition is satisfied, i.e., when $\omega_R<m\Omega_{\rm H}$, the superradiant mechanism sets in and the spacetime
is unstable, $\omega_I>0$, with an instability time scale given by $\tau\equiv 1/\omega_I$.

A numerical search of the parameter space shows that the peak growth rate for the instability is around $\omega_I\sim 3\times 10^{-4}$
at $r_+/L\sim 0.07$ for a nearly extremal BH. For $r_+/L>0.15$ there are no signs of unstable modes.

\begin{figure}[ht]
\centering
\begin{tabular}{cc}
 \includegraphics[width=.48\textwidth]{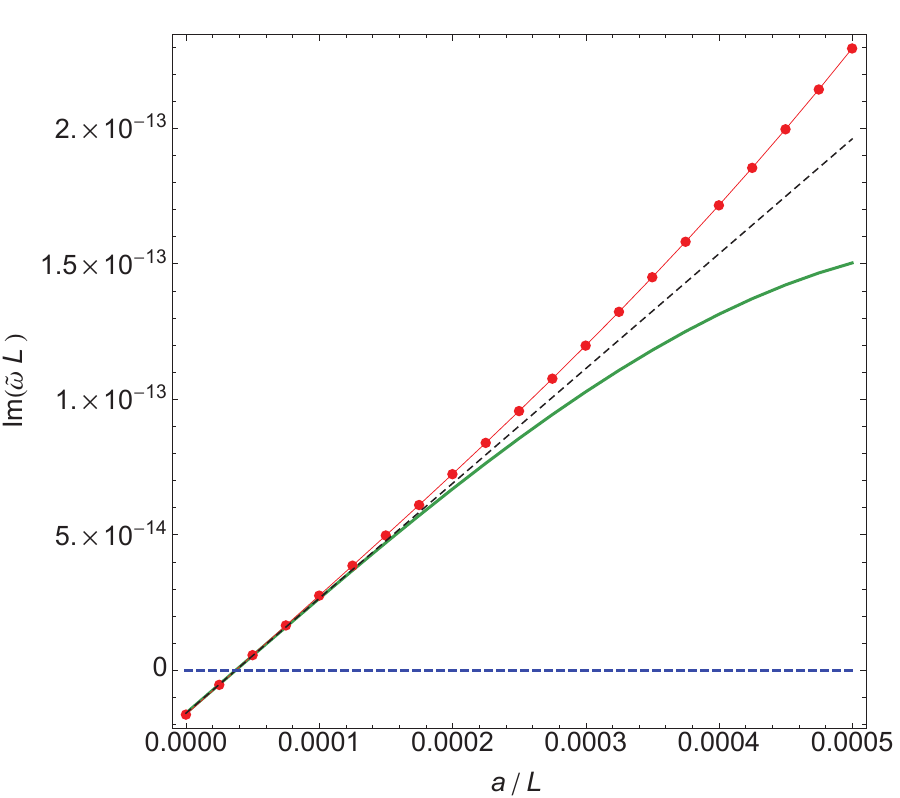}&
 \includegraphics[width=.48\textwidth]{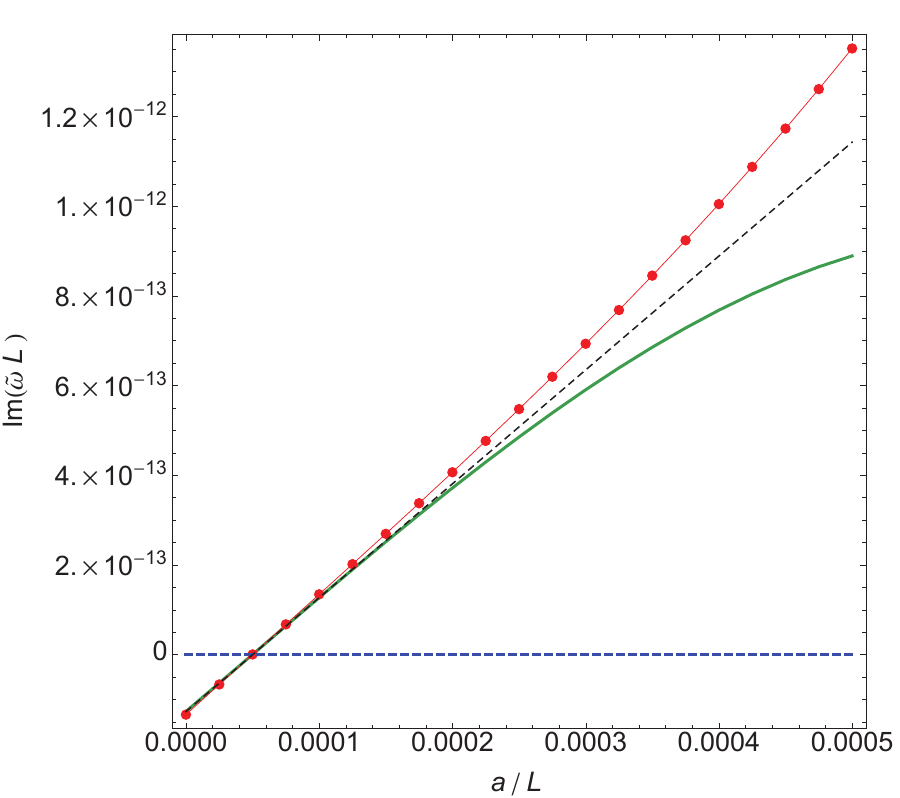}
\end{tabular}
\caption{
Imaginary part of the QNM frequency as a function of the rotation parameter $a/L$, for fixed horizon radius $r_+/L=0.005$, for $l=2$ gravitational scalar ({\it Right Panel}) and vector modes ({\it Left Panel}). 
Here $\omega=\Sigma\tilde{\omega}$.
The red dots are numerical points. The green curve is the numerical solution of the matching transcendental equation \eqref{FreqMatchingl2}, while the dashed black curve is the approximated analytical solution \eqref{SnormalM} or \eqref{VnormalM} of \eqref{FreqMatchingl2}. In both figures there is a critical rotation where $\omega_I=0$ and $\omega_R-m\Omega_{\rm H}\simeq 0$ to within $0.01\%$. For lower rotations the QNMs are damped and with $\omega_R-m\Omega_{\rm H}>0$, while for higher rotations we have unstable superradiant modes with $\omega_R-m\Omega_{\rm H}<0$.
From Ref.~\cite{Cardoso:2013pza}.}\label{Fig:matchNum}
\end{figure}  
Gravitational~\cite{Cardoso:2013pza} and EM perturbations~\cite{Wang:2015goa,Wang:2015fgp} can be handled in a similar 
way\footnote{A comprehensive discussion of the acceptable boundary conditions for gravitational fluctuations is 
presented in Refs.~\cite{Dias:2013sdc,Cardoso:2013pza,Sullivan:2017agx}.};
these perturbations have two degrees of freedom which have traditionally been termed gravitational vector (or 
Regge-Wheeler or odd) perturbations and gravitational scalar (or Zerilli or even) perturbations, and similarly for EM 
perturbations.
For small $r_+/L$, a similar matched asymptotic expansion technique can be used~\cite{Cardoso:2013pza}. For the lowest gravitational harmonic $l=2$, the characteristic frequencies satisfy 
\begin{eqnarray}\label{FreqMatchingl2}
&& \hspace{-3.2cm} 
i (-1)^{L \tilde{\omega} +1} L^{-5}\left(r_+-\frac{a}{r_+^2}\right)^5 L \tilde{\omega}  \left(L^2 \tilde{\omega} ^2-1\right)\left(L^2 \tilde{\omega} ^2-4\right) \Gamma (5-2 i \varpi ) \nonumber\\
&&   \hspace{3cm} +5400 \left[\varepsilon+(-1)^{L \tilde{\omega} }\right] \Gamma (-2 i \varpi )=0\,,
\end{eqnarray}
where $\varepsilon=1$ describes gravitational scalar modes while $\varepsilon=-1$ represents gravitational vector modes (both with the boundary conditions corresponding to a non-deformed AdS boundary~\cite{Cardoso:2013pza}).
Note also that Ref.~\cite{Cardoso:2013pza} uses a slightly different coordinate system with time coordinate $\tilde{t}$
and characteristic frequency $\tilde{\omega}$. In the Boyer-Lindquist coordinates we adopt here, the characteristic frequencies are $\omega=\Sigma\tilde{\omega}$, where $\Sigma$ was defined in Eq.~\eqref{metric parameters}.

An approximate analytic solution (valid in the limit $a/L\ll r_+/L \ll 1$) of the transcendental equations above is
\begin{eqnarray}
&&  \hspace{-1.7cm} 1) \quad \hbox{Scalar modes:} \quad  \tilde{\omega}_I  L\simeq \frac{16 }{15 \pi }\left[-\frac{3 r_+^6}{L^6}+\frac{ m \,a \,r_+^4}{L^5}\left(1+15 (5 \gamma -7) \,\frac{r_+^2}{L^2}\right)\right] +\cdots \,, \label{SnormalM}
\\ 
&& \hspace{-1.7cm}  2) \quad \hbox{Vector modes:} \quad  \tilde{\omega}_I  L\simeq \frac{96}{15 \pi }\left[-\frac{4 r_+^6}{L^6}+\frac{m\, a\, r_+^4}{L^5}\left(1+\frac{80 (5 \gamma -7)}{3}\,\frac{r_+^2}{L^2}\right)\right] +\cdots \,,\label{VnormalM}   
\end{eqnarray}
where $\gamma\simeq 0.577216$ is the Euler-Mascheroni constant. The overall behavior is identical to that of scalar fields. For both scalar and vector modes the imaginary part of the frequency is negative at $a=0$, consistent with the fact that QNMs of Schwarzschild-AdS are always damped~\cite{Berti:2009kk,Cardoso:2001bb}. However, as $a/L$ increases, $\omega_I$ increases. As in the previous cases, at the critical rotation where the crossover occurs, i.e. $\omega_I=0$, one has $\omega_R-m\Omega_{\rm H}\simeq 0$ to within $0.01\%$. For smaller rotations one has  $\omega_R-m\Omega_{\rm H}>0$ and for higher rotations one has  $\omega_R-m\Omega_{\rm H}<0$ and $\omega_I>0$. Therefore, the instability which is triggered at large rotation rates has a superradiant origin since the superradiant factor becomes negative precisely when the QNMs go from damped to unstable.
These analytical matching results provide also a good testbed check to our numerics. Indeed we find that our analytical and numerical results have a very good agreement in the regime of validity of the matching analysis. This is demonstrated  in Fig. \ref{Fig:matchNum} where we plot the numerical and analytical results for the fundamental $l=2$ scalar and vector modes.

Finally, note that the strength of the scalar or vector gravitational instabilities can be orders of magnitude higher than the strength of the same superradiant instability sourced by a scalar field perturbation~\cite{Cardoso:2004hs,Uchikata:2009zz,Sullivan:2017agx}.
The maximum growth rate for the scalar and vector superradiant instability is of order $L\omega_I \sim 0.032, 0.058$ respectively at $(r_+/L,a/L)\sim (0.445,0.589), (0.32,0.386)$ (for further details see Ref.~\cite{Cardoso:2013pza}); the peak growth rates are therefore
substantially larger than those for scalar field fluctuations, as might be anticipated. Indeed the maximum growth rates are two orders of magnitude larger than in the scalar case, as might be expected
from the corresponding two orders of magnitude difference in superradiant amplification factors.

Direct evolutions in the time-domain were recently reported for scalar fields yielding instability time scales consistent with the frequency-domain analysis~\cite{Cardoso:2015fga}.
Finally, rigorous growth-rate estimates for generic initial data are provided in Ref.~\cite{Shlapentokh-Rothman:2013ysa}.

This mechanism is likely to render other rotating black objects in asymptotically AdS spacetimes unstable.
One example of such objects are rotating black rings, recently discussed in Ref.~\cite{Figueras:2014dta}.
The above results were generalized to scalar and Maxwell perturbations with generic boundary conditions~\cite{Wang:2015goa,Ferreira:2017tnc}, and also to higher-dimensional spacetimes~\cite{Delice:2015zga,Li:2016kws}; mirror-like conditions in higher-dimensional and asymptotically AdS spacetimes were studied in Ref.~\cite{Huang:2016zoz}.

\paragraph{Three-dimensional BHs in AdS}
The only exception to this rule are $(2+1)$-dimensional BHs, whose spectra shares some similarities to those of Kerr BHs.
Studies of the so-called rotating BTZ BH spacetime or ``squashed'' versions present in modified theories of gravity 
thereof show that these geometries are {\it stable} for the restricted set of boundary conditions~\cite{Birmingham:2001pj,Ferreira:2013zta}. Note that in three dimensional 
GR there are no gravitational degrees of freedom, and that stability results refer only to scalar or EM fluctuations. Notwithstanding, it is possible to show that superradiance and associated instabilities are present for other, more general boundary conditions~\cite{Iizuka:2015vsa,Dappiaggi:2017pbe}; for Robin boundary conditions it is possible to construct BH solutions surrounded by scalar clouds~\cite{Ferreira:2017cta}.

\paragraph{The end-state of the instability and new BH solutions}
Small, rapidly spinning BHs in AdS are unstable. Where does the instability drives the system to? For such confining geometries, the final state cannot be a Kerr-AdS BH: energy and angular momentum conservation guarantee that the BH would have exactly the same parameters as the initial state, hence it would be unstable.
Furthermore, the BH is amplifying low-frequency radiation which can not penetrate the horizon.
We are thus led to the conclusion that the final state of the instability must be a rotating BH surrounded by a bosonic ``cloud'',
generically a very dynamic spacetime due to GW emission induced by the cloud.

In certain cases, it is possible to suppress GW emission by considering contrived matter content, as it was done in Ref.~\cite{Dias:2011at} where the authors have explicitly constructed
an AdS BH with scalar hair, albeit in five-dimensional spacetimes. The action considered includes 2 complex scalar fields in five dimensions,
\be
S=\int d^5 x \sqrt{-g}\left[R+\frac{12}{L^2}-2\left|\nabla \vec{\Pi}\right|^2\right]\,,
\ee
with
\be
\vec{\Pi}=\Pi e^{-i\omega t+i\psi}\left\{ 
\begin{array}{l}
             \sin{\left(\theta/2\right)}e^{-i\phi/2}\\
             \cos{\left(\theta/2\right)}e^{-i\phi/2}
\end{array}
\right\}\,.
\ee
With the ansatz
\be
ds^2=-fgdt^2+dr^2/f+r^2\left[h\left(d\psi+\frac{\cos\theta}{2}d\phi-\Omega dt\right)^2+\frac{d\theta^2+\sin^2\theta d\phi^2}{4}\right]\,,
\ee
then all metric coefficients $f,g,h,\Omega$ and the field $\Psi$ are real functions of a radial coordinate $r$.
Notice that such ansatz is special in the sense that even though the scalars are dynamical, the stress-tensor
\be
T_{ab}=\partial_a \vec{\Pi}^*\partial_b\vec{\Pi}+\partial_a \vec{\Pi}\partial_b\vec{\Pi}^*-g_{ab}\partial_c\vec{\Pi}\partial^c\vec{\Pi}^*\,,
\ee
has the same symmetry as the metric. It is then possible to find five-dimensional AdS BHs with scalar hair by simply solving a set of coupled ODEs~\cite{Dias:2011at}. The BHs are neither
stationary nor axisymmetric, but are invariant under a single Killing field which is tangent
to the null generators of the horizon. These solutions can then be viewed either as the end-state of
the superradiant instability, or as interpolations between (equal angular momenta) Myers-Perry-AdS BHs and rotating boson stars in AdS.
In a phase diagram, these solutions bifurcate from the threshold of the superradiant instability of the original Myers-Perry BH.

More general solutions representing the end-point of superradiant instabilities, without the assumptions above, are thought to exist~\cite{Cardoso:2006wa,Cardoso:2009nz}; in fact, two such solutions have recently been studied~\cite{Iizuka:2015vsa,Dias:2015rxy} and they underline the role of superradiance in a vast set of physical phenomena including in the construction of novel BH solutions.

The nonlinear evolution of the superradiant instability of spinning BHs in four-dimensional AdS space was recently 
investigated in Ref.~\cite{Chesler:2018txn,Ishii:2020muv}. Evidence for the formation of a solution with a single 
helical Killing vector was reported. The evolution displays a multi-stage process with distinct superradiant 
instabilities. To ascertain the final fate of the system, longer duration nonlinear simulations must be performed.

%%%%%%%%%%%%%%%%%%%%%%%%%%%%%%%%%%%%%%%%%%%%%%%%%%%%%%%%%%%%%%%%%%%%%%%%%%%%%%%%%%%%%%%%%%%%%%%%%%%%%%%%%%%%%%%%%%%%%%%%%%
\subsubsection{Instabilities of charged AdS black holes: superradiance, spontaneous symmetry breaking and holographic 
superconductors}\label{sec:breaking}
%%%%%%%%%%%%%%%%%%%%%%%%%%%%%%%%%%%%%%%%%%%%%%%%%%%%%%%%%%%%%%%%%%%%%%%%%%%%%%%%%%%%%%%%%%%%%%%%%%%%%%%%%%%%%%%%%%%%%%%%%%

As might be anticipated, charged BHs in AdS are also unstable through superradiance, in line with what we discussed in Sec.~\ref{sec:chargedbombs} for charged BHs in a cavity. 
The setup describing a spherically symmetric, charged BH and a charged scalar -- generalizing to AdS 
asymptotics the charged BH bomb discussed in Sec.~\ref{sec:chargedbombs},
was also studied nonlinearly~\cite{Bosch:2016vcp,Dias:2016pma}. The results show strong evidence of a final stationary spacetime 
where a BH is in equilibrium with a charged condensate in its exterior, in line with the asymptotically flat 
case~\cite{Sanchis-Gual:2015lje,Sanchis-Gual:2016tcm}. Together with the follow-up of the instability of spinning BHs discussed above, these studies show that 
the instability exists and that there is a rich structure in its development and final state. Superradiant instabilities arise also in other spacetimes with
AdS or dS asymptotics~\cite{Gonzalez:2017shu,Burikham:2017gdm}.

In fact, instabilities had been studied extensively in the context of the gauge-gravity duality and prompted a flurry 
of activity on so-called holographic superconductors and superfluids~\cite{Hartnoll:2008vx,Hartnoll:2009sz}. Curiously, 
the connection of this phenomenon to superradiance was initially almost unnoticed, and has been realized only some years 
after the original proposal (see Refs.~\cite{Hartnoll:2011fn,Dias:2011tj,Basu:2010uz,Dias:2016pma}). We present here a unified 
picture of this problem.

Instabilities of charged BHs in AdS have been studied in Ref.~\cite{Gubser:2008px} under a different guise, namely with the aim to provide a holographic dual description of a spontaneous symmetry-breaking mechanism at finite temperature. Ref.~\cite{Gubser:2008px} considered an Abelian Higgs theory in four-dimensional curved spacetime, which is given by action~\eqref{eq:MFaction} with a massless Maxwell field.
A solution of the theory above is a RN-AdS BH (cf. Eq.~\eqref{RNLambda}) endowed with an electric potential\footnote{As discussed in Ref.~\cite{Gubser:2008px}, the electric potential at the horizon should vanish to ensure regularity of the one-form $\Phi dt$.} $\Phi=Q/r-Q/r_+$ and a vanishing scalar field. A small scalar fluctuation on this background is governed by the Klein-Gordon equation with an effective mass term given by
\begin{equation}
 m^2_{\rm eff}=-\frac{q^2\Phi^2}{f(r)}\,, 
\end{equation}
where $f(r)$ is given in Eq.~\eqref{RNLambda} and for simplicity we have neglected the actual mass term $\mu_S$ of the scalar field, whose role is not crucial in this analysis. Thus, the effective mass squared is negative outside the horizon. If $q$ is sufficiently large, the negative potential well can produce unstable modes. 
Such modes only exist when the spacetime is asymptotically AdS and have no analog in flat space\footnote{A similar mechanism occurs also for neutral fields with nonminimal couplings~\cite{Cadoni:2009xm}. However, in that case the instability occurs also in asymptotically-flat spacetime~\cite{Gubser:2005ih} and does not have a holographic interpretation in terms of spontaneous symmetric breaking. In fact, this mechanism is akin to superradiant instabilities triggered by nonminimal couplings, as those discussed in Sec.~\ref{sec:nonminimal}.}. 
In fact, there are two different mechanisms at play~\cite{Dias:2010ma,Dias:2011tj,Dias:2016pma,Dias:2018zjg}, only one of which is associated with superradiance. 
One (nonsuperradiant) mechanism is related to the near-horizon geometry of the extremal RN BH which is described by 
${\rm AdS}_2$. When $m^2_{\rm eff}<m_{\rm BL}^2$ (where $m_{\rm BL}$ is the Breitenlohner-Freedman bound of the 
near-horizon $AdS_2$ geometry), the mode is effectively space-like and produces a tachyonic instability. Such 
instability also exists for non-extremal BHs although it requires larger values of $q$. On the other hand the second, 
superradiant, mechanism is related to the fact that charged scalar perturbations can be superradiantly amplified by the 
RN BH, the energy being trapped by the AdS boundary, which provides the arena for the instability. In fact, the 
linearized analysis is equivalent to that presented in Sec.~\ref{sec:superradiance_charged}. This mechanism is akin to 
the BH bomb and requires confinement due to the AdS boundary. Therefore, it only exists in global AdS and not for planar 
RN-AdS black branes~\cite{Basu:2010uz,Dias:2011tj}.

\begin{figure}
\epsfig{file=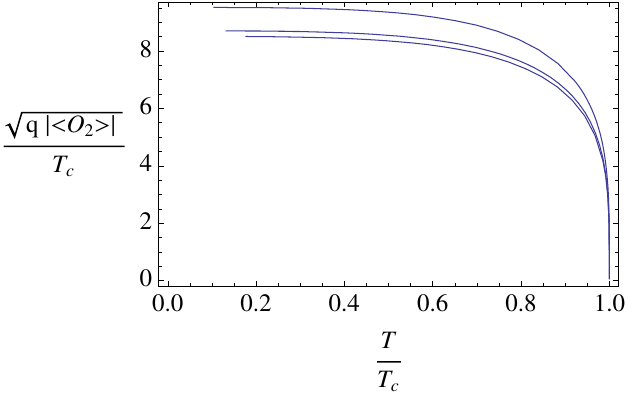,width=0.48\textwidth,angle=0,clip=true}
\epsfig{file=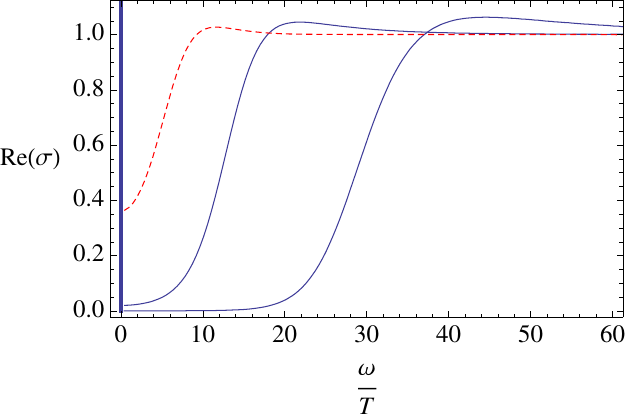,width=0.48\textwidth,angle=0,clip=true}
\caption{Left panel: the scalar operator of the boundary theory dual to the Abelian Higgs model~\eqref{eq:MFaction} (with massless Maxwell field) which is related to the asymptotic behavior of the scalar hair through the AdS/CFT dictionary. A hairy BH geometry branches off the RN-AdS solution and exists only below a certain critical temperature $T_c\sim\sqrt{\rho}$, where $\rho$ is the charge density. The behavior at $T\sim T_c$ shows that the phase transition is of second order. Different curves correspond to various values of the scalar charge $q$.
Right panel: the electric conductivity of the dual theory in the superfluid phase at various temperatures (decreasing from left to the right). A Dirac delta function appears at $\omega=0$ and there is a frequency gap at small temperatures.
From Ref.~\cite{Hartnoll:2008kx}.
}
\label{fig:HSC}
\end{figure}

\paragraph{Spontaneous symmetry breaking and holographic superconductors}
In the context of the gauge-gravity duality, this instability has far-reaching consequences, as it signals the onset of a phase transition towards a hairy BH configuration that breaks the $U(1)$ symmetry of the initial RN-AdS solution. In a quantum field theory, such spontaneous symmetry breaking (akin to the Higgs mechanism) is associated to superfluidity and the scalar condensate is associated to Cooper pairs~\cite{Hartnoll:2009sz,Hartnoll:2011fn}.
This same mechanism is at play in the Abelian Higgs model as was demonstrated in the seminal work~\cite{Hartnoll:2008vx} 
where a ``holographic superconductor'' was constructed as the nonlinear endstate of the superradiant instability. At 
small temperatures, the RN-AdS BH becomes unstable through superradiance and spontaneously develops a 
spherically-symmetric scalar hair. This is in agreement with our previous analysis, and only small BHs are unstable 
through this mechanism; in addition, planar RN-AdS black branes are {\it stable}~\cite{Dias:2011tj}.
The scalarized phase is energetically favored at low temperatures and corresponds to a nonvanishing expectation value of a scalar operator ${\cal O}_2$ living on the boundary, as shown in the left panel of Fig.~\ref{fig:HSC}. 

The behavior of the scalar condensate near the critical temperature signals a second-order phase transition. Other 
properties of the dual phase such as the existence of a gap in the conductivity, infinite DC conductivity, the existence 
of Cooper-like pairs and a Meissner-like effect, can all be studied by solving the linear response of the hairy BH 
solutions to scalar and EM perturbations~\cite{Hartnoll:2009sz}. An example is presented in the right panel of 
Fig.~\ref{fig:HSC}, showing the conductivity of the superfluid phase.

\begin{figure}[htbp]
  \centering
  \epsfig{file=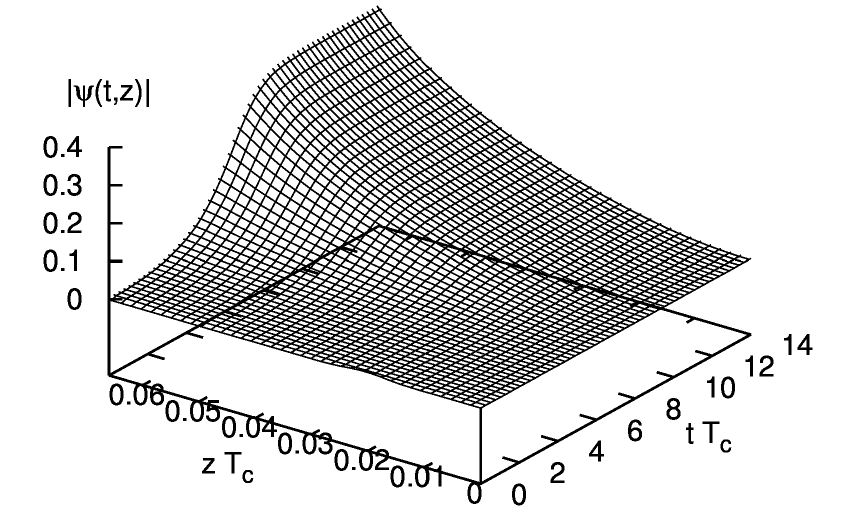,width=0.48\textwidth,angle=0,clip=true}
  \epsfig{file=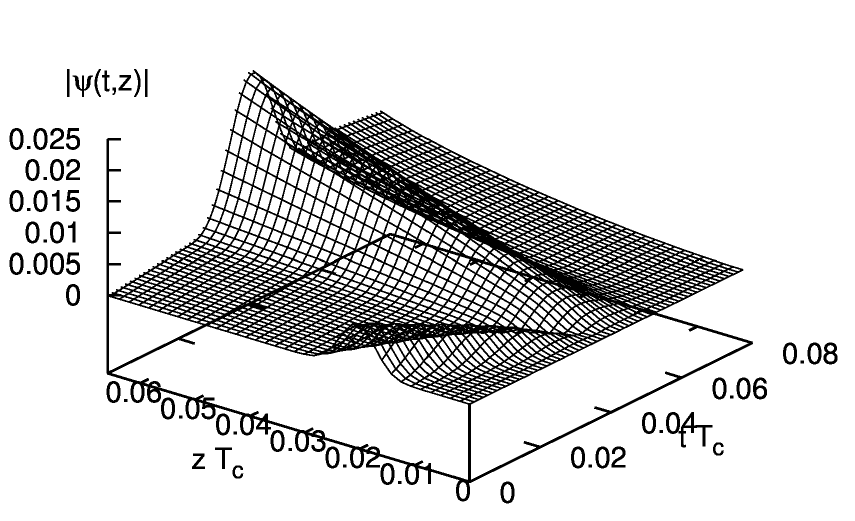,width=0.48\textwidth,angle=0,clip=true}
  \caption{\label{fig:HSC_evolution}
Example of the evolution of the scalar field on $(t,z)$-plane starting with a perturbed RN-AdS solution in the theory. Left panel: the time interval $0 \leq tT_c \leq 14$ is shown, where $T_c$ is the critical temperature for the phase transition. Because of the instability of the RN-AdS BH,  the scalar density grows exponentially for $tT_c\lesssim 6$ and for $tT_c\gtrsim 6$, the scalar density approaches a stationary configuration.
Right panel: the same evolution at initial times, $0 \leq tT_c \leq 0.08$. The wave packet is reflected by the AdS boundary at $t\simeq 0.04$ and most of it is absorbed at the BH horizon within $tT_c\lesssim 0.06$. From Ref.~\cite{Murata:2010dx}.
}
\end{figure}

The results in Ref.~\cite{Hartnoll:2008vx} triggered a flurry of activity in this field that goes well beyond the scope of this work (for a somehow outdated review see~\cite{Horowitz:2010gk}). Relevant to our discussion is the analysis of Ref.~\cite{Murata:2010dx}, in which nonequilibrium processes in the holographic superfluid phase and the energy extraction from the normal phase described by the (planar, in contrast to Ref.~\cite{Bosch:2016vcp} which studied spherical topology BHs) RN-AdS BH have been investigated through time evolutions. An example of such evolution is described in Fig.~\ref{fig:HSC_evolution}.

It is also interesting to mention the case of a charged massive fermion coupled to Einstein-Maxwell theory in AdS. As 
previously discussed, Pauli exclusion principle implies that fermions cannot condensate and, in turn, superradiance does 
not occur. From the holographic perspective, the quantum state will not have a coherent phase and the $U(1)$ symmetry is 
unbroken (cf. Ref.~\cite{Hartnoll:2011fn} for a review). While classical fermionic instabilities are prevented, 
Schwinger pair production of fermions can occur for sufficiently light fermions, in analogy to the bosonic case. The 
result of this process is the population of a Fermi sea delimited by a Fermi surface outside the BH, giving rise to 
so-called ``electron stars''~\cite{Hartnoll:2010gu} which are the (planar, AdS) cousins of astrophysical NSs.

%%%%%%%%%%%%%%%%%%%%%%%%%%%%%%%%%%%%%%%%%%%%%%%%%%%%%%%%
\subsection{Massive bosonic fields}\label{sec:mass}
%%%%%%%%%%%%%%%%%%%%%%%%%%%%%%%%%%%%%%%%%%%%%%%%%%%%%%%%
So far we have discussed two classes of BH-bomb systems: BHs enclosed in a reflecting cavity and BHs in asymptotically AdS spacetimes. The former are highly idealized and unrealistic configurations, whereas the latter --~although of great theoretical interest especially in the context of the gauge-gravity duality~-- are of little relevance for astrophysical BHs. 

Fortunately, sometimes ``nature provides its own mirrors''~\cite{Press:1972zz,Cardoso:2004nk}. A massive bosonic field naturally confines low-frequency radiation due to a Yukawa-like suppression $\sim e^{-\mu r}/r$ where $\mu$ is the mass term. Thus, it was suspected since the 1970s~\cite{Press:1972zz,Damour:1976kh} that superradiance triggers instabilities in spinning BH geometries.

This section is devoted to the superradiant instability of spinning BHs triggered by massive bosonic fields in asymptotically-flat spacetime, a topic that has recently flourished with numerous developments in the last few years. The busy reader will find a unified discussion of such instabilities in Sec.~\ref{sec:massive_unified}. We anticipate that superradiant instabilities triggered by massive bosons are relevant (i.e. their time scale is sufficiently short) only when the gravitational coupling $M\mu_i\lesssim1$. We recall that the physical mass of the fields is $m_{i}=\mu_{i}\hbar$, where $i=S,V,T$ for scalar, vector and tensor fields, respectively. In Planck units ($G=c=\hbar=1$), the following conversions are useful
%%%
\begin{equation}
 1\,{\rm eV} \approx 7.5\times 10^9 M_\odot^{-1}\approx 5.1 \times 10^{9}\, {\rm km}^{-1}\approx 1.5\times10^{15}\,{\rm s}^{-1}\,. \label{conversion}
\end{equation}
%%%
so that $\mu_i M\sim 0.75$ when $M\sim M_\odot$ and $\mu_i\sim 10^{-10}{\rm eV}$, i.e. for a ultralight bosonic field.

%%%%%%%%%%%%%%%%%%%%%%%%%%%%%%%%%%%%%%%%%%%%%%%%%%%%%%%%%%%%%%%%%%%%%%%%%%%%%%%%%%%%%%%%%%%%%%%%%%%%%%%%%%%%%%%%%%%%%%
\subsubsection{The zoo of light bosonic fields in extensions of the Standard Model}\label{sec:ultralight_motivations}
%%%%%%%%%%%%%%%%%%%%%%%%%%%%%%%%%%%%%%%%%%%%%%%%%%%%%%%%%%%%%%%%%%%%%%%%%%%%%%%%%%%%%%%%%%%%%%%%%%%%%%%%%%%%%%%%%%%%%%
All observed elementary particles are either fermions or bosons, according to the statistics they obey, which in turn determines whether they have half-integer or integer spin, respectively. Because superradiance does not occur for fermionic fields, here we are interested in massive bosons. 
All observed elementary bosons are all either massless or very massive, such as the $W$ and $Z$ bosons and the 
recently-discovered Higgs boson, whose masses are of the order $m\sim 100~{\rm GeV}$. As we discuss below, the condition 
$\mu_i M\lesssim1$ sets the range of mass $\hbar\mu_i$ which is phenomenologically relevant for a given BH mass $M$. A 
hypothetical boson with mass in the electronvolt range would trigger a sufficiently strong instability only for light 
BHs with masses $M\sim 10^{20}{\rm g}$. Although the latter could be formed in the early universe as ``primordial'' 
BHs~\cite{1971MNRAS.152...75H,1966AZh....43..758Z,1974MNRAS.168..399C} (see Ref.~\cite{Carr:2009jm} for 
a review) and are also promising dark-matter candidates~\cite{Carr:2016drx}, here we focus mostly on massive BHs, i.e. 
those with masses ranging from a few solar masses to billions of solar masses.

Superradiant instabilities of such massive BHs require ultralight bosonic fields in order to have astrophysically relevant time scales.
Such bosons are almost ubiquitous in extensions of the Standard Model of particle physics and in various extensions of GR.
The prototypical example of a light boson is the Peccei-Quinn axion~\cite{PecceiQuinn} introduced as a possible resolution for the strong CP problem in QCD, i.e. the observed suppression of CP violations in the Standard Model despite the fact that, in principle, the nontrivial vacuum structure of QCD allows for large CP violations. The Peccei-Quinn mechanism is based on a global symmetry, whose spontaneous breaking is associated to a new particle, the axion~\cite{1978PhRvL..40..223W,Wilczek:1977pj}. The axion should acquire a small mass due to nonlinear instanton effects in QCD and its mass is theoretically predicted to be below the electronvolt scale. 
For a massive BH with $M\sim 5 M_\odot$, axions with mass of the order of $10^{-11}{\rm eV}$ would have a superradiant 
coupling $\mu_S M\sim 0.4$, so that superradiant instabilities are potentially important. In addition to solve the 
strong CP problem, light axions are also interesting candidates for cold dark 
matter~\cite{Fairbairn:2014zta,Marsh:2014qoa,Marsh:2015xka,Bertone:2019irm,Braaten:2019knj,OHare:2020wah}. For example, using a fundamental discrete symmetry, two-axion models have been constructed where the QCD axion solves the strong-CP problem, and an ultralight axion provides the dominant form of dark matter~\cite{Kim:2015yna}.

Other light bosons, such as familions~\cite{Wilczek:1982rv} and Majorons~\cite{Chikashige1981265}, emerge from the spontaneous breaking of the family and lepton-number symmetries, respectively.
A common characteristic of these light bosons is that their coupling to Standard-Model particles is suppressed by the energy scale that characterizes the symmetry breaking, so that it is extremely challenging to detect these fields in the laboratory. Thus, massive BHs are probably the best candidates to investigate the putative effects of light bosons in a range which is complementary to searches using cosmological observations~\cite{Hlozek:2014lca}.

Furthermore, a plenitude of ultralight bosons might arise from moduli compactification in string theory. In the ``axiverse'' scenario, multiples of light axion-like fields can populate the mass spectrum down to the Planck mass, $m_{\rm P}\sim10^{-33}~{\rm eV}$, and can provide interesting phenomenology at astrophysical and cosmological scales~\cite{Arvanitaki:2009fg,Mehta:2020kwu}.

Light bosonic fields with spin are also a generic feature of extensions of the Standard Model. For example, massive vector fields (``dark photons''~\cite{Ackerman:mha,Nakayama:2019rhg}) arise in the so-called hidden $U(1)$ sector~\cite{Goodsell:2009xc,Jaeckel:2010ni,Camara:2011jg,Goldhaber:2008xy,Hewett:2012ns}. 
On the other hand, coupling massive spin-2 fields to gravity is a much more involved problem from a theoretical standpoint, but progress in this direction has been recently done in the context of nonlinear massive gravity and bimetric theories (cf. Refs.~\cite{Hinterbichler:2011tt,deRham:2014zqa} for reviews). A light massive graviton modifies the gravitational interaction at long distances and is a natural alternative to explain the accelerating expansion of the Universe.

In addition to fundamental bosonic fields, effective scalar degrees of freedom arise naturally due to nonminimal couplings or in several modified theories of gravity~\cite{Berti:2015itd}.
For example, in so-called scalar-tensor theories, the gravitational interaction is mediated by a scalar field in addition to the standard massless graviton. The no-hair theorems of GR extend to scalar-tensor gravity under certain conditions~\cite{Sotiriou:2011dz} so that GR BHs are also the unique vacuum, stationary solution of these theories. However, if the scalar field is massive such BHs would be unstable due to the superradiant instability. Due to a correspondence between scalar-tensor theories and theories which replace the Einstein-Hilbert term by a generic function of the Ricci curvature (so-called $f(R)$ gravity~\cite{Berti:2015itd}), effective massive scalar degrees of freedom are also present in these theories and trigger superradiant instabilities~\cite{Hersh:1985hz}\footnote{Interestingly, in the context of $f(R)$ gravity the effective scalar field is related to the scalar curvature of the metric, which grows exponentially through superradiance. This suggests that, at variance with the case of real massive fields in which the final state is likely a Kerr BH with lower spin (as discussed in the rest), the end-state of superradiant instabilities in $f(R)$ gravity might be different from a Kerr BH~\cite{Hersh:1985hz}. If such theories are to satisfy the no-hair theorem~\cite{Sotiriou:2011dz}, the end state of the instability should be a non-stationary solution.}. 

The phenomenological implications of superradiant instabilities triggered by light bosons are discussed in Sec.~\ref{sec:bounds_mass}, here we simply consider the mass of the boson (either $m_S$, $m_V$ or $m_T$ for spin-0, -1 and -2 particles, respectively) to be a free parameter of the model.

%%%%%%%%%%%%%%%%%%%%%%%%%%%%%%%%%%%%%%
\subsubsection{Massive scalar fields}
%%%%%%%%%%%%%%%%%%%%%%%%%%%%%%%%%%%%%%
The simplest and best understood case of superradiant instability triggered by massive bosons is the case of a massive probe scalar field propagating on a fixed Kerr geometry. 
The existence of this instability was originally suggested by Damour, Deruelle and Ruffini~\cite{Damour:1976kh} and has been thoroughly investigated by several authors since then.

The linearized dynamics is governed by the massive Klein-Gordon equation
%%%
\begin{equation}
 [\square-\mu_S^2]\Psi=0\,, \label{massiveKG}
\end{equation}
%%%
where the D'Alembertian operator is written on the Kerr metric and $\mu_S$ is the scalar mass term (we recall that we use Planck units henceforth; the physical mass $m_S$ of the field reads $m_S=\mu_S \hbar$). In the Teukolsky formalism~\cite{Teukolsky:1972my,Teukolsky:1973ha}, Eq.~\eqref{massiveKG} can be separated by use of spin-0 spheroidal wavefunctions~\cite{Berti:2005gp} as discussed in Sec.~\ref{sec:Teukolsky} for the massless case. 
The ODE for the angular part is identical to the massless case after the redefinition $\omega^2\to\omega^2-\mu_S^2$, whereas the potential of the radial equation gets a further contribution proportional to $\mu_S^2 \Delta  r^2$.
%%%%%%%%%%%%%%%%%%%%%%%%%%%%%%
\paragraph{Analytical results}
%%%%%%%%%%%%%%%%%%%%%%%%%%%%%%
The crucial parameter regulating the interaction between the geometry and the massive scalar is the gravitational coupling $M\mu_S$, which is just the ratio between the gravitational radius of the BH and the Compton wavelength of the field. In the scalar case analytical results are available in the $M\mu_S\ll1$ and $M\mu_S\gg1$ limits.

For small $M\mu_S$, it can be shown that the eigenvalue problem
admits a hydrogenic-like solution\footnote{Higher-order 
corrections to this formula have been computed in 
Ref.~\cite{Baumann:2018vus}. The eikonal, large $l$ limit is shown in Ref.~\cite{Eperon:2019viw} for arbitrary $M\mu_S$.}~\cite{Detweiler:1980uk,Rosa:2009ei,Pani:2012bp} with $\lambda\sim l(l+1)$ and
\begin{equation}
\label{omegaDetweiler}
\omega\sim \mu_S-\frac{\mu_S}{2}\left(\frac{M\mu_S}{l+n+1}\right)^2+\frac{i}{\gamma_{nlm}M}\left(\frac{am}{M}-2\mu_S 
r_+\right)(M\mu_S)^{4l+5}\,, \qquad M\mu_S\ll1\,,
\end{equation}
where $n=0,1,2...$ is analog to the principal quantum number in the hydrogen atom and $\gamma_{nlm}$ is a coefficient that depends on $(n,l,m)$; $\gamma_{011}=48$ for the dominant unstable mode~\cite{Pani:2012bp} (this result corrects the prefactor found in the original reference, for further details we refer the reader to the appendix of Ref.~\cite{Pani:2012bp}). Note that the QNMs are complex, $\omega=\omega_R+i\omega_I$, unless the superradiant condition is saturated. This happens when 
\begin{equation}
 a=a_{\rm crit}\approx \frac{2\mu_S M r_+}{m}\,.\label{acrit}
\end{equation}
%%%%
Because of the time dependence of the field, when $a>a_{\rm crit}$ the imaginary part of the modes is positive and the instability time scale can be defined as $\tau\equiv1/\omega_I$. In this case, the field grows exponentially in time, $\Psi\sim e^{t/\tau}$. The instability time scale depends on the coupling $\mu_S M$, on the spin $a/M$ and on the mode numbers $(l,m,n)$. The strongest instability occurs for $l=m=1$, $n=0$ and for highly-spinning BHs. 

In the same limit, $M\mu_S\ll1$, the eigenfunctions can be written in terms of Laguerre polynomials~\cite{Detweiler:1980uk,Yoshino:2013ofa}
\be
\psi(\mu_S,a,M,r)\propto\tilde{r}^l e^{-\tilde{r}/2}L_{n}^{2l+1}(\tilde{r})\,,\label{eigenfunctionDetweiler}
\ee
and $\psi$ becomes a universal function of the dimensionless quantity $\tilde{r}=2r M {\mu_S}^2/(l+n+1)$~\cite{Brito:2014wla}.
For the single most unstable mode, $l=m=1$ and $n=0$, the eigenfunctions simplify to $\psi\propto\tilde{r} e^{-\tilde{r}/2}$.

In the opposite regime, $M\mu_S\gg1$, the instability is exponentially suppressed. By using a WKB analysis,  Zouros and Eardley found that the shortest time scale reads~\cite{Zouros:1979iw}
%%%
\begin{equation}
 \tau \sim 10^7 M e^{1.84 M \mu_S} \qquad M\mu_S\gg1 \,.\label{Zouros}
\end{equation}

It can be shown that the super-radiant instability regime is bounded by the relation
\be
\mu_S<\sqrt{2}\,m\Omega\,,
\ee
and that the upper bound can be approached arbitrarily close in the eikonal regime, $M\mu\gg 1$~\cite{Hod:2012zza}.

Note that for a solar mass BH and a field of mass $m_S\sim 1$~eV, the parameter
$M\mu_S\sim10^{10}$ and the instability time scale is much larger than the age of the
universe. Therefore, the case $M\mu_S\gg1$ has little phenomenological relevance. Below we discuss a more interesting case, when the gravitational coupling is of order unity, $M\mu_S\lesssim1$.

%%%%%%%%%%%%%%%%%%%%%%%%%%%%%%%%%%%%
\paragraph{Numerical results}
%%%%%%%%%%%%%%%%%%%%%%%%%%%%%%%%%%%%
Exact results for any value of $M\mu_S$ and $a/M$ can be obtained by solving the problem numerically.
This was originally done in Ref.~\cite{Cardoso:2005vk} and a very complete analysis of the instability can be found in Ref.~\cite{Dolan:2007mj} which used an extension of the continued-fraction method~\cite{Berti:2009kk} to compute the unstable modes\footnote{The spectrum of massive scalar perturbations of the Kerr metric contains both stable QNMs and quasibound states, which are localized near the BH~\cite{Dolan:2007mj,Rosa:2011my,Pani:2012bp}. The quasibound states are those that become unstable in the superradiant regime.}.

\begin{figure}
\epsfig{file=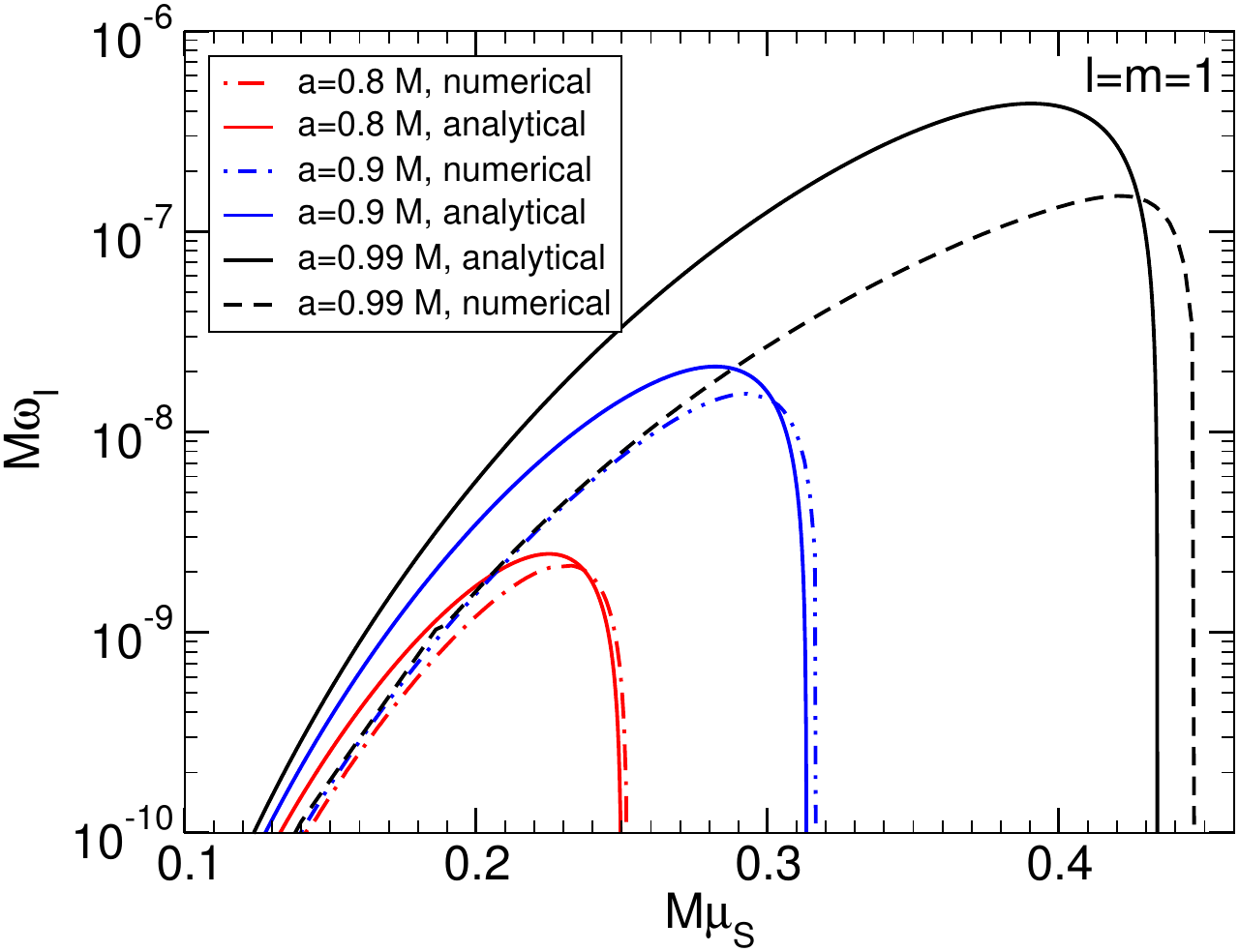,width=0.48\textwidth,angle=0,clip=true}
\epsfig{file=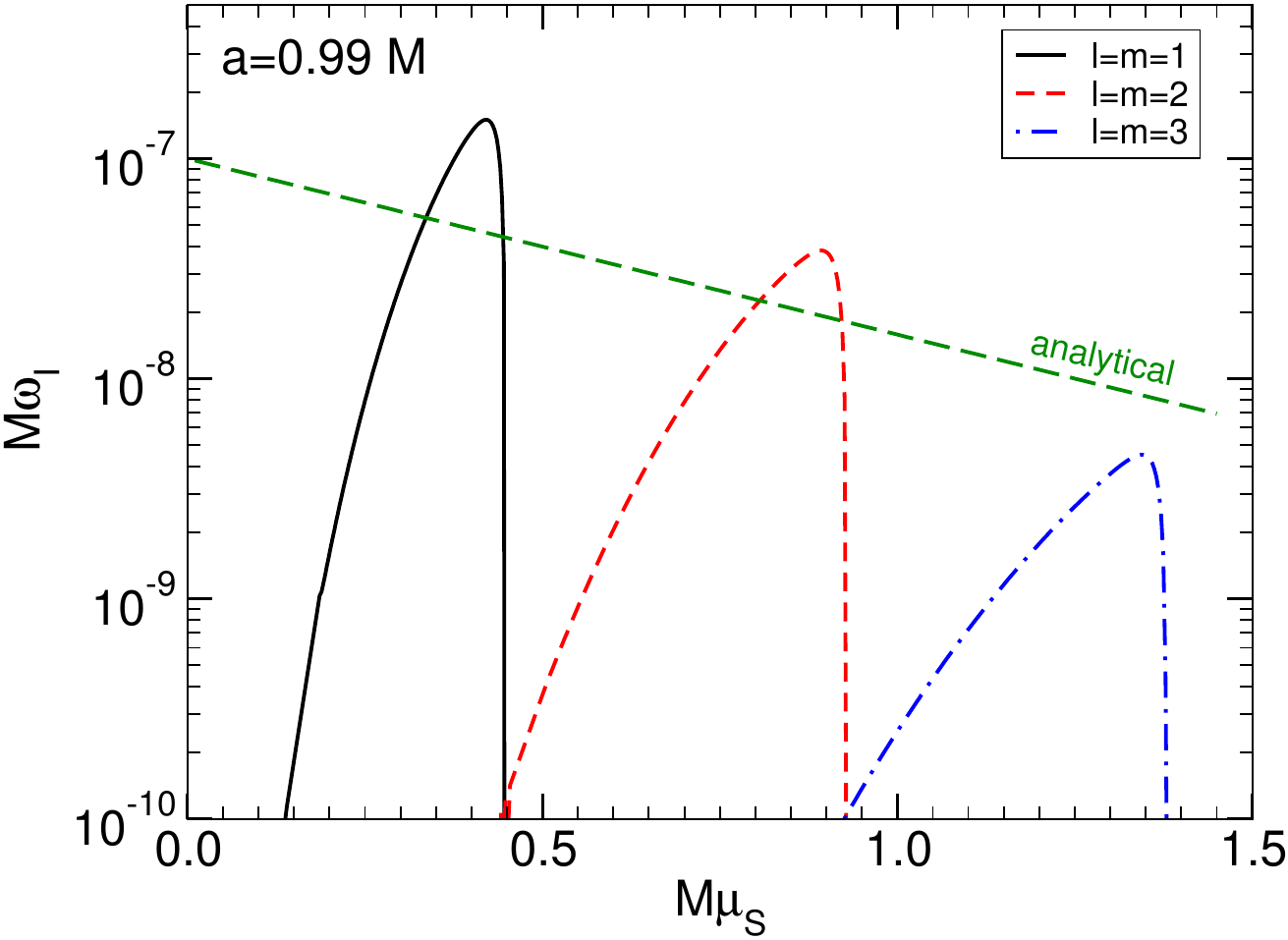,width=0.48\textwidth,angle=0,clip=true}
 \caption{
 Left: Superradiant instability for the fundamental ($n=0$) $l=m=1$ modes as a function of the gravitational coupling $M\mu_S$ and for various BH spin [see publicly available {\scshape Mathematica}\textsuperscript{\textregistered} notebook in Appendix~\ref{app:codes}]. The dotted lines shows Detweiler's approximation~\eqref{omegaDetweiler} \cite{Detweiler:1980uk} (with the coefficient corrected as in Ref.~\cite{Pani:2012bp}), valid in the limit $M\mu_S \ll 1$. Although not shown, the instability terminates (i.e. $\omega_I\to0$) when the superradiance condition is not satisfied.
Right: The same for the fundamental $l=m=1$, $l=m=2$ and $l=m=3$ modes. The fastest growth occurs for the $l=m=1$ state at $M \mu_S \approx 0.42$, with $a = 0.999 M$. The dotted line shows Zouros and Eardley's approximation~\cite{Zouros:1979iw}, valid when $M\mu \gg 1$ (cf. Eq.~\eqref{Zouros}).}
\label{fig:instabilityscalar}
\end{figure}

Some representative results are displayed in Fig.~\ref{fig:instabilityscalar}, which shows $\omega_I$ as a function of the gravitational couplings for various parameters.
The instability corresponds to $\omega_I>0$, which occurs when $\omega_R<m\Omega_{\rm H}$, i.e. when the mode satisfies the superradiance condition~\eqref{eq:superradiance_condition}. As expected, faster rotation leads to shorter growth time scales. Furthermore, for a given $l$, the mode with the faster growth rate has $m=l$, and clearly the axisymmetric mode with $m=0$ is stable. As in the analytical case, the dominant unstable mode has $l=m=1$ and $n=0$. For this mode the shortest instability time scale is approximately
%%%
\begin{equation}
 \tau_S\equiv\tau\sim 6.7\times 10^6 M\sim \left(\frac{M}{10^6M_\odot}\right){\rm yr}\,,\label{tauscalar}
\end{equation}
%%%
and occurs when $M\mu_S\sim 0.42$, corresponding to a light scalar field of mass
%%%
\begin{equation}
 \mu_S\sim 0.42 M^{-1}\sim 5.6\times 10^{-17} \left(\frac{10^6 M_\odot}{M}\right){\rm eV}\,.
\end{equation}

The exact numerical results can be used to check the validity of the analytical approximation when $\mu_S M\ll1$. It turns out that the spectrum~\eqref{omegaDetweiler} and the eigenfunctions~\eqref{eigenfunctionDetweiler} are a valid approximation of the exact results even for moderately large coupling (roughly up to $\mu_S M\lesssim0.2$) and even at large BH spin.

More recently, massive Klein-Gordon perturbations of Kerr BHs were also investigated through a time-domain analysis. This was done in Ref.~\cite{Witek:2012tr} by adapting a $3+1$ code, whereas subsequently an elegant decomposition in spherical harmonics was used to reduce the Klein-Gordon equation to an infinite set of hyperbolic partial differential equations for perturbations with different harmonic indices, which can then be solved with a $1+1$ code~\cite{Dolan:2012yt}. The results of these works are in remarkably good agreement with the frequency-domain analysis. Furthermore, Ref.~\cite{Witek:2012tr} provides an explanation for an apparent discrepancy between
time and frequency domain calculations of the instability growth rates as obtained in Ref.~\cite{Strafuss:2004qc}. This 
is related to an interference effect between different overtones that will be discussed in the context of massive vector 
fields below (cf. Fig.~\ref{fig:ProcaNum}).

%%%%%%%%%%%%%%%%%%%%%%%%%%%%%%%%%%%%%%%%%%%%%%%%%%%%%%%%%%%%%%%%%%
\paragraph{The end-state of the instability and new BH solutions}
%%%%%%%%%%%%%%%%%%%%%%%%%%%%%%%%%%%%%%%%%%%%%%%%%%%%%%%%%%%%%%%%%%
Unlike the AdS case discussed in Section~\ref{sec:AdS}, massive fields can confine only low-frequency radiation. The issue of the final state of the instability is discussed in detail in Sections~\ref{sec:nonlinear} and~\ref{sec:cloudproperties}. We anticipate here that --~because of the no-hair theorems ensuring that axisymmetric vacuum solutions of GR in asymptotically-flat spacetime are described by the Kerr geometry~\cite{HawkingBook,Heusler:1995qj,Sotiriou:2011dz,Graham:2014ina}~-- the final state of the superradiant instability of a Kerr BH will still be a Kerr BH with smaller mass and spin. An important counterexample is provided by the hairy BHs discussed in Sec.~\ref{sec:hair}. In that case, similarly to the AdS case, the metric remains stationary even if the scalar field oscillates in time~\cite{Herdeiro:2014goa,Herdeiro:2017phl}. This time dependence in the matter sector circumvents the hypothesis of the no-hair theorem and, at the same time, prevents GW emission.

%%%%%%%%%%%%%%%%%%%%%%%%%%%%%%%%%%%
\paragraph{Massive charged scalars}
%%%%%%%%%%%%%%%%%%%%%%%%%%%%%%%%%%%
Massive charged scalars propagating on a Kerr-Newman background were studied (both analytically and numerically) in Ref.~\cite{Furuhashi:2004jk}, which found that the instability growth rate also depends on the coupling $qQ$, where $q$ and $Q$ are the charges of the field and of the BH, respectively. For a given value of the BH spin the shortest instability time scale is comparable to that of the neutral case, although it occurs for different values of $\mu_S M$ and with $qQ\neq0$.

Because the BH-bomb effect occurs also for minimally-coupled, charged scalar perturbations of a static, charged BH in a
cavity (cf. Sec.~\ref{sec:chargedbombs}), one might be tempted to conclude that a similar instability exist also when 
the cavity's surface is replaced by a massive perturbation. However, unlike their rotating counterpart, 
asymptotically-flat charged BHs were shown to be stable against massive charged scalar perturbations. This is due to the 
fact that the conditions required in order to trigger the superradiant instability (existence of bound states in the 
superradiant regime) are incompatible~\cite{Hod:2013eea,Hod:2013nn}, and rotation is a necessary ingredient~\cite{Hod:2016bas}. The same absence of superradiant instability was recently proven for charged BHs in low-energy effective string theory~\cite{Zhang:2015jda}; rotation seems to be a necessary ingredient~\cite{Siahaan:2015xna}. Exceptions to this rule may arise when new couplings are present, changing the equations of motion and effectively a potential well for an instability to set in~\cite{Kolyvaris:2018zxl}.

%%%%%%%%%%%%%%%%%%%%%%%%%%%%%%%%%%%%%%%%%%%%%%%%%%%%%%%%%%%%%%%%%%%%%
\paragraph{Superradiance and effective mass in more generic theories}
%%%%%%%%%%%%%%%%%%%%%%%%%%%%%%%%%%%%%%%%%%%%%%%%%%%%%%%%%%%%%%%%%%%%%
A particularly important aspect of modified theories of gravity, is that some couplings can either mimic a mass term, or include features which are able to trap and amplify massless fields. A generic theory of scalar fields is ``Horndeski'' theory~\cite{Horndeski:1974wa,Berti:2015itd}. It displays at least two important features: i. it gives rise to instability of charged spacetimes, even in the absence of rotation~\cite{Kolyvaris:2018zxl}; ii. 
some of the couplings are equivalent to an effective mass term, thus leading to the possibility of studying such theories based almost exclusively on what one learns from
minimally coupled scalars on Kerr geometries~\cite{Tattersall:2019pvx}.
Finally, we should mention that there are superradiant instabilities which do not require a mass term to confine the field and make it grow. 
For example, even massless and charged scalar fields are unstable around a charged BH in de Sitter spacetime~\cite{Zhu:2014sya,Konoplya:2014lha,Destounis:2019hca}; the precise mechanism is not well understood, but it seems that the cosmological constant is responsible for creating the necessary potential well, where the superradiant mode can grow.

%%%%%%%%%%%%%%%%%%%%%%%%%%%%%%%%%%%%%%%%%%%%%%%%%%%%%%
\subsubsection{Massive vector fields}\label{ref:proca}
%%%%%%%%%%%%%%%%%%%%%%%%%%%%%%%%%%%%%%%%%%%%%%%%%%%%%%

While superradiant instabilities of spinning BHs against massive scalar perturbations have a relatively long 
history~\cite{Press:1972zz,Damour:1976kh,Cardoso:2004nk,Cardoso:2005vk,Dolan:2007mj,Rosa:2009ei,Cardoso:2011xi}, the 
case of massive bosonic fields with nonvanishing spin (i.e. massive vector and tensor fields) has been investigated much 
more recently. This is due to technical difficulties that were only recently overcome, as we now discuss.

%%%%%%%%%%%%%%%%%%%%%%%%%%%%%%%%%%%%%%%%%%%%%%%%%%%%%%%%%%%%
\paragraph{The issue of the separability of Proca equations}
%%%%%%%%%%%%%%%%%%%%%%%%%%%%%%%%%%%%%%%%%%%%%%%%%%%%%%%%%%%%

The equation governing massive vector (spin-1) fields is the Proca equation
\begin{equation}
\nabla_\sigma F^{\sigma\nu}=\mu_V^2 A^\nu\,,\label{proca}
\end{equation}
where $F_{\mu\nu}=\partial_\mu A_\nu-\partial_\nu A_\mu$, $A_\mu$
is the vector potential and we will focus again on the case in which the differential operator is written on the Kerr 
background. 
Maxwell's equations are recovered when
$\mu_V=m_V/\hbar=0$, where $m_V$ is the mass of the vector field.  Note
that, as a consequence of Eq.~\eqref{proca}, the Lorenz condition
$\nabla_\mu A^\mu=0$ is automatically satisfied, i.e. in the massive
case there is no gauge freedom and the field $A_\mu$ propagates
$2s+1=3$ degrees of freedom~\cite{Rosa:2011my}.

Initial studies of Proca fields on a BH geometry were restricted to the nonrotating
case~\cite{Gal'tsov:1984nb,Herdeiro:2011uu,Rosa:2011my,Konoplya:2005hr} and clearly fail to describe the superradiant 
regime. The main reason is
that the Proca equation~\eqref{proca} does not seem to be separable in the Kerr background by using the standard 
Teukolsky approach. 
Recently, the problem has been solved by various complementary approaches (in chronological order):
%%%
\begin{itemize}
 \item Through semi-analytical techniques in 
the slow-rotation limit~\cite{Pani:2012vp,Pani:2012bp} (briefly discussed in Appendix~\ref{app:HT}, cf. 
Ref.~\cite{Pani:2013pma} for a review on the slow-rotation approximation);
 %%%%
 \item Through a fully numerical time evolution of the Proca system, both in a fixed Kerr 
background~\cite{Witek:2012tr} and fully nonlinearly~\cite{East:2017ovw,East:2018glu} (see also 
Sec.~\ref{sec:nonlinear});
 %%%%
 \item Through a fully numerical frequency-domain computation, solving a system of elliptic partial differential 
equations~\cite{Cardoso:2018tly,Baumann:2019eav};
 %%%%
 \item Through a novel technique~\cite{Frolov:2018pys} to \emph{separate} the Proca equations on a Kerr 
background~\cite{Frolov:2018ezx}, resulting in system of coupled ordinary differential equations that can be solved with 
standard methods~\cite{Dolan:2018dqv}. This method was extended to include Kerr-Newman and Kerr-Sen backgrounds~\cite{Cayuso:2019ieu}.
 \end{itemize}
%%% 

The latter (and most recent) approach is the optimal one to compute the instability spectrum accurately.

%%%%%%%%%%%%%%%%%%%%%%%%%%%%%%
\paragraph{Analytical results}
%%%%%%%%%%%%%%%%%%%%%%%%%%%%%%
Even before proving separability of the Proca equation on a Kerr background, analytical results were obtained in the 
small-coupling (or ``Newtonian'') regime, $\mu_V M\ll1$ using matched-asymptotics 
techniques~\cite{Endlich:2016jgc,Baryakhtar:2017ngi,Baumann:2019eav}. To the leading order, the spectrum of unstable 
modes is
%%%
\begin{eqnarray}
 \omega_R^2 &\sim& \mu_V^2\left[1-\left(\frac{M\mu_V}{l+n+S+1}\right)^2\right]\,,\label{Proca_wR}\\
 M\omega_I  &\sim& 2\gamma_{Sl}r_+\left(m\Omega_{\rm H}-\omega_R \right) (M\mu_V)^{4l+5+2S}\,,\label{Proca_wI}
\end{eqnarray}
%%%
where $n\geq0$ is the overtone number, $l$ is the total angular momentum number, $m$ is the azimuthal number, and $S$ 
is the polarization; $S=0$ for axial modes and $S=\pm1$ for the two 
classes of polar modes
\footnote{\label{foot:notation}
Note that different harmonic decompositions have been used in the literature. In the one 
used in Refs.~\cite{Endlich:2016jgc,Baryakhtar:2017ngi}, the angular harmonics $\bm{Y}^{\ell,jm}$ are described by 
three quantum 
numbers $j$, $\ell$ and $m$. The quantum number $j$ specifies the total angular momentum of the perturbation (that is here denoted with $l$), $\ell$ describes the orbital angular momentum (that in our notation is $l+S$), and $m$ is the azimuthal number. Regularity of these harmonics requires that the integer $m$ must be such that $|m| \leq j$ and that $\ell$ is a nonnegative integer that is constrained to take one of three values, namely $\ell \in\{j - 1, j, j + 1\}$. Each one of these three $\ell$'s represents one of the $3$ families of perturbations that we call here $S=0$ and $S=\pm1$. The above decomposition is valid in the static case. For a spinning background, this separation ansatz and associated harmonic decomposition are no longer valid. However, we still have three sectors of perturbations that are continuously
connected to the three ``polarizations'' families in the static limit. Accordingly, if we wish so, we can designate
these three families by the $\ell = j - 1$, $\ell = j$, and $\ell = j + 1$ families respectively for $S=-1$, $S=0$ and $S=1$.}, and the 
coefficient $\gamma_{Sl}$ depends on $S$ and 
$l$~\cite{Baryakhtar:2017ngi,Baumann:2019eav}. The most unstable mode is $S=-1$, $l=1$ and yields 
$\gamma_{-11}=4$~\cite{Baryakhtar:2017ngi}. Interestingly, Eq.~\eqref{Proca_wR} predicts a degeneracy for modes with 
the same value of $l+n+S$ when $M\mu_V\ll 1$, which is akin to the degeneracy in the spectrum of the hydrogen atom.

Massive vector perturbations of rotating BHs are expected to induce a stronger
superradiant instability than in the scalar case because, as previously discussed, superradiance is stronger for EM 
waves. This is confirmed by Eq.~\eqref{Proca_wI} which shows that for the dominant unstable mode (with $l=m=1$, $n=0$, 
and even parity with $S=-1$) the strongest instability should occur on a time scale
\begin{equation}
\tau\equiv \tau_V\sim\frac{M(M\mu_V)^{-7}}{4
(a/M-2 \mu_V r_+)}\,,\label{tau_vector} 
\end{equation}
to be compared with the scalar case, $\tau_S\sim\frac{48M(M\mu_V)^{-9}}{a/M-2 \mu_V r_+}$, cf. 
Eq.~\eqref{omegaDetweiler}.
Roughly speaking, the shortest instability time scale against vector polar perturbations
is of order $\tau_V\sim 0.0025 ({M}/{M_\odot})\,{\rm s}$, i.e. some orders of magnitude shorter than 
in the scalar case.

%%%%%%%%%%%%%%%%%%%%%%%%%%%%%%
\paragraph{Numerical results}
%%%%%%%%%%%%%%%%%%%%%%%%%%%%%%

\begin{figure}%[htpb!]
\begin{center}
\begin{tabular}{ccc}
\epsfig{file=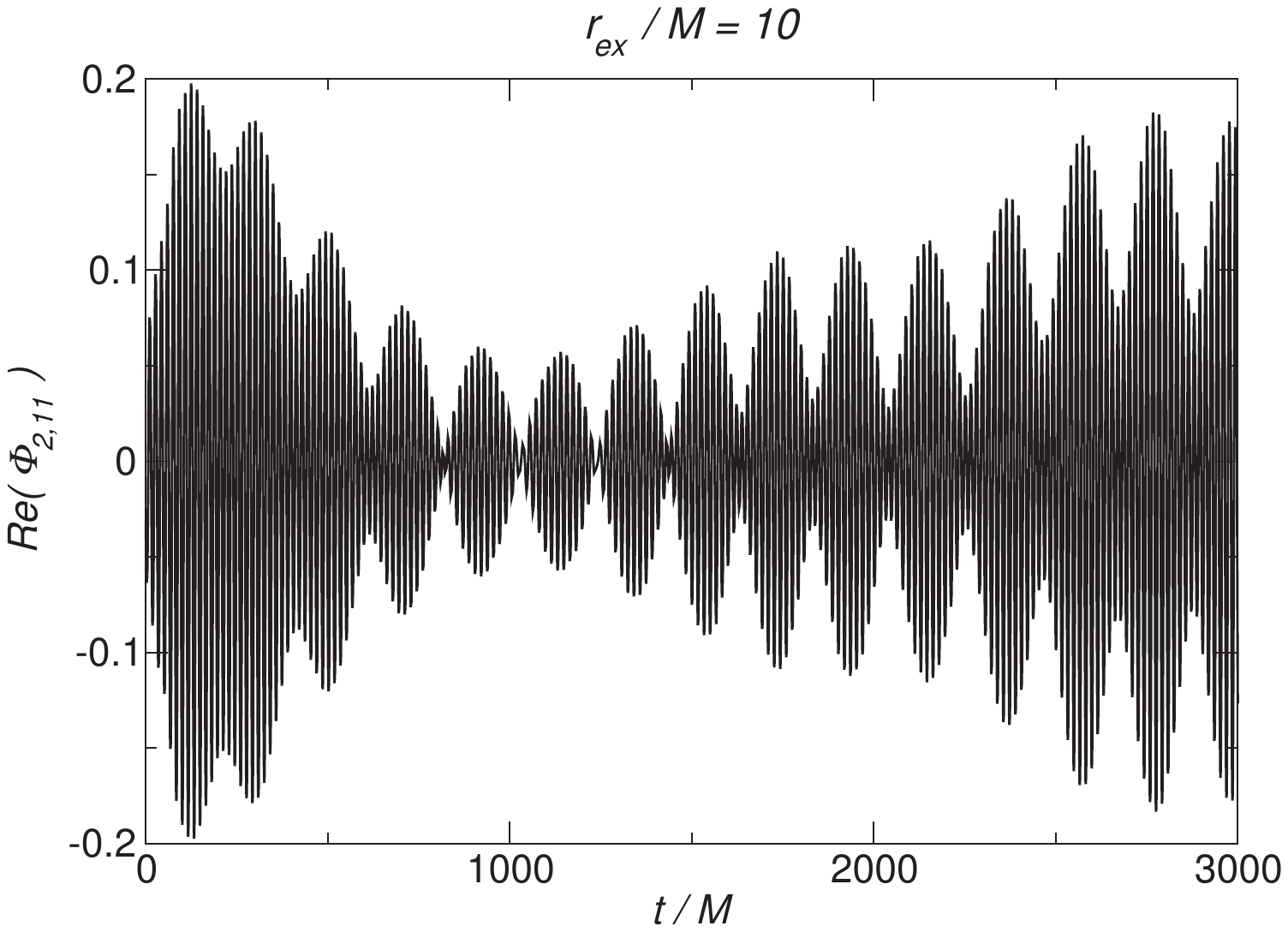,width=0.31\textwidth,angle=0,clip=true} &
\epsfig{file=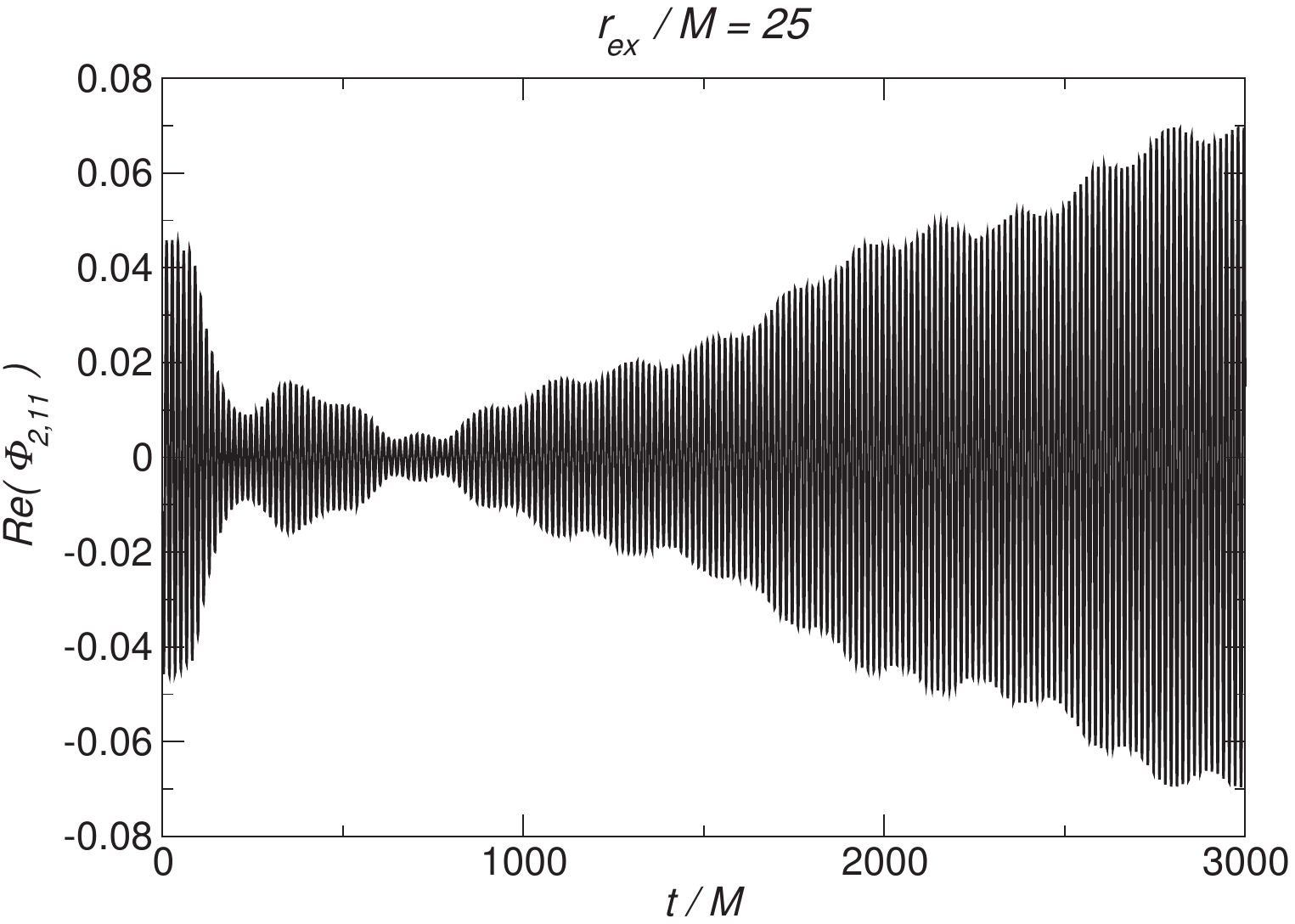,width=0.31\textwidth,angle=0,clip=true} &
\epsfig{file=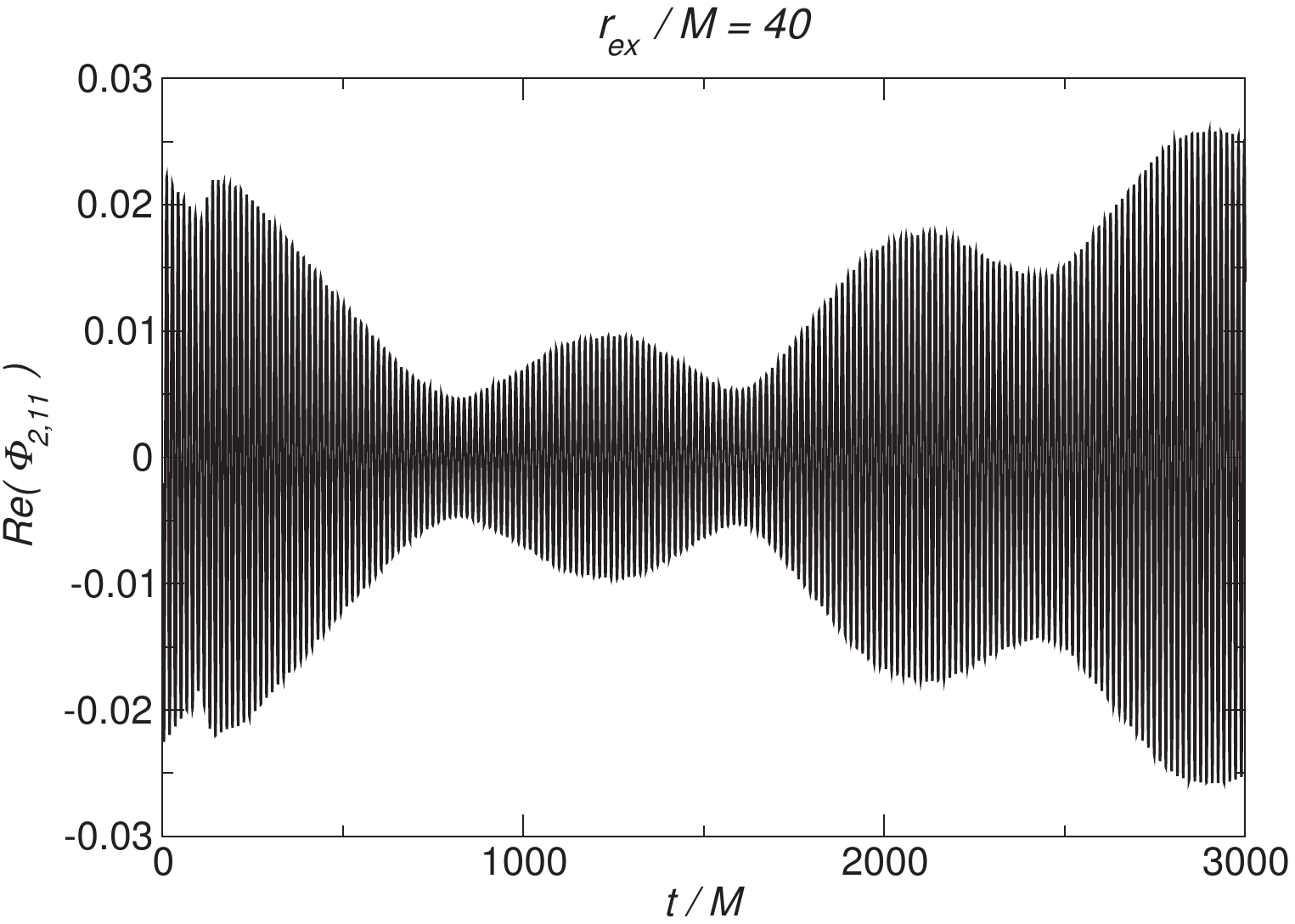,width=0.31\textwidth,angle=0,clip=true} \\
\epsfig{file=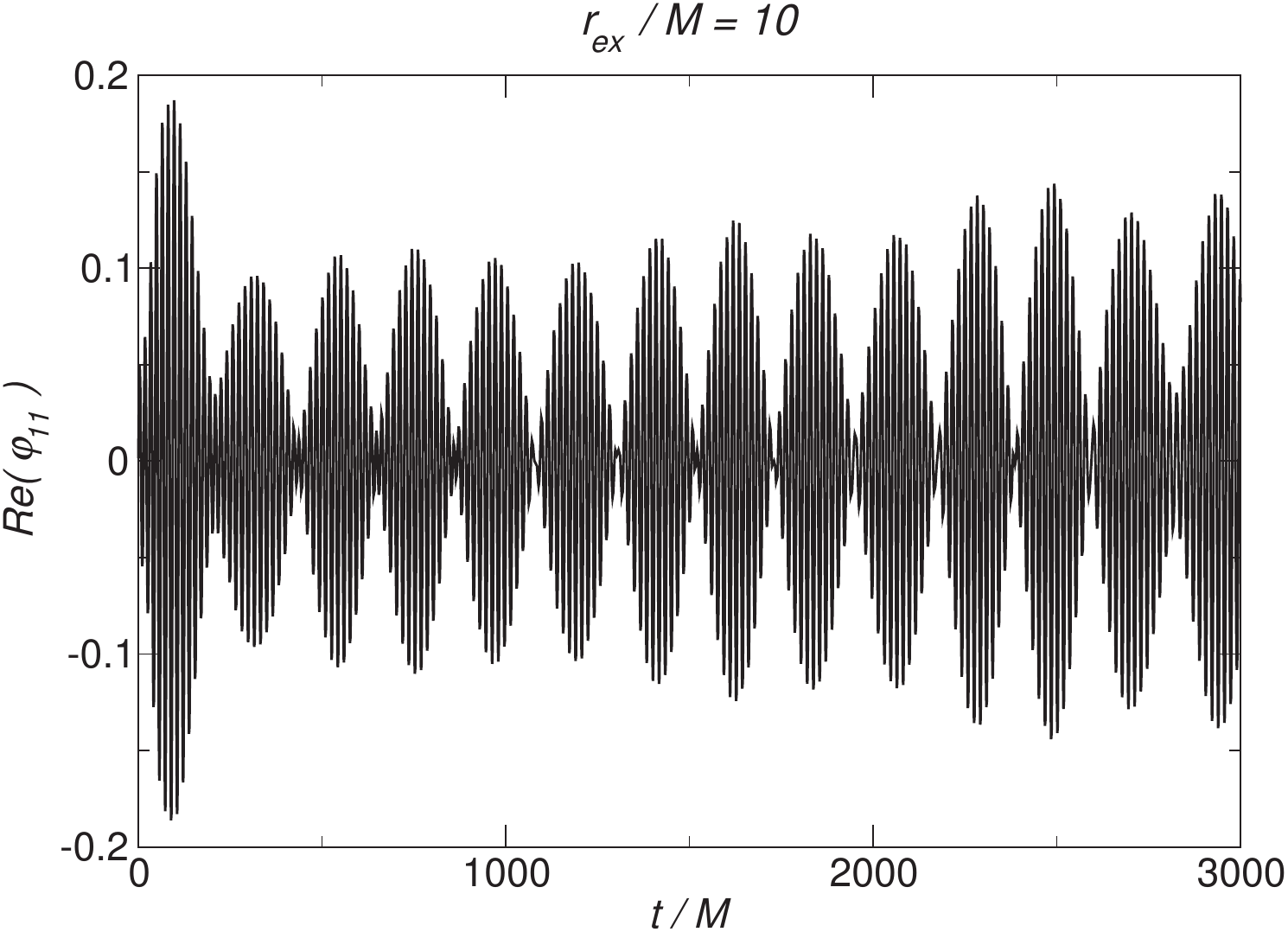,width=0.31\textwidth,angle=0,clip=true} &
\epsfig{file=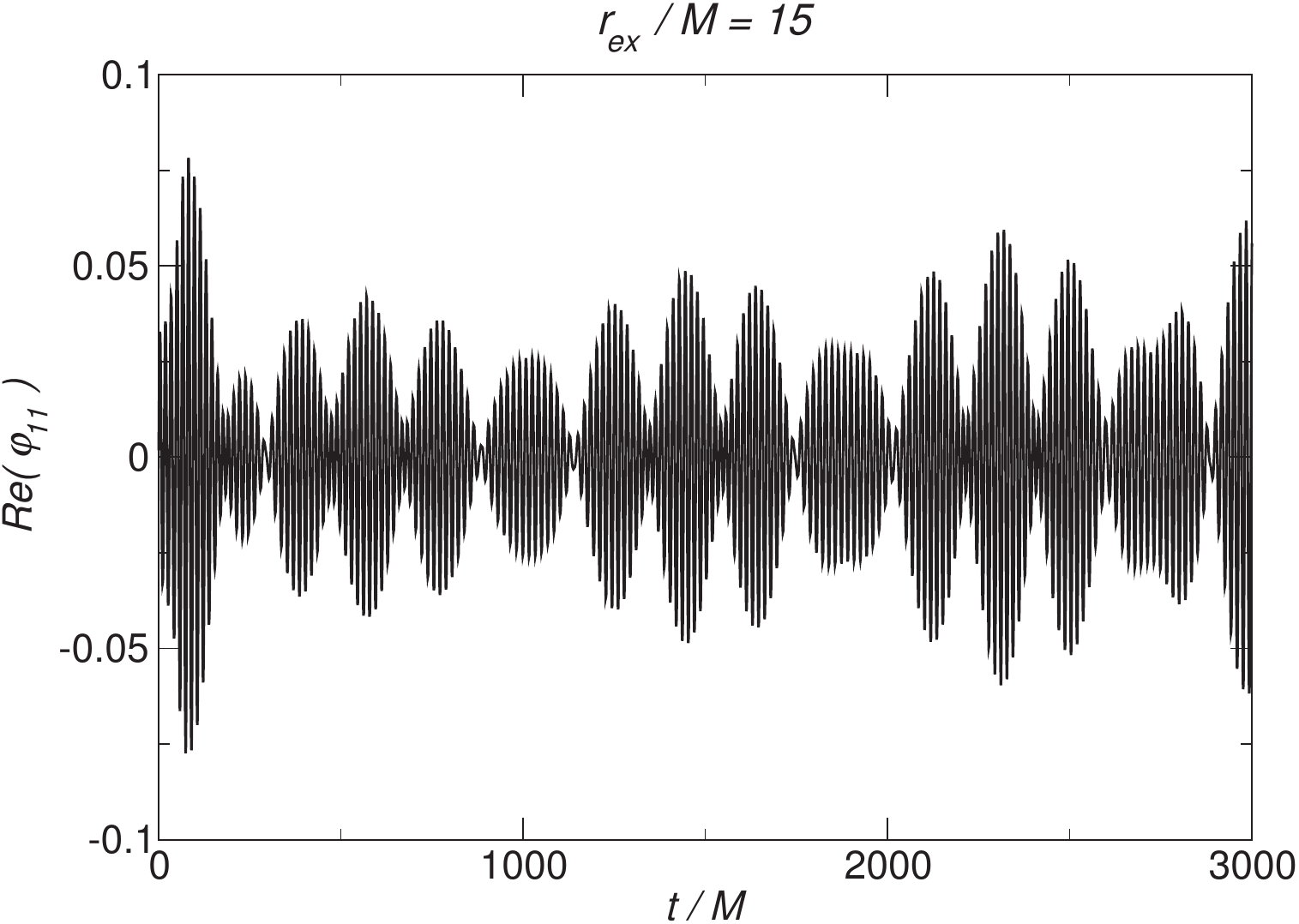,width=0.31\textwidth,angle=0,clip=true} &
\epsfig{file=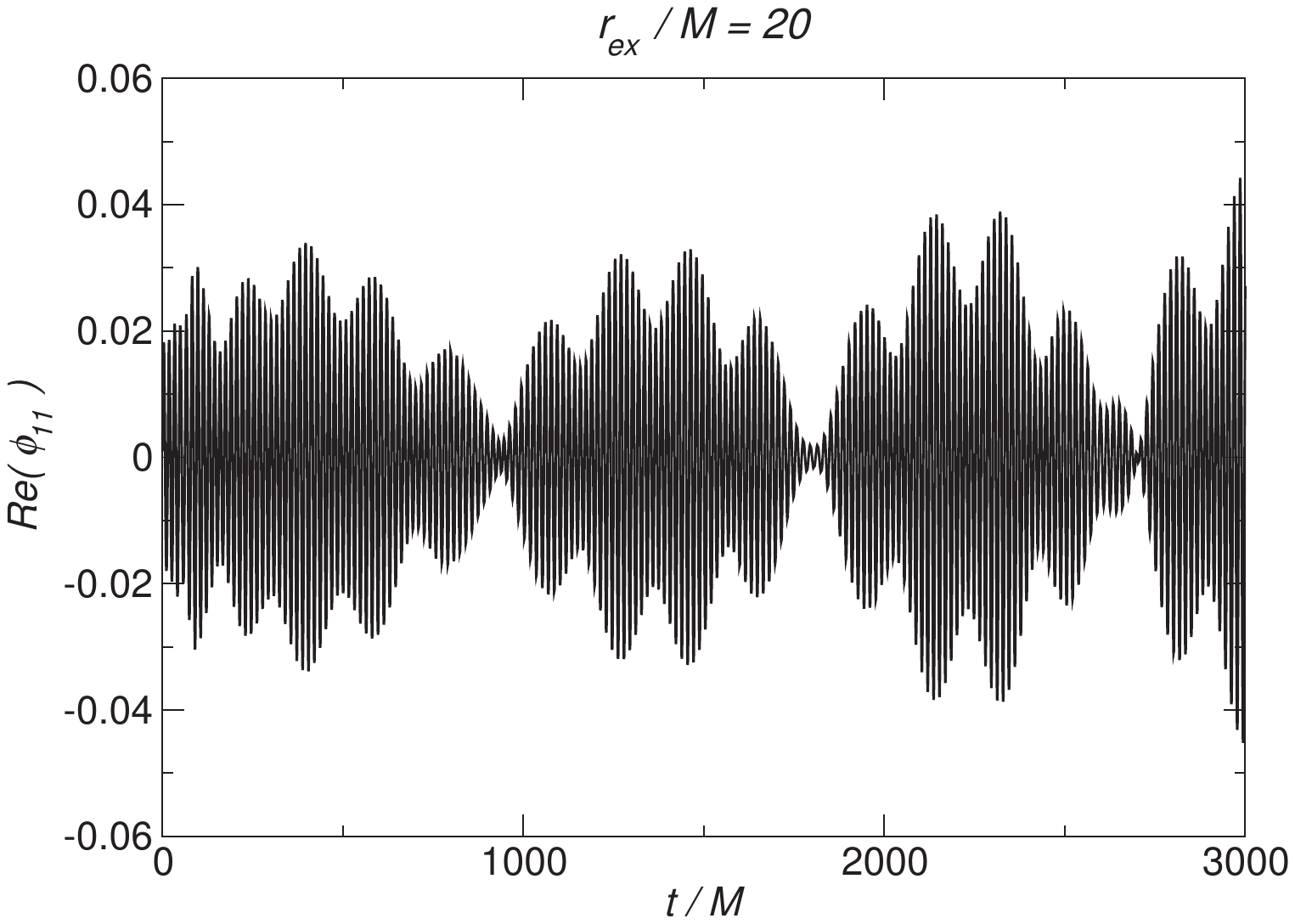,width=0.31\textwidth,angle=0,clip=true} 

\end{tabular}
\end{center}
\caption{\label{fig:ProcaNum}
Time evolution of the Proca field with gravitational coupling $M\mu_V=0.40$
in Kerr background with $a/M=0.99$, at different extraction radii.
The $l=m=1$ mode of the Newman-Penrose scalar $\Phi_2$ (upper panels)
and of the scalar component $\varphi$ (lower panels) are shown. From Ref.~\cite{Witek:2012tr}.
}
\end{figure}

As previously mentioned, early numerical results on the Proca instability of Kerr BHs include a slow-rotation 
approximation up to second order in the spin~\cite{Pani:2012vp,Pani:2012bp} and a fully-numerical analysis of the Proca 
equation on a fixed Kerr geometry~\cite{Witek:2012tr}. In the latter case it was also shown that generic initial date 
excite several overtones (i.e., modes with different 
principal quantum number $n$). Because these modes all have similar frequencies $\omega_R$ (see Eq.~\eqref{Proca_wR}) 
and very long time scales, the overall waveform shows beating patterns~\cite{Witek:2012tr}. An example of this effect is shown in Fig.~\ref{fig:ProcaNum}.

More recently, numerical results have been extended both in the 
time domain~\cite{East:2018glu} and in the frequency domain~\cite{Cardoso:2018tly,Baumann:2019eav}. A breakthrough occurred in 2018 
when the Proca equation was shown to be separable~\cite{Frolov:2018ezx}, reducing the problem to one of the same 
complexity of the scalar case.

\begin{figure}
\begin{center}
 \includegraphics[width=10cm]{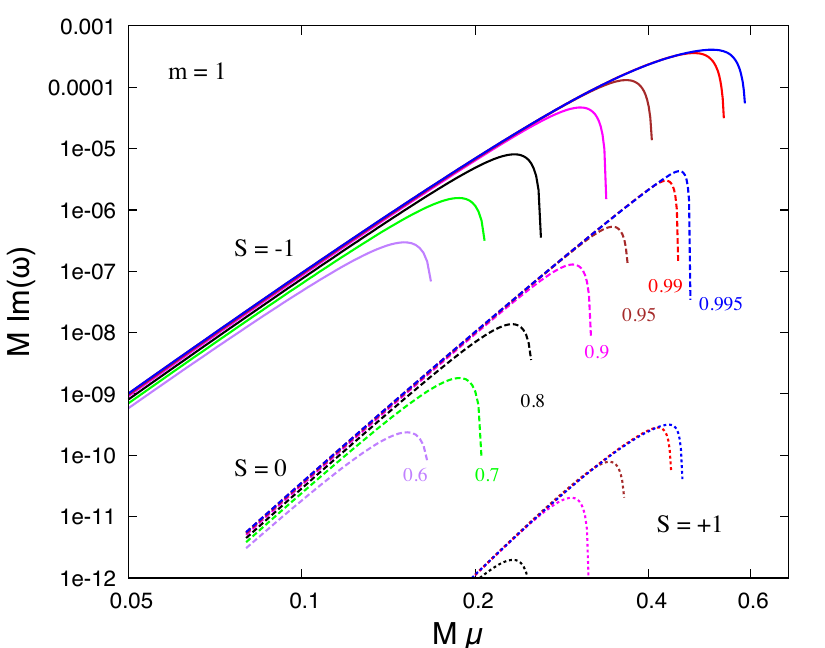}
\end{center}
\caption{
Growth rate of the fundamental $m = 1$ modes of the Proca field, for the three polarizations 
$S = -1$ (solid), $S = 0$ (dashed) and $S = +1$ (dotted), and for various BH spins. The vertical axis shows 
$M\omega_I$ on a logarithmic scale, whereas the horizontal axis shows $M \mu_V$. Taken from \cite{Dolan:2018dqv}.
}
\label{fig:Proca}
\end{figure}

Figure~\ref{fig:Proca} shows results obtained by solving the separable equations~\cite{Dolan:2018dqv}. The results 
agree perfectly~\cite{Frolov:2018ezx} with those obtained without separability~\cite{Cardoso:2018tly}.
A useful fit for the most unstable ($-S=m=1$ and $n=0$) mode as a 
function of the dimensionless spin $\chi=a/M$ and coupling $\mu_V M$ is given by~\cite{Cardoso:2018tly}	
\begin{eqnarray}
M\omega_R &\simeq& 
M\mu_V\left(1+\alpha_1M\mu_V+\alpha_2(M\mu_V)^2+\alpha_3(M\mu_V)^3\right)\,, 
\label{fitProcaR}\\
M\omega_I &\simeq& \beta_0\, (M\mu_V)^7\left(1+\beta_1M\mu_V+\beta_2(M\mu_V)^2\right) \left(\chi- 2 
\omega_R r_+\right) \,, \label{fitProcaI}
\end{eqnarray}
where 
%%%%
\begin{eqnarray}
 \alpha_i &=& \sum_{p=0}^4 A_p^{(i)}(1-\chi^2)^{p/2}\,,\qquad \beta_i =\sum_{p=1}^4 
B_p^{(i)}(1-\chi^2)^{p/2}+\sum_{p=0}^4 C_p^{(i)} \chi^{p}\,, \label{alphabetafit}
\end{eqnarray}
%%%%
and the coefficients $A_p^{(i)}$, $B_p^{(i)}$ and $C_p^{(i)}$ are given in Table~\ref{tab:fitProca}. 
In a large region of the parameter space the precision of the fit is at least $0.2\%$ ($50$\%) for the real (imaginary) 
part of the frequency.

\begin{table}
\centering
\caption{Fitting parameters appearing in Eq.~\eqref{alphabetafit}. See~\cite{Cardoso:2018tly} for 
details of the fit and for a comparison with numerical results.}\label{tab:fitProca}
 \begin{tabular}{cccccc}
  \hline
  \hline
  & $p=0$ & $p=1$ & $p=2$ & $p=3$& $p=4$\\
 \hline
  $A_p^{(1)}$ & $+0.142$ 	& $-1.170$	& $+3.024$	& $-3.234$	& $+1.244$ \\
  $A_p^{(2)}$ & $-1.298$ 	& $+6.601$	& $-15.21$	& $+14.47$	& $-5.070$ \\
  $A_p^{(3)}$ & $+0.726$ 	& $-8.516$	& $+15.43$	& $-11.15$	& $+3.277$   \\
  $B_p^{(1)}$ & -- 		& $-27.76$	& $+114.9$	& $-311.1$	& $+177.2$   \\
  $B_p^{(2)}$ & -- 		& $-14.05$ 	& $+20.78$	& $-36.84$	& $+58.37$  \\
  $B_p^{(3)}$ & -- 		& $+14.78$ 	& $-4.574$	& $-248.5$	& $+108.1$ \\
  $C_p^{(1)}$ & $+48.86$	& $-8.430$	& $+45.66$	& $-132.8$	& $+52.48$     \\
  $C_p^{(2)}$ & $-31.20$	& $+32.52$	& $-73.50$	& $+161.0$	& $-91.27$  \\
  $C_p^{(3)}$ & $+189.1$	& $-85.32$	& $+388.6$	& $-1057$	& $+566.1$       \\
  \hline
  \hline
  \end{tabular}
\end{table}

%%%%%%%%%%%%%%%%%%%%%%%%%%%%%%%%%%%%%%%%%%%%%%%%%%%%%%%%%%%%%%%%%%%%%%
\subsubsection{Massive tensor fields\label{sec:spin2SR}} 
%%%%%%%%%%%%%%%%%%%%%%%%%%%%%%%%%%%%%%%%%%%%%%%%%%%%%%%%%%%%%%%%%%%%%%
Massive tensor fields cannot be trivially coupled to gravity. The development of a consistent theory of massive spin-2 
fields has an interesting history on its own and we refer the reader to the recent 
reviews~\cite{Hinterbichler:2011tt,deRham:2014zqa}.

At the linear level it is known, since the works of Fierz and Pauli, that there is a unique ghost- and tachyon-free mass term that preserves Lorentz invariance and describes the five polarizations of a massive spin-2 field on a flat background~\cite{Fierz:1939ix}.
On a curved spacetime the most general linearized field equations describing a massive spin-2 field read 
\be
\label{eqmotioncurved}
\left\{
\begin{array}{l}
\bar\Box h_{\mu\nu}+2 \bar R_{\alpha\mu\beta\nu} h^{\alpha\beta}-\mu_{T}^2 h_{\mu\nu}=0\,,\\
\mu_T^2\bar\nabla^{\mu}h_{\mu\nu}=0\,,\\
\left(\mu_{T}^2-{2\Lambda}/{3}\right)h=0\,.
\end{array}\right.
\ee
At the linear level these equations are only consistent if we assume the background to be a vacuum solution of Einstein's equations with a cosmological constant $\Lambda$, 
so that $\bar{R}=4\Lambda$, $\bar{R}_{\mu\nu}=\Lambda \bar{g}_{\mu\nu}$ and barred quantities refer to the background. Interestingly, the same equations can also describe the propagation of a massive graviton in nonlinear massive gravity and in bimetric gravity in some special configurations~\cite{Brito:2013wya}.
These equations were shown to lead to a superradiant instability of Kerr BHs in these theories~\cite{Brito:2013wya,Brito:2013yxa,Brito:2020lup}\footnote{
In fact, even the Schwarzschild spacetime is unstable against a spherically symmetric mode in these theories.
The instability of the Schwarzschild metric against massive spin-2 perturbations was first discovered in Ref.~\cite{Babichev:2013una}, where it was shown that the mass term for a massive spin-2 field can be interpreted as a Kaluza-Klein
momentum of a four-dimensional Schwarzschild BH extended into a flat higher dimensional spacetime.
Such ``black string'' spacetimes are known to be unstable against long-wavelength perturbations,
a mechanism known as the Gregory-Laflamme instability \cite{Gregory:1993vy,Kudoh:2006bp}, which in turn is the analog of a Rayleigh-Plateau instability of fluids~\cite{Cardoso:2006ks,Camps:2010br}. Based on these results, Ref.~\cite{Babichev:2013una} pointed out that massive tensor perturbations on a Schwarzschild BH in massive gravity and bimetric theories would generically give rise to a (spherically symmetric) instability. The unstable mode is absent in partially massless gravity~\cite{Brito:2013yxa} and in solutions of bimetric theories other than the bi-Schwarzschild solution~\cite{Babichev:2014oua}. The former case corresponds to the Higuchi bound $\mu_T^2=2\Lambda/3$, so that the scalar equation in~\eqref{eqmotioncurved} becomes an identity and the scalar modes does not propagate.}.

Around a Kerr BH there exist long-lived bound states which follow the same kind of hydrogenic-like 
scaling~\eqref{Proca_wR} and \eqref{Proca_wI} observed for massive bosons with lower spin~\cite{Brito:2013wya,Brito:2020lup}. In addition to these modes, a 
new polar dipole mode was found~\cite{Brito:2013wya}. This mode was shown to be isolated, does not follow the same 
small-mass behavior, and does not have any overtone. For this mode, the real part is  smaller than the mass of the 
spin-2 field, and in the region $M\mu_T\lesssim 0.4$ is very well fitted by
\be
\label{polar_di_Re}
\omega_R/\mu_T\approx 0.72(1-M\mu_T)\,.
\ee
In the limit $M\mu_T\ll 1$, and for the static case, the imaginary part was found to scale as~\cite{Brito:2013wya}
\be
\omega_I/\mu_T\approx -(M\mu_T)^{3}\,.
\ee
This mode has the largest binding energy ($\omega_R/\mu_T-1$) among all couplings $M\mu_T$, much higher than the ground states of the 
scalar, Dirac, and vector fields. 

\begin{figure*}[htb]
\begin{center}
% \begin{tabular}{c}
\epsfig{file=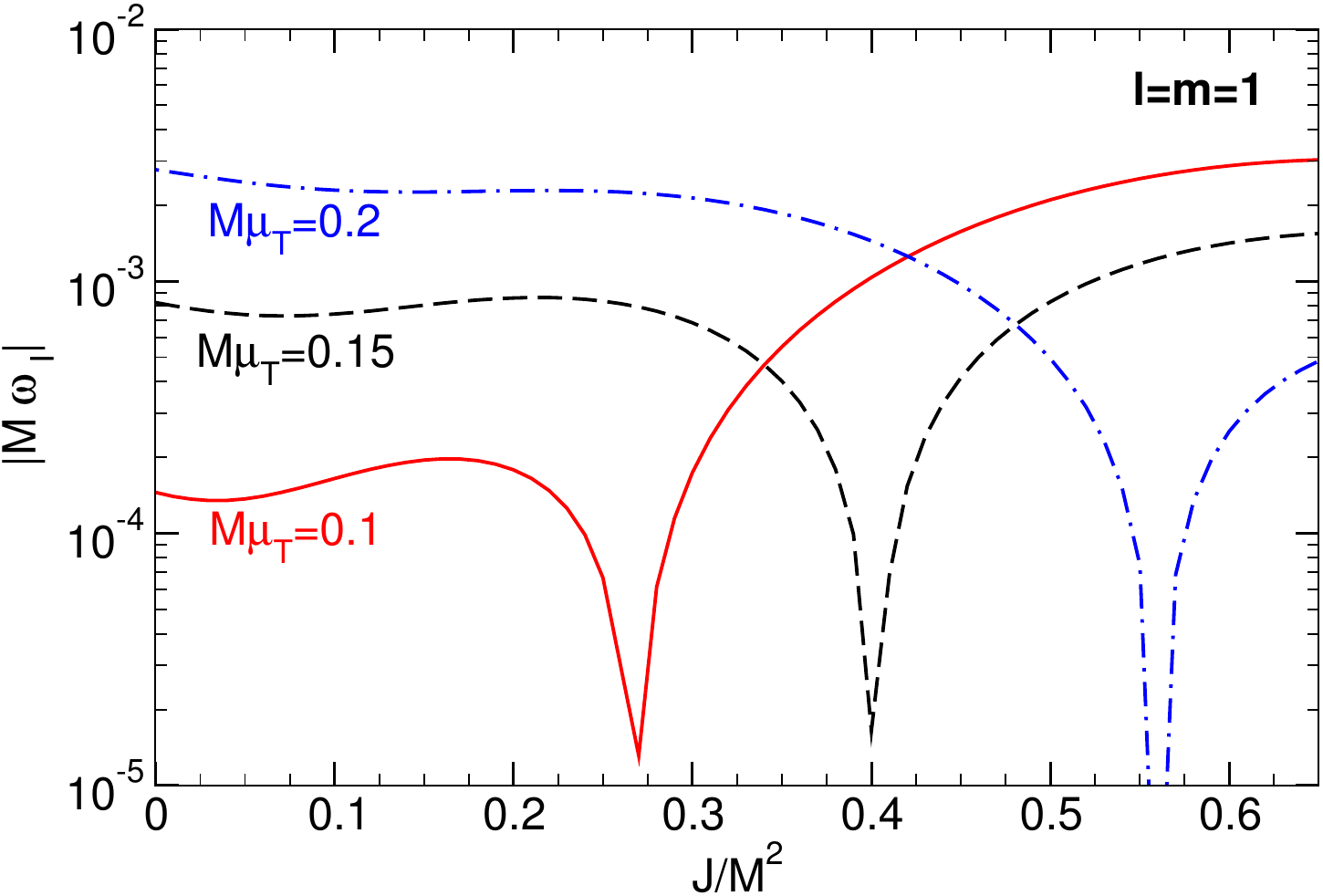,width=0.48\textwidth,angle=0,clip=true}
\epsfig{file=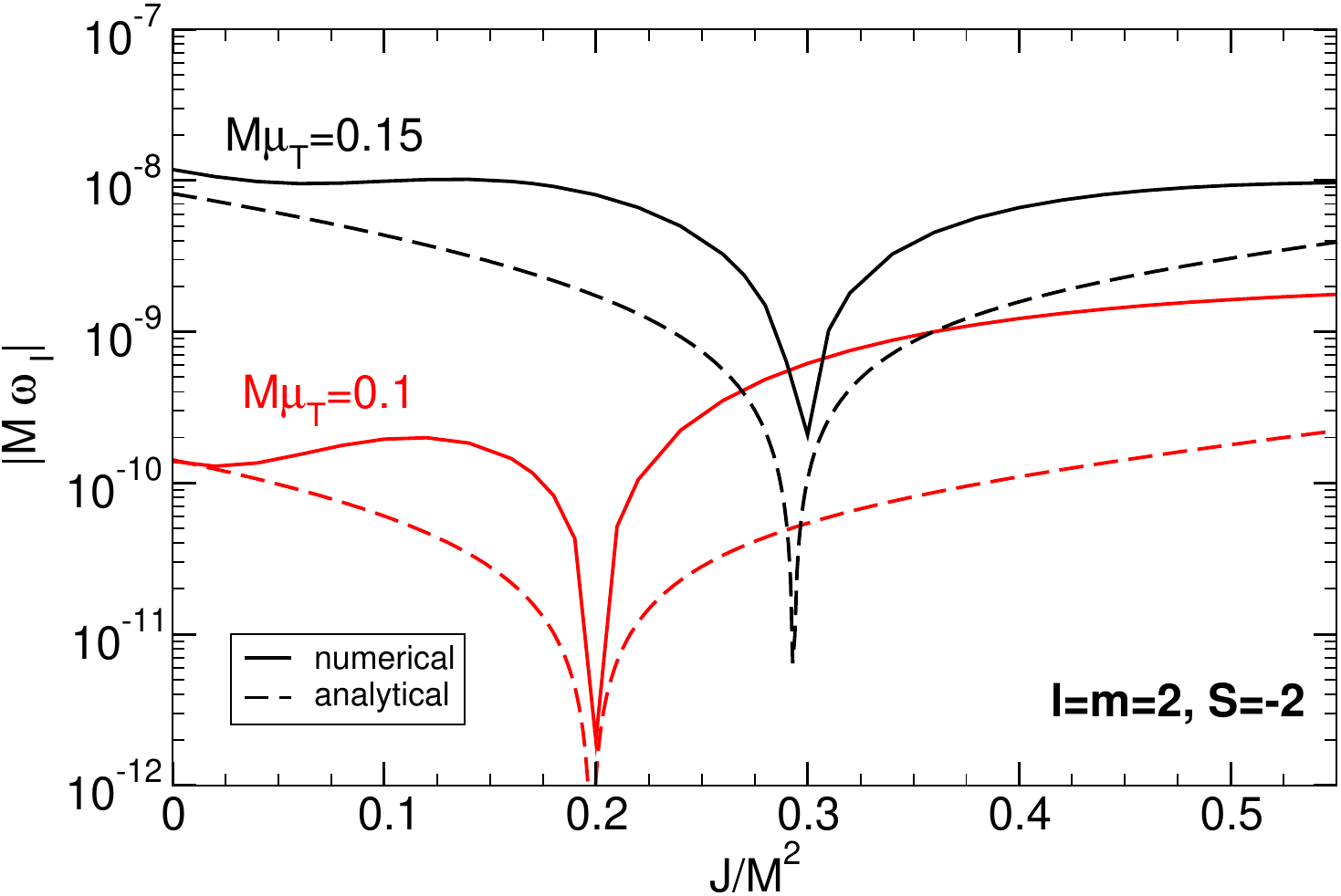,width=0.48\textwidth,angle=0,clip=true}
% \end{tabular}
\caption{Absolute value of the imaginary part of the polar quasibound modes as a function of the BH rotation rate $a/M$ for different values of $l$ and $m$ and different values of the mass coupling $\mu_T M$. 
Left panel: polar dipole mode for $l=m=1$, computed numerically using an expansion to first-order in the dimensionless spin $a/M$, valid for any value of the mass coupling $\mu_T M$~\cite{Brito:2013wya}. Right panel: polar mode $l=m=2$, $S=-2$, comparing the numerical results (solid lines) of Ref.~\cite{Brito:2013wya}, which are valid for $a/M\ll 1$ and any value of $\mu_T M$, with the analytical estimates (dashed lines) of Ref.~\cite{Brito:2020lup}, which are valid for $\mu_T M \ll 1$ and any value of the spin.
For any mode with $m\geq 0$, the imaginary part crosses the axis and become unstable when the superradiance condition $\omega_R<m\Omega_{\rm H}$ is met.
\label{fig:spin2}}
\end{center}
\end{figure*}

Since no separable ansatz of the system of equations~\eqref{eqmotioncurved} in a Kerr background has been found so far, 
to investigate the superradiant instability, Ref.~\cite{Brito:2013wya} adopted a slow-rotation approximation to first order in the spin, whereas Ref.~\cite{Brito:2020lup} showed that in the $M\mu_T\ll 1$ limit the unstable hydrogenic modes can be computed analytically for any spin. A representative example of the unstable modes is presented in Fig.~\ref{fig:spin2}, where it is shown that the growth timescale of the dipole polar mode is very small even for small couplings $M\mu_T$.
Indeed, the time scale for this unstable mode is~\cite{Brito:2013wya}
\begin{equation}
 \tau_{T}\sim \frac{M(M\mu_T)^{-3}}{\gamma_{{\rm polar}}(a/M-2r_+\omega_R)}\,,
\end{equation}
where $\gamma_{{\rm polar}}\sim{\cal O}(1)$. This time scale is four orders of magnitude shorter than the corresponding Proca field instability and, in fact, it is the shortest instability time scale of a four dimensional, asymptotically-flat GR BH known to date.

%%%%%%%%%%%%%%%%%%%%%%%%%%%%%%%%%%%%%%%%%%%%%%%%%%%%%%%%%%%%%%%%%%%%%%%%%%%%%%%%%%%%%%%%%%%%%%%%%%%%%%%%%%%%%%%%%%%%%
\subsubsection{A unified picture of the linearized superradiant instability of massive bosonic fields} 
\label{sec:massive_unified}
%%%%%%%%%%%%%%%%%%%%%%%%%%%%%%%%%%%%%%%%%%%%%%%%%%%%%%%%%%%%%%%%%%%%%%%%%%%%%%%%%%%%%%%%%%%%%%%%%%%%%%%%%%%%%%%%%%%%%
The results presented in the previous sections for spin-0, spin-1 and spin-2 fields suggest the following unified picture describing the superradiant instability of massive bosonic fields, of mass $\mu$, around a spinning BH. For any bosonic field propagating on a spinning BH, there exists a set of quasibound states whose frequency satisfies the superradiance condition $\omega_R<m\Omega_{\rm H}$. These modes are localized at a distance from the BH which is governed by the Compton wavelength $1/\mu$ and decay exponentially at large distances. In the small gravitational coupling limit, $M\mu\ll1$ (where $\mu$ denotes the mass of the field), the spectrum of these modes resembles that of the hydrogen atom: 
\be
\omega_R/\mu \sim 1-\frac{(M\mu)^2}{2(l+S+1+n)^2}\,, \label{hydrogenic}
\ee
where $l$ is the total angular momentum of the state with spin projections $S=-s,-s+1,\ldots,s-1,s$, $s$ being the spin of the field.

Again in the small gravitational coupling limit, to leading order in $\alpha$, a detailed matched-asymptotic 
calculation 
yields~\cite{Detweiler:1980uk,Baryakhtar:2017ngi,Baumann:2019eav,Brito:2020lup}
\begin{equation}
%  \omega_I\propto (m\Omega_{\rm H}-\omega_R)(M\mu)^{\eta} \qquad \eta=4l+2S+5\,. \label{wIslope}
\omega_I=-\frac{1}{2}C^{(s)}_{lS}\frac{{\cal P}_{lm}(\chi)}{{\cal P}_{lm}(0)}\alpha^{4l+2S+5}
(\omega_R-m\Omega_{\rm H})\,, \label{wIslope}
\end{equation}
where 
\begin{equation}
 {\cal P}_{lm}(\chi)=(1+\Delta)\Delta^{2l}\prod_{q=1}^l\left(1+4M^2\left(\frac{\omega_R-m\Omega_{\rm 
H}}{q\kappa}\right)^2\right)
\end{equation}
is proportional to the BH absorption probability~\cite{Starobinski:1973,Starobinski2:1973} (see 
Sec.~\ref{sec:super_anavsnum}), $\Delta=\sqrt{1-\chi^2}$, and $\kappa=\frac{\Delta}{1+\Delta}$.
The constants $C^{(s)}_{lS}$ are given in Table~\ref{tab:inst}.

In the nonspinning case ($\Omega_{\rm H}=0$), the decay rate of these modes is $\omega_I/\mu\propto -(M\mu)^{4l+2S+5}$.
For spinning geometries, $\omega_I$ changes sign in the superradiant regime. Indeed, 
when $\omega_R<m\Omega_{\rm H}$ the imaginary part becomes positive and $\omega_I$ corresponds to the growth rate of 
the field ($\tau\equiv\omega_I^{-1}$ being the instability time scale). According to Eq.~\eqref{wIslope}, the shortest 
instability time scale occurs for $l=1$, $S=-1$ and for $l=2$, $S=-2$. The only exception to the 
scaling~\eqref{hydrogenic} and 
\eqref{wIslope} is given by the (non-hydrogenic) dipole polar mode of a spin-2 field, whose frequency is given by 
Eq.~\eqref{polar_di_Re} 
and the scaling of the imaginary part is similar to Eq.~\eqref{wIslope}, but with $\omega_I\propto (m\Omega_{\rm 
H}-\omega_R)(M\mu)^{3}$.

Analytical results in the small-coupling limit are in good agreement with the exact numerical results in the scalar and vector case, where the linearized problem has been fully 
solved~\cite{Detweiler:1980uk,Dolan:2007mj,Baryakhtar:2017ngi,Baumann:2019eav,Dolan:2018dqv}. The case of spin-$2$ 
perturbations is 
less explored but --~when a comparison between analytical and numerical results is available~-- the agreement turns out 
to be very good~\cite{Brito:2013wya,Brito:2020lup}.

\begin{table}\label{tab:inst}
\begin{center}
 \caption{Coefficients appearing in Eq.~\eqref{wIslope} for the imaginary part of the most relevant massive spin-$s$ 
bosonic unstable modes of a Kerr BH. In our notation $l$ is the total angular momentum of the state with spin 
projections $S=-s,-s+1,\ldots,s-1,s$, $s$ being the spin of the field. (In the notation of 
Refs.~\cite{Endlich:2016jgc,Baryakhtar:2017ngi}, $j=l$ and $\ell=l+S$, see footnote~\ref{foot:notation}.)}
\vspace{0.2cm}
 \begin{tabular}{c|cc|c}
  \hline
  \hline
  $s$ 	& $l$ 	& 	 $S$ 	& 	$C^{(s)}_{lS}$	\\
  \hline
  \multirow{3}{*}{$0$} 	& $1$	&	 $0$	&	$1/6$	\\
   	& $2$	&	 $0$	&	$128/885735$		\\
   	& $3$	&	 $0$	&	$1/32256000$	\\
  \hline
  \multirow{3}{*}{$1$} 	& $1$	&	 $-1$	&	$16$	\\
   	& $2$	&	 $-1$	&	$1/54$		\\
   	& $3$	&	 $-1$	&	$128/22143375$	\\
  \hline
  \multirow{3}{*}{$2$} 	& $2$	&	 $-2$	&	$128/45$	\\
   	& $1$	&	 $0$	&	$10/9$		\\
   	& $3$	&	 $-2$	&	$4/4725$	\\
   	& $1$	&	 $1$	&	$640/19683$	\\
  \hline
  \hline
 \end{tabular}
\end{center}
\end{table}

%%%%%%%%%%%%%%%%%%%%%%%%%%%%%%%%%%%%%%%%%%%%%%%%%%%%%%%%%%%%%%%%%%%%%%%%%%%%%%%%%%%%%%%%%%%%%%%
\subsubsection{Superradiant instabilities of massive bosonic fields: nonlinear evolutions} 
\label{sec:nonlinear}
%%%%%%%%%%%%%%%%%%%%%%%%%%%%%%%%%%%%%%%%%%%%%%%%%%%%%%%%%%%%%%%%%%%%%%%%%%%%%%%%%%%%%%%%%%%%%%%
Although the instability of massive bosonic fields around rotating BHs is very well understood at the linear level, the 
nonlinear development of the instability has received much less attention and successful simulations of its first stages 
have only recently been performed~\cite{East:2017ovw,East:2018glu}. These simulations are in fact extremely challenging, 
mainly due to the large separation of scales between the instability time scale $1/\omega_I$ and the oscillation period 
of the field $1/\omega_R$. In addition, the problem must, in general, be solved using full (3+1)-simulations, 
further increasing the required computational 
cost~\cite{Okawa:2014nda,Zilhao:2015tya,East:2017ovw,East:2018glu}.

As we just saw, the instability time scale can be significantly shorter for vector fields when compared to scalar fields. With this is mind, Ref.~\cite{East:2017ovw} reported the first successful numerical study of the nonlinear evolution of the instability for a Proca field. They simulated the time evolution of a BH with initial mass $M_0$ and initial spin $a=0.99 M_0$  surrounded by a complex Proca field with initial energy $\sim 10^{-3} M_0$ and an azimuthal dependence of the form $e^{i \varphi}$ for which the stress-energy tensor is axisymmetric, allowing them to use a $(2+1)$ numerical domain.

\begin{figure}[hbt]
\begin{center}
\begin{tabular}{cc}
\epsfig{file=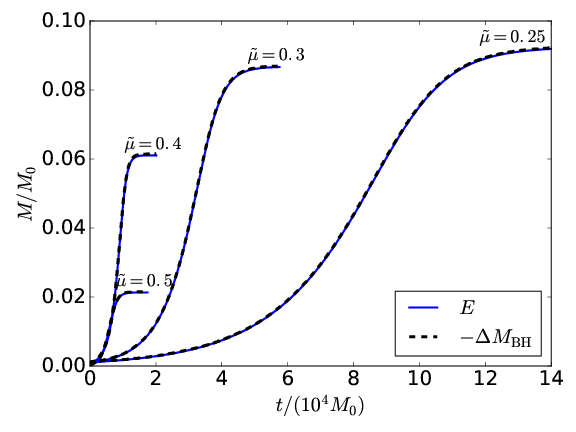,width=0.5\textwidth,angle=0,clip=true}&
\epsfig{file=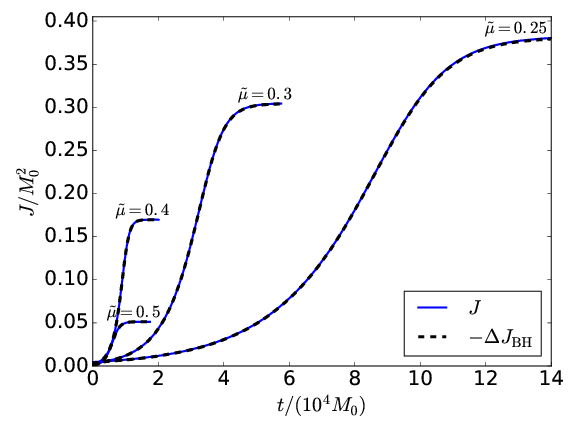,width=0.5\textwidth,angle=0,clip=true}
\end{tabular}
\caption{The evolution of the energy (left) and angular momentum (right) of the Proca field as a function of time (blue solid lines) for a BH with initital mass $M_0$ and initial spin $a=0.99M_0$, for different values of the boson mass $\tilde{\mu}\equiv M_0\mu$. For comparison the mass and angular momentum lost by the BH is also shown (black dashed lines). The curves overlap perfectly showing that the energy lost by the BH is entirely transferred to the Proca field. From~\cite{East:2017ovw}.\label{Fig:cloud_evol}}
\end{center}
\end{figure}

Figure~\ref{Fig:cloud_evol} shows the evolution of the energy and angular momentum of the Proca field as a function of time for different values of the boson mass. Initially, the field energy and angular momentum undergo an exponential growth, as predicted by the linear analysis, with the corresponding changes in the BH mass and spin closely tracking this growth, suggesting that all the mass and spin extracted from the BH is transferred to the Proca field. However, at late times the instability growth rate slowly decreases until the energy and angular momentum saturate to a given value. 

\begin{figure}[hbt]
\begin{center}
\begin{tabular}{c}
\epsfig{file=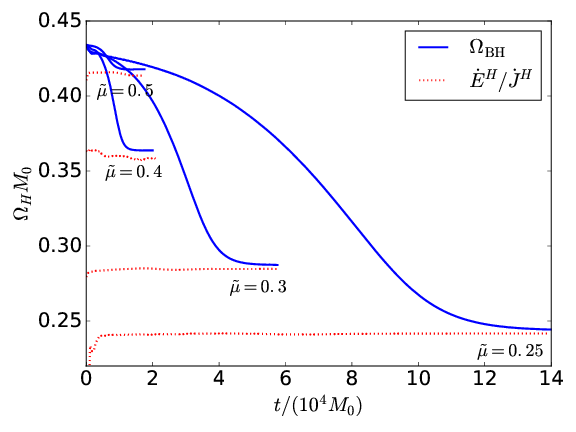,width=0.7\textwidth,angle=0,clip=true}
\end{tabular}
\caption{Evolution of the BH's horizon angular frequency $\Omega_{\rm H}$ along with the ratio between the Proca energy and angular momentum flux at the event horizon. For a field dominated by a single mode with frequency $\omega_R$, one has $\dot{E}^H/\dot{J}^H\sim \omega_R/m$ (c.f. Appendix~\ref{appendix_energyangularmomentum}). For $m=1$ these results show that the instability saturates when $\Omega_{\rm BH}$ approaches the frequency $\omega_R$ of the most unstable mode. From~\cite{East:2017ovw}.\label{Fig:BH_evol}}
\end{center}
\end{figure}

As shown in Fig.~\ref{Fig:BH_evol}, the instability saturates when the ratio between the energy and angular momentum flux of the field at the event horizon $\dot{E}^H/\dot{J}^H$ approaches the horizon's angular frequency $\dot{E}^H/\dot{J}^H\sim \Omega_{\rm H}$. This is consistent with the field being dominated by a single mode with frequency $\omega_R$ and azimuthal quantum number $m=1$ (cf. Appendix~\ref{appendix_energyangularmomentum}). In fact, as expected from Eq.~\eqref{Proca_wI}, when $\dot{E}^H/\dot{J}^H< \Omega_{\rm H}$ the Proca field is unstable and extracts energy and angular momentum from the BH. Due to the BH's spin loss, $\Omega_{\rm H}$ decreases up to the point where $\dot{E}^H/\dot{J}^H\sim \Omega_{\rm H}$, at which point superradiance stops and the instability saturates. 

These simulations show that the superradiant instability leads to the formation of a long-lived boson cloud around the BH. In the process, up to $\sim 10 \%$ of the BH mass can be extracted by the boson field~\cite{East:2017ovw,Herdeiro:2017phl}. The solutions found at the end of these simulation were shown to match very well~\cite{Herdeiro:2017phl} the stationary hairy BH solutions constructed in Ref.~\cite{Herdeiro:2016tmi}, consisting of a rotating BH surrounded by a complex Proca hair oscillating with a frequency $\omega= m\Omega_{\rm H}$ but with a stationary stress-energy tensor (see Section~\ref{sec:hair} for more details).


\begin{figure}[hbt]
\begin{center}
\begin{tabular}{cc}
\epsfig{file=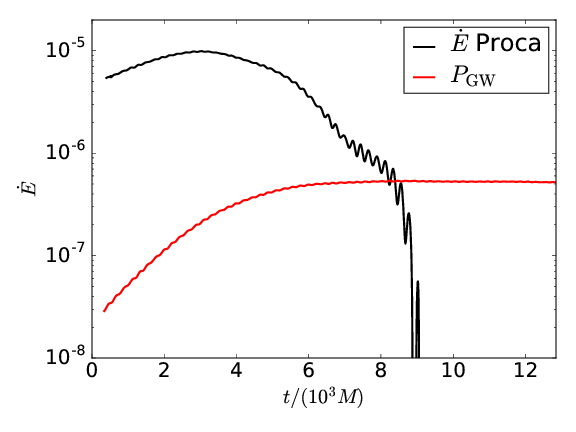,width=0.5\textwidth,angle=0,clip=true}
\epsfig{file=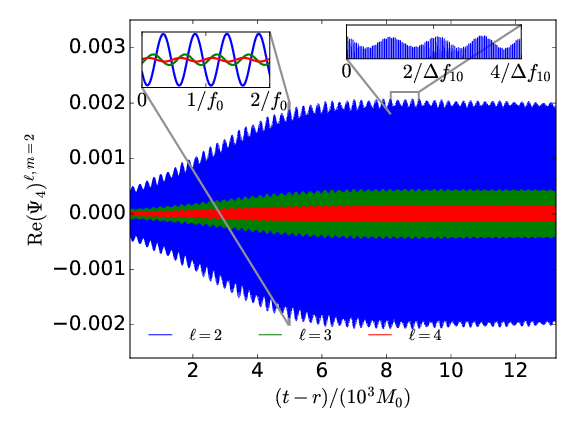,width=0.5\textwidth,angle=0,clip=true}
\end{tabular}
\caption{Left: The time derivative of the energy carried by the Proca field (black line) and the GW power (red lines) emitted to infinity. Right: Leading-order spherical harmonic components of the real part of $\Psi_4$ emitted by a Proca field with mass $\mu=0.4/M_0$ around a BH with initital mass $M_0$ and initial spin $a=0.99M_0$. The $2\leq\ell\leq 4$, $m=2$ components are shown. The left-hand inset shows that the signal is predominantly at twice the frequency of the
most unstable mode, while the right-hand inset shows the variation in the amplitude due to the beating between the first and second most unstable modes. From~\cite{East:2018glu}.\label{Fig:GW_emission}}
\end{center}
\end{figure}

These results were extended to real Proca fields in Ref.~\cite{East:2018glu} with no restrictions on the symmetries of the problem. It was shown that, just like the axisymmetric simulations discussed above, the initial unstable growth and saturation agrees with the linear analysis predictions. However, unlike the axisymmetric case, due to the oscillations of the Proca field and the non-axisymmetric profile of the field, GWs are emitted in the process. As can be seen in  Fig.~\ref{Fig:GW_emission} after the superradiant growth saturates, the field energy and angular momentum steadily decrease on a time scale $t_{\rm GW}\sim 10^5 M_0$. 

During the exponential growth of the cloud, GW emission is suppressed and therefore does not hinder the superradiant growth, as can be seen in the left panel of Fig.~\ref{Fig:GW_emission}. However, after the instability saturates, the cloud steadily dissipates by emitting GWs. As we will discuss in more detail in Section~\ref{sec:astrophysics}, this GW signal can be understood as due to the annihilation of Proca particles in the cloud. The right panel of Fig.~\ref{Fig:GW_emission} shows that the GW signal emitted is predominantly quadrupolar and the dominant frequency is given by twice the frequency of the dominant unstable mode of the Proca field $\omega_{\rm GW} \approx 2\omega_R$. In addition, on top of the GW signal with frequency $\omega_{\rm GW} \approx 2\omega_R$, small modulations in the GW amplitude can also be observed. These modulations have a smaller frequency than the signal coming from annihilation and can be attributed to the interference of the dominant unstable mode ($l=m=1,\, S=-1,\, n=0$) with a smaller amplitude overtone ($n=1$).

These numerical simulations provide a very simple picture for the evolution of the instability (to be 
discussed in detail later): a nonspherical bosonic cloud grows near the BH on a instability 
time scale, extracting energy and angular momentum until superradiance stops and on much larger time scales the cloud is 
slowly dissipated through GW emission. These simulations only explored the $m=1$ mode instability. As we already 
discussed, higher $m$ modes are also unstable whenever $\Omega_{\rm H}>\omega_R/m$. Therefore, modes with $m>1$ are 
expected to further spin down the BH even after the $m=1$ mode instability saturates. However, the much longer 
instability time scales for these modes (cf. Eq.~\eqref{Proca_wI}) makes it challenging to simulate their evolution. 
Further work is needed to fully understand the final state of the instability.

Although these results were obtained by performing fully nonlinear simulations of the Einstein-Proca equations, the results are in excellent agreement with the predictions made through an adiabatic evolution of the instability~\cite{Brito:2014wla} that we discuss in more detail in below. We will also see in Section~\ref{sec:bounds_mass} that the formation of a boson cloud around spinning BHs and subsequent dissipation through GW emission leads to several observational consequences that can be used to either detect ultralight bosons or constrain their existence.

%%%%%%%%%%%%%%%%%%%%%%%%%%%%%%%%%%%%%%%%%%%%%%%%%%%%%%%%%%%%%%%%%%%%%%%%%%%%%%%%%%%%%%%%%%%%%%%
\subsubsection{Anatomy of scalar clouds around spinning black holes}\label{sec:cloudproperties}
%%%%%%%%%%%%%%%%%%%%%%%%%%%%%%%%%%%%%%%%%%%%%%%%%%%%%%%%%%%%%%%%%%%%%%%%%%%%%%%%%%%%%%%%%%%%%%%
%
\begin{figure}[htb]
\begin{tabular}{cc}
\includegraphics[width=7cm]{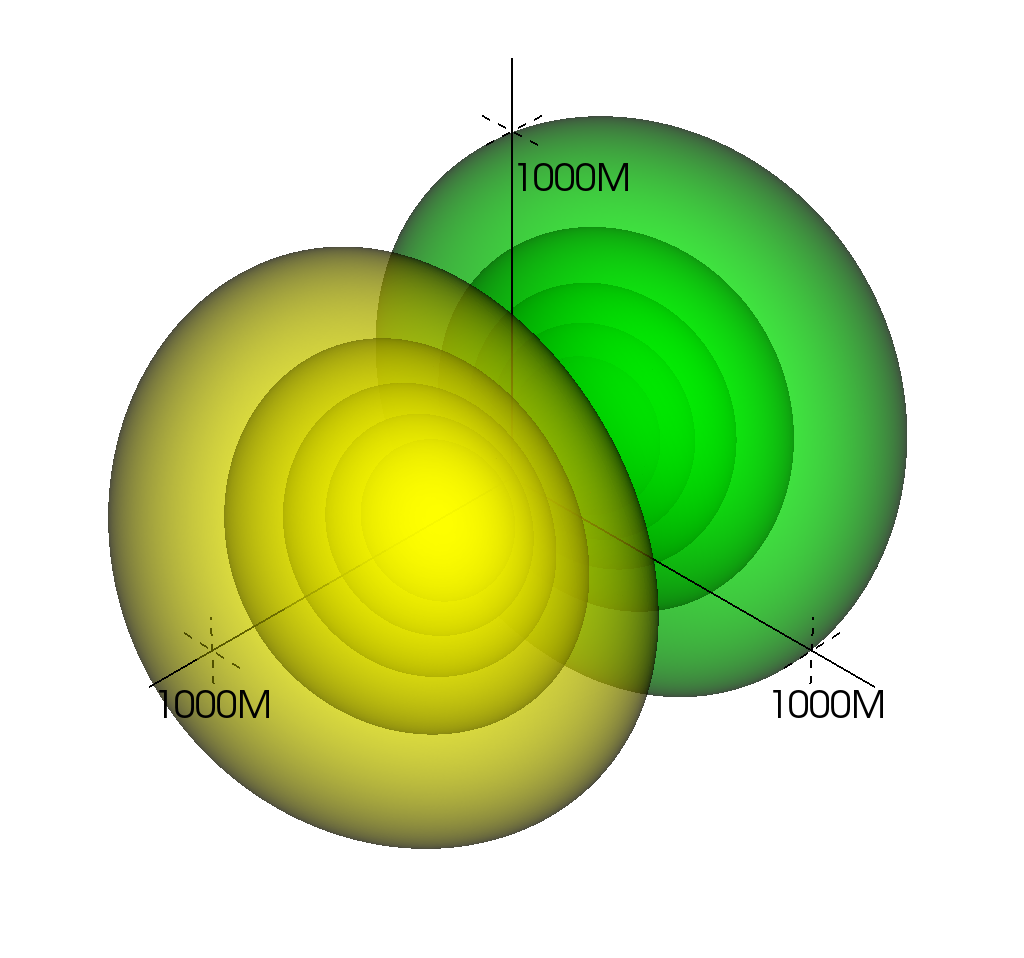}&
\includegraphics[width=7cm]{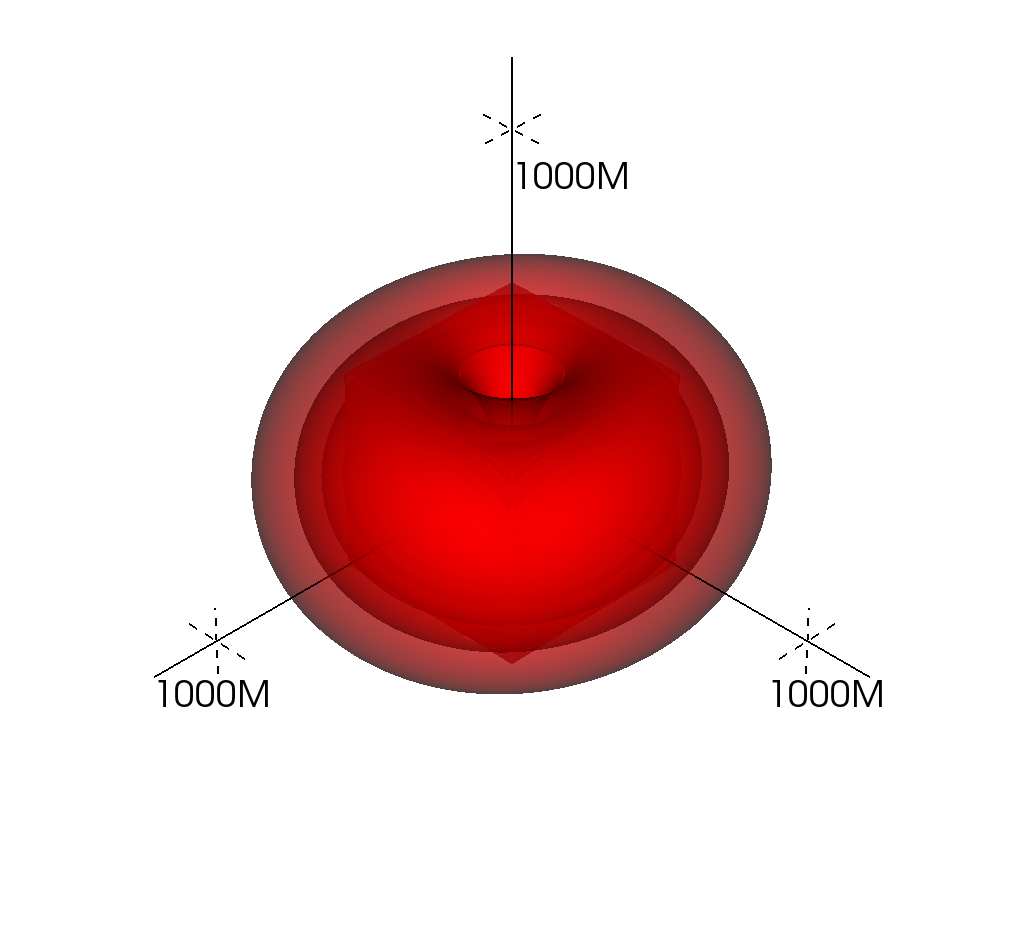}
\end{tabular}
\caption{Field (left) and energy density (right) distribution along the equatorial plane for a coupling $M\mu=0.1$ and a dipolar configuration around a non-spinning BH. The initial data is of the form \eqref{eq:FrequencySpec}, in particular it is described by
\eqref{scalar}. The field is dipolar, as expected, whereas the energy density at the equator is almost -- but not exactly -- symmetric along the rotation axis. The lengthscale of these images is of order $100M$. From Ref.~\cite{Cardoso:2020hca}.
\label{fig:a0_Mc0_mu01_snapshot}
}
\end{figure}
The linear and nonlinear results discussed above can be summarized as follows.
%%%%%%%%%%%%%%%%%%%%%%%%%%%%%%%%%%%%%%%%%%%%%%%%%%%%%%%%%%%%%%%%%%%%%%%
\paragraph{The superradiant stage.}
%%%%%%%%%%%%%%%%%%%%%%%%%%%%%%%%%%%%%%%%%%%%%%%%%%%%%%%%%%%%%%%%%%%%%%%
As the superradiant instability develops, two time scales are important: the superradiant time scale and the GW emission time scale. Neglect first GW emission. Then the system asymptotes to a stationary state where superradiance ``shuts-off,'' and a cloud forms outside the BH. Focus on scalar fields for simplicity. As we mentioned, in the non-relativistic limit, the scalar cloud is governed by an equation which is formally equivalent to Schr\"{o}dinger's equation with a Coulomb potential, controlled by a single parameter
\be
\alpha\equiv M\mu_S \, .
\ee
This can be seen by making the standard \textit{ansatz} for the dynamical evolution of $\Phi$~\cite{Mendes:2016vdr,Baumann:2018vus}
\beq
\Phi\left(t,\bf{r}\right)=\frac{1}{\sqrt{2\mu_S}}\left(\psi\left(t,\bf{r}\right)e^{-i\mu_S t}+\psi^*\left(t,\bf{r}\right)e^{i\mu_S t}\right)\,,
\eeq
where $\psi$ varies on time scales much larger than $1/\mu_S$. Then, one can re-write Eq.~\eqref{eq:MFEoMScalar} as
\beq
i \frac{\partial}{\partial t}\psi&=&\left(-\frac{1}{2\mu_S}\nabla^2-\frac{\alpha}{r}\right)\psi \, ,
\eeq
where we kept only terms of order $\mathcal{O}\left(r^{-1}\right)$ and linear in $\alpha$.

The eigenstates of the system are hydrogenic-like, with an adapted ``fine structure constant'' $\alpha$ and ``reduced Bohr radius'' $a_0$~\cite{Detweiler:1980uk,Arvanitaki:2010sy,Baumann:2018vus}. They were shown in Eq.~\eqref{eigenfunctionDetweiler} and can be normalized as
\beq
\psi_{nlm}&=&\,e^{-i\left(\omega_{nlm}-\mu_S\right)t}\,R_{nl}\left(r\right)Y_{lm}\left(\theta,\phi\right) \, ,\\
R_{nl}\left(r\right)&=&C\left(\frac{2r}{\tilde{n} a_0}\right)^lL_{n}^{2l+1}\left(\frac{2r}{\tilde{n}a_0}\right)e^{-\frac{r}{\tilde{n}a_0}} \, , \nonumber\\
\tilde{n}\equiv l+n+1\,,\quad a_0&=&1/\left(\mu_S\alpha\right) \, ,\quad C=\sqrt{\left(\frac{2}{\tilde{n}a_0}\right)^3\frac{\left(n\right)!}{2\tilde{n}\left(\tilde{n}+l\right)!}} \,,\label{eq:FrequencySpec}
\eeq
where $L_{n}^{2l+1}$ is a generalized Laguerre polynomial. To lowest order in $M\mu_S$, the eigenvalues are given by Eq.~\eqref{omegaDetweiler}. Up to terms of order $\alpha^5$ one finds~\cite{Baumann:2019eav}
\be
\omega_{nlm}=\mu_S\left(1-\frac{\alpha^2}{2\tilde{n}^2}-\frac{\alpha^4}{8\tilde{n}^4}+\frac{\left(2l-3\tilde{n}+1\right)\alpha^4}{\tilde{n}^4\left(l+1/2\right)}\right)\,.
\ee

One estimate for the ``size'' of the scalar cloud is the expectation value of the radial coordinate for a given state,
\be
\left< r \right> = \frac{\int_0^{\infty}dr\,r^3 R_{nl}^2\left(r\right)}{\int_0^{\infty}dr\,r^2 R_{nl}^2\left(r\right)} = \frac{a_0}{2}\left(3\tilde{n}^2-l(l+1)\right) \, . \label{peak}
\ee
Thus, for most couplings the cloud extends well beyond the horizon, where rotation effects can be neglected and where a flat-space approximation is justified.

If a light fundamental bosonic field exists in the universe, any small fluctuation will grow around a spinning BH.
For scalars, the dominant mode is dipolar. This state will grow linearly in the way we described previously, and will extract rotational energy away from the BH and deposit it in such a state. At leading order, the geometry is described by the Kerr spacetime and the scalar evolves in this fixed background: the size of the cloud is large and the backreaction in the geometry is mostly negligible~\cite{Brito:2014wla}. In the quasi-adiabatic approximation (and focusing on the $l=m=1$ fundamental mode), the cloud is stationary and described by
\beq
R&=&A_0 \tilde{r}e^{-\tilde{r}/2} \cos\left(\varphi-\omega_Rt\right)\sin\vartheta\,, \label{scalar}\\
\tilde{r}&\equiv& \frac{2M\mu_S^2}{\tilde{n}}r \,,
\eeq
where the amplitude $A_0$ can be expressed in terms of the mass $M_S$ of the scalar cloud through~\cite{Brito:2014wla}
\begin{equation}\label{amplitude}
A_0^2=\frac{1}{32\pi}\left(\frac{M_S}{M}\right) \alpha^4\,. 
\end{equation}
The spatial distribution of such a field and the corresponding energy density is shown in Fig.~\ref{fig:a0_Mc0_mu01_snapshot}.

The amount of energy in the cloud depends on the initial BH spin $a_i$ and on the field mass $\mu_S$. In an ideal process for energy extraction, the process develops adiabatically with constant BH area, and the final spin is related to the initial one via
$2M_i(M_i+\sqrt{M_i^2-a_i^2})=2M_f(M_f+\sqrt{M_f^2-a_f^2})$. For the most extreme example of an initially extremal BH being spun down by a very light field such that $a_f\sim 0$, one finds $M_f=\sqrt{2}M_i/2$. Thus, the cloud can contain up to $29\%$ of the BH energy, within a factor 3 of what has so far been seen numerically.

Following the development of the instability in a fully nonlinear evolution is extremely challenging because of the time scales involved: $\tau_{\rm BH}\sim M$ is the light-crossing time, $\tau_S\sim 1/\mu_S$ is the typical oscillation period of the scalar cloud and $\tau \sim M/(M\mu_S)^9$ is the dominant instability time scale in the small-$M\mu_S$ limit. As previously discussed, in the most favorable case for the instability, $\tau\sim 10^6\tau_S$ is the minimum evolution time scale required for the superradiant effects to become noticeable in the scalar case. Thus, current nonlinear evolutions (which typically last at most $\sim 10^3 \tau_S$~\cite{Okawa:2014nda}) have not yet probed the development of the instability, nor the impact of GW emission.
However, as we mentioned, nonlinear evolutions of complex vector fields have been done and confirm the above description~\cite{East:2018glu}.

%%%%%%%%%%%%%%%%%%%%%%%%%%%%%%%%%%%%%%%%%%%%%%%%%%%%%%%%%%%%%%%%%%%%%%%
\paragraph{The GW emission stage.}
%%%%%%%%%%%%%%%%%%%%%%%%%%%%%%%%%%%%%%%%%%%%%%%%%%%%%%%%%%%%%%%%%%%%%%%
A nonspherical monochromatic cloud as in Eq.~\eqref{scalar} will emit GWs with frequency $ 2\pi/\lambda\sim 2\omega_R\sim 2\mu_S$, the wavelength $\lambda$ being in general \emph{smaller} than the size of the source, $r_{\rm cloud}\sim\langle r\rangle$. Thus, even though the cloud is nonrelativistic, the quadrupole formula does not apply because the emission is incoherent~\cite{Arvanitaki:2010sy,Yoshino:2013ofa,Brito:2014wla,Yoshino:2015nsa}. However, due to the separation of scales between the size of the cloud and the BH size for $\mu_S M\ll 1$, the GW emission can be analyzed taking the source to lie in a nonrotating (or even flat~\cite{Yoshino:2013ofa}) background.

\begin{figure}[t]
\centering
\includegraphics[width=0.75\textwidth]{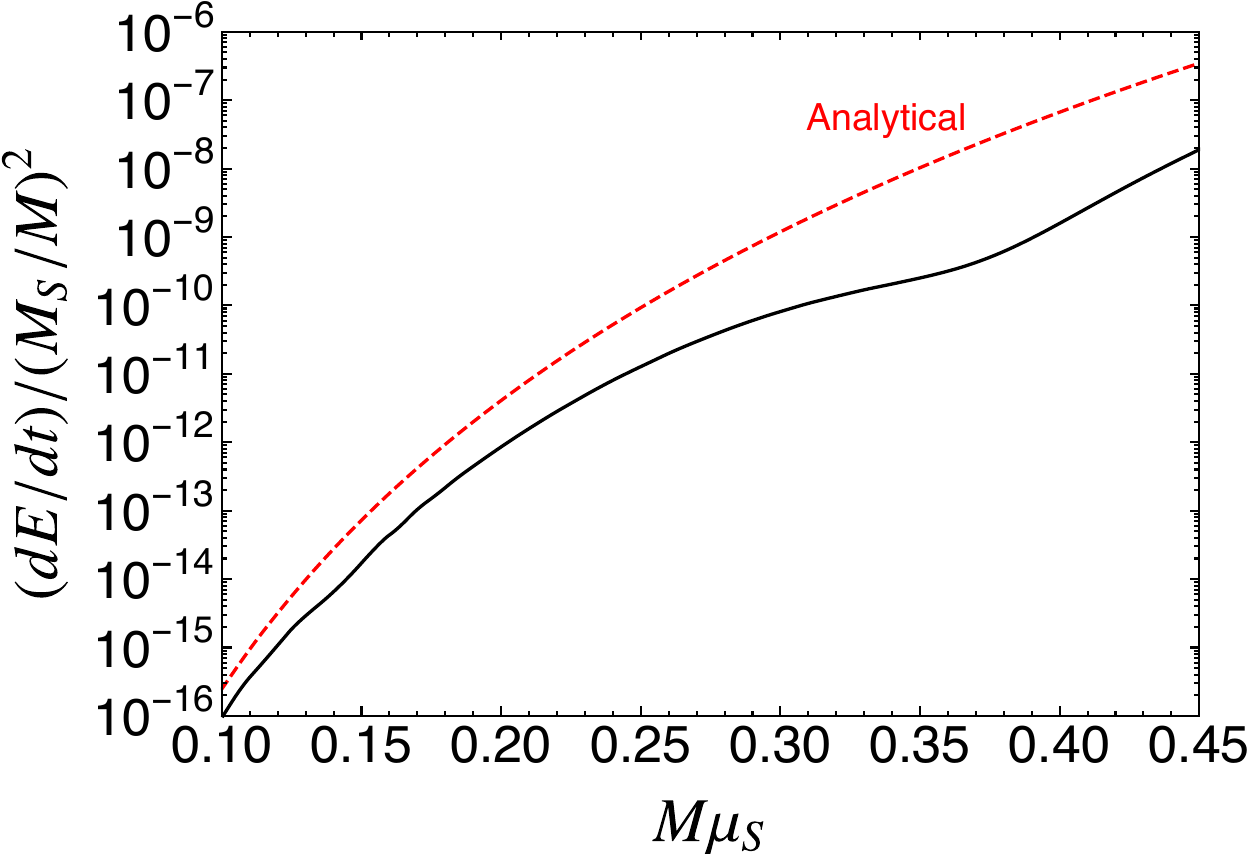}
\caption{Fully relativistic GW flux for a dipolar $l=m=1$ scalar mode (black solid line) as a function of the coupling $M\mu_S$, computed at the threshold at which this mode is marginally stable $\Omega_{\rm H}=\omega_R$. For comparison we also show the curve predicted by the analytical formula [cf. Eq.~\eqref{dEdtF}] valid in the small $M\mu_S$ (dashed red line). Adapted from Ref.~\cite{Brito:2017zvb}.
\label{fig:GW_flux}
}
\end{figure}

By performing a fully relativistic analysis within the Teukolsky formalism, Refs.~\cite{Brito:2014wla,Brito:2017zvb} found that the energy and angular-momentum fluxes of gravitational radiation emitted by a dipolar $l=m=1$ scalar cloud read
\begin{eqnarray}
 \dot{E}_{\rm GW}&=&\frac{484+9 \pi ^2}{23040}\left(\frac{M_S^2}{M^2}\right)(M\mu_S)^{14}\,, \label{dEdtF}\\
 \dot{J}_{\rm GW}&=&\frac{1}{\omega_R} \dot{E}_{\rm GW}\,. \label{dJdtF}
\end{eqnarray}
This result has been obtained for small values of the coupling $M\mu_S$ and by neglecting spin effects, i.e. by considering a Schwarzschild background. The latter is a well-motivated assumption, because the cloud is localized away from the horizon, when spin effects are negligible. The energy flux above is in agreement with a previous analysis~\cite{Yoshino:2013ofa} except for a different prefactor in Eq.~\eqref{dEdtF} due to the fact that Ref.~\cite{Brito:2014wla} considered a Schwarzschild background, whereas Ref.~\cite{Yoshino:2013ofa} considered a flat-metric approximation. This analytical result is an \emph{upper bound} relative to the exact numerical flux, as we show in Fig.~\eqref{fig:GW_flux}, the latter being valid for any $\mu_S$ and any BH spin~\cite{Yoshino:2013ofa}. Therefore, using Eq.~\eqref{dEdtF} to estimate the energy loss in GWs is a very conservative assumption, since the GW flux is generically smaller.

\paragraph{Multiple modes}

The evolution discussed above considered the case of a single ($l=m=1$) mode. Multiple modes were discussed in detail in Ref.~\cite{Ficarra:2018rfu}. The evolution of multimode data is much richer and depends strongly on the initial ``seed'' energy of the perturbation, on the relative amplitude between modes, and on the gravitational coupling. If the seed energy is smaller than a few percent of
the BH mass --~as in most realistic cases including an instability triggered by quantum fluctuations~-- the effect of nonsuperradiant modes is negligible and the effect of higher multipoles ($l>1$) can be easily included independently of the $l=m=1$. As discussed in Sec.~\ref{sec:nospinningBHs}, although modes with $l>1$ are typically much less unstable than the dipole, owing to the superradiant condition $\omega_R<m\Omega$ modes with higher values of $m$ might exist when the dipole mode $l=m=1$ is not superradiantly unstable (see Fig.~\ref{fig:holes} below).
On the other hand, if the seed energy is a few percent of
the BH mass, a BH surrounded by a mixture of superradiant and nonsuperradiant modes with comparable amplitudes might not undergo a superradiant unstable phase, depending on the value of the boson mass. This case is relevant only when the BH is initially surrounded by a non-negligible scalar environment, or if it is formed out of the coalescence of two BHs merging with their own scalar clouds. Finally, multiple-mode initial data are useful to make a quantitative comparison between the quasi-adiabatic evolution (to be discussed in more detail in Sec.~\ref{sec:simplifiedmodel} below) and numerical-relativity simulations, since the latter often make use of multimode initial data~\cite{Okawa:2014nda,East:2017mrj,East:2018glu}.

%%%%%%%%%%%%%%%%%%%%%%%%%%%%%%%%%%%%%%%%%%%%%%%%%%%%%%%%%%%%%%%%%%%%%%%%%%%%%%%%%%%%%%%%%%%%%%%
\subsubsection{Circumventing the no-hair theorem with superradiance}\label{sec:hair}
%%%%%%%%%%%%%%%%%%%%%%%%%%%%%%%%%%%%%%%%%%%%%%%%%%%%%%%%%%%%%%%%%%%%%%%%%%%%%%%%%%%%%%%%%%%%%%%

A series of works established what is now known as uniqueness theorems in GR:
regular, stationary BHs in Einstein-Maxwell theory are extremely simple
objects, being characterized by three parameters only: mass, angular momentum and electric 
charge~\cite{Bekenstein:1996pn,Carter:1997im,Heusler:1998ua,Chrusciel:2012jk,uniqueness,Cardoso:2016ryw}.
Because of quantum and classical discharge effects, astrophysical BHs are thought to be neutral to a very good
approximation~\cite{Gibbons:1975kk,Goldreich:1969sb,Ruderman:1975ju,Blandford:1977ds,Barausse:2014tra,Cardoso:2016olt}.
Therefore the geometry of astrophysical BHs in GR is simply described by the two-parameter Kerr 
metric~\eqref{metricKerrLambda}. 
On the other hand, NSs --~the most compact, nonvacuum objects that exist~-- cannot be more massive than $\sim 
3M_\odot$~\cite{Rhoades:1974fn};
taken together, these two results imply that any observation of a compact object with mass larger than $\sim 3M_\odot$ 
must belong to the Kerr family. Therefore tests of strong-field gravity targeting
BH systems aim at verifying the ``Kerr hypothesis'' in various ways~\cite{Berti:2015itd,Cardoso:2016ryw}.

We just saw that when (electro-vacuum) GR is enlarged to include minimally coupled, massive scalar fields, Kerr BHs 
may become superradiantly unstable.
For {\it real} scalars, such phenomenon leads to a bosonic cloud around the BH, whose nonzero quadrupole moment results in
periodic GW emission, described above. Thus, the end-state is thought to be a Kerr BH
with lower spin~\cite{Witek:2012tr,Okawa:2014nda,Cardoso:2013krh}, as dictated by the uniqueness theorems.
All the linear and nonlinear results available confirm this picture.

%%%%%%%%%%%%%%%%%%%%%%%%%%%%%%%%%%%%%%%%%%%%%%%%%%%%%%%%%%%%%%%%%%%%%%%%%%%%%%%%%%%%%%%%%%%%%%%%%%%%%%%%%%%%%%%%%%%
%\subsubsection{Circumventing the no-hair theorem with complex scalars}
%%%%%%%%%%%%%%%%%%%%%%%%%%%%%%%%%%%%%%%%%%%%%%%%%%%%%%%%%%%%%%%%%%%%%%%%%%%%%%%%%%%%%%%%%%%%%%%%%%%%%%%%%%%%%%%%%%%
However, there is a subtle way of circumventing the hypothesis of the uniqueness theorem. Namely, the scalar field could 
be time dependent but in such a way that the geometry remains stationary. This requires that the stress-energy tensor of 
the scalar field shares the same symmetries of the metric, similarly to the AdS case discussed in 
Sec.~\ref{sec:KerrAdS}. Having such stationary configuration is impossible for a single real scalar field, but for a 
\textit{complex} scalar field with time dependence $\Psi(t,\mathbf{x})=e^{-i\omega t}\psi(\mathbf{x})$, it is possible 
precisely when
the frequency saturates the superradiant condition~\eqref{eq:superradiance_condition}, i.e. when
\begin{equation}
 \omega=m\Omega_{\rm H} \,. \label{cloudcond}
\end{equation}
This is easily seen from the analytic formula \eqref{omegaDetweiler} together with the flux
result~\eqref{flux_scalar}. Consequently, there is no scalar field flux through the horizon as long as
\eqref{cloudcond} is obeyed and the field is allowed to be complex.

This argument suggests the existence of \textit{asymptotically-flat rotating BHs with complex scalar hair}.
In fact, the argument parallels the discussion of hairy solutions in asymptotically AdS spacetimes, discussed in 
Sec.~\ref{sec:KerrAdS}. Such solutions in asymptotically flat spacetimes were indeed found
and studied in the limit that the BH is extremal~\cite{Hod:2012px,Hod:2013zza}.
The solutions in full generality were found numerically in Ref.~\cite{Herdeiro:2014goa}, and formally shown to exist later~\cite{Chodosh:2015oma}; a detailed discussion on their 
construction and physical properties can be found in Ref.~\cite{Herdeiro:2015gia}.
The ultimate physical reason for the existence of a stationary geometry endowed with an oscillating scalar field is that 
GW emission is halted due to cancellations in the stress-energy tensor, which
becomes independent on the time and azimuthal variables, thus avoiding
GW emission and consequent angular momentum losses.

The fact that the condition~\eqref{cloudcond} for the existence of hairy BHs lies
precisely at the threshold of the superradiant condition~\eqref{eq:superradiance_condition} arises from the fact that 
\textit{real frequency} bound states are possible if and only if Eq.~\eqref{cloudcond} is satisfied. The hairy BHs found 
in Ref.~\cite{Herdeiro:2014goa} can be thought of as
nonlinear extensions of the linear bound states, when the backreaction of the scalar condensate on the metric is 
included (see also Ref.~\cite{Herdeiro:2014ima}).

The minimally coupled hairy solutions are described by the following ansatz~\cite{Herdeiro:2014goa} 
\begin{align}
\label{ansatz_hairy}
ds^2&=e^{2F_1}\left(\frac{dR^2}{N }+R^2 d\theta^2\right)+e^{2F_2}R^2 \sin^2\vartheta (d\varphi-W dt)^2-e^{2F_0} N 
dt^2\,, \nonumber \\
\Psi&=\phi(r,\vartheta)e^{i(m\varphi-\omega t)}\,,
\end{align} 
where $N\equiv 1-{R_H}/{R}$, the parameter $R_H$ being the location of the event horizon. The five functions of 
$(R,\vartheta)$, $F_0,F_1,F_2,N,\phi$, are obtained by solving numerically a system of nonlinear, coupled PDEs, with 
appropriate boundary conditions that
ensure both asymptotic flatness and regularity at the horizon; the latter requirement implies 
condition~\eqref{cloudcond}. 

The solutions form a five-parameter family described by the ADM mass $M$, the ADM angular momentum $J$, the Noether 
scalar charge $Q$ (which roughly measures the amount of scalar hair outside the horizon), and by two discrete 
parameters: the azimuthal harmonic index $m$ and the node number $n$ of the scalar field~\cite{Herdeiro:2014goa}. One 
may regard $n=0$ as the fundamental configuration and $n\ge 1$ as excited states. 
Remarkably, these solutions interpolate between a Kerr BH when $q\equiv Q/2J=0$ and a rotating boson 
stars~\cite{Yoshida:1997qf,Kleihaus:2005me} when $q=1$. The latter are (horizonless)
gravitating solitons, that are discussed in Sec.~\ref{sec:BHmimickers} in the context of so-called ``BH mimickers''. 
Because the scalar charge $Q$ is a free parameter, the solutions found in Ref.~\cite{Herdeiro:2014goa} corresponds to 
hairy spinning BHs with primary hair (in contrast to BH solutions with secondary hair, in which the scalar charge is 
fixed in terms of other parameters, such as the mass~\cite{Berti:2015itd}).

\begin{figure}[H]
\centering
\includegraphics[width=0.8\textwidth]{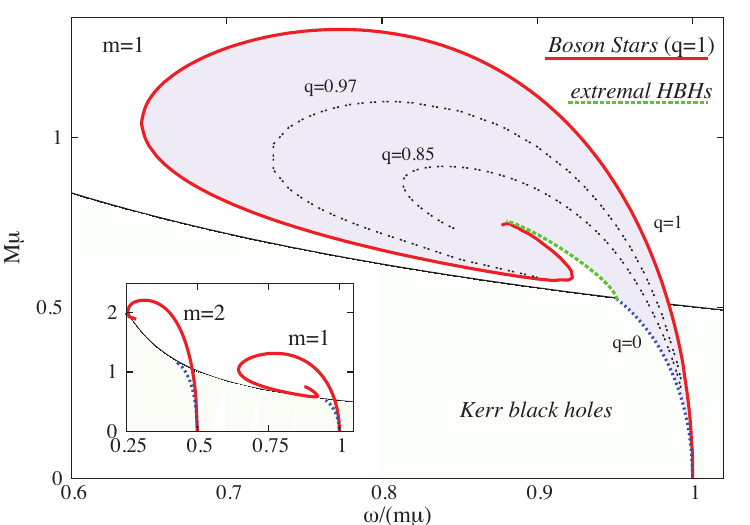}\\
\caption{The $M$-$\omega$ parameter space of hairy BHs with a complex scalar field for $n=0$ and $m=1$. These solutions 
exist in the shaded blue region. The black solid curve corresponds to extremal Kerr BHs and nonextremal Kerr BHs exist 
below it. For $q\equiv Q/2J=0$, the domain of existence connects to Kerr solutions (dotted blue line). For $q=1$, hairy 
BHs reduce to boson stars (red solid line). The final line that delimits the domain of existence of the hairy BHs 
(dashed green line) corresponds to extremal BHs, i.e. with BHs with zero temperature. The inset shows the boson star 
curves for $m=1,2$. Units in the axes are normalized to the scalar field mass $\mu$. Adapted from 
Ref.~\cite{Herdeiro:2014goa}.}
\label{hairy-pm}
\end{figure}

Figure~\ref{hairy-pm} shows the parameter space for the ground-state ($n=0$) solutions with 
$m=1$~\cite{Herdeiro:2014goa}. Interestingly, uniqueness in the $(M,J)$ subspace is broken because there is a region in 
which hairy BHs and the Kerr solution coexist with the same values of mass and angular momentum.  However, no two 
solutions were found with the same
$(M,\,J,\,q)$~\cite{Herdeiro:2014goa}. In the region of nonuniqueness, hairy BHs have
\textit{larger} entropy than the corresponding Kerr BHs. Therefore, the
former cannot decay into the latter adiabatically.

\begin{figure}[ht]
\centering
\includegraphics[width=0.8\textwidth]{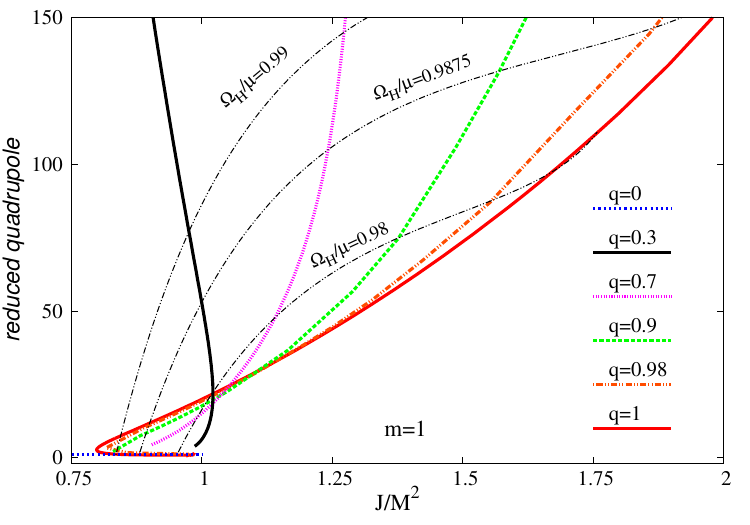}\\
\caption{Quadrupole moment of a hairy BH normalized by its Kerr value for the same mass and angular momentum and as a 
function of the dimensionless spin $J/M^2$. Several lines of constant $\Omega_{\rm H}$ (dashed black) and $q\equiv Q/2J$ 
are displayed. From Ref.~\cite{Herdeiro:2014goa}.}
\label{fig:hairyQ}
\end{figure}

As found in Ref.~\cite{Herdeiro:2014goa}, the quadrupole moment and the angular frequency at the ISCO can differ 
significantly for hairy BHs, as compared to the standard Kerr
values. This is shown in Fig.~\ref{fig:hairyQ}.

In fact, in one corner of the parameter space these solutions can be interpreted as Kerr BHs perturbed by a small scalar 
field (and whose quadrupole moment is close to that of Kerr), whereas in the opposite corner they describe a small BH 
within a large boson star. In the latter case the properties of the solutions are resemble those of a stellar 
configuration rather than those of a BH.

Since these solutions also possess an ergoregion they are expected to be unstable against the superradiant instability. Indeed it was shown in Ref.~\cite{Ganchev:2017uuo} that solutions with azimuthal number $m=1$ are unstable against small perturbations with azimuthal number $m>1$. For small scalar hair the time scale of the instability was shown to be almost identical to those of a massive scalar field on a fixed Kerr background, as should be expected. Non-linear hairy solutions with $m>1$ were studied in detail in Ref.~\cite{Delgado:2019prc} showing that consecutive $m$ families can share solutions with the same global mass and spin, and when this occurs higher $m$ are always entropically favoured, therefore suggesting that the superradiant instability can trigger the migration of solutions with lower $m$ to the ones with larger $m$, in an approximately adiabatic way. It should be noted however,
that such instability has a time scale which can be extremely large~\cite{Degollado:2018ypf}.

Finally, hairy BHs have a richer structure of ergoregions
than Kerr. For example, besides the ergosphere of Kerr-like configurations (cf. Fig.~\ref{fig:ERKerr}) also 
\emph{ergosaturn} can form in a certain region of parameter space~\cite{Herdeiro:2014jaa}.

Nonlinear, hairy solutions were also extended to encompass rotating, charged geometries~\cite{Benone:2014ssa,Delgado:2016jxq}, whereas in Refs.~\cite{Hod:2014baa,Hod:2014sha} these solutions were constructed and analyzed analytically at linear level. 
Solutions with complex Proca hair were also constructed and shown to share very similar properties to the 
solutions presented above~\cite{Brito:2015pxa,Herdeiro:2016tmi}. These solutions also emerge as metastable states during the numerical evolution of complex Proca fields~\cite{East:2018glu}.

%%%%%%%%%%%%%%%%%%%%%%%%%%%%%%%%%%%%%%%%%%%%%%%%%%%%%%%%%%%%%%%%%%%%%%%
\paragraph{Other hairy solutions and the role of tidal dissipation.}
%%%%%%%%%%%%%%%%%%%%%%%%%%%%%%%%%%%%%%%%%%%%%%%%%%%%%%%%%%%%%%%%%%%%%%%
Generalizations were soon found that encompass hairy BHs with self-interacting scalar fields.
For example, in Ref.~\cite{Herdeiro:2014pka} the authors studied a complex massive scalar field with quartic plus hexic 
self-interactions, dubbed nonlinear $Q$-clouds.
Without the self-interactions, it reduces to the hairy solutions just described and correspond to zero modes of the 
superradiant instability. Non-linear $Q$-clouds, on the other hand, are also in synchronous rotation with the BH 
horizon; but they exist on a 2-dimensional subspace, delimited by a minimal horizon angular velocity and by an 
appropriate existence line, wherein the nonlinear terms become irrelevant and the $Q$-cloud reduces to a linear cloud. 
Thus, $Q$-clouds provide an example of scalar bound states around Kerr BHs which, generically, are not zero modes of the 
superradiant instability. Note that self-interaction terms only become important in the nonlinear regime:
accordingly, it could be anticipated that nonlinear solutions exist (where the nonlinear terms play the role of an 
effective mass term) despite not corresponding to any superradiant bound state in the linear regime. Indeed, Ref.~\cite{Herdeiro:2015tia} constructed fully non-linear hairy solutions for a complex massive scalar field with a quartic potential, showing that solutions exist even for very large self-interactions.

Other hairy solutions were also found in higher dimensional, asymptotically flat spacetime~\cite{Brihaye:2014nba};
the construction parallels that of AdS spacetime (see Section~\ref{sec:KerrAdS} and Ref.~\cite{Dias:2011at}) and 
consists on finding
rotating BHs with scalar hair and a regular horizon, within five dimensional Einstein's gravity minimally coupled to a 
complex, massive scalar field doublet. They are described by their mass $M$, two equal angular momenta and a conserved 
Noether charge $Q$, measuring the scalar hair. For vanishing horizon size the solutions reduce to five dimensional boson 
stars. In the limit of vanishing Noether charge density, the scalar field becomes point-wise arbitrarily small and the 
geometry becomes, locally, arbitrarily close to that of a specific set of Myers-Perry BHs (the higher-dimensional 
versions of the Kerr solution~\cite{Myers:1986un}); but there remains a global difference with respect to the latter, 
manifest in a finite mass gap. Thus, the scalar hair never becomes a linear perturbation of the Myers-Perry geometry. 
This is a qualitative difference when compared to Kerr BHs with scalar hair~\cite{Herdeiro:2014goa}. Whereas the 
existence of the latter can be anticipated in linear theory, from the existence of scalar bound states on the Kerr 
geometry (i.e. scalar clouds), the hair of these Myers-Perry BHs is intrinsically nonlinear.

An aspect that deserves to be highlighted is condition~\eqref{cloudcond} for stationary solutions, which holds even when 
the hairy solution cannot easily be mapped onto a linearly, superradiantly unstable spacetime. This condition is tightly 
connected to tidal dissipation, in turn associated with superradiance, as we explained in
Section ~\ref{sec:tides} (see also Refs.~\cite{Cardoso:2012zn,Brito:2012gw}). In summary, if the scalar ``cloud'' does 
not obey Eq.~\eqref{cloudcond}, tidal forces (of gravitational or other nature) will act and the system cannot possibly 
be in equilibrium. This fact is reminiscent of the phenomenon of ``tidal locking'' that occurs, for instance, in the 
Earth-Moon system~\cite{Cardoso:2012zn}.

%%%%%%%%%%%%%%%%%%%%%%%%%%%%%%%%%%%%%%%%%%%%%%%%%%%%%%%%%%%%%%%%%%%%%%%
\paragraph{Formation of hairy solutions and bounds on bosonic fields.}
%%%%%%%%%%%%%%%%%%%%%%%%%%%%%%%%%%%%%%%%%%%%%%%%%%%%%%%%%%%%%%%%%%%%%%%
In parallel with the open problem of stability of the hairy BHs discussed above, a relevant question is the mechanism of 
formation of such solutions. 
Formation scenarios based on collapse or Jeans-like instability arguments are hard to devise. Indeed, if the collapsing 
matter does not possess any scalar charge, it is reasonable to expect that collapse would form a Kerr BH, which might 
eventually migrate towards a hairy BH solution through superradiant amplification of a scalar-field fluctuation. 
However, should these solutions arise from a superradiant instability of 
the Kerr metric, the energy-density of the scalar field is negligible and the geometry would be very well described by 
the Kerr solution~\cite{Brito:2014wla}. In other words, superradiant instabilities require a Kerr BH to start with, and 
they can at most produce ``light'' scalar clouds, i.e. condensates which backreact very weakly on the geometry. The 
physical reason is that superradiance can only extract a finite amount of mass from the BH (at most $29\%$ of the 
initial BH mass~\cite{Begelman:2014aea}), and therefore the scalar cloud can only grow to a limited value. 
Although it is unlikely that configurations that deviate considerably from Kerr can arise from the evolution of 
initially isolated Kerr BHs, they may arise as the end-state of some other initial conditions, most likely involving a 
large scalar field environment; for instance they could arise from the collapse of ordinary stars inside a large 
boson-star environment.

Finally, the putative existence of hairy BH solutions as the end-state of the superradiant instability does {\it not} 
invalidate
any of the results related to BH spin down, which will be discussed in detail in Sections~\ref{sec:evolution} and \ref{sec:bounds_mass}. The reason is that hairy BHs lie 
along the $\omega=m\Omega\sim\mu$ line.
In other words, for a Kerr BH to evolve towards a hairy BH it will necessarily loose angular momentum, in the same way 
as Kerr BHs do, as discussed in Section~\ref{sec:astrophysics}.

%%%%%%%%%%%%%%%%%%%%%%%%%%%%%%%%%%%%%%%%%%%%%%%%%%%%%%%%%%%%%%%%%%%%%%%%%%%%%%%%%%%%%%%%%%%
\subsection{Tidal effects induced by a companion star or black hole}\label{sec:cloud_tides}
%%%%%%%%%%%%%%%%%%%%%%%%%%%%%%%%%%%%%%%%%%%%%%%%%%%%%%%%%%%%%%%%%%%%%%%%%%%%%%%%%%%%%%%%%%%
%
\begin{figure}[htb]
\begin{center}
\begin{tabular}{c}
\includegraphics[width=10cm]{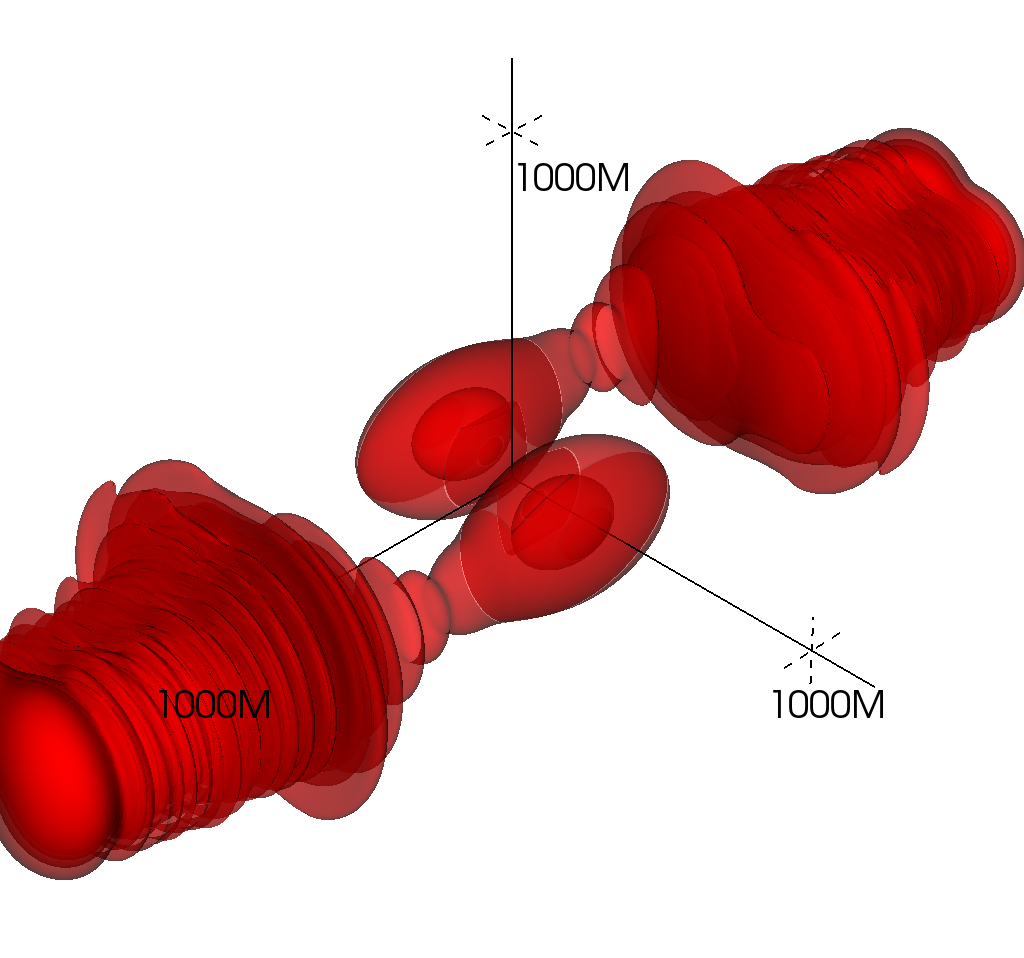}
\end{tabular}
\caption{Snapshot of a tidally disrupting cloud. The snapshot depicts the energy density along the equator of a scalar 
cloud which was set initially around a non-spinning BH. 
In the absence of a companion mass, the energy density is almost spherical and remains so for thousands of dynamical 
time scales. Here, the simulation starts with one symmetric initial scalar energy distribution, but in the presence of a star
of mass $M_c=0.1$ at a distance $R=100$. The gravitational coupling $M\mu_{\rm S}=0.1$.
The snapshot is taken after $7000M$ and is leading to disruption of the cloud. Taken from Ref.~\cite{Cardoso:2020hca}.
\label{fig:a0_Mc01_mu01_snapshot}}
\end{center}
\end{figure}
The presence of a companion affects the growth and development of superradiant instabilities,
since it destroys the axi-symmetry of the problem. This loss of symmetry leads to mode-mixing
and to a wealth of new effects~\cite{Baumann:2018vus}. The effects of the companion can been modelled as follows.
Take a Kerr BH surrounded by a superradiant cloud, and assumed it to cause negligible backreaction in the spacetime. The 
geometry is therefore kept fixed, in the sense that the scalar field never backreacts on the geometry. This working 
hypothesis holds true for most of the situations of interest~\cite{Brito:2014wla}. A companion of mass $M_c$ is now 
present, at a distance $R$, and located at $\theta=\theta_c,\phi=\phi_c$ in the BH sky. The companion induces a change 
$\delta ds_{\rm tidal}^2$ in the geometry. Thus, the spacetime geometry is described by
\be
ds^2=ds^2_{\rm Kerr}+\delta ds_{\rm tidal}^2\,.
\ee
For the tidal perturbation induced by the companion, the literature as focused so far only on the leading-term 
non-spinning approximation~\cite{Cardoso:2017cfl}
\beq
\delta ds^2&=&\sum_m r^2{\cal E}_{2m}Y_{2m}(\theta,\phi)(f^2dt^2+dr^2+(r^2-2M^2)d\Omega^2)\nonumber\\
{\cal E}_{2m}&=&\frac{8\pi}{5}\frac{M_c}{R^3}Y^*_{2m}(\theta_c,\phi_c)\,,%\quad {\cal B}_{2m}= 
\mathcal{O}\left(\frac{v}{c}\right) \, ,
\eeq
where $f=1-2M/r$. This expression deals only with electric-type tides, since they dominate over magnetic type terms.
This approximation is not accurate close to the BH horizon, where spin effects change the tidal description. However, 
for all the parameters considered here the cloud is localized sufficiently far-away that such effects ought to be very 
small.

The tidally-corrected geometry above is now ready to be used in the equation of motion for the scalar field. As a 
result, one finds a wave-like equation similar to \eqref{teu_radial}, which can be put in the form 
$d^2R/dr_*^2+\left(\omega^2-V\right)R=0$,
with an effective potential $V=V_0+\delta V$, where $V_0$ is the potential in pure Kerr. This form is apt for a 
perturbative treatment, familiar from quantum mechanics~\cite{Landau:QM}.

In particular, the companion will cause transition between levels to an extent which can easily be computed. A thorough 
study of this problem and associated level transitions was done in Refs.~\cite{Baumann:2018vus,Berti:2019wnn,Cardoso:2020hca}.
There are two important effects on the cloud itself, generically: the presence of a companion excites higher overtones, 
causing the cloud to ``puff up,'' since overtones have a larger spatial distribution. Tidal effects also cause mixing 
between angular modes, causing octupolar and higher modes to be excited and leading to a different cloud spatial 
profile.

The tidally distorted cloud also backreacts on the companion, in a way very similar to that of Earth tides acting on the 
moon, described in detail in Section~\ref{sec:tides}: transferal of angular momentum from the cloud to the companion 
occurs via the gravitational interaction only and may lead to companions stalled (``floating'') at some critical radius where tides are 
resonantly excited~\cite{Cardoso:2011xi,Ferreira:2017pth,Zhang:2018kib}. The mechanism in fact transfers angular momentum from the BH to the companion, using the cloud as intermediary; there is a one-to-one correspondence with the Earth-Moon system of Section~\ref{sec:tides}, where the cloud plays the role of oceans. Such floating orbits are described in more 
detail below, in the context of a charged companion. For purely gravitational interactions, we refer the reader to 
Ref.~\cite{Ferreira:2017pth} for a Newtonian analysis and to Refs.~\cite{Cardoso:2011xi,Zhang:2018kib,Baumann:2019ztm} for a full 
 relativistic and thorough analysis. For example, for angular momentum exchanged via the $l=1,m=\pm 1$ mode, floating occurs at
at an orbital frequency $M\Omega=a\mu_{S}(M\mu_{S})^5/12$~\cite{Zhang:2018kib,Baumann:2019ztm}.

For large tidal fields, one expects the scalar configuration to be disrupted. A star of mass $M_*$, radius $R_*$, in the 
presence of a companion of mass $M_c$ at distance $R$ is on the verge of disruption if -- up to numerical factors of 
order unity,
\be
2M_c/R^3=M_*/R_*^3\,.
\ee
For configurations where the mass in the scalar cloud is a fraction of that of the BH, $M_*=M$ and its radius is of the 
order of $R_*=5/(M_{\rm BH}\mu_{S}^2)$ (for the fundamental dipolar mode, see Section~\ref{sec:cloudproperties}). As 
such, one finds the critical moment 
\be
\left(\frac{M_c}{R^3}\right)_{\rm crit}\approx\frac{(M\mu_{S})^6}{250M^2}\,.\label{tde_scaling}
\ee
Numerical simulations are consistent with this behavior. Figure~\ref{fig:a0_Mc01_mu01_snapshot} shows the equatorial 
distribution of energy density for a cloud with a mass coupling $M\mu_{S}=0.1$ in the neighborhood of a star with 
mass $0.1M$ at a distance $R=100M$.
The result is compatible with disruption of the cloud, which quickly looses mass to asymptotic infinity. The scaling 
with the parameters of the system is consistent with the previous predictions.

Note that, at the verge of tidal disruption by a companion, the binary itself is emitting GWs at a rate
\be
\dot{E}_{\rm binary}= \frac{32}{5}\frac{M_c^2M^3}{R^5}\,,
\ee
where we assume the companion to be much lighter than the BH. The GW flux emitted by the cloud-BH system scales as Eq.~\eqref{dEdtF}~\cite{Yoshino:2013ofa,Brito:2014wla,Brito:2017zvb}
\beq
\dot{E}_{\text{cloud}}\sim \frac{1}{50}\left(\frac{M_S}{M}\right)^2\left(M\mu_S\right)^{14} \,.
\eeq
Thus, GW emission by the binary dominates the signal whenever
\be
\frac{M_c}{M_S}\gtrsim (M_S/M)^5(5M\mu_S/2)^{12}\,,
\ee
with $M_S$ the mass in the scalar cloud~\cite{Yoshino:2013ofa,Brito:2014wla,Brito:2017zvb}. Therefore, in the context of GW emission and detection, for all practical purposes disruption will not affect our ability to probe the system:
if it was visible via monochromatic emission by the cloud before disruption, it will be seen after disruption as a binary.

%%%%%%%%%%%%%%%%%%%%%%%%%%%%%%%%%%%%%%%%%%%%%%%%%%%%%%%%%%%%%%%%%%%%%%%%%%%%%%%%%%%%%%%%%%%%%%%%%%%%%%
\subsection{Superradiant instabilities and other absorption channels}\label{sec:super_couplings}
%%%%%%%%%%%%%%%%%%%%%%%%%%%%%%%%%%%%%%%%%%%%%%%%%%%%%%%%%%%%%%%%%%%%%%%%%%%%%%%%%%%%%%%%%%%%%%%%%%%%%%
%%%%%%%%%%%%%%%%%%%%%%%%%%%%%%%%%%%%%%%%%%%%%%%%%%%%%%%%%%%%%%%%%%%%%%%%%%%%%%%
\subsubsection{Axionic couplings and bursts of light\label{sec:axion_coupling}}
%%%%%%%%%%%%%%%%%%%%%%%%%%%%%%%%%%%%%%%%%%%%%%%%%%%%%%%%%%%%%%%%%%%%%%%%%%%%%%%
We have given a complete description of how minimally coupled, massive fields evolve
in the spacetime containing a spinning BH, but otherwise isolated. Generically, an instability ensues through extraction
of energy from the spinning BH, and deposition onto the bosonic field, which ``condensates'' outside the horizon.
However, it is likely that such fundamental fields are coupled to other, possibly standard-model, fields.

As a simple example, consider the theory described by a charged massive scalar coupled to the Maxwell sector, 
Eq.~\eqref{eq:MFaction}.
A vacuum Kerr spacetime with vanishing scalar and vector $\Psi, A_\mu$ solves the equations of motion. Thus, a neutral 
spinning BH will trigger a scalar instability.
As the scalar grows, and as is clear from the equations of motion, it induces a non-vanishing vector field, potentially 
affecting the development and growth of the scalar condensate.

An example of interest to particle physics and where new effects appear concerns axionic couplings~\cite{Anastassopoulos:2017ftl}. Consider the theory 
where a massive, real scalar $\Psi$ with possible axionic couplings to a vector (through the coupling constant 
$k_{\rm a}$) and scalar couplings to the Maxwell invariant through a coupling constant $k_{\rm s}$,
\beq
\label{eq:MFaction2}
{\cal L}=\frac{R}{k}- \frac{1}{4} F^{\mu\nu} F_{\mu\nu} - \frac{1}{2} g^{\mu\nu} \partial_{\mu}\Psi\partial_{\nu} 
\Psi-\frac{\mu^{2}_{\rm S}}{2} \Psi\Psi - \frac{k_{\rm a}}{2} \Psi \,^{\ast}F^{\mu\nu} F_{\mu\nu}-\frac{\left(k_{\rm 
s}\Psi\right)^p}{4} F^{\mu\nu} F_{\mu\nu} \,,
\eeq
with $p=1,2$ popular choices~\cite{Stadnik:2015kpa,Olive:2007aj}. Again, the mass of the scalar $\Psi$ is given by $m_S = \mu_S \hbar$, 
$F_{\mu\nu} \equiv
\nabla_{\mu}A_{\nu} - \nabla_{\nu} A_{\mu}$ is the Maxwell tensor. Here, the dual $\,^{\ast}F^{\mu\nu} \equiv 
\frac{1}{2}\epsilon^{\mu\nu\rho\sigma}F_{\rho\sigma}$ and we use the definition $\epsilon^{\mu\nu\rho\sigma}\equiv
\frac{1}{\sqrt{-g}}E^{\mu\nu\rho\sigma}$ where $E^{\mu\nu\rho\sigma}$ is
the totally anti-symmetric Levi-Civita symbol with $E^{0123}=1$.
The quantities $k_{\rm a}, k_{\rm s}$ are constants.
Depending on the parity transformation of the 
(pseudo)scalar, coupling to the Maxwell sector is realized through ${\bf E}\cdot {\bf B}$ (pseudo-scalar, setting 
$k_s=0$) or ${\bf E}^2- 
{\bf B}^2$ (scalar, setting $k_a=0$) invariant. 
We set $k_s=0$, but the extension is trivial~\cite{Ikeda:2019fvj,Boskovic:2018lkj,Sen:2018cjt}.
We get the following equations of motion for the theory above:
%
%\begin{widetext}
\begin{subequations}
\label{eq:MFEoMgen2}
\begin{align}
\label{eq:MFEoMScalar2}
&\left(\nabla^{\mu}\nabla_{\mu} - \mu^{2}_{\rm S} \right) \Psi =\frac{k_{\rm a}}{2} \,^{\ast}F^{\mu\nu} F_{\mu\nu}\,,\\
\label{eq:MFEoMVector2}
&\nabla^{\nu} F_{\mu\nu} = - 2 k_{\rm a} \,^{\ast}F_{\mu\nu} \nabla^{\nu} \Psi\,.
\end{align}
\end{subequations}
For any astrophysically interesting scenario, the matter content around BHs is not significant, and as we explained
one can take the background Kerr geometry as a good approximation.

It is important to note that the equations of motion are solved by $A_\mu=0, \Psi=0$ but also by $A_\mu=0$ and $\Psi$
described by a superradiant state around Kerr. Thus, one expects that small scalar fluctuations initially grow through 
the superradiant instability. Although $A_\mu=0$ is always a solution, it is an unstable one. This can be seen most 
easily in a Minkowski background and for a uniform, coherent oscillating background axion state of amplitude $\Psi_0$. 
A 
simple analysis shows that the equation of motion for the vector is of Mathieu type, and that $A_\mu=0$ is an unstable 
state. There are unstable modes at frequency $\omega=n\mu_S/2$, with $n \in \mathbb{N}$. At the lowest 
frequency, the 
vector grows at a rate $A_\mu\sim e^{\lambda t}$ with $\lambda =k_a\mu_{\rm S}\Psi_0/2$.

Consider now a scalar cloud of finite extent around a spinning BH. The flat space analysis may be expected to hold, 
provided that the photon produced through the instability interacts before exiting the cloud. Thus, the cloud 
lengthscale $1/(M\mu_{\rm S}^2)$ (Section~\ref{sec:cloudproperties}) now controls
the process and introduces a cutoff coupling $k_a^{\rm crit}$ below which no instability develops. The threshold 
coupling constant was found to be~\cite{Ikeda:2019fvj,Boskovic:2018lkj}
\be\label{eq:EMinst_crit}
k_a^{\rm crit}\gtrsim 2\left(\frac{M}{M_S}\right)^{1/2}(M\mu_{\rm S})^{-2}\,,
\ee
where $M_S$ is the total mass in the scalar cloud and $M$ the BH mass.

\begin{figure}[tb]
\begin{center}
\begin{tabular}{c}
\epsfig{file=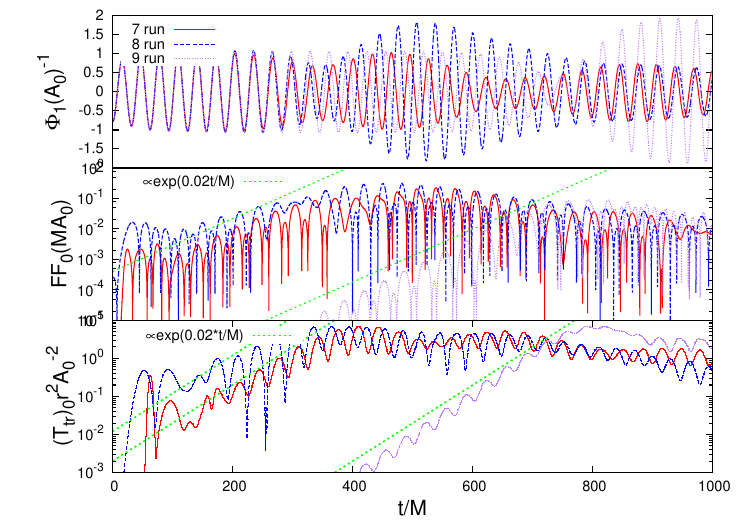,width=10cm,angle=0,clip=true}
\end{tabular}
\caption{
Time evolution of a small vectorial fluctuation around an axionic cloud of amplitude $A_0$ (the overall amplitude in 
\eqref{eigenfunctionDetweiler}). Here, $\Phi_{1}$ (upper panel) is the dipolar component of the axion, $FF_{0}$ (middle 
panel) is the monopolar component of the Maxwell invariant $F_{\mu\nu}F^{\mu\nu} $at $r=20M$. 
We also show the energy flux (bottom panel) for an axion with mass $M\mu_{\rm S}=0.2$ around a BH with $a=0.5M$. The 
coupling constant is super-critical with $k_{\rm a}A_0=0.3$.
The initial EM profile is described by an amplitude and width $(E_0/A_0,w/M)=(10^{-3},5),(10^{-3},20),(10^{-4},5)$ for 
run $7, 8, 9$ respectively, and $r_0=40$. The overall behavior and growth rate of the instability
at large time scales are insensitive to the initial conditions. Taken from Ref.~\cite{Ikeda:2019fvj}.
\label{Kerr_axion}}
\end{center}
\end{figure}
These predictions describe well the numerical simulations. As an example, Fig.~\ref{Kerr_axion} shows
the evolution of a small vector fluctuation around a scalar cloud. For different initial disturbances,
the growth rate is very similar, and agrees with the simple estimate above.

The vector instability occurs as a burst of radiation, that decreases the axion amplitude, lowering it to sub-critical 
values. On a long, superradiant time scale, the scalar is replenished and further, periodic bursts occur.
In other words, the coupling between the axion and the Maxwell sector effectively limits the maximum amount of mass in 
the scalar cloud. For sufficiently large coupling constants, cloud growth can be enormously suppressed. This is a generic consequence of couplings to other channels, as we will see below. For effects of plasma on this mechanism, see Refs.~\cite{Sen:2018cjt,Boskovic:2018lkj}.

%%%%%%%%%%%%%%%%%%%%%%%%%%%%%%%%%%%%%%%%%%%%%%%%%%%%%%%%%%%%%%%%%%%%%%%%%%%%%%%%%%%%%%%%%%%
\subsubsection{Nonlinear self-interactions and ``bosenova'' explosions}\label{sec:bosenova}
%%%%%%%%%%%%%%%%%%%%%%%%%%%%%%%%%%%%%%%%%%%%%%%%%%%%%%%%%%%%%%%%%%%%%%%%%%%%%%%%%%%%%%%%%%%

In the same way that the field $\Psi$ interacts with the Maxwell field through an axionic coupling, non-perturbative 
effects can also produce self-interaction terms. A widely used theory considers QCD axions with periodic potential
\beq
\label{eq:MFaction_bosenova}
{\cal L}&=&\frac{R}{k}- \frac{1}{2} g^{\mu\nu} \partial_{\mu}\Psi\partial_{\nu} \Psi-U(\Psi) \,,\\
U(\Psi)&=& f_a^2 \mu^2\left[1-\cos\left(\frac{\Psi}{f_a}\right)\right] \,, \label{axion_pot}
\eeq
parameterized by a mass $\mu$ and coupling $f_a$, the axion decay 
constant~\cite{Kodama:2011zc,Yoshino:2012kn,Marsh:2015xka}. 
The axion decay constant depends on the model, in some models it is of the order of the GUT scale, $f_a\approx 
10^{16}{\rm GeV}$.
Note that at small amplitudes, $U\sim \mu^2\Psi^2/2$ and one recovers the theory of a simple minimally coupled massive 
field.

The equation of motion for the scalar is,
\be
\nabla^{\mu}\nabla_{\mu}\Psi - \mu^{2}f_a\sin\left(\Psi/f_a\right)=0\,.
\ee
When the scalar is small, the superradiant instability proceeds as it does for a minimally coupled massive field.
However, when $\Psi^2\sim 6 f_a^2$, the next term in the Taylor expansion of the sin becomes comparable to the mass 
term, and new effects appear. In particular, modes carrying angular momentum down the horizon and off the cloud become 
relevant.
At the level of the equation of motion, such phenomenology is similar to the previous discussion on couplings to the 
Maxwell sector:
new physics appears each time one term in the equations of motion becomes comparable to the others.
Thus, one can anticipate that nonlinear effects become important whenever the mass in the cloud $M_S$ (see 
Section~\ref{sec:cloudproperties}) is of order
\be\label{eq:crit_bosenova}
M_S\sim 192\pi M\frac{f_a^2}{(M\mu)^4}\,.
\ee
Here, one assumes starting with fundamental $l=m=1$ mode and evaluates the field at its peak, according to 
Eq.~\eqref{amplitude}. This prediction should be compared with simulations done for $M\mu=0.4$ around a highly spinning 
BH~\cite{Kodama:2011zc,Yoshino:2012kn}, which indicate a dependence $M_S\approx 1600(f_a/m_{\rm P})^2 M$, with $m_{\rm 
P}$ being the Planck mass~\cite{Yoshino:2012kn,Kodama:2011zc,Yoshino:2014wwa}. The dependence on the coupling parameter 
$M\mu$ is untested thus far. This process was dubbed ``bosenova'' in analogy with a similar phenomenon occurring in 
condensed-matter systems. For sufficiently strong self-interactions ($f_a\ll m_{\rm P}$) bosenovae can happen during the 
superradiant growth and before extracting all the BH spin as allowed by the superradiant condition. For example, if 
$f_a$ corresponds to the GUT scale, $f_a\approx 10^{16}{\rm GeV}$, the bosenova occurs when $M_S\gtrsim 0.16 M$. As 
shown in the evolutions presented Sec.~\ref{sec:evolution}, the scalar cloud can typically attain such fraction of the 
BH mass under conservative assumptions (cf. Fig.~\ref{fig:evolution} and Ref.~\cite{Brito:2014wla}), and therefore the 
effects of bosenovae can have interesting phenomenological applications.
During the bosenova, a fraction of the cloud energy is absorbed by the BH, whereas the rest is emitted in a GW burst 
(see Sec.~\ref{sec:GWs}), leaving just a small fraction of the cloud bound to the 
BH~\cite{Yoshino:2012kn,Kodama:2011zc,Yoshino:2014wwa,Arvanitaki:2010sy}. 
This reduces the size of the cloud and the effects of nonlinearities. After the first collapse, the cloud is replenished 
through superradiance until the next bosenova possibly occurs (assuming the conditions are such that nonlinearities can 
become important before superradiant extraction is exhausted). This superradiance-bosenova cycle repeats until all 
available BH spin is exhausted. Thus, at variance with annihilation and level transition, the signal from bosenova 
explosions is a \emph{periodic} emission of bursts, whose separation depends on the fraction of the cloud which remains 
bound to the BH after each subsequent collapse. 

An analytical or semi-analytical description of superradiant growth in the presence of nonlinearities is challenging, but necessary to understand the generality of bosenova-like processes;
a recent renormalization group analysis~\cite{Omiya:2020vji} suggests that indeed, relativistic axion clouds, an explosive bosenova may be inevitable. This analysis was complemented with a detailed study of quartic interactions~\cite{Baryakhtar:2020gao}.
%\begin{flushleft}
%\end{flushleft}

Finally, Ref.~\cite{Mocanu:2012fd} have modeled the dynamics of the axion cloud by a simple cellular
automaton, showing that the process exhibits self-organized criticality.
%%%%%%%%%%%%%%%%%%%%%%%%%%%%%%%%%%%%%%%%%%%%%%%%%%%%%%%%%%%%%%%%%%%%%%%%%%%%%%%%%%%
\subsubsection{Plasma-triggered superradiant instabilities}\label{plasma-triggered}
%%%%%%%%%%%%%%%%%%%%%%%%%%%%%%%%%%%%%%%%%%%%%%%%%%%%%%%%%%%%%%%%%%%%%%%%%%%%%%%%%%%
Since the early days of BH superradiance it was speculated that plasma could provide a natural confining mechanism for 
low-frequency photons~\cite{teukolskythesis,Press:1972zz}. Indeed, even standard 
photons interacting with the plasma acquire an effective mass given (in Planck units) by the plasma 
frequency~\cite{Sitenko:1967,Kulsrud:1991jt}
\begin{equation}
\omega_p=\sqrt{\frac{4\pi e^2 n}{m_e}}\,,\label{plasma_freq}
\end{equation}
where $n$ is the electron density and $m_e$ and $e$ are the electron mass and charge, respectively. 
% As a consequence of the modified dispersion relation, Maxwell equations within the plasma in flat spacetime read

Consider a spinning BH surrounded by a plasma. If the total mass of the surrounding matter is sufficiently small, its 
gravitational backreaction is negligible and the background spacetime is uniquely described by the Kerr metric.
In order to incorporate the modified dispersion relation in an effective way, Maxwell equations within the plasma could be modified as~\cite{Kulsrud:1991jt}:
%%%
\begin{equation}
\nabla_\sigma F^{\sigma\nu}=\omega_p^2 A^\nu\,.\label{procaplasma}
\end{equation}
%%%
% The equation above is also valid in curved spacetime as long as the background is slowly varying compared to $\omega_p^{-1}$ and the density gradient is small compared to the gravitational field~\cite{Kulsrud:1991jt}.
This description is valid for a hot and ultrarelativistic plasma, whereas in the cold-plasma approximation (typically more relevant in accretion disks) Maxwell equations take a more complicated form~\cite{Cannizzaro:2020uap}.
When the plasma density is constant and homogeneous, Eq.~\eqref{procaplasma} coincides with Proca equation~\eqref{proca}, where the plasma frequency can be identified with the mass of the vector field. More generically, the plasma density might have a nontrivial radial and angular profile. In this case the instability can be investigated semi-analytically by using the methods developed in Refs.~\cite{Pani:2012vp,Pani:2013pma,Dima:2020rzg} or by a fully numerical analysis.

In the Proca-like approximation and at the linearized level, superradiant instabilities triggered by plasma were 
analyzed in Ref.~\cite{Pani:2013hpa}, where it was shown that they are relevant for small primordial BHs in the early 
universe (see also Sec.~\ref{sec:astro_plasma}). It has also been suggested that plasma-induced 
superradiant instabilities could explain observations of ``fast radio bursts''~\cite{Conlon:2017hhi}. A numerical 
investigation of superradiant instabilities triggered by plasmas and accretion disks was recently performed in 
Ref.~\cite{Dima:2020rzg}.

However, the above studies are based on a number of strong assumptions that have been recently challenged.
In addition to the breakdown of the Proca approximation~\cite{Cannizzaro:2020uap}, an important caveat with any claims 
about the role of plasmas in superradiance concerns backreaction and nonlinear effects. A growing EM field will easily 
change the plasma distribution around a compact object, and potentially even blow the plasma away from its vicinity; 
likewise, the full Maxwell's system includes nonlinear corrections in the electric field amplitude which can 
make the plasma transparent for some frequencies~\cite{1970PhFl...13..472K,1971PhRvL..27.1342M,Cardoso:2020nst}. 
The impact of these effects on the development of superradiant instabilities was recently studied, and shown to be important~\cite{Cardoso:2020nst}. In particular, nonlinear effects
make the plasma transparent and can effectively quench the instability growth.
%%%%%%%%%%%%%%%%%%%%%%%%%%%%%%%%%%%%%%%%%%%%%%%%%%%%%%%%%%%%%%%%%%%%%%%%%%
\subsection{Superradiance in scalar-tensor theories}\label{sec:nonminimal}
%%%%%%%%%%%%%%%%%%%%%%%%%%%%%%%%%%%%%%%%%%%%%%%%%%%%%%%%%%%%%%%%%%%%%%%%%%
Nonminimal couplings can produce effective mass terms in the perturbation equations and confine radiation, thus giving rise to superradiant instabilities akin to the ones discussed above for massive bosonic fields. Here we discuss superradiant instabilities in a modified theory of gravity, namely scalar-tensor theories.

%%%%%%%%%%%%%%%%%%%%%%%%%%%%%%%%%%%%%%%%%%%%%%%%%%%%%%%%%%%%%%%%%%%%%%%%%%%%%%%%%
\subsubsection{Spontaneous superradiant instabilities in scalar-tensor theories}
%%%%%%%%%%%%%%%%%%%%%%%%%%%%%%%%%%%%%%%%%%%%%%%%%%%%%%%%%%%%%%%%%%%%%%%%%%%%%%%%%
As discussed in Sec.~\ref{sec:super_ST}, the presence of matter may drastically affect the superradiant amplification of scalar waves in scalar-tensor theories~\cite{Cardoso:2013fwa,Cardoso:2013opa}. Indeed, the Klein-Gordon equation for a massless scalar field acquires an effective, spacetime-dependent mass term $\mu_{\rm eff}$ proportional to the trace of the stress-energy tensor. 

When $\mu_{\rm eff}^2>0$, a ``spontaneous superradiant instability'' might be present for rotating BHs, similarly to the case of massive Klein-Gordon fields previously discussed. 
Focusing on separable solutions of the Klein-Gordon equation with $\Phi=\Psi(r)S(\vartheta)e^{-i\omega t+im\varphi}$, Refs.~\cite{Cardoso:2013fwa,Cardoso:2013opa} found that if the (trace of the stress-energy tensor of the) matter profile has the general form
\be
T(r,\vartheta)\sim 2\frac{{\cal F}(\vartheta)+{\cal G}(r)}{a^2+2r^2+a^2\cos2\vartheta}\,,\label{separable2}
\ee
then the scalar acquires an effective mass $\mu_{\rm eff}^2\sim \mu_0^2+T$, {\it and} the Klein-Gordon equation is separable, where
$\mu_0$ is the original, constant, mass of the scalar~\cite{Cardoso:2013fwa,Cardoso:2013opa}. In this case, the scalar perturbations reduce to the following coupled system of equations:
\beq
&&\frac{(\sin\vartheta S')'}{\sin\vartheta}+\left[
a^2\left(\omega^2-\mu_0^2\right)\cos^2\vartheta-\frac{m^2}{\sin^2\vartheta}-
{\cal F}+\lambda \right]S=0,\nn\\
&&\Delta 
\frac{d}{dr}\left(\Delta\frac{d\Psi}{dr}\right)+\left[
K^2-\Delta\left({\cal G}+r^2\mu_0^2+\lambda\right)\right]\Psi=0\,,\nn
\eeq
where $\Delta$, $K$ and $\lambda$ have been defined in Sec~\ref{sec:Teukolsky}, whereas $\mu_0$ is a ``bare'' mass that will be set to zero in the following, because we are interested in an effective mass term that vanishes at large distances.

\begin{center}
\begin{figure}[ht]
\begin{center}
\begin{tabular}{cc}
\epsfig{file=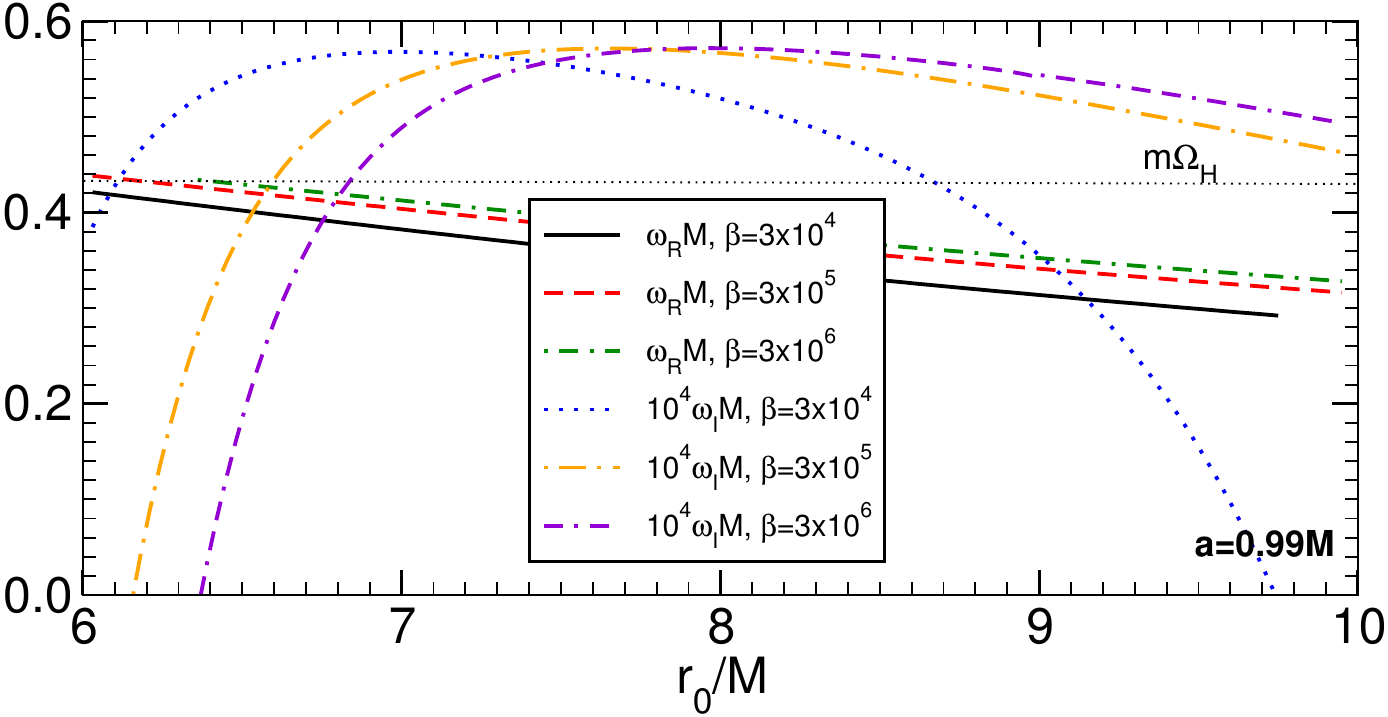,width=0.48\textwidth,angle=0,clip=true}&
\epsfig{file=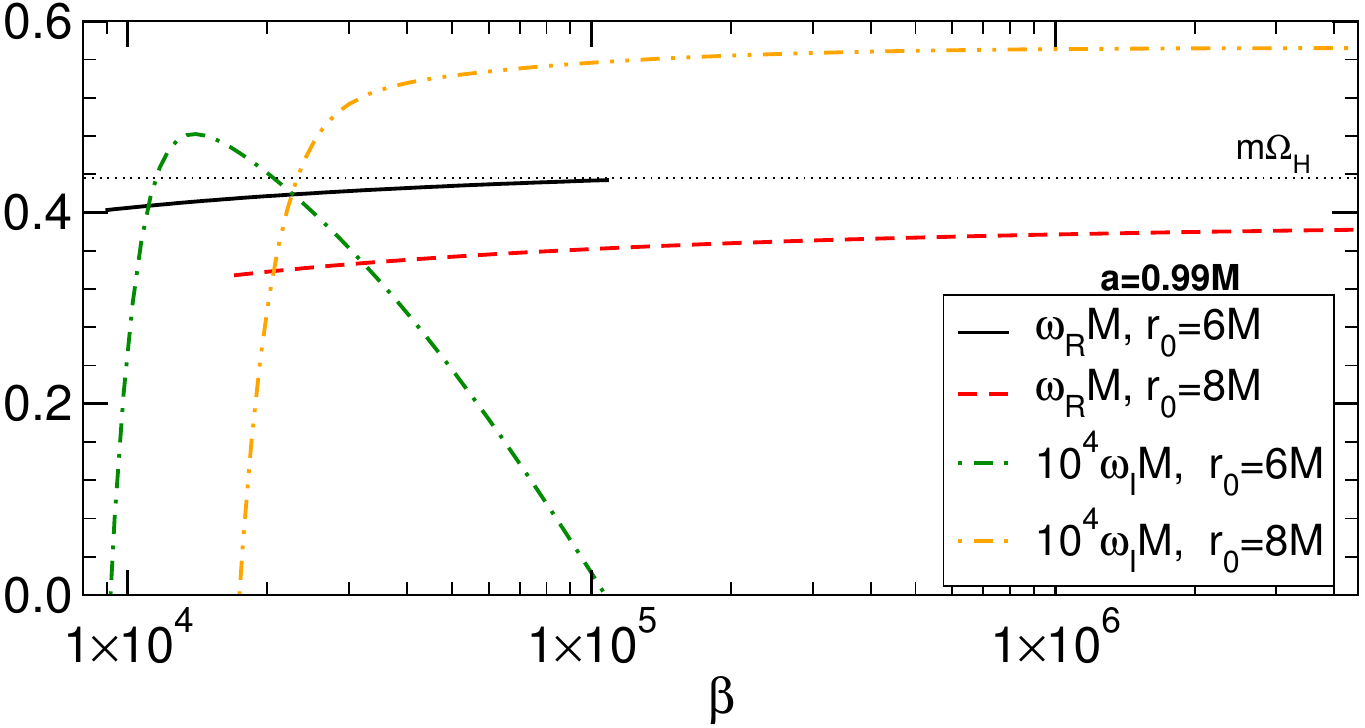,width=0.48\textwidth,angle=0,clip=true}
\end{tabular}
\end{center}
\caption{\label{fig:modelG}
Superradiant instability for a matter profile characterized by Eq.~\eqref{separable2} with $\mu_0=0$, ${\cal F}=0$ and ${\cal G}=\Theta[r - r_0]\beta(r-r_0)/r^3$, where $\beta$ parametrizes the strength of the scalar-tensor coupling. For large
$\beta$ the system behaves as a BH enclosed in a cavity with radius $r_0$. Curves are truncated when the modes become stable. From Ref.~\cite{Cardoso:2013fwa}.
}
\end{figure}
\end{center}
A representative case is summarized in Fig.~\ref{fig:modelG} for a matter profile characterized by $\mu_0=0$, ${\cal F}=0$ and ${\cal G}=\Theta[r - r_0]\beta(r-r_0)/r^3$, where $\beta$ parametrizes the strength of the scalar-tensor coupling. Even though the effective mass term vanishes at large distances, the instability is akin to the
original BH bomb, i.e. a spinning BH enclosed by a mirror located at $r=r_0$: as discussed in Sec.~\ref{sec:mirror},
for small $r_0$ there is no instability, as the natural frequencies of this system scale like $1/r_0$ and are outside the superradiant
regime. It is clear from Fig.~\ref{fig:modelG} that this is a superradiant phenomenon, as the instability is quenched
as soon as one reaches the critical superradiance threshold. At fixed large $r_0/M$, and for {\it any} sufficiently large $\beta$, the instability time scale $\omega_I^{-1}$ is roughly constant. In agreement with the simpler BH bomb system, a critical $\beta$ corresponds to a critical barrier height which is able to reflect radiation back. After this point increasing $\beta$ further is equivalent to a further increase of the height of the barrier and has no effect on the instability.

Although spontaneous superradiant instabilities seem to be a generic feature of scalar-tensor theories~\cite{Cardoso:2013opa}, so far they have been investigated only through the ansatz~\eqref{separable2}, i.e. when the equations are separable. Further investigation is necessary in order to understand realistic configurations 
such as accretion disks. In that case, methods such as those used in Ref.~\cite{Pani:2012vp,Pani:2012bp,Witek:2012tr,Dolan:2012yt} would be required.

Finally, spontaneous superradiant instabilities of Kerr-de Sitter BHs in scalar-tensor theories and the role of a positive cosmological constant were recently investigated~\cite{Zhang:2014kna}.

%%%%%%%%%%%%%%%%%%%%%%%%%%%%%%%%%%%%%%%%%%%%%%%%%%%%%%%%%%%
\subsubsection{Floating orbits} \label{sec:bounds_floating}
%%%%%%%%%%%%%%%%%%%%%%%%%%%%%%%%%%%%%%%%%%%%%%%%%%%%%%%%%%%
%
\begin{figure}[thb]
\begin{center}
\begin{tabular}{c}
\epsfig{file=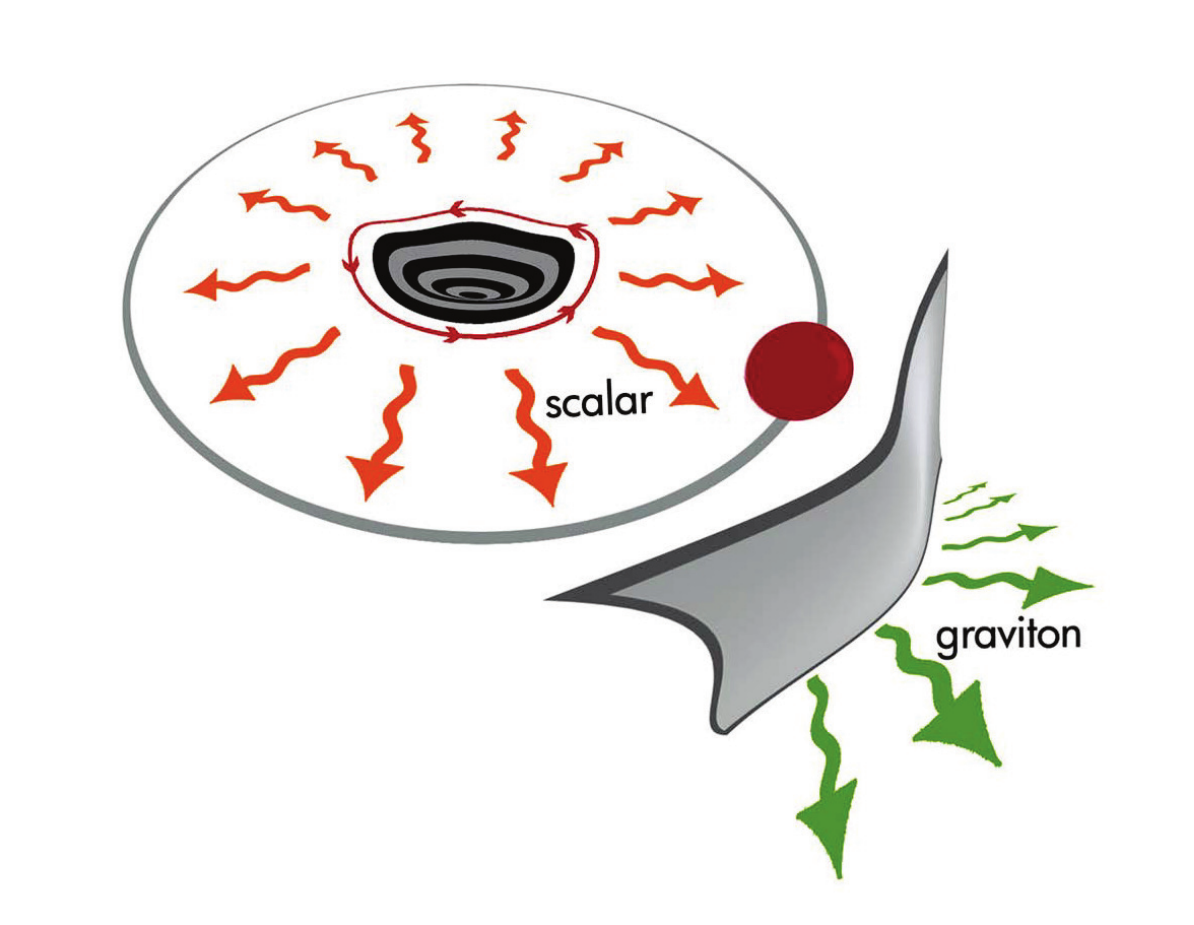,width=0.9\textwidth,angle=0,clip=true}
\end{tabular}
\caption{Pictorial description of floating orbits. An orbiting body excites superradiant scalar modes close to the BH 
horizon which are
prevented from escaping to infinity due to their being massive (represented by the gray ``wall''). Since the scalar 
field is massive, the flux at infinity consists solely of gravitational radiation. From Ref.~\cite{Cardoso:2011xi}.
\label{fig:floating1}}
\end{center}
\end{figure}
\begin{figure}[thb]
\begin{center}
\begin{tabular}{c}
\epsfig{file=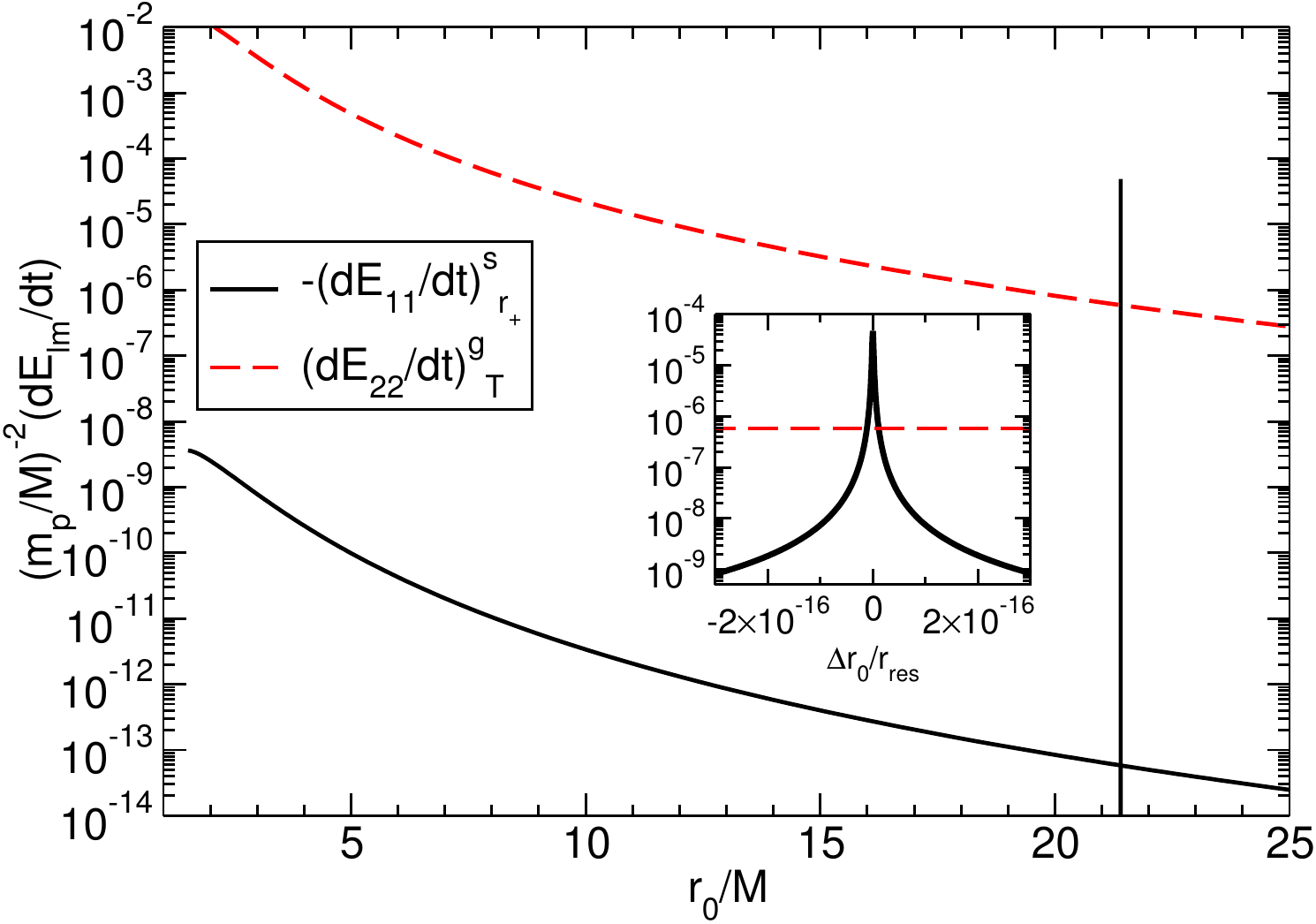,width=0.9\textwidth,angle=0,clip=true}
\end{tabular}
\caption{Dominant fluxes of scalar and gravitational energy ($l=m=1$ and $l=m=2$, respectively) for $\mu_s M=10^{-2}$, 
$\alpha=10^{-2}$ and $a=0.99M$. The inset is a zoom around resonance. From Ref.~\cite{Cardoso:2011xi}.
\label{fig:floating2}}
\end{center}
\end{figure}
Although not directly related to superradiance --~when the bosonic field is coupled to matter~-- new effects related to 
stimulated emission may be triggered, modifying the inspiral dynamics of compact binaries~\cite{Cardoso:2011xi,Yunes:2011aa,Alsing:2011er,Fujita:2016yav}. Figure~\ref{fig:floating1} illustrates one such process: a point particle of mass $m_p$ orbits a supermassive BH on a quasi-circular orbit of Boyer-Lindquist radius $r_0$.
The point particle is coupled to a scalar field through the trace of its stress-energy tensor ${\cal T}$, yielding the 
equation of motion for the scalar field,
\be
\left[\square-\mu_s^2\right]\Phi=\alpha {\cal T}\label{EQBD2b}\,.
\ee
the coupling $\alpha$ is related to the specific theory under consideration~\cite{Cardoso:2011xi,Cardoso:2013fwa}.
Within a perturbation framework, for small masses $m_p$ the scalar field $\Phi$ is small and its backreaction in the 
geometry can be neglected. In other words, the particle follows a geodesic in the spacetime of a rotating BH, emitting scalar 
and GWs of frequency proportional to the orbital frequency of the circular geodesic.

The power emitted as gravitational radiation can be estimated with the use of the quadrupole formula to be
$\dot E^{g}_\infty={32}/{5}\left({r_0}/{M}\right)^{-5}m_p^2/M^2$. This is the power at spatial infinity
in the low-frequency regime, the flux through the horizon being negligible for large orbital radii.
The scalar flux at infinity can be computed in the low-frequency regime,
\be
\dot E^{s}_{\infty}=\frac{\alpha^2M^2}{12\pi}\frac{\left(1-\mu_s^2r_0^3/M\right)^{3/2}}{r_0^4}m_p^2
\Theta(\Omega_p-\mu_s)\,,\label{dipole}
\ee
where $\Theta(x)$ is the Heaviside function. As expected, for orbital radii large enough that the orbital frequency 
$\Omega_p<\mu_s$, scalar radiation is quenched at spatial infinity.
However, we learned in Section~\ref{sec:mass} that the Kerr spacetime admits the existence of {\it superradiant} 
resonances at
\be
\omega_{\rm res}^2=\mu_s^2-\mu_s^2\left(\frac{\mu_s M}{l+1+n}\right)^2\,,\quad n=0,1,...\label{om_resonance}
\ee

Thus, one might expect enhanced scalar flux {\it at the horizon} close to these resonances. Indeed, 
Fig.~\ref{fig:floating2} shows that the flux of (scalar) energy
at the horizon is greatly enhanced close to these resonances. We can estimate the peak flux close to the resonant 
frequencies for large distances and for $l=m=1$,
\be
\dot{E}^{s, {\rm peak}}_{r_+}\sim -\frac{3 \alpha ^2  \sqrt{\frac{r_0}{M}}m_p^2 M}{16\pi 
r_+\left(M^2-a^2\right)\left(\frac{a}{2r_+}-(\frac{M}{r_0})^{3/2}\right){\cal F}}\,,
\label{flux_ana_fin2}
\ee
with ${\cal F}=1+4 P^2$. The scalar flux at the horizon {\it grows} in magnitude with $r_0$ and it is negative, due to 
superradiance, at sufficiently large distances. Eccentric but equatorial orbits do not affect the overall picture~\cite{Fujita:2016yav}.

Thus, for any $\mu_s M\ll1$, there exists a frequency $\omega_{\rm res}\lesssim\mu_s$ for which the total flux $\dot 
E^{s}_\infty+\dot E^{s}_{r_+}+\dot E^{g}_\infty+\dot E^{g}_{r_+}=0$, because the negative scalar flux at the horizon is 
large enough to compensate for the other positive contributions. These points are called {\it floating orbits}, because 
an energy balance argument suggests that at these locations the small point particle does not inspiral (neither inwards nor outwards). All the energy lost at  infinity under GWs is provided entirely by the rotational energy of the BH.
Under ideal conditions, floating would stop only when the peak of the scalar flux at the horizon is too small to 
compensate for the gravitational flux, $|\dot E^{g}|>|\dot E_\text{peak}^{s}|$.

Floating orbits are not possible in vacuum GR~\cite{Kapadia:2013kf}. Thus, they are a smoking-gun of new physics; the orbital 
frequency at which the particle stalls exactly matches the mass of the putative fundamental field, making BHs ideal self-tuned 
``particle detectors''. The existence of floating orbits manifests itself in a sizeable and detectable dephasing of GWs, with respect to pure GR waveforms~\cite{Cardoso:2011xi,Yunes:2011aa,Alsing:2011er,Barausse:2014tra,Fujita:2016yav}.

The above analysis assumes the orbiting pointlike to carry some nonzero charge, with which it interacts with the field and therefore with the spinning BH. However, purely gravitational exchange angular momentum is also possible and gives rise to similar phenomena~\cite{Ferreira:2017pth,Zhang:2018kib,Baumann:2019ztm}.

%%%%%%%%%%%%%%%%%%%%%%%%%%%%%%%%%%%%%%%%%%%%%%%%%%%%%%%%%%%%%%%%%%%%%%%%%%%
\subsection{Superradiant instability from stars}\label{sec:instabilitystar}
%%%%%%%%%%%%%%%%%%%%%%%%%%%%%%%%%%%%%%%%%%%%%%%%%%%%%%%%%%%%%%%%%%%%%%%%%%%%
As discussed in Sec.~\ref{sec:SRNS} rotating conducting stars, or stellar magnetospheres in axionic physics can amplify radiation due to superradiance. Hence, in 
analogy with the BH case, the mass term for the Proca field can lead to superradiant 
instabilities in conducting stars. The parameter space of the QNM spectrum is large and complicated, since --~even for 
fixed ``quantum'' numbers $(l,m,n)$~-- it still depends on four dimensionless parameters, namely ($\mu_V M, M/R, 
\Omega/\Omega_K$, $\sigma M$), where we remind that $M$, $R$, and $\Omega$ are the mass, radius, and angular velocity 
of the star, respectively, and $\Omega_K=\sqrt{M/R^3}$ is the Keplerian angular velocity, and $\sigma$ is the 
conductivity.

In the axial case, the numerical results for the $l=m=1$ fundamental unstable mode are well approximated in the small 
$\mu_VM$ limit and to linear order in $\Omega/\Omega_K$ by~\cite{Cardoso:2017kgn}
\begin{eqnarray}
\omega_R^2&\sim&\mu_V^2\left(1-\frac{\mu_V^2M^2}{8}\right)\,,\\
\omega_I&\sim&-\left[\alpha_1\frac{\sigma M}{\alpha_2+(\sigma M)^{3/2}}\right](\mu_VM)^8 (\mu_V-m\Omega)\,, 
\label{wIaxial}
\end{eqnarray}
where $\alpha_i$ are dimensionless constants that depend on the compactness and also on $\Omega$ since the combination 
$ 
\Omega/\sigma$ is not necessarily small. Besides the prefactor in square brackets in Eq.~\eqref{wIaxial}, the 
functional 
form of the superradiantly unstable modes is the same as that found for a 
BH~\cite{Pani:2012bp,Pani:2012vp,Rosa:2011my,Witek:2012tr}.
The dependence of the prefactor in Eq.~\eqref{wIaxial} on $\sigma$ and $M/R$ are presented in the right panel of
Fig.~\ref{fig:SRstar}, which displays a linear behavior in $\sigma$ at small conductivities and a 
$\sim\sigma^{-1/2}$ behavior at large conductivities. The dependence on the compactness is monotonic at 
small conductivities, but it is more complicated at large conductivities, in line with the aforementioned behavior of 
the amplification factor of massless fields.
A simple model that shares many features with these numerical results was presented in Ref.~\cite{Cardoso:2017kgn}.
Bounds on ultralight bosons using the superradiant instability of pulsars~\cite{Cardoso:2017kgn,Kaplan:2019ako} are 
discussed in Sec.~\ref{sec:bounds_pulsar} below. A discussion of superradiant instabilities triggered by axion-photon couplings
can be found in Refs.~\cite{Boskovic:2018lkj,Day:2019bbh} and may play an important role in magnetars~\cite{Day:2019bbh}. Although unrelated to superradiance, we should add that stars with magnetic fields
can naturally develop a non-trivial axionic profile~\cite{Garbrecht:2018akc} due to couplings with the standard model, such as the one in Section~\ref{sec:axion_coupling}.

%%%%%%%%%%%%%%%%%%%%%%%%%%%%%%%%%%%%%%%%%%%%%%%%%%%%%%%%%%%%%%%%%%%%%%%%%
\subsection{Black holes immersed in a magnetic field}\label{sec:magnetic}
%%%%%%%%%%%%%%%%%%%%%%%%%%%%%%%%%%%%%%%%%%%%%%%%%%%%%%%%%%%%%%%%%%%%%%%%%
Magnetic fields can also confine radiation and work as ``natural'' mirrors. Strong magnetic fields are believed to exist around astrophysical BHs, mainly supported by accretion disks. Realistic astrophysical BHs are in general very complex systems which involve the coupling of gravity to the surrounding accretion disk and magnetic field. However some approximate solutions have been found that can give an accurate qualitative, and in some cases quantitative, description of stationary magnetized BH solutions. 

The first approximate solution to be found describes a test uniform magnetic field in a Kerr background~\cite{Wald:1974np}. In addition to this solution, there exists a class of exact ``Ernst'' solutions of the Einstein--Maxwell equations, which describe BHs immersed in a uniform magnetic field~\cite{Ernst:1976BHM}. These solutions are not asymptotically flat. At infinity the Ernst solutions resemble another solution of the Einstein--Maxwell found by Melvin~\cite{Melvin:1963qx,Melvin:1965zza} and further studied by Thorne~\cite{PhysRev.139.B244}, describing a uniform magnetic field held together by its own gravitational pull. Much like AdS spacetime which behaves as a covariant box for perturbations (cf. Section~\ref{sec:AdS}), the Melvin solution also admits normal modes~\cite{Brito:2014nja}, because the asymptotic boundary of the Melvin solution is able to confine perturbations. The model introduced in Section~\ref{sec:model} then predicts that a rotating BH immersed in this spacetime should be superradiantly unstable.  

\begin{figure}[hbt]
\begin{center}
%\begin{tabular}{c}
\epsfig{file=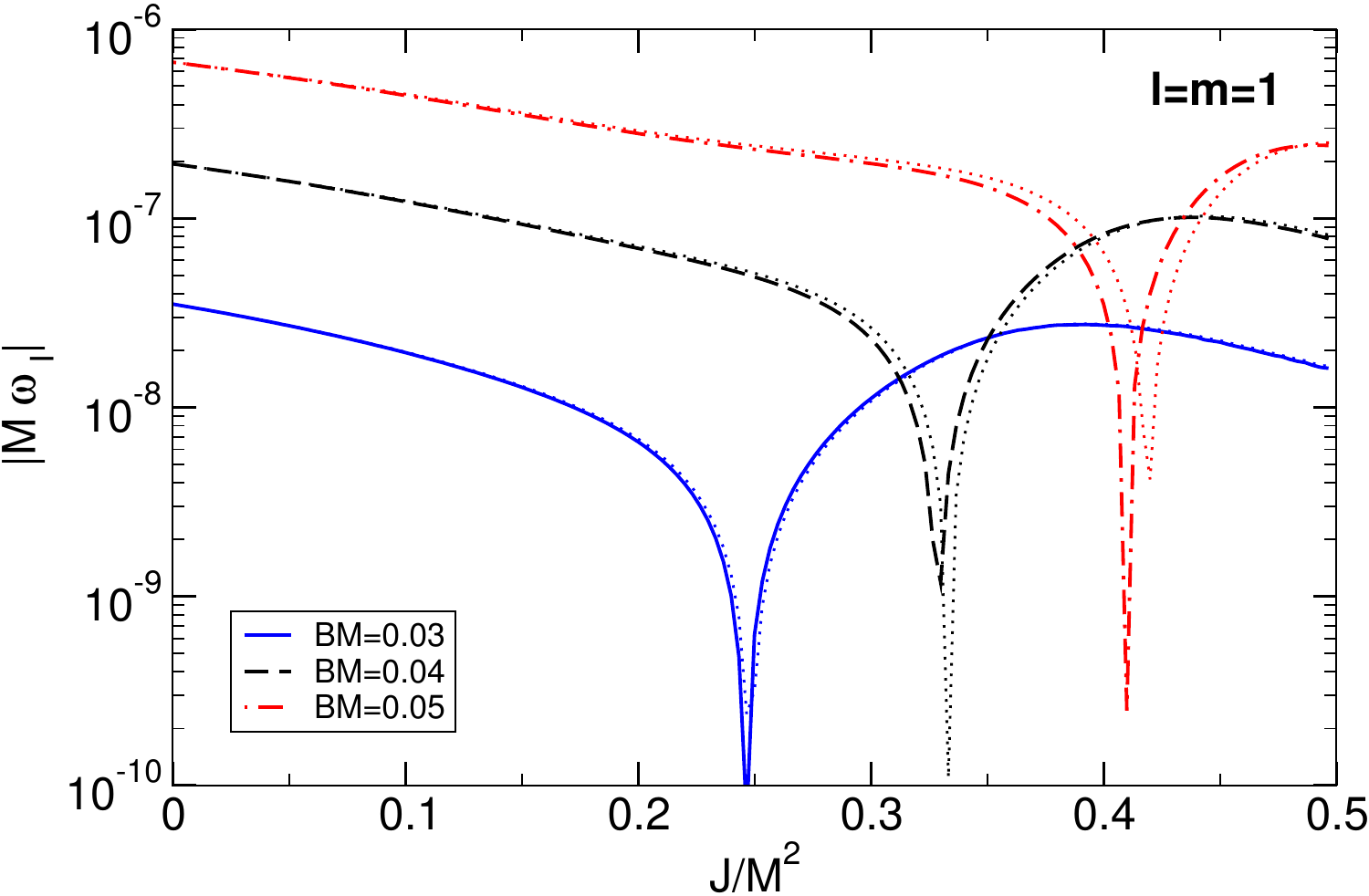,width=0.7\textwidth,angle=0,clip=true}
%\end{tabular}
\caption{Imaginary part of the fundamental modes of a Kerr--Newman--Ernst BH with Wald's charge $q=-2a M B$, computed at second order in the rotation, as a function of the BH rotation $J/M^2$, for $l=m=1$, and different values of the magnetic field. The dotted thinner lines correspond to a magnetized BH without charge $q=0$. The only effect of the charge is to change the superradiance threshold. From \cite{Brito:2014nja}.\label{Fig:super}}
\end{center}
\end{figure}

Similarly to massive vector and tensor perturbations of a Kerr background (cf. Section~\ref{sec:mass}), perturbations do 
not separate in the Ernst backgrounds. Due to this difficulty, up to date no study of the gravito-EM perturbation of 
this solution has been performed. However scalar field perturbations have been studied by several 
authors~\cite{Galtsov:1978ag,Konoplya:2007yy,Konoplya:2008hj,Brito:2014nja}. This was first done in 
Ref.~\cite{Galtsov:1978ag}, who found that in a $Br\ll 1$ expansion (with $B$ being the magnetic field strength and $r$ 
the radial coordinate, both in geometric units) the massless scalar field equation~\eqref{massiveKG} is separable and is 
equivalent to a massive scalar perturbation propagating on a Schwarzschild or Kerr metric with an effective mass 
$\mu_{\rm eff}=Bm$, where $m$ is the field's azimuthal number. This was further developed in 
Refs.~\cite{Konoplya:2007yy,Konoplya:2008hj} who showed that the magnetic field triggers the same superradiant 
instability associated to massive fields. However, this approximation becomes inaccurate at distances comparable to or 
larger than $\sim 1/B$.
To handle the problem of non-separability, Ref.~\cite{Brito:2014nja} used a slow-rotation approximation (cf. Section~\ref{ref:proca}) and methods introduced in Ref.~\cite{Dolan:2012yt} to study in full detail scalar perturbations of the Ernst solutions without any approximation in the magnetic field strength $B$. In particular, they studied perturbations around the most generic of these solutions, a magnetized version of the Kerr-Newman metric, and found that in this background, the mode spectrum reads
\beq
&&\omega_R\sim \left[0.75n+1.2m+0.25l+0.7\right]B+{\cal O}(B^3)\,,\\
&&\omega_I M\sim \gamma\left(a m/M-\frac{2\omega_R r_+}{1+8B^2M^2-16B^4M^4}\right)\left(BM\right)^{2(l+1)}\label{wI_magnetic}\,.
\eeq
This estimate was computed including Wald's result for the charge induction~\cite{Wald:1974np} caused by the  magnetic field, which implies that to have a vanishing total electric charge at infinity a rotating BH should acquire a non-zero charge $q=-2a M B$. It is clear that the instability time scale can be orders of magnitude smaller than the one estimated using the $Br\ll 1$ approximation of Refs.~\cite{Galtsov:1978ag,Konoplya:2007yy,Konoplya:2008hj}, in terms of an effective mass $\mu_{\rm eff}=Bm$ (cf. Eq.~\eqref{omegaDetweiler}). An example of the instability growth rate for the Kerr--Newman--Ernst BH is shown in Fig.~\ref{Fig:super}. 

The model presented in Section~\ref{sec:model} suggests that magnetized Kerr--Newman BHs should also be unstable against 
gravito-EM perturbations. The same model predicts that the instability growth rate should follow the same scaling as 
scalar perturbations~\eqref{wI_magnetic}. Moreover, since superradiant extraction is more efficient for gravitational 
and EM perturbations (cf. Section~\ref{sec:super_anavsnum}) we expect them to trigger a slightly stronger 
instability. This generic instability of BHs surrounded by magnetic fields can be used to impose intrinsic limits on the 
strength of magnetic fields around rotating BHs as we discuss in more detail in Section~\ref{sec:bounds_magnetic}.

%

%%%%%%%%%%%%%%%%%%%%%%%%%%%%%%%%%%%%%%%%%%%%%%%%%%%%%%%%%%%%%%%%%%%%%%%%%%%%%%%%%%%%%%
\subsection{Superradiant instability of black holes surrounded by conducting rings}\label{sec:rings}
%%%%%%%%%%%%%%%%%%%%%%%%%%%%%%%%%%%%%%%%%%%%%%%%%%%%%%%%%%%%%%%%%%%%%%%%%%%%%%%%%%%%%%

%
\begin{figure}
\begin{center}
\begin{tabular}{cc}
\epsfig{file=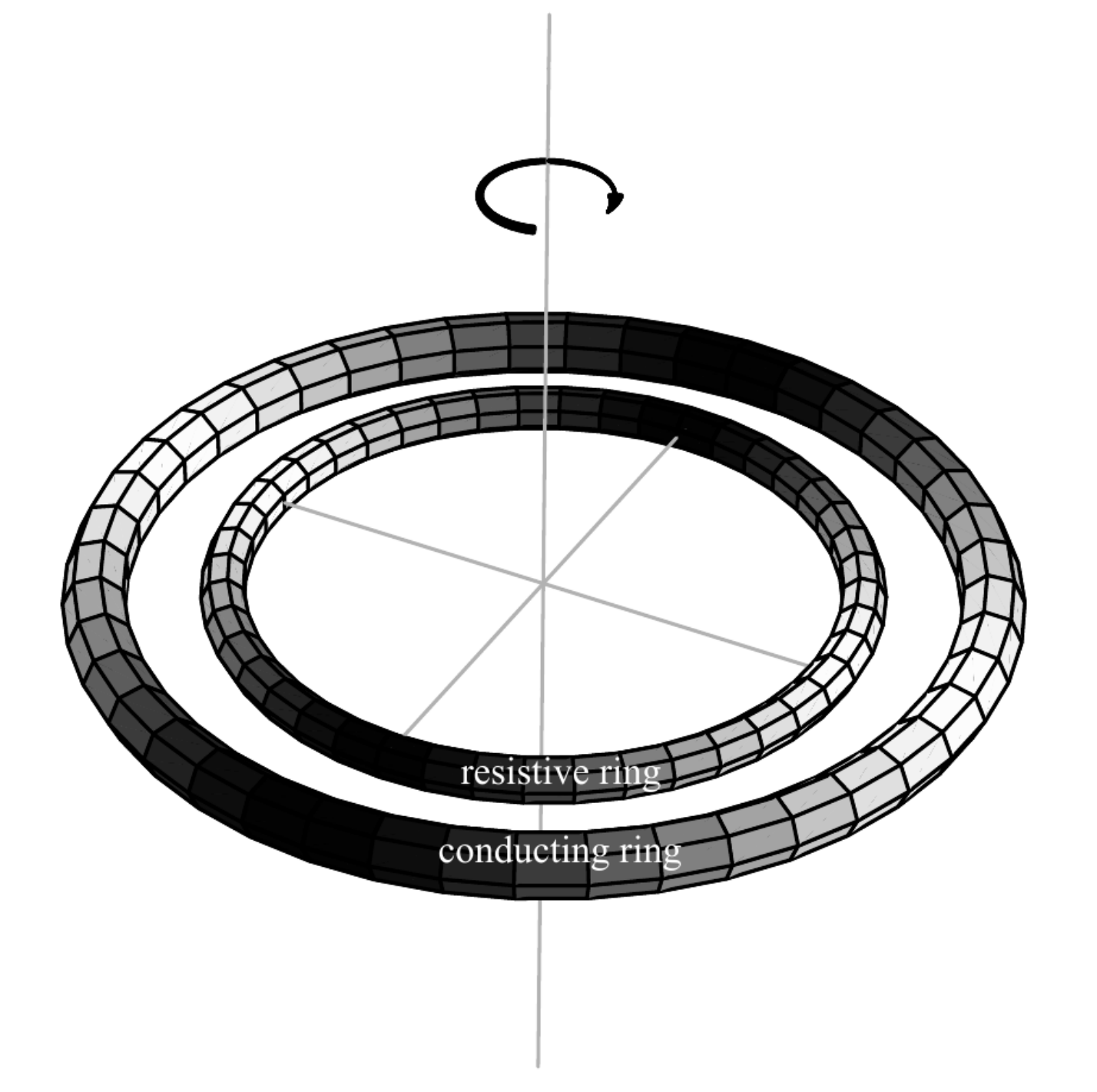,width=0.495\textwidth,angle=0,clip=true}&
\epsfig{file=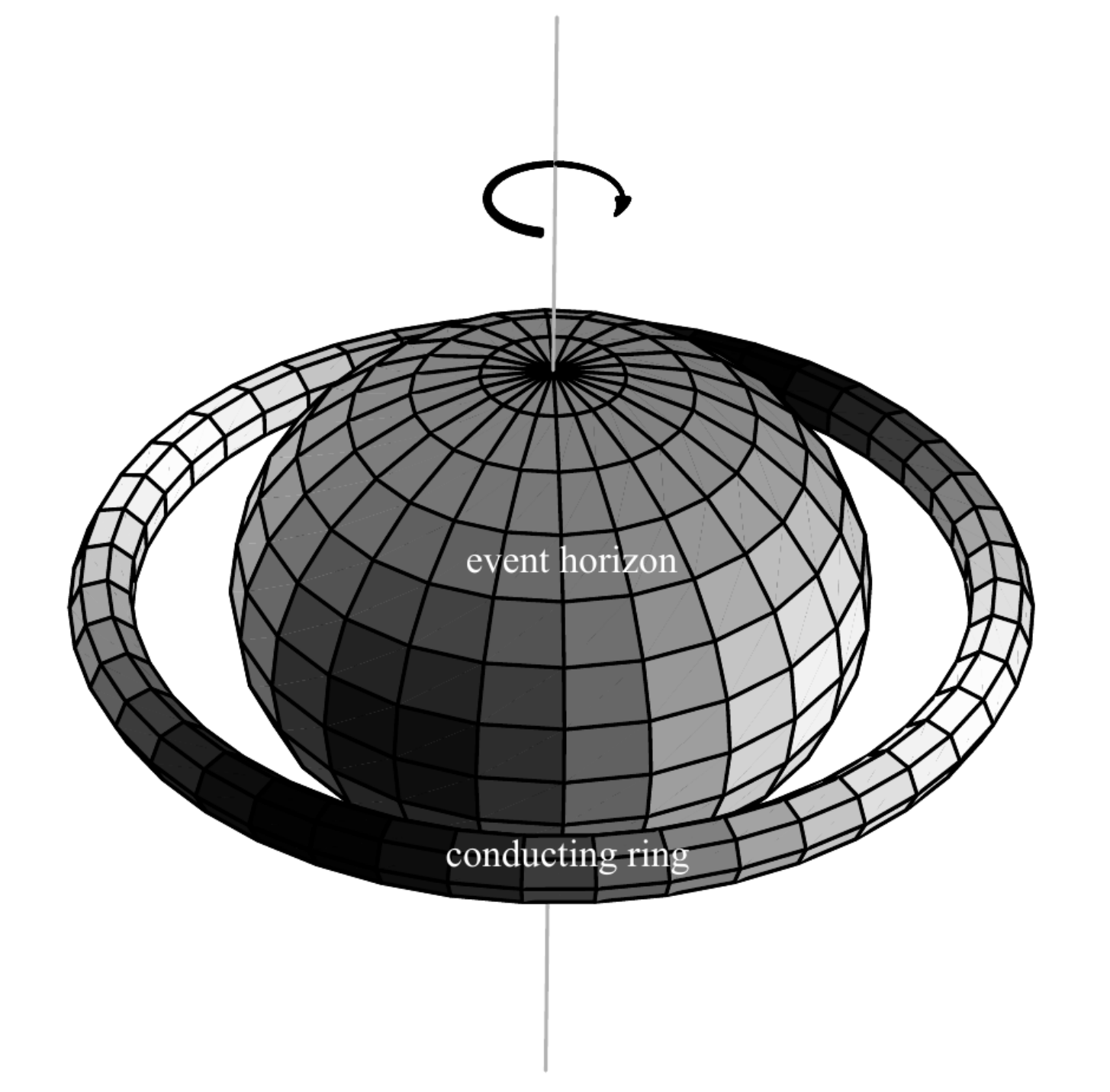,width=0.495\textwidth,angle=0,clip=true}
\end{tabular}
\caption{Left: Table-top model for superradiant amplification by two conducting rings. The inner resistive ring rotates at relativistic speed, whereas the outer ring is a conductor and might be nonrotating. Rotational energy is extracted from the resistive ring and may be larger than radiative losses to infinity, yielding exponential growth of the stored field energy. 
Shading shows schematically the location of positive and negative charge in an $m=2$ unstable mode. 
Right: the conjectured BH analog of the table-top model, where the resistive rotating ring is replaced by a Kerr BH.
Shading shows the charge density on the ring, and the image charge density on the horizon. 
From Ref.~\cite{PressRing}.
\label{fig:ring}}
\end{center}
\end{figure}

An interesting toy model of superradiant-triggered energy extraction in astrophysical systems was proposed by Press~\cite{PressRing}. As depicted in Fig.~\ref{fig:ring}, the model consists of two coaxial rings, the inner of which is resistive and rotates around the common axis of symmetry, whereas the outer one is a conductor (which we take to be nonspinning for simplicity). The astrophysical analog of such toy model (right panel of Fig.~\ref{fig:ring}) is obtained by replacing the inner ring by a Kerr BH, the event horizon playing the role of the rotating resistor\footnote{The membrane paradigm assigns an electrical resistance of $\sim 377 \,{\rm Ohm}$ to the horizon~\cite{MembraneParadigm}.}.

In the two-ring model, the electric field is computed by solving Maxwell equations in terms of retarded potentials~\cite{PressRing,Jackson}. The key point of the derivation is to recognize that Ohm's law $J=\sigma E_\varphi$ (where $J$, $\sigma$ and $E$ are the electric current on the ring, the conductivity and the $\varphi$ component of the electric field, respectively) must be applied in the matter rest frame of each ring element. Using Lorentz transformations into the inner rotating ring frame yields
\begin{equation}
 {J_1'}^{\hat\varphi}=\sigma {E'}^{\hat \varphi} \quad \to \quad \gamma\left(1-\frac{m\Omega}{\omega}\right)J_1=\sigma E_1^{\hat{\varphi}}\,,
\end{equation}
where a prime denotes the ring rest frame, the hatted index is the orthonormal tetrad component~\cite{PressRing}, $\gamma$ is the Lorentz factor associated to the inner ring angular velocity $\Omega$, i.e. $\gamma=(1-v^2)^{-1/2}$ where $v$ is the linear velocity. Note the superradiant factor emerging in the equation above when Ohm's law is written in the inertial frame.

%%%%%%%%%%%%%%%%%%%%%%%%%%%%%%%%%%%%%%%%%%%%%%%%%%%%%%%%%%%%%%%%%%%%%%%%%%%%%%%%%%%%%
\subsection{Kaluza-Klein mass: superradiant instabilities in higher dimensions}
%%%%%%%%%%%%%%%%%%%%%%%%%%%%%%%%%%%%%%%%%%%%%%%%%%%%%%%%%%%%%%%%%%%%%%%%%%%%%%%%%%%%%

For higher-dimensional BH spacetimes, instabilities are the rule rather than the exception. For example, black strings and black branes are unstable against long wavelength modes along the flat dimension. This is known as the Gregory-Laflamme instability~\cite{Gregory:1993vy,Gregory:1994bj} (see also Sec.~\ref{sec:mass} for the relation between this instability and the instability of the Kerr BH family against massive spin-2 fluctuations~\cite{Babichev:2013una,Brito:2013wya}). As another example, for $D\geq 6$ dimensions where no upper bound on the rotation of Myers-Perry BHs exists, a Gregory-Laflamme-like instability renders ultra-spinning BHs unstable~\cite{Emparan:2003sy,Dias:2009iu,Dias:2010eu,Dias:2010maa,Dias:2011jg}. 

Besides these instabilities, spinning black branes in $D=d+n$ dimensions (and black strings for the particular case $n=1$) are unstable against massless bosonic fields due to the superradiant instability when $d=4$~\cite{Cardoso:2004zz,Cardoso:2005vk,Ishibashi:2015rya}. Spinning black branes in $D=4+n$ have the form
\be\label{black_brane}
ds^2=ds^2_{\rm Kerr}+dx^jdx_j\,, \quad (j=1,2,\ldots,n)\,,
\ee
where Kerr stands for the 4$D$ Kerr geometry given in~\eqref{metricKerrLambda} with $\Lambda=0$. With the ansatz $\Psi=e^{-i\omega t+im\varphi+i\mu_jx^j}S_{0lm}(\vartheta)\psi(r)$,
the {\it massless} Klein-Gordon equation~\eqref{massiveKG} in the background~\eqref{black_brane} results in the decoupled Teukolsky equations for a scalar field with effective mass $\mu_S^2\equiv \sum_i \mu_i^2$. It can be shown that, more generally, certain gravitational fluctuations are also governed by the massive Klein-Gordon or Maxwell equations~\cite{Ishibashi:2015rya}.
Thus the propagation of  massless fields around the geometry above is equivalent to the propagation of a massive field in the vicinity of the 4$D$ Kerr BH, the mass of field being played by the ``Kaluza-Klein'' momenta along the flat dimensions. Since Kerr BHs are unstable against massive bosonic fields, the black brane~\eqref{black_brane} is also unstable. Surprisingly, this is only true if $d=4$~\cite{Cardoso:2005vk}. For $d>4$ there are no  stable bound orbits for massive particles~\cite{Cardoso:2008bp}, which in terms of wave propagation means that there is no well in the the effective potential, and thus there are no (quasi)-bound states. As discussed in Section~\ref{sec:mass}, this is a fundamental property needed to trigger the superradiant instability. Similar arguments were used to show that large doubly spinning black rings in $D=5$~\cite{Dias:2006zv}\footnote{Black rings have topology $S^1 \times S^{D-3}$ unlike Myers-Perry BHs which have topology $S^{D-2}$. The first 5D black ring was found by Emparan and Reall~\cite{Emparan:2001wn,Emparan:2006mm}.} are unstable. That this geometry must be unstable was realized from the fact that in the large-radius limit they reduce to boosted Kerr black strings, which are unstable due to the reasons stated above. The superradiant instability for massive scalar fields around boosted Kerr black strings was recently studied~\cite{Rosa:2012uz}.

%%%%%%%%%%%%%%%%%%%%%%%%%%%%%%%%%%%%%%%%%%%%%%%%%%%%%%%%%%%%%%%%%%%%
\subsection{Ergoregion instability}\label{sec:ergoregioninstability}
%%%%%%%%%%%%%%%%%%%%%%%%%%%%%%%%%%%%%%%%%%%%%%%%%%%%%%%%%%%%%%%%%%%%
We argued in Section~\ref{sec:penrosegeneral} that the standard Penrose process and superradiance from spinning BHs are two distinct phenomena: the former only requires the existence of an ergoregion, whereas the latter requires the existence of a horizon. For stationary and axisymmetric BHs, this distinction is superfluous because the existence of an ergoregion implies that of a horizon (cf. proof in Sec.~\ref{sec:ERhor}). However, an interesting effect occurs for those geometries that possess an ergoregion but not a horizon: the so-called \emph{ergoregion instability}~\cite{1978CMaPh..63..243F,Sato:1978ue}. The mechanism is simple: a negative-energy fluctuation in the ergoregion is forced to travel outwards; at large distances only positive-energy states exist, and energy conservation implies that the initial disturbance gives rise to a positive fluctuation at infinity plus a larger (negative-energy)
fluctuation in the ergoregion. Repetition of the process leads to a cascading instability. The only way to prevent such cascade from occurring is by absorbing the negative energy states, which BHs do efficiently (and hence Kerr BHs are stable against massless fields), but horizonless objects must then be unstable\footnote{The only exception to this rule and argument may occur if the ergoregion extends all the way to infinity as in certain non-asymptotically flat geometries~\cite{Dias:2009ex,Dias:2012pp}; we thank \'Oscar Dias for drawing our attention to this point.}.

This instability was discovered by Friedman while studying ultracompact slowly-rotating stars with an ergoregion~\cite{1978CMaPh..63..243F,Vilenkin:1978uc,Sato:1978ue}, with subsequent work quantitatively describing 
the unstable modes for a scalar field propagating on a slowly-rotating metric in the large-$l$ limit~\cite{CominsSchutz}. 
This approach has been extended in subsequent work~\cite{1996MNRAS.282..580Y,Cardoso:2007az,Chirenti:2008pf}.
Most notably, Ref.~\cite{1996MNRAS.282..580Y} extended the analysis to the case of small multipoles $(l,m)$, finding that the instability time scale is much shorter. Finally, Ref.~\cite{Kokkotas:2002sf} studied axial gravitational modes (but again only to first order in the spin), by neglecting the coupling to polar modes that arises in the slow-rotation limit. They find that the time scale can be of the order of the seconds/minutes
depending on the compactness of the star. A discussion of these results and their connection to the CFS instability and the r-mode instability is given in Sec.~\ref{sec:astrophysics}.

However, these works are based on an initial assumption which is not fully consistent, because they consider a slowly-rotating, perfect-fluid star including some terms to second order in the rotation but neglecting others (see below). Although this approximation is expected to be reliable for very compact stars~\cite{CominsSchutz}, no consistent treatment of the ergoregion instability has been developed to date. Below, based on recent developments in the study of perturbations of slowly rotating spacetimes~\cite{Kojima:1992ie,1993ApJ...414..247K,Pani:2012bp,Pani:2013pma}, we give the first fully-consistent treatment of this problem.
%%%%%%%%%%%%%%%%%%%%%%%%%%%%%%%%%%%%%%%%%%%%%%%%%%%%%%%%%%%%%%%%%%%%%%%%%%%%%%%%%%%%%%%%%%%%%%%%%%%%%%
\subsubsection{Ergoregion instability of rotating objects: a consistent approach \label{ergo_consistent}}
\begin{figure}
\begin{center}
\begin{tabular}{cc}
\epsfig{file=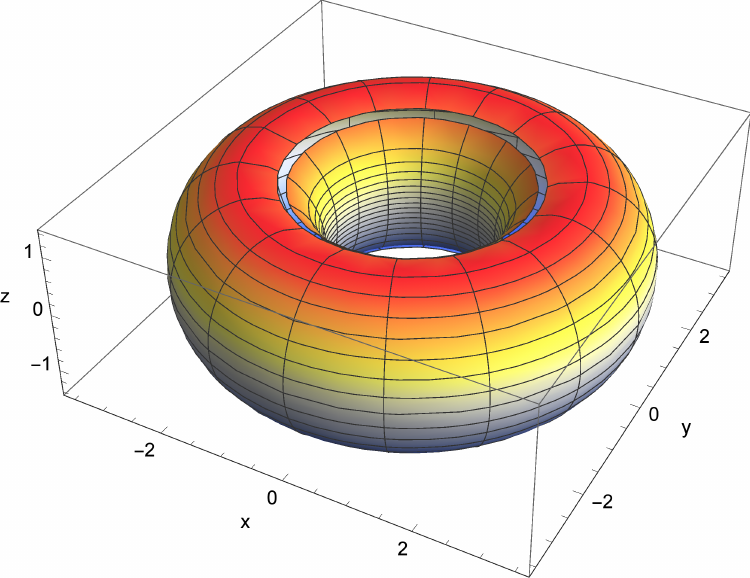,width=0.5\textwidth,angle=0,clip=true}
\end{tabular}
\caption{The toroidal ergoregion of a NS with APR equation of state and spinning at the mass-shedding limit, $\Omega=\Omega_K\equiv\sqrt{M/R^3}$, for a mass slightly above the maximum value (to be compared with Fig.~\ref{fig:ERKerr} for a Kerr BH).
The coordinates $(x,y,z)$ are Cartesian-like coordinates obtained from $(r,\th,\varphi)$ of the line element~\eqref{metricHT2b}.
\label{fig:ER}}
\end{center}
\end{figure}

%%%%%%%%%%%%%%%%%%%%%%%%%%%%%%%%%%%%%%%%%%%%%%%%%%%%%%%%%%%%%%%%%%%%%%%%%%%%%%%%%%%%%%%%%%%%%%%%%%%%%%%
The technical details of this computation are given in Appendix~\ref{app:HT} and in a publicly available {\scshape Mathematica}\textsuperscript{\textregistered} notebook (cf. Appendix~\ref{app:codes}).
Our starting point is the line element~\eqref{metricHT2b}. To second order in the spin, the ergosphere condition $g_{tt}=0$ becomes
\begin{equation}
e^\nu (1+2h_0)=[r^2\varpi^2\sin^2\th+e^\nu h_2(3\sin^2\th-2)]\,.\label{ergosphere}
\end{equation}
The solution to Eq.~\eqref{ergosphere} is topologically a torus. Thus, to characterize the ergoregion it is necessary to include the second-order terms $h_0$ and $h_2$. All previous analysis of the ergoregion instability neglected such terms, based on the fact that for a very compact object $e^\nu\sim0$ and the terms proportional to $h_0$ and $h_2$ should be subdominant relative to the term proportional to $\varpi^2$. However, it is easy to show that this approach would give the wrong result for the ergosphere. For example, in the particular case of a Kerr BH, Eq.~\eqref{ergosphere} is solved by
\begin{equation}
r_{\rm ergo}=2M-\frac{a^2}{4M}\cos2\vartheta+{\cal O}\left(\frac{a^4}{M^4}\right)\,,\label{ergoKerr2nd}
\end{equation}
which agrees with the exact result to second order in the spin\footnote{Note that the metric~\eqref{metricHT2b} is not written in Boyer-Lindquist coordinates, so the ergoregion location does not coincide with that given in Eq.~\eqref{ergoKerr}.}. On the other hand, neglecting the second-order terms $h_0$ and $h_2$ in Eq.~\eqref{ergosphere} would give the wrong result, $r_{\rm ergo}=2M\left(1+\frac{a^2}{2M^2}\sin^2\vartheta\right)+{\cal O}\left(\frac{a^4}{M^4}\right)$, i.e. the ergoregion would always be larger than the Schwarzschild radius, in clear contrast with the correct result~\eqref{ergoKerr2nd}. Clearly, computing the ergoregion of slowly-rotating spacetimes requires to go at least to second order in the rotation. The formalism to construct slowly-rotating geometries has been developed by Hartle \& Thorne and is described in Appendix~\ref{app:HT}. The ergoregion of a compact rotating star, computed by solving Einstein's equations to second order in the angular momentum and using Eq.~\eqref{ergosphere}, is shown in Fig.~\ref{fig:ER}.

\begin{figure}
\begin{center}
\begin{tabular}{cc}
 \epsfig{file=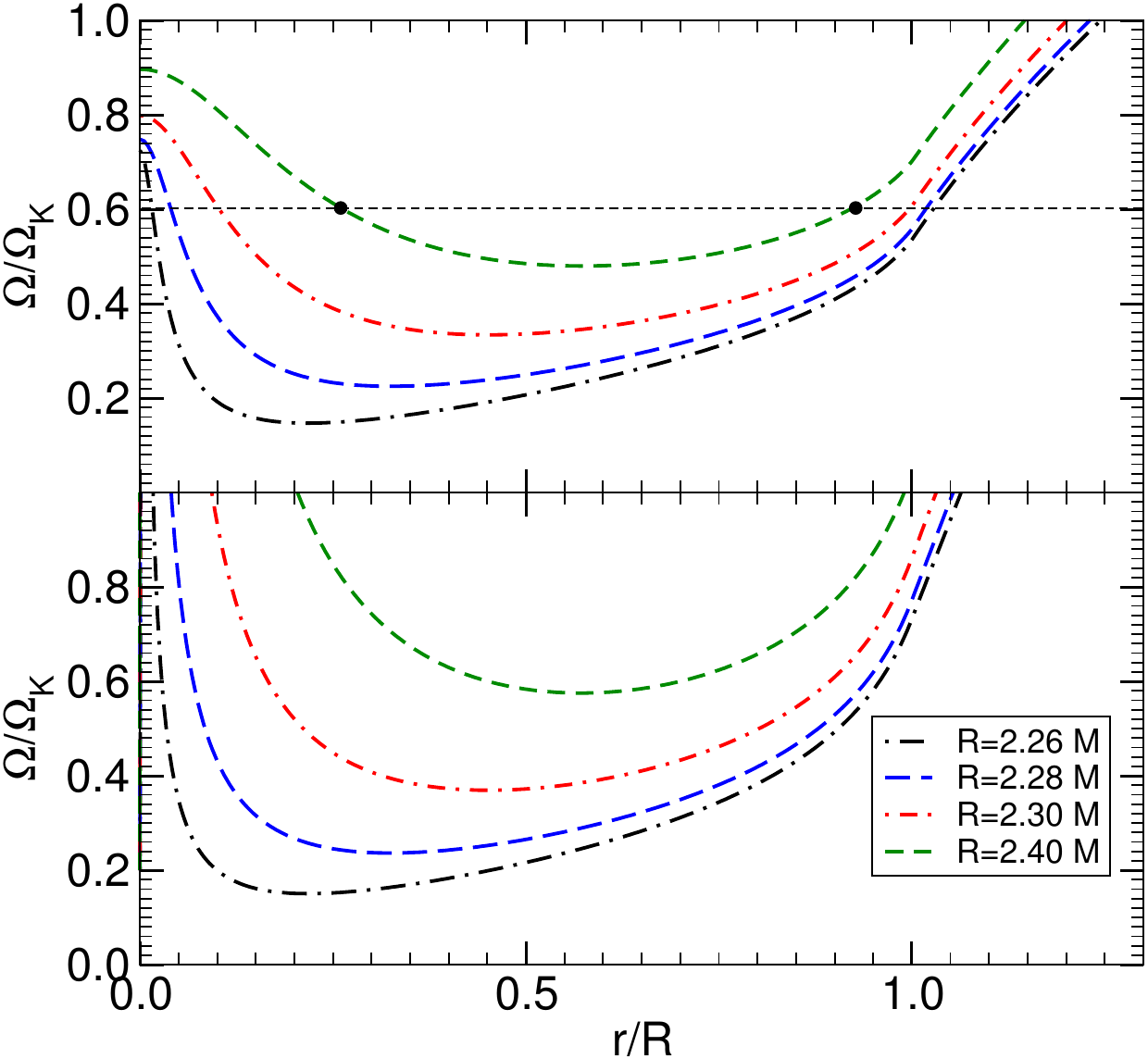,width=0.7\textwidth,angle=0,clip=true}
\end{tabular}
\caption{Size of the ergoregion on the equatorial plane of a constant-density star with various compactnesses for the consistent second-order case (top panel) and for the inconsistent case obtained neglecting $h_0$ and $h_2$ in Eq.~\eqref{ergoKerr2nd} (bottom panel). For a given rotational frequency $\Omega$ and a given compactness, the boundaries of the ergoregion are the intersections between the corresponding curve and the horizontal line. For example in the consistent case with $R=2.40 M$ and $\Omega\sim0.6\Omega_K$, the ergoregion extends between the two black markers, $0.25\lesssim r/R\lesssim0.95$.
\label{fig:ERsize}}
\end{center}
\end{figure}

In Figure~\ref{fig:ERsize}, we show the size of the ergoregion for a constant-density star (whose metric in the static case is given in Eqs.~\eqref{fstar} and~\eqref{Bstar}) for the consistent second-order case (top panel) and for the inconsistent case obtained neglecting $h_0$ and $h_2$ in Eq.~\eqref{ergoKerr2nd} (bottom panel). For a given rotational frequency $\Omega$, the boundaries of the ergoregion are the intersections between each curve and the horizontal line. The two cases can differ substantially, especially as the compactness decreases. In particular, two striking differences appear: (i) in the consistent case the ergoregion extends to the center of the star, while it disappears in the inconsistent case, and (ii) in the consistent case the ergoregion can extend well beyond the radius of the star. Overall, the inconsistent result tends to underestimate the size of the ergoregion.

%%%

The spectrum of perturbations of spinning geometries is generically involved, due the coupling between modes with opposite parity and different harmonic index $l$. Nonetheless, within a slow-rotation approach, certain classes of perturbations can be studied consistently by neglecting such couplings~\cite{Kojima:1992ie,ChandraFerrari91,Pani:2012bp,Pani:2013pma}. For example, for perturbations of a perfect-fluid star to first order in the spin, the following master equation can be derived:
\begin{equation}
\frac{d^2\Psi}{dr_*^2}+\left[\omega^2-2m\omega\varpi-e^\nu\left(\frac{l(l+1)}{r^2}+\eta \frac{2M(r)}{r^3}+4\pi(P-\rho)\right)\right]\Psi(r)=0\,,\label{master_scalar_star_1order}
\end{equation}
where $dr/dr_*=e^{(\nu-\lambda)/2}$ and $\eta=-3,1$ for gravitational-axial and probe-scalar perturbations, respectively. For an ultracompact star with an ergoregion, the former and the latter perturbations were studied in Ref.~\cite{Kokkotas:2002sf} and \cite{CominsSchutz,1996MNRAS.282..580Y}, respectively, finding a family of unstable modes\footnote{We remark that Refs.~\cite{CominsSchutz,1996MNRAS.282..580Y} studied scalar perturbations propagating in the toy-model metric~\eqref{ds2rotapprox}.}. The instability growth rate increases with the spin of the object, is typically maximum for $l=m=1$ modes, and is also larger for gravitational perturbations than for scalar modes.

\begin{figure}
\begin{center}
\begin{tabular}{c}
 \epsfig{file=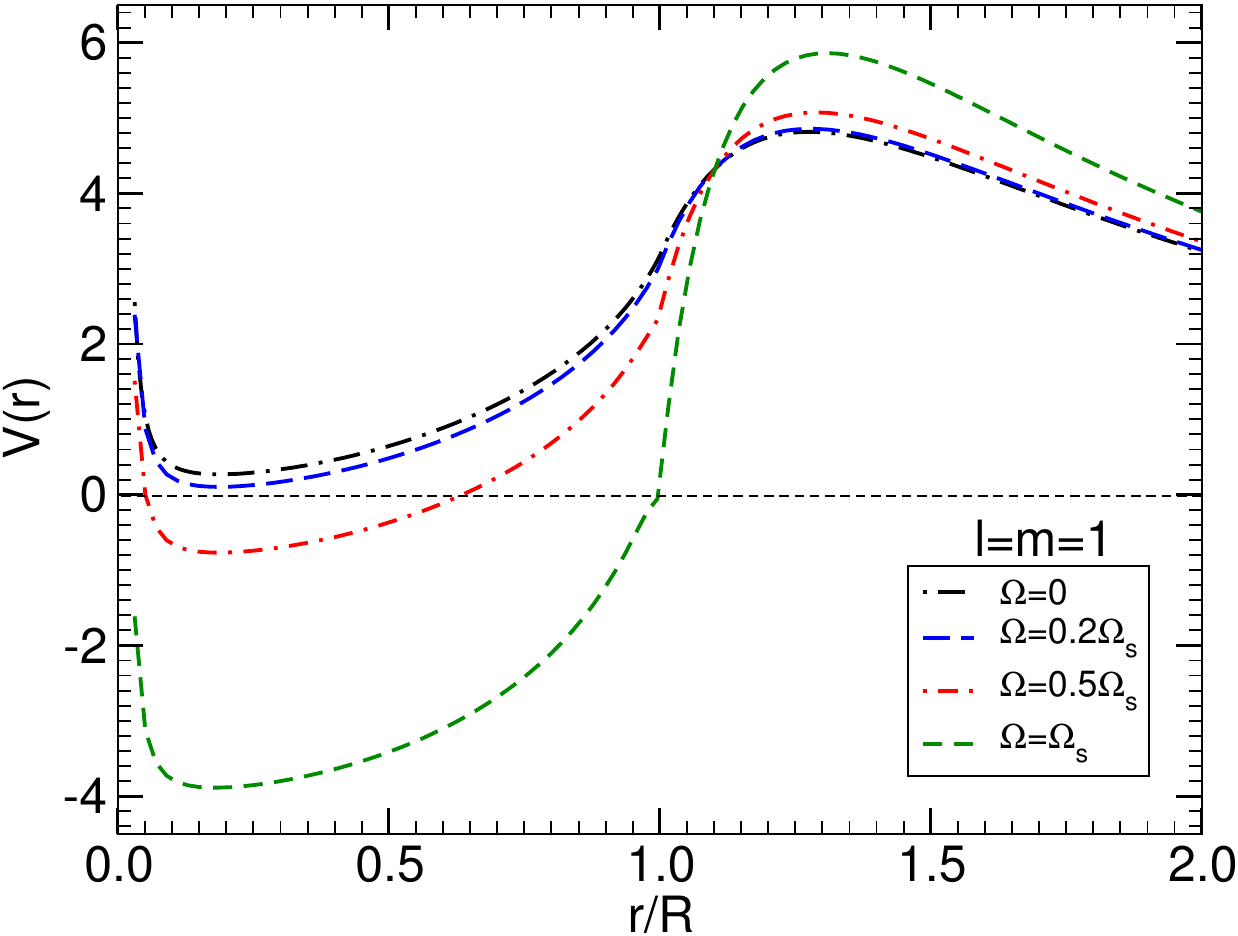,width=0.6\textwidth,angle=0,clip=true}
\end{tabular}
\caption{The potential $V$ for $\omega=0$ as defined in Eq.~\eqref{V_ER} for a constant-density spinning star with $R\sim 2.26 M$ for different values of the angular velocity. As the mass-shedding limit $\Omega\to\Omega_K$ is approached, the potential develops a deeper negative well. Note that $V$ becomes negative only because of second-order corrections and is positive when $\Omega=0$, although its minimum almost crosses the real axis, giving rise to long-lived modes in the nonspinning case, cf. Sec.~\ref{sec:ERlonglived}.
\label{fig:VscalarER}}
\end{center}
\end{figure}

Nonetheless, our previous analysis shows that --~to treat the ergoregion instability consistently~-- one has to include a background geometry to second-order in the spin. Here we consider the simplest case of a probe scalar field that propagates on the background of a spinning NS. The perturbation equations to second order in the spin are derived in Appendix~\ref{app:HT}, the final result is the master equation
\begin{equation}
\frac{d^2\Psi}{dr_*^2}+\left[\omega^2-2m\omega\varpi-V\right]\Psi(r)=0\,, \label{master_scalar_star_2order}
\end{equation}
where 
\begin{equation}
 V=e^\nu\left(\frac{l(l+1)}{r^2}+ \frac{2M(r)}{r^3}+4\pi(P-\rho)+V_2(\omega)\right)\,, \label{V_ER}
\end{equation}
and $V_2$ is a second-order quantity in the spin that is a cumbersome function of the background metric coefficients appearing in~\eqref{metricHT2b}, of the pressure $P$ and the density $\rho$, and of their derivatives. Indeed, because $V_2$ contains second radial derivatives of $\rho$, solving the corresponding eigenvalue problem is quite challenging. For this reason, here we consider a constant-density star which simplifies the problem considerably. The effective potential $V$ for this case is shown in Fig.~\ref{fig:VscalarER} for various spin rates.

We have solved the eigenvalue problem associated to Eq.~\eqref{master_scalar_star_2order} on the background of a constant-density spinning star to second order in the angular velocity. The background problem is solved in the interior by requiring continuity of the metric functions at the star radius $R$\footnote{Note that, because of the absence of Birkhoff theorem in the spinning case, the exterior geometry is not a slowly-spinning Kerr metric.}. For the scalar perturbations, the fact that $\rho=\rho_c={\rm const}$ in the interior and $\rho=0$ in the exterior produces discontinuities in $V$ at the star's radius, which can be taken into account by suitable junction conditions for the derivative of the scalar field. At the stellar radius we impose $\psi_-=\psi_+$ and $\partial_r\psi_+=\partial_r\psi_- -\Delta V \psi_-/(1-2M/r)^2$, where $\Delta V=V_+-V_-$ and we defined $A_\pm=\lim_{\epsilon\to0} A(R\pm\epsilon)$.

\begin{figure}
\begin{center}
\begin{tabular}{c}
 \epsfig{file=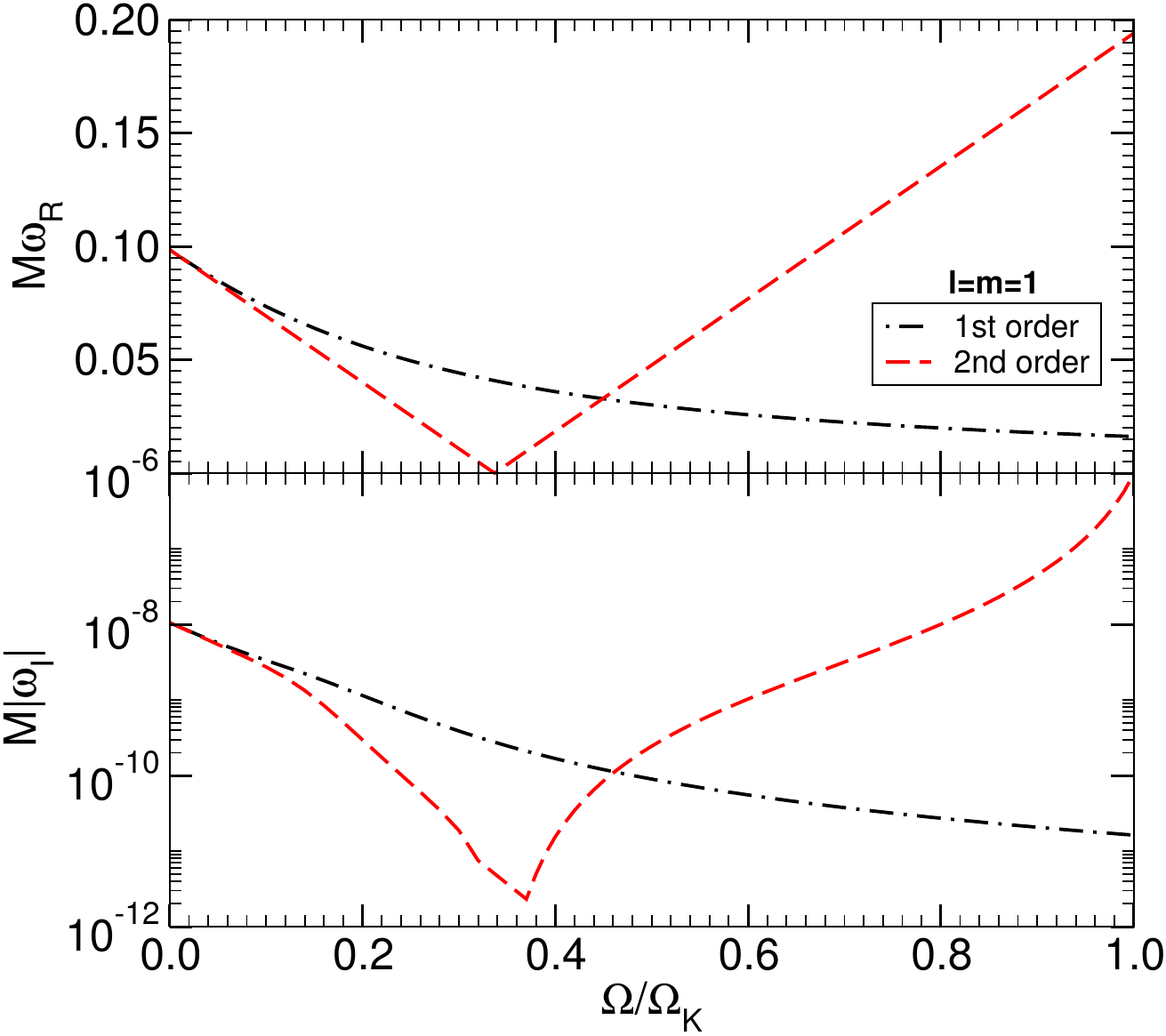,width=0.6\textwidth,angle=0,clip=true}
\end{tabular}
\caption{Real and imaginary parts of the fundamental $l=m=1$ mode for a constant-density star with ultrahigh compactness, $R\sim 2.26M$, as a function of the angular velocity of the star normalized by the mass shedding limit. Note that the vertical axis of the bottom plot shows the absolute value of $\omega_I$ and is in a log scale. The first order result fails to capture the instability ($\omega_I>0$, rightmost part of the plots) because in this case the background geometry does not possess an ergoregion. To second order, the threshold of the instability corresponds to a zero crossing of both $\omega_R$ and $\omega_I$, see text for details.
\label{fig:ER_modes}}
\end{center}
\end{figure}

The fundamental modes of the system are shown in Fig.~\ref{fig:ER_modes} for a constant-density star with ultrahigh compactness, $R\sim 2.26 M$, whose effective potential is shown in Fig.~\ref{fig:VscalarER}. We present both first-order and second-order computations. As expected, these two cases are in agreement with each other for small angular velocities, but they are dramatically different when $\Omega\gtrsim 0.1\Omega_K$. Indeed, while the modes remain stable to first order in the spin, they become unstable to second order. Interestingly, the threshold of the instability corresponds (within numerical accuracy) to a zero crossing also of the \emph{real} part of the mode. In Fig.~\ref{fig:ER_modes}, we focus only on $\omega_R>0$ by exploiting the symmetry of the field equations under $m\to-m$ and $\omega\to-\omega$.

The fact that the second-order terms play such an important role in the stability analysis can be understood by the fact that the ergoregion of the spacetime appears only at the second order. Indeed, while our results are generically in qualitative agreement with previous analysis~\cite{CominsSchutz,1996MNRAS.282..580Y,Cardoso:2007az,Chirenti:2008pf,Kokkotas:2002sf}, it is important to note that in all cases the latter have been obtained by including \emph{some} (but not all) second-order terms. Should all second-order terms be neglected, no unstable mode would be found. The results in Fig.~\ref{fig:ER_modes} represent the first fully-consistent computation of the ergoregion instability for a spinning compact star. The phenomenology of this instability is discussed in detail in Sec.~\ref{sec:ERphenom_stars}.

%

%%%%%%%%%%%%%%%%%%%%%%%%%%%%%%%%%%%%%%%%%%%%%%%%%%%%%%%%%%%%%%%%%%%%%%%%%%%%%%%%%%%%%%%
\subsubsection{Ergoregion instability and long-lived modes} \label{sec:ERlonglived}
%%%%%%%%%%%%%%%%%%%%%%%%%%%%%%%%%%%%%%%%%%%%%%%%%%%%%%%%%%%%%%%%%%%%%%%%%%%%%%%%%%%%%%%

The underlying origin of the ergoregion instability is the existence of long-lived modes in ultracompact spacetimes in the static limit; these modes are very slowly damped and can become unstable when rotation is included. This has been first discussed in the eikonal limit~\cite{CominsSchutz} and it has been recently put on a firmer basis in Ref.~\cite{Cardoso:2014sna}.

\begin{figure}[t]
\begin{center}
\begin{tabular}{cc}
\epsfig{file=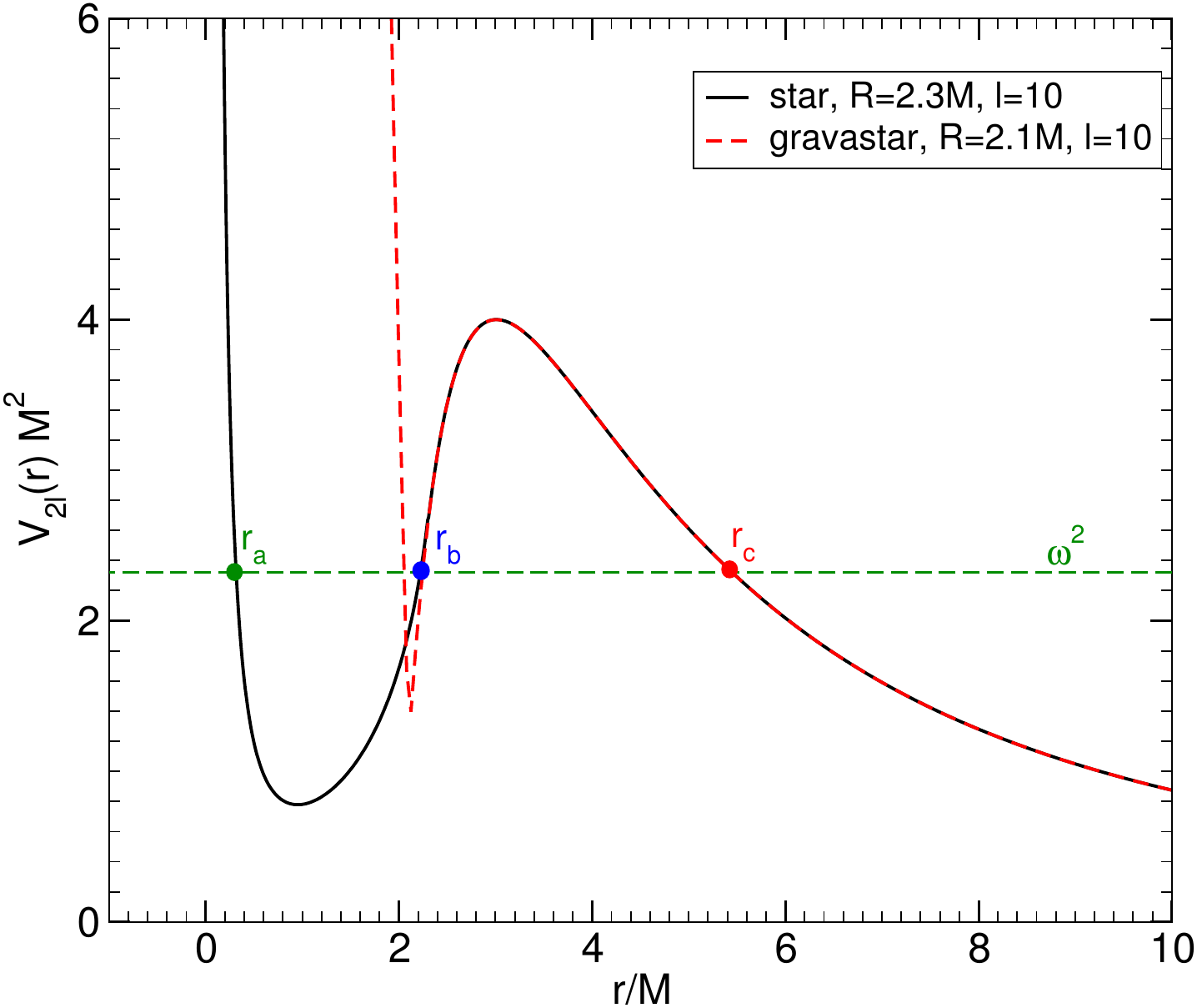,width=8.5cm,angle=0,clip=true}
\end{tabular}
\end{center}
\caption{\label{fig:LRpotential}
Examples of the potential governing linear perturbations of a static ultracompact star. The black solid line and the red dashed line correspond to $l=10$ gravitational axial perturbations of a uniform star with $R=2.3 M$ and of a gravastar with $R=2.1M$, respectively.}
\end{figure}

Such long-lived modes exist in ultracompact spacetimes which possess a light ring (i.e. an unstable circular orbit as in the Schwarzschild case) but not a horizon~\cite{Cardoso:2014sna,Keir:2014oka}.
The reason for that is explained in Fig.~\ref{fig:LRpotential} (cf. also Fig.~\ref{fig:VscalarER} above), which shows the effective potential~\eqref{potentialmaster} (cf. Appendix~\ref{app:WKB} for details) corresponding to two models of static ultracompact objects: a constant-density star with compactness $M/R\sim0.435$ (black solid curve) and of a thin-shell gravastar\footnote{Thin-shell gravastars~\cite{Mazur:2001fv} are discussed in Sec.~\ref{sec:BHmimickers} in the context of so-called ``BH mimickers''.} with compactness $M/R\sim 0.476$ (dashed red curve), respectively\footnote{Other regular geometries which possess a light-ring are the perfect-fluid stellar objects with multiple necks discussed in Refs.~\cite{Karlovini:2000zu,Karlovini:2000xd,Karlovini:2000rn}.}. 
Because the radius of these objects is smaller than the light-ring location of the external Schwarzschild spacetime, $r=3M$, the effective potential develops a maximum at that location. Furthermore, the centrifugal potential near the center of these objects is responsible for the existence of a further~\emph{stable} null circular orbit in the object's interior. This corresponds to the minimum shown in Fig.~\ref{fig:LRpotential}, where the long-lived modes are localized~\cite{Cardoso:2014sna}. These modes (sometimes dubbed ``s-modes'' in the context of ultracompact stars~\cite{ChandraFerrari91}) are computed in the WKB approximation in Appendix~\ref{app:WKB} and they agree quite well with exact numerical results (cf. Fig.~\ref{fig:star} and Ref.~\cite{Cardoso:2014sna}).

The dependence of the frequency and damping time of these long-lived modes to instability as functions of the spin in connection to the ergoregion has been first discussed in Ref.~\cite{CominsSchutz}, which considers an approximate line element
\begin{equation}
 ds^2=-F(r)dt^2+B(r) dr^2+r^2 d\theta^2+r^2 \sin^2\theta(d\phi-\varpi(r) dt)^2\,. \label{ds2rotapprox}
\end{equation}
Although not being a solution of Einstein's equations coupled to a fluid, this metric should approximate the exact geometry describing a spinning star in the case of slow rotation and high compactness, as we discussed. In such approximate metric, the ergoregion is defined by
\begin{equation}
\varpi(r)\sin\theta>\frac{\sqrt{F(r)}}{r}\,.
\end{equation}
In the eikonal limit, the Klein-Gordon equation in the background~\eqref{ds2rotapprox} can be written in the form~\cite{CominsSchutz}
\begin{equation}
\psi''+m^2 \frac{B}{F}(\bar{\omega}+V_+)(\bar{\omega}+V_-)\psi=0\,, \label{KGrot}
\end{equation}
where $\bar{\omega}=\omega/m$ is a rescaled frequency, $m$ is the azimuthal number associated to the axisymmetry of the background, and
%%%
\begin{equation}
 V_\pm=-\varpi\pm \frac{\sqrt{F}}{r}\,,
\end{equation}
%%%
are the effective potentials that describe the motion of (counter-rotating for the plus sign and co-rotating for the minus sign) null geodesics in the equatorial plane of the geometry~\eqref{ds2rotapprox}.

Now, the boundary of the ergoregion (if it exists) corresponds to two real roots of $V_+=0$ and $V_+<0$ inside the ergoregion. Because $V_+\to+\infty$ at the center and attains a positive finite value in the exterior, it is clear that the ergoregion must contain a point in which $V_+$ displays a (negative) local minimum. This simple argument shows the important result that the presence of an ergoregion in a horizonless object implies the existence of \emph{stable} counter-rotating photon orbits~\cite{Cardoso:2014sna}. 

Furthermore, Eq.~\eqref{KGrot} supports unstable modes, whose computation is briefly presented in Appendix~\ref{app:WKB} in the WKB approximation. In the eikonal limit, the instability time scale depends exponentially on the azimuthal number, 
\begin{equation}
\tau_{\rm ergo} \sim 4\alpha e^{2\beta m}\,, \label{tau_ergo_eikonal}
\end{equation}
where $\alpha$ and $\beta$ are two positive constants~\cite{CominsSchutz} (cf. Appendix~\ref{app:WKB}). The instability can be understood from the fact that the corresponding modes are localized near the stable photon orbit, which is situated within the ergosphere, and are confined within the star. This confinement provides the arena for the instability to grow through the negative-energy states that are allowed within the ergoregion~\cite{1978CMaPh..63..243F}. Likewise, this argument also explains why spinning BHs --~that also possess a light ring and an ergoregion~-- are linearly stable, because the presence of the horizon forbids the existence of trapped modes.
Together with superradiant instabilities of rotating BHs, the ergoregion instability provides a general mechanism that turns rotating compact objects unstable when placed in a cavity or in AdS spacetimes~\cite{Ishii:2020muv}.

%%%%%%%%%%%%%%%%%%%%%%%%%%%%%%%%%%%%%%%%%%%%%%%%
\subsubsection{Ergoregion instability in fluids}
%%%%%%%%%%%%%%%%%%%%%%%%%%%%%%%%%%%%%%%%%%%%%%%%
%  
\begin{figure*}
\includegraphics[width=16cm]{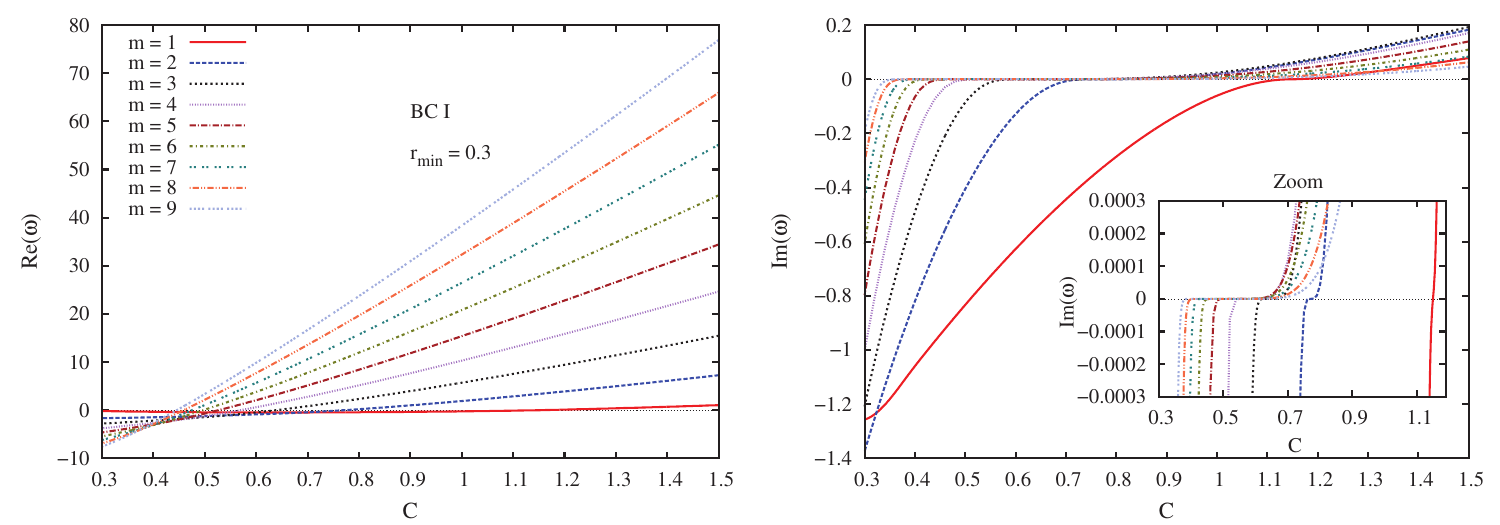}\\
\includegraphics[width=16cm]{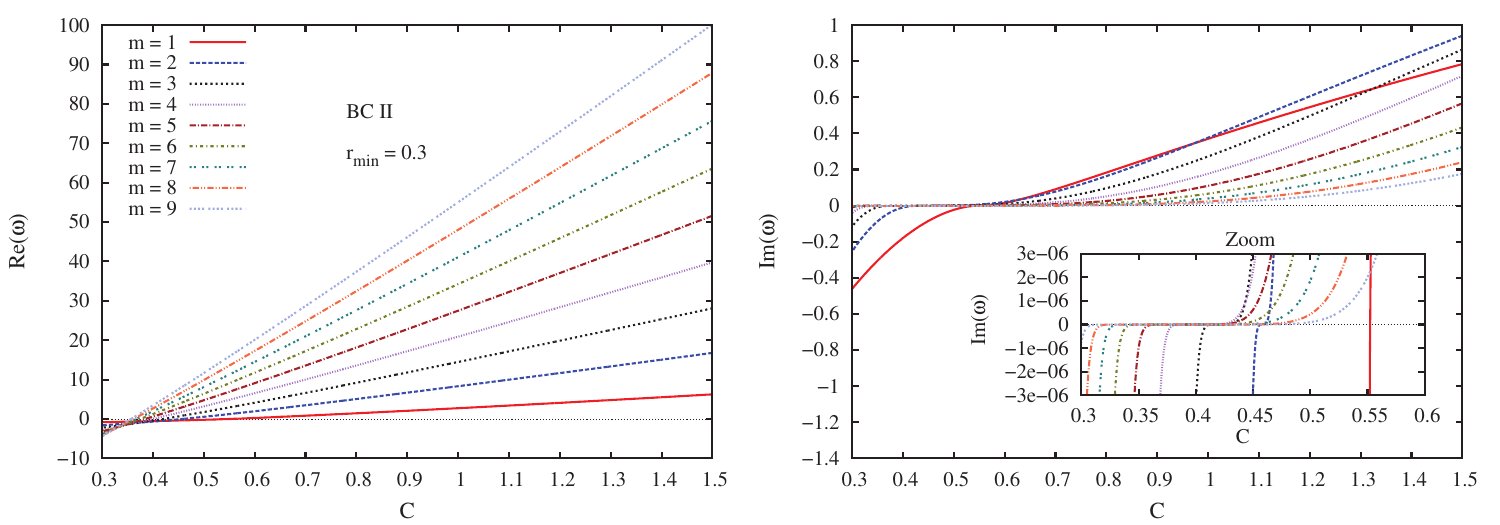}
\caption{Real (left) and imaginary (right) components of the fundamental QNM frequencies, plotted as a function of $C$, for $r_{\rm min}=0.3$ and different values of  $m$.
The top plots correspond to Dirichlet-like boundary conditions, whereas the bottom plots correspond to Neumann-like boundary conditions. Note the striking similarity with Fig.~\ref{fig:ER_modes}. From Ref.~\cite{Oliveira:2014oja}.
}
\label{fig-stable1}
\end{figure*}
\begin{figure*}
\includegraphics[width=16cm]{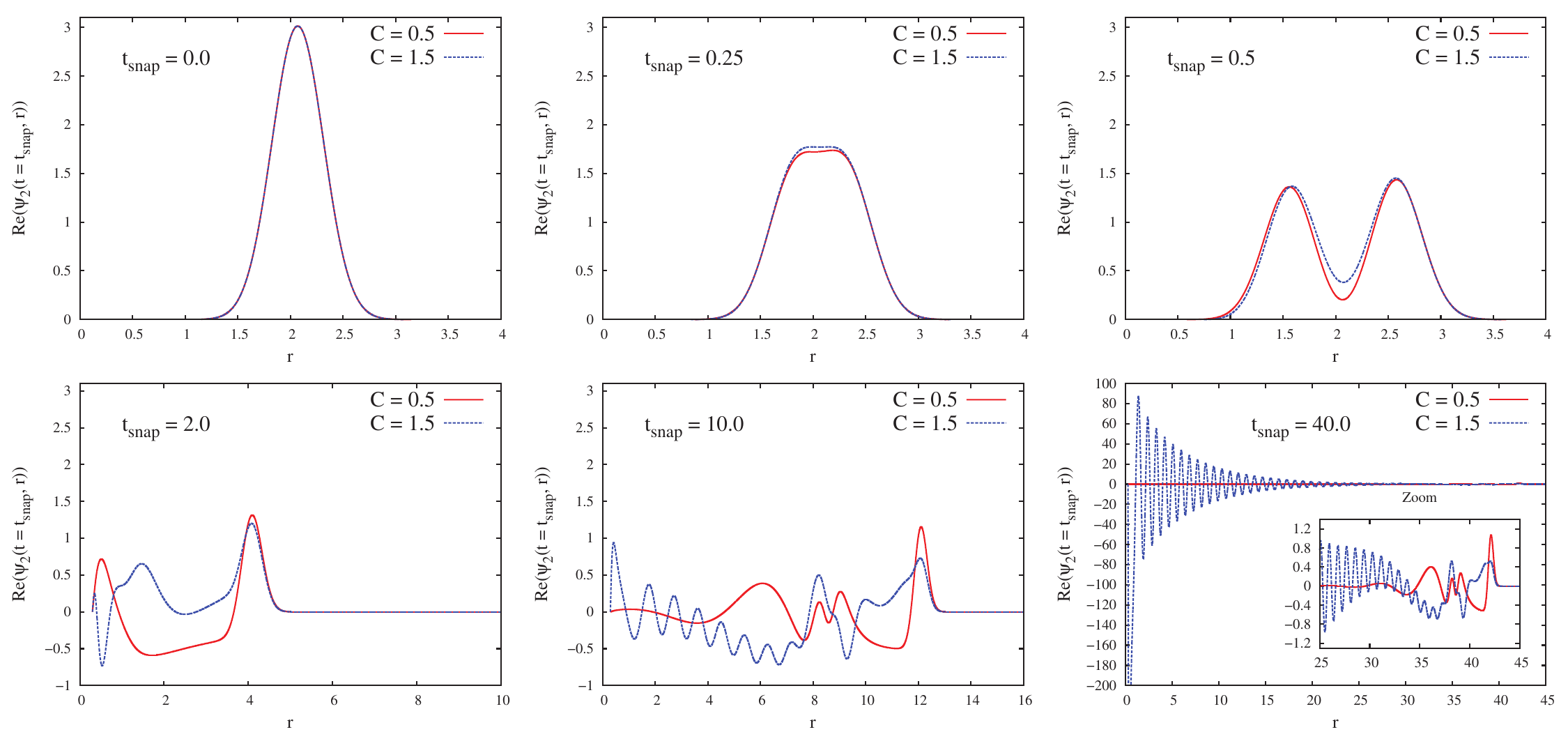}
\caption{Snapshots of the radial profiles of $\text{Re}(\psi_{m}(t, r))$ for azimuthal number $m = 2$, circulations $C=0.5$ (stable case) and $C=1.5$ (unstable case). Dirichlet-like boundary conditions are imposed at $r_{\rm min}=0.3$. From Ref.~\cite{Oliveira:2014oja}.}
\label{fig_snapshots}
\end{figure*} 
In the context of acoustic geometries introduced in Section~\ref{super_discontinuity} and expanded in Sections~\ref{sec:lab} and \ref{sec:SR_analog},
sound waves propagate in moving fluids as a massless scalar field in curved spacetime, with an effective geometry dictated by the background fluid flow.
There are simple acoustic setups with instabilities that can be framed in the language of curved spacetime as ergoregion instabilities~\cite{Oliveira:2014oja,Hod:2017lho}.

Let us focus again on the two-dimensional fluid flow of Section~\ref{sec:SR_analog}, but consider a specific flow with vanishing radial speed ($A=0$ in Eq.~\eqref{metric2}),
the so-called the {\it hydrodynamic vortex}, whose line element is
\beq
ds^2=-c^{2}\left(1-\frac{C^2}{c^2r^2} \right) dt^2+dr^2-2Cdtd\theta+r^2d\theta^2+dz^2\,.
\label{vortex}
\eeq
This effective spacetime presents an ergoregion with outer boundary at $r_{\rm ergo} = C/c$, which coincides with the circle at which the (absolute value of the) background flow velocity equals the speed of sound $c$. Henceforth we set the speed of the sound equal to unity ($c=1$).

The {\it background} velocity diverges at the origin as $1/r$, signaling a physically singular behavior.
Possible experimental setups can be mimicked by imposing boundary conditions at a finite
position $r=r_{\rm min}$, the precise form of which depend on the specific experimental apparatus.
 Assume therefore that an infinitely long cylinder of radius $r_{\rm min}$
is placed at the center of our coordinate system. The cylinder is made of a certain material with acoustic impedance $Z$~
\cite{Lax:1948}. Low-impedance materials correspond to Dirichlet-type boundary conditions on the master variable (see Section~\ref{sec:lab}) and, for completeness, we also consider Neumann-type conditions~\cite{Oliveira:2014oja}).

Together with Sommerfeld conditions at large distance, the problem is an eigenvalue problem for the possible frequencies, the 
solution of which is shown in Fig.~\ref{fig-stable1} for a specific cylinder position at $r_{\rm min}=0.3$ as a function of rotation rate $C$.
Notice that our generic arguments in favor of an ergoregion instability predict that the geometry is unstable as long as the cylinder position is within the ergosurface.
In other words, as long as $C>0.3$. Figure~\ref{fig-stable1} shows that indeed the large-$m$ threshold of the instability asymptotes to $C=0.3$, as can be seen from Fig.~\ref{fig-stable1}, and as anticipated from our discussion. 
The striking similarity between Fig.~\ref{fig-stable1} and Fig.~\ref{fig:ER_modes} is also remarkable. Indeed, in this analog geometry we recover all the qualitative features previously discussed for ultracompact stars. In particular, in both cases at the threshold for the instability the frequency of the mode has a zero crossing and the imaginary part of the mode has an inflection point.
Further insights into the onset of the instability were derived in Ref.~\cite{Hod:2014hda}.

The results also indicate (cf. Fig.~\ref{fig-stable1}) that all modes $m>5$ are unstable for $r_{\rm min}=0.3$ and circulation $C=0.5$. Moreover, at fixed inner boundary location $r_{\rm min}$ and fixed $m$ the instability gets stronger for larger $C$, as might also be anticipated. All the numerical results fully support the statement that the presence of an ergoregion without event horizon gives rise to instabilities.  
A complementary facet of the instability is shown in snapshots of the evolution, as those depicted in Fig.~\ref{fig_snapshots}. These snapshots compare the evolution of a stable ($C=0.5$) and unstable ($C=1.5$) configuration, both for $m=2$, and show clearly how the instability develops inside the ergoregion and close to the inner boundary at $r_{\rm min}=0.3$. Notice the scale in the last snapshot, and how the field decays in space but grows in time.

As might be expected in a centuries-old field, similar instabilities were reported decades ago in fluid dynamics, within that specific field's language.
Broadbent and Moore have conducted a thorough study of stability of rotating fluids, but imposing slightly different boundary conditions~\cite{Broadbent:1979}. In line with our findings, they uncover an instability for compressible fluids related also to sound wave amplification (note that incompressible fluids were also analyzed by Lord Kelvin and were found to be marginally stable~\cite{Kelvin}).
The evidence that the hydrodynamic vortex is an unstable system and that the instabilities are directly related to the existence of an ergoregion
together with the absence of an event horizon agrees with the prediction in Ref.~\cite{1978CMaPh..63..243F}. This confirmation further strengthens the similarities between effective spacetimes in fluids and BHs.

The simple physical description of such ergoregion instabilities was recently used to understand the dynamical instability of multiply quantized vortices in spatially homogeneous atomic Bose--Einstein condensates~\cite{Giacomelli:2019tvr}. In particular, an analysis of the role of boundary conditions at large distance in the linearized Bogoliubov problem showed that this is a dispersive version of the ergoregion instability of rotating spacetimes with respect to scalar field perturbations.

%%%%%%%%%%%%%%%%%%%%%%%%%%%%%%%%%%%%%%%%%%%%%%%%%%%%%%%%%%%%%%%%%%%%%%%%%%%%%%%%%%%%%%%%%%%%%%%%%%%%%%%%
\subsubsection{Ergoregion instability and Hawking radiation}
%%%%%%%%%%%%%%%%%%%%%%%%%%%%%%%%%%%%%%%%%%%%%%%%%%%%%%%%%%%%%%%%%%%%%%%%%%%%%%%%%%%%%%%%%%%%%%%%%%%%%%%%
As we mentioned in Section~\ref{sec:KerrCFT}, string theory has made great progress in understanding the microphysics of BHs. In particular, for certain (nearly) supersymmetric BHs, one is able to show that the Bekenstein-Hawking entropy $S_{\rm BH} = A/4$, as
computed in the strongly-coupled supergravity description, can be reproduced in a weakly-coupled $D-$brane description
as the degeneracy of the relevant microstates~\cite{Strominger:1996sh}. The AdS/CFT correspondence~\cite{Maldacena:1997re,Gubser:1998bc,Witten:1998qj} allows
further insights into these issues by providing a dictionary relating the geometric description of the physics in the near-horizon
region with the physics of a dual conformal field theory. In particular, the AdS/CFT
indicates that Hawking evaporation should be a unitary process, in keeping with the basic tenets of quantum theory.
The discussion of BHs in the context of the AdS/CFT correspondence makes it evident that the path integral
over geometries in the bulk may include multiple saddle-points, i.e., several classical supergravity solutions~\cite{Witten:1998zw}. Another point that was realized early on is that the geometric description of individual microstates would not have a horizon~\cite{Myers:1997qi}. These ideas were incorporated by Mathur and colleagues in a radical revision of the stringy description
of BHs, the ``fuzzball'' proposal~\cite{Mathur:2005zp,Mathur:2014nja}. They argue that each of the CFT microstates corresponds to a separate spacetime geometry with no horizon. The BH is dual to an ensemble of such microstates and so the BH geometry only emerges in a coarse-grained description which ``averages'' over the BH microstate geometries.

In a fuzzball microstate the spacetime ends just outside the horizon (because compact directions ``cap-off''~\cite{Mathur:2014nja})
thus avoiding issues like the information paradox in BH physics. However, it seemingly introduces an unexpected problem: if the horizon is 
not the traditional one, how is it possible to recover traditional BH thermodynamics like the Hawking radiation rate?
Surprisingly, for the few microstates known explicitly -- which rotate and possess an ergoregion -- it was shown that the Hawking radiation rate can be {\it exactly} reproduced from the ergoregion instability~\cite{Chowdhury:2007jx,Chowdhury:2008bd} (because these effective geometries have no horizon, spin will in general give rise to an ergoregion hence an instability~\cite{Cardoso:2005gj}).

%%%%%%%%%%%%%%%%%%%%%%%%%%%%%%%%%%%%%%%%%%%%%%%%%%%%%%%%%%%%%%%%%%%%%%%%%%%%%%%%%%%%%%%%%%%%%%%%%%%%
\subsection{Black-hole lasers and superluminal corrections to Hawking radiation}\label{sec:BHlasers}
%%%%%%%%%%%%%%%%%%%%%%%%%%%%%%%%%%%%%%%%%%%%%%%%%%%%%%%%%%%%%%%%%%%%%%%%%%%%%%%%%%%%%%%%%%%%%%%%%%%%

A completely different, semi-classical realization of the BH-bomb mechanism was put forward in Ref.~\cite{Corley:1998rk}. In this model, one considers Hawking radiation from a geometry with an outer and an inner horizon and in the presence of high-energy modifications that change the dispersion relation $\omega(k)$ of photons at high frequencies\footnote{The example considered in Ref.~\cite{Corley:1998rk} was inspired by analogue
BH models and, as should be clear from Section~\ref{super_discontinuity}, the geometry only plays the role of a spectator. The laser effect occurs in analogue models as well as in true, gravitational BHs (for example, in the RN geometry).}. 

For a geometry with a single (event) horizon, Hawking radiation is rather insensitive to high-energy modifications, producing the classical thermal spectrum~\cite{Hawking:1974sw} at frequencies much lower than the new scale. However, in the presence of two horizons and if the dispersion relation is superluminal, the negative-energy partners of Hawking quanta are able to bounce back and return to the outer horizon on a superluminal trajectory.
Indeed, the origin of the laser effect can be attributed to the closed trajectories followed by the negative Killing frequency partners of Hawking quanta, which can bounce between the two horizons due to the modified dispersion relation.
If the quanta are fermions, they suppress Hawking radiation, whereas if they are bosons they stimulate a secondary emission which is correlated to the original radiation, unlike in the usual Hawking effect. The process sustains itself as in the BH-bomb mechanism (and, in fact, as in the stimulated emission of a laser), the role of the mirror being played by the ergoregion between the two horizons which allows for superluminal bouncing trajectories with negative energies (see Ref.~\cite{Corley:1998rk} for details). 
A thorough mode analysis of the BH laser effect shows that
it is described in terms of frequency eigenmodes that are spatially bound. The spectrum contains a discrete and finite
set of complex frequency modes which appear in pairs and which encode the laser effect~\cite{Coutant:2009cu,Coutant:2014wga}. 
Related, zero-frequency ``undulation'' modes were dealt with in Refs.~\cite{Coutant:2012zh,Coutant:2012mf}.

The BH laser is a dynamical instability, the origin of which can be traced back to the 
negative energy states behind the outer horizon, and which work in fact as an ergoregion for 
the modes ``living'' there. One can then naturally associate the BH laser instability with a superradiant
instability~\cite{Coutant:2009cu,Coutant:2014wga}.

%%%%%%%%%%%%%%%%%%%%%%%%%%%%%%%%%%%%%%%%%%%%%%%%%%%%%%%%%%%%%%%%%%%%%%%%%%%%%%%%%%%%%%%%%%%%%%%%%%%%%%%%%%%%%%%%
\subsection{Black holes in Lorentz-violating theories: nonlinear instabilities}\label{sec:BH_Lorentz_violating}
%%%%%%%%%%%%%%%%%%%%%%%%%%%%%%%%%%%%%%%%%%%%%%%%%%%%%%%%%%%%%%%%%%%%%%%%%%%%%%%%%%%%%%%%%%%%%%%%%%%%%%%%%%%%%%%%
A related instability is thought to occur for BHs in Lorentz-violating theories~\cite{Eling:2007qd,Blas:2011ni}.
In these theories, BH solutions can exist (see e.g. Refs.~\cite{Giannios:2005es,Eling:2006ec,Barausse:2013nwa} or the overview~\cite{Berti:2015itd}) with multiple, nested horizons, one for each maximal speed of propagation in the
theory. Each horizon traps the corresponding species of
field excitations. Consider two particles, with different propagation speeds, and therefore two horizons.
In this framework, the region between the two horizons
is classically accessible to the faster particle and it is a classically inaccessible ergoregion for the slower one.
If these particles are now allowed to interact gravitationally, it is {\it possible}
that an energy transfer occurs from the slower to the faster particle, resulting in a nonlinear ergoregion instability.
Hints of nonlinear instabilities were discussed in Ref.~\cite{Blas:2011ni}, but it is not clear whether they are related to
this particular mechanism.

%%%%%%%%%%%%%%%%%%%%%%%%
\subsection{Open issues}
%%%%%%%%%%%%%%%%%%%%%%%%
Superradiant (or ``BH bomb'') instabilities are a fascinating and rapidly growing topic. Here we list some of the most 
urgent open questions related to this problem at the time of writing.

\begin{itemize}

\item While superradiant instabilities of spinning BHs have been studied thoroughly for long time, it has 
been only recently that a similar understanding for massive vector fields has been 
reached~\cite{Pani:2012vp,Pani:2012bp,Witek:2012tr,Baryakhtar:2017ngi,East:2017mrj,East:2018glu,Frolov:2018pys,
Frolov:2018ezx}. Massive spin-$2$ perturbations are still largely unexplored.  
In this case, the only available results have considered either an expansion at first-order in the 
spin~\cite{Brito:2013wya} or a small gravitational coupling approximation~\cite{Brito:2020lup}. It would also be very 
interesting to understand the nature of the 
``special'' mode found in Ref.~\cite{Brito:2013wya} (see discussion in Sec.~\ref{sec:spin2SR}).
Finally, it is unknown whether the system of 
equations~\eqref{eqmotioncurved} is separable in a Kerr background. Can techniques similar to those recently used in 
the Proca case~\cite{Frolov:2018pys,Frolov:2018ezx} be used for spin-two massive fields?

\item It was shown that RN-dS BHs are unstable under 
spherically-symmetric charged scalar perturbations~\cite{Zhu:2014sya,Konoplya:2014lha,Destounis:2019hca}\footnote{Higher dimensional RN-dS 
were shown to be unstable in $D\geq 
7$ dimensions against gravito-EM perturbations~\cite{Konoplya:2008au,Cardoso:2010rz,Konoplya:2013sba}. However this 
instability is of different nature.}. Given the fact that asymptotically flat RN BHs are stable against these 
perturbations, this is a quite surprising and still not very well understood result. In Ref.~\cite{Konoplya:2014lha} it 
was shown that a necessary condition for the instability to occur is that the field's frequency $\omega_R$ satisfies:
\be
\frac{qQ}{r_c}<\omega_R<\frac{qQ}{r_+}\,.
\ee
This is exactly the superradiant condition for this spacetime  (cf. Eq.~\eqref{superradiance_dS}), which suggests that the instability is of superradiant nature. However, it was also found that not all the superradiant modes are unstable and only the monopole $l=0$ suffers from this instability. The instability only occurs at small values of the coupling $q Q\lesssim 1$, as long as $qQ\gg \mu M$, where $\mu$ is the mass of the scalar field, and disappears when $\Lambda\to 0$.  
The end-state of the instability is still an open-problem, but the fact that the system is not confined, unlike in the RN-AdS case (see Sec.~\ref{sec:breaking}), makes it likely that the instability will extract charge and mass from the BH, evolving to a stable region in the parameter space. 
A detailed understanding of charged BHs in asymptotically dS spacetimes may also prove fundamental to explore the Strong Cosmic Censorship Conjecture:
it was shown recently that near-extremal geometries damp fluctuations fast enough that Cauchy horizons are stable, opening the door to the lack of determinism in GR~\cite{Cardoso:2017soq,Luna:2018jfk,Dias:2018ufh,Dias:2018etb}.

\item Shibata and Yoshino found that rapidly singly-spinning higher-dimensional BHs with spherical topology are 
unstable against non-axisymmetric perturbations (the so-called ``bar''--mode 
instability)~\cite{Shibata:2009ad,Shibata:2010wz} in $D=5,6,7,8$ dimensions (see also~\cite{Dias:2014eua}). This was 
extended to equal angular momenta Myers-Perry BHs in odd dimensions in Ref.~\cite{Hartnett:2013fba} and analytically 
studied in the large-$D$ limit in Ref.~\cite{Emparan:2014jca}; these unstable BHs will emit gravitational radiation and 
consequently spin down and decrease their mass~\cite{Shibata:2010wz}. The area theorem (cf. Section~\ref{areaimpliesSR}) 
then requires that the unstable modes should satisfy the superradiant condition~\eqref{eq:superradiance_condition}, 
which indicates that the instability is of superradiant nature. However not all the superradiant modes are unstable and 
unlike the superradiant instability discussed in this Section, this instability is not due to confinement. A complete 
comprehension of the physical mechanism behind this instability is still an open problem.

\item An interesting open question is the effect of rotation in the \emph{outer} disk of the two-ring model discussed in Sec.~\ref{sec:rings}, for example to investigate possible resonant effects when both rings are spinning. Likewise, the BH analog of the two-ring model proposed by Press~\cite{PressRing}, namely a Kerr BH surrounded by a conductive disk -- in particular whether such system is unstable or not, and on which time scales -- has not been studied yet.

\item One of the important missing studies concerns a detailed investigation of the ergoregion instability of ultracompact spinning NSs or other compact objects.
Rapidly or (consistently built, see Section~\ref{ergo_consistent}) slowly-spinning stars are all basically uncharted 
territory (but see Ref.~\cite{Tsokaros:2019mlz} for a recent development showing that stars with an ergoregion can be 
stable at least for a dynamical time).
Gravitational perturbations of slowly-spinning NSs can, in principle, be computed by extending recently-developed perturbative methods to second order in the spin (including the star structure~\cite{Hartle:1967he} and its perturbations~\cite{Kojima:1992ie,1993ApJ...414..247K,Pani:2013pma}).

\item Massive fermions near a Kerr BH form bound states that, rather than inducing an instability as in the bosonic case, condense and form a Fermi sea which extends outside the ergosphere~\cite{Hartman:2009qu}. This analysis has been performed in the WKB limit and hints at possible important nonlinear effects in the behavior of fermion fields. Whether or not such systems can trigger superradiant instabilities at the nonlinear level is unclear. In a different but related vein, Ref.~\cite{Matsas:2007bj} opened the possibility of overspinning a RN BH by quantum tunneling;
such possibility was later argued to be ruled out, and that cosmic censorship conjecture is actually respected in this situation~\cite{Hod:2008kq}. The physical mechanism is a quantum version of superradiance, which protects the integrity of the BH horizon by spontaneously emitting low-energy ($\omega<m\Omega$) fermions. The final destiny of charged BHs is still unclear, as quantum effects may still play an important role~\cite{Richartz:2011vf} (and references therein).

\item As we discussed in Section~\ref{sec:magnetic}, BHs in strong magnetic fields are unstable. Because these are confining geometries, the lesson from AdS spacetimes (see Section~\ref{sec:KerrAdS})
implies that non-axially symmetric BH solutions should exist. These would be interesting to construct, even if only numerically.

\item One of the most exciting topics concerns the experimental verification of the theoretical results.
Rotational superradiance was recently seen in the laboratory using fluids~\cite{Torres:2016iee}, as we summarized in Section~\ref{sec:lab2}.
Important open issues concern validation of the previous results by independent teams, or measurement of superradiance in other setups, such as ``photon fluids''
~\cite{Prain:2019jqk}. The realization of the ergoregion instability or of instabilities triggered by massive fluctuations is specially interesting, in light of all
the important phenomenology and physics associated.

\item The superradiant instability triggered by plasma (Sec.~\ref{plasma-triggered}) is poorly studied and 
several interesting extensions are worth investigating. For example, the dynamics of an EM field in a plasma is certainly more involved than the Proca equation (with an effective mass given by the plasma frequency 
$\omega_p$), even at the linear level~\cite{Cannizzaro:2020uap}. Furthermore, relativistic and nonlinear 
effects (in particular, backreaction of growing EM fields on the plasma distribution and on the transparency of plasmas at certain frequencies) might affect the propagation and trapping of photons and they should be included in more sophisticated analyses~\cite{1970PhFl...13..472K,1971PhRvL..27.1342M,Cardoso:2020nst,Blas:2020kaa}.

\item An intriguing mechanism to trigger instabilities in astrophysical systems concerns the ergoregion instability in {\it fluids}, such
as accretion disks around gravitational BHs. In an analogue description, sound waves in these systems are described by an effectively-curved background geometry~\cite{Das:2004wf,Das:2006an,Chaverra:2015aya}. When the accretion disk velocity surpasses the local {\it sound} speed, an acoustic ergoregion appears, presumably giving rise to ergoregion instabilities.
As far as we are aware, these phenomena have not been explored.

\item We mentioned in Section~\ref{sec:BH_Lorentz_violating} that nonlinear ergoregion instabilities are thought to occur for BHs with multiple horizons in Lorentz-violating theories. Explicit examples do not exist yet.

\item Superradiance of self-interacting fields, or fields with nontrivial dispersion relations have hardly been 
explored, with a noteworthy (but one-spatial dimensional) toy-model~\cite{Richartz:2012bd}. Recently, the role of 
nonlinearities in the superradiant instability triggered by massive scalars has been discussed in 
Ref.~\cite{Fukuda:2019ewf} but further studies are needed.

\end{itemize}

%%%%%%%%%%%%%%%%%%%%%%%%%%%%%%%%%%%%%%%%%%%%%%%%%%%%%%%%%%%%%%%%%%%%%%%%%%
\clearpage
\newpage
\section{Black hole superradiance in astrophysics}\label{sec:astrophysics}
%%%%%%%%%%%%%%%%%%%%%%%%%%%%%%%%%%%%%%%%%%%%%%%%%%%%%%%%%%%%%%%%%%%%%%%%%%
BHs are one of the most striking predictions of Einstein's GR, or of any relativistic theory of gravity~\cite{Berti:2015itd}. Since Schmidt's identification
of the first quasar~\cite{Schmidt:1963}, large consensus in the astronomy community has mounted that nearly all galactic 
centers harbor a supermassive BH and that compact objects with mass above $\sim 3M_\odot$ should be BHs (we discuss some alternatives to 
this paradigm in Sec.~\ref{sec:BHmimickers} below; see Ref.~\cite{Cardoso:2019rvt} for a recent review). Indeed, strong 
evidence exists that astrophysical BHs with masses ranging from few solar masses to billions of solar masses are 
abundant objects.

BHs with masses in the range $\sim 5-30\,M_\odot$ have been indirectly observed through the x-ray
emission from their accretion disk, whose inner edge can be associated
to their innermost stable circular orbit (ISCO). Since 2015, BHs with
masses up to $\sim 10^2\,M_\odot$ have been detected in compact binary coalescence GW event by laser
interferometers~\cite{LIGOScientific:2018mvr}. To date, the heaviest BH detected by LIGO-Virgo is the remnant of
GW170729, with mass $M\approx 80\,M_\odot$. In 2019, during the third observation run of
the LIGO-Virgo interferometers, binary BH mergers have been detected on a weekly basis.

GR's uniqueness theorems imply a very strong prediction: all isolated, vacuum BHs in the Universe are described by the 
two-parameter Kerr family. Not only does this implies that BHs are perfect testbeds for strong-gravity effects due to 
their simplicity, but it also means that observing any deviation from this ``Kerr paradigm'' --~a goal within the reach 
of current~\cite{LIGO,VIRGO} and future~\cite{Doeleman:2008qh,KAGRA,ET,ELISA} GW and 
EM~\cite{Lu:2014zja,GRAVITY,Akiyama:2019cqa} facilities~-- would 
inevitably imply novel physics beyond GR~\cite{Cardoso:2019rvt}. Finally, the equivalence principle guarantees that gravity couples universally to matter.
Altogether, these properties suggest that predictions based on gravitational effects of extra fields around BHs should be very solid.

\begin{figure}[ht]
\begin{center}
\begin{tabular}{c}
\epsfig{file=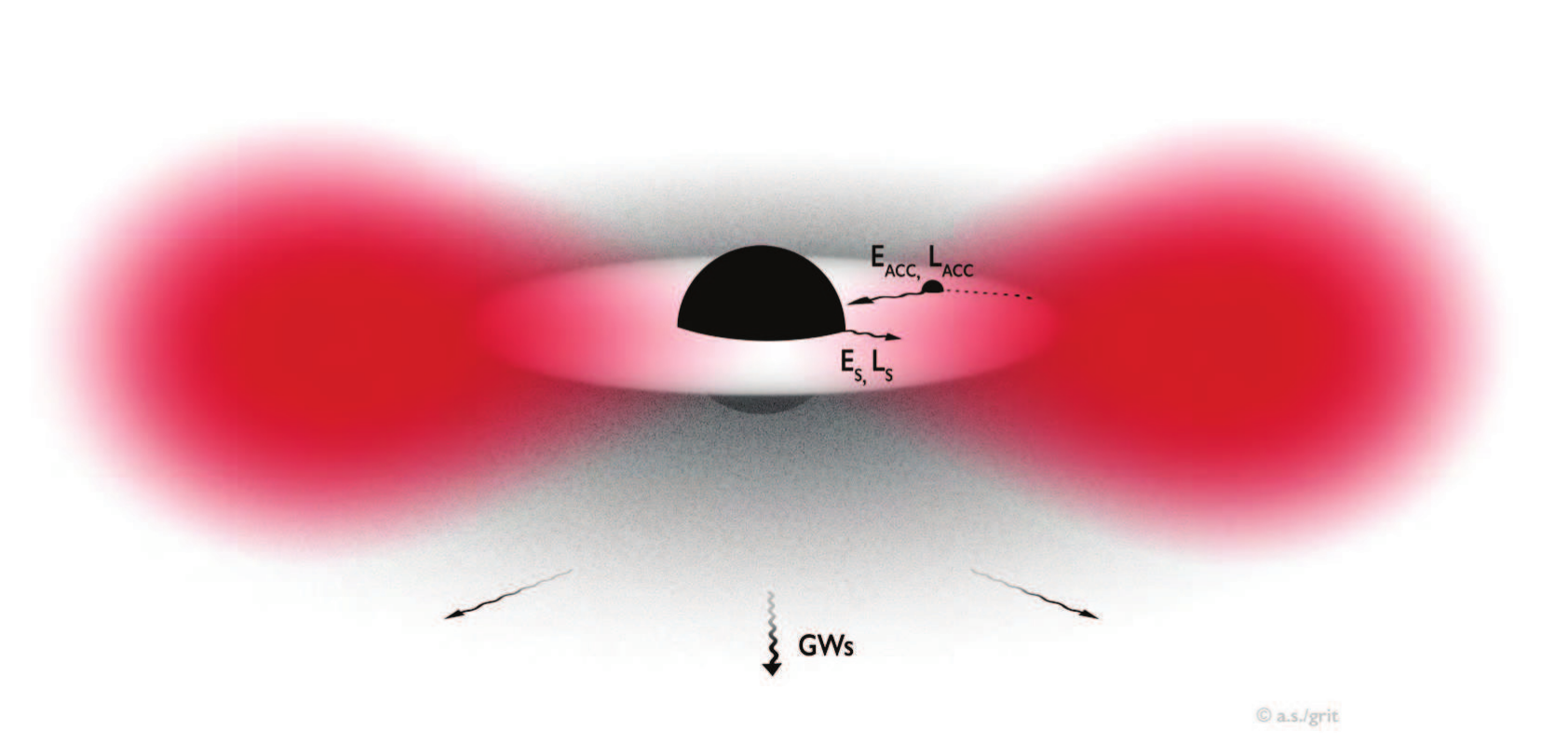,width=0.7\textwidth,angle=0,clip=true}
\end{tabular}
\end{center}
\caption{\label{fig:draw}
Pictorial description of a bosonic cloud around a spinning BH in a realistic astrophysical environment. The BH loses energy $E_S$ and angular momentum $L_S$ through superradiant extraction of scalar waves and emission of GWs, while accreting gas from the disk, which transports energy $E_{\rm ACC}$ and angular momentum $L_{\rm ACC}$. Notice that accreting material is basically in free fall after it reaches the ISCO. A scalar cloud would be localized at a distance $\sim 1/M\mu_S^2>2M$.}
\end{figure}
%

%%%%%%%%%%%%%%%%%%%%%%%%%%%%%%%%%%%%%%%%%%%%%%%%%%%%%%%%%%%%%%%%%%%%%%%%%%%%%%%%%%%%%%%%%%%%%%%%%%%%%%%%%%%%%%%%%%%%%%%%
\subsection{Evolution of superradiant instabilities for astrophysical BHs}\label{sec:evolution}
%%%%%%%%%%%%%%%%%%%%%%%%%%%%%%%%%%%%%%%%%%%%%%%%%%%%%%%%%%%%%%%%%%%%%%%%%%%%%%%%%%%%%%%%%%%%%%%%%%%%%%%%%%%%%%%%%%%%%%%%
We saw in the previous chapter that quantum or classical fluctuations of \emph{any} massive bosonic field can trigger a superradiant instability of the Kerr metric, whose time scale $\tau$ can be extremely short when compared to typical astrophysical time scales. For a BH with mass $M$, the shortest instability time scale (occurring for an optimal value of the coupling is $\tau\sim \left(M/10^6 M_\odot\right){\rm yr}$ for a ultralight scalar~\cite{Cardoso:2005vk,Dolan:2007mj,Pani:2012vp,Witek:2012tr,Brito:2013wya}, and shorter for vector~\cite{Pani:2012vp,Pani:2012bp,Witek:2012tr} and tensor fields~\cite{Brito:2013wya} for which superradiance is more efficient (cf. Sec.~\ref{sec:mass}).

As we discussed in Sec.~\ref{sec:mass}, the nonlinear development of the superradiant instability leads to the formation of a nonspherical bosonic cloud near the BH on a time scale $\tau$ that extracts energy and angular momentum from the BH until superradiance stops. On even longer time scales the cloud is slowly dissipated through GW emission~\cite{Arvanitaki:2009fg,Arvanitaki:2010sy, Witek:2012tr,Okawa:2014nda,Cardoso:2013krh,Brito:2014wla,Yoshino:2015nsa,Brito:2017zvb,East:2017ovw,East:2018glu}. During the evolution, the BH acquires an effective ``hair'' as pictorially depicted in Fig.~\ref{fig:draw}. This process effectively occurs for all unstable modes, but during the finite lifetime of a BH only the most unstable modes grow through superradiance.

In recent years superradiant instabilities have been used \emph{to turn astrophysical BHs into effective particle detectors}, by exploiting the prediction that putative ultralight bosons (cf. Sec.~\ref{sec:ultralight_motivations}) would make such massive BHs superradiantly unstable, in disagreement with current observations of spinning BHs. In addition, the GWs emitted by the cloud would lead to very specific quasi-monochromatic signals that would be a smoking-gun feature of the existence of ultralight bosons. These exciting possibilities are discussed in Sec.~\ref{sec:bounds_mass} below. 
However, before venturing in the astrophysical implications of superradiant instabilities, we need to assess how effects such as GW emission and gas accretion affect the development of the process.

%%%%%%%%%%%%%%%%%%%%%%%%%%%%%%%%%%%%%%%%%%%%%%%%%%%%%%%%%%%%%%%%%%%%%%%%%
\subsubsection{A simplified model including gas accretion}\label{sec:simplifiedmodel}
%%%%%%%%%%%%%%%%%%%%%%%%%%%%%%%%%%%%%%%%%%%%%%%%%%%%%%%%%%%%%%%%%%%%%%%%%%
The evolution of the superradiant instability including the effect of GW emission and of gas accretion can be studied with a simple quasi-adiabatic, fully-relativistic model~\cite{Brito:2014wla}. For simplicity, let us focus on the case of a (generically complex) massive scalar field minimally coupled to gravity.
The starting point of the analysis is the action~\eqref{eq:MFaction} with vanishing gauge field.
As previously discussed following the development of the instability in a fully nonlinear evolution is extremely challenging because of the time scales involved: 
$\tau_{\rm BH}\sim M$ is the light-crossing time, $\tau_S\sim 1/\mu_S$ is the typical oscillation period of the scalar cloud, and $\tau \sim M/(M\mu_S)^9$ is the instability time scale in the small-$M\mu_S$ limit. Even in the most favorable case for the instability $\tau\sim 10^6\tau_S$ is the minimum evolution time scale required for the superradiant effects to become noticeable. 
This requires extremely long and time-consuming simulations.
However, in such configuration the system is suitable for a \emph{quasi-adiabatic approximation}: over the dynamical time scale of the BH the scalar field can be considered almost stationary and its backreaction on the geometry can be neglected as long as the scalar energy is small compared to the BH mass~\cite{Brito:2014wla}.

At leading order, the geometry is described by the Kerr spacetime and the scalar evolves in this fixed background. For small mass couplings $M\mu_S$, the spectrum of the scalar perturbations admits the hydrogenic-like solution~\eqref{omegaDetweiler}, whereas the eigenfunctions are given in Eq.~\eqref{eigenfunctionDetweiler}~\cite{Detweiler:1980uk,Yoshino:2013ofa}. Their typical length scale is given by Eq.~\eqref{peak} and thus extends well beyond the horizon, where rotation effects can be neglected. 
The analytical result is a good approximation to the numerical eigenfunctions for moderately large couplings, $\mu_S M\lesssim0.2$, even at large BH spin~\cite{Brito:2014wla}. 

In the quasi-adiabatic approximation (and focusing on the $l=m=1$ fundamental mode), the cloud is stationary and described by Eq.~\eqref{scalar},
where the amplitude $A_0$ can be expressed in terms of the mass $M_S$ of the scalar cloud through Eq.~\eqref{amplitude}.
This dipolar cloud will emit GWs with frequency $ 2\pi/\lambda\sim 2\omega_R\sim 2\mu_S$. As previously discussed, the emission is incoherent so the quadrupole formula does not apply~\cite{Arvanitaki:2010sy,Yoshino:2013ofa,Brito:2014wla,Yoshino:2015nsa}. 
By performing a fully relativistic analysis within the Teukolsky formalism, Ref.~\cite{Brito:2014wla,Brito:2017zvb} found that the energy and angular-momentum fluxes of gravitational radiation emitted from the cloud are given by Eqs.~\eqref{dEdtF} and \eqref{dJdtF}.

%%%%%%%%%%%%%%%%%%%%%%%%%%%%%
\paragraph{Gas accretion}
%%%%%%%%%%%%%%%%%%%%%%%%%%%%%
Astrophysical BHs are not in isolation but surrounded by matter fields in the form of gas and plasma. On the one hand, addition of mass and angular momentum to the BH via accretion competes with superradiant extraction. On the other hand, a slowly-rotating BH which does not satisfy the superradiance condition might be spun up by accretion and become superradiantly unstable precisely \emph{because} of angular momentum accretion. Likewise, for a light BH whose coupling parameter $\mu_S M$ is small, superradiance might be initially negligible but it can become important as the mass of the BH grows through gas accretion. It is therefore crucial to include accretion in the treatment of BH superradiance.

Reference~\cite{Brito:2014wla} considered a conservative and simple model in which mass accretion occurs at a fraction of the Eddington rate (see e.g.~\cite{Barausse:2014tra}):
\begin{equation}
 \dot M_{\rm ACC} \equiv f_{\rm Edd} \dot M_{\rm Edd}\sim 0.02 f_{\rm Edd} \frac{M(t)}{10^6 M_\odot} M_\odot {\rm yr}^{-1}\,.\label{dotMaccr}
\end{equation}
%%%
The formula above assumes an average value of the radiative efficiency $\eta\approx0.1$, as required by Soltan-type arguments, i.e. a comparison between the luminosity of active galactic nuclei and the mass function of BHs~\cite{LyndenBell:1969yx,Soltan:1982vf}. The Eddington ratio for mass accretion, $f_{\rm Edd}$, depends on the details of the accretion disk surrounding the BH and it can be of order unity (or even larger for ultraluminous sources) for quasars and active galactic nuclei, whereas it is typically much smaller for quiescent galactic nuclei (e.g. $f_{\rm Edd}\sim 10^{-9}$ for SgrA$^{*}$). If we assume that mass growth occurs via accretion through Eq.~\eqref{dotMaccr}, the
BH mass grows exponentially with $e$-folding time given by a fraction $1/f_{\rm Edd}$ of the Salpeter time scale
%%%%
\begin{equation}
 \tau_{\rm Salpeter}=\frac{\sigma_T}{4\pi m_p}\sim4.5\times 10^7~{\rm yr}\,, \label{Salpeter}
\end{equation}
%%%%
where $\sigma_T$ is the Thompson cross section and $m_p$ is the proton mass. Therefore, the minimum time scale for the BH spin to grow via gas accretion is roughly $\tau_{\rm ACC}\sim \tau_{\rm Salpeter}/f_{\rm Edd}\gg \tau_{\rm BH}$ and also in this case the adiabatic approximation is well justified.

Regarding the evolution of the BH angular momentum through accretion, Ref.~\cite{Brito:2014wla} made the conservative assumption that the disk lies on the equatorial plane and extends down to the ISCO. 
If not, angular momentum increase via accretion is suppressed and superradiance becomes (even) more dominant.
Ignoring radiation effects\footnote{In the absence of superradiance the BH would reach extremality in finite time, whereas radiation effects set an upper bound of $a/M\sim 0.998$~\cite{Thorne:1974ve}. To mimic this upper bound in a simplistic way, a smooth cutoff in the accretion rate for the angular momentum can be introduced~\cite{Brito:2014wla}. This cutoff merely prevents the BH to reach extremality and does not play any role in the evolution.}, the evolution equation for the spin reads~\cite{Bardeen:1970zz}
\begin{equation}
\dot J_{\rm ACC} \equiv \frac{L(M,J)}{E(M,J)} \dot M_{\rm ACC}\,,\label{dotJaccr}
\end{equation}
%%%
where $L(M,J)=2M/(3\sqrt{3})\left(1+2 \sqrt{3 r_{\rm ISCO}/{M}-2}\right)$ and $E(M,J)=\sqrt{1-2M/3r_{\rm ISCO}}$ are the angular momentum and energy per unit mass, respectively, of the ISCO of the Kerr metric, located at $r_{\rm ISCO}=r_{\rm ISCO}(M,J)$ in Boyer-Lindquist coordinates.

%%%%%%%%%%%%%%%%%%%%%%%%%%%%%%%%%%%%%%%%%%%%%%%%%%%%%%%%%%%%%%%%%%%%%%%%%%%%%%%%%%
\paragraph{Growth and decay of bosonic condensates around spinning black holes}
%%%%%%%%%%%%%%%%%%%%%%%%%%%%%%%%%%%%%%%%%%%%%%%%%%%%%%%%%%%%%%%%%%%%%%%%%%%%%%%%%%
The evolution of the cloud is governed by a simple set of differential equations~\cite{Brito:2014wla}.
Energy and angular momentum conservation requires that
%%%%
\begin{eqnarray}
 \dot{M}+\dot{M}_S&=&-\dot{E}_{\rm GW}+\dot M_{\rm ACC}\,,\\
 \dot{J}+\dot{J}_S&=&-\frac{1}{{\mu_S}}\dot{E}_{\rm GW}+\dot J_{\rm ACC}\,,
\end{eqnarray}
%%%%
where $M_S$ and $J_S$ are the mass and the angular momentum of the scalar cloud, we have neglected the subdominant contributions of the mass of the disk and of those GWs that are absorbed at the horizon, and we have approximated the local mass and angular momentum by their ADM counterparts. The latter approximation is valid as long as backreaction effects are small, as we discuss below. The system is closed by two further equations
%%%%
\beq
\dot{M}&=& -\dot{E}_S+\dot M_{\rm ACC} \,,\\
\dot{J}&=& -\frac{1}{{\mu_S}}\dot{E}_S +\dot J_{\rm ACC}\,,
\eeq
which describe the superradiant extraction of energy and angular momentum and the competitive effects of gas accretion at the BH horizon. In the equations above we have introduced the scalar energy flux that is extracted from the horizon through superradiance,
\be
\dot{E}_S=2M_S\omega_I\,,
\ee
where $M\omega_I=\frac{1}{48}({a/M-2{\mu_S} r_+})(M{\mu_S})^9$ for the $l=m=1$ fundamental mode. These equations assume that the scalar cloud is not directly (or only very weakly)  coupled to the disk.

\begin{figure*}[ht]
\begin{center}
\begin{tabular}{ccc}
\epsfig{file=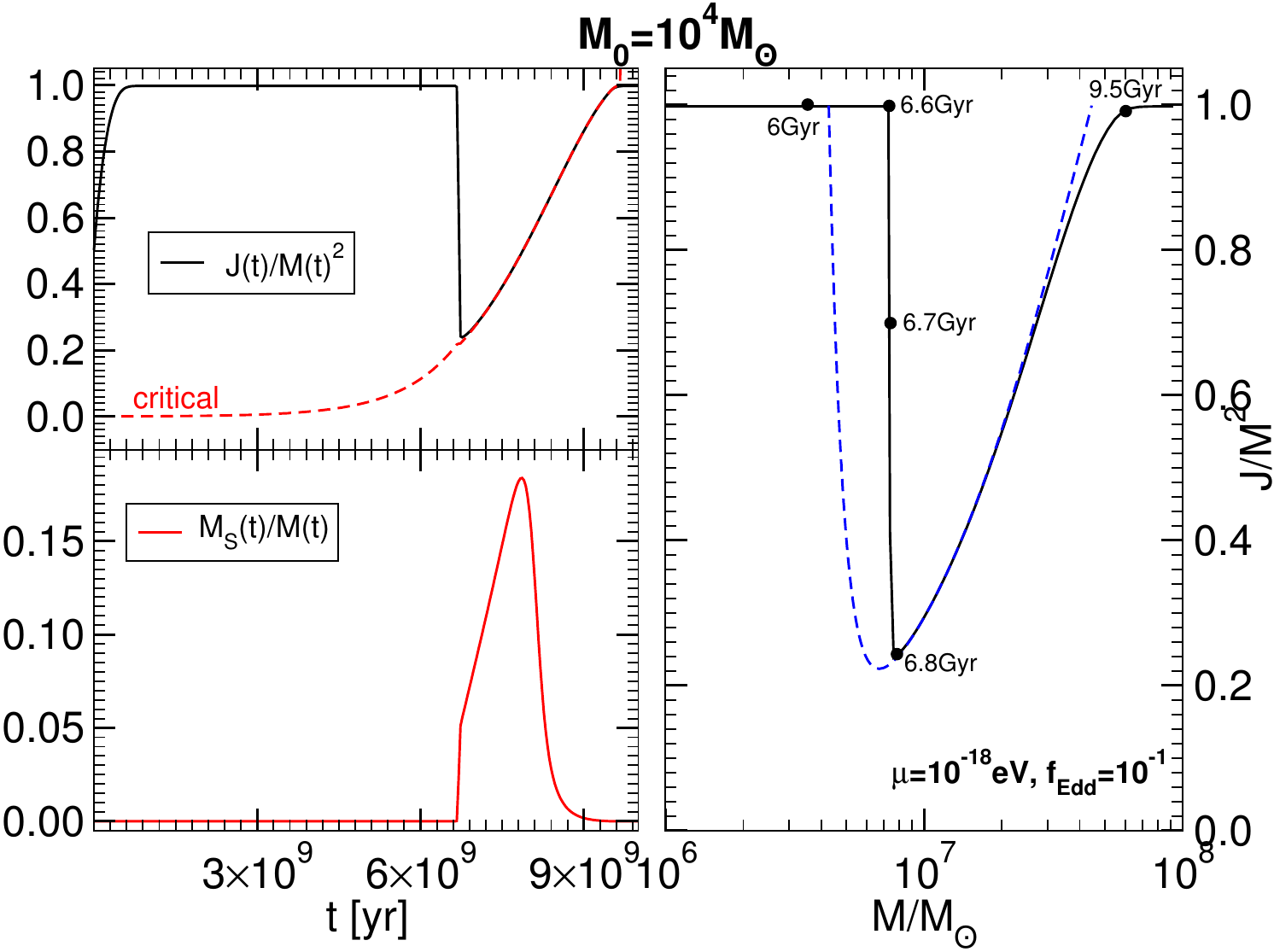,width=0.49\textwidth,angle=0,clip=true}&\hspace{0.3cm}&
\epsfig{file=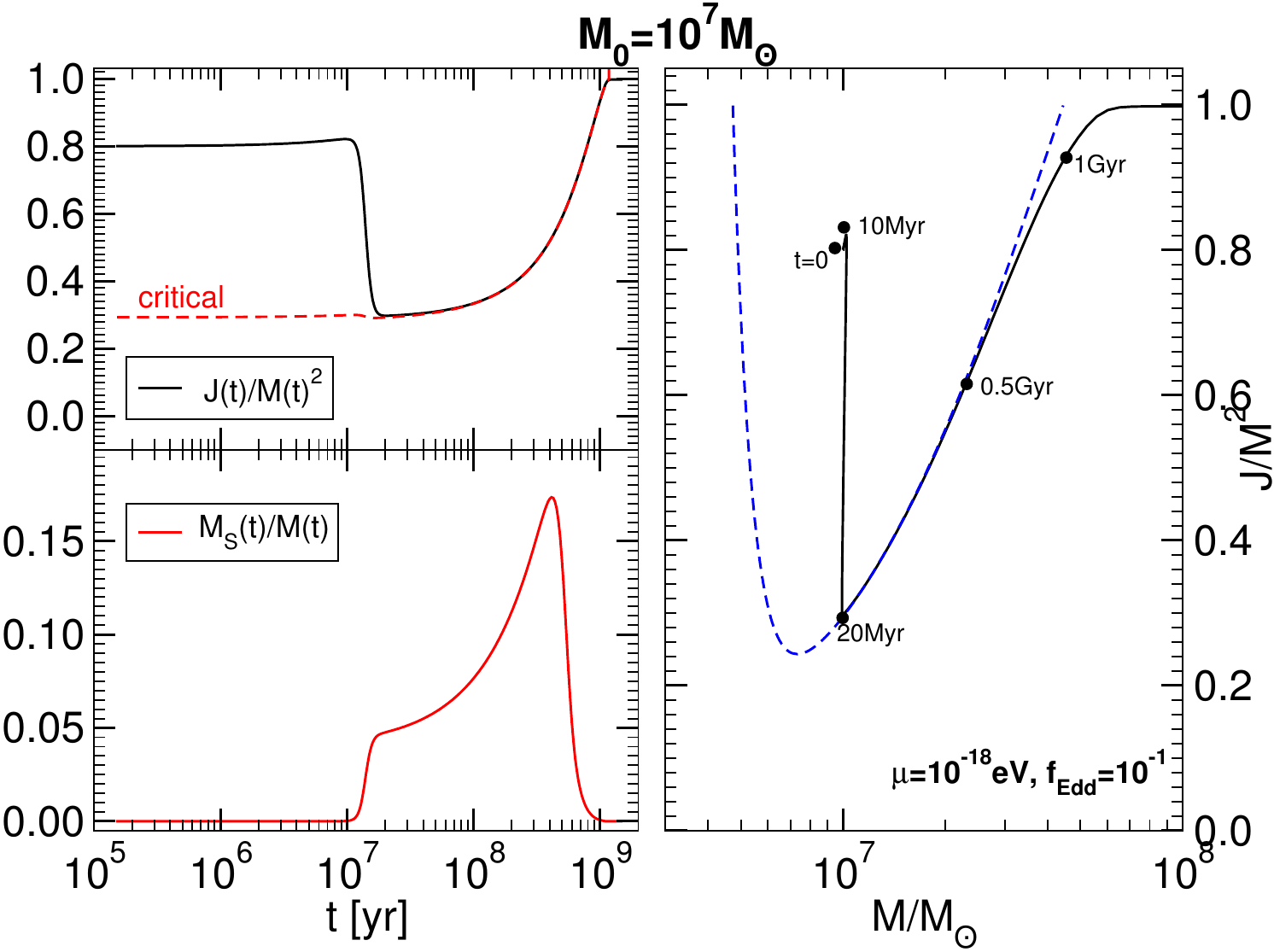,width=0.49\textwidth,angle=0,clip=true}
%%%%%
\end{tabular}
\end{center}
\caption{\label{fig:evolution}
Evolution of the BH mass and spin, and of the scalar cloud mass due to superradiance, accretion of gas and emission of GWs. The two sets of plots show two different cases. In Case I (left set) the initial BH mass $M_0=10^4 M_\odot$ and the initial BH spin $J_0/M_0^2=0.5$. The BH enters the instability region at about $t\sim 6\,{\rm Gyr}$, when its mass $M\sim10^7 M_\odot$ and its spin is quasi-extremal. The set of plots on the right shows Case II, in which $M_0=10^7M_\odot$ and $J_0/M_0^2=0.8$, and the evolution starts already in the instability region for this scalar mass $\mu=10^{-18}{\rm eV}$. For both cases, the left top panels show the dimensionless angular momentum $J/M^2$ and the critical superradiant threshold $a_{\rm crit}/M$ (cf. Eq.~\eqref{acrit}); the left bottom panels show the mass of the scalar cloud $M_S/M$ (note the logarithmic scale in the x-axis for Case II); and the right panels show the trajectory of the BH in the \emph{Regge plane}~\cite{Arvanitaki:2010sy} during the evolution. The dashed blue line denotes the depleted region as estimated by the linearized analysis, i.e. it marks the threshold at which $\tau\sim\tau_{\rm ACC}$. From Ref.~\cite{Brito:2014wla}.
}
\end{figure*}

Representative results for the evolution of the system are shown in Fig.~\ref{fig:evolution} where we consider the scalar-field mass $\mu_S=10^{-18}{\rm eV}$ and mass accretion near the Eddington rate, $f_{\rm Edd}=0.1$. We consider two cases: (I) the left set of plots corresponds to a BH with initial mass $M_0=10^4 M_\odot$ and initial spin $J_0/M_0^2=0.5$, whereas (II) the right set of plots corresponds to $M_0=10^7 M_\odot$ and $J_0/M_0^2=0.8$.

In Case I, superradiance is initially negligible because $\mu_S M_0\sim 10^{-4}$ and superradiant extraction is suppressed. Thus, the system evolves mostly through gas accretion, reaching extremality ($J/M^2\sim0.998$) within the time scale $\tau_{\rm ACC}\sim 10\,\tau_{\rm Salpeter}$. At about $t\sim 6\,{\rm Gyr}$, the BH mass is sufficiently large that the superradiant coupling $\mu_S M$ becomes important. This corresponds to the BH entering the region delimited by a dashed blue curve in the Regge plane~\cite{Arvanitaki:2010sy} shown in Fig.~\ref{fig:evolution} for Case I. At this stage superradiance becomes effective very quickly: a scalar cloud grows exponentially near the BH (left bottom panel), while
mass and angular momentum are extracted from the BH (left top panel). This abrupt phase lasts until the BH spin reaches the critical value $a_{\rm crit}/M$ [cf Eq.~\eqref{acrit}] and superradiance halts. Because the initial growth is exponential, the evolution does not depend on the initial mass and initial spin of the scalar cloud as long as the latter are small enough, so that in principle also a quantum fluctuation would grow to a sizeable fraction of the BH mass in finite time. 

Before the formation of the scalar condensate, the evolution is the same regardless of GW emission and the only role of accretion is to bring the BH into the instability window.  After the scalar growth, the presence of GW dissipation and accretion produces two effects: (i) the scalar condensate loses energy through the emission of GWs, as shown in the left bottom panel of Fig.~\ref{fig:evolution} [the signatures of this GW emission are discussed in Sec.~\ref{sec:GWs} below]; (ii) gas accretion returns to increase the BH mass and spin.

However, because accretion restarts in a region in which the superradiance coupling $\mu_S M$ is nonnegligible, the ``Regge trajectory'' $J(t)/M(t)^2\sim a_{\rm crit}/M$ (cf. Eq.~\eqref{acrit}) is an attractor for the evolution and the BH ``stays on track'' as its mass and angular momentum grow. For Case I, this happens between $t\sim 6.8{\rm Gyr}$ and $t\sim9.5\,{\rm Gyr}$, i.e. the Regge trajectory survives until the spin reaches the critical value $J/M^2\sim 0.998$ and angular momentum accretion saturates.

A similar discussion holds true also for Case II, presented in the right set of plots in Fig.~\ref{fig:evolution}. In this case, the BH starts already in the instability regime, its spin grows only very little before superradiance becomes dominant, and the BH angular momentum is extracted in about $10\,{\rm Myr}$. After superradiant extraction, the BH evolution tracks the critical value $a_{\rm crit}/M$ while the BH accretes over a time scale of $1\,{\rm Gyr}$. 

%%%%%%%%%%%%%%%%%%%%%%%%%%%%%%%%%%%%%%%%%%%%%%%%%%%%%%%%%%%%%%%%%%%%%%%%%%%%%%%
\subsubsection{Superradiant instabilities imply no highly-spinning black holes}\label{sec:nospinningBHs}
%%%%%%%%%%%%%%%%%%%%%%%%%%%%%%%%%%%%%%%%%%%%%%%%%%%%%%%%%%%%%%%%%%%%%%%%%%%%%%%

While GW emission is always too weak to affect the evolution of the BH mass and spin (nonetheless being responsible for the decay of the scalar condensate, as shown in Fig.~\ref{fig:evolution}), accretion 
plays a more important role. From Fig.~\ref{fig:evolution}, it is clear that accretion produces two effects. First, for BHs which initially are not massive enough to be in the superradiant instability region, accretion can bring them to the instability window by feeding them mass as in Case~I. Furthermore, when $J/M^2\to a_{\rm crit}/M$ the superradiant instability is exhausted, so that accretion is the only relevant process and the BH inevitably spins up again. This accretion phase occurs in a very peculiar way, with the dimensionless angular momentum following the trajectory $J/M^2\sim a_{\rm crit}/M$ over very long time scales.

Therefore, a very solid prediction of BH superradiance is that supermassive BHs would move on the Regge plane following the bottom-right part of the superradiance threshold curve. The details of this process depend on the initial BH mass and spin, on the scalar mass $\mu_S$ and on the accretion rate. A relevant problem concerns the \emph{final} BH state at the time of observation; namely, given the observation of an old BH and the measurement of its mass and spin, would these measurements be compatible with the evolution depicted in Fig.~\ref{fig:evolution}? 
\begin{figure}[ht]
\begin{center}
\begin{tabular}{c}
\epsfig{file=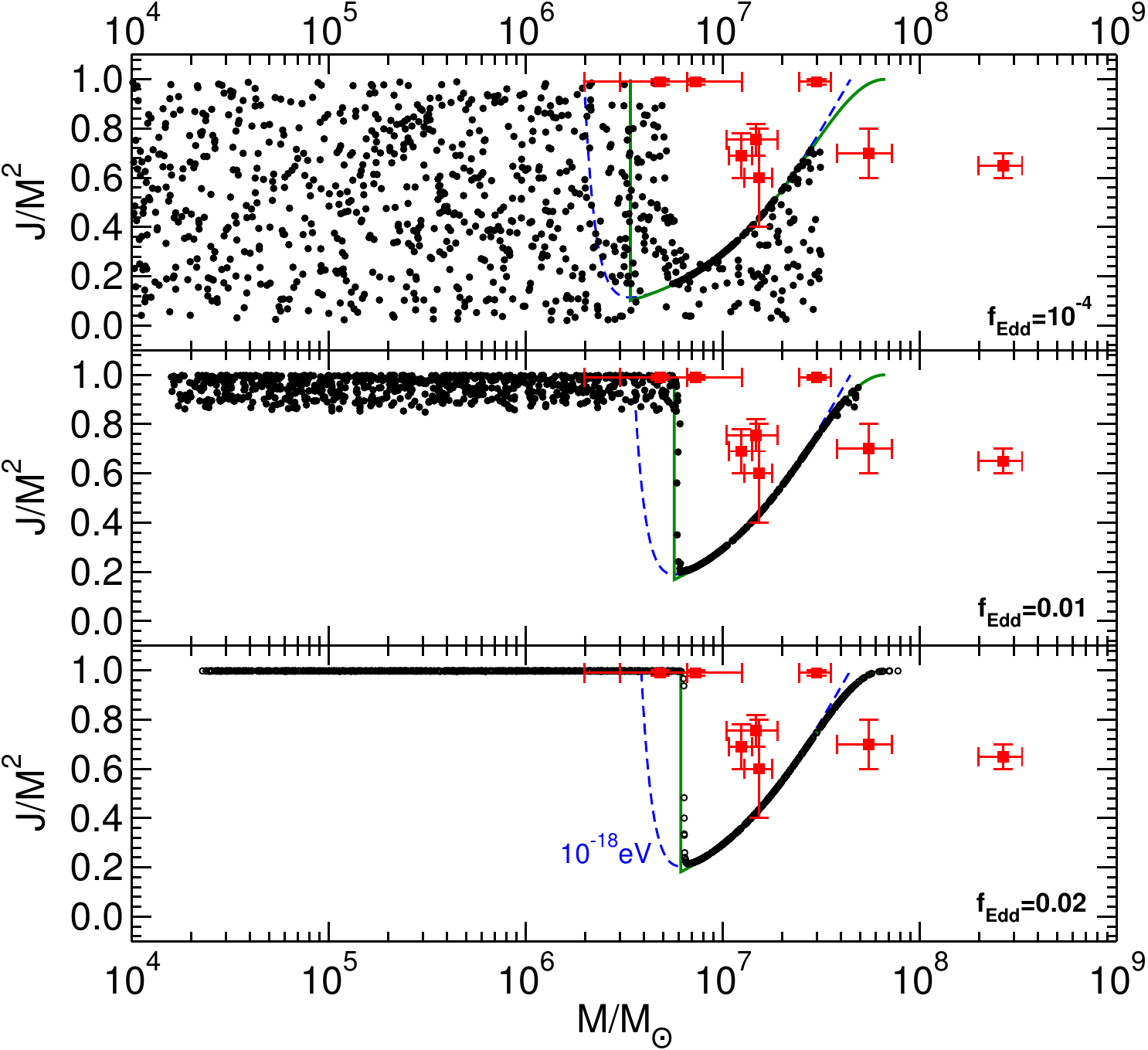,width=0.65\textwidth,angle=0,clip=true}
\end{tabular}
\end{center}
\caption{\label{fig:ReggeMC}
The final BH mass and spin in the Regge plane for initial data consisting of $N=10^3$ BHs with initial mass and spin randomly distributed between $\log_{10}M_0\in[4,7.5]$ and $J_0/M_0^2\in[0.001,0.99]$. The BH parameters are then extracted at $t=t_F$, where $t_F$ is distributed on a Gaussian centered at $\bar t_{F}\sim 2\times 10^9{\rm yr}$ with width $\sigma=0.1\bar t_{F}$. We considered $\mu_S=10^{-18}{\rm eV}$. The dashed blue line is the prediction of the linearized analysis obtained by comparing the superradiant instability time scale with the accretion time scale, $\tau\approx\tau_{\rm Salpeter}/f_{\rm Edd}$, whereas the solid green line denotes the region defined through Eq.~\eqref{region}. Old BHs do not populate the region above the green threshold curve. The experimental points with error bars refer to the supermassive BHs listed in Ref.~\cite{Brenneman:2011wz}. From Ref.~\cite{Brito:2014wla}.
}
\end{figure}

This problem is addressed in Fig.~\ref{fig:ReggeMC}, which shows the final BH mass and spin in the Regge plane~\cite{Arvanitaki:2009fg,Arvanitaki:2010sy} (i.e. a BH mass-spin diagram) for $N=10^3$ Monte Carlo evolutions. We consider a scalar field mass $\mu=10^{-18}{\rm eV}$ and three different accretion rates $f_{\rm Edd}$ (which, we recall, is defined as the fraction of mass accretion rate relative to the Eddington limit) and, in each panel, we superimpose the bounds derived from the linearized analysis, i.e. the threshold line when the instability time scale equals the accretion time scale (cf. Sec.~\ref{sec:bounds_mass} below for details). As a comparison, in the same plot we include the experimental points for the measured mass and spin of some supermassive BHs listed in Ref.~\cite{Brenneman:2011wz}.

Various comments are in order. First, it is clear that the higher the accretion rate the better the agreement with the linearized analysis. This seemingly counter-intuitive result can be understood by the fact that higher rates of accretion make it more likely to find BHs that have undergone a superradiant instability phase over our observational time scales. In fact, for high accretion rates it is very likely to find supermassive BHs precisely on the ``Regge trajectory''~\cite{Arvanitaki:2009fg,Arvanitaki:2010sy} given by $J/M^2\sim a_{\rm crit}/M$ (cf. Eq.~\eqref{acrit}).

Furthermore, for any value of the accretion rate, we always observe a depleted region (a ``hole'') in the Regge plane~\cite{Arvanitaki:2009fg,Arvanitaki:2010sy}, which is not populated by old BHs. While the details of the simulations might depend on the distribution of initial mass and spin, this qualitative result is very solid and is a generic feature of the evolution. For the representative value $\mu_S=10^{-18}{\rm eV}$ adopted here, the depleted region is incompatible with observations~\cite{Brenneman:2011wz}. Similar results would apply for different values\footnote{Note that, through Eq.~\eqref{dotMaccr}, the mass accretion rate only depends on the combination $f_{\rm Edd} M$, so that a BH with mass $M=10^6 M_\odot$ and $f_{\rm Edd}\sim10^{-3}$ would have the same accretion rate of a smaller BH with $M=10^{4} M_\odot$ accreting at rate $f_{\rm Edd}\sim10^{-1}$. Because this is the only relevant scale for a fixed value of $\mu_S M$, in our model the evolution of a BH with different mass can be obtained from Fig.~\ref{fig:evolution} by rescaling $f_{\rm Edd}$ and $\mu_S$.} of $\mu_S$ in a BH mass range such that $\mu_S M\lesssim1$. Therefore, as discussed in Refs.~\cite{Arvanitaki:2009fg,Arvanitaki:2010sy,Pani:2012vp,Brito:2013wya} and reviewed in Sec.~\ref{sec:bounds_mass} below, observations of massive BHs with various masses can be used to rule out various ranges of the boson-field mass $\mu_S$.

Finally, Fig.~\ref{fig:ReggeMC} suggests that when accretion and GW emission are properly taken into account, the holes in the Regge plane are smaller than what naively predicted by the relation $\tau\approx\tau_{\rm ACC}$, i.e. by the dashed blue curve in Fig.~\ref{fig:ReggeMC}. Indeed, a better approximation for the depleted region is~\cite{Brito:2014wla,Ficarra:2018rfu}  
%%%
\begin{equation}
 \frac{J}{M^2}\gtrsim \frac{a_{\rm crit}}{M}\sim 4\mu M \quad  \cup \quad M \gtrsim \left({\frac{96}{\mu ^{10} \tau_{\rm ACC}}}\right)^{1/9} \,,\label{region}
\end{equation}
%%%%
whose boundaries are shown in Fig.~\ref{fig:ReggeMC} by a solid green line. These boundaries correspond to the threshold value $a_{\rm crit}$ (cf. Eq.~\eqref{acrit}) for superradiance and to a BH mass which minimizes the spin for which $\tau\approx \tau_{\rm ACC}$, for a given $\mu$~\cite{Pani:2012bp}. As shown in Fig.~\ref{fig:ReggeMC}, the probability that a BH populates this region is strongly suppressed as the accretion rate increases.

\begin{figure}[thb]
\begin{center}
\begin{tabular}{cc}
 \epsfig{file=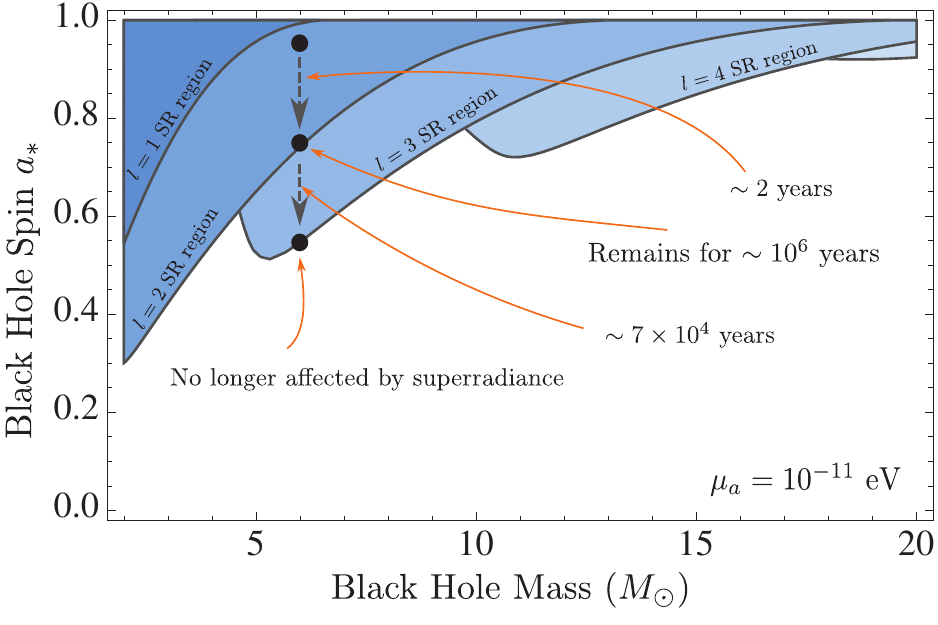,width=0.8\textwidth,angle=0,clip=true}
\end{tabular}
\caption{Holes in the Regge plane for an ultralight scalar with mass $\mu_S=10^{-11}{\rm eV}$ for various multipoles ($l=1,...,5$). Shaded regions correspond to BH parameters which would result in spindown within $10^6{\rm yr}$. The description refers to a representative evolution of a BH with $M=6 M_\odot$ and initial spin $a/M=0.95$. From Ref.~\cite{Arvanitaki:2014wva}.
  \label{fig:holes} }
\end{center}
\end{figure}

Although the instability is strongly suppressed for higher multipoles, the first few $(l,m)$ modes (and not only 
the dipole with $l=m=1$) can contribute to the depleted region in the Regge plane~\cite{Arvanitaki:2010sy}. Because the 
superradiance condition depends on the azimuthal number $m$, for certain parameters it might occur that the modes with 
$l=m=1$ are stable, whereas the modes with $l=m=2$ are unstable, possibly with a superradiant extraction stronger than 
accretion. When this is the case, the depleted region of the Regge plane is the union of various 
holes~\cite{Arvanitaki:2009fg,Arvanitaki:2010sy}, as shown in the schematic evolution depicted in Fig.~\ref{fig:holes} for a massive scalar field.

%%%

Finally, note that in the case of multiple modes with large initial amplitude the shape of the Regge gaps is more complex, mostly due to the absorption of nonsuperradiant modes~\cite{Ficarra:2018rfu}.

%%%%%%%%%%%%%%%%%%%%%%%%%%%%%%%%%%%%%%%%%%%%%%%%%%%%%%%%%%%%%%%%%%%%%%
\subsubsection{Summary of the evolution of superradiant instabilities}
%%%%%%%%%%%%%%%%%%%%%%%%%%%%%%%%%%%%%%%%%%%%%%%%%%%%%%%%%%%%%%%%%%%%%%

Because the results discussed above play an important role for the analysis of the next sections, it is relevant to summarize here the main features of the evolution of superradiant instabilities:

\begin{itemize}

 \item The mass of the cloud remains a sizeable fraction of the BH mass over cosmological times, so that such systems can be considered as (quasi)-stationary hairy BHs for any astrophysical purpose. Nonetheless, the energy-density in the scalar field is negligible because the cloud typically extends over very large distances. Therefore, the geometry is very well described by the Kerr metric during the entire evolution.
  
 \item The role of gas accretion is twofold. On the one hand, accretion competes against superradiant extraction of mass and angular momentum. On the other hand accretion may produce the optimal conditions for superradiance, for example by increasing the BH spin before the instability becomes effective or by ``pushing'' the BH into the instability region in the Regge plane. 
 
 \item A very generic prediction of BH superradiant instabilities is the existence of \emph{holes in the Regge plane}. For mass accretion near the Eddington rate, such depleted regions are very well described by Eq.~\eqref{region}, which refines the estimate obtained just by comparing the instability time scale against a typical accretion time scale (cf. Sec.~\ref{sec:bounds_mass} below). A more sophisticated analysis --~including radiative effects and the geometry of the disk~-- would be important to refine the bounds previously derived~\cite{Arvanitaki:2010sy,Kodama:2011zc,Pani:2012vp,Brito:2013wya}. 
 
\end{itemize}

Finally, the above discussion focused on the scalar case. However, the main qualitative features of the evolution of the 
superradiant instability extend also to the vector case. In that case, as discussed above, the case the superradiant instability is shorter and so is the GW emission time scale. The phenomenological consequences of these features are discussed in the next section.

%%%%%%%%%%%%%%%%%%%%%%%%%%%%%%%%%%%%%%%%%%%%%%%%%%%%%%%%%%%%%%%%%%%%%%%%%%%%%%%%%%%%
\subsection{Astrophysical black holes as particle detectors} \label{sec:bounds_mass}
%%%%%%%%%%%%%%%%%%%%%%%%%%%%%%%%%%%%%%%%%%%%%%%%%%%%%%%%%%%%%%%%%%%%%%%%%%%%%%%%%%%%
As we already alluded to in the previous section, the instabilities discussed in Sec.~\ref{sec:bombs} have important astrophysical implications that arise from the surprising connections between strong-field gravity and particle physics. 
One generic prediction of these instabilities is that --~over the superradiance time scale~-- isolated, massive BHs 
should not spin above the superradiant threshold. In other words, superradiant instabilities set an upper bound on the 
BH spin which is smaller than the theoretical Kerr bound for the absence of naked singularities. Another prediction is a 
peculiar emission of GWs through various channels, as discussed below. These effects have been investigated in the 
context of stringy axions and ultralight 
scalars~\cite{Arvanitaki:2009fg,Arvanitaki:2010sy,Kodama:2011zc,Yoshino:2012kn,Yoshino:2013ofa,Brito:2014wla,Arvanitaki:2014wva,
Arvanitaki:2016qwi,Brito:2017zvb,Brito:2017wnc,Cardoso:2018tly,Stott:2018opm,Davoudiasl:2019nlo} (with bounds being 
complementary to those from cosmological observations~\cite{Bozek:2014uqa,Hlozek:2014lca}), light vector 
fields~\cite{Pani:2012vp,Baryakhtar:2017ngi,Cardoso:2018tly,Davoudiasl:2019nlo}, and light tensor field, the latter 
also related to massive gravity and bimetric theories~\cite{Brito:2013wya,Brito:2020lup}.

Here, we present an overview of astrophysical signatures that can be used to either constrain the existence of new light particles or constrain their existence. As discussed, for a bosonic field of mass $\mu$, the only parameter regulating the strength of the gravitational coupling to a BH of mass $M$ is the dimensionless combination $\mu M$. The instability time scale is minimized when $2\mu M\sim 1$, i.e. when the Compton wavelength of the bosonic field is roughly comparable to the size of the BH. However, the details of the process depend on the nature of the bosonic field. As discussed in Sec.~\ref{sec:bombs}, for a given coupling $\mu M$ the instability time scale is shorter for bosonic fields with spin. % due to spin-spin interactions.

%%%%%%%%%%%%%%%%%%%%%%%%%%%%%%%%%%%%%%%%%%%%%%%%%%%%%%%%%%%%%%%%%%%%%%%%%%%%%%%%%%%%%%%%%%%%%%%%%%%%%%%%%%%%%%
\subsubsection{Bounds on the mass of bosonic fields from gaps in the Regge plane}\label{sec:bounds_mass_regge}
%%%%%%%%%%%%%%%%%%%%%%%%%%%%%%%%%%%%%%%%%%%%%%%%%%%%%%%%%%%%%%%%%%%%%%%%%%%%%%%%%%%%%%%%%%%%%%%%%%%%%%%%%%%%%%

A very generic and solid prediction of BH superradiant instabilities is the existence of holes in the Regge plane, as discussed in Sec.~\ref{sec:evolution}. 
The estimates for the instability time scale, together with reliable spin measurements for BHs in various mass ranges,
can be used to impose stringent constraints on the allowed mass
of ultralight 
bosons~\cite{Arvanitaki:2009fg,Arvanitaki:2010sy,Pani:2012vp,Brito:2013wya,Arvanitaki:2014wva,Arvanitaki:2016qwi,
Baryakhtar:2017ngi,Brito:2017zvb,Stott:2018opm,Cardoso:2018tly,Davoudiasl:2019nlo,Ng:2020ruv}. 

\paragraph{Measuring the BH spin}
BH spin is routinely obtained from the EM spectrum using reliable proxies for the position of the ISCO~\cite{Middleton:2014sma}. Both the (mass-independent) shape of the iron K${\alpha}$ line seen in reflection~\cite{Fabian:2000nu} and the thermal emission from the inner edge of the disc~\cite{Zhang:1997dy} --~assumed to be the ISCO~\cite{Shafee:2007sa,Penna:2010hu}~-- are commonly employed for both stellar mass BHs in binaries and supermassive BHs in active galactic nuclei (although the latter's use is a more recent development~\cite{Done:2011at}). While important caveats exist for both traditional approaches, convincing evidence for truncation at the ISCO comes from the consistent position of the inner disc edge in LMC X-3 from modeling of the thermal dominant state, providing a remarkably stable spin value over a baseline of $26~{\rm yr}$~\cite{Steiner:2010kd}. While not as well sampled (due to the source rarely entering the requisite thermal dominant state), Cygnus X-1 also shows a remarkably stable spin value over $14~{\rm yr}$ from the same approach~\cite{Gou:2009ks}.
These observations imply that at the moment LMC X-3 and Cygnus X-1 are not undergoing a superradiant instability \emph{at least} over a time scale of $26~{\rm yr}$ and $14~{\rm yr}$, respectively. This fact was used in Ref.~\cite{Cardoso:2018tly} to put \emph{direct} constraints on the mass of ultralight scalar and vector fields, as discussed below.
More stringent (albeit less direct) constraints come from comparing the instability time scale against a typical accretion time scale, that we estimate here to be the Salpeter time scale given in Eq.~\eqref{Salpeter}.

A novel approach to measure the masses and spins of astrophysical BHs comes from GW astronomy. Binary BHs are arguably the cleanest gravitational sources so measurements of the mass and spin of the binary components should be less affected by systematics than in the EM case.
While the spins of the primary and secondary objects in the coalescence events detected by LIGO/Virgo in the first two 
observation runs are affected by large uncertainties and are (marginally) compatible with zero spin for most sources 
(but see~\cite{LIGOScientific:2018mvr,Venumadhav:2019lyq} for a few events in which the effective spin of the binary is 
nonzero and \cite{Abbott:2020niy} for O3 events), future detections will provide more stringent constraints on the 
individual spins, at the level of $30\%$~\cite{TheLIGOScientific:2016pea}.
More precise measurement will come from the LISA space mission~\cite{Audley:2017drz}. LISA will be able to measure the mass and spin of binary BH components out to cosmological distances. Depending on the mass of BH seeds in the early Universe, LISA will also detect intermediate mass BHs and measure their mass and spin, thus probing the existence of
light bosonic particles in a large mass range (roughly $m_s\sim 10^{-13}$--$10^{-16}$~eV) that is inaccessible to EM observations of stellar and supermassive BHs and to Earth-based GW detectors.
\begin{figure}
\begin{center}
\includegraphics[width=0.65\textwidth]{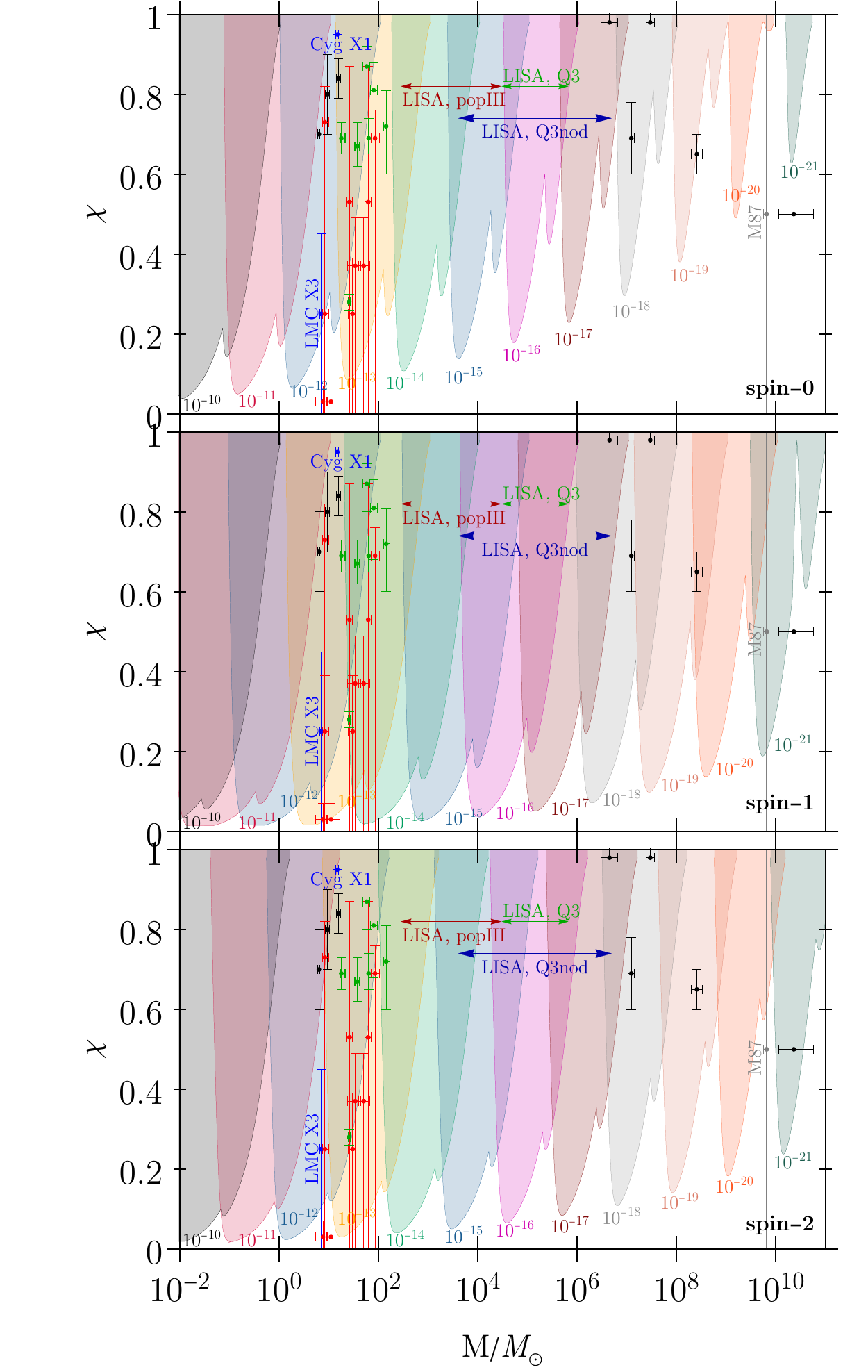}\\
\caption{
Exclusion regions in the BH spin-mass diagram obtained from the superradiant 
instability of Kerr BHs against massive bosonic fields for the two most unstable modes.
The top, middle, and bottom panels refer to scalar (spin-$0$), vector (spin-$1$), and tensor (spin-$2$) fields, 
respectively.
For each mass of the field (reported in units of eV), the separatrix corresponds to an instability time scale 
equal to the Salpeter time $\tau_S=2\times 10^7 {\rm\, yr\,}$. 
% PAOLO: i quote half od the Sapter time because am using Gamma=2*OmegaI
%
Note that the rightmost part of each curve is universal, $a\sim a_{\rm crit}$ [cf. Eq.~\eqref{acrit}], i.e. it does 
not depend on the spin of the field.
The meaning of the markers is explained in the main text.
BHs lying above each of these curves would be unstable on an observable time scale, and therefore each point can in 
principle rule out a range of the boson field masses. 
Adapted from 
Refs.~\cite{Arvanitaki:2014wva,Baryakhtar:2017ngi,Brito:2017zvb,Cardoso:2018tly,Brito:2020lup}.
In the spin-2 case we do not show results for the ``special'' dipolar mode~\cite{Brito:2013wya}, which might in 
principle provide stronger constraints but has been only computed to first order in the spin.
\label{fig:ReggePlane}
}
\end{center}
\end{figure}

\paragraph{Bounds from BH-spin measurements}
In order to quantify
the dependence of the boson mass bounds on the mass and spin of
astrophysical BHs, in Fig.~\ref{fig:ReggePlane} we show exclusion regions
in the BH Regge plane.
More precisely, using the results summarized in Sec.~\ref{sec:massive_unified}, we plot 
contours corresponding to an instability
time scale of the order of the Salpeter time for different masses
of the bosonic field and considering the unstable modes with the largest growth rate. From top to bottom, the three 
panels refer to a spin-0, spin-1, and spin-2 field, respectively.
The figure shows that observations of astrophysical BHs with $M_\odot\lesssim
M\lesssim 10^{10}M_\odot$ spinning above a certain threshold would exclude a wide range of boson-field masses. Because superradiance is 
stronger for bosonic fields with spin, the exclusion windows are wider as the spin of the field increases, and they also 
extend almost down to $J\sim0$ for spin-1 and spin-2 bosons. This feature is important because current spin measurements 
might be affected by large systematics. Nonetheless, it is clear from Fig.~\ref{fig:ReggePlane} that almost any BH spin 
measurement would exclude a considerable range of masses.

In Fig.~\ref{fig:ReggePlane}, black data points denote EM
estimates of stellar or supermassive BH spins obtained using either the
K$\alpha$ iron line or the continuum fitting
method~\cite{Brenneman:2011wz,Middleton:2015osa}, whereas red data points are LIGO-Virgo 90\% confidence levels for the 
spins of the primary and secondary BHs in (a selection of) the merger events detected to 
date~\cite{LIGOScientific:2018mvr,LIGOScientific:2020stg,Abbott:2020khf,Abbott:2020tfl}, including the recent data from the second GW transient catalogue~\cite{Abbott:2020niy}. 
Although current LIGO-Virgo measurements are still not precise enough to measure the 
spins from the merging BHs and give any meaningful constraints, by combining information from multiple detections and 
with the expected improvement of future ground-based detectors~\cite{Sathyaprakash:2019yqt} it is expected that 
constraints coming from the observation of merging stellar-mass BHs will improve significantly~\cite{Arvanitaki:2016qwi, 
Baryakhtar:2017ngi,Ng:2019jsx,Ng:2020ruv}. On the other hand, as discussed in Ref.~\cite{Brito:2017zvb}, LISA will have 
the ability 
to measure BH masses and spins with a very good precision~\cite{Klein:2015hvg}, in a range of masses largely 
complementary to the ones coming from ground-based GW detector and EM observations. The range of the projected LISA 
measurements using three different population models for supermassive BH growth (popIII, Q3 and Q3-nod 
from~\cite{Klein:2015hvg}) are denoted by arrows in Fig.~\ref{fig:ReggePlane}. 
Green points are the $90\%$ confidence levels for the mass-spin of a selection of the GW coalescence 
remnants detected to 
date~\cite{LIGOScientific:2018mvr,LIGOScientific:2020stg,Abbott:2020khf,Abbott:2020tfl,Abbott:2020niy}. 
While those events cannot be used to constrain the Regge plane (because the 
observation time scale is much shorter than $\tau_{\rm inst}$), they identify targets of merger follow-up 
searches~\cite{Arvanitaki:2014wva,Arvanitaki:2016qwi,Baryakhtar:2017ngi,Isi:2018pzk,Ghosh:2018gaw} (see 
Sec.~\ref{sec:GWs}). This is particularly important in the spin-$1$ and spin-$2$ cases, since $\tau_{\rm inst}$ can be 
as small as a fraction of seconds for typical remnants in the LIGO/Virgo band.

Instead of using the Salpter time as a reference time scale, more direct constraints would come from comparing 
$\tau_{\rm inst}$ 
against the baseline (typically ${\cal O}(10\,{\rm yr})$) during which the spin of certain BH candidates is 
measured to be constant~\cite{Cardoso:2018tly}, as it is the case for LMC X-3~\cite{Steiner:2010kd} and 
Cyg~X-1~\cite{Gou:2009ks}, shown in the panels of Fig.~\ref{fig:ReggePlane} by blue points. 
These sources could confidently exclude the range $m_b\in(10^{-11},10^{-13})$.

Note that, for a single BH observation, superradiant instabilities can only exclude a \emph{window} in the mass range 
of the fields, as shown in Fig.~\ref{fig:ReggePlane}. For each BH observation, the upper limit comes from the fact that 
when $M\mu\ll1$ the time scale grows with some power of $1/(\mu M)$ and eventually the instability is ineffective on 
astrophysical time scales. The lower limit comes from the fact that the instability exists only when the superradiant 
condition is satisfied, and this imposes a constraint on $\mu$ for a given azimuthal number $m$\footnote{As we already 
discussed, as $m$ increases, larger values of $\mu$ are allowed in the instability region and virtually any value of 
$\mu$ gives some unstable mode in the eikonal ($l,m\gg1$) limit. However, the instability is highly suppressed as $l$ 
increases so that, in practice, only the first few allowed values of $l=m$ correspond to an effective instability.}. 
Indeed, the rightmost part of the curves shown in Fig.~\ref{fig:ReggePlane} for fixed $\mu$ is universal and arises 
from saturation of the superradiant condition, $a\sim a_{\rm crit}$, where $a_{\rm crit}$ is given in 
Eq.~\eqref{omegaDetweiler}. Such condition does not depend on the spin of the field, so the upper bounds arising from 
the Regge gaps are the same for scalar, vector and tensor fields.

By combining different BH observations in a wide 
range of BH masses and assuming\footnote{Some recent observations of ultraluminous X-ray sources suggest that these 
sources contain  intermediate-mass BHs (e.g.~\cite{2014Natur.513...74P,Lin:2018dev}), suggesting that the BH mass 
spectrum might be populated continuously from few solar masses to billions of solar masses.} that spinning BHs exist in 
the entire mass range $M_\odot\lesssim M\lesssim 10^{10} M_\odot$, one would be able to constrain approximately the entire range 
\begin{equation}
 10^{-21}{\rm eV}\lesssim m_b\lesssim 10^{-10}{\rm eV}\,,\label{eq_bounds_entire}
\end{equation}
where the best upper bound would come from the lightest massive BHs (with $M\approx 
5M_\odot$~\cite{Middleton:2015osa}), whereas the best lower bound would come from the heaviest supermassive BHs for 
which spin measurements are reliable e.g. the BH candidate Fairall~9~\cite{Schmoll:2009gq}. Note that in the 
$10^{-21}\,{\rm eV}$ range ultralight bosons are also compelling dark-matter 
candidates~\cite{Hui:2016ltb}.

This lower bound could be further improved if the supermassive BH in M87$^{*}$ is confirmed to have a large spin, as 
recently suggested~\cite{Akiyama:2019fyp,Tamburini:2019vrf} (single gray point in Fig.~\ref{fig:ReggePlane}). In 
fact, Ref.~\cite{Davoudiasl:2019nlo} showed that together with the BH mass measurement $M\sim 6.5 \times 10^9 
M_{\odot}$ coming from the imaging of M87$^{*}$'s shadow by the Event Horizon Telescope~\cite{Akiyama:2019cqa}, this 
source alone rules out scalar field masses $2.8\times 10^{-21}\,{\rm eV}<m_S<4.6\times 10^{-21}\,{\rm eV}$ and vectors 
with masses $8.5\times 10^{-22}\,{\rm eV}<m_V<4.6\times 10^{-21}\,{\rm eV}$ at $1\sigma$ level. In addition, if the 
largest known supermassive BHs with $M\simeq 2\times 10^{10} M_\odot$ \cite{McConnell:2011mu,2012arXiv1203.1620M} were 
confirmed to have nonzero spin, we could get even more stringent bounds.

Aside from constraining the existence of these particles, an even more exciting prospect is to use these arguments to actually \emph{detect} ultralight bosons and measure their properties. For massive scalar fields this was studied in Refs.~\cite{Arvanitaki:2016qwi,Ng:2019jsx,Fernandez:2019qbj} for stellar-mass BHs detected in ground-based GW detectors and in~\cite{Brito:2017zvb} for massive BHs detected by LISA. It was shown that with the expected number of merging BHs that will be observed by aLIGO and LISA, the mass of the boson could be inferred with accuracies between $\sim 5\%$ to $50\%$, in case such particles exist in nature.

Under certain conditions, the constraint~\eqref{eq_bounds_entire} (see also Table~\ref{tab:bounds} below) on massive spin-2 fields also applies to massive gravitons propagating on a Kerr BH~\cite{Brito:2013wya}, and sets a stringent bound on the mass of the graviton~\cite{PDG}. Similarly, such bounds can also be translated in a bound on the photon mass~\cite{Pani:2012vp}, although in this case the effect of the coupling between photons and accreting matter on the instability needs to be assessed (see Ref.~\cite{Pani:2012vp} for a discussion). A more rigorous analysis should be performed to assess whether plasma interactions (see Sec.~\ref{sec:astro_plasma}) can affect the bounds discussed above in the case of massive photons.

%%%%%%%%%%%%%%%%%%%%%%%%%%%%%%%%%%%%%%%%%%%%%%%%%%%%%%%%%%%%%%%%%%%%%%
\subsubsection{Gravitational-wave signatures} \label{sec:GWs}
%%%%%%%%%%%%%%%%%%%%%%%%%%%%%%%%%%%%%%%%%%%%%%%%%%%%%%%%%%%%%%%%%%%%%%

Upcoming precise spin measurements of massive BHs~\cite{GRAVITY,Lu:2014zja,Brito:2017zvb} will be useful to refine the 
bounds discussed above. However, a very different phenomenology can be probed through detection of GWs that are possibly 
emitted by bosonic clouds around spinning 
BHs~\cite{Arvanitaki:2009fg,Arvanitaki:2010sy,Yoshino:2013ofa,Brito:2014wla,Arvanitaki:2014wva,Yoshino:2014wwa,
Arvanitaki:2016qwi,Baryakhtar:2017ngi,East:2017mrj,Brito:2017zvb,Brito:2017wnc,Isi:2018pzk,Ghosh:2018gaw,East:2018glu,
Siemonsen:2019ebd,Ng:2020jqd}.

As first discussed in Refs.~\cite{Arvanitaki:2009fg,Arvanitaki:2010sy,Yoshino:2012kn,Arvanitaki:2014wva}, a bosonic condensate around a spinning BH as the one depicted in Fig.~\ref{fig:draw} would emit GWs through different channels, which are discussed below. When the specific analysis applies to a generic massive scalar, vector or tensor, we will denote its mass by $\mu_S$, $\mu_V$ or $\mu_T$, respectively, or more generically as $\mu$ if it applies to all cases, whereas $\mu_a$ and $f_a$ will specifically refer to axions as in Eq.~\eqref{axion_pot}.  
%%%

\paragraph{Direct continuous GW emission}
As we discussed in Sec.~\ref{sec:cloudproperties} the formation of a nonspherical monochromatic boson cloud anchored on a spinning BH leads to emission of GWs with frequency $\sim \omega_R/\pi$. In a particle-like description of the interaction, such waves can be interpreted as arising from the annihilation of the boson field to produce gravitons~\cite{Arvanitaki:2010sy}. 

Detailed relativistic computations find that for scalar, vector, and tensor\footnote{In the spin-2 case there 
exists a special dipolar mode~\cite{Brito:2013wya} that does not follow the behavior of Eq.~\eqref{eq:gw_flux}.} fields 
the emitted GW flux for the $l$ multipole and spin polarization $S$ scales 
as~\cite{Yoshino:2013ofa,Brito:2014wla,Ficarra:2018rfu,Baryakhtar:2017ngi,Brito:2020lup}
%%%%
\begin{equation}\label{eq:gw_flux}
 \dot{E}_{\rm GW}\propto \left(\frac{M_S}{M}\right)^2 (\mu M)^{4l+4S+10}\,,
\end{equation}
%%%%
where $M_S$ denotes the cloud's total mass.
This result is valid only when $\mu M\ll1$, but it approximates the scaling of the exact results reasonably well also for moderately large values of the coupling which can be computed numerically by solving the Teukolsky equation~\cite{Yoshino:2013ofa,Brito:2017zvb,Siemonsen:2019ebd} (cf. Fig.~\ref{fig:GW_flux}).

This radiation is \emph{monochromatic} with frequency
%%%
\begin{equation}
 f_{\rm GW}\sim \omega_R/\pi \sim 5\,{\rm kHz}\left(\frac{\mu \hbar}{10^{-11}{\rm eV}}\right)\,, \label{eq:frequency_SR}
\end{equation}
%%%
with a typical duration for the dominant mode roughly given by~\cite{Yoshino:2013ofa,Brito:2017zvb,Baryakhtar:2017ngi,Siemonsen:2019ebd,Brito:2020lup}
%
%\begin{equation} \label{eq:tgw_ann}
%\tau_{\rm GW}^{S} \approx 6.5 \times 10^{4}\, {\rm yr} \left(\frac{M}{10\, M_{\odot}}\right) \left(\frac{0.1}{M\mu_S}\right)^{15}\hspace{-2pt} \frac{M}{a}\,,\quad \tau_{\rm GW}^{V,T}\approx 1\, {\rm day} \left(\frac{M}{10\, M_{\odot}}\right) \left(\frac{0.1}{M\mu_{V,T}}\right)^{11}\hspace{-2pt} \frac{M}{a}\,,
%\end{equation}
%
\begin{eqnarray}
&&\tau_{\rm GW}^{S} \approx 1.3 \times 10^{5}\, {\rm yr} \left(\frac{M}{10\, M_{\odot}}\right) \left(\frac{0.1}{M\mu_S}\right)^{15}\hspace{-2pt} \left(\frac{0.5}{\chi_i-\chi_f}\right)\,, \label{eq:tgw_sca}\\
&&\tau_{\rm GW}^{V,T}\approx 2\,\, {\rm days} \left(\frac{M}{10\, M_{\odot}}\right) \left(\frac{0.1}{M\mu_{V,T}}\right)^{11}\hspace{-2pt} \left(\frac{0.5}{\chi_i-\chi_f}\right)\,,\label{eq:tgw_vec}
\end{eqnarray}
for the scalar, vector and tensor case, respectively, and where $\chi_{i}\,(\chi_{f})$ is the dimensionless BH spin at the beginning (end) of the superradiant growth.
Thus, BH-boson condensates are \emph{continuous GW sources},
like pulsars for LIGO or verification binaries for LISA. There are, however, two notable differences: (i) depending on the value of $M\mu$, the GW emission time scale 
$\tau_{\rm GW}$ for vectors and tensors can be significantly shorter than the observation time, resulting in an impulsive signal; (ii)~for the tensor case, at variance with the massive scalar and vector field case, GW emission for the dominant hydrogenic mode is mostly \emph{hexadecapolar} and not quadrupolar~\footnote{This is because the signal is produced by a spinning quadrupolar field and not by a spinning dipolar field~\cite{Brito:2020lup}. The hexadecapolar nature of the radiation implies that the signal vanishes along the BH spin axis, at variance with the quadrupolar case, for which it is maximum in that direction.}.

The frequency of the signal~\eqref{eq:frequency_SR} is only weakly dependent on the BH mass and spin, and since one needs $2\mu M\sim \mathcal{O}(1)$ for the boson occupation number to grow sufficiently fast through superradiance, ground-based detectors would be sensitive to the presence of bosonic clouds around stellar-mass BHs, whereas space-based detectors are sensitive to signatures of clouds around supermassive BHs.

In the $M\mu\ll 1$ limit and considering only the dominant unstable hydrogenic modes, the GW strain amplitude at its peak (using the definition in Ref.~\cite{Zhu:2020tht}) can be approximated by~\cite{Yoshino:2013ofa,Arvanitaki:2014wva,Brito:2014wla,Brito:2017zvb,Baryakhtar:2017ngi,Siemonsen:2019ebd,Brito:2020lup}:
\begin{eqnarray}\label{eq:h0_approx}
&& h^{S} \approx 5\times 10^{-27} \left(\frac{M}{10 M_{\odot}}\right)
\left(\frac{M\mu_S}{0.1}\right)^7 \left(\frac{\rm Mpc}{d}\right) 
\left(\frac{\chi_i - \chi_f}{0.5}\right)\,,\\
&& h^{V,T} \approx 10^{-23} \left(\frac{M}{10 M_{\odot}}\right)
\left(\frac{M\mu_{V,T}}{0.1}\right)^5 \left(\frac{\rm Mpc}{d}\right) 
\left(\frac{\chi_i - \chi_f}{0.5}\right)\,,
\end{eqnarray}
for a source at luminosity distance $d$. Although strictly valid only when $M\mu\ll 1$, these expressions give a reasonably good order of magnitude estimate also when $M\mu\gtrsim 0.1$ (see e.g. Ref.~\cite{Isi:2018pzk} for an explicit comparison against numerical results for the scalar case).

\begin{figure}
\begin{center}
\includegraphics[width=0.6\textwidth]{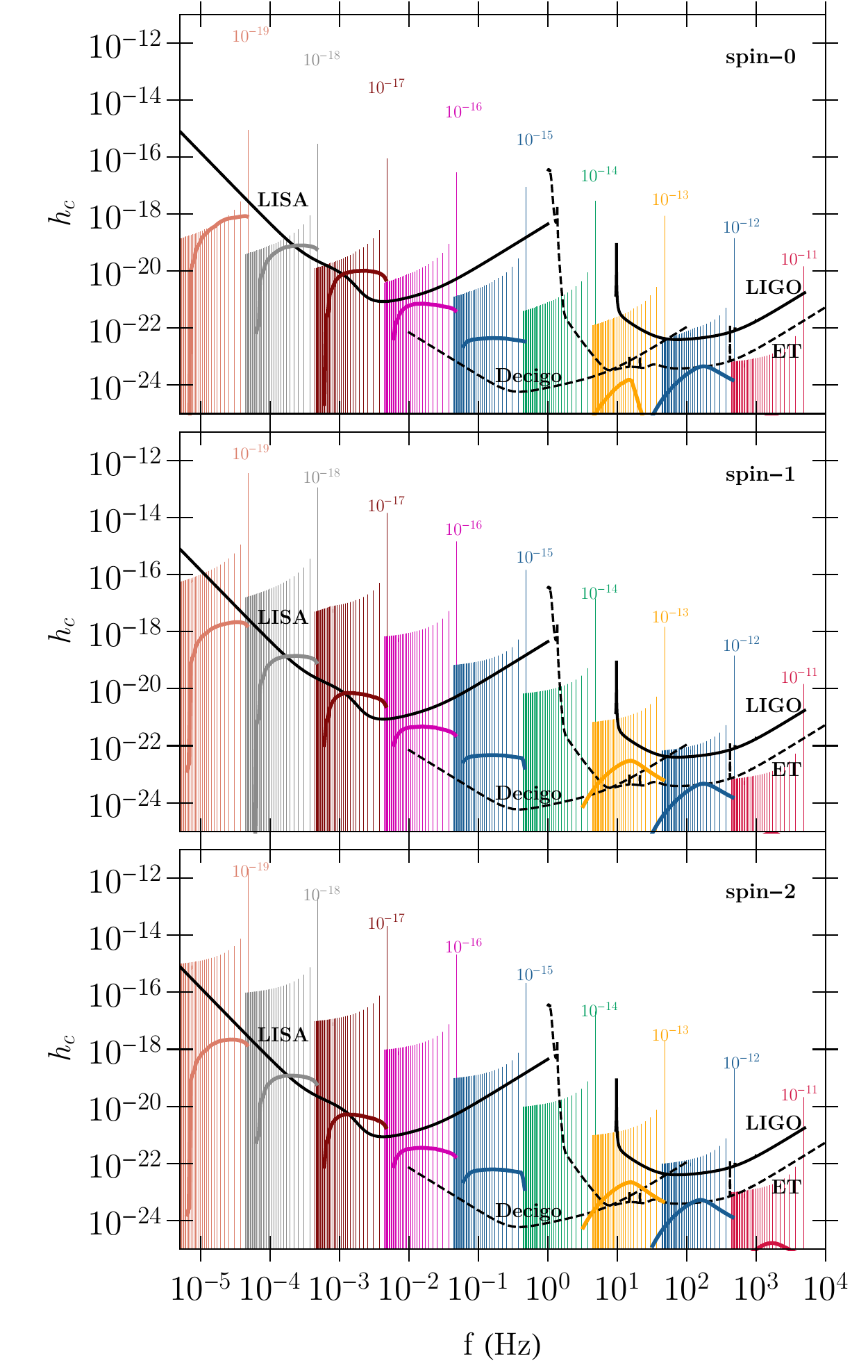}
\caption{GW characteristic strain (thin colored lines), as defined by Eq.~\eqref{eq:char_strain}, produced by BH-boson condensates (for spin-$0$, spin-$1$, and spin-$2$ fields). The GW strain amplitude is computed at its peak value and we consider the emission from the dominant unstable hydrogenic mode, an initial BH spin $\chi_i=0.9$ and an observation time $T_{\rm obs}=4\,{\rm yr}$. Each (nearly vertical) line shows the characteristic strain for a given boson mass $m_b$, computed at redshift $z\in(0.001,10)$ (from right to left, in steps of $\delta z=0.3$), with 
$M\mu/m$ varying in the superradiant range $(0, M\Omega_{\rm H})$ along each line (we use $m=1$ for spin-0 and spin-1 fields and $m=2$ for spin-2 fields). Different colors correspond to different boson masses $m_b$. For comparison we also show the characteristic noise strain amplitude (defined as $\sqrt{f S_n(f)}$, with $S_n(f)$ the noise PSD) of Advanced LIGO at design sensitivity~\cite{Aasi:2013wya} and the sky-averaged characteristic noise strain of LISA~\cite{Audley:2017drz,Cornish:2018dyw} (black thick curves). 
Thick colored lines show the stochastic GW background produced by the population of extra-galactic BH-boson systems under optimistic assumptions~\cite{Brito:2017wnc,Brito:2017zvb}. The characteristic noise strain of DECIGO~\cite{Kawamura:2006up} and the Einstein Telescope (ET)~\cite{Hild:2010id} (dashed lines) are also shown for reference. Extended from~\cite{Brito:2017wnc,Brito:2020lup}.  \label{sensitivity}}
\end{center}
\end{figure}

To estimate the detectability of these sources for different GW detectors it is useful to define the characteristic strain amplitude of the GW signal as~\cite{Moore:2014lga}
\begin{equation}
h_c=\sqrt{N_{\rm cycles}}\,h\,, 
\label{eq:char_strain}
\end{equation}
%%%
where the approximate number of GW cycles in a given detector's frequency band can be estimated as $N_{\rm cycles}\sim \min[\sqrt{f T_{\rm obs}},\sqrt{ f_{\rm GW} \tau_{\rm GW}}]$ for a signal with detector's frame frequency $f=f_{\rm GW}/(1+z)$, and considering an observation time $T_{\rm obs}$ for a source located at cosmological redshift $z$.
The characteristic GW strain amplitude of these signals is shown in Fig.~\ref{sensitivity} for $\chi_i=0.9$ and several values of the boson mass, BH mass and cosmological redshift. We compare it against the expected noise power spectral density (PSD) of LISA and Advanced LIGO at design sensitivity considering an observation time of 4 years. Also shown in thick solid curves is the estimated stochastic background, for the most optimistic models of Refs.~\cite{Brito:2017wnc,Brito:2017zvb}, from the annihilation GW signal emitted by the whole BH population too faint to be detected as individual sources. We discuss this background in more detail below. With the addition of third-generation ground-based detectors, such as the Einstein Telescope~\cite{Hild:2010id} and Cosmic Explorer~\cite{Evans:2016mbw,Essick:2017wyl}, or a detector sensitive in the deci-Hz such as DECIGO~\cite{Kawamura:2006up}, bosons with masses in the range $10^{-19}\,{\rm eV} \lesssim m_b \lesssim 10^{-11}\,{\rm eV}$ could be detectable by GW observatories.

\begin{figure}
\begin{center}
\includegraphics[width=0.65\textwidth]{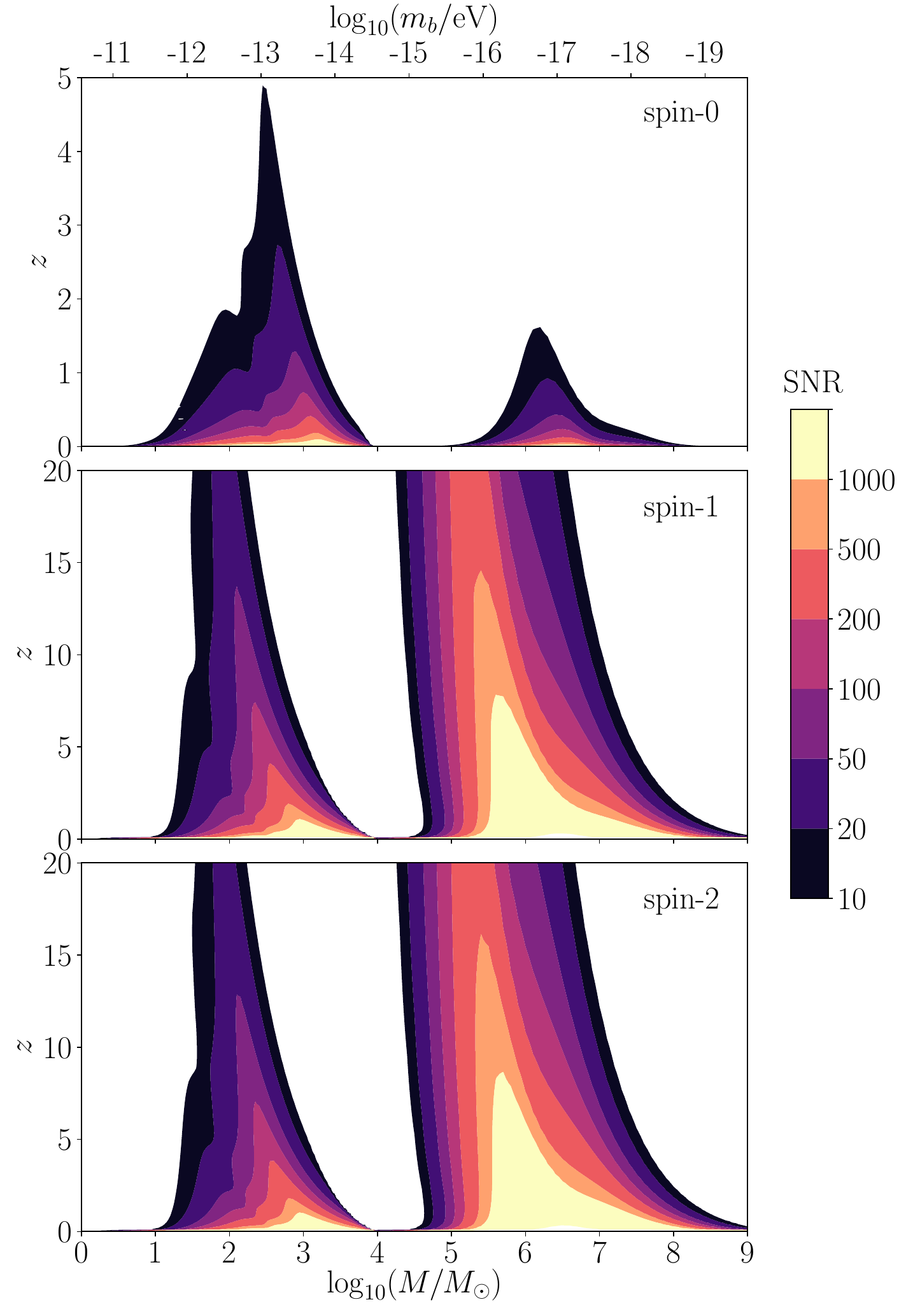}
\caption{Contours of constant angle-averaged SNR for annihilation GW signals detected with the Einstein Telescope and LISA, for scalar (top), vector (middle) and tensor (bottom) condensates, as a function of the source-frame BH mass and cosmological redshift. We considered $M\mu=0.2$, $\chi_i=0.9$ and an observation time $T_{\rm obs}=4 {\rm yr}$. The sources detectable by the Einstein Telescope and LISA correspond to BHs with masses in the range $\sim [10,10^4]\, M_{\odot}$ and $\sim [10^4,10^9]\, M_{\odot}$, respectively.
For simplicity we estimated the SNR using ${\rm SNR}\propto h_c/\sqrt{f S_n(f)}$ (with a proportionality factor taking into account the angle-averaged signal and triangular shape of LISA and the Einstein Telescope, see e.g. ~\cite{Berti:2005ys,Regimbau:2012ir}) and neglected the possible confusion noise from the stochastic GW background (see Fig.~\ref{sensitivity}). We computed the GW amplitude and duration time using analytical estimates for the GW flux and assumed that the signal is detected starting from its peak amplitude.
Notice the different scale for the redshift in the spin-0 case.
\label{fig:waterfall}}
\end{center}
\end{figure}

Quite interestingly, LISA and third generation ground-based detectors have the potential to detect sources at very high 
cosmological redshifts, especially for vector and tensor fields~\cite{Brito:2020lup}.  This is better illustrated in the 
``waterfall''~\cite{Audley:2017drz} plot in Fig.~\ref{fig:waterfall}, where we show the typical angle-averaged signal-to-noise ratio (SNR) for sources detected by ET and LISA, corresponding to BHs with masses in the range $\sim 
[10,10^4]\, M_{\odot}$ and $\sim [10^4,10^9]\, M_{\odot}$, respectively.
For vector and tensor fields, the continuous GW signal could be detected even when $z\approx 20$ 
or higher, if bosons in the right mass range exist, although the exact detection horizons will depend on the sensitivity of the search methods to these type signals~\cite{Isi:2018pzk}.
In practice, the range of potentially detectable boson masses through direct GW detections depend on the formation rate of spinning BHs. This was studied in Refs.~\cite{Arvanitaki:2014wva,Brito:2017wnc,Brito:2017zvb,Zhu:2020tht} for scalar bosons and in~\cite{Baryakhtar:2017ngi} for vector bosons in the context of all-sky searches with Advanced LIGO and with LISA. Depending on the boson mass (and to a lesser degree its spin) and the astrophysical BH population, up to ${\cal O}(10^4)$ signals could be detected, assuming that boson fields with masses in the optimal range exist. 
%
%The prospect for detection of GWs emitted by scalar clouds in blind all-sky searches with the Advanced LIGO-Virgo experiments~\cite{LIGO,VIRGO} and with an LISA-like mission~\cite{AmaroSeoane:2012km,Audley:2017drz} was studied in Refs.~\cite{Arvanitaki:2014wva,Brito:2017wnc,Brito:2017zvb}.
%%%
\begin{figure}[thb]
\begin{center}
\begin{tabular}{c}
 \epsfig{file=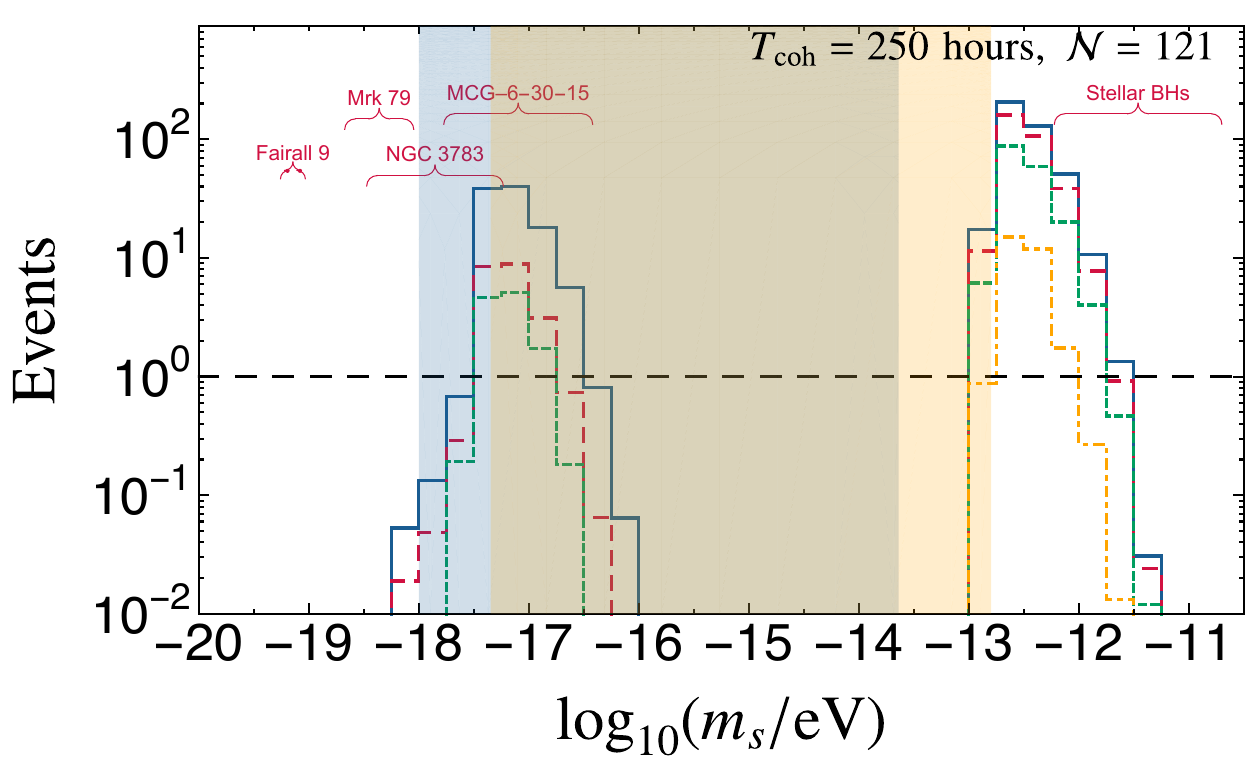,width=0.8\textwidth,angle=0,clip=true}
 \end{tabular}
 \caption{ 
Expected number of scalar field annihilation events observable with aLIGO at design sensitivity~\cite{LIGO} and for the space-based detector LISA~\cite{AmaroSeoane:2012km,Audley:2017drz} as a function of the scalar field mass, assuming that the search is done with a semi-coherent method, where the signal is divided in 121 coherent segments with time length $T_{\rm coh}=250$ hours. The different lines correspond to different assumptions about the astrophysical models that bracket optimistic to pessimistic estimates. Shaded areas correspond to exclusion regions that could be placed by a 4-year LISA mission from massive BH spin measurements (cf. Sec.~\ref{sec:bounds_mass_regge}).  For reference we also show with brackets the constraints
  that can be placed by current spin measurements of massive/stellar-mass BHs~\cite{Brenneman:2011wz,Arvanitaki:2014wva} as discussed in Sec.~\ref{sec:bounds_mass_regge}. See Refs.~\cite{Brito:2017zvb,Brito:2017wnc} for details. Similar estimates were first computed in Ref.~\cite{Arvanitaki:2014wva} and more recently also in Ref.~\cite{Zhu:2020tht}.\label{fig:GW_annih} }
\end{center}
\end{figure}
%
%Using recent mass distributions for stellar BHs and supermassive BHs, Refs.~\cite{Arvanitaki:2014wva,Brito:2017wnc,Brito:2017zvb} estimated an expected number of signals as large as ${\cal O}(10^4)$, assuming that scalar fields with masses in the optimal range for a given BH exist. 
A summary of these results is shown in Fig.~\ref{fig:GW_annih} for scalar fields. The range of scalar masses that is detectable is complementary to the one that could be excluded by BH spin measurements (cf. Sec.~\ref{sec:bounds_mass}). Similar estimates were obtained for massive vector fields in Ref.~\cite{Baryakhtar:2017ngi}.

A detailed study of the detectability of the ensemble of boson annihilation GW signals emitted by a population of isolated Galactic BHs was done more recently~\cite{Zhu:2020tht}. It was shown that, with the sensitivity of the latest all-sky searches for continuous GWs~\cite{Palomba:2019vxe,Dergachev:2019wqa,Dergachev:2019oyu,Pisarski:2019vxw}, up to $\sim 10^3$ ($\sim 10^2$) signals could have already been detected, assuming a population of stellar mass BHs between 5 and 30 $M_{\odot}$ and a uniform distribution $\chi\in [0,1]$ ($\chi \in [0,0.5]$) for the BH spin. Negative results can, in principle, be turned into a constraint on ultralight scalar fields; however, for signals that can be numerous and very loud in a small frequency range -- in particular for scalar fields with masses in the range $[3,10]\times 10^{-13}$ eV -- the sensitivity of the current methods to search for continuous GW signals needs to be reassessed in order to properly interpret null results.

Aside from blind all-sky searches, Refs.~\cite{Arvanitaki:2016qwi,Baryakhtar:2017ngi} proposed that these GW signals could also be detected by performing targeted follow-up searches of known binary BH mergers remnants. The prospects of performing targeted searches~\cite{Arvanitaki:2016qwi,Baryakhtar:2017ngi,Isi:2018pzk,Ghosh:2018gaw} is especially promising for planned third-generation GW detectors, such as the Einstein Telescope~\cite{Hild:2010id} and Cosmic Explorer~\cite{Evans:2016mbw,Essick:2017wyl}. Dedicated pipelines to search for these signals in LIGO data are actively being developed and implemented for both blind all-sky~\cite{DAntonio:2018sff,Palomba:2019vxe,Dergachev:2019wqa,Dergachev:2019oyu,Zhu:2020tht} and targeted searches~\cite{Isi:2018pzk,Sun:2019mqb} but no GW signal consistent with a boson annihilation has been detected so far. 

Using LIGO O2 data, it as been proposed that the absence of signals in generic all-sky searches on LIGO and Virgo data constrains the existence of scalar fields in the mass range $\sim [1.1$~--~$4.0]\times 10^{-13}$ ($[1.2$~--~$ 1.8]\times 10^{-13}$) eV assuming the existence of a speculative population of young heavy BHs ($M\gtrsim 30 M_{\odot}$) in the Milky Way born with high spins $\sim 0.998$ (moderate spins $\sim 0.6$)~\cite{Palomba:2019vxe}. In principle, current null results already disfavor the existence of scalar bosons in the range $[2,25]\times 10^{-13}$ eV~\cite{Zhu:2020tht}. However, as mentioned above, those constraints should be reassessed by properly taking the account the overall expected GW signal from the population of galactic BHs.
 
%%%%%%%%%%%%%%%%%%%%%%%%%%%%%%%%%%%%%%
\paragraph{GWs from level transitions}
 %%%%%%%%%%%%%%%%%%%%%%%%%%%%%%%%%%%%%
Because the boson condensate has a hydrogenic-like spectrum (cf. Eq.~\eqref{omegaDetweiler}), GW emission can also occur from level transitions between states with the same harmonic indices ($l,m$) but different overtone numbers $n$, similarly to photon emission through atomic transitions. 
This process occurs when the growth rate of some $n>0$ mode is stronger than that of the fundamental $n=0$, as this can happen for high values of $(l,m)$~\cite{Arvanitaki:2009fg,Arvanitaki:2010sy,Arvanitaki:2014wva,Siemonsen:2019ebd} (a detailed analysis of this effect is presented in Ref.~\cite{Yoshino:2015nsa} for scalar fields and Ref.~\cite{Siemonsen:2019ebd} for vector fields). In the GW signal this phenomenon shows up as a beating pattern due to the close frequencies of the excited and group-state modes, as can be seen in the numerical simulations of Fig.~\ref{fig:ProcaNum} and~\ref{Fig:GW_emission}.
The frequency of the emitted graviton is given by the frequency difference between the excited ($\tilde{n}_e>\tilde{n}_g$) state and the ground state $\tilde{n}_g$\footnote{Here $\tilde{n}$ denotes the principal quantum number given by $\tilde{n}=l+S+n+1$.},
\begin{equation}
 \omega_{\rm trans}\sim\frac{\mu}{2}(M \mu)^2\left(\frac{1}{\tilde{n}_g^2}-\frac{1}{\tilde{n}_e^2}\right)\,,
\end{equation}
and the corresponding wavelength is usually much longer than the size of the system. Therefore, in this case the quadrupole formula is valid~\cite{Arvanitaki:2010sy}. 
In this approximation, the dominant GW flux for a scalar field (corresponding to a transition between the $l=m=4,n=1$ and $l=m=4,n=0$ modes) reads~\cite{Arvanitaki:2010sy,Arvanitaki:2014wva}
%%%%
\begin{equation}
\dot{E}_{\rm GW}\sim {\cal O}(10^{-6} \div 10^{-8})\frac{M_gM_e}{M^2}(\mu_S M)^8\,,
\end{equation}
%%%%
where $M_g$ and $M_e$ are the mass in the ground and excited states, respectively.

\begin{figure}[thb]
\begin{center}
\begin{tabular}{cc}
 \epsfig{file=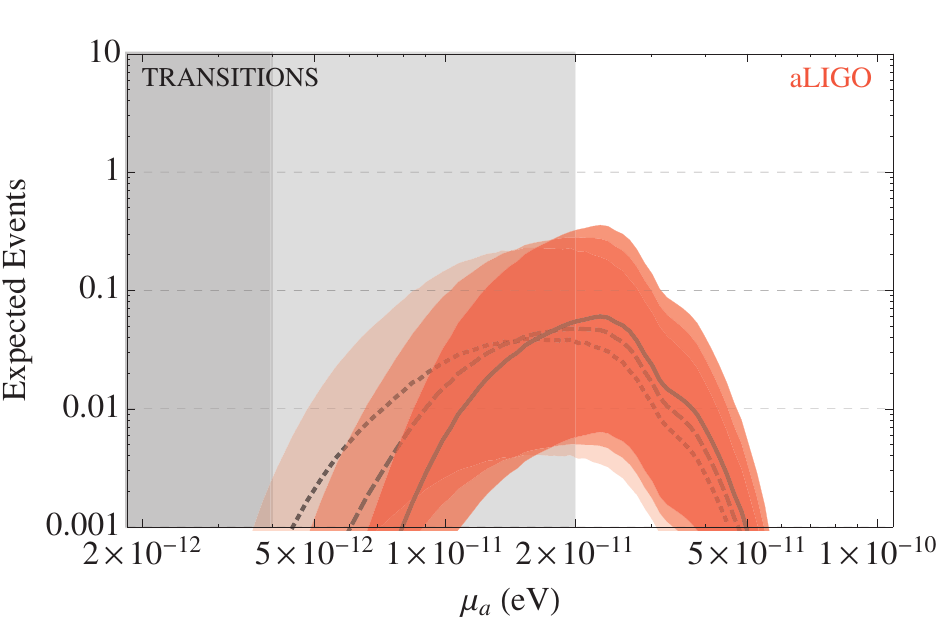,width=0.8\textwidth,angle=0,clip=true}
\end{tabular}
\caption{Expected number of events for GWs emitted thorough level transitions with aLIGO at design sensitivity~\cite{LIGO} as a function of the axion mass (but the results are also valid for a generic scalar field). The vertical shaded regions are disfavored by BH spin measurements assuming the QCD axion coupling strength (light gray) or stronger (dark grey). For stronger axion couplings nonlinear effects become important and extrapolations of the linearized results are not reliable (cf. Sec.~\ref{sec:super_couplings}). Each of the three bands corresponds to varying the BH mass distribution width including optimistic and pessimistic estimates of astrophysical uncertainties. See Ref.~\cite{Arvanitaki:2014wva} for details.
  \label{fig:GW_trans} }
\end{center}
\end{figure}

Although this is usually tiny, the GW strain is enhanced by the occupation number of the two levels, which grow exponentially through superradiance. Also in this case the signal is monochromatic. For the dominant transition of a scalar field, the typical frequency is
%%%
\begin{equation}
 f_{\rm trans}\sim 13\,{\rm Hz}\left(\frac{\mu_S \hbar}{10^{-11}{\rm eV}}\right)^3\left(\frac{M}{5M_\odot}\right)\,, \label{eq:frequency_SR_trans}
\end{equation}
%%%
which falls in the sensitivity bands of advanced ground-based detectors for a boson with mass about $10^{-11}{\rm eV}$ around a stellar-mass BH with $M\sim 5M_\odot$, whereas it falls within LISA milliHerz band for a boson with mass about $10^{-15}{\rm eV}$ around a supermassive BH with $M\sim 10^5 M_\odot$. 
The number of transition events for aLIGO/aVirgo as estimated in Ref.~\cite{Arvanitaki:2014wva} for a massive scalar field is shown in Fig.~\ref{fig:GW_trans}. For space-based detectors, the peak of sensitivity falls in the range of intermediate-mass BHs, for which precise mass distributions are lacking. This affects the event estimates, but it is promising that the reach radius for transition signals of a LISA-like detector would extend up to hundred megaparsec~\cite{Arvanitaki:2014wva}. 

%%%
%

\begin{figure}[thb]
\begin{center}
\begin{tabular}{c}
\includegraphics[width=0.8\textwidth]{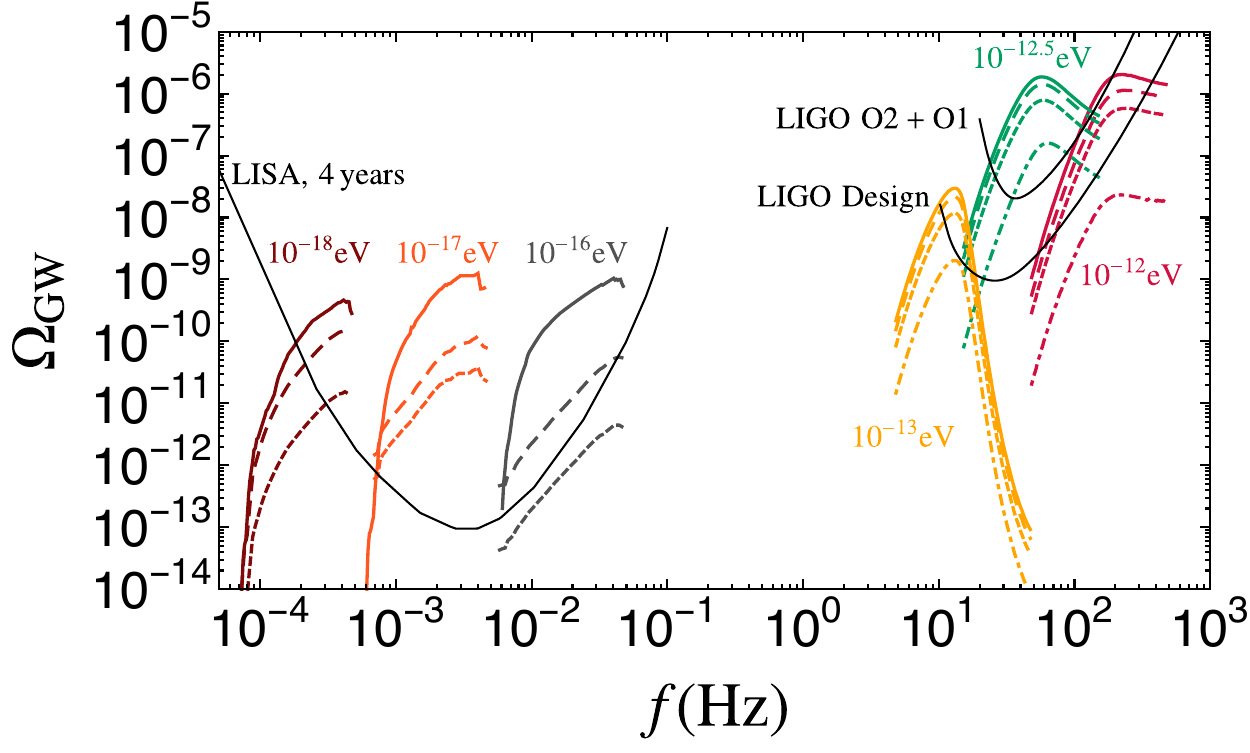}
\end{tabular}
\caption{Stochastic background from boson annihilation GW signals in the LISA and LIGO bands.
  For each boson mass the the different predicted signals correspond to different astrophysical models described in Ref.~\cite{Brito:2017wnc}. The black lines are the power-law integrated curves of~\cite{Thrane:2013oya}, computed using predicted
  noise PSDs for LISA~\cite{Audley:2017drz}, advanced LIGO at design sensitivity~\cite{LIGOScientific:2019vic}, and upper limits from LIGO's first two observing runs (O1+O2)~\cite{LIGOScientific:2019vic}. A power law stochastic background which lies tangent to one of these curves would be detectable with a 2$\sigma$ significance. Adapted from~\cite{Brito:2017wnc}. \label{fig:background}}
\end{center}
\end{figure}

%%%%%%%%%%%%%%%%%%%%%%%%%%%%%%%%%%%%%%
\paragraph{Stochastic GW background}
%%%%%%%%%%%%%%%%%%%%%%%%%%%%%%%%%%%%%%
In addition to GW events that can be detected as individual sources, one expects the existence of an even larger number of sources too faint to be detected individually. The incoherent superposition of all these sources produces a stochastic GW background that was first computed in~\cite{Brito:2017wnc,Brito:2017zvb} for the GW signal emitted by scalar annihilations. The background can be characterized by its (dimensionless) 
energy spectrum
\begin{equation}
\Omega_{\rm gw}(f)=\frac{1}{\rho_c}\frac{d\rho_{\rm gw}}{d\ln f}\,, \label{OmegaGW}
\end{equation}
where $\rho_{\rm gw}$ is the background's energy density, $f$ the GW frequency measured at the detector
and $\rho_c$ the critical density of the Universe at the present time. The predicted stochastic background is shown in Fig.~\ref{fig:background}.  The maximum frequency of the background is determined by the boson mass, whereas its amplitude is mostly determined by the astrophysical population of BHs. 
For the most optimistic astrophysical models, LIGO could detect or rule out scalar masses in the range $2\times 10^{-13}\,{\rm eV} \lesssim m_S \lesssim 10^{-12}\,{\rm eV}$ whereas LISA would be sensitive to scalar masses in the range $5\times 10^{-19}\,{\rm eV}\lesssim m_S \lesssim 5\times 10^{-16}\,{\rm eV}$, and similar ranges should apply for vector and tensor fields.

Dedicated searches in Advanced LIGO's first observing runs found no evidence for such stochastic signal~\cite{Tsukada:2018mbp,Tsukada:2020lgt}, excluding scalar fields with masses in the range $2.0\times 10^{-13}\,{\rm eV}< m_S < 3.8\times 10^{-13} \,{\rm eV}$ at 95\% credibility~\cite{Tsukada:2018mbp} and vector fields in the range $0.8\times 10^{-13}\,{\rm eV}< m_V < 6.0\times 10^{-13} \,{\rm eV}$~\cite{Tsukada:2020lgt}, under optimistic assumptions about the BH population.

%%%%%%%%%%%%%%%%%%%%%%%%%%%%%%%%%%%%%%%%%%%%%%%%
\paragraph{GW bursts from bosenova explosions}
 %%%%%%%%%%%%%%%%%%%%%%%%%%%%%%%%%%%%%%%%%%%%%%%
As we discussed in Sec.~\ref{sec:bosenova} when nonlinear terms are taken into account, additional effects can arise. For example, for an axion-like field described by the sine-Gordon potential~\eqref{axion_pot} nonlinear effects become important when the mass in the cloud reaches a critical mass (cf. Eq.~\eqref{eq:crit_bosenova}) and a bosenova explosion can occur. At variance with annihilation and level transition signals, the signal from bosenova explosions is a \emph{periodic} emission of bursts, whose separation depends on the fraction of the cloud which remains bound to the BH after each subsequent collapse.

The typical frequency of a single bosenova burst is~\cite{Yoshino:2012kn,Kodama:2011zc,Arvanitaki:2014wva}
%%%
\begin{equation}
 f_{\rm bn}\sim 30\,{\rm Hz}\left(\frac{16 r_{\rm cloud}}{t_{\rm bn}}\right)\left(\frac{\mu_a M}{0.4 l}\right)^2 \left(\frac{10 M_\odot}{M}\right)\,,
\end{equation}
%%%
where $t_{\rm bn}$ is the infall time and $r_{\rm cloud}$ is the characteristic size of the cloud as given in Eq.~\eqref{peak}. For example, a burst from a $10 M_\odot$ BH would last approximately one millisecond and, as the result of multiple subsequent explosions, there can be various spikes separated by a quiet period of approximately $300\, {\rm s}$~\cite{Arvanitaki:2014wva}. A quadrupole estimate of the GW strain for such signal yields~\cite{Yoshino:2012kn,Kodama:2011zc,Arvanitaki:2014wva}
%%%
\begin{equation}
 h\sim 10^{-21} \left(\frac{{\rm kpc}}{d}\right)\left(\frac{\epsilon}{0.05}\right) \left(\frac{16 r_{\rm cloud}}{t_{\rm bn}}\right)^2  \left(\frac{\mu_a M}{0.4 l}\right) \left(\frac{M}{10 M_\odot}\right)\left(\frac{f_a}{f_a^{\rm max}}\right)^2\,,
\end{equation}
%%%
where $\epsilon$ is the fraction of the cloud falling into the BH (typically $\epsilon\approx 5\%$~\cite{Yoshino:2012kn}), $f_a^{\rm max}$ is the largest coupling for which bosenova occurs and $d$ is the distance of the source from the detector.

%%%%%%%%%%%%%%%%%%%%%%%%%%%%%%%%%%%%%%%%%%%%%%%%%%%%
\subsubsection{Electromagnetic signatures}
%%%%%%%%%%%%%%%%%%%%%%%%%%%%%%%%%%%%%%%%%%%%%%%%%%%% 
In addition to the GW signatures discussed above, direct EM signatures of ultralight bosons around astrophysical BHs have also been proposed. Those can be divided in two categories: (i) direct imaging of sources close to supermassive BHs or through spectral properties of X-rays emission by accreting BHs~\cite{Cunha:2015yba,Cunha:2016bpi,Vincent:2016sjq,Cunha:2019ikd,Franchini:2016yvq,Ni:2016rhz}; (ii) bursts of light from compact objects, or polarimetric measurements of light from the BH vicinity~\cite{Rosa:2017ury,Ikeda:2019fvj,Boskovic:2018lkj,Plascencia:2017kca,Chen:2019fsq} (cf. Sec.~\ref{sec:axion_coupling}). We now discuss these signatures in more detail.

\paragraph{Black-hole shadows} 

The presence of a dense boson cloud around astrophysical BHs would affect the geodesic structure of the spacetime surrounding the BH. Therefore, any deviation from a vacuum BH spacetime is encoded in the way light rays are lensed in the neighborhood of the BH, and in particular in the properties of the ``shadow'' that the BH casts. 

The shadows of spinning BHs surrounded by scalar clouds (cf. Sec.~\ref{sec:hair}) were first studied in Refs.~\cite{Cunha:2015yba,Cunha:2016bpi,Vincent:2016sjq}. It was shown that for very heavy clouds the shadow can be considerably different from their vacuum Kerr BH counterparts, and measuring this difference is potentially in the reach of current and future very long baseline interferometric observations of supermassive BHs, such as the recent observations of M87$^{*}$ by the Event Horizon Telescope~\cite{Akiyama:2019cqa}. However, for bosonic clouds formed dynamically due to the superradiant instability, the prospect of directly imaging deviations from Kerr is more challenging~\cite{Cunha:2015yba,Vincent:2016sjq,Cunha:2019ikd,Creci:2020mfg}. As we discussed in Sec.~\ref{sec:hair} the superradiant instability BH can only extract a limited amount of energy, therefore producing bosonic clouds which backreact weakly on the geometry~\cite{Brito:2014wla}. In fact, it was shown in Ref.~\cite{Cunha:2019ikd} that the Event Horizon Telescope observations are not yet precise enough to constrain the possibility that the M87$^{*}$ supermassive BH is surrounded by superradiantly-produced bosonic cloud.

\begin{figure}[thb]
\begin{center}
\begin{tabular}{ccc}
\includegraphics[width=0.25\textwidth]{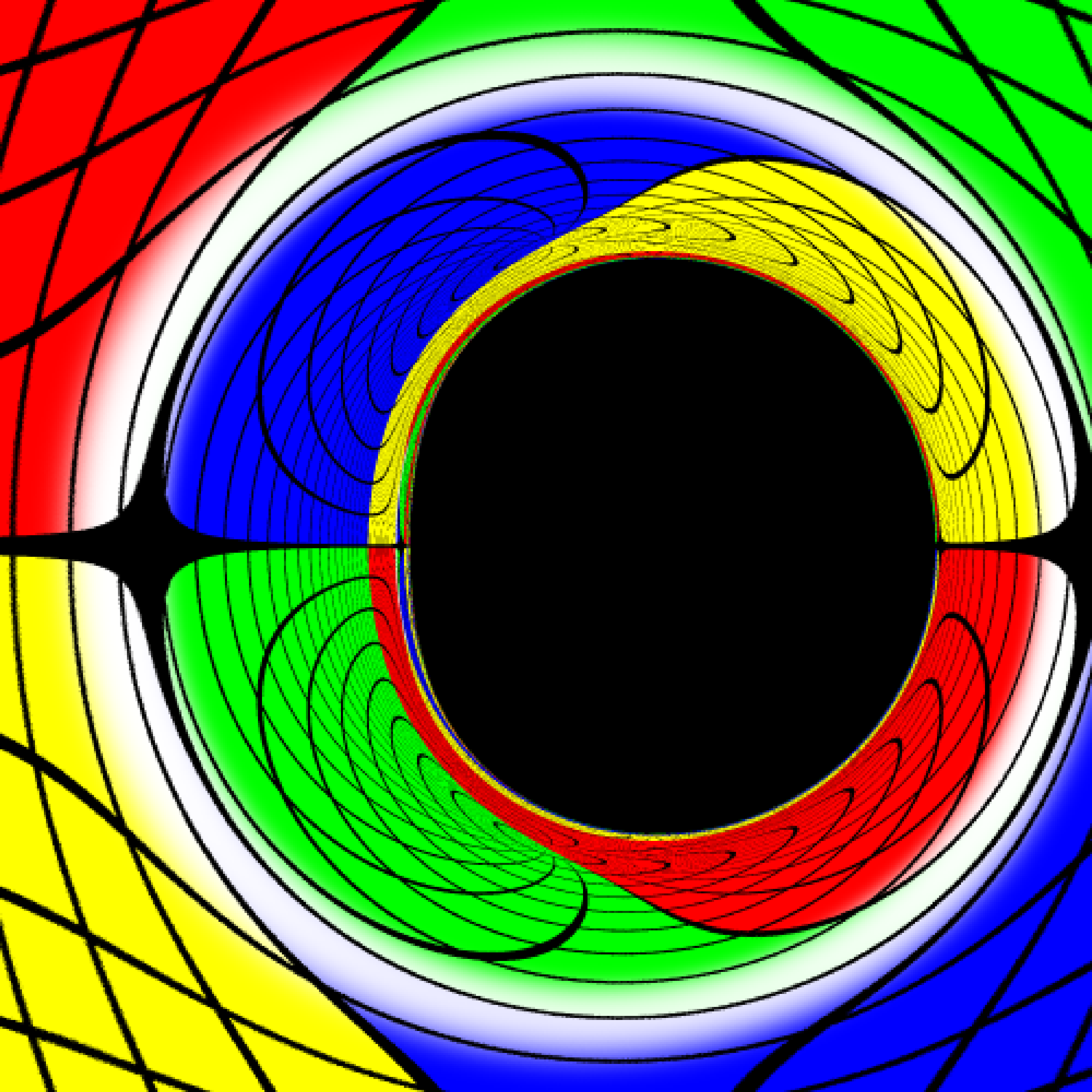} & \includegraphics[width=0.25\textwidth]{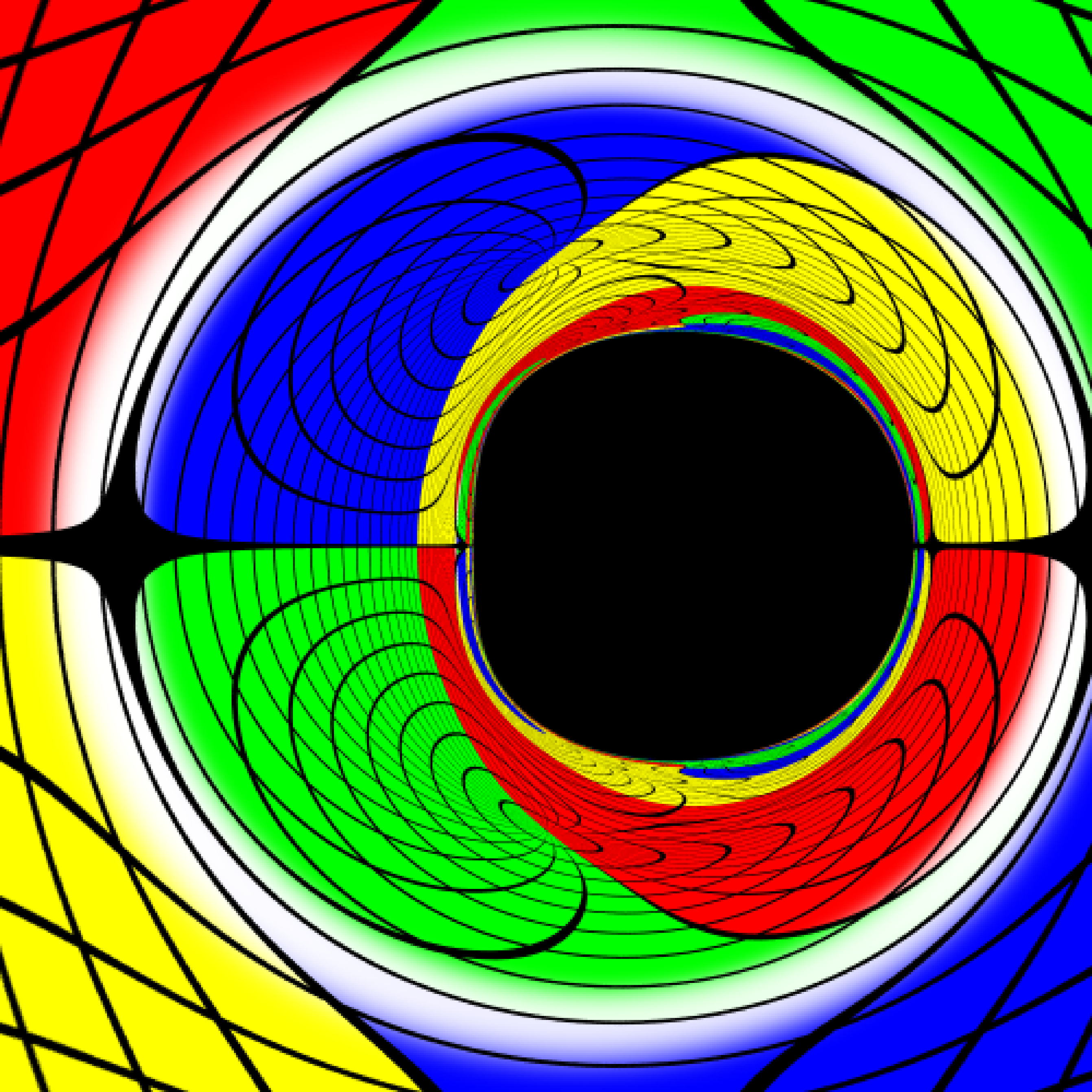}& \includegraphics[width=0.25\textwidth]{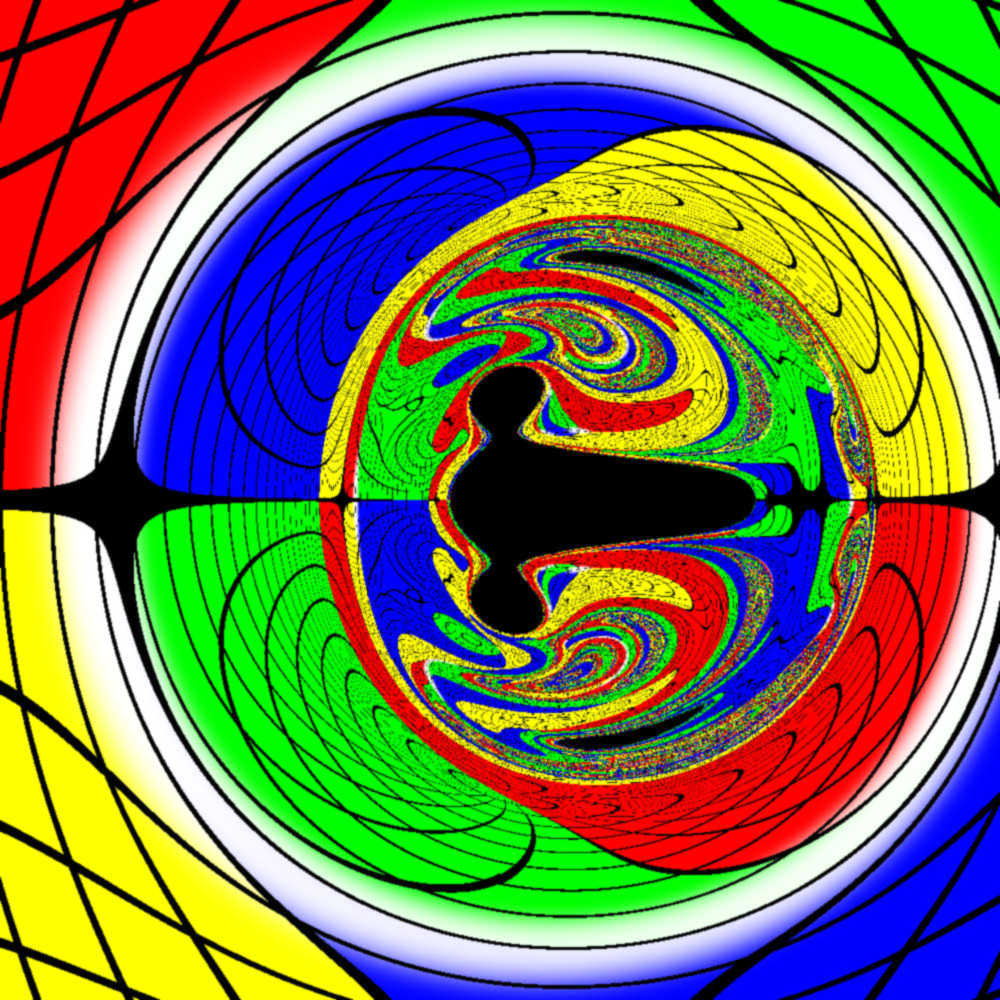}
\\
\includegraphics[width=0.25\textwidth]{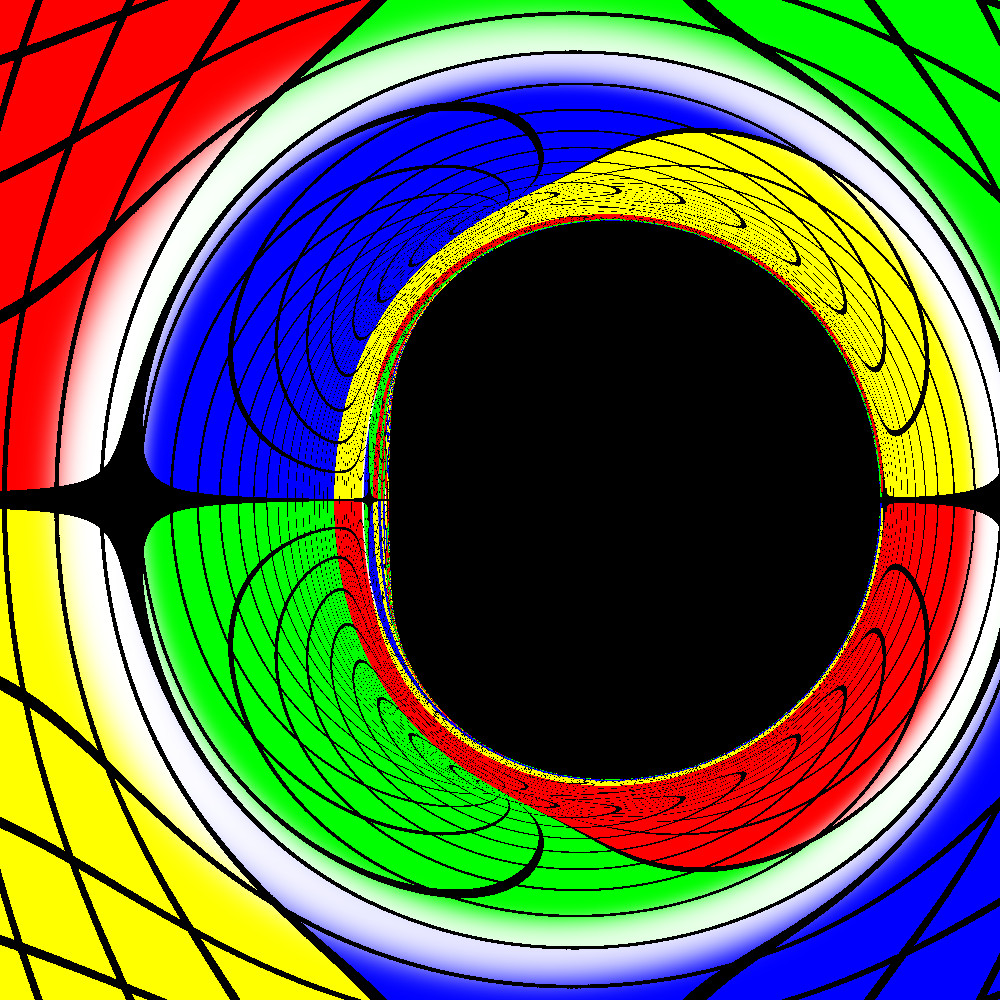} & \includegraphics[width=0.25\textwidth]{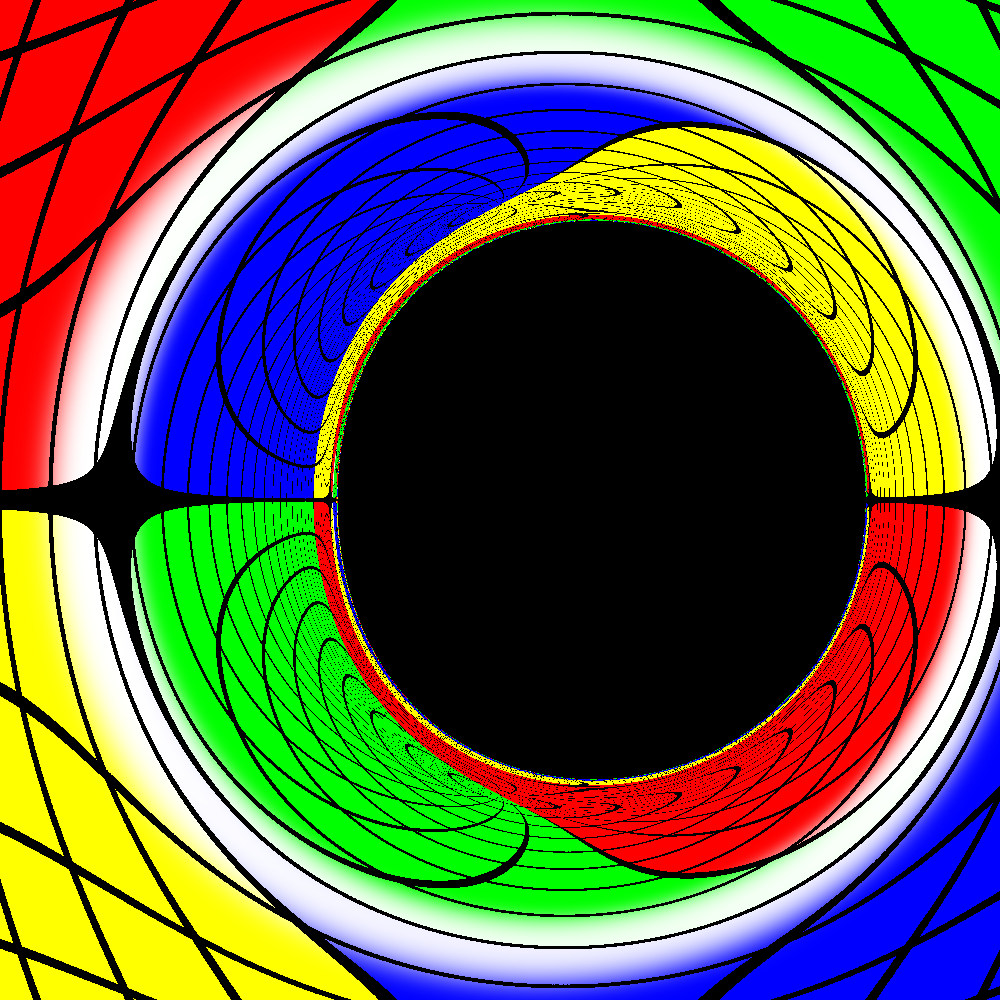}& \includegraphics[width=0.25\textwidth]{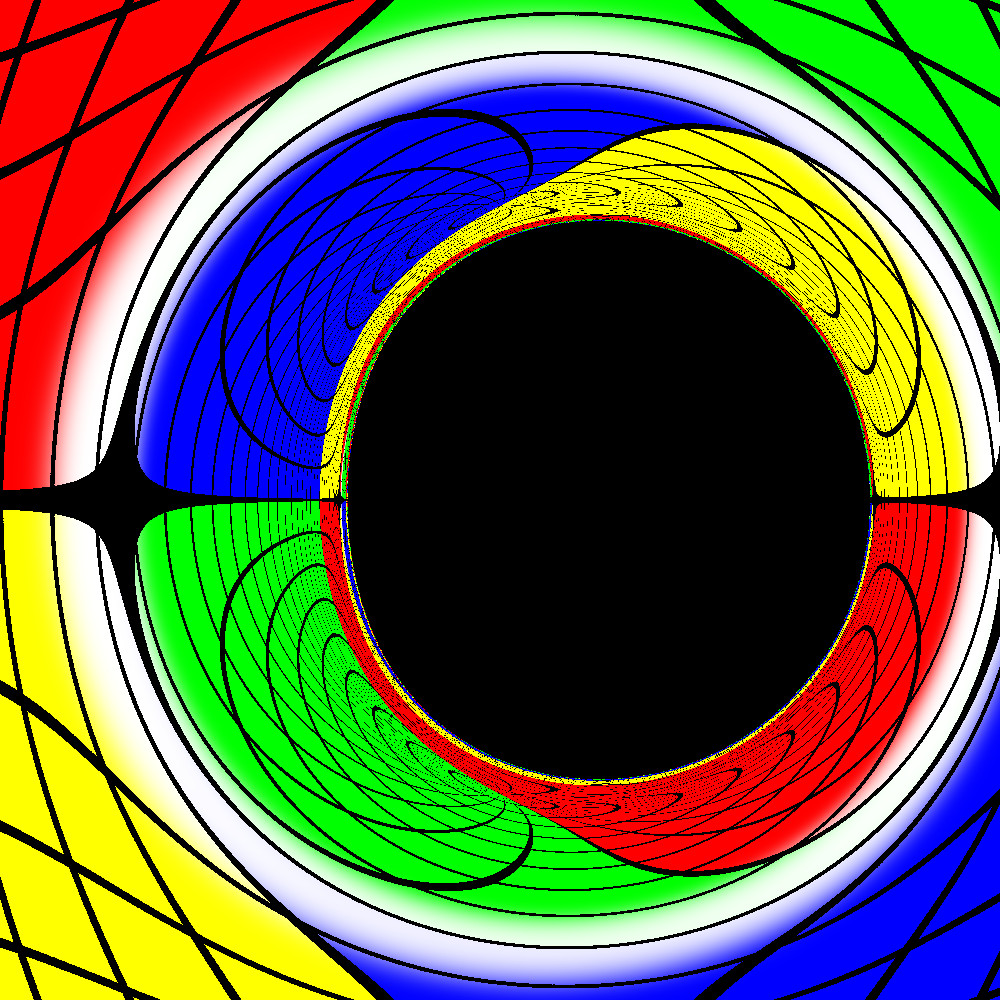}
\end{tabular}
\caption{Top row: Shadows of spinning BHs surrounded by a dipolar scalar cloud. From left to right we show the shadow of a configuration where the scalar cloud contains approximately $5\%$, $75\%$ and $98.2\%$ of the total ADM mass, respectively. Bottow row: For reference we also show the shadow of a Kerr BH with the same ADM mass and spin for each configuration shown in the top row. From left to right the dimensionless spin is given by $\chi \simeq 0.999$, $\chi \simeq 0.85$  and $\chi \simeq 0.894$, respectively. To help visualizing the extreme gravitational lensing in the vicinity of the BH, the celestial sphere is divided in four different patches as represented by the different colors. See Ref.~\cite{Cunha:2015yba} for details.
 \label{fig:shadows}}
\end{center}
\end{figure}

Examples of shadows for spinning BHs surrounded by scalar clouds are shown in Fig.~\ref{fig:shadows}. Three different configurations illustrate the differences between light and heavy clouds. For comparison we also show the shadow of a Kerr BH with comparable mass and spin. As clearly illustrated, the shadow of light clouds is similar to the one produced by a Kerr BH with comparable parameters. However for very heavy scalar configurations the structure of the shadow can be quite distinct from that of an isolated Kerr BH. 

%%%%%%%%%%%%%%%%%%%%%%%%%%%%%%%%%%%%%%%%%%%%%%%%%%%%%%%%%%%%%%%%%
\paragraph{Quasi-Periodic-Oscillations and Iron K$\alpha$ line} 
%%%%%%%%%%%%%%%%%%%%%%%%%%%%%%%%%%%%%%%%%%%%%%%%%%%%%%%%%%%%%%%%%
In addition to signatures in the BH shadow, deviations in the geodesic structure due to superradiant clouds have also been proposed to be potentially observable through the observation of ``quasi-periodic-oscillations'' (QPOs) in the X-ray spectrum of accreting BHs~\cite{Franchini:2016yvq}. High frequency QPOs are thought to originate from the innermost regions of the accretion disk and to be associated with the epicyclic frequencies of the geodesic  motion of time-like particles. Therefore QPOs provide a direct measure of the spacetime structure around the BH and could constrain the presence of superradiant clouds. The iron K$\alpha$ line, expected in the reflection spectrum of accreting BHs, can also be used to constrain the presence of a structure in the BH vicinity~\cite{Ni:2016rhz}, since such structure would affect the spacetime geodesics probed by the iron spectrum.

A detailed study of the possible constraints from QPOs was done in Ref.~\cite{Franchini:2016yvq} whereas prospects for iron K$\alpha$ line measurement were studied in Ref.~\cite{Ni:2016rhz}. Both studies concluded that heavy bosonic clouds would be easily distinguishable from a vacuum Kerr BH, but measuring the effect of ``light'' superradiantly-produced bosonic clouds is not feasible with current data.

%%%%%%%%%%%%%%%%%%%%%%%%%%%%%
\paragraph{Black-hole lasers}
%%%%%%%%%%%%%%%%%%%%%%%%%%%%%
As we discussed  in Sec.~\ref{sec:axion_coupling}, if the scalar field that populates the boson cloud couples to photons, the system can become unstable against the emission of EM bursts when the mass of the cloud reaches a critical value set by Eq.~\eqref{eq:EMinst_crit}~\cite{Ikeda:2019fvj,Boskovic:2018lkj}. 
For the axionic couplings discussed in Sec.~\ref{sec:axion_coupling} the EM waves emitted during the process are mostly monochromatic with frequencies given by~\cite{Rosa:2017ury,Sen:2018cjt,Ikeda:2019fvj,Boskovic:2018lkj}
\be
f_{\rm EM}\sim \frac{\omega_R}{4\pi} \sim 1200\,{\rm Hz}\left(\frac{m_S}{10^{-11}{\rm eV}}\right)\,.
\ee
For bosons with masses $m_S\lesssim 10^{-11}$ eV, corresponding to the range of masses for which astrophysical BHs with masses $M\gtrsim 10 M_{\odot}$ can become superradiantly unstable (cf.~\ref{sec:bounds_mass}), the frequency of these EM waves would be smaller than the typical plasma frequency of the interstellar medium~\cite{Sen:2018cjt,Boskovic:2018lkj}, thus making their direct detection unlikely. However it has been suggested that the properties of the EM bursts could be linked to ``Fast Radio Bursts'' if a QCD axion with mass $m_S\sim 10^{-5}$ eV and primordial BHs with masses around $M\sim 10^{-6} M_{\odot}$ exist in our Universe~\cite{Rosa:2017ury}. This process could also leave imprints in the CMB~\cite{Blas:2020nbs}.

Finally, even if not directly observable, this process effectively limits the maximum amount of energy that superradiance can extract before the EM burst sets in and therefore would directly affect the constraints from GW searches (cf. Sec.~\ref{sec:GWs}). 

\paragraph{Polarization-dependent propagation}

The pseudo-scalar nature of the axion-photon interaction, described by the equations of motion~\eqref{eq:MFEoMScalar2} and~\eqref{eq:MFEoMVector2}, also affects the polarization properties of EM waves propagating through an axion background~\cite{Plascencia:2017kca,Chen:2019fsq,Chigusa:2019rra}.

Due to the axion interaction the cloud behaves like an optically active medium, i.e. a medium in which left- and right-circular polarizations propagate at different speeds. To see this, consider for simplicity an EM plane wave propagating along the $z$-axis with wavelength much smaller than the length scale over which the background axion field changes appreciably, such that space derivatives of the axion field can be neglected. We assume that the wave propagates in a region sufficiently far from the BH such that the spacetime is approximately flat. Within this approximation the field equations~\eqref{eq:MFEoMVector2} admit plane-wave solutions of the form~\cite{Harari:1992ea}
\begin{equation}
\mathbf{A}_{\pm} \approx A_{0} \cos\left[\omega t-z\left(\omega \pm k_{\rm a}\frac{\partial \Psi}{\partial t}\right)\right] \mathbf{e}_x \mp \sin\left[\omega t-z\left(\omega \pm k_{\rm a}\frac{\partial \Psi}{\partial t}\right)\right]\mathbf{e}_y\,,
\end{equation}
where $\pm$ denotes right- and left-handed circularly polarized EM waves and we expanded the phase up to linear order in $k_{\rm a}$. In particular, due to the pseudo-linear nature of the axion coupling the two EM circular polarizations propagate with different phase velocities, a property known as birefringence. For a linearly-polarized EM wave propagating through an axion background this effect induces a relative phase shift between the two circularly polarized EM components. This leads to a rotation of the plane of polarization of the wave while it propagates through the axion background, in analogy with the Faraday rotation effect that occurs for linearly-polarized EM waves propagating in a magnetized plasma. 

Although birefringence is not unique to axion clouds, a smoking-gun feature of the axion birefringence would be the existence of a time-dependent angle of rotation of the plane of polarization, which would oscillate with period $\sim 1/\mu_S$. In Refs.~\cite{Plascencia:2017kca,Chen:2019fsq} this effect was proposed as a means to detect or constrain the presence of an axion cloud around supermassive BHs. In particular, Ref.~\cite{Chen:2019fsq} showed that polarimetric measurements by the Event Horizon Telescope of the supermassive BH M87$^*$ (Sgr A$^*$) could detect or rule out the presence of an axion with masses around $m_S\sim 10^{-20}$ eV ($m_S\sim 10^{-17}$ eV) with decay constant $k_{\rm a} \gtrsim \mathcal{O}(10^{-16})$ GeV$^{-1}$.

In addition, Ref.~\cite{Plascencia:2017kca} also studied how the axion interaction affects the bending of an EM wave passing through the axion cloud. Let us assume that the cloud's self-gravity can be neglected and, again, that the length scale over which the axion field changes is much larger than the wavelength of the EM wave. Under those assumptions, and considering propagation in the equatorial plane of a non-spinning BH, Ref.~\cite{Plascencia:2017kca} showed that in the eikonal limit the geodesic equation~\eqref{geodesics:kerr} gets modified to: 
\be
\dot{r}^2=E^2-\frac{L^2}{r^2}\left(1-\frac{2M}{r} \right) \pm 2 k_{\rm a}E\frac{\partial \Psi}{\partial t}\,,
\ee
where again $\pm$ corresponds to right- and left-handed circularly polarized EM waves. Since the two polarizations follow different photon orbits when passing through the cloud, they experience a different deflection angle. This angular splitting was computed in Ref.~\cite{Plascencia:2017kca}. For axion masses between $10^{-18}\,{\rm eV} \lesssim m_S\lesssim 10^{-12}\,{\rm eV}$ and decay constants in the range $10^{-18}\,{\rm GeV}^{-1}\lesssim k_{\rm a} \lesssim 10^{-12}\,{\rm GeV}^{-1}$ they found that the angular splitting could take values of the order ${\cal O}(10^{-8} \div 1)$ arcsec, depending on the axion mass, decay constant and field's amplitude. A proper investigation of whether this effect can be detected with future radio telescopes remains to be done, however as noted in~\cite{Plascencia:2017kca} angular resolutions of the order ${\cal O}(10^{-3})$ arcsec are achievable using very-long-baseline interferometry techniques, making a detection potentially achievable.

%%%%%%%%%%%%%%%%%%%%%%%%%%%%%%%%%%%%%%%%%%%%%%
\subsubsection{Signatures in binary systems}
%%%%%%%%%%%%%%%%%%%%%%%%%%%%%%%%%%%%%%%%%%%%%%
The constraints discussed so far assume that the BH-cloud system can be considered in isolation. In the presence of a companion star or BH a wealth of novel effects can occur that lead to very distinct observational signatures, as we now discuss. 

\paragraph{Tidal effects} 
As we discussed in Sec.~\ref{sec:cloud_tides}, the tidal force exerted on the boson by a companion orbiting body can excite additional modes in the cloud which can lead to a very rich dynamical evolution of the system. In particular, the cloud's tidal deformation induced by the companion triggers a mixing between modes that leads to an efficient transfer of angular momentum between the cloud and the orbiting object~\cite{Zhang:2018kib,Baumann:2019ztm}. For weak tidal perturbations, this mixing is in general too small to significantly affect the orbital motion. However, as first shown in Ref.~\cite{Baumann:2018vus}, the effect can be enhanced when the orbital frequency $\Omega$ matches the energy split $\Omega=(\omega_a-\omega_b)/(m_a-m_b)$ between two states $\psi_a$ and $\psi_b$ for which the tidal perturbation induces a non-zero transition probability (see~\cite{Berti:2019wnn,Zhang:2019eid} for an extension to eccentric and hyperbolic orbits). As shown in~\cite{Zhang:2018kib,Baumann:2018vus}, depending on the excited modes and the orientation of the orbit, the angular momentum transfer can make the orbit shrink faster than in vacuum, or even be strong enough to compensate for the angular momentum loss caused by GW emission and temporarily stall the orbit (cf. Sec.~\ref{sec:bounds_floating}). Considering for example a mixing between two levels with principal quantum number $\tilde{n}_{a}$ and $\tilde{n}_{b}$ one finds that the GWs emitted by the binary system during such resonances have frequency~\cite{Baumann:2018vus,Zhang:2018kib,Baumann:2019ztm}:
\begin{equation}
f_{\rm res}=\frac{\Omega}{\pi}=0.2\,{\rm Hz}\frac{1}{|\Delta m|}\left(\frac{60 M_{\odot}}{M}\right)\left(\frac{M\mu}{0.07}\right)^3\left|\frac{1}{\tilde{n}_a^2}-\frac{1}{\tilde{n}_a^2}\right|\,,
\end{equation}
where $\Delta m=m_a-m_b$ is the difference between the azimuthal numbers of the two modes. Ref.~\cite{Baumann:2019ztm} showed that these resonances can lead to a significant dephasing of the GW signal when compared to a binary system without a cloud, therefore suggesting that they can leave detectable imprints in the GW signal emitted by the binary, especially for systems observable through the future space-based GW detector LISA~\cite{Audley:2017drz}.

The structure of the cloud would also be imprinted in the GW signal of a binary system through its multipole moments or through its tidal Love numbers that could be significantly different than in the case where no cloud is present~\cite{Baumann:2018vus,Baumann:2019ztm}.
The mode-mixing discussed above would induce a strong time-dependence on these quantities, especially at frequencies close to the resonances~\cite{Baumann:2018vus,Baumann:2019ztm}. These signatures would leave interesting imprints in the GW signal emitted by the binary system, in particular Ref.~\cite{Baumann:2019ztm} suggested that they could be used to distinguish scalar and vector clouds, however a detailed study of the  detectability of these effects has not yet been done. 

\paragraph{Tidal disruption}
The discussion above assumes that the tidal field induced by the companion is sufficiently weak such that it can be treated perturbatively, but for sufficiently large tidal fields the boson cloud will be disrupted (cf. Section~\ref{sec:cloud_tides}). The results of Sec.~\ref{sec:cloud_tides} allows us to estimate when such disruptions can occur for known BHs. For example, let us consider the Cygnus X-1 system and the BH at the center of our galaxy, SgrA$^*$. 

Cygnus X-1 is a binary system composed of a BH of mass $M\sim 15 M_\odot$ and a companion with $M_c\sim 20 M_\odot$ at a distance $R\sim 0.2 \,{\rm UA} \sim 3\times 10^{10}  \,{\rm m}$~\cite{Orosz_2011}. With these parameters, we find a tidal moment $M_cM^2/R^3 \sim 5\times 10^{-19}$ (see Section~\ref{sec:cloud_tides}). For it to sit at the critical tide for disruption, Eq.~\eqref{tde_scaling}, one needs $M\mu_S\sim 2\times 10^{-3}$. The time scale for the growth of a scalar cloud via superradiance is given in Eq.~\eqref{wIslope} and is of order $\sim (M\mu_S)^9 M$, and therefore the growth timescale for a cloud that can be tidally disrupted is too large to be meaningful for scalar fields, but it could be potentially important for vector fields that instead can grow on much smaller timescales $\sim (M\mu_V)^7 M$. The tide is also small enough that it should not be affecting any of the constraints derived from the possible non-observation of GWs from the system~\cite{Yoshino:2014wwa,Sun:2019mqb}.

On the other hand, at the center of our galaxy there is a supermassive BH of mass $\sim 4\times 10^6 M_{\odot}$ with known companions~\cite{Abuter:2018drb,Naoz:2019sjx}. For the closest known star, S2, with a pericenter distance of $\sim 1400M$ we find $M_cM^2/R^3\sim 2\times 10^{-15}$, or a critical coupling $M\mu \sim 9\times 10^{-3}$ (we assume $M_c\sim 20 M_{\odot}$ but the result above is only mildly dependent on the unknown mass of S2). This is now a potential source of tidal disruption for interesting coupling parameters, and will certainly affect the estimates using pure dipolar modes to estimate GW emission.
Tidal disruptions could also be a potential source of GWs, although the detailed GW signal emitted during the process has not been computed so far.

%%%%%%%%%%%%%%%%%%%%%%%%%%%%%%%%%%%%%
\paragraph{Self-gravity of the cloud}
%%%%%%%%%%%%%%%%%%%%%%%%%%%%%%%%%%%%% 
The presence of a spatially extended boson cloud around supermassive BHs can also significantly influence the motion of small compact objects and stars orbiting them~\cite{Ferreira:2017pth,Hannuksela:2018izj,Bar:2019pnz,Amorim:2019hwp}. For example, compared to the vacuum case, the boson cloud changes the gravitational potential felt by the orbiting bodies which is potentially measurable through precise measurements of the orbit of stars around SgrA$^*$~\cite{Ferreira:2017pth,Bar:2019pnz,Amorim:2019hwp}. This would also be imprinted in the GWs emitted by small compact objects orbiting supermassive BHs~\cite{Ferreira:2017pth,Hannuksela:2018izj},  one of the most promising GW sources for LISA~\cite{Audley:2017drz}.

In addition to this effect, it was also shown in Ref.~\cite{Ferreira:2017pth} that the presence of a non-axially symmetric scalar field profile causes orbiting bodies to precess at a rate that depends solely on the parameters governing the scalar field. The field also causes variations of other orbital parameters, such as the orbital nodes and the orbital eccentricity~\cite{Ferreira:2017pth,Amorim:2019hwp}. These effects add to the general relativistic ones and therefore the presence of a cloud can be inferred by carefully measuring the orbital parameters of satellites around supermassive BHs~\cite{Ferreira:2017pth,Amorim:2019hwp}. In fact, as shown in~\cite{Amorim:2019hwp} precise measurements by the GRAVITY experiment~\cite{GRAVITY} of the S2 star could constrain the fractional mass of a possible scalar cloud around SgrA$^*$ to be $\lesssim 1\%$ for scalar masses in the range $10^{-20}\, {\rm eV} \lesssim m_S\lesssim 10^{-18}\, {\rm eV}$. 

The presence of a boson cloud can also give rise to resonant orbits when the orbital frequency matches the characteristic frequencies of the system~\cite{Ferreira:2017pth}, which includes one corotation resonance, where a large number of orbiting bodies will tend to pile up, and two Lindbad resonance where the scalar field can exchange angular momentum with the orbiting body~\cite{Ferreira:2017pth}, which can potentially cause floating orbits (cf. Sec.~\ref{sec:bounds_floating} and discussion above).

Finally, the dissipation of the cloud due to the monochromatic GWs discussed in Sec.~\ref{sec:GWs} can also significantly affect the orbital motion of a binary system. It was shown in Ref.~\cite{Kavic:2019cgk} that this energy loss tends to push the companion outwards (similarly to the floating orbits discussed above and in Sec~\ref{sec:bounds_floating}) and this could be used to detect scalar masses in the range $10^{-15}{\rm eV}\lesssim m_S\lesssim 3\times 10^{-12}{\rm eV}$ through the precise timing of the orbital period of pulsar-BH binaries.

%%%%%%%%%%%%%%%%%%%%%%%%%%%%%%%%%%%%%%%%%%%%
\paragraph{Gravitational drag and accretion} 
%%%%%%%%%%%%%%%%%%%%%%%%%%%%%%%%%%%%%%%%%%%%
In addition to the effects discussed above, a binary system evolving in a boson environment will  be influenced by at least two additional effects: accretion and dynamical friction. As the bodies move through the medium, they accrete material while simultaneously exerting a gravitational pull on the surrounding environment. These effects leave an imprint on the inspiral dynamics and, in consequence, the GW phasing. The effect of dynamical friction and accretion for boson clouds as been studied in~\cite{Ferreira:2017pth,Zhang:2019eid} (see also~\cite{Macedo:2013qea} for related studies). Those effects are especially important close to the peak of the scalar configuration, where they can dominate over gravitational radiation reaction~\cite{Zhang:2019eid}.

The cumulative effect of these effects over a large number of orbits can leave a clear imprint in the gravitational waveform, measurable as a phase shift in the GW signal relative to the inspiral in vacuum. In particular, Ref.~\cite{Zhang:2019eid} showed that the effect of dynamical friction could be particularly important for stellar mass BHs sweeping over the LISA band. 

In addition, the transfer of angular momentum induced by the dynamical friction caused by the cloud could also increase 
the probability of a BH dynamically capturing other compact objects therefore increasing the formation rate of BH 
binaries~\cite{Zhang:2019eid}.

%%%%%%%%%%%%%%%%%%%%%%%%%%%%%%%%%%%%%%%%%%%%%%%%%%%%%%%%%%%%%%%%%%%%%%%%%%%%%%%%%%%%%%%%%%%%%%%%%%%%%%%%%%%%%%%%%%%
\subsection{Bounds on ultralight particles from pulsar timing}\label{sec:bounds_pulsar}
%%%%%%%%%%%%%%%%%%%%%%%%%%%%%%%%%%%%%%%%%%%%%%%%%%%%%%%%%%%%%%%%%%%%%%%%%%%%%%%%%%%%%%%%%%%%%%%%%%%%%%%%%%%%%%%%%%%
%
The observation of an isolated compact object with spindown time scale $\tau_{\rm spindown}$ would exclude superradiant instabilities for that system, at least on instability time scales $\tau<\tau_{\rm spindown}$. Therefore, spinning compact objects for which a (possibly small) spindown rate can be measured accurately are ideal candidates to put constraints on superradiant mechanisms, especially those triggered by ultralight bosons.

Unfortunately, direct measurements of the spin derivative of BHs are not available\footnote{As previously discussed, observations of LMC X-3 and Cygnus X-1 are consistent with a constant spin on a time scale of the order of ten years, which gives a very mild bound compared to the one discussed in this section.}, so that constraints on superradiant instabilities using BH mass and spin measurements are only meaningful in a statistical sense, as discussed above. On the other hand, \emph{both} the spin and the spindown rate of \emph{pulsars} are known with astonishing precision through pulsar timing (cf., e.g., Ref.~\cite{lorimer2005handbook}). For several sources, the rotational frequency is moderately high, $f_{\rm spin}=\Omega/(2\pi)\simeq (500-700)\,{\rm Hz}$, and the spindown time scale can be extremely long, $\tau_{\rm spindown} = \Omega/(\dot\Omega)\simeq 10^{10}\,{\rm yr}$. As an example, the ATNF Pulsar Catalogue~\cite{catalogue,Manchester:2004bp} contains $398$ ($40$) pulsars for which $\tau_{\rm spindown}>2\times 10^9\,{\rm yr}$ ($\tau_{\rm spindown}>2\times 10^{10}\,{\rm yr}$).

\begin{figure}[th]
\begin{center}
\begin{tabular}{c}
  \includegraphics[width=0.7\textwidth]{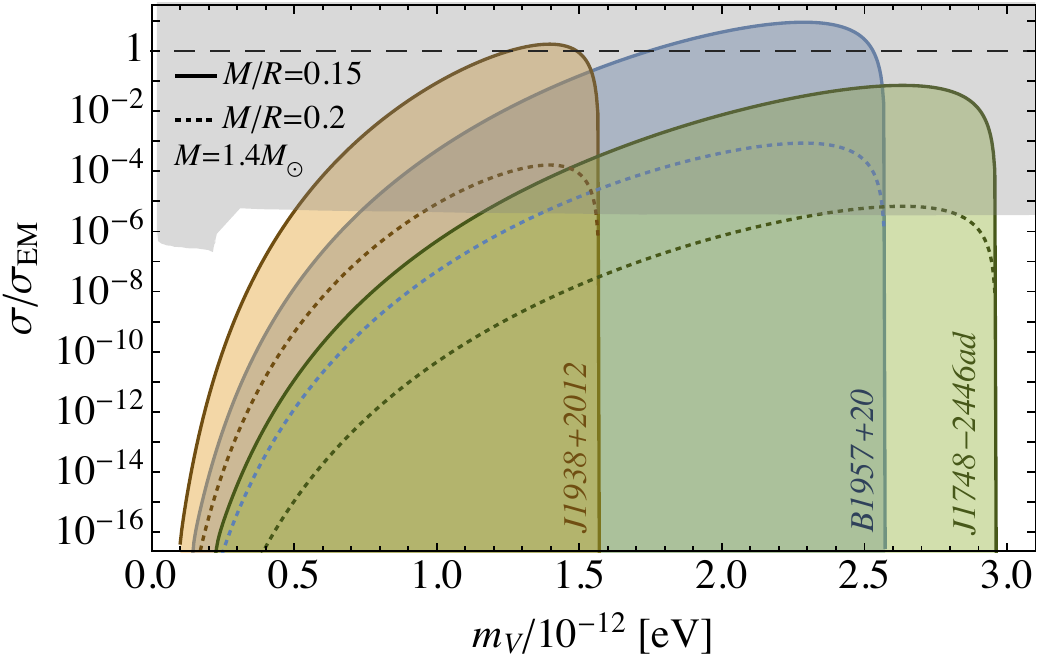}
\end{tabular}
\caption{Exclusion plots in the $\sigma/\sigma_{\rm{EM}}$ vs. $m_V$ plane for the model given by action~\ref{action}. The bounds are obtained from the measurements of spin and spindown rate of pulsars J1938+2012 (orange)~\cite{Stovall:2016unz} and J1748-2446ad (green)~\cite{Hessels:2006ze}, and of the pulsar binary B1957+20 (blue)~\cite{Arzoumanian:1993qt}. In all cases we assumed $M= 1.4M_\odot$ and two values of the compactness, namely $M/R= 0.15$ (solid) and $M/R= 0.2$ (dotted). %
The shaded areas correspond to regions excluded by the superradiant instability because $\tau<\tau_{\rm spindown}$ for a given pulsar (i.e., the pulsar is observed to spin down at much longer rate than that predicted by the superradiant instability in that region of the parameter space). 
The horizontal dashed line indicates when $\sigma$ has the same value than the standard electric conductivity, i.e. $\sigma$ $\sigma=\sigma_{\rm EM}$.
We only display the region where $\sigma\gg \Omega$. In the opposite limit, the instability time scale grows as $\tau\sim1/\sigma$ [cf.\ Eq.~\eqref{wIaxial}] and eventually $\tau>\tau_{\rm spindown}$ for sufficiently small $\sigma$. 
The shaded gray region is excluded from distortions of the CMB blackbody from $\gamma\to X$ photon depletion~\cite{Mirizzi:2009iz}. Taken from Ref.~\cite{Cardoso:2017kgn}.
} \label{fig:pulsar}
\end{center}
\end{figure}

In Fig.~\ref{fig:pulsar}, we show the excluded regions in the \emph{conductivity vs dark-photon mass plane} for the model given by action~\eqref{action}, where the conductivity can be directly related to the fractional hidden charge of the electron~\cite{Cardoso:2017kgn}.
The bound is obtained by imposing $\tau<\tau_{\rm spindown}$ for three known sources, namely pulsars J1938+2012~\cite{Stovall:2016unz} and J1748-2446ad~\cite{Hessels:2006ze}, and pulsar binary B1957+20~\cite{Arzoumanian:1993qt}. The first one is representative of a pulsar with an exceptionally long spindown time scale ($\tau_{\rm spindown}\simeq1.1\times 10^{11}\,{\rm yr}$), but with a moderately large spin ($f_{\rm spin}\simeq 380\,{\rm Hz}$, which corresponds to $\Omega/\Omega_K\approx0.28$ assuming $M=1.4 M_\odot$ and $M/R=0.15$). 
The second one is the fastest pulsar known to date ($f_{\rm spin}\simeq 716\,{\rm Hz}$, corresponding to $\Omega/\Omega_K\approx0.53$ for $M=1.4 M_\odot$ and $M/R=0.15$), but only an upper bound on its spin derivative is available, from which we infer $\tau_{\rm spindown}>7.6\times 10^7\,{\rm yr}$.
The last one is representative of a pulsar with very large spin ($f_{\rm spin}\simeq 622\,{\rm Hz}$, which corresponds to $\Omega/\Omega_K\approx0.46$ again assuming $M=1.4 M_\odot$ and $M/R=0.15$), but moderately long spindown time scale ($\tau_{\rm spindown}\simeq3\times 10^{9}\,{\rm yr}$).
Furthermore, because our fits for $\alpha_1$ and $\alpha_2$ appearing in Eq.~\eqref{wIaxial} are independent of $\Omega$ only for $\Omega/\sigma\ll1$, in Fig.~\ref{fig:pulsar} we show only values of the conductivity which satisfy $\sigma\gg \Omega$. 

The exclusion plot shown in Fig.~\ref{fig:pulsar} is obtained as follows.
For a given measurement of the spin frequency of a pulsar, $f_{\rm spin}$, we can estimate $\Omega$ and compute the instability time scale as a function of $\sigma$ and $\mu_V$ through Eq.~\eqref{wIaxial}. Furthermore, the measurement of a spindown time scale for a pulsar, $\tau_{\rm spindown}$, implies that a faster spindown rate caused by the superradiant instability would be incompatible with observations.
Thus, imposing $\tau<\tau_{\rm spindown}$ yields an excluded region in the $\sigma$-$m_V$ plane. Fastly spinning pulsars constrain the rightmost part of the $\sigma$-$m_V$ diagram because the instability requires $\mu_V\sim\omega_R <m\Omega$. On the other hand, pulsars with longer spindown time scale correspond to higher threshold lines in the leftmost part of the $\sigma$-$m_V$ diagram.

The above results were obtained for the model~\eqref{action}, but similar analysis should be valid also in other cases, provided the ultralight bosons are coupled to the material of the star. For example, using a toy model for the coupling~\cite{Cardoso:2015zqa}, Ref.~\cite{Kaplan:2019ako} recently studied the superradiant instability of millisecond pulsars due to the interaction between ultralight scalar fields and neutrons.

%%%%%%%%%%%%%%%%%%%%%%%%%%%%%%%%%%%%%%%%%%%%%%%%%%%%%%%%%%%%
\subsection{Summary of the bounds on ultralight bosons from superradiant instabilities} \label{sec:bounds}
%%%%%%%%%%%%%%%%%%%%%%%%%%%%%%%%%%%%%%%%%%%%%%%%%%%%%%%%%%%%

Table~\ref{tab:bounds} summarizes current constraints on ultralight bosons coming from the absence of effects predicted 
by the superradiant instability of compact objects. 
For each measurement we provide the bounds on the mass of a scalar, vector, and tensor field, and also provide the 
relevant references in which these bounds have been derived. For a discussion on the caveats for each bound we refer to
the corresponding reference. 
In some cases, only bounds on scalars (and sometimes on vectors) have been explicitly derived. When a bound is not 
available in the literature, we derived it independently using the results discussed in the previous sections, 
especially in Sec.~\ref{sec:massive_unified}, and other results in the references. These new bounds are marked in 
Table~\ref{tab:bounds} with an asterisk. 

Note that, in the tensor (spin-2) case, we only consider hydrogenic modes (using the results of 
Ref.~\cite{Brito:2020lup}) and neglect the ``special'' dipolar mode~\cite{Brito:2013wya}, which might in principle 
provide stronger constraints but has been only computed to first order in the spin. As shown in Table~\ref{tab:inst}, 
the time scale for the most unstable hydrogenic tensor mode has the same scaling with the coupling than the most 
unstable vector mode, but a larger prefactor by a factor $\approx 5.6$. This explains why the lower bounds shown in 
Table~\ref{tab:bounds} are always less stringent for spin-$2$ relative to the spin-$1$ case.

%%%%%%%%%%%%%%%%%%%%%%%%%%%%%%%%%%%%%%%
\begin{table}
 \tiny
 \begin{tabular}{cccc}
  \hline
  \hline
 & excluded region (in eV)                         & source            & references\\
    \hline
  %%%%%
{\bf *} &  $5.2\times10^{-13}<{m}_S<6.5\times 10^{-12}$       & \multirow{3}{*}{Direct bounds from 
absence of spin down in Cyg~X-1.}  &
\multirow{3}{*}{\cite{Cardoso:2018tly,Brito:2020lup}} \\
{\bf *} &  $1.1\times10^{-13}<{m}_V<8.2\times10^{-12}$    &   & \\
{\bf *} &  $2.9\times10^{-13}<{m}_T<9.8\times10^{-12}$    &   & \\
  \hline
  %%%%%
&  $3.8\times 10^{-14}<{m}_S<2\times 3.4\times10^{-11}$ & \multirow{6}{*}{Indirect bounds from BH mass-spin measurements.}  
&
\multirow{3}{*}{\cite{Arvanitaki:2014wva,Pani:2012vp,Brito:2013wya,Baryakhtar:2017ngi}} \\
&   $5.5\times 10^{-20}<{m}_S<1.3\times 10^{-16}$ &    &\\
&   $2.5\times 10^{-21}<{m}_S<1.2\times 10^{-20}$ &    &\\
 &  $6.2\times 10^{-15}<{m}_V<3.9\times10^{-11}$     &   &  
\multirow{3}{*}{\cite{Brito:2017zvb,Stott:2018opm,
Cardoso:2018tly,Brito:2020lup,Stott:2020gjj,Unal:2020jiy}}\\
 &  $2.8\times 10^{-22}<{m}_V<  1.9\times 10^{-16}$ &    &\\
 &  $2.2\times 10^{-14}<{m}_T<  2.8\times10^{-11}$  &   & \\
 &  $1.8\times 10^{-20}<{m}_T<  1.8\times 10^{-16}$ &    &\\
 &  $6.4\times 10^{-22}<{m}_T<  7.7\times 10^{-21}$ &    &\\
  \hline 
&  $1.2\times 10^{-13}<{m}_S<1.8\times 10^{-13}$   & \multirow{4}{*}{Null results from blind all-sky searches for 
continuous GW signals.}  & \multirow{4}{*}{\cite{Palomba:2019vxe,Zhu:2020tht}} \\
&  $2.0\times 10^{-13}<{m}_S<2.5\times 10^{-12}$     &   & \\
&  ${m}_V$: NA     &   & \\
&  ${m}_T$: NA     &   & \\
  %%%%%
  \hline
  %%%%%
&  $6.4\times 10^{-13}<m_S<8.0\times 10^{-13}$     & \multirow{3}{*}{Null results from searches for continuous GW 
signals from Cygnus X-1.}  &
\multirow{3}{*}{\cite{Sun:2019mqb,Yoshino:2014wwa}} \\
&  ${m}_V$: NA     &   & \\
&  ${m}_T$: NA     &   & \\
    \hline
  %%%%%
&  $2.0\times 10^{-13}<m_S<3.8\times 10^{-13}$    & \multirow{3}{*}{Negative searches for a GW background.}  &
\multirow{3}{*}{\cite{Brito:2017wnc,Brito:2017zvb,Tsukada:2018mbp,Tsukada:2020lgt}} \\
&  $0.8\times 10^{-13}\,{\rm eV}< m_V < 6.0\times 10^{-13} \,{\rm eV}$     &   & \\
&  ${m}_T$: NA     &   & \\
      \hline
  %%%%%
&  $5\times 10^{-13}<{m}_S<3\times 10^{-12}$     & \multirow{3}{*}{Bounds from pulsar timing.}  &
\multirow{3}{*}{\cite{Cardoso:2017kgn,Kaplan:2019ako}} \\
&  ${m}_V\sim10^{-12}$     &   & \\
&  ${m}_T$: NA     &   & \\
  \hline
    %%%%%
&  $2.9\times 10^{-21}<m_S<4.6\times 10^{-21}$ & \multirow{3}{*}{Bounds from mass and spin measurement of M87 with 
EHT.} 
 &
\multirow{3}{*}{\cite{Davoudiasl:2019nlo,Stott:2020gjj}} \\
& $8.5\times 10^{-22}<m_V<4.6\times 10^{-21}$     &   & \\
& $7.2\times 10^{-22}<m_T<2.5\times 10^{-20}$     &   & \\
  \hline
  \hline
 \end{tabular}
 \caption{Current constraints on ultralight bosons coming from 
the absence of effects predicted by the superradiant instability of compact objects. These bounds will be constantly 
updated online~\cite{webpage} as new results come out (``NA'' means that the bound is not available yet, whereas an 
asterisk stands for bounds originally derived in this work based on previous results). The 
quoted bounds are typically at $1\sigma$ significance level, but some of them are affected by caveats discussed in the 
corresponding reference. For a recent detailed analysis on bounds from BH mass-spin measurements, see Refs.~\cite{Stott:2020gjj,Unal:2020jiy}.
}\label{tab:bounds}
\end{table}
%%%%%%%%%%%%%%%%%%%%%%%%%%%%%%%%%%%%%%%

%%%%%%%%%%%%%%%%%%%%%%%%%%%%%%%%%%%%%%%%%%%%%%%%%%%%%%%%%%%%
\subsection{Plasma interactions} \label{sec:astro_plasma}
%%%%%%%%%%%%%%%%%%%%%%%%%%%%%%%%%%%%%%%%%%%%%%%%%%%%%%%%%%%%
Already in his PhD thesis, Teukolsky proposed that plasmas could be used as mirrors to trigger superradiant 
instabilities~\cite{teukolskythesis,Press:1972zz}. This expectation is based on the fact that the frequency of 
amplified radiation is much smaller than the plasma frequency $\omega_p^{-1}$ (cf. Eq.~\eqref{plasma_freq}) of the 
interstellar medium, therefore photons scattered by a BH in vacuum would be reflected by a spherically-symmetric plasma 
distribution.

Modelling the plasma effect with the effective Proca equation~\eqref{procaplasma}, Ref.~\cite{Pani:2013hpa} studied 
plasma-triggered superradiant instabilities in the context of small primordial BHs in the early 
universe~\cite{Carr:2009jm}. In that case the BHs are surrounded by a mean cosmic electron density that depends on the 
redshift, which translates to a (weakly) time-dependent plasma frequency through Eq.~\eqref{plasma_freq}.
It was also argued that the same plasma confinment can take place for stellar-origin BHs due to the low plasma 
frequency of the interstellar medium~\cite{Conlon:2017hhi}.

However, as mentioned in Sec.~\ref{plasma-triggered} the above studies are based on some strong assumptions, most 
notably on the validity of the effective Proca equation~\eqref{procaplasma} and on neglecting relativistic and 
nonlinear effects. As recently realized, the correct field equations governing the photon-plasma interaction are 
different from Proca even at the linear level~\cite{Cannizzaro:2020uap} and, most importantly, nonlinear and 
relativistic effects (relevant during the development of the superradiant instability) make the plasma 
transparent, quenching the instability~\cite{Cardoso:2020nst,Blas:2020kaa}. It would be interesting to study the 
problem in its full glory by performing a detailed numerical simultation of the full Maxwell-plasma system. 

Another important limitation concerns the role of the geometry of the accretion flow, since a nonspherical plasma would 
be less efficient in confining radiation, even at the linear level. In particular, a spinning BH drags plasma in its 
vicinity so that the density profile is expected to be nonspherical. As an example of superradiance stimulated 
amplification in a realistic setting, Ref.~\cite{VanPutten:1999vda} studied superradiant confinement in a toroidal 
magnetosphere around a Kerr BH, arguing that the repeated amplification of EM (with time scales of the order of the 
second for stellar-mass BHs) might be a model for periodic $\gamma$-ray bursts.

\subsection{Intrinsic limits on magnetic fields} \label{sec:bounds_magnetic}
%%%%%%%%%%%%%%%%%%%%%%%%%%%%%%%%%%%%%%%%%%%%%%%%%%%%%%%%%%%%%%%%%%%%%%%%%%%%

In Sec.~\ref{sec:magnetic} we showed that rotating BHs immersed in a magnetic field are unstable against superradiant modes. In complete analogy with the discussion of Sections~\ref{sec:evolution} and~\ref{sec:bounds_mass}, due to this instability, the energy density of the radiation in the region $r\lesssim 1/B$, with $B$ the magnetic field strength, would grow exponentially in time at the expense of the BH angular momentum, with the end state being a spinning BH with a spin set by the superradiant threshold\footnote{As was pointed out in Ref.~\cite{Brito:2014nja}, for the (unrealistic) Ernst metric in which radiation cannot escape, the end state is most likely a rotating BH in equilibrium with the outside radiation, similarly to the asymptotically AdS case discussed in Sec.~\ref{sec:KerrAdS}. However, in realistic situations part of the radiation will escape to infinity, reducing the BH spin (see discussion below).}. This implies an upper bound on the spin of magnetized BHs, again leading to holes in the BH Regge plane (cf. Sec.~\ref{sec:evolution}). This was used in Ref.~\cite{Brito:2014nja} to put intrinsic limits on magnetic fields around astrophysical BHs.

\begin{figure}[t]
\begin{center}
%\begin{tabular}{c}
\epsfig{file=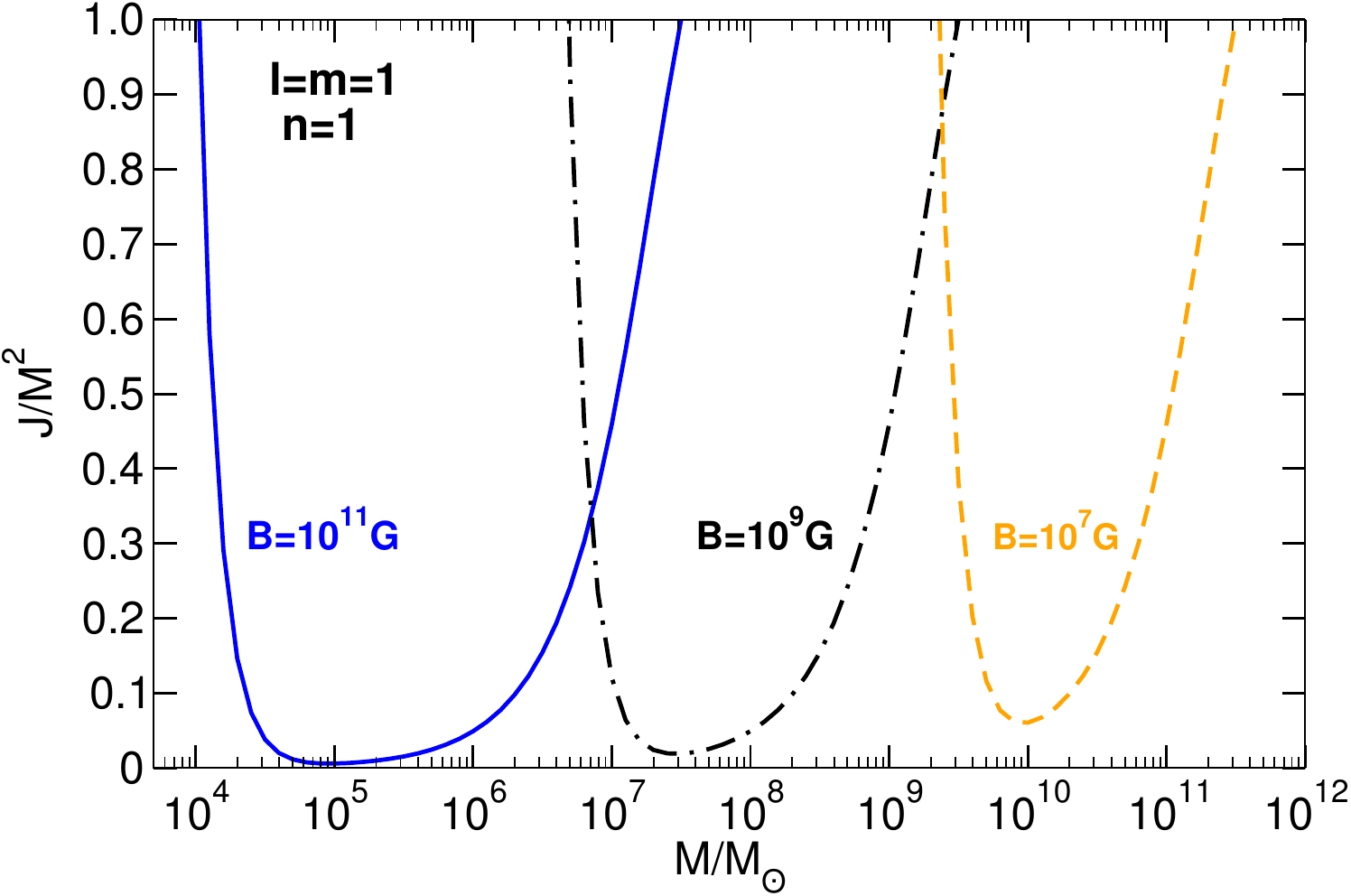,width=0.7\textwidth,angle=0,clip=true}
%\end{tabular}
\caption{Contour plots in the BH Regge plane corresponding to an instability time scale shorter than $\tau_{\rm Salpeter}\sim 4.5\times 10^7{\rm yr}$ for different values of the magnetic field strength $B$ for modes with $l=m=n=1$. BHs lying above each of these curves would be unstable on an observable time scale. The threshold lines are obtained using Eq.~\eqref{wI_magnetic} in the range $10^{-4}\lesssim B/B_M\lesssim 1$. From Ref.~\cite{Brito:2014nja}.
\label{fig:Regge_magnetic}}
\end{center}
\end{figure}

In Fig.~\ref{fig:Regge_magnetic} we show the BH Regge plane with contour curves corresponding to an instability time scale $1/\omega_I$, given by Eq.~\eqref{wI_magnetic}, of the order of the Salpeter time. % Spin measurements of supermassive BHs would allow us to locate data points on the Regge plane, thus excluding a whole range of possible magnetic fields. 
Since the contours extend almost up to $J/M^2\sim 0$, one interesting consequence of these results is that essentially \emph{any} observation of a spinning supermassive BH (even with spin as low as $J/M^2\sim 0.1$) would provide some constraint on $B$. However, these observations can possibly exclude only very large values of $B$. For example a putative observation of a supermassive BH with $M\sim 10^9 M_\odot$ and $J/M^2\gtrsim 0.5$ can potentially exclude the range $10^7 {\rm Gauss}\lesssim B\lesssim 10^9 {\rm Gauss}$\footnote{The strength of the magnetic field can be measured defining the characteristic magnetic field $B_M=1/M$ associated to a spacetime curvature of the same order of the horizon curvature. In physical units this is given by $B_M\sim 2.4\times 10^{19} \left(M_{\odot}/M\right) {\rm Gauss}$.}.

Although these results are only valid when $B/B_M\lesssim 1$, this limit includes the most interesting region of the parameter space. Indeed, the strongest magnetic fields around compact objects observed in the Universe are of the order of $10^{13}$--$10^{15} {\rm Gauss}$~\cite{McGill} and, in natural units, this value corresponds to $B/B_M\sim 10^{-6}$--$10^{-4}$. For astrophysical BHs, a reference value for the largest magnetic field that can be supported in an accretion disk is given by $B\sim 4\times 10^8 \left(M/M_\odot\right)^{-1/2}{\rm Gauss}$~\cite{ReesAGN} so that the approximation $B\ll B_M$ is well justified.

The main caveat of these bounds is that they were obtained using the Ernst metric which, as we discussed in Sec.~\ref{sec:magnetic}, is not asymptotically flat, but instead describes a BH immersed in a magnetic field which is supported by some form of matter at infinity. In most realistic models it is expected that the Ernst metric is a relatively good approximation for the geometry of astrophysical BHs only up to a cutoff distance associated with the matter distribution.
Considering that the accretion disk is concentrated near the innermost stable circular orbit, this would imply that these results can be trusted only when $B/B_M\gtrsim 0.1$~\cite{Brito:2014nja}, which is a very large value for typical massive BHs. On the other hand, the Ernst metric is more accurate to describe configurations in which the disk extends much beyond the gravitational radius, as is the case in various models (cf. Refs.~\cite{lrr-2013-1,Barausse:2014tra}). In this case, however, the magnetic field will not be uniform and the matter profile has to be taken into account.  While the simplistic analysis of Ref.~\cite{Brito:2014nja} can provide the correct order of magnitude for the instability, a more refined study is needed to assess the validity of these results in the full range of $B$.

%%%%%%%%%%%%%%%%%%%%%%%%%%%%%%%%%%%%%%%%%%%%%%%%%%%%%%%%%%%%%%%%%%%%%%%%%%%%%%
\subsection{Phenomenology of the ergoregion instability} \label{sec:ERphenom}
%%%%%%%%%%%%%%%%%%%%%%%%%%%%%%%%%%%%%%%%%%%%%%%%%%%%%%%%%%%%%%%%%%%%%%%%%%%%%%
The ergoregion instability discussed in Sec.~\ref{sec:ergoregioninstability} has important phenomenological 
implications. Indeed, building on the results by Friedman~\cite{1978CMaPh..63..243F} that a horizonless object with an 
ergoregion is unstable, a series of more recent 
works~\cite{Cardoso:2007az,Cardoso:2008kj,Pani:2010jz,Chirenti:2008pf,Moschidis:2016zjy,Cardoso:2014sna,Maggio:2017ivp,
Maggio:2018ivz} have established that --~under certain assumptions~-- this instability rules out extremely compact NSs 
and various exotic alternatives to BHs.

%%%%%%%%%%%%%%%%%%%%%%%%%%%%%%%%%%%%%%%%%%%%%%%%%%%%%%%%%%%%%%%%%%%%%%%%%%%%%%%%%%%%%%%
\subsubsection{Ergoregion instability of ultracompact stars} \label{sec:ERphenom_stars}
%%%%%%%%%%%%%%%%%%%%%%%%%%%%%%%%%%%%%%%%%%%%%%%%%%%%%%%%%%%%%%%%%%%%%%%%%%%%%%%%%%%%%%%

\begin{figure}[t]
\begin{center}
%\begin{tabular}{c}
\epsfig{file=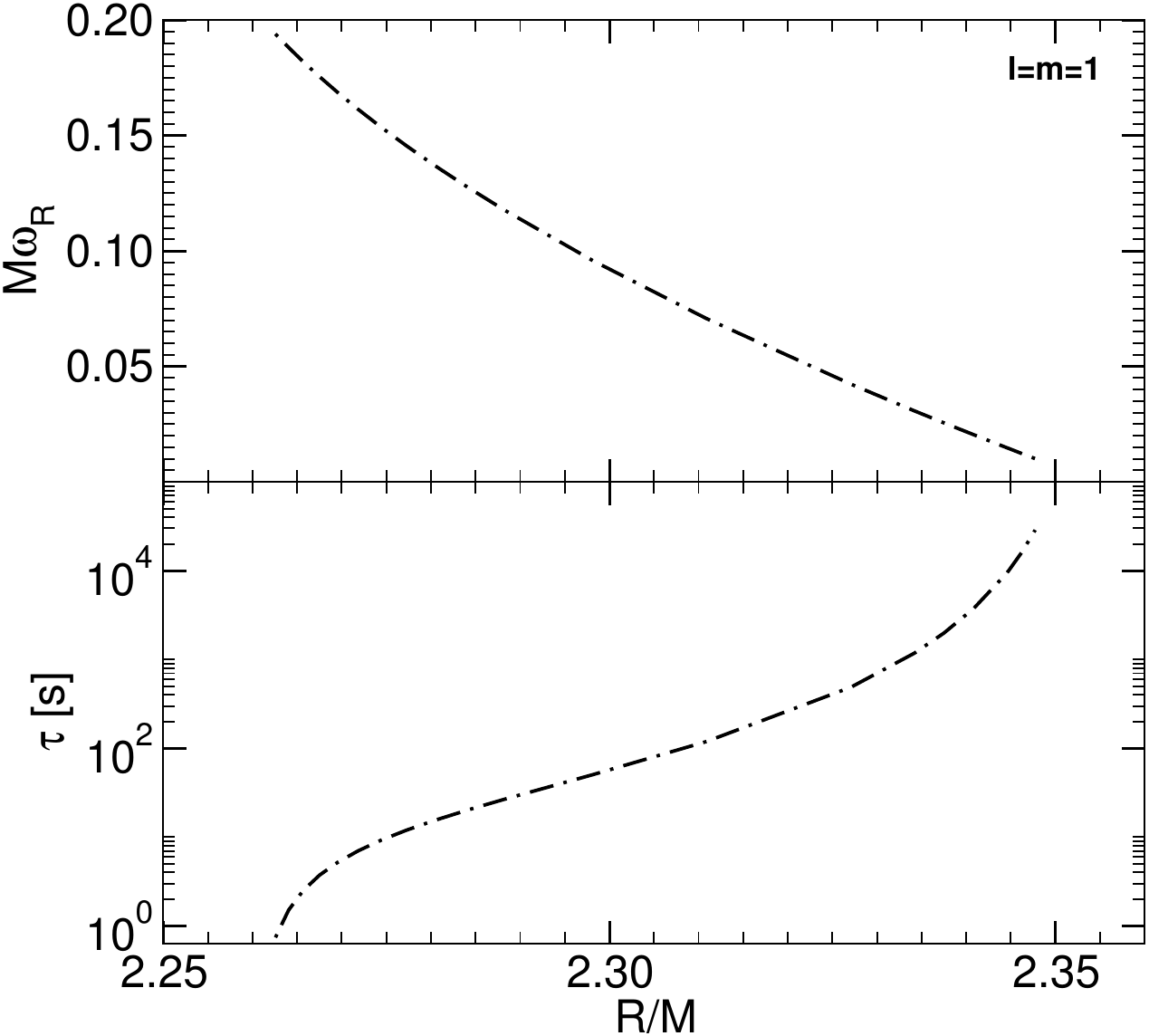,width=0.7\textwidth,angle=0,clip=true}
%\end{tabular}
\caption{The frequency (top panel) and the time scale (bottom panel) of the fundamental $l=m=1$ unstable mode of a constant-density NS as functions of the stellar compactness $R/M$. The $l=m=1$ mode turns stable at about $R\sim 2.35 M$. Although not shown, higher multipoles with $l=m\gg1$ will remain unstable until $R\lesssim 3M$~\cite{CominsSchutz,1996MNRAS.282..580Y} (cf. also Fig.~\ref{fig-stable1} where higher unstable multipoles are shown in a different system.).
\label{fig:ER_compactness}}
\end{center}
\end{figure}

As shown in Fig.~\ref{fig:ER_modes}, the time scale of the ergoregion instability of a compact spinning star can be as short as $\tau_{\rm ER}\sim 10^7 M$ (although this requires an extrapolation to $\Omega\sim\Omega_K$ beyond the slowly-rotating regime). For a compact star with $M\approx 1.4M_\odot$, this corresponds to a short time scale of the order of seconds.
A relevant question concerns the dependence of the instability on the compactness of the star and on its equation of state. A representative example is shown in Fig.~\ref{fig:ER_compactness}, which presents the frequency and time scale of the fundamental $l=m=1$ mode as functions of the stellar compactness $R/M$ for a constant-density star, whose pressure is given in terms of the constant density in Eq.~\eqref{Pstar}.

The instability time scale grows very rapidly as the compactness decreases and the $l=m=1$ mode turns stable at 
$R\approx 2.35 M$. This result is valid to second order in the spin, a consistent treatment was described in 
Sec.~\ref{ergo_consistent}. On the other hand, by neglecting some of the second-order terms in the perturbations 
equations, various works have explored the dependence of the time scale on $l$, $m$, and on the stellar compactness. 
Yoshida and Eriguchi have presented a detailed analysis~\cite{1996MNRAS.282..580Y}, showing that various $l=m$ modes can 
become unstable. 
The WKB analysis by Comins and Schutz~\cite{CominsSchutz} shows that in the eikonal ($l=m\to\infty$) limit an unstable mode appears precisely when an ergoregion is formed, although the time scale is exponentially long, cf. Eq.~\eqref{tau_ergo_eikonal}. This is also confirmed by our previous analysis of the ergoregion instability in analogue-gravity systems, see Fig.~\ref{fig-stable1}.

Clearly, the instability is phenomenologically relevant only if the associated time scale is shorter than the age of the 
star. Indeed, \emph{dynamically stable} (i.e., with $\tau_{\rm ER}\gg M$) spinning stars with an ergoregion can exist 
also in realistic scenarios~\cite{Tsokaros:2019mlz} (see below).

In addition, the compactness of a NS is bounded from above by the requirement that the speed of sound in the stellar 
interior is smaller than the speed of light. This causality bound translates into the lower constraint $R\gtrsim 3 
M$~\cite{Lattimer:1990zz,Glendenning:1992vb} on the NS radius. This seems to exclude the ergoregion instability because, 
as we have shown, the latter is associated with long-lived modes which exist only for ultracompact stars with $R\lesssim 
3M$ in the nonspinning limit\footnote{
Recently, Ref.~\cite{Hod:2014ena} showed that long-lived modes necessarily exist for matter configurations whose trace of the stress-energy tensor is positive (or zero). For a perfect-fluid star, this requires $P>\rho/3$, where $P$ and $\rho$ are the NS pressure and density. This is an extreme configuration which is unlikely to exist in ordinary stars, but it might occur in other models of ultracompact objects, as those discussed in Sec.~\ref{sec:BHmimickers}.}. Furthermore, causality also constrains the maximum angular velocity of a spinning NS~\cite{Koranda:1996jm}, thus disfavoring the formation of an ergoregion.

However, it is likely that an ergoregion forms for NSs whose radius is larger than $3M$ if the star is fastly-spinning. 
In addition, differential rotation seems to help. The first NSs with an ergoregion (stable on dynamical time scales) 
were recently constructed for a compressible, causal equation of state~\cite{Tsokaros:2019mlz}. Ergoregion-instability 
time scales for such objects have not been systematically studied yet.

%%%%%%%%%%%%%%%%%%%%%%%%%%%%%%%%%%%%%%%%%%%%%%%%%%%%%%%%%%%%%%%%%%%%%%%%%%%%%%%%%%%%%%%%%%%%%%%%%%%%%%%%%%%%%%%%%
\subsubsection{Testing the black-hole paradigm: extreme compact objects} \label{sec:BHmimickers}
%%%%%%%%%%%%%%%%%%%%%%%%%%%%%%%%%%%%%%%%%%%%%%%%%%%%%%%%%%%%%%%%%%%%%%%%%%%%%%%%%%%%%%%%%%%%%%%%%%%%%%%%%%%%%%%%%
BHs in GR have a remarkable property: being vacuum solutions of Einstein's field equations they do not depend on any external scale and, therefore, can exist in any size (or, equivalently, with any mass).
Dark objects as compact and massive as BHs but that do not possess an event horizon go under the generic name of ``BH 
mimickers'', ``Exotic Compact Objects'', or generically ``Extreme Compact 
Objects''~(ECOs) (see Refs.~\cite{Cardoso:2017cqb,Cardoso:2019rvt,Carballo-Rubio:2018jzw} for some overviews).
Notwithstanding, ordinary matter 
--~even when in extreme conditions~-- cannot support the enormous self-gravity of a massive and ultracompact object. 
For example, NSs --~the densest material objects known in the Universe~-- cannot sustain masses larger 
than\footnote{Even constant-density NSs have a maximum compactness which is smaller than the BH limit $M/R=1/2$. 
Inspection of Eq.~\eqref{fstar} shows that $M/R\leq 4/9$ to ensure regularity of the geometry. More realistic equations 
of state yield a maximum mass and a maximum compactness.} $\approx 3M_{\odot}$. Therefore, supporting the self-gravity 
of an ECO requires (at least!) exotic matter, or quite drastic modifications to GR: this is the price to pay to avoid 
dealing with event horizons.

There are strong motivations to study ECOs as alternatives to ordinary 
BHs, or as new ``species'' of compact objects that might co-exist in the universe along with ordinary BHs and 
NSs~\cite{Cardoso:2017cqb,Cardoso:2019rvt}. Despite the growing evidence in favor of the \emph{BH paradigm}, a definite 
proof that massive compact objects are endowed with a horizon is fundamentally impossible. On the contrary, the 
observation of a surface would be a bullet-proof indication that compact dark objects have 
star-like properties (see e.g. Ref.~\cite{Narayan:2005ie}). Such tests are extremely challenging to 
perform in the electromagnetic window~\cite{Abramowicz:2002vt,Cardoso:2017cqb,Cardoso:2019rvt}.
On the other hand, GW-based tests are already providing strong constraints and the whole field will improve 
significantly in the next years.
%

%%%%%%%%%%%%%%%%%%%%%%%%%%%%%%%%%%%%%%%%%
\paragraph{ECO taxonomy}
%%%%%%%%%%%%%%%%%%%%%%%%%%%%%%%%%%%%%%%%%
Within GR, Buchdhal's theorem~\cite{Buchdahl:1959zz} sets an upper bound on the maximum compactness of a 
self-gravitating compact object. This implies that ECOs can exist either within GR if some of the assumptions of 
the theorem are 
violated (e.g., staticitiy, perfect-fluid matter, isotropy) or if the underlying theory of gravity is not 
GR~\cite{Cardoso:2019rvt}. 

While the list of ECO models is ever growing, the most popular models are reviewed and discussed in Ref.~\cite{Cardoso:2019rvt}. 
%%%%%%%%%%%%%%%%%%%%%%%%%%%%%%%%%%%%%%%%%%%%%%%%%%%%%%%%%%%
\paragraph{Diagnostics for ECOs}
%%%%%%%%%%%%%%%%%%%%%%%%%%%%%%%%%%%%%%%%%%%%%%%%%%%%%%%%%%%%
Observational signatures of ECOs can be divided in two categories: {\rm EM tests} and {\rm GW tests}. The former 
includes EM counterparts of accretion and two-body processes~\cite{Broderick:2005xa,Broderick:2007ek,Lu:2017vdx}, 
mostly in the $X$-ray and radio band (including the 
recently observed shadow of the supermassive BH candidate in M87~\cite{Akiyama:2019cqa,Vincent:2015xta,Cunha:2018gql}). 
The latter includes:
%%%
\begin{itemize}
\item Tests of the inspiral phase: BH no-hair theorem tests using multipole 
moments~\cite{Krishnendu:2017shb,Raposo:2018xkf,Krishnendu:2018nqa,Pani:2015tga}, absence/presence of an 
horizon from tidal heating~\cite{Maselli:2017cmm,Datta:2019euh,Datta:2019epe}, measurement of the tidal 
deformability~\cite{Cardoso:2017cfl,Sennett:2017etc,Maselli:2017cmm,Pani:2019cyc} (so-called ``tidal Love 
numbers''), resonance excitation during the 
inspiral~\cite{Pani:2010em,Macedo:2013qea,Macedo:2013jja,Zhang:2018kib,Zhang:2019eid};
\item Signatures of matter or of an effective surface during the 
merger~\cite{Kesden:2004qx,Macedo:2013qea,Pani:2009ss,Barausse:2014tra,Cardoso:2014sna,Yunes:2009ke,Yunes:2016jcc}; 
possible EM counterpart associated with it;
\item Ringdown tests using the oscillation modes of the merger remnant~\cite{Berti:2005ys,Berti:2009kk,Berti:2006qt};
\item Inspiral-merger-ringdown consistency tests~\cite{TheLIGOScientific:2016src,Cabero:2017avf}
\item Post-merger tests using GW echoes~\cite{Cardoso:2016rao,Cardoso:2016oxy}.
\end{itemize}
%%%
A detailed discussion of these signatures and of the corresponding bounds on ECOs is given in 
Ref.~\cite{Cardoso:2019rvt}. Below we shall focus only on those aspects which are directly related to superradiance.

%%%%%%%%%%%%%%%%%%%%%%%%%%%%%%%%%%%%%%%%%%%%%%%%%%%%%%%%%%%%%%%%%%%%
\paragraph{Ergoregion instabilities of ECOs and their phenomenology}
%%%%%%%%%%%%%%%%%%%%%%%%%%%%%%%%%%%%%%%%%%%%%%%%%%%%%%%%%%%%%%%%%%%%
%
Very compact objects can develop negative-energy regions once spinning, and become unstable. Such instability affects 
any horizonless geometry with an ergoregion, similarly to the aforementioned case of spinning ultracompact 
stars~\cite{Friedman:1978wla,Kokkotas:2002sf,Moschidis:2016zjy,Cardoso:2007az,Oliveira:2014oja,Maggio:2017ivp,
Vicente:2018mxl}. As discussed in Sec.~\ref{sec:ergoregioninstability}, the only way to prevent this 
instability from occurring is by absorbing the negative-energy modes. BHs can absorb them efficiently (and hence Kerr 
BHs are stable against massless fields), but horizonless objects --~at least if perfectly-reflecting~-- must then be 
unstable.

The ergoregion instability of various boson stars and ``gravastars'' (objects with a de Sitter interior~\cite{Mazur:2001fv}) 
models has been studied in some detail in Ref.~\cite{Cardoso:2007az,Chirenti:2008pf}, showing that unstable modes generically exist. 
While gravastars have been studied only in the slowly rotating limit, numerical solutions of highly-spinning boson stars are available\footnote{The 
angular momentum of a boson star is quantized~\cite{Liebling:2012fv}; this prevents performing a standard slow-rotation 
approximation. Furthermore, there are indications that spinning scalar boson stars (at variance with their vector counterpart known as Proca 
stars~\cite{Brito:2015pxa}) are also subject to another (nonsuperradiant) type of instability, at least in the absence 
of self-interactions~\cite{Sanchis-Gual:2019ljs}.}. For a given compactness of the order of $M/R\sim 1/2$, gravastars 
and boson stars develop an ergoregion when spinning above a certain threshold. As for ultracompact stars, also in this 
case the instability arises from long-lived modes that exist when these objects possess a light-ring, which typically 
happens when $R\lesssim 3M$ in the nonspinning limit.

The ergoregion instability of other objects with ergoregions (e.g. ``superspinars'' and wormholes) was studied in Ref.~\cite{Cardoso:2008kj}, showing 
that similar results hold. Because in this case the exact form of the geometry is unknown, the stability analysis has 
been performed by imposing Dirichlet boundary conditions at the excision surface at $r=r_0$. The latter should 
approximate the boundary conditions required by a hard surface at the would-be horizon location. A detailed analysis of 
the instability of Kerr-like objects with different boundary conditions, was performed in 
Refs.~\cite{Pani:2010jz,Maggio:2017ivp,Maggio:2018ivz}.

\begin{figure}[th]
	\centering		
\includegraphics[width=.7\textwidth]{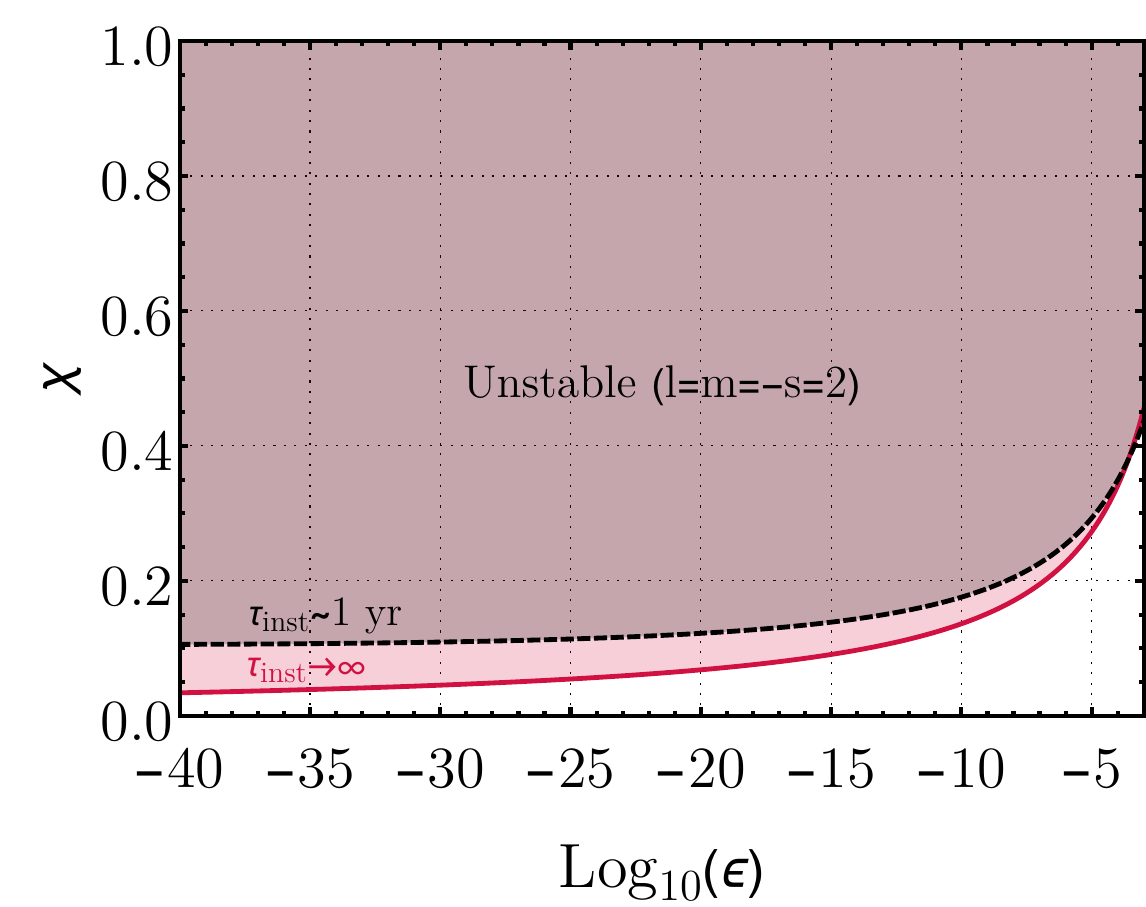}
\caption{Exclusion plot in the $\chi-\epsilon$ plane due to the ergoregion instability of a perfectly-reflecting ECO, 
assumed to be described by the Kerr geometry in their exterior, $r>r_0=r_+(1+\epsilon)$. Shaded areas represent regions 
where the object is unstable against $l=m=2$ gravitational perturbations with a time scale equal to or shorter than 
$\tau_{\rm inst}$. 
For details see Refs.~\cite{Cardoso:2019rvt,Maggio:2017ivp,Maggio:2018ivz}.}
\label{fig:exclusion}
\end{figure}

More recently, Refs.~\cite{Maggio:2017ivp,Maggio:2018ivz} studied in detail ultracompact \emph{Kerr-like ECOs}, with radius $r_0=r_+(1+\epsilon)$ (where $\epsilon\ll1$), 
and whose exterior geometry ($r>r_0$) is described by the Kerr metric. The properties of the object's interior and surface can be parametrized in terms of 
boundary conditions at $r=r_0$, in particular by a complex and (possibly) frequency-and-spin-dependent
reflection coefficient, ${\cal R}$~\cite{Mark:2017dnq,Maggio:2017ivp}.
The ergoregion instability of Kerr-like ECOs has been studied for scalar, electromagnetic, and gravitational 
perturbations --~both numerically and analytically in the low-frequency regime~-- for a variety of boundary conditions 
at $r=r_0$.
The overall summary of these studies is that the instability time scale depends strongly on the spin and on the 
compactness of the objects, which can be parametrized in terms of $\epsilon$. The ergoregion-instability 
timescale can very long~\cite{Friedman:1978wla,Cardoso:2007az,Maggio:2017ivp}. For concreteness, for gravastars with 
$\epsilon\sim 0.1-1$ the ergoregion is absent even for moderately high spin~\cite{Chirenti:2008pf}. However, at least 
for perfectly-reflecting Kerr-like ECOs in the $\epsilon\to0$ limit, the critical spin above which the object is 
unstable is very low~\cite{Maggio:2018ivz}.
As for the case of superradiance, the instability removes angular momentum from the spinning body until the latter 
reaches the critical spin for which the instability is absent~\cite{Barausse:2018vdb}. This implies that highly-spinning 
ECOs should be unstable over relatively short time scales in a large portion of the spin-compactness plane, as shown in 
Fig.~\ref{fig:exclusion}.

%%%%%%%%%%%%%%%%%%%%%%%%%%%%%%%%%%%%%%%%%%%%%%%%%%%%%%%%%%%%%%%%%%%%%%%%%%%%%%%%%%%%%%%%%%%%
\paragraph{Ergoregion instability for partially-absorbing objects} \label{sec:ERpartial}
%%%%%%%%%%%%%%%%%%%%%%%%%%%%%%%%%%%%%%%%%%%%%%%%%%%%%%%%%%%%%%%%%%%%%%%%%%%%%%%%%%%%%%%%%%%%
%  
The above discussion assumed that the ECOs are \emph{perfectly reflecting}.
A possible way to quench the instability is by absorbing the negative-energy modes trapped within the ergoregion, 
similarly to the BH case.

A first attempt to understand the role of partial absorption in the superradiant energy extraction was discussed in 
Refs.~\cite{Maggio:2017ivp,Maggio:2018ivz}, which showed that the instability is totally quenched if the absorption 
rate at the ECO surface is at least equal to the maximum superradiance amplification. For highly spinning 
objects, this
requires at least $0.4\%$ absorption rate for scalar fields, but up to $100\%$ absorption rate for
gravitational perturbations and almost maximal spins. While these numbers
reduce to $\lesssim 0.1\%$ for spins $\chi\lesssim0.7$, they are still several orders of magnitude larger than
what achievable with viscosity from nuclear matter. 
Nonetheless, certain models of quantum BHs seem to have reflectivity properties such that the ergoregion instability is 
absent~\cite{Wang:2019rcf,Oshita:2019sat}.

Finally, given its long timescales, it is possible that the instability can be efficiently quenched by some dissipation 
mechanism of nongravitational nature, although this effect would be 
model-dependent~\cite{Maggio:2017ivp,Maggio:2018ivz}. 
Unfortunately, the effect of viscosity in ECOs is practically unknown~\cite{Cardoso:2014sna,Guo:2017jmi},
and so are the timescales involved in putative dissipation mechanisms that might quench this instability. 

It is also possible that, when spinning, a partially-absorbing object can support quasi-trapped superradiant modes 
with $\omega_R<m\Omega$, which might lead to an instability similar to that of massive bosonic fields around Kerr 
BHs. To better understand this point, let us consider 
Eq.~\eqref{reflectivity} for an object with reflectivity ${\cal R}_{\rm surface}$ at its surface. In such case, ${\cal 
O}={\cal T}{\cal R}_{\rm surface}$ and the scattering reflectivity 
coefficient reads
%%%%%
\begin{equation}
 |{\cal R}|^2=|{\cal I}|^2-\frac{\omega-m\Omega}{\omega}|{\cal T}|^2\left(1-|{\cal R}_{\rm 
surface}|^2\right)\,,\label{reflectivity2}
\end{equation}
%%%
where for simplicity we assumed $k_\infty=\omega$ and $k_H=\omega-m\Omega$, with $\Omega$ being 
the angular velocity at the surface. 
If the object is only partially reflecting then $|{\cal R}_{\rm surface}|<1$ 
and superradiance would occur whenever $\omega<m\Omega$ (regardless of the caveat discussed in Sec.~\ref{sec:nohorSR} 
related to possible instabilities). In more 
realistic configurations ${\cal R}_{\rm surface}={\cal R}_{\rm surface}(\omega)$ (and it would be, in general, a 
complex function). If an object has a reflectivity ${\cal R}_{\rm surface}(\omega)$ in the \emph{static} limit, when 
spinning the effective reflectivity can acquire a superradiant term $\omega\to \omega-m\Omega$. This happens, for 
instance, to the reflectivity coefficient of a BH, see Eq.~\eqref{sigma0} and 
Eq.~\eqref{absor_wIm_rot}\footnote{The dependence of $|{\cal R}_{\rm surface}|^2$ on the combination 
$\omega-m\Omega$ should actually be linear, hence the 
change of sign in the superradiant regime. Heuristically this can be understood as follows. Let us consider a compact 
object such that the effective potential vanishes at its surface. The field near 
the surface is $\Psi\sim e^{-i (\omega-m\Omega) r_*}$ and the energy flux depends on linear partial derivatives with 
respect to the tortoise coordinate, $\partial_{r_*}\Psi$, which brings a \emph{linear} $\omega-m\Omega$ dependence.}.

%%%%%%%%%%%%%%%%%%%%%%%%%%%%%%%%%%%%%%%%%%%%%%%%%%%%%%%%%%%%
\paragraph{Nonlinear instability}
%%%%%%%%%%%%%%%%%%%%%%%%%%%%%%%%%%%%%%%%%%%%%%%%%%%%%%%%%%%%
Linearized gravitational fluctuations of any 
nonspinning ECO are extremely long-lived and decay no faster than 
logarithmically~\cite{Keir:2014oka,Cardoso:2014sna,Eperon:2016cdd,Eperon:2017bwq}. 
Being trapped between the center of the
object and the light ring, and being localized near a second, stable null geodesic~\cite{Cardoso:2014sna,Cunha:2017qtt}, 
the long damping time of 
these modes has led to the conjecture that ECOs are \emph{nonlinearly} unstable and may 
evolve through a Dyson-Chandrasekhar-Fermi type of mechanism~\cite{Dyson:1893:1,Dyson:1893:2,1953ApJ...118..116C} at the 
nonlinear level~\cite{Keir:2014oka,Cardoso:2014sna}.

\begin{figure}[t]
\begin{center}
\begin{tabular}{cc}
\epsfig{file=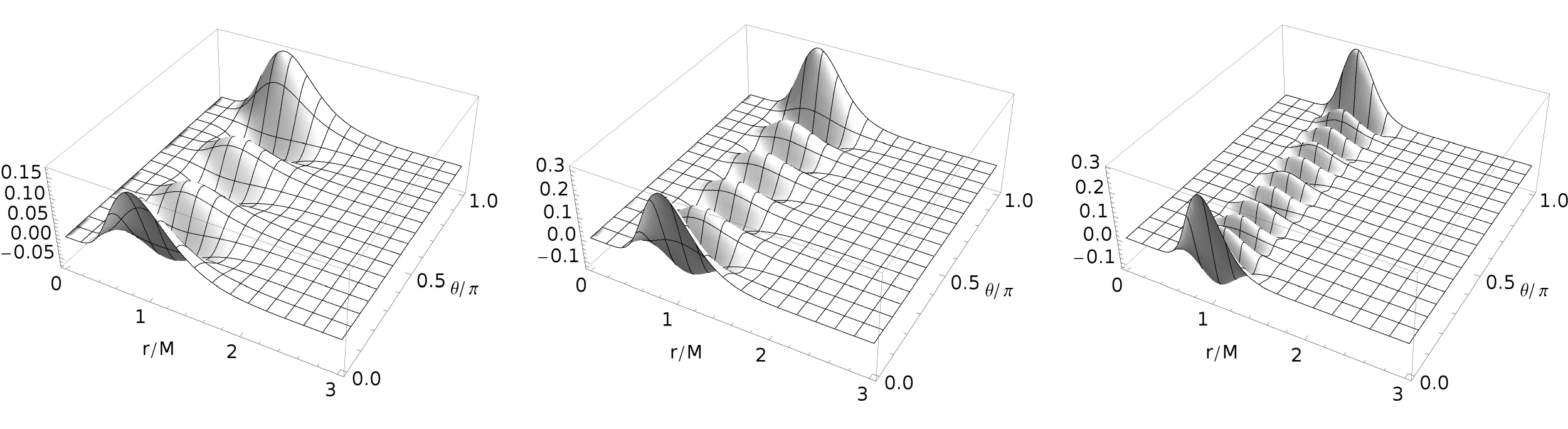,width=0.95\textwidth,angle=0,clip=true}
\end{tabular}
\end{center}
\caption{\label{fig:3D}
Scalar eigenfunctions of a static, ultracompact star with radius $R=2.3M$ for $m=0$ and $l=6,10,20$ (from left to the right).
The eigenfunctions have a typical width that scales as $l^{-1}$ in the angular direction and a width in the radial direction
that depends on the model, but typically ranges between $l^{-0.4}-l^{-0.8}$. Therefore, the ``aspect ratio'' of the perturbation $\sim l^{0.6}-l^{0.2}$ grows in the large-$l$ limit and the perturbation becomes more and more elongated along the radial direction. See Ref.~\cite{Cardoso:2014sna} for details.
}
\end{figure}

To understand this mechanism, it is illustrative to inspect the eigenfunctions of the linearized problem. An example is 
shown in Fig.~\ref{fig:3D} for the case of a ultracompact star (qualitatively similar results hold for other BH 
mimickers). 
As the multipolar index $l$ increases, the eigenfunctions becomes more and more elongated along the radial direction. If we assume for simplicity that the perturbations are axisymmetric ($m=0$), these elongated, long-lived modes are unstable against the same ``Dyson-Chandrasekhar-Fermi'' 
mechanism that affects thin cylinders or rings of matter~\cite{Dyson:1893:1,Dyson:1893:2,1953ApJ...118..116C,Cardoso:2006sj}. 
The minimum growth time scale of this instability scales as $\tau_{\rm DCF} \sim \delta\rho^{-1/2}$, where $\delta\rho$ is the density fluctuation.
The requirement that nonlinearities take over is that $\tau_{\rm DCF}$ be much smaller than the lifetime of linear fluctuations. Because the latter grows exponentially with $l$ (cf. Eq.~\eqref{tau_ergo_eikonal}), it is easy to show that fragmentation becomes important already at moderately small values of $l$ even for $\delta\rho/\rho \sim 10^{-16}$ or smaller~\cite{Cardoso:2014sna}. 

The fragmentation of the linear eigenfunction leads to a configuration which consists on a spherically symmetric core surrounded by droplets of the star fluid, whose sizes are much smaller than that of the original object~\cite{Cardoso:2014sna} (see also nonlinear results for fragmentation of black strings~\cite{Lehner:2010pn}). It is easy to see that these smaller droplets, although of the same material as the original star, are much less compact because they are much smaller and are therefore expected to be themselves stable.
Likewise, the core of the star is also less compact and stable. On longer time scales, these droplets re-arrange and fall into the core, and the process continues. The dynamical picture looks like that of a ``boiling'' fluid, and radiates a nonnegligible amount of radiation. Exact calculations have not been performed yet but, if this scenario is correct, a sizable fraction of the object's initial mass can disperse to infinity, possibly reducing the compactness of the final object to values which no longer allow for the existence of light rings. 
In an alternative scenario, nonlinear interactions over the ultralong lifetime of the unstable modes may lead to the formation of small BHs close to the stable light ring~\cite{Cardoso:2014sna}.

%%%%%%%%%%%%%%%%%%%%%%%%%%%%%%%%%%%%%%%%%%%%%%%%%%%%%%%%%%%%
\paragraph{Do light rings imply black holes?}
%%%%%%%%%%%%%%%%%%%%%%%%%%%%%%%%%%%%%%%%%%%%%%%%%%%%%%%%%%%%
To summarize, ultracompact objects with $R\lesssim 3M$ might be plagued by various instabilities. When these objects 
are 
almost perfectly reflecting and spinning sufficiently fast, they suffer from the ergoregion instability at the linear 
level. Even when they are only slowly spinning or static, long-lived modes trapped by the light-ring can become 
unstable at the nonlinear level. In the latter case, the instability can lead to fragmentation (thus reducing the 
object's compactness) or to gravitational collapse (thus forming a BH). In both cases, the instability can be 
sufficiently strong to be dynamically effective. As recently pointed out~\cite{Hod:2014ena}, 
exotic matter configurations with $T^\mu_\mu\equiv T >$ 0 are necessarily 
characterized by the existence of long-lived modes. 
Altogether, these results give further theoretical support to the BH hypothesis~\cite{Cardoso:2019rvt}: the mere 
observation of a light ring --~a much simpler task than the observation of the horizon~-- might thus be seen as a 
compelling (albeit indirect) evidence for the existence of BHs.

%%%%%%%%%%%%%%%%%%%%%%%%%%%%%%%%%%%%%%%%%%%%%%%%%%%%%%%%%%%%
\paragraph{Superradiance and gravitational-wave echoes}
%%%%%%%%%%%%%%%%%%%%%%%%%%%%%%%%%%%%%%%%%%%%%%%%%%%%%%%%%%%%
A powerful discriminator for the presence of \emph{structure} at the horizon scale is the post-merger GW 
signal~\cite{Cardoso:2016rao,Cardoso:2016oxy}. When ECOs are very compact, this signal displays the same prompt 
ringdown as observed for BHs, but followed by a repeated and modulated train of pulses, known as GW 
echoes~\cite{Cardoso:2016oxy,Abedi:2016hgu}. Echoes have been studied with several approaches, most notably time- and 
frequency-domain 
templates~\cite{Abedi:2016hgu,Abedi:2017isz,Nakano:2017fvh,Testa:2018bzd,Maggio:2019zyv,Micchi:2019yze}, using transfer 
functions~\cite{Mark:2017dnq,Bueno:2017hyj,Testa:2018bzd}, Dyson series~\cite{Correia:2018apm}, or resonance 
methods~\cite{Conklin:2017lwb,Conklin:2019fcs} (see 
Refs.~\cite{Westerweck:2017hus,Tsang:2018uie,Nielsen:2018lkf,Lo:2018sep,
Wang:2019rcf,Uchikata:2019frs,Oshita:2020dox,Abedi:2020ujo} for a discussion of the detectability of this effect in 
actual LIGO/Virgo data).

Interestingly, superradiance is also relevant for echoes. Indeed, the frequency content of the echo signal decreases in 
time for each subsequent echo, so that eventually the frequency can be small enough to satisfy the superradiant 
condition, $\omega<m\Omega$~\cite{Nakano:2017fvh,Maggio:2019zyv}. This produces interesting effects and resonances in 
the transfer function~\cite{Maggio:2019zyv}.
In the case of perfectly-reflecting ECOs, one might be tempted to conclude that the echo amplitude will grow in time 
due to the superradiant/ergoregion instability. While this is true, the time scale for this process is parametrically 
longer than the echo delay time, and likely plays a negligible role in actual searches for echoes~\cite{Maggio:2019zyv}.
On the other hand, if spinning compact objects are ECOs, this effect might produce a strong stochastic background from 
all unresolved sources which would be detectable with current and future interferometers~\cite{Barausse:2018vdb}.

%%%%%%%%%%%%%%%%%%%%%%%%%%%%%%%%%%%%%%%%%%%%%%%%%%%%%%%%%%%%
\paragraph{Tidal heating \& superradiance as discriminators for horizons}
%%%%%%%%%%%%%%%%%%%%%%%%%%%%%%%%%%%%%%%%%%%%%%%%%%%%%%%%%%%%
Another strong diagostic for horizons which is directly related to superradiance is tidal 
heating~\cite{Hartle:1973zz,Hughes:2001jr} (see Sec.~\ref{sec:tides}).
As we have discussed, a spinning BH absorbs radiation of frequency $\omega>m\Omega_{\rm H}$, 
but amplifies radiation of smaller frequency due to superradiance. In this respect, BHs are dissipative systems which 
behave just like a Newtonian viscous fluid~\cite{Damour_viscous,Poisson:2009di,Cardoso:2012zn}. 

For low-frequency circular binaries, the energy flux associated to tidal heating at the 
horizon, $\dot E_H$, corresponds to the rate of change of the BH 
mass~\cite{Alvi:2001mx,Poisson:2004cw} (see also Eq.~\eqref{tidesNewt}),
\begin{eqnarray}
 \dot M = \dot E_H \propto \frac{\Omega_{\rm orb}^5}{M^2}(\Omega_{\rm orb}-\Omega_{\rm H})\,, \label{Edot}
\end{eqnarray}
where $\Omega_{\rm orb}\ll 1/M$ is the orbital angular velocity (related to the linear velocity $v$ by $v=(M\Omega_{\rm 
orb})^{1/3}$) and the (positive) prefactor 
depends on the masses and spins of the two bodies. Thus, tidal heating is stronger for 
highly spinning bodies relative to the nonspinning by a factor $\sim \Omega_{\rm H}/\Omega_{\rm orb}\gg1$.

The energy flux~\eqref{Edot} leads to a potentially observable phase shift of GWs emitted 
during the inspiral, which can be parametrized by higher-order post-Newtonian (PN) corrections to the 
phase~\cite{Blanchet:2006zz}. Absorption at the horizon introduces a $2.5{\rm PN}\times\log v$ (resp., a $4{\rm 
PN}\times\log v$) correction to the GW phase of spinning (resp., nonspinning) binaries, relative to the leading 
term~\cite{Maselli:2017cmm}. This effect has been used to estimate the constraints coming from the detection of 
supermassive spinning binaries with future detectors~\cite{Maselli:2017cmm,Datta:2019euh}. More recently, the effect of 
tidal heating in extreme-mass-ratio inspirals~\cite{Hughes:2001jr} has been shown to lead to exquisite constraints on the effective 
reflectivity of an ECO~\cite{Datta:2019epe} (see also Ref.~\cite{Datta:2020rvo}).

%%%%%%%%%%%%%%%%%%%%%%%%%%%%%%%%%%%%%%%%%%%%%%%%%%%%%%%%%%%%%%%%%%%%%
\subsection{Superradiance \& relativistic jets } \label{sec:BZ}
%%%%%%%%%%%%%%%%%%%%%%%%%%%%%%%%%%%%%%%%%%%%%%%%%%%%%%%%%%%%%%%%%%%%%
Relativistic jets emitted by astrophysical sources are one of the most interesting and mysterious phenomena in our 
Universe. The most powerful jets are seen in active galactic nuclei (AGNs), and are believed to be the result of 
accretion of matter by supermassive BHs~\cite{LyndenBell:1969yx}. AGNs are the most powerful sources in the Universe, 
making it very hard to conceive viable models for their production without invoking very compact objects. Although the 
first AGNs (such as quasars and radio galaxies) were discovered four decades ago, the engine powering these events is 
still largely unknown. The energy needed for the acceleration of these relativistic outflows of matter is widely 
believed to either come from gravitational binding energy and/or from the object's rotational energy. In the first case, 
accretion of matter onto the BH leads to a transfer of gravitational binding energy to particles which are tossed away 
along the rotational axis of the BH (see e.g. Ref.~\cite{Blandford:1982di} for such a process). Other mechanisms, akin 
to superradiance or to the Penrose process, make use of the rotational energy of the BH. This is the case of the 
Blandford-Znajek (BZ) mechanism~\cite{Blandford:1977ds} which occurs for BHs immersed in magnetic fields (see also e.g. 
Refs.~\cite{Komissarov:2008yh,Lasota:2013kia,Kinoshita:2017mio} for a discussion on the relationship between the BZ mechanism and 
superradiance or the Penrose process\footnote{Interestingly the analogy between the BZ process and superradiance might be more than just an analogy. In fact, Ref.~\cite{Noda:2019mzd} recently showed that the BZ process can be interpreted as the long wavelength limit of the superradiant scattering from Alfv\'{e}nic waves in the plasma.}). In this mechanism the magnetic field lines, which are anchored in the accretion 
disk, are twisted due to the frame dragging effect near the rotating BH (see Sec.~\ref{sec:ZAMO}), thus increasing the 
magnetic flux.
Similar to the Earth-Moon system discussed in Sec.~\ref{sec:tides}, due to dissipative effects, this can lead to energy 
transfer from the BH to the magnetic field~\cite{MacDonald:1982zz}. This energy is then used to accelerate the 
surrounding plasma and to power a jet collimated along the BH rotational axis. In general both the accretion process and 
the BZ mechanism might contribute to the energy released in the jets, making it difficult to prove from numerical 
simulations that the latter mechanism is at work, but recent general relativistic magnetohydrodynamic (GRMHD) 
simulations seem to indicate that this is indeed the 
case~\cite{Tchekhovskoy:2012up,McKinney:2012vh,Penna:2013rga,Lasota:2013kia}.

%%%%%%%%%%%%%%%%%%%%%%%%%%%%%%%%%%%%%%%%%
\subsubsection{Blandford-Znajek process~\label{sec:BF}}
%%%%%%%%%%%%%%%%%%%%%%%%%%%%%%%%%%%%%%%%%

%
\begin{figure}[ht]
\begin{center}
\begin{tabular}{c}
\epsfig{file=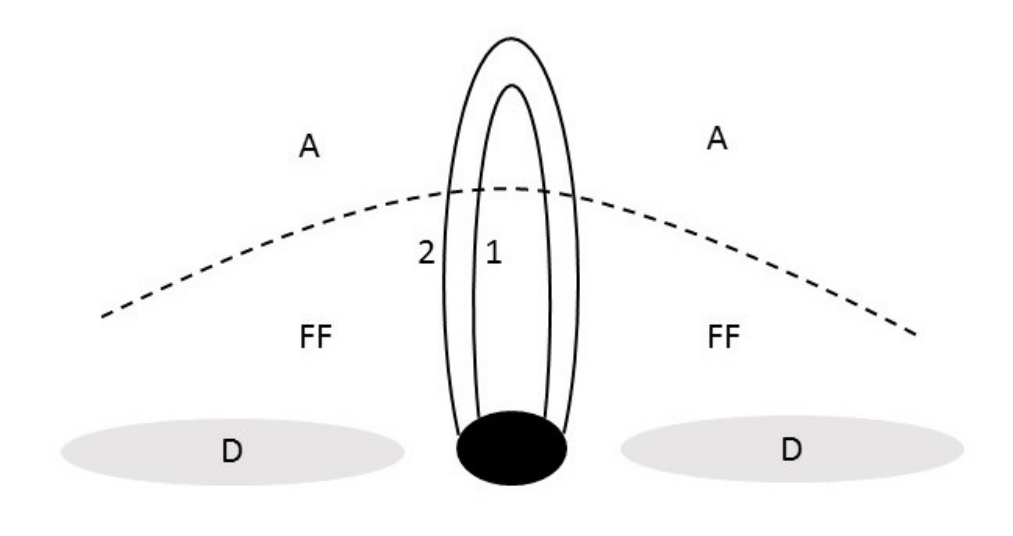,width=0.8\textwidth,angle=0,clip=true}
\end{tabular}
\end{center}
\caption{Pictorial description of the magnetosphere surrounding a BH in the BZ mechanism. The solid lines denote 
electric equipotential surfaces. The magnetosphere is composed of three different regions: a region (D) which includes 
the accretion disk and the horizon, where the field is degenerate, i.e., the electric field is perpendicular to the 
magnetic field, but not force-free. This last condition is required for these regions to be able to anchor the magnetic 
field; a region (FF) where the field is force-free. In this region the current flows along equipotential surfaces; an 
acceleration region (A) in which the field is neither degenerate nor force free. In region (A) the equipotential 
surfaces close up, and the energy extracted from the BH is used to accelerate charged particles. In regions (D) and (A) 
the current can flow across the equipotential surfaces. Reproduction of figure from 
Ref.~\cite{MacDonald:1982zz}.\label{fig:BZ}}
\end{figure}

In the BZ solution a Kerr BH is considered to be immersed in a stationary axisymmetric force-free 
magnetosphere~\cite{Blandford:1977ds}. 
In Ref.~\cite{Blandford:1977ds} it was argued that in analogy with what happens in pulsars, a rotating BH would trigger 
an electron-positron pair cascade just outside the accretion disk and the horizon where the plasma is rarefied, 
establishing an approximately force-free magnetosphere\footnote{A condition for this to happen is that initially there 
is a small electric field component parallel to the magnetic field (note that this is a Lorentz invariant condition). In 
Ref.~\cite{Wald:1974np}, this was shown to occur for rotating BHs immersed in a magnetic field.}. In Fig.~\ref{fig:BZ} 
we depict the region where this force-free magnetosphere is localized plus the other regions that characterize the 
magnetosphere. Region (FF) is where the transfer of energy takes place. This energy is then deposited in region (A) 
where particles are accelerated.

To describe the force-free magnetosphere, in addition to Maxwell's equations with a source 
\be\label{maxwell_source}
\nabla_{\nu}F^{\mu\nu}=\frac{J^{\mu}}{\epsilon_0}\,,
\ee
the EM field must satisfy the following three conditions
\be\label{force_free}
F_{\mu\nu}J^{\nu}=0\,,\qquad
\,^{\ast}F^{\mu\nu}F_{\mu\nu}=0\,,\qquad
F_{\mu\nu}F^{\mu\nu}>0\,,
\ee
where $\,^{\ast}F^{\mu\nu} \equiv \frac{1}{2}\epsilon^{\mu\nu\rho\sigma}F_{\rho\sigma}$ is Maxwell's tensor dual (we use 
the definition $\epsilon^{\mu\nu\rho\sigma}\equiv
\frac{1}{\sqrt{-g}}E^{\mu\nu\rho\sigma}$ where $E^{\mu\nu\rho\sigma}$ is the totally anti-symmetric Levi-Civita symbol 
with $E^{0123}=1$), $\epsilon_0$ is the vacuum permittivity and $J^{\mu}$ is the current generated by the 
electron-positron plasma. The first condition implies --~assuming that the vector potential has the same symmetries 
(axisymmetry and stationarity) than the BH spacetime~-- that the magnetic field lines lie along surfaces of constant 
$A_{\varphi}$. On the other hand, if the second condition is satisfied but not the third one can always find a local 
inertial frame where the EM field is purely electric.
From these equations, it also follows that one can define a function $\Omega_{\rm EM}(r,\vartheta)$ as
\be
\Omega_{\rm EM}(r,\vartheta)=-\frac{A_{t,r}}{A_{\varphi,r}}=-\frac{A_{t,\vartheta}}{A_{\varphi,\vartheta}}\,, 
\label{omegaBZ}
\ee
which can be interpreted as being the ``angular velocity'' of the EM field as will become clear below.

The field equations must also be supplemented with appropriate boundary conditions at the horizon and at infinity. At 
the horizon it was shown in Ref.~\cite{1977MNRAS.179..457Z} that regularity implies (assuming $A_{\varphi}$ to be 
finite) 
\be
\frac{\Delta\sin\vartheta}{\rho^2}F_{r\vartheta}=\frac{2M r_+\left(\Omega_{\rm EM}-\Omega_{\rm 
H}\right)}{r_+^2+a^2\cos^2\vartheta}A_{\varphi,\vartheta}(r_+,\vartheta)\,,
\ee
where in the force-free approximation, $F_{r\vartheta}$ can be shown to be a function of $A_{\varphi}$ only. On the 
other hand, the boundary conditions at infinity are not unique but they can be chosen, e.g., by matching the field to 
known flat-space solutions.

%%%%%%%%%%%%%%%%%%%%%%%%%%%%%%%%%%%%%%%%%
%\paragraph{Extraction of energy}
%%%%%%%%%%%%%%%%%%%%%%%%%%%%%%%%%%%%%%%%%

The factor $\Omega_{\rm EM}-\Omega_{\rm H}$ appearing in the boundary conditions above (compare it with the 
superradiant 
condition~\eqref{eq:superradiance_condition}) already suggests that stationary axisymmetric solutions of the 
inhomogeneous Maxwell's equations~\eqref{maxwell_source} in a Kerr background are akin to a superradiance-like process. 
In fact the conserved radial EM energy and angular momentum fluxes at the horizon are given by~\cite{Blandford:1977ds}
\beq
\delta E_{\rm hole}^r&\equiv&-T_{\mu}^{\phantom{\mu}r}\xi^{\mu}_{(t)}=\Omega_{\rm EM}(\Omega_{\rm EM}-\Omega_{\rm 
H})\left(\frac{A_{\varphi,\vartheta}}{r_+^2+a^2\cos^2\vartheta}\right)^2(r_+^2+a^2)\epsilon_0\,,\label{BZ_energy}\\
\delta J_{\rm hole}^r&\equiv&T_{\mu}^{\phantom{\mu}r}\xi^{\mu}_{(\varphi)}=\frac{\delta E_{\rm hole}^r}{\Omega_{\rm 
EM}}\,,\label{BZ_angular}
\eeq
and thus when $0<\Omega_{\rm EM}<\Omega_{\rm H}$ there is a net radial negative energy and angular momentum flux 
$\delta 
E_{\rm hole}^r<0$, $\delta J_{\rm hole}^r<0$  at the horizon, i.e., energy and angular momentum are extracted from the 
BH. From Eq.~\eqref{BZ_angular} one sees that the function $\Omega_{\rm EM}$ can indeed be interpreted as the ``angular 
velocity'' of the EM field.

By deriving specific solutions for the EM field, it is possible to construct the function $\Omega_{\rm EM}$ through 
Eq.~\eqref{omegaBZ} explicitly. Particularly important are the split monopole, and the paraboidal magnetic field 
solutions found perturbatively in the slowly-rotating limit~\cite{Blandford:1977ds}. In these cases, $\Omega_{\rm 
EM}=\Omega_{\rm H}/2$ and $\Omega_{\rm EM}\approx 0.38\Omega_{\rm H}$, respectively (see e.g. 
Ref.~\cite{Gralla:2014yja} 
for a recent summary of these solutions and also 
Refs.~\cite{Lupsasca:2014pfa,Lupsasca:2014hua,Zhang:2014pla,Li:2014bta} 
for recent exact solutions found around extreme Kerr BHs). Recently Ref.~\cite{Yang:2014zva} studied the linear 
stability of the monopole solution and their results suggest that the solution is mode stable. 
%%%
%%%%
In fact, force-free simulations 
(e.g.~\cite{Komissarov:2004ms,Komissarov:2008yh,McKinney:2006sc,Tchekhovskoy:2008gq,Palenzuela:2010xn}) and recent GRMHD 
simulations seem to indicate that magnetic fields generated by accretion disks have large split monopole 
components~\cite{Tchekhovskoy:2012up,McKinney:2012vh,Penna:2013rga} suggesting that the BZ mechanism should occur in 
fully dynamical setups. 

%%%%%%%%%%%%%%%%%%%%%%%%%%%%%%%%%%%%%%%%%%%%%%%%%%%%%%%%%%%%%%%%%%%
\subsubsection{Blandford-Znajek process and the membrane paradigm}
%%%%%%%%%%%%%%%%%%%%%%%%%%%%%%%%%%%%%%%%%%%%%%%%%%%%%%%%%%%%%%%%%%%

The understanding of the physics behind the BZ mechanism was at the origin of a new paradigm to describe BHs, the 
so-called \emph{membrane paradigm}. This paradigm uses a $3+1$ spacetime decomposition in which the BH event horizon is 
regarded as a two-dimensional surface residing in a three-dimensional space, while the region inside the horizon is 
``thrown away'' from the picture since it is causally disconnected from any observer outside the horizon\footnote{The 
use of a $3+1$ spacetime decomposition was mainly useful to write the equations in a more familiar form for the 
astrophysics community. In fact most of the work done in this area in the last decades has been done using this 
formalism. Recently the GR community has regained interest in the subject and some remarkable effort has been done to 
develop a fully covariant theory of force-free magnetospheres around rotating BHs~\cite{Gralla:2014yja}.}. This surface 
can be shown to behave as an electrically charged viscous fluid with finite surface electrical resistivity, entropy and 
temperature. In this picture the interaction of the membrane with the rest of the Universe is then governed by 
well-known physical laws for the horizon's fluid, such as the Navier-Stokes equation, Ohm's law, tidal force equations 
and the laws of thermodynamics. Originally all quantities were computed in the ZAMO frame (see Sec.~\ref{sec:kerr}) in 
relation to which electric and magnetic fields are defined and physical laws are formulated, although the membrane 
paradigm has also been reformulated in a covariant form in Ref.~\cite{Parikh:1997ma}. For stationary (or static) BH 
spacetimes the membrane paradigm is fully equivalent to the standard spacetime approach as long as one is only 
interested in physics occurring outside the horizon. The teleological nature of the paradigm makes it more challenging 
to study time-dependent problems although some cases involving weakly perturbed non-stationary spacetimes have 
successfully been studied~\cite{MembraneParadigm}. For astrophysical purposes this paradigm has been quite successful to 
describe and understand relativistic phenomena in BH spacetimes (see Ref.~\cite{MembraneParadigm} for a pedagogical 
introduction and a compilation of works which led to the full formulation of the membrane paradigm. See also 
Ref.~\cite{Parikh:1997ma} for a derivation of the membrane paradigm starting from an action principle).

In the membrane paradigm, one can understand how the BZ mechanism works through an analogy with the tidal acceleration 
effect (see Sec.~\ref{sec:tides})~\cite{MacDonald:1982zz}. Taking an infinitesimal tube of magnetic flux $\delta \psi$ 
in the force-free region (for example a tube with walls given by surfaces 1 and 2 of Fig.~\ref{fig:BZ}) and which 
intersects the hole, it is possible to show that the torque exerted by the membrane on this tube 
is~\cite{MacDonald:1982zz}
\be
-\frac{d\delta J}{dt}=\frac{\Omega_{\rm H}-\Omega_{\rm EM}}{4\pi}g_{\varphi\varphi}B_{\bot}\delta\psi\,,
\ee
where $B_{\bot}$ is the magnetic field perpendicular to the membrane as seen by the ZAMO's observer and 
$g_{\varphi\varphi}$ is to be taken at the horizon. The power transmitted to the flux tube due to this torque is then
\be
P=-\Omega_{\rm EM}\frac{d\delta J}{dt}=\Omega_{\rm EM}\frac{\Omega_{\rm H}-\Omega_{\rm 
EM}}{4\pi}g_{\varphi\varphi}B_{\bot}\delta\psi\,.
\ee
This torque and power are transmitted through the tube up to region A, where angular momentum gets gradually deposited 
into charged particles. 
A direct comparison with Eqs.~\eqref{torquetides} and ~\eqref{tidesGeneral} shows that from the point of view of the 
ZAMO's observer this is indeed a analogous process to tidal acceleration, and thus completely analogous to 
superradiance.

\begin{figure}[ht]
\begin{center}
\begin{tabular}{c}
\epsfig{file=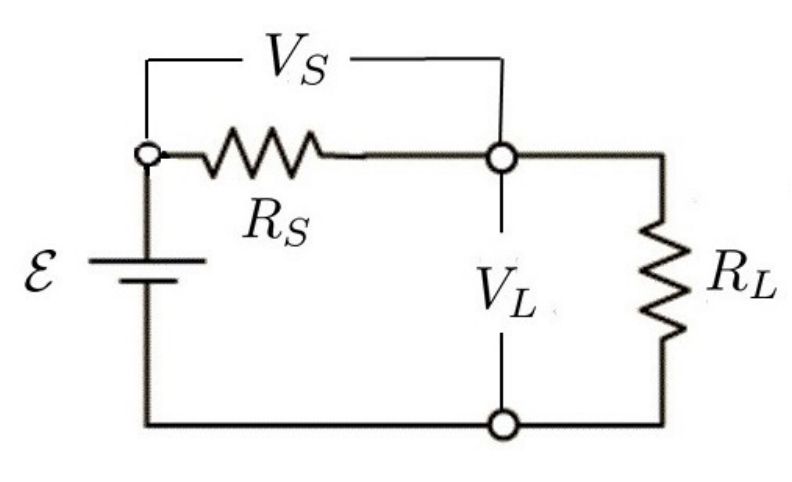,width=0.5\textwidth,angle=0,clip=true}
\end{tabular}
\end{center}
\caption{Circuit analogy of the BZ mechanism in which a battery transfers energy to a load. A battery of electromotive 
force $\mathcal{E}$ with internal resistance $R_S$ drives a current $I$ through the load resistance $R_L$ (which could 
be for example an electric light). Maximum power transfer is attained when $R_L=R_S$.\label{fig:battery}}
\end{figure}

The membrane picture also suggests an analogy between the BZ mechanism and the circuit of Fig.~\ref{fig:battery}, in 
which energy is transferred from a battery (the BH) to a load (the acceleration region A)~\cite{MacDonald:1982zz}. The 
current flowing along the resistance $R_S$ produces a potential drop $V_S$, while at $R_L$ it produces a potential drop 
$V_L$ such that the electromotive force of the battery is given by $\mathcal{E}=V_S+V_L$. From Ohm's law the current $I$ 
flowing along the circuit is given by
\be
I=\frac{\mathcal{E}}{R_S+R_L}\,,
\ee
while the power dissipated in the load is given by
\be
P_L=I^2R_L=\frac{\mathcal{E}^2}{R_S^2/R_L+2R_S+R_L}\,.
\ee
On the other hand the efficiency of this process, defined by the ratio of the power dissipated in the load to the total 
power generated by the source, reads
\be\label{efficiency_circuit}
\eta=\frac{1}{R_S/R_L+1}\,.
\ee
Although the efficiency has its maximum when $R_L\gg R_S$, the maximum power output at the load is obtained when 
$R_L=R_S$. Note that in this case only half of the energy is really transferred to the load, the other half being 
dissipated as heat due to the source internal resistance. On the other hand, if $R_L\ll R_S$, then most of the power 
output is dissipated as heat at the source, whereas if $R_L\gg R_S$ the current $I$ generated at the source will be very 
low and thus the power transferred to the load will be very small, even though the efficiency will tend to $100\%$.

In the BZ process case, the current flowing from surface 1 to 2 of Fig.~\ref{fig:BZ} in the horizon membrane's produces 
a potential drop $\delta V_H$ due to the membrane internal resistance $R_H$, given by~\cite{MacDonald:1982zz}
\be\label{battery_potential}
\delta V_H=I \delta R_H=\frac{(\Omega_{\rm H}-\Omega_{\rm EM})\delta\psi}{2\pi}\,.
\ee
where $\delta R_H$ is related to $R_H$ through~\cite{MacDonald:1982zz}
\be\label{battery_resistance}
\delta R_H=R_H\frac{\delta\psi}{4\pi^2 g_{\varphi\varphi} B_{\bot}}\,.
\ee
The potential drop in region A can be thought as being due to a resistance $\delta R_A$, and it can be shown to be given 
by~\cite{MacDonald:1982zz}
\be\label{load_potential}
\delta V_A=I \delta R_A=\frac{\Omega_{\rm EM}\delta\psi}{2\pi}\,,
\ee
where it is assumed that the acceleration region A is sufficiently far away such that frame dragging effects are 
negligible.
Using Eqs.~\eqref{battery_potential},~\eqref{battery_resistance} and~\eqref{load_potential}, the ratio between the 
potentials in the acceleration region and at the horizon are then given by
\be\label{ratio_resistance}
\frac{\delta V_A}{\delta V_H}=\frac{\delta R_A}{\delta R_H}=\frac{1}{\Omega_{\rm H}/\Omega_{\rm EM}-1}\,.
\ee
By comparison with Eq.~\eqref{efficiency_circuit}, one can define the efficiency of the BZ mechanism by 
$\eta=\Omega_{\rm EM}/\Omega_{\rm H}$~\cite{Blandford:1977ds}\footnote{This is not to be confused with the jet 
efficiency, defined by $\eta_{\rm jet}=\left<L_{\rm jet}\right>/\big<\dot{M}\big>$ where $\left<L_{\rm jet}\right>$ is 
the time-average jet luminosity and $\big<\dot{M}\big>$ is the time-average rate of matter accretion by the BH. 
Recently, efficiencies up to $\eta_{\rm jet}\sim 300\%$ have been obtained in GRMHD 
simulations~\cite{Tchekhovskoy:2011zx,Tchekhovskoy:2012up,McKinney:2012vh,Lasota:2013kia} which is a strong indication 
that the BZ mechanism is at work.}.
The sum of the potential drops is equal to the total electromotive force $\mathcal{E}=\delta V_H+\delta V_A$ around a 
closed loop that passes along the horizon from surface 1 to 2, then up the surface 2 poloidally to the region A in which 
it crosses to surface 1 again and then back down to the horizon. Thus, the total current $I$ and the total power 
transmitted $P$ to the acceleration region are given by
\beq
I&=&\frac{\mathcal{E}}{\delta R_A+\delta R_H}=\frac{1}{2}\left(\Omega_{\rm H}-\Omega_{\rm 
EM}\right)g_{\varphi\varphi}B_{\bot}\delta \psi\,,\\
P&=&\delta R_A I^2=\Omega_{\rm EM}\frac{\left(\Omega_{\rm H}-\Omega_{\rm 
EM}\right)}{4\pi}g_{\varphi\varphi}B_{\bot}\delta \psi\,.
\eeq
Maximum power transmission then implies $\Omega_{\rm EM}=\Omega_{\rm H}/2$. From Eq.~\eqref{ratio_resistance}, this 
happens when $R_A=R_H$ and $\delta V_A=\delta V_H$, which corresponds to the condition obtained from the circuit 
analogy. In Ref.~\cite{MacDonald:1982zz} it was argued that the configuration $\Omega_{\rm EM}=\Omega_{\rm H}/2$ would 
be likely to be achieved in a dynamical setup due to the backreaction of charged particles onto the field lines. In fact 
recent GRMHD simulations seem to obtain $\Omega_{\rm EM}/\Omega_{\rm H}\approx 0.3\text{ -- }0.4$, in agreement with 
this analysis~\cite{McKinney:2012vh,Penna:2013rga}.
  	
A key ingredient for this analogy to work is to understand the physical origin behind the electromotive force 
$\mathcal{E}$ driving the current $I$. The membrane paradigm suggests an analogy with Faraday's unipolar inductor. 
Consider a rotating conducting disk, which can be idealized as a perfect conductor, immersed in a uniform magnetic field 
perpendicular to the rotational axis of the disk. Due to the rotational motion of the disk through the magnetic field 
there is a radial Lorentz force on the free charges in the disk, which in turn produces a potential difference between 
the center and the boundary of the disk. On the other hand, due to the magnetic field, this current feels a Lorentz 
force opposite to the rotational motion of the disk, producing a reaction torque on the conductor which will make it 
slow down in analogy with the BZ mechanism. Completing this circuit with a wire attached at the boundary and the center 
of the disk, one can effectively use the disk as a battery. 
This is in fact the mechanism behind the electromotive force developed by rotating magnetized 
stars~\cite{Goldreich:1969sb,Ruderman:1975ju} and planets~\cite{1969ApJ...156...59G}. However, as was pointed out in 
Refs.~\cite{Komissarov:2004ms,Komissarov:2008yh} the membrane paradigm suggests that the horizon plays a similar role 
to 
the surface of a magnetized rotating star, hiding the role played by the ergosphere. Unlike the surface of a disk in 
which an electromotive force can indeed drive an electric current, Einstein's equivalence principle tells us that the 
BH 
horizon is not a physical surface where electrics current can flow\footnote{However from the point of view of \emph{BH 
complementarity} introduced in Ref.~\cite{Susskind:1993if}, the membrane is real as long as the observers remain 
outside 
the horizon, but fictitious for observers who jump inside the BH. Since neither observer can verify a contradiction 
between each other, the two are complementary in the same sense of the wave-particle duality.}. In 
Ref.~\cite{Komissarov:2004ms} the author showed that inside the ergoregion there are no stationary axisymmetric 
solutions
of the Einstein-Maxwell equations, describing a EM field supported by a remote source, that satisfy both the second and 
third conditions of Eq.~\eqref{force_free} along the magnetic field lines (see also 
Ref.~\cite{Toma:2014kva,Ruiz:2012te}). This implies that near a rotating BH there are no stationary solutions with a 
completely screened electric field. This is in fact a purely gravitational effect caused by the dragging of inertial 
frames near the BH. Although the force-free approximation is for all purposes a good approximation for the 
magnetosphere 
near a rotating BH, it fails to predict that current sheets must form inside the ergoregion, where a strong enough 
unscreened electric field perpendicular to the magnetic field must persist in order to sustain the potential drop along 
the magnetic field lines. On the other hand, in the region where the force-free approximation holds, it is the residual 
component of the electric field parallel to the magnetic field that drives the poloidal 
currents~\cite{Komissarov:2004ms}.

%%%%%%%%%%%%%%%%%%%%%%%%%%%%%%%%%%%%%%%%%%%%%%%%%%%%%%%%%%%%%%%%%%%%%%%%%%%%%%%%
\subsection{Superradiance, CFS instability, and r-modes of spinning stars}
%%%%%%%%%%%%%%%%%%%%%%%%%%%%%%%%%%%%%%%%%%%%%%%%%%%%%%%%%%%%%%%%%%%%%%%%%%%%%%%%
Another important astrophysical process that bears some resemblance with superradiant phenomena is the 
Chandrasekhar-Friedman-Schutz (CFS) instability of spinning NSs driven by gravitational radiation. This instability was 
discovered by Chandrasekhar in 1970 while studying Maclaurin spheroids~\cite{Chandrasekhar:1992pr}. In 1978, Friedman 
and Schutz extended the analysis to the case of compressible, perfect-fluid stars and explained the instability in an 
elegant way~\cite{1978ApJ...221..937F}. In fact, such instability is very generic and occurs whenever a mode that is 
retrograde in a frame corotating with the star appears as prograde to a distant inertial observer (see 
Refs.~\cite{Stergioulas:2003yp,Andersson:2006nr,RotatingRelativisticStars} and references therein).

\begin{figure}[ht]
\begin{center}
\begin{tabular}{c}
\epsfig{file=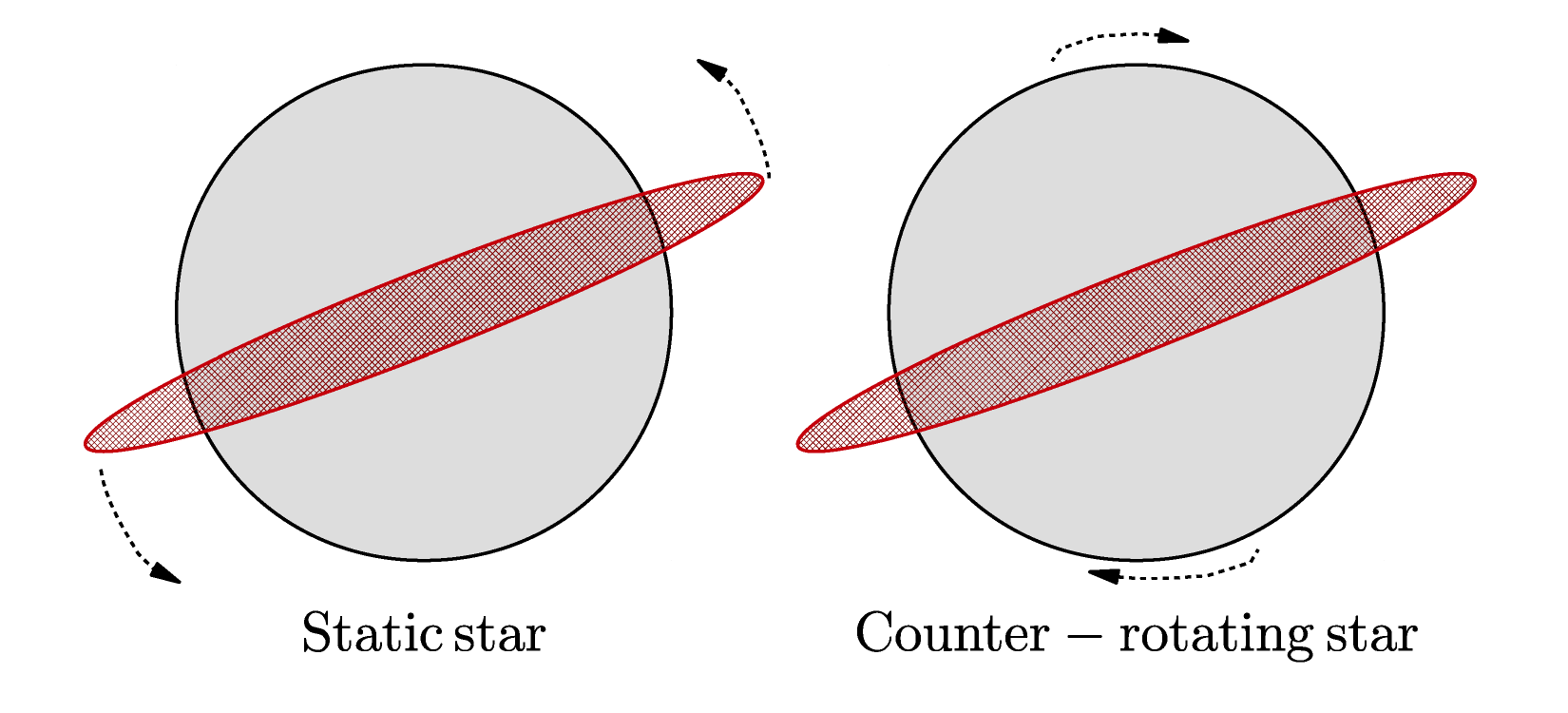,width=0.75\textwidth,angle=0,clip=true}
\end{tabular}
\end{center}
\caption{\label{fig:CFS}
Illustration of the CFS instability as seen from the laboratory frame. In the left panel a bar-like mode of the fluid in 
a static star rotates counter-clockwise. This perturbation tends to increase the angular momentum of the star. Because 
the perturbation carries away positive angular momentum through GWs, it also reduces its amplitude. In the right panel 
the star rotates clockwise (with rotational axis perpendicular to the plane of the figure) such that, in the laboratory 
frame, the phase velocity of the mode vanishes and so does the emission of GWs. For a slightly higher stellar spin, the 
mode would appear to rotate \emph{clockwise} and it would emit GWs with \emph{negative} angular momentum. This negative 
value is subtracted from the (positive) angular momentum of the perturbation, which therefore increases in amplitude. 
The larger the perturbation grows, the larger is the angular momentum radiated in GWs, thus producing a positive 
feedback.}
\end{figure}

The mechanism for the instability is depicted in Fig.~\ref{fig:CFS}. In the left panel we show a stable configuration: a 
fluid perturbation of a static star with phase velocity $\omega/m$ moving counter-clockwise. Within our axis 
conventions, this perturbation carries a \emph{positive} angular momentum and also emits positive angular momentum 
through GWs. The angular momentum emitted in GWs has to be subtracted by that of the perturbation, whose amplitude 
consequently decreases. However, a drastically different picture emerges when the star rotates (right panel of 
Fig.~\ref{fig:CFS}). 
In such case the sign of the angular momentum carried by GWs depends only on the relative motion of the perturbation 
with respect to the laboratory frame, whereas the sign of the angular momentum of the perturbation depends only on the 
motion of the mode relative to the star. 
Therefore, as the star rotates faster and faster in clockwise direction, the counter-clockwise mode starts rotating more 
slowly as viewed from the laboratory frame, decreasing the rate of angular momentum emission in GWs, but not its 
intrinsic angular momentum, which remains roughly the same as in the nonrotating case. For some critical angular 
velocity, the phase velocity of the mode will vanish and the mode will freeze relative to the laboratory (as shown in 
the right panel of Fig.~\ref{fig:CFS}). For a slightly higher stellar rotation rate, the initially counter-clockwise 
mode rotates in the \emph{clockwise} sense, thus emitting \emph{negative} angular momentum through GWs. This emission 
has to be compensated by an \emph{increase} of the (positive) angular momentum of the perturbation, which therefore 
increases in amplitude. The larger the perturbation grows, the larger is the angular momentum radiated in GWs, and the 
instability ensues.
The instability evolves on a secular time scale, extracting angular momentum from the star via GW emission, unless it is 
suppressed by other mechanisms, such as viscosity.

This qualitative picture already shows some similarity with the fact that superradiant modes within the ergoregion 
appear to be prograde to a distant inertial observer but are in fact retrograde in a frame corotating with the BH.
To put this in more quantitative terms, let us consider Newtonian stars within the Lagrangian perturbation framework 
developed in Ref.~\cite{1978ApJ...221..937F}. We consider a normal mode (i.e. ignoring GW dissipation) of the star in 
the form $\xi=\hat\xi e^{-i(\omega t-m\varphi)}$. In such case, the canonical energy and angular momentum of the mode 
are related as~\cite{1978ApJ...221..937F,Andersson:2006nr}
%%%
\begin{equation}
 E_c=\frac{\omega}{m} J_c\,, \label{energyCFS}
\end{equation}
%%%
which resembles Eq.~\eqref{delta_J}, as expected for the perturbation of an axisymmetric object. When the star rotates 
with angular velocity $\Omega>0$, the canonical angular momentum must also satisfy the 
inequality~\cite{1978ApJ...221..937F,Andersson:2006nr}
%%%
\begin{equation}
 \frac{\omega-m\Omega-\Omega}{m}\leq \frac{J_c/m^2}{\langle \hat\xi,\rho\hat\xi\rangle} \leq 
\frac{\omega-m\Omega+\Omega}{m}\,.
\end{equation}
%%%
where $\rho$ is the fluid density and the angular parenthesis denote the inner product over the volume of the star. The 
equation above shows that, in the static $\Omega\to0$ limit, corotating modes (with $\omega/m>0$) must have $J_c>0$, 
whereas counter-rotating modes have $J_c<0$. From Eq.~\eqref{energyCFS}, this implies $E_c>0$ and therefore the modes 
are stable. However, when the star rotates in the opposite direction relative to the mode phase velocity, an initially 
counter-rotating mode can become corotating as discussed before. When this happens $E_c$ can change sign and the mode 
becomes unstable when $\omega\leq m\Omega$ (in the laboratory frame), with the inequality saturated for marginally 
stable modes. Therefore, it is clear that the CFS instability requires the existence of modes satisfying the 
superradiant condition~\eqref{eq:superradiance_condition}.

The relativistic framework to study this instability was developed in a series of papers during the 
1970s~\cite{1975ApJ...200..204F,Friedman:1978hf,1978ApJ...221..937F}, the crucial additional ingredient being the 
emission of GWs generated by fluid and spacetime perturbations of the star. These works confirmed the Newtonian 
analysis, finding that a mode becomes unstable at the point where its phase velocity vanishes in the inertial frame, 
i.e. when $\omega/m=\Omega$ (see Refs.~\cite{Stergioulas:2003yp,Andersson:2006nr} for some reviews on the important of 
the CFS instability in astrophysics).

%%%%%%%%%%%%%%%%%%%%%%%%%%%%%%%%%%%%%%%%%%%%%%%%%%%%%%%%
\paragraph{The r-mode instability of rotating stars}
%%%%%%%%%%%%%%%%%%%%%%%%%%%%%%%%%%%%%%%%%%%%%%%%%%%%%%%%
Some axial fluid modes of static, Newtonian stars (as well as the axial gravitational modes of relativistic stars) are 
degenerate at zero frequency. Therefore,  even in the nonrotating case such modes are marginally stable towards the CFS 
instability. As soon as rotation is turned on, these \emph{r-modes} become unstable for arbitrarily small rotation 
rates~\cite{Andersson:1997xt} (cf. Ref.~\cite{Andersson:2000mf} for a review).

To first order in the stellar spin, the frequency of the r-modes in the inertial frame reads
\begin{equation}
 \omega=m\Omega\left(1-\frac{2}{l(l+1)}\right)\,.
\end{equation}
Therefore, modes with positive phase velocity, $\omega/m>0$, relative to the laboratory frame have always a 
\emph{negative} phase velocity $\omega/m-\Omega<0$ in the star comoving frame for any value of $l$ and $\Omega$ (the 
special case of $l=1$ fluid perturbations is marginally stable to first order in the spin). This is confirmed by the 
canonical energy of these modes which, to first order in $\Omega$, reads~\cite{Andersson:2006nr}
%%%
\begin{equation}
 E_c=A(\omega+m\Omega)(\omega-m\Omega)\,,
\end{equation}
%%%
where $A>0$ is a constant depending on the amplitude of the modes, the harmonic index $l$ and on the stellar density. 
Therefore, to first order in the spin the instability occurs when $\omega<m\Omega$, i.e. when the superradiant 
condition~\eqref{eq:superradiance_condition} holds. Such analogy remains valid also to second order in $\Omega$ in the 
large-$l$ limit.

%%%%%%%%%%%%%%%%%%%%%%%%%%%%%%%%%%%%%%%%%%%%%%%%%%%%%%%%%%%%%%%%%%%%%%%%%%%%%%%%%%%%%
\subsection{Open issues}
%%%%%%%%%%%%%%%%%%%%%%%%%%%%%%%%%%%%%%%%%%%%%%%%%%%%%%%%%%%%%%%%%%%%%%%%%%%%%%%%%%%%%

\begin{itemize}

\item The impact of nonlinearities on the bounds discussed in Sec.~\ref{sec:bounds_mass} has not been fully explored. Nonlinearities might slow down or even saturate the superradiant growth of bosonic clouds, thus making the constraints derived from BH superradiance less stringent. On the other hand, nonlinear effects similar to the bosenova~\cite{Yoshino:2012kn,Kodama:2011zc} can provide novel smoking guns for bosonic condensates around astrophysical BHs.

\item Plasma-triggered superradiant instabilities have been studied in Ref.~\cite{Pani:2013hpa} but mostly for homogeneous configurations. It would be interesting to extend such analysis to more realistic matter profiles around a spinning BH, for example by extending perturbative~\cite{Pani:2013pma} or fully-numerical~\cite{Witek:2012tr} methods (see Ref.~\cite{VanPutten:1999vda} for a related analysis). Such studies can be important for scalar-tensor theories where accretion disks are able to provide tachyonic modes to a fundamental scalar~\cite{Cardoso:2013fwa,Dima:2020rzg}.

\item As mentioned above, a systematic study of the ergoregion instability in realistic models of spinning NSs is still lacking. It would be interesting to compute the time scale and check whether the instability can be used to rule out some region of the NS mass-radius-spin parameter space.

\item Wormholes are interesting alternatives to the BH paradigm. Traversable wormholes are predicted in GR for matter 
that violates the null energy condition~\cite{Visserbook}. When rotating, such objects are expected to be unstable 
because of the ergoregion instability, but a detailed computation, together with a discussion of possible astrophysical 
implications, is not available yet.

\item The role of the horizon in the BZ 
mechanism is still unclear and whether it is necessary for the process to occur is still a matter of debate. In fact, 
some recent numerical simulations~\cite{Ruiz:2012te} seem to indicate that the ergosphere alone is sufficient for the 
process to occur.

\item  As was pointed out in Ref.~\cite{Penna2014}, recent GRMHD simulations studying the BZ mechanism suggest that the 
magnetosphere leading to BH jets has a large split-monopole 
component~\cite{Tchekhovskoy:2012up,McKinney:2012vh,Penna:2013rga}; however a simple explanation for why the system 
tends to this solution is still missing. 

\item Recent numerical simulations suggest that BHs carrying linear momentum~\cite{Neilsen:2010ax} and coalescing BH--BH 
or BH--NS binary systems can also power jets~\cite{Palenzuela:2010nf}. Although some work has been done to understand 
the mechanism behind these jets~\cite{Lyutikov:2011vca,McWilliams:2011zi}, a complete theoretical understanding is still 
needed.

\item In the context of the BH analog of the two-ring model discussed in Sec.~\ref{sec:rings}, it is important to 
understand whether such mechanism (or extensions therein) can be used to power gamma-ray burst, as discussed in 
Ref.~\cite{PressRing}. More in general, a purely-superradiant model for gamma-ray burst production has not yet been 
developed.

\end{itemize}

%%%%%%%%%%%%%%%%%%%%%%%%%%%%%%%%%%%%%%%%%%%%%%%%%%%%%%%%
\clearpage
\newpage
\section{Conclusions \& Outlook} \label{sec:Conclusion}
%%%%%%%%%%%%%%%%%%%%%%%%%%%%%%%%%%%%%%%%%%%%%%%%%%%%%%%%
Energy extraction through superradiance is ubiquitous in physics and appears in essentially any dissipative system under different guises. In fact, we have discussed how superradiance can be understood in simple kinematical terms. In flat spacetime the most common superradiant phenomenon is Cherenkov emission, but many classical and quantum systems can be turned into superradiant amplificators. Sound and surface waves can be amplified in a variety of settings that can be easily devised in the laboratory.

In gravitational theories, superradiance is intimately connected to tidal acceleration, 
even at Newtonian level. Relativistic gravitational theories predict the existence of BHs, gravitational vacuum 
solutions with two special properties: the first is the defining feature of a BH, the horizon. But spinning BHs possess 
also ergoregions, where matter is forced to co-rotate with the BH and which provide a coupling between the matter and 
the background spacetime. Ergoregions allow superradiance to occur in BH 
spacetimes, and to extract energy from vacuum even at the classical level. 
When semiclassical effects are taken into 
account, superradiance occurs also in static configurations, as in the case of Hawking radiation from a Schwarzschild 
BH. The efficiency of superradiant scattering of GWs by a spinning (Kerr) BH can be larger than $100\%$ and this 
phenomenon is deeply connected to other important mechanisms associated to spinning compact objects, such as the Penrose 
process, the ergoregion instability, the Blandford-Znajek effect, and the CFS instability. 
Rotational superradiance has been recently observed in the laboratory in analog BH systems, and is associated with a number of interesting effects and instabilities
in astrophysics, which potential observable imprints. We have presented a unified treatment of BH superradiant phenomena including charged 
BHs, higher dimensions, nonasymptotically flat spacetimes, analog models of gravity, and theories beyond GR.

An important point of our analysis is the role played by the event horizon and by the ergoregion in energy-extraction 
processes, such as superradiance, tidal acceleration, and the Penrose process. 
The ergosphere (allowing negative energy states in its interior) is responsible for energy 
amplification. As we have shown, the existence of 
an event horizon in stationary and axisymmetric spacetimes implies that of an ergoregion, so the existence 
of dissipation and negative-energy states are indissolubly connected to each other.
However, superradiance in GR is not a prerogative of BHs, but it can also occur in conducting stars and in exotic, 
horizonless compact objects; in the latter case it is associated with instabilities of these objects.

One of the most interesting applications of BH superradiance is the possibility of tapping the amplified radiation 
through various mechanisms of confinement, thus producing a ``BH bomb'' instability. We have discussed various of such 
confining mechanisms, including reflecting surfaces, AdS boundaries, massive fields, magnetic fields, and other 
nonminimal interactions. Superradiant instabilities of charged BHs in AdS spacetimes provide a holographic dual 
description of a spontaneous symmetry-breaking mechanism at finite temperature, and are associated with a phase 
transition between RN-AdS BHs and a novel hairy BH which is the ground state at low temperatures. 
Novel hairy BH solutions branch off the superradiant threshold in the AdS case and in asymptotically-flat spacetimes with massive complex scalars. These solutions can be interpreted as the nonlinear extension of linear bound states of frequency saturating the superradiant condition~\eqref{eq:superradiance_condition}, and give rise to stationary hairy BHs that interpolate between Kerr BHs and boson stars. These hairy BHs evade the no-hair theorem of GR and might play an important role in astrophysical tests of the Kerr hypothesis.

The study of superradiant instabilities triggered by light bosons has flourished in recent years, because of the 
exciting connections between BH superradiance and particle physics. We have provided a unified picture of the 
state-of-the-art in this field and have described the evolution of these instabilities in a Kerr BH. Superradiant 
instabilities of massive bosons have important phenomenological effects, being associated to very peculiar EM and GW 
emission from astrophysical BHs and NSs. The effects we have discussed (formation of bosonic condensates 
near spinning BHs, lack of highly-spinning BHs, emission of peculiar monochromatic GWs and dipolar scalar waves) are currently being investigated to 
constrain ultralight bosons arising in various extensions of the Standard Model, to rule out dark-matter candidates, and 
to study various astrophysical effects in the strong-curvature regime. Table~\ref{tab:bounds} summarizes the state-of-the-art,
an astonishing set of constraints on ultralight fields, obtained using astrophysical BHs!

BH superradiance has been discovered almost 50 years ago, but it is nowadays more alive than ever. Not only new 
exciting theoretical developments have been recently discovered, but recent EM and GW facilities 
are currently searches for direct evidence of BH superradiance, thus providing a new tool to test 
gravitational interactions and particle physics in curved spacetime. 

We concluded the first edition of this book with this list of open problems:
\begin{quote}
 Among the most urgent open problems are the fully nonlinear evolution of the superradiant instability, the stability 
 of hairy BHs, EM and GW tests of bosonic condensates around massive BHs, observing superradiance in analog-gravity 
 models in the laboratory, understanding completely the holographic dual of superradiant states and their microscopic 
description.
\end{quote}
It is amusing to note that, in the five years past the first edition, essentially all these outstanding open problems 
have been solved. Nonetheless, as we have discussed, this area is still very active and we are positive that novel 
exciting developments will be uncovered in the near future.

%%%%%%%%%%%%%%%%%%%%%%%%%%%%%%%%%%%%%%%%%%%%%%%%%%%%%%%%
\section*{Acknowledgments}
%%%%%%%%%%%%%%%%%%%%%%%%%%%%%%%%%%%%%%%%%%%%%%%%%%%%%%%%
We thank Ana Sousa Carvalho for kindly preparing the illustrations for us.
We are indebted to Asimina Arvanitaki, Jacob Bekenstein, \'Oscar Dias, Sam Dolan, Roberto Emparan, Sean Hartnoll, Shahar Hod, Luis Lehner, Carlos Palenzuela, Robert Penna, Silke Weinfurtner for useful comments on a preliminary draft of this manuscript, and especially to Jo\~ao Rosa for comments and for comparing our superradiant amplification factors with his code. We thank Carlos Herdeiro and Pedro Cunha for kindly providing us the images for Fig.~\ref{fig:shadows}.
We are also much indebted to Lorenzo Annulli, Enrico Barausse, Emanuele Berti, Mateja Boskovic, \'Oscar Dias, Roberto 
Emparan, Shrobana Ghosh, Leonardo Gualtieri, Carlos Herdeiro, Taishi Ikeda, Maximiliano Isi, Luis Lehner, Hideo Kodama, 
Akihiro Ishibashi, Jos\'e Lemos, Avi Loeb, Hirotada Okawa, Frans Pretorius, Thomas Sotiriou, Ulrich Sperhake, Ling Sun, 
Rodrigo Vicente, Helvi Witek, Shijun Yoshida, Hirotaka Yoshino, and Tien-tien Yu for many and interesting discussions 
throughout the years.

R.B. acknowledges financial support from the European Union's Horizon 2020 research and innovation programme under the Marie Sk\l odowska-Curie grant agreement No. 792862.
V.C. acknowledges financial support provided under the European Union's H2020 ERC 
Consolidator Grant ``Matter and strong-field gravity: New frontiers in Einstein's 
theory'' grant agreement no. MaGRaTh--646597. V. C. is indebted to Waseda University for warm hospitality while this work was being finalized.
P.P. acknowledges financial support provided under the European Union's H2020 ERC, Starting 
Grant agreement no.~DarkGRA--757480, and under the MIUR PRIN and FARE programmes (GW-NEXT, CUP:~B84I20000100001).
The authors would like to acknowledge networking support by the GWverse COST Action 
CA16104, ``Black holes, gravitational waves and fundamental physics'' and 
support from the Amaldi Research Center funded by the MIUR program "Dipartimento di 
Eccellenza" (CUP: B81I18001170001).
We thank FCT for financial support through Project~No.~UIDB/00099/2020.
We acknowledge financial support provided by FCT/Portugal through grant PTDC/MAT-APL/30043/2017.
%

%%%%%%%%%%%%%%%
\appendix
%%%%%%%%%%%%%%%%%%%%%%%%%%%%%%%%%%%%%%%%%%%%%%%%%%%%%%%%%%%%
\section{List of publicly available codes}\label{app:codes}
%%%%%%%%%%%%%%%%%%%%%%%%%%%%%%%%%%%%%%%%%%%%%%%%%%%%%%%%%%%%
The numerical and analytical methods used in this work have been implemented in ready-to-be-used  {\scshape Mathematica}\textsuperscript{\textregistered} notebooks, which are publicly available~\cite{webpage}.
Here we give a short description of them:
%%%
\begin{itemize}
 \item \href{http://centra.tecnico.ulisboa.pt/network/grit/files/amplification-factors/}{\bf superradiance charge.nb}: Amplification factors of the superradiant scattering of a charged wave off a spherically-symmetric or a slowly-rotating BH with generic metric.
 \item \href{http://centra.tecnico.ulisboa.pt/network/grit/files/amplification-factors/}{\bf superradiance spin.nb}: Amplification factors of the superradiant scattering of a neutral bosonic wave of generic spin off a Kerr BH, obtained by solving the Teukolsky equations. 
 \item \href{http://centra.tecnico.ulisboa.pt/network/grit/files/ringdown/}{\bf Kerr massive scalar bound states.nb}: solves the eigenspectrum of unstable modes of a Kerr BH under massive scalar perturbations through Leaver's continued fraction method.
 \item \href{http://centra.tecnico.ulisboa.pt/network/grit/files/nonlinear/}{\bf HartleThorne.nb}: (i) computes and solves Einstein's equations for a rotating self-gravitating perfect-fluid to second order in the spin and  (ii) derives in detail the procedure to separate the Klein-Gordon equation in this background.
 %%%%
\end{itemize}
%%%
Some data presented in the main text are also available on the webpage~\cite{webpage}. For example, the data files contained in the file \href{http://centra.tecnico.ulisboa.pt/network/grit/files/amplification-factors/}{\bf superradiance spin.nb} provide the dependence of the amplification factor $Z_{slm}(\omega)$ for a Kerr BH in the entire parameter space. The number of digits in the tables is not indicative of the precision; our tests indicate a precision of roughly one part in $10^5$.
%%%%%%%%%%%%%%%%%%%%%%%%%%%%%%%%%%%%%%%%%%%%%%%%%%%%%%%%

%%%%%%%%%%%%%%%%%%%%%%%%%%%%%%%%%%%%%%%%%%%%%%%%%%%%%%%%%%%%%%%%%%
\section{Analytic computation of the amplification coefficients}\label{appendix_super_ana}
%%%%%%%%%%%%%%%%%%%%%%%%%%%%%%%%%%%%%%%%%%%%%%%%%%%%%%%%%%%%%%%%%%%
In this section we compute the cross section of a Kerr BH for generic spin. We will follow Refs.~\cite{Starobinski:1973,Starobinski2:1973}. We assume that the Compton wavelength of the particle is much bigger than the gravitational size of the BH, i.e., $\omega M\ll 1$. We also consider the slowly rotating regime $a\omega\ll 1$. 

To solve the radial equation~\eqref{teu_radial} we use a matching procedure, dividing the spacetime in two overlapping regions, the near-region $r-r_+\ll 1/\omega$, and the far-region $M\ll r-r_+$.

Changing variables to
\be
x=\frac{r-r_+}{r_+-r_-}\,\,,
\ee
equation~\eqref{teu_radial} is approximately given by
\be\label{ana_radial}
x(1+x)^2\frac{d^2R}{dx^2}+(s+1)x(x+1)(2x+1)\frac{dR}{dx}+\left[k^2x^4+2iskx^3-\lambda x(x+1)-is Q(2x+1)+Q^2\right]R=0\,,
\ee
where $Q=\frac{\omega-m\Omega_{\rm H}}{4\pi T_H}$, $4\pi T_H=(r_+-r_-)/r_+^2$ and $k=\omega(r_+-r_-)$.

\paragraph{(i) Near-region solution}

In this region we consider $kx\ll 1$ such that equation~\eqref{ana_radial} is then approximately given by
\be\label{near_radial}
x(1+x)^2\frac{d^2R}{dx^2}+(s+1)x(x+1)(2x+1)\frac{dR}{dx}+\left[Q^2-is Q(2x+1)-\lambda x(x+1)\right]R=0\,.
\ee

The most general solution to Eq.~\eqref{near_radial}, satisfying the ingoing boundary condition is given by
\be
R=A_1 x^{-s-i Q}(x+1)^{-s+iQ} F(\alpha,\beta,\gamma,-x)\,,
\ee
\bea
\gamma&=&1-s-2iQ\,,\\
\alpha&=&-l-s\,,\\
\beta&=&l-s+1\,.
\eea

The large $x$ behavior is 
\be\label{sol_near}
R\sim A_1\left[x^{l-s}\frac{\Gamma(\gamma)\Gamma(\beta-\alpha)}{\Gamma(\gamma-\alpha)\Gamma(\beta)}+
x^{-l-1-s}\frac{\Gamma(\gamma)\Gamma(\alpha-\beta)}{\Gamma(\alpha)\Gamma(\gamma-\beta)}\right]\,.
\ee

\paragraph{(ii) Far-region solution}

In the asymptotic region Eq.~\eqref{ana_radial} is approximately given by
\be\label{near_radial2}
\frac{d^2R}{dx^2}+\frac{2(1+s)}{x}\frac{dR}{dx}+\left(k^2+\frac{2isk}{x}-\frac{\lambda}{x^2}\right)R=0\,,
\ee
The solution of this equation can be written in terms of the confluent hypergeometric function 
\be\label{sol_confluent}
R=C_1e^{-i kx}x^{l-s}U(l-s+1,2l+2,2ikx)+C_2e^{-ik x}x^{-l-s-1}U(-l-s,-2l,2ikx)\,.
\ee
Expanding for small $kx\ll 1$, we obtain
\be\label{sol_far}
R\sim C_1x^{l-s}+C_2x^{-l-s-1}\,.
\ee
Matching~\eqref{sol_near} and~\eqref{sol_far} we get
\bea
C_1&=&A_1\frac{\Gamma(1-s-2iQ)\Gamma(2l+1)}{\Gamma(l-s+1)\Gamma(l+1-2iQ)}\,,\\
C_2&=&A_1\frac{\Gamma(1-s-2iQ)\Gamma(-1-2l)}{\Gamma(-l-2iQ)\Gamma(-l-s)}\,.
\eea

When $r\to \infty$ and in the low-frequency limit, the solution of~\eqref{teu_radial} behaves as
\be\label{inf_sol}
R_{slm}\sim \mathcal{I}_s\frac{e^{-i\omega r}}{r}+\mathcal{R}_s\frac{e^{i\omega r}}{r^{2s+1}}\,, {\rm as} \quad r \to \infty\,.
\ee

To compute the fluxes at infinity, we must relate the $C_1$ and $C_2$ with $\mathcal{I}_s$ and $\mathcal{R}_s$. Expanding~\eqref{sol_confluent} at infinity and matching to~\eqref{inf_sol} we find
\bea
\mathcal{I}_s&=&\frac{1}{\omega}\left[k^{l+1+s}\frac{C_2 (-2i)^{l+s} \Gamma (-2 l)}{\Gamma (-l+s)}+k^{s-l}\frac{C_1 (-2 i)^{s-l-1} \Gamma (2 l+2)}{\Gamma (l+s+1)}\right]\,,\\
\mathcal{R}_s&=&\omega ^{-2 s-1} \left[k^{l+1+s}\frac{C_2 (2i)^{l-s} \Gamma (-2 l)}{\Gamma (-l-s)}+k^{s-l}\frac{C_1 (2 i)^{-l-s-1} \Gamma (2
   l+2)}{\Gamma (l-s+1)}\right]\,.
\eea
To obtain the fluxes one can use the trick proposed in Ref.~\cite{Page:1976df}: solve equation~\eqref{teu_radial} replacing $s$ by $-s$, with the asymptotic behavior of $R_{-slm}$ given by
\be
R_{-slm}\sim \mathcal{I}_{-s}\frac{e^{-i\omega r}}{r}+\mathcal{R}_{-s}\frac{e^{i\omega r}}{r^{-2s+1}}\,.
\ee
Making use of the symmetries of the radial equation,
%%%%%%
\beq\label{teu_sym}
R_{s\,l\,m\,\omega}(r)&=&(-1)^{m} R^*_{s\,l\,-m\,-\omega}(r)\,,\nn\\
R^*_{s\,l\,m\,\omega}(r)&=&\Delta^{-s} R_{-s\,l\,m\,\omega}(r)\,,
%
% S^*_{s\,l\,m\,\omega}(\vartheta,\varphi)&=&(-1)^{m} S_{-s\,l\,-m\,-\omega}(\vartheta,\varphi)\,,\nn\\
% S_{s\,l\,m\,\omega}(\pi-\vartheta,\varphi+\pi)&=&S_{-s\,l\,m\,\omega}(\vartheta,\varphi)\,,
\eeq
the absorption coefficient can then be computed using
\be
Z_{slm}=\frac{dE_{\rm out}}{dE_{\rm in}}-1=\left|\frac{\mathcal{R}_s\mathcal{R}_{-s}}{\mathcal{I}_s\mathcal{I}_{-s}}\right|-1\,.
\ee
After some algebra one finally finds~\eqref{sigma} (see also the Appendix of Ref.~\cite{Page:1976df} for details).

%%%%%%%%%%%%%%%%%%%%%%%%%%%%%%%%%%%%%%%%%%%%%%%%%%%%%%%%%%%%%%%%%%%%%%%%%%%
\section{Angular momentum and energy\label{appendix_energyangularmomentum}}
%%%%%%%%%%%%%%%%%%%%%%%%%%%%%%%%%%%%%%%%%%%%%%%%%%%%%%%%%%%%%%%%%%%%%%%%%%%
Consider a stationary and axially symmetric spacetime with Killing vector fields $\xi_{(t)}^{\mu}\equiv \partial^{\mu} t$ and $\xi_{(\varphi)}^{\mu} \equiv \partial^{\mu} \varphi$. For a stress-energy tensor $T_{\mu\nu}$ the conserved energy flux vector is given by
\be
\epsilon^{\mu}=-T^{\mu}_{\phantom{\mu}\nu}\xi^{\nu}_{(t)}\,,
\ee
and the conserved angular momentum flux vector by
\be
l^{\mu}=T^{\mu}_{\phantom{\mu}\nu}\xi^{\nu}_{(\varphi)}\,.
\ee
Thus over a hypersurface $\mathbf{d}\Sigma_{\mu}$ the energy and angular momentum fluxes are
\be
\delta E=\epsilon^{\mu} \mathbf{d} \Sigma_{\mu}\,, \qquad \delta J=l^{\mu} \mathbf{d} \Sigma_{\mu}\,.
\ee
Over a spherical surface $\mathbf{d}\Sigma_{\mu} \equiv n_{\mu}r^2 dt d\Omega$, where $n_{\mu}$ is the radial outgoing normal vector to the surface, we then have
\be
\frac{\delta J}{\delta E}=-\frac{T^{r}_{\phantom{r}\varphi}}{T^{r}_{\phantom{r}t}}\,.
\ee

Considering a scalar field $\Phi(t,r,\vartheta,\varphi)=f(r,\vartheta)e^{-i\omega t+im\varphi}$ with the scalar stress-energy tensor
\be
T_{\mu\nu}=\Phi_{,\mu}\Phi_{,\nu}-\frac{1}{2}g_{\mu\nu}\Phi_{\alpha}\Phi^{\alpha}\,,
\ee
one finds
\be
\frac{\delta J}{\delta E}=\frac{m}{\omega}\,.
\ee

This applies for generic fields (photons, gravitons, \dots) as can be seen, by using the EM stress-energy 
tensor or using the effective stress-energy tensor for linearized GWs~\cite{Misner:1974qy}. We can also see it using the 
following simple argument~\cite{Bekenstein:1973mi}. At infinity the wave is composed of many quanta each with energy 
$E=\hbar \omega$ and angular momentum in the $\varphi$ direction $J=\hbar m$. Thus the ratio of the total angular 
momentum to the total energy carried by the wave across a sphere must be $m/\omega$.

%%%%%%%%%%%%%%%%%%%%%%%%%%%%%%%%%%%%%%%%%%%%%%%%%%%%%%%%%%%%%%%
\subsection{Energy and angular momentum fluxes at the horizon}
%%%%%%%%%%%%%%%%%%%%%%%%%%%%%%%%%%%%%%%%%%%%%%%%%%%%%%%%%%%%%%%
The energy flux at the horizon, as measured  at infinity, is given by
\be
\delta E_{\rm hole}=-T_{\mu}^{\phantom{\mu}\nu}\xi^{\mu}_{(t)}d^3\Sigma_{\nu}\,,
\ee
where $\xi^{\mu}_{(t)}\equiv \partial^{\mu}t$ is the time Killing vector of the Kerr metric and $\Sigma_{\mu}$ is the 3--surface element of the hole given by
\be
d^3\Sigma_{\mu}=n_{\mu}2Mr_+\sin\vartheta d\vartheta d\varphi dt\,,
\ee
with the normal vector $n_{\mu}$ in the inward direction. Likewise we can define a conserved angular momentum flux associated with the axial Killing vector $\xi^{\mu}_{(\varphi)}\equiv \partial^{\mu}\varphi$,
\be
\delta J_{\rm hole}=T_{\mu}^{\phantom{\mu}\nu}\xi^{\mu}_{(\varphi)}d^3\Sigma_{\nu}\,.
\ee
On the horizon we have
\be
n^{\mu}=\xi^{\mu}_{(t)}+\Omega_{\rm H}\xi^{\mu}_{(\varphi)}\,,
\ee
thus for any wave that enters the BH we obtain
\be\label{energy_hole}
\frac{d^2E_{\rm hole}}{dtd\Omega}-\Omega_{\rm H} \frac{d^2J_{\rm hole}}{dtd\Omega}=2Mr_+ T^{\mu\nu}n_{\mu}n_{\nu}\,.
\ee
Because of energy conservation, an angular momentum increment $\delta J$ is related to an energy increment $\delta E\equiv \delta M$ by Eq.\eqref{delta_J}~\cite{Bekenstein:1973mi}.
Inserting this in~\eqref{energy_hole} gives
\be
\frac{d^2E_{\rm hole}}{dtd\Omega}=\frac{\omega}{k_H}2Mr_+ T^{\mu\nu}n_{\mu}n_{\nu}\,.
\ee
%

%%%%%%%%%%%%%%%%%%%%%%%%%%%%%%%%%%%%%%%%%%%%%%%%%%%%%%%%%%%%%%%%%%%%%%%%%%%%
\section{Electromagnetic fluctuations around a rotating black hole enclosed in a mirror\label{app:EM_BCs}}
%%%%%%%%%%%%%%%%%%%%%%%%%%%%%%%%%%%%%%%%%%%%%%%%%%%%%%%%%%%%%%%%%%%%%%%%%%%%
%Equations for linearized Maxwell field perturbations in curved spacetimes can be obtained along the lines of the scalar field. To separate the
%angular dependence we need {\it vector spherical harmonics}~\cite{Nollert:1999ji,Ruffini,edmonds}. 

Consider the evolution of a Maxwell field in a Schwarzschild background with metric given by
\begin{equation}
ds^{2}= -f(r) dt^{2}+ \frac{dr^{2}}{f(r)}+r^{2}(d\vartheta^{2}+\sin^2\vartheta d\varphi^{2})\,,
\label{sch_lineelement}
\end{equation}
where, $f(r)=1-2M/r$ and $M$ is the BH mass. The perturbations are governed by Maxwell's equations:
\begin{equation}
{F^{\mu\nu}}_{;\nu}=0 \quad, F_{\mu\nu}=A_{\nu,\mu}-A_{\mu,\nu}\,,
\label{maxwell}
\end{equation}
where a comma stands for ordinary derivative and a semi-colon
for covariant derivative. Since the background is spherically symmetric,
we can expand $A_{\mu}$ in 4-dimensional vector spherical harmonics (see~\cite{Ruffini}):

{\small
\be
A_{\mu}(t,r,\vartheta,\varphi)=\sum_{l,m}\left[\left(
 \begin{array}{c} 0 \\ 0 \\
 a^{lm}(t,r)\bar{\bm{S}}_{lm}\end{array}\right)
+\left(\begin{array}{c}f^{lm}(t,r)Y_{lm}\\h^{lm}(t,r)Y_{lm} \\
 k^{lm}(t,r) \bar{\bm{Y}}_{lm}\end{array}\right)\right]\,,
\label{expansion}
\ee
}
with the vector spherical spherical harmonics given by,
\be
\bar{\bm{Y}}^\intercal_{lm}=\left(\partial_\vartheta Y_{lm}, \partial_\varphi Y_{lm}\right)\,,\quad
\bar{\bm{S}}^\intercal_{lm}=\left(\frac{1}{\sin\vartheta}\partial_\varphi Y_{lm}, -\sin\vartheta\partial_\vartheta Y_{lm}\right)\,,
\ee
and where $Y_{lm}$ are the usual scalar spherical harmonics, $m$ is the azimuthal number and $l$ the angular quantum number. The first term in the right-hand side has parity $(-1)^{l+1}$,
and the second term has parity $(-1)^{l}$. We shall call the former the axial modes and the latter the polar modes. 

Upon defining
\be 
\Upsilon^{lm}=\frac{r^2}{l(l+1)}\left(\partial_t h^{lm}-\partial_r f^{lm}\right)\,,\label{upsilon}
\ee
and inserting~\eqref{expansion} into Maxwell's equations~\eqref{maxwell}, and after some algebra, we get the following system of equations
\beq
& & \frac{\partial^{2} a^{lm}(t,r)}{\partial r_*^{2}} + \left\lbrack -\frac{\partial^{2}}{\partial
t^{2}}-V(r)\right\rbrack a^{lm}(t,r)=0 \,,\\
& &\frac{\partial^{2}\Upsilon^{lm}(t,r)}{\partial r_*^{2}} + \left\lbrack -\frac{\partial^{2}}{\partial
t^{2}}-V(r)\right\rbrack \Upsilon^{lm}(t,r)=0 \,,\\
& & V=f\frac{l(l+1)}{r^2}\,.\label{potentialmaxwell}
\eeq
If we assume a time dependence $a^{lm}\,,\Upsilon^{lm}\propto e^{-i\omega t}$, the equation for
EM perturbations of the Schwarzschild geometry takes the form
\be
\frac{\partial^{2}\Psi}{\partial r_*^{2}} + \left\lbrack \omega^2-V\right\rbrack \Psi=0
\,,\label{wavemaxwell}
\ee
where the tortoise coordinate is defined through $dr/dr^*=f(r)$, $\Psi=a^{lm}$ for axial modes and $\Psi=\Upsilon$ for polar modes. The
potential $V$ appearing in equation~\eqref{wavemaxwell} is given by Eq~\eqref{potentialmaxwell}.

Let us now assume we have a spherical conductor at $r=r_m$. The conditions to be satisfied are then that the electric/magnetic field as seen by an observer at rest with respect to the conductor has no tangential/parallel components, $E_{\vartheta}\propto F_{\vartheta \,t}=0,\,E_{\varphi}\propto F_{\varphi \,t},\,B_{r}\propto F_{\varphi \,\vartheta}=0$.
This translates into
\be
\partial_t a^{lm}(t,r_m)=0\,,\quad f^{lm}(t,r_m)-\partial_t k^{lm}(t,r_m)=0\,.
\ee
Using Maxwell's equations~\eqref{maxwell}, we get the relation
\be
f^{lm}(t,r_m)-\partial_t k^{lm}(t,r_m)=\frac{f}{l(l+1)}\partial_r\left(r^2\partial_r f^{lm}-r^2\partial_t h^{lm}\right)\,.
\ee
Finally, using Eq.~\eqref{upsilon} we get
\be
\partial_r\Upsilon=0\,.
\ee
In other words, the boundary conditions at the surface $r=r_m$ are $\Psi=0$ and $\partial_r\Psi=0$ for axial and polar 
perturbations respectively. This can be used to easily compute the EM modes inside a resonant cavity in flat space. 
Taking $M=0$ in Eq.~\eqref{wavemaxwell} we find the exact solution
\be
\Psi=\sqrt{r}\left[C_1 J_{l+1/2}(r \omega)+C_2 Y_{l+1/2}(r \omega)\right]\,,
\ee
where $C_{i}$ are constants and $J_n(r\omega)$ and $Y_n(r\omega)$ are Bessel functions of the first and second kind, respectively. Imposing regularity at the origin $r=0$ implies $C_2=0$. The Dirichlet boundary condition $\Psi=0$ at $r=r_m$, which can easily be shown to correspond to the \emph{transverse electric} modes (modes with $E_r=0$)~\cite{Jackson}, then gives
\be\label{TE_modes}
\omega_{\rm TE}=\frac{j_{l+1/2,n}}{r_m}\,,
\ee
where $j_{l+1/2,n}$ are the zeros of the Bessel function $J_{l+1/2}$ and $n$ is a non-negative integer. On the other hand the eigenfrequencies for the Neumann boundary condition $\partial_r\Psi=0$, which corresponds to the \emph{transverse magnetic} modes (modes with $B_r=0$)~\cite{Jackson}, can be computed solving
\be
\left\{\partial_r\left[\sqrt{r}J_{l+1/2}(r\omega)\right]\right\}_{r=r_m}=\frac{(l+1)J_{l+1/2}(r_m\omega)-r_m\omega J_{l+3/2}(r_m\omega)}{\sqrt{r_m}}=0\,.
\ee
Defining $\tilde{j}_{l+1/2,n}$ as being the zeroes of $\partial_r\left[\sqrt{r_m}J_{l+1/2}(r_m\omega)\right]$ we find
\be\label{TM_modes}
\omega_{\rm TM}=\frac{\tilde{j}_{l+1/2,n}}{r_m}\,.
\ee
The eigenfrequencies for $l=1$ and $n=0$ are shown in Fig.~\ref{Fig:BQNM} where we see that when $r_m\gg M$, the real part of the quasinormal frequencies of a BH enclosed in a mirror asymptotically reduces to the flat space result.  
One can write down a relation between the Regge-Wheeler function $\Psi$~\cite{Chrzanowski:1975wv,Ori:2002uv,Hughes:2000pf} and the Teukolsky radial function $R$ (cf. Eq.~\eqref{teu_eigen}) given by
\begin{eqnarray}
\frac{\Psi}{r(r^2-2Mr)^{s/2}}&=&\left(r\sqrt{\Delta}\right)^{|s|}\mathcal{D}_-^{|s|}\left(r^{-|s|}R\right)\,,\, s<0,\nn\\
\frac{\Psi}{r(r^2-2Mr)^{s/2}}&=&\left(\frac{r}{\sqrt{\Delta}}\right)^{s}\mathcal{D}_+^{s}\left[\left(\frac{r^2-2Mr}{r}\right)^{s}R\right]\,,\, s>0,\nn\\\label{hughes}
\end{eqnarray}
where $\mathcal{D}_{\pm}=d/dr\pm i \omega/f$. Using these relations and Teukolsky's radial equation~\eqref{teu_radial}, one finds that the Dirichlet and the Neumann boundary conditions for $\Psi$, correspond to the Robin boundary conditions for the radial function $R$ given respectively by
\beq
\partial_r R_{-1}&=&\frac{r-2M+i r^2\omega}{r(r-2M)}R_{-1}\,,\label{cond1}\\
\partial_r R_{-1}&=&\frac{r\omega[2M+r(-1-i r \omega)]-i l(l+1) (2M-r)}{(2M-r)r^2\omega}R_{-1}\,.\label{cond2}
\eeq

After having understood the nonrotating case, below we turn to the rotating case. The main difficulty relies in 
describing the EM physical quantities in terms of the Newman-Penrose quantities. We will show that doing so, will allow 
us to generalize the conditions~\eqref{cond1} and~\eqref{cond2}.

%%%%%%%%%%%%%%%%%%%%%%%%%%%%%%%%%%%%%%%%%%%%%%%%%%%%%%%%%%%%%%%%%%%%%%%%%%%%
\paragraph{Newman-Penrose approach.}
%%%%%%%%%%%%%%%%%%%%%%%%%%%%%%%%%%%%%%%%%%%%%%%%%%%%%%%%%%%%%%%%%%%%%%%%%%%%
In the Newman-Penrose formalism, the EM field is characterized by three complex scalars from which one can obtain the 
electric and magnetic field. The details of this procedure are not important for us here so we refer the reader to 
Ref.~\cite{King:1977}. In the frame of a ZAMO observer (cf. Section~\ref{sec:kerr}), the relevant electric and magnetic 
field components read~\cite{King:1977}
\beq
E_{(\vartheta)}&=&\left[\frac{\Delta^{1/2}(r^2+a^2)}{\sqrt{2}\rho^* A^{1/2}(r^2+a^2\cos^2\vartheta)}\left(\frac{\phi_0}{2}-\frac{\phi_2}{\rho^2\Delta}\right)+{\rm c.c.}\right]
-\frac{2 a \Delta^{1/2}}{A^{1/2}}\sin\vartheta\, {\rm Im}(\phi_1)\,,\nn\\
E_{(\varphi)}&=&\left[-i\Delta^{1/2}\rho\left(\frac{\phi_0}{2\sqrt{2}}+\frac{\phi_2}{\sqrt{2}\rho^2\Delta}\right)+{\rm c.c.}\right]\,,\nn\\
B_{(r)}&=&\left[\frac{a\sin\vartheta}{\sqrt{2}\rho A^{1/2}}\left(\phi_2-\Delta\rho^2\frac{\phi_0}{2}\right)+{\rm c.c.}\right]
+2\frac{r^2+a^2}{A^{1/2}}{\rm Im}(\phi_1)\,,
\eeq
where $\rho=-(r-i a\cos\vartheta)^{-1}$, $A=(r^2+a^2)^2-a^2\Delta\sin^2\vartheta$ and $\Delta=r^2-2Mr+a^2$.

If we assume a conducting spherical surface surrounding the BH at $r=r_m$, then Maxwell's equations require that
$E_{(\vartheta)}=E_{(\varphi)}=B_{(r)}=0$ at $r=r_m$ and we are left with the boundary conditions at the conductor:
\be
{\rm Re}\left(\rho\Phi_0\right)=\frac{{\rm Re}\left(\rho\Phi_2\right)}{\Delta}\,,\quad
{\rm Im}\left(\rho\Phi_0\right)=-\frac{{\rm Im}\left(\rho\Phi_2\right)}{\Delta}\,,\quad
{\rm Im}(\phi_1)=0\,,
\ee
where we defined $\Phi_0=\phi_0$ and $\Phi_2=2\rho^{-2}\phi_2$ . This can be simplified to
\be\label{bc_NP}
|\Phi_0|^2=\frac{|\Phi_2|^2}{\Delta^2}\,.
\ee
%
%
%Using $\rho=-(r+ia\cos\vartheta)/(r^2+a^2\cos^2\vartheta)$ we see that to satisfy Eq.~\eqref{bc_NP} one needs to satisfy the two conditions
%
%\beq
%r\Phi_0&=&r\frac{\Phi_2^*}{\Delta}\,,\label{bc_1}\\
%\frac{ia\cos\vartheta}{r^2+a^2\cos^2\vartheta}\Phi_0&=&-\frac{ia\cos\vartheta}{r^2+a^2\cos^2\vartheta}\frac{\Phi_2^*}{\Delta}\,,\label{bc_2}
%\eeq
%
We use the decomposition
\beq\label{decom1}
\Phi_0&=&\sum_{l m}\int \,d\omega e^{-i\omega t+im\varphi}R_{s\,l\,m\,\omega}S_{s\,l\,m\,\omega}(\vartheta)\,,\nonumber\\
\Phi_2&=&\sum_{l m}\int \,d\omega e^{-i\omega t+im\varphi}R_{-s\,l\,m\,\omega}S_{-s\,l\,m\,\omega}(\vartheta)\,,
\eeq
where the radial and the angular function, $R$ and $S$, satisfy Teukolsky's Eqs.\eqref{teu_radial} and~\eqref{spheroidal}, respectively.
The functions $R_{s=1}$ can be written as a linear combination of $R_{s=-1}$, and $S_{s=-1}$ can be written as a linear combination of $S_{s=1}$ and its derivative through the Starobinski-Teukolsky identities~\cite{Teukolsky:1974yv,1973ZhETF..65....3S,Staro2}
\be\label{ST_iden}
\mathcal{D}_0\mathcal{D}_0{}R_{-1}=B R_{1}\,,\quad
\mathcal{L}_0\mathcal{L}_1S_{1}=B S_{-1}\,,
\ee
where $B=\sqrt{(A_{-1lm}+a^2\omega^2-2am\omega)^2+4ma\omega-4a^2\omega^2}$ and the linear operators are given by
\be
\mathcal{D}_0=\partial_r-i\frac{K}{\Delta}\,,\quad
\mathcal{L}_n=\partial_{\vartheta}+m\csc\vartheta-a\omega\sin\vartheta+n\cot\vartheta.
\ee
%
%Furthermore, from Teukolsky's equations one can derive the following identities~\cite{Chrzanowski:1975wv}
%
%\be\label{iden}
%R^*_{s\,l\,m\,\omega}= (-1)^{m} R_{s\,l\,-m\,-\omega}\,,\quad
%
%S^*_{s\,l\,m\,\omega}=(-1)^{m} S_{-s\,l\,-m\,-\omega}\,.
%\ee
%
Finally, replacing~\eqref{decom1} and \eqref{ST_iden} in eq.~\eqref{bc_NP} and integrating~\eqref{bc_NP} over the sphere we find the following conditions for the two polarizations:
\beq
&&\partial_r R_{-1}=\frac{-i\Delta\left[\pm B+A_{-1lm}+\omega  \left(a^2 \omega -2 a m+2 i
   r\right)\right]}{2 \Delta\left(a^2 \omega -a m+r^2 \omega \right)}R_{-1}\nonumber\\
&&		+\frac{\left(a^2 \omega -a m+r^2 \omega \right) \left(2 i a^2 \omega -2 i a m+2 M+2 i r^2 \omega+ \partial_r\Delta-2  r\right)}{2 \Delta\left(a^2 \omega -a m+r^2 \omega \right)}R_{-1}\,,
\eeq
where we integrated out the angular dependence using the normalization condition~\eqref{sphe_norm}. This is the result shown in Section~\ref{sec:bombs}. Note that to for $a=0$ we recover the condition~\eqref{cond1} when using the minus sign, while for the plus sign we recover the condition~\eqref{cond2}. 

%%%%%%%%%%%%%%%%%%%%%%%%%%%%%%%%%%%%%%%%%%%%%%%%%%%%%%%%%%%%%%%%%%%%%%%%%%%%%%%
\section{Hartle-Thorne formalism for slowly-rotating spacetimes and perturbations}\label{app:HT}
%%%%%%%%%%%%%%%%%%%%%%%%%%%%%%%%%%%%%%%%%%%%%%%%%%%%%%%%%%%%%%%%%%%%%%%%%%%%%%%
In this Appendix we summarize the formalism originally developed by Hartle and Thorne~\cite{Hartle:1967he} to construct slowly-rotating stars and that developed by Kojima~\cite{Kojima:1992ie,1993ApJ...414..247K} to include generic nonspherical perturbations (see also Refs.~\cite{Pani:2012bp,Brito:2013wya} for extensions and \cite{Pani:2013pma} for a review.). 
%%%%%%%
\subsection{Background} 
%%%%%%%
Let us start by considering the most general stationary axisymmetric spacetime (we also assume circularity, see Sec.~\ref{sec:ERhor} and Ref.~\cite{Chandra}) 
%%%%
\begin{equation}
 ds^2_0=g_{tt}dt^2+g_{rr}dr^2+2g_{t\varphi}dtd\varphi+g_{\th\th}d\vartheta^2+g_{\varphi\varphi}d\varphi^2\,,\label{genericspinningmetric}
\end{equation}
where $g_{tt}$, $g_{rr}$, $g_{t\varphi}$, $g_{\th\th}$ and $g_{\varphi\varphi}$ are functions of $r$ and $\vartheta$ only. Assuming slow rotation, we construct a perturbative expansion in the angular momentum $J$ (or in some other parameter linear in $J$, which characterizes the rotation rate). To second order in rotation, the metric above can be expanded as~\cite{Hartle:1967he}
%%%%
\begin{eqnarray}
 d\tilde{s}^2&&=-e^\nu\left[1+2\epsilon^2\left(h_0+h_2 P_2\right)\right]dt^2+\frac{1+2\epsilon^2(m_0+m_2P_2)/(r-2M)}{1-2M/r}dr^2\nn\\
 &&+r^2\left[1+2\epsilon^2(v_2-h_2)P_2\right]\left[d\vartheta^2+\sin^2\vartheta(d\varphi-\epsilon\varpi dt)^2\right]\,, \label{metricHT2b}
\end{eqnarray}
%%%
where $P_2=P_2(\cos\vartheta)=(3\cos^2\vartheta-1)/2$ is a Legendre
polynomial. The radial functions $\nu$ and $M$ are of zeroth order in
rotation, $\varpi$ is of first order, and $h_0$, $h_2$,
$m_0$, $m_2$, $v_2$ are of second order. 

We consider a perfect fluid coupled to gravity with a barotropic equation of state $P=P(\rho)$, where $P$ and $\rho$ are the pressure and the density of the fluid, respectively. Under an infinitesimal rotation both $P$ and $\rho$ transform as scalars. As shown in~\cite{Hartle:1967he,Hartle:1968si}, in order to perform a valid perturbative expansion it is necessary to transform the radial coordinate in such a way that the deformed density in the new coordinates coincides with the unperturbed density at the same location. It can be shown that this transformation is formally equivalent to working in the original coordinates but expanding the pressure and the density as
\begin{eqnarray}
 P&\equiv&P_0+\Delta P=P_0+(\rho_0+P_0)(p_0+p_2 P_2)\,, \label{P}\\
 \rho&\equiv&\rho_0+\Delta\rho=\rho_0+(\rho_0+P_0)\frac{\pa\rho_0}{\pa P_0}(p_0+p_2 P_2)\,,\label{rho}
\end{eqnarray}
where $P_0$ and $\rho_0$ denote the corresponding quantities in the nonrotating case. Finally, the stress-energy tensor is the standard one, 
%%%
\begin{equation}
 T^{\mu\nu}=(P+\rho)u^\mu u^\nu+g^{\mu\nu}P\,,
\end{equation}
%%%
where $u^\mu$ is the fluid four-velocity. By plugging the decompositions above into the
gravitational equations $G_{\mu\nu}=8\pi T_{\mu\nu}$, and by solving the
equations order by order in the spin, we obtain a system of ODEs for the rotating background, which can be solved by standard methods~\cite{Hartle:1967he,Hartle:1968si,Pani:2014jra}.

%%%%%%%%%%%%%%%%%%%%%%%%%%%
\subsection{Perturbations of a slowly-rotating object}
%%%%%%%%%%%%%%%%%%%%%%%%%%%
Perturbations of slowly rotating and oscillating compact objects can be studied by perturbing the solution discussed above.
Scalar, vector and tensor field equations in the background
metric~\eqref{metricHT2b} can be linearized in the field perturbations.
Any perturbation function $\delta X$ can be expanded in a complete basis of spherical harmonics, similarly to the static case. Schematically, in the frequency domain we have
%%%%
\begin{equation}
\delta X_{\mu_1\dots}(t,r,\vartheta,\varphi)=
\delta X^{(i)}_{l m}{\cal Y}_{\mu_1\dots}^{l m\,(i)}e^{-i\omega t}\,,
\label{expa}
\end{equation}
%%%
where ${\cal Y}_{\mu_1\dots}^{l m\,(i)}$ is a basis of scalar,
vector or tensor harmonics, depending on the tensorial nature of the
perturbation $\delta X$. As in the spherically symmetric case, the perturbation variables $\delta X^{(i)}_{l m}$ can
be classified as ``polar'' or ``axial'' depending on their behavior
under parity transformations.

The linear response of the system is fully characterized by a coupled system of ODEs in the perturbation functions $\delta X^{(i)}_{l  m}$.  
In the case of a spherically symmetric background,
perturbations with different values of $(l,\,m)$, as well as
perturbations with opposite parity, are decoupled. In a rotating,
axially symmetric background, perturbations with different values of
$m$ are still decoupled but perturbations with different values of
$l$ are not.

To second order, the perturbation equations read schematically as (cf. Ref.~\cite{Pani:2013pma} for details)
\begin{eqnarray}
0&=&{\cal A}_{l}+\tilde a m \bar{\cal A}_{{l}}+\tilde{a}^2 \hat{{\cal A}}_l+\tilde a ({\cal Q}_{{l}}\tilde{\cal P}_{l-1}+{\cal Q}_{l+1}\tilde{\cal P}_{l+1})\nn\\
&+&\tilde{a}^2 \left[{\cal Q}_{l-1} {\cal Q}_l \breve{{\cal A}}_{l-2} + {\cal Q}_{l+2} {\cal Q}_{l+1} \breve{{\cal A}}_{l+2} \right]+{\cal O}(\tilde{a}^3)\,,\label{eq_axial}\\
%%%%%
0&=&{\cal P}_{l}+\tilde a m \bar{\cal P}_{{l}}+\tilde{a}^2 \hat{{\cal P}}_l+\tilde a ({\cal Q}_{{l}}\tilde{\cal A}_{l-1}+{\cal Q}_{l+1}\tilde{\cal A}_{l+1})\nn\\
&+&\tilde{a}^2 \left[{\cal Q}_{l-1} {\cal Q}_l \breve{{\cal P}}_{l-2} + {\cal Q}_{l+2} {\cal Q}_{l+1} \breve{{\cal P}}_{l+2} \right]+{\cal O}(\tilde{a}^3)\,,\label{eq_polar}
\end{eqnarray}
%%%%%
where $\tilde{a}=a/M$, ${\cal Q}_l=\sqrt{\frac{l^2-m^2}{4l^2-1}}$ and the
coefficients ${\cal A}_l$ and ${\cal P}_l$ (with various
superscripts) are \emph{linear} combinations of axial and polar
perturbation variables, respectively.

The structure of Eqs.~\eqref{eq_axial}--\eqref{eq_polar} is interesting.
In the limit of slow rotation there is a Laporte-like
``selection rule''~\cite{ChandraFerrari91}. Perturbations with a given parity and index $l$ are coupled to: (i)
perturbations with \emph{opposite} parity and index $l\pm1$ at
order $\epsilon_a$; (ii) perturbations with \emph{same} parity and
\emph{same} index $l$ up to order $\epsilon_a^2$; (iii)
perturbations with \emph{same} parity and index $l\pm2$ at order
$\epsilon_a^2$. 
Furthermore, from the definition of ${\cal Q}_l$ it follows that ${\cal Q}_{\pm m}=0$, and
therefore if $|m|=l$ the coupling of perturbations with index
$l$ to perturbations with indices $l-1$ and $l-2$ is
suppressed. This general property is usually called~\cite{ChandraFerrari91} ``propensity
rule'' in atomic theory, and states that transitions $l\to l+1$
are strongly favored over transitions $l\to l-1$. Indeed, the slow-rotation technique is well known in quantum mechanics and the coefficients ${\cal Q}_{l}$ are related to the
usual Clebsch-Gordan coefficients.

\subsubsection{Scalar perturbations of a slowly-rotating star}

The formalism above can be applied to any type of perturbations of a generic stationary and axisymmetric background. The simplest example is a probe scalar field governed by the Klein-Gordon equation~\eqref{KG} and propagating on the fixed geometry~\eqref{metricHT2b}. The entire procedure is performed in a publicly available {\scshape Mathematica}\textsuperscript{\textregistered} notebook, cf. Appendix~\ref{app:codes}.

We start by the standard decomposition of the scalar field in spherical harmonics in Fourier space,
%%%
\begin{equation}
 \Phi=\sum_{l m}\int d\omega\, \Psi_l(r)Y^l(\vartheta,\varphi) e^{-i\omega t}\,.
\end{equation}
%%%
By plugging this equation into~\eqref{KG} and using the line element~\eqref{metricHT2b}, we obtain the following equation in schematic form:
\begin{equation}
 A_{l} Y^l+ \hat A_l \cos^2\vartheta Y^l+\tilde{B}_l\cos\th\sin\th Y^l_{,\th}=0\,,\label{eq_expY}
\end{equation}
%%%
where a sum over $(l,m)$ is implicit, and the explicit form of the radial coefficients
$A_l$, $\hat{A}_l$ and $\tilde B_l$ is given in the notebook. The coefficients $\hat{A}_l$ and $\tilde{B}_l$ are proportional to terms quadratic in the spin, so they vanish to first order. Indeed, to first order the equation reduces to $A_l=0$ which can be explicitly written as in Eq.~\eqref{master_scalar_star_1order}.

The separation of Eq.~\eqref{eq_expY} can be achieved by using the
identities~\cite{Kojima:1992ie}
\begin{eqnarray}
\cos\th Y^{l}&=&{\cal Q}_{l+1}Y^{l+1}+{\cal Q}_{l}Y^{l-1}\,,\nn\\
\sin\th \partial_\vartheta Y^{l}&=&
{\cal Q}_{l+1}l Y^{l+1}-{\cal Q}_{l}(l+1)Y^{l-1}\,,\nn\\
%%%%
\cos^2\th Y^{l}&=&\left({\cal Q}_\lp^2 + {\cal Q}_l^2\right)Y^l+{\cal Q}_\lp {\cal Q}_\lpp Y^\lpp + {\cal Q}_l {\cal Q}_\lm Y^\lmm\,,\nn\\
%%%%
\cos\th\sin\th \pa_\th Y^{l}&=&\left(l{\cal Q}_\lp^2 -(l+1){\cal Q}_l^2\right)Y^l+{\cal Q}_\lp {\cal Q}_\lpp l Y^\lpp- {\cal Q}_l {\cal Q}_\lm (l+1) Y^\lmm\,,\nn
\end{eqnarray}
%%%
and so on, as well as the orthogonality property of scalar spherical harmonics.
The result reads
%%%%
\begin{eqnarray}
&& A_l+{\cal Q}_{l+1}^2[\hat{A}_l+l\tilde{B}_l]+{\cal Q}_{l}^2[\hat{A}_l-(l+1)\tilde{B}_l]\nn\\
&&+{\cal Q}_{l-1}{\cal Q}_{l} [\hat{A}_{l-2}+(l-2)\tilde{B}_{l-2}]+{\cal Q}_{l+2}
{\cal Q}_{l+1} [\hat{A}_{l+2 }-(l+3)\tilde{B}_{l+2}]=0\label{final_schematic}\,.
\end{eqnarray}
%%%%

Therefore, at second order,
perturbations with harmonic index $l$ are coupled to perturbations
with $l\pm2$. Crucially, this coupling does not
contribute to the eigenfrequencies to second order~\cite{Pani:2012bp,Pani:2013pma}. 
Therefore, for given values of $l$ and $m$, the eigenspectrum of the scalar perturbations is governed by a single ODE:
\begin{equation}
A_l+{\cal Q}_{l+1}^2[\hat{A}_l+l\tilde{B}_l]+{\cal Q}_{l}^2[\hat{A}_l-(l+1)\tilde{B}_l]=0\,.\label{eqSCAL2}
\end{equation}
%%%
In the online notebook \url{HartleThorne.nb} we show that the equation above reduces to~\eqref{master_scalar_star_2order} and we give the explicit form of $V_2$, which is too involved to be reproduced here.

%%%%%%%%%%%%%
\section{WKB analysis of long-lived and unstable modes of ultracompact objects} \label{app:WKB}
%%%%%%%%%%%%%%%%
As discussed in Sec.~\ref{sec:ERlonglived}, ultracompact objects have two light rings~\cite{Cardoso:2014sna}. From a point of view of massless
fields, which propagate as null particles in the eikonal regime, the light rings effectively confine
the field and give rise to long-lived modes, which may become unstable if they form within the ergoregion. Here we perform a WKB analysis of these trapped modes.

Let us first discuss static, spherically symmetric spacetimes described by a line element given in Eq.~\eqref{ds2rotapprox} with $\varpi=0$.
Various classes of perturbations of this geometry are described by a master equation of the form~\eqref{wave} where $V_{\rm eff}=\omega^2-V_{sl}(r)$, and the effective potential for wave propagation reads~\cite{Cardoso:2014sna}
\begin{eqnarray}
V_{sl}(r)&=&f\left[\frac{l(l+1)}{r^2}+\frac{1-s^2}{2r B}\left(\frac{f'}{f}-\frac{B'}{B}\right)+8\pi(p_{\rm rad}-\rho) \delta_{s2}\right]\,,\label{potentialmaster}
\end{eqnarray}
and the prime denotes derivative with respect to the coordinate $r$, which is related to the tortoise coordinate $r_*$ through $dr/dr_*=\sqrt{f/B}$. In the potential~\eqref{potentialmaster} $l\geq s$, $s=0,1$ for test Klein-Gordon and Maxwell fields, respectively, whereas $s=2$ for axial perturbations of a (generically anisotropic) fluid in GR (where $p_{\rm rad}=T_r^r$ and $\rho=-T_t^t$ are the radial pressure and the energy density of the fluid, respectively).

Figure~\ref{fig:LRpotential} shows an example of $V_{sl}(r)$ for two representative ultracompact objects: the so-called gravastar model discussed in Sec.~\ref{sec:BHmimickers} and a constant density star which, in the static case, is described by the line element~\eqref{ds2rotapprox} with $\varpi=0$ and
\begin{eqnarray}
F(r)&=&\frac{1}{4 R^3}\left(\sqrt{R^3-2Mr^2}-3R\sqrt{R-2M}\right)^2\,, \label{fstar}\\
B(r)&=&\left(1-\frac{2 M r^2}{R^3}\right)^{-1}\,, \label{Bstar}
\end{eqnarray}
where $R$ is the radius of the star. The pressure is given by
\begin{equation}
 P(r)=\rho_c\frac{\sqrt{3-8 \pi  R^2 \rho_c}-\sqrt{3-8 \pi  r^2 \rho_c}}{\sqrt{3-8 \pi  r^2 \rho_c}-3 \sqrt{3-8 \pi  R^2 \rho_c}}\,, \label{Pstar}
\end{equation}
where $\rho_c=3M/(4\pi R^3)$ is the density of the uniform star.

This potential $V_{sl}(r)$ shares many similarities with the geodesic potential to which it reduces in the eikonal limit~\cite{Cardoso:2008bp}:
it has a local maximum, diverges at the origin  and is constant at infinity.
Because the potential necessarily develops a local minimum, it is possible to show that in the eikonal limit ($l\gg1$) the spectrum contains long-lived modes whose damping time grows exponentially with $l$~\cite{Festuccia:2008zx,Berti:2009wx,Cardoso:2008bp}. 
\begin{figure}[t]
\begin{center}
\begin{tabular}{cc}
\epsfig{file=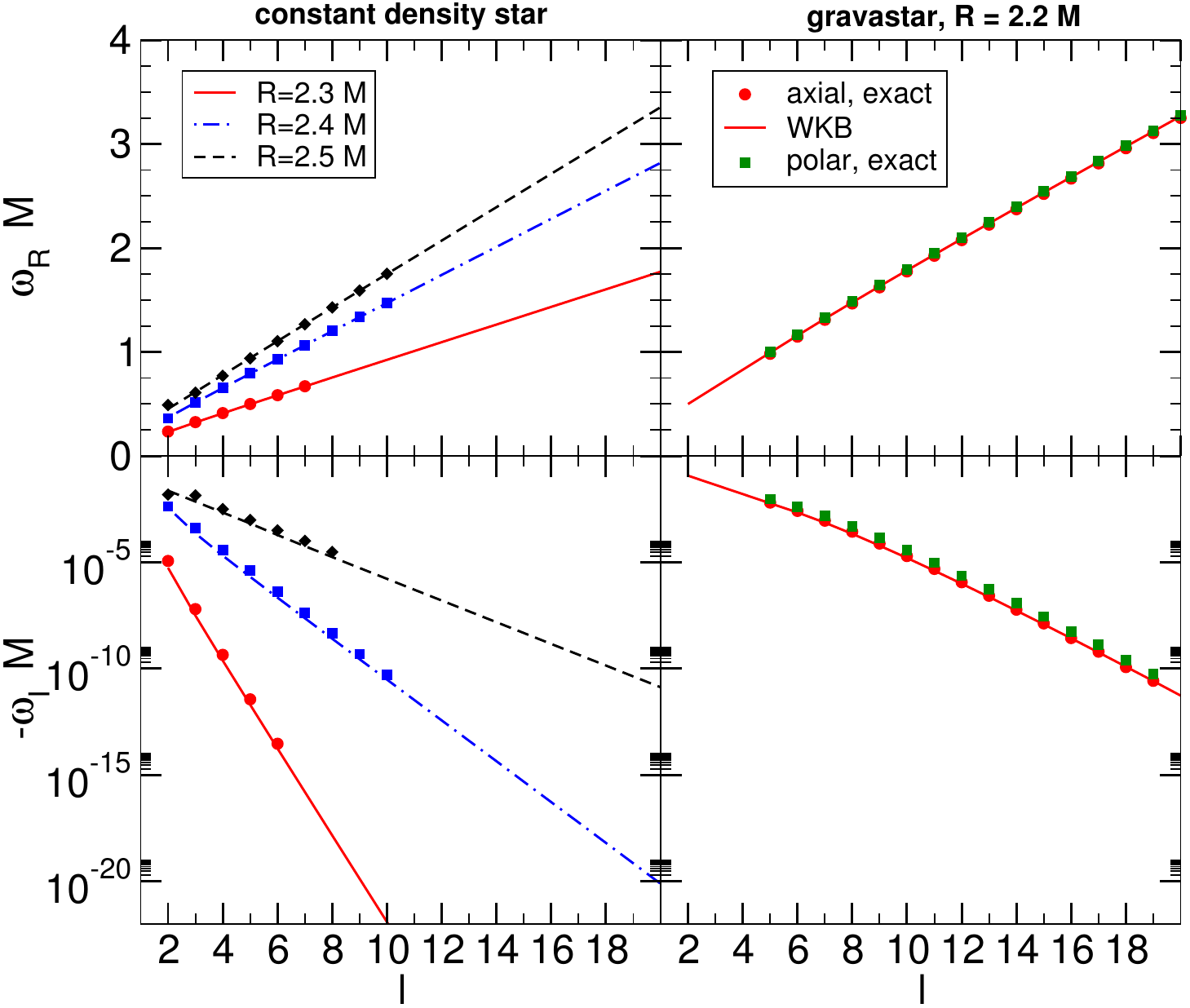,width=8.5cm,angle=0,clip=true}
\end{tabular}
\end{center}
\caption{\label{fig:star}
Real and imaginary parts of the long-lived modes of a uniform star for different compactness (left panels) and for a gravastar with $R=2.2M$ (right panels). The lines are the WKB results, whereas markers show the numerical results obtained in Ref.~\cite{Cardoso:2014sna} by using direct integration or continued fractions. For uniform stars we show gravitational axial modes, whereas for gravastar we show both axial modes (red circles) and gravitational polar modes with $v_s=0.1$ (green squares), where $v_s$ is related to the speed of sound on the shell~\cite{Pani:2009ss}. See Ref.~\cite{Cardoso:2014sna} for details.
}
\end{figure}
To first order in the eikonal limit, the potential can be approximated as $V_{sl}(r)\sim l^2f/r^2$. Let us define $r_a$, $r_b$ and $r_c$ to be the three real turning points of $\omega_R^2-V_{sl}(r)=0$ as shown in Fig.~\ref{fig:LRpotential} for the black solid curve. When such turning points exist, the real part of the frequency of a class of long-lived modes is given by the WKB condition:
\begin{equation}
\int_{r_a}^{r_b}\frac{dr}{\sqrt{f/B}}\sqrt{\omega_R^2-V_{sl}(r)}=\pi\left (n+1/2\right),\label{bohrsommerfeld}
\end{equation}
where $n$ is a positive integer.
The imaginary part of the frequency $\omega_I$ of these modes is 
\begin{equation}
\omega_I=-\frac{1}{8\omega_R\gamma} \exp\left[-2\int_{r_b}^{r_c}\frac{dr}{\sqrt{f/B}} \sqrt{V_{sl}(r)-\omega_R^2}\right]\,,\\ \label{omI}
\end{equation}
where
\begin{eqnarray}
\gamma&=& \int_{r_a}^{r_b}\frac{dr}{\sqrt{f/B}} \frac{\cos^2\chi(r)}{\sqrt{\omega_R^2-V_{sl}(r)}} \,,\qquad \chi(r)= -\frac{\pi}{4}+\int_{r_a}^{r}\frac{dr}{\sqrt{f/B}} \sqrt{\omega_R^2-V_{	sl}(r)}\,.
\end{eqnarray}
By expanding Eqs.~\eqref{bohrsommerfeld} and~\eqref{omI}, one can show that, to leading order in the eikonal limit, the mode frequency reads
\begin{equation}
\omega\sim a\, l -i\, b\, e^{-c l}\qquad l\gg1\,, \label{omega}
\end{equation}
where $a$, $b$ and $c$ are positive constants. By expanding Eq.~\eqref{bohrsommerfeld} near the minimum of the potential displayed in Fig.~\ref{fig:LRpotential}, it is possible to show that~\cite{Cardoso:2014sna}
\begin{equation}
 a\sim\Omega_{\rm LR2}\equiv\frac{\sqrt{f(r_{\rm LR2})}}{r_{\rm LR2}}\,,
\end{equation}
where $\Omega_{\rm LR2}$ is the angular velocity of the \emph{stable} null geodesic at the light-ring location $r=r_{\rm LR2}$. Note that the damping time of these modes is exponentially large, so that they are arbitrarily long-lived in the large-$l$ limit. In Fig.~\ref{fig:star}, we compare the long-lived modes computed through the above WKB formula with the exact numerical result~\cite{Cardoso:2014sna} for two representative ultracompact objects, showing a quite good agreement in the large-$l$ limit.

Practically, the long-lived modes of a static ultracompact object are metastable and it is reasonable to expect that they can turn unstable when rotation is included. In the slow-rotation limit one may consider a probe scalar field propagating on the approximate spinning geometry~\eqref{ds2rotapprox}; the Klein-Gordon equation in the eikonal limit reduces to Eq.~\eqref{KGrot}, which is suitable for a WKB analysis similarly to the nonrotating case~\cite{CominsSchutz,Cardoso:2007az}. By defining $W=\frac{B(r)}{f(r)}\left (\bar{\omega}-V_+\right )\left (\bar{\omega}-V_-\right )$, the quasi-bound unstable modes are determined by
\begin{equation}
m\int_{r_a}^{r_b}\sqrt{W(r)}dr =\frac{\pi}{2}+n\pi\,,\quad n=0\,,1\,,2\,,\dots
\end{equation}
and their characteristic time scale can be computed through
\begin{equation}
\tau=4\exp{\left[2m\int_{r_b}^{r_c}\sqrt{|W|}dr\right]}\int_{r_a}^{r_b}\frac{d}{d\bar{\omega}}
\sqrt{W}dr\,,
\end{equation}
where $r_a$, $r_b$ are solutions of $V_+=\bar{\omega}$ and $r_c$ is determined by the condition
$V_-=\bar{\omega}$. This result agrees with Eq.~\eqref{tau_ergo_eikonal} quoted in the main text.
As discussed in Sec.~\ref{sec:ergoregioninstability}, the long-lived modes become unstable (i.e. their imaginary part changes sign) above a critical spin and precisely when an ergoregion forms in the geometry~\cite{CominsSchutz,Cardoso:2007az,Cardoso:2014sna}.

%%%%%%%%%%%%%%%%%%%%%%%%%%%%%%%%%%%%%%%%%%%%%%%%%%%%%%%%

%\bibliography{ref}  
\bibliographystyle{myutphys}
\bibliography{superradiance_ref}
\end{document}